%% file: main2.tex
\documentclass[10pt,twoside,hidelinks,openright]{book}
\usepackage[utf8]{inputenc}
\usepackage[T1]{fontenc}
\usepackage{amsmath}
\usepackage{amsfonts}
\usepackage{amssymb}
\usepackage{graphicx}
\usepackage[english]{babel}
\usepackage{csquotes}
\usepackage[a4paper]{geometry}
\usepackage{caption}
\usepackage{subcaption}
\usepackage{fancyhdr}
\usepackage{nameref}
\usepackage{hyperref}
\usepackage{multirow}
\usepackage{appendix}
\usepackage{setspace}
\usepackage{titlesec}

\spacing{1.25}

\titlespacing{\chapter}{0pt}{0pt}{0pt}
\titlespacing{\section}{0pt}{0pt}{0pt}
\titlespacing{\subsection}{0pt}{0pt}{0pt}
\titlespacing{\subsubsection}{0pt}{0pt}{0pt}

\usepackage[doi=true,isbn=false,url=true,giveninits=true,sorting=none,style=nature,backend=biber]{biblatex}
\begin{document}

\author{Memoria presentada por \\\textbf{David Cintas González}\\para optar al grado de \\Doctor en Física\\ \\ \\ \\Grupo de Física Nuclear y Astropartículas\\Área de Física Atómica, Molecular y Nuclear\\Departamento de Física Teórica\\Centro de Astropartículas y Física de Altas Energías (CAPA)\\\textbf{UNIVERSIDAD DE ZARAGOZA}\\ \\}
\title{\textbf{New strategies to improve the sensitivity of the ANAIS-112 experiment at the Canfranc Underground Laboratory}}

\frontmatter
\maketitle
\thispagestyle{empty}
\newpage
\input{curiosidad.tex}

\newpage
\thispagestyle{empty}
\setlength{\parskip}{0mm}
\fancyhead[LE]{\emph{Contents}}
\fancyhead[RO]{\emph{Contents}}
\tableofcontents
\setlength{\parskip}{5mm}
\newpage
\thispagestyle{empty}

\mainmatter

\input{Introduction.tex}
\input{ANAIS-112-New2.tex}
\newpage
\thispagestyle{empty}
\input{ASim2.tex}
\input{QF2.tex}
\newpage
\thispagestyle{empty}
\input{SiPMIntro.tex}
\newpage
\thispagestyle{empty}
\input{SiPM_frio2.tex}
\input{SiPMsZgzNew2.tex}

\newpage
\thispagestyle{empty}
\fancyhead[LE]{\emph{Summary and conclusions}}
\fancyhead[RO]{\emph{Summary and conclusions}}
\input{Conclusions.tex}

\selectlanguage{spanish}
\fancyhead[LE]{\emph{Resumen y conclusiones}}
\fancyhead[RO]{\emph{Resumen y conclusiones}}
\input{Conclusiones.tex}

\appendix
\selectlanguage{english}
\newpage
\thispagestyle{empty}
\input{ANEXO1.tex}
\newpage
\thispagestyle{empty}
\input{ANEXO2.tex}

\fancyhead[LE]{\emph{Agradecimientos}}
\fancyhead[RO]{\emph{Agradecimientos}}
\addcontentsline{toc}{chapter}{Agradecimientos}
\selectlanguage{spanish}
\input{Agradecimientos.tex}

\newpage
\thispagestyle{empty}

\fancyhead[LE]{\emph{Acknowledgements}}
\fancyhead[RO]{\emph{Acknowledgements}}
\addcontentsline{toc}{chapter}{Acknowledgements}
\selectlanguage{english}
\input{Acknowledgements.tex}

\newpage
\thispagestyle{empty}

\backmatter
\fancyhead[LE]{\emph{Bibliography}}
\fancyhead[RO]{\emph{Bibliography}}
\addcontentsline{toc}{chapter}{Bibliography}
\printbibliography

\end{document}

%% file: curiosidad.tex
\chapter*{}
\begin{flushright}
\textit{La curiosidad rompe barreras}
\end{flushright}

%% file: Introduction.tex
\chapter{Introduction}\label{Chapter:Intro}
\fancyhead[LE]{\emph{Chapter \thechapter. \nameref{Chapter:Intro}}}

The most recent measurements indicate that the universe is spatially-flat, with an accelerated expansion, and is composed of 69\% dark energy, 5\% of ordinary matter and 26\% non-baryonic dark matter~\cite{Planck:2018vyg} (Section~\ref{Section:Intro_StdModel}). Although the existence of this non-baryonic dark matter is strongly supported by numerous cosmological and astrophysical observations, its nature is still unknown. Different candidates have been proposed over the years. Concerning particle dark matter candidates, they have to be searched for in theories beyond the Standard Model of Particle Physics (SM)~\cite{Baudis:2016qwx} (Section~\ref{Section:Intro_DM}). Among them, Weakly Interacting Massive Particles (WIMPs) could be detected by experiments specifically designed for their search by production in colliders, indirect detection and direct detection (Section~\ref{Section:Intro_Detection}). Despite the intense effort made by the scientific community in this search, no experiment has reported an unequivocal positive signal from dark matter to date. Within this context, the DAMA/LIBRA experiment, which uses NaI(Tl) crystals and is located at the Gran Sasso National Laboratory (LNGS), in Italy, has been reporting since the 1990s the observation of an annual modulation in its detection rate compatible with that expected for dark matter particles with a high statistical significance~\cite{DAMA:2008jlt,DAMA:2010gpn,Bernabei:2013xsa,Bernabei:2018jrt}. This observation, on the other hand, has not been reproduced by other experiments using different targets and technologies, reason why an experiment using the same target, as ANAIS, is required for a model-independent comparison (Section~\ref{Section:Intro_Scintillators}).

\section{Brief introduction to Cosmology}\label{Section:Intro_StdModel}
\fancyhead[RO]{\emph{\thesection. \nameref{Section:Intro_StdModel}}}

Cosmology is the study of the origins, structure, and evolution of the Universe. It aims to provide a comprehensive description of the largest structures and global dynamics of the Universe, and to answer fundamental questions about its nature.

The human understanding of the Universe has evolved significantly over time. In the past, humans could only observe the Universe with the naked eye. Many ancient cultures believed that the Earth was the center of the Universe, and that the Sun, Moon, planets and stars all revolved around it. The ancient Greeks improved this understanding. Hipparchus made the first known catalog of stars, measuring the location of over 850 stars. He also introduced the concept of magnitude to describe the brightness of stars. Greeks also measured the movements of the planets with high accuracy, developing complex models to explain their behaviour. Based on the observations, two different models were proposed for the universe: the Aristarchus model (with the Sun at the center) and the Ptolemy model (with the Earth at the center). This last one was the most accepted over the centuries.

In the 16th century, the Polish astronomer Copernicus recovered the heliocentric model of the solar system, and tried to explain the movements of the planets observed supposing that they follow circular orbits. At the same time, he introduced the Copernican principle: there are not privileged observers in the Universe and then, the Earth was a planet like the others. This model was later expanded by Galileo, who used a telescope to observe the planets. His observations of the phases and orbit elongations of Mercury and Venus indicated that they orbit the Sun, and the discovery of the moons of Jupiter were a proof that not all the celestial bodies orbit the Earth. Later on, the mathematician Johannes Kepler used the Tycho Brahe measurements of the movements of the planets to propose three laws that described them with very high accuracy. These laws were based on the premise that they follow elliptical orbits around the Sun, instead of circular. However, they described but did not explain the planets motion.

Newton revolutionized the understanding of the Universe when he introduced the laws of motion and gravity. He demonstrated that the force of gravity acting between two objects depends on their masses and the distance between them. He also supposed that the laws that rule the movement of the objects in the Earth are the same for all the bodies in the Universe. The laws of motion and the law of universal gravitation provided a framework for the study of the cosmos.

The development of Einstein's Theory of General Relativity in 1915 was a turning point in the history of the cosmology~\cite{GALE1996279}. This theory introduced the concept of space-time, and showed that the curvature of space is determined by the distribution of matter and energy. This new theory challenged many of the previous beliefs about the nature of the Universe, and paved the way for new ideas about its origins and structure, thus giving rise to modern cosmology. The Einstein's field equations were
\begin{equation}\label{eq:Relativity}
	R_{\mu\nu} - \frac{1}{2} R g_{\mu\nu} = \frac{8\pi G}{c^4} T_{\mu\nu},
\end{equation}
where $g_{\mu\nu}$ is the metric tensor and defines the geometry of the space-time, $R_{\mu\nu}$ is the Ricci tensor, $R$ is the Ricci scalar and $T_{\mu\nu}$ is the energy/momentum tensor. When this theory was applied to the study of the evolution of the universe as a whole~\cite{LEMAÎTRE1931,LEMAÎTRE1950}, it was found that it should expand or contract under the only presence of matter. However, as for that time the universe was believed to be static and unchanging, Einstein introduced in the equations an additional term, being $\Lambda$ the cosmological constant, that acted as a repulsive force that would balance the collapse of the universe:
\begin{equation}\label{eq:Relativity2}
	R_{\mu\nu} - \frac{1}{2} R g_{\mu\nu} + \Lambda g_{\mu\nu} = \frac{8\pi G}{c^4} T_{\mu\nu},
\end{equation}
This history changed in 1929, when Edwin Hubble studied distant galaxies. From the red-shift observed in their light emissions he concluded that they were all moving away from the Earth~\cite{Hubble:1929,Hubble:1931}. This was a revolution: as the universe was expanding, the cosmological constant seemed unnecessary. Einstein called the introduction of this constant in the general relativity as "his greatest blunder". Further observations showed that the velocities of the galaxies relative to us were directly proportional to their distances, following:
\begin{equation}
	v = H_0 \cdot r,
\end{equation}
where $H_0$ is the Hubble parameter, $v$ is the galaxy velocity and $r$ is the distance to the observer. This is known today as the Hubble-Lema\^itre's law.

Viewing this expansion from the light of the cosmological principle lead to understand that every galaxy is moving away from everything else and that the space-time itself is expanding. Under this consideration, moving back in time, the Universe should have been smaller and hotter. This hypothesis was later known as the Big Bang Theory.

This theory states that the very early Universe would have consisted of a hot soup of elementary particles (such as quarks, leptons, photons, etc) in thermal equilibrium. However, while the Universe expanded and cooled down, the different types of particles would decouple from the rest of the universe contents at a time dependent on its mass and coupling strength. In particular, the "freeze out" would occur when the Universe reaches a temperature low enough to make the production rate for a type of particle fall down. Then, those particles could annihilate while their density is high enough before the expansion of the Universe dilutes them and the particle number freezes. This thermal history is very successful to explain the evolution of the Universe. Quarks formed baryons, and baryons formed light nuclei in a well understood process known as primordial nucleosynthesis~\cite{Gamow:1946eb,Alpher:1948ve,Alpher:1948gsu,Alpher:1949sef}.

The Universe was a plasma consisting of electrons, light nuclei and photons in equilibrium until the temperature fell down below 3000~K. Then, approximately 380000~years after the Big Bang, the electrons recombined with the light nuclei, forming the first atoms, and the Universe became neutral, i.e. transparent to photons. Photons decoupled from the rest of the Universe contents and have been cooling down and red-shifting while the Universe expanded. Nowadays these photons are filling the Universe with an average temperature of 2.7~K and we call them Cosmic Microwave Background (CMB) radiation. This radiation was first observed in 1964 by Arno Penzias and Robert Wilson, who were awarded the Nobel Prize in Physics for their discovery~\cite{Penzias:1965apj}.

\begin{figure}[h!]
	\begin{center}
		\includegraphics[width=\textwidth]{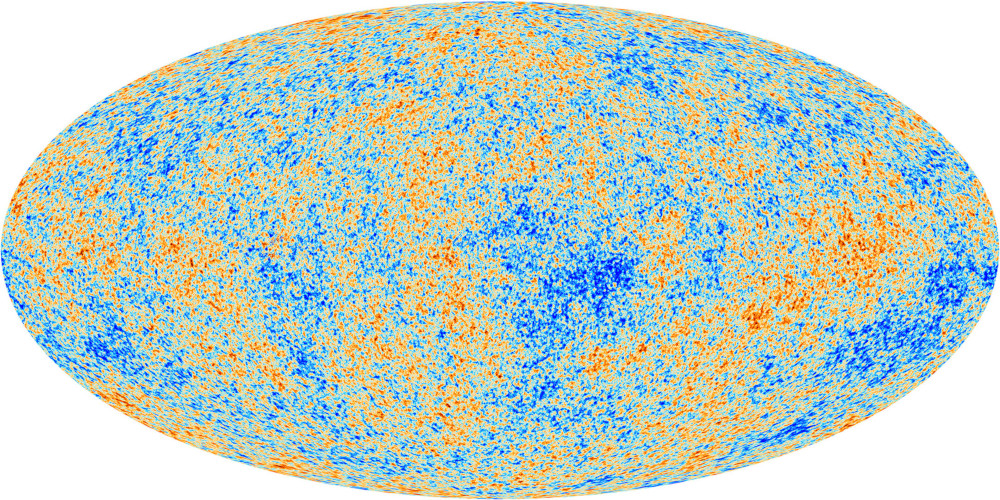}
		\caption{\label{PlanckCMB}All-sky image based on over 1500~days of data from the Planck satellite. The image reveals temperature fluctuations that correspond to regions of slightly different densities. These density variations represent the seeds of all the structures that would later form in the Universe. The signal from our own galaxy was removed from the data using the multifrequency data. Image from~\cite{PlanckCMB}.}
	\end{center}
\end{figure}

From the 60's, the CMB has been mapped with higher and higher precision, as it gives very valuable information of the early times of the universe. The most recent and accurate measurement of CMB fluctuations was produced by the Planck satellite~\cite{Planck:2018vyg,Planck:2019nip}, which provided a high-resolution map of the temperature distribution of the Universe at the recombination time (shown in Figure~\ref{PlanckCMB}), with the warmer shown in red, and the colder shown in blue. This was an improvement of the previous measurements of the WMAP satellite~\cite{WMAP:2012fli,WMAP:2012nax}.

The first characteristic observed of the CMB is its huge homogeneity, with anisotropies of the order of 1~over~10$^5$, meaning that regions of the space that were not causally connected at the recombination time had the same temperature. To solve this issue, the cosmic inflation theory was proposed by Alan Guth in 1981~\cite{Guth:1980zm}. This theory proposes that the universe presented a period of exponential expansion lasting for 10$^{-36}$~s in the first 10$^{-32}$~s after the Big Bang, thus producing thermally connected regions much larger than those allowed by the light cone. Quantum fluctuations of the density of the primordial fluid at the microscopic scale would have grown up very fast into cosmic scale inhomogeneities~\cite{Linde:1981mu}. These inhomogeneities are the anisotropies observed in the CMB. They would have acted as seeds for the formation of structures under gravitational collapse, leading to the formation of galaxies and galaxy clusters~\cite{Starobinsky:1982ee,Mukhanov:1990me}.

However, the amount of baryonic matter in the universe is not able to reproduce the structures we observe in the present universe starting from the CMB anisotropies. But it is required another kind of matter that does not interact with the electromagnetic radiation. A matter that, for historical reasons, has been known as dark matter.

The first evidences of dark matter were found in the beginning of the 20th century. In 1924, Jan Hendrik Oort analyzed the velocity distribution of the stars near the Sun and found that the gravitational potential produced by the known stars was not sufficient to retain the stars in the galactic disk, as most of them had velocities higher than the escape velocity. This could be explained if large amount of mass in the galaxy was not found in luminous stars, but gas or other "dark" astronomical objects as Oort suggested~\cite{Oort}.

The next solid evidence for the existence of dark matter came from Fritz Zwicky's study of the Coma cluster in 1933~\cite{Zwicky:1933gu}. Using the virial theorem, Zwicky obtained that the averaged mass of the galaxies was more than two orders of magnitude larger than that obtained with the luminosity. Since then, other numerous observations have provided further evidence for the existence of dark matter, which, in addition, is different from the conventional (baryonic) matter.

\begin{figure}[h!]
	\begin{center}
		\includegraphics[width=\textwidth]{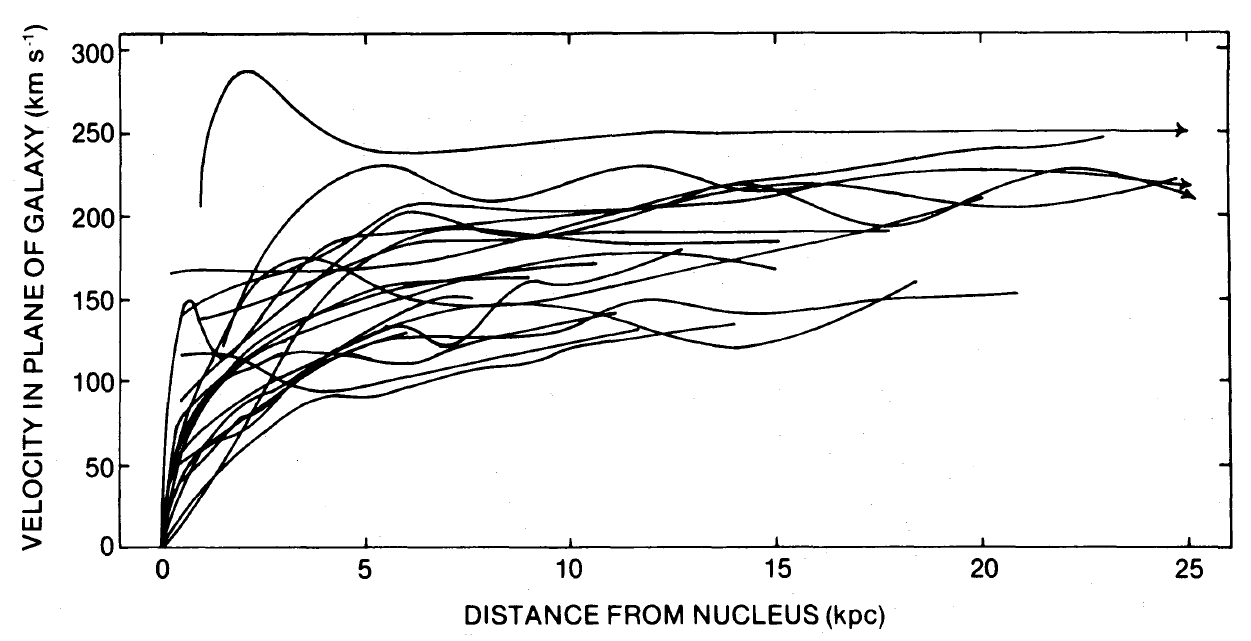}
		\caption{\label{RotationCurves}Galaxy rotation curves measured and published in~\cite{Rubin:1978kmz}.}
	\end{center}
\end{figure}

One of the most compelling pieces of evidence for the existence of dark matter comes from the study of galactic rotation curves, which are the rotational velocities of objects in a galaxy, such as stars and gas clouds, as a function of their distance from the center of the galaxy. According to Newton's law of gravity, the rotational velocities $v(r)$ of these objects should decrease with the distance $r$ from the center of the galaxy as:
\begin{equation}
	v(r) = \sqrt{\frac{G M(r)}{r}},
\end{equation}
in the case the galaxy could be considered an spherical mass distribution, $M(r)$. Although spiral galaxies are not spherical distributions, a decrease of the rotational velocity beyond the visible galactic radius is expected. Observations made by Vera Rubin, Kent Ford and Albert Bosma in the 70's~\cite{Rubin:1978kmz} showed that the rotational velocities of objects in galaxies remained roughly constant at large distances from the galactic center, at more than 10 times the radius of the visible galactic disk, $R_{opt}$, (as it is shown in Figure~\ref{RotationCurves}). The most widely accepted explanation for this observed behaviour is the existence of a halo made of dark matter that extends much farther than the visible galaxy and has a mass distribution $M(r) \propto r$.

\begin{figure}[h!]
	\begin{center}
		\includegraphics[width=0.75\textwidth]{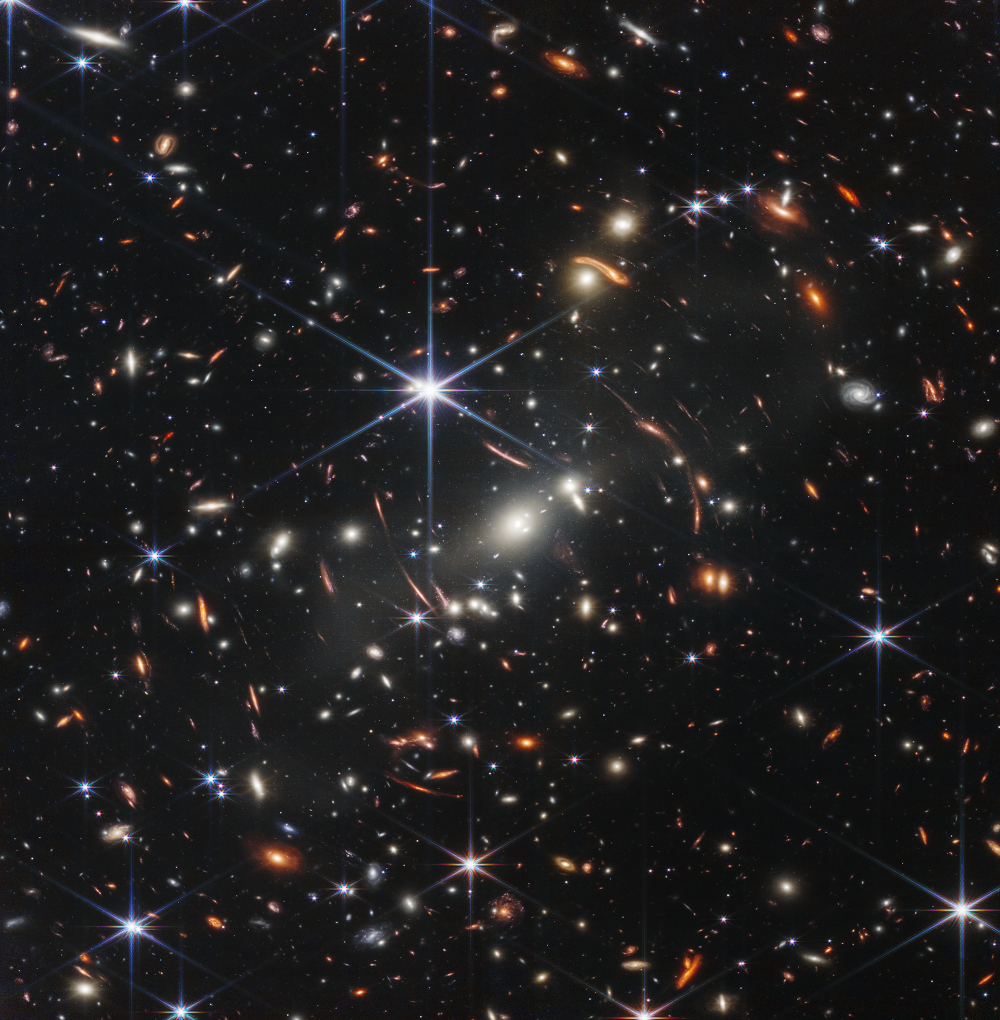}
		\caption{\label{JamesWebb}Near-infrared image from NASA’s James Webb Space Telescope of the galaxy cluster SMACS-0723. Images formed by the gravitational lensing effect are observed, for instance, the orangish arc structures found around the brightest galaxy, near the center of the image. Image from~\cite{JamesWebbTelescope}.}
	\end{center}
\end{figure}

The Theory of General Relativity predicts that the mass curves the space-time, which lenses the light. This lensing effect can be measured to estimate the amount of mass in a system, as for example a cluster of galaxies~\cite{Tyson2009}. This effect can be observed in Figure~\ref{JamesWebb}, an impressive image taken by the new space telescope James Webb~\cite{JamesWebbTelescope}. The clusters' mass estimates resulting from these lensing measurements are by far larger than the visible mass they contain.

\begin{figure}[h!]
	\begin{center}
		\includegraphics[width=0.75\textwidth]{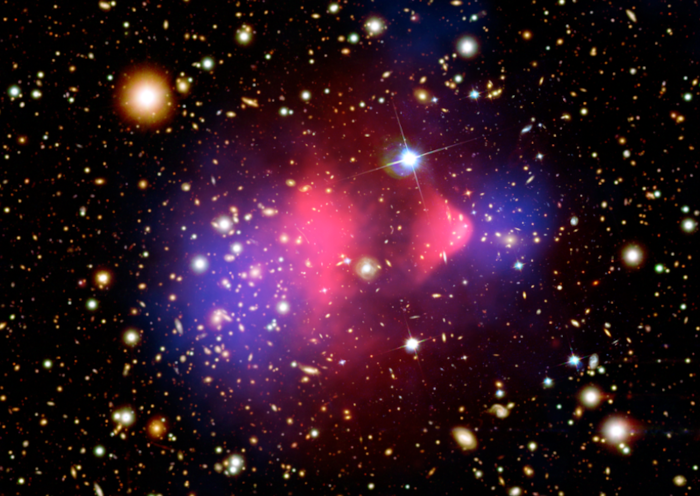}
		\caption{\label{BulletCluster}This composite image shows the Bullet cluster, also known as 1E 0657-556. The image combines data from the Chandra x-ray telescope, which detects hot gas in the cluster (shown in pink), with optical data from the Magellan and Hubble Space telescopes, which show the galaxies in the cluster. The blue region of the image shows the mass distribution in the cluster, as deduced from observations using gravitational lensing. This image illustrates the separation of dark matter and hot gas in the cluster. Image from~\cite{Bullet}.}
	\end{center}
\end{figure}

The study of clusters of galaxies has also provided a very important result by combining information from optical emissions, x-rays emitted from the hot gas, and the reconstruction of the cluster mass distribution by analysing the gravitational lensing. In the case of the Bullet cluster, shown in Figure~\ref{BulletCluster}~\cite{Bullet}, the distributions of hot gas and matter are very different, and can only be understood if this cluster is the result of a collision between two galaxy clusters. When this collision happened, the gas of each cluster slowed down due to the electromagnetic interaction. On the other hand, the matter in form of galaxies in the clusters had a very low probability to collide, and both clusters passed through each other. However, if the dark matter forming large halos around the galaxies would have consisted of particles with a large interaction probability, then like the gas, the dark matter would have slowed down in the collision, and the distribution of matter in the Bullet cluster would have been very different from the observed one. The Bullet cluster can only be understood if the dark matter consists of a very weakly interacting particle~\cite{Clowe:2003tk}. This evidence for dark matter brings into discussion different arguments with respect to previous evidences, as it is not relying on gravitational effects.

It is possible to extract some properties that the dark matter must fulfill to explain all these observations~\cite{Bergstrom:2000pn}. First, this matter must not emit, absorb or reflect any electromagnetic radiation. To enable the formation of galaxies and large-scale structures, it must be "cold", meaning that the velocity of the particles should have been non-relativistic at the time of the structure formation. Its self-interaction and the interaction with baryonic matter must be very weak, as it does not dissipate energy through collisions~\cite{Bertone:2010zza}. Although nowadays the nature of dark matter remains unknown, several particle candidates have been proposed, some of them reviewed in Section~\ref{Section:Intro_DM}. Moreover, diverse experimental techniques have been developed to identify its nature, as it is overviewed in Section~\ref{Section:Intro_Detection}.

During the 90's some observatories were focused on the detection of Ia supernovae at very long distances. These supernovae have a luminosity curve very well characterized, which allows to use them as standard candles. By measuring the effective magnitude as a function of the red-shift of a large number of supernovae, the Supernova Cosmological Project (SCP)~\cite{SupernovaCosmologyProject:1998vns,SupernovaCosmologyProject:2003dcn,SupernovaCosmologyProject:2008ojh} and High-z Supernova Search Team (HZT)~\cite{SupernovaSearchTeam:1998fmf,SupernovaSearchTeam:2003cyd,SupernovaSearchTeam:2004lze} confirmed, in 1998, that the universe is currently in an accelerated expansion. This acceleration can be accounted for by introducing in the Einstein field equations the cosmological constant term (Equation~\ref{eq:Relativity2}). This term can be seen as a contribution to the universe energy density, behaving in a different way than matter does, and it is referred to as dark energy. The nature of this energy is still unknown. It could be related with the vacuum energy associated to quantum fluctuations or to the evolution of some scalar field, known as quintessence~\cite{Caldwell:2000wt,Wang:1999fa}. However, successful dark energy models are still to be found.

The most widely accepted model for the Universe is the Lambda-Cold-Dark-Matter ($\Lambda$CDM) model, known as the Standard Cosmological Model~\cite{Carroll:2001,Peebles:2003} and it is able to explain all the previously referred observations of the Universe in the different spatial scales and evolution times. According to this model, the Universe is made up of four fundamental components: dark energy ($\Lambda$)~\cite{Copeland:2006wr}, non-baryonic Cold Dark Matter (CDM)~\cite{Arbey:2021}, ordinary matter (baryons and leptons) and radiation (photons). The Cosmological principle implies that the Universe expansion/contraction is described by a single  parameter: the scale factor $a(t)$. The metric used in the Einstein field equations to describe the dynamics of the Universe is known as the Friedmann-Lema\^itre-Robertson-Walker (FLRW) metric, and includes a curvature parameter $k$, determined by the matter/energy sources. The line element is:
\begin{equation}
	ds^2 = c^2 dt^2 - a^2(t) \left(\frac{dr^2}{1-kr^2} +r^2 d\theta + r^2 \sin^2\theta d\phi^2 \right), 
\end{equation}
The possible values of the parameter $k$ are +1 (closed universe), 0 (flat universe) or -1 (open universe). Considering the universe as an homogeneous fluid with a pressure $P$ and density $\rho$, the corresponding energy-momentum tensor is 
\begin{equation}
	T_{\mu\nu} = \left(\rho+\frac{P}{c^2}\right)u_\mu u_\nu + Pg_{\mu\nu},
\end{equation}
where $u_\mu$ is the four-velocity of the fluid element. It can be introduced in the Einstein's field equations (Equation~\ref{eq:Relativity2}), to obtain the two Friedmann equations:
\begin{equation}
	H^2(t) \equiv \left(\frac{\dot{a}}{a}\right)^2 = \frac{8\pi G}{3} \rho_{tot} - \frac{kc^2}{a^2},
\end{equation}
and
\begin{equation}
	\frac{\ddot{a}}{a} = -\frac{4\pi G}{3c^2} \left(\rho_{tot} c^2 +3P\right),
\end{equation}
where $H(t)$ is the Hubble's parameter, which measures the expansion rate of the Universe at each time of its evolution ($H_0$ at present) and $\rho_{tot}$ is the total mass/energy density, which can be expressed as sum of the different contributions: matter $\rho_{m}$, radiation $\rho_r$ and dark energy $\rho_{\Lambda} = \frac{\Lambda}{8\pi G}$ densities. Then, the temporal evolution of each energy density $i$ is
\begin{equation}
	\frac{d{\rho_i}}{dt} = -3\left(\rho_i + \frac{P_i}{c^2}\right) \frac{\dot{a}}{a}.
\end{equation}
The solution for this density assuming perfect fluid equations of state
\begin{equation}
	P_i = \omega_i \rho_i c^2,
\end{equation}
is different for each component:
\begin{equation}
	\rho_i(t) \propto a(t)^{-3(\omega_i+1)}.
\end{equation}
For matter $\omega_m = 0$, for radiation $\omega_r = 1/3$, which results in the corresponding red-shift of the energy density. For dark energy $\omega_{\Lambda} < -1/3$ ($=-1$ for the cosmological constant, with a possible a time dependence $\omega_{\Lambda}(t)$ in other dark energy models). The critical density can be defined as the density corresponding to $k$ = 0 (flat universe), being:
\begin{equation}
	\rho_c = \frac{3H_o^2}{8\pi G}
\end{equation}
Therefore, if $\rho_{tot} > \rho_c$ then the universe is closed, if $\rho_{tot} = \rho_c$ the universe is flat and if $\rho_{tot} < \rho_c$ the universe is open. Then, a dimensionless density parameter, $\Omega_i$, can be defined for every component $i$ as the energy density relative to the critical density
\begin{equation}
	\Omega_i = \frac{\rho_i}{\rho_c},
\end{equation}
which in case of a flat Universe, indicates the energy contribution of each component.

\begin{figure}[h!]
	\begin{center}
		\includegraphics[width=0.75\textwidth]{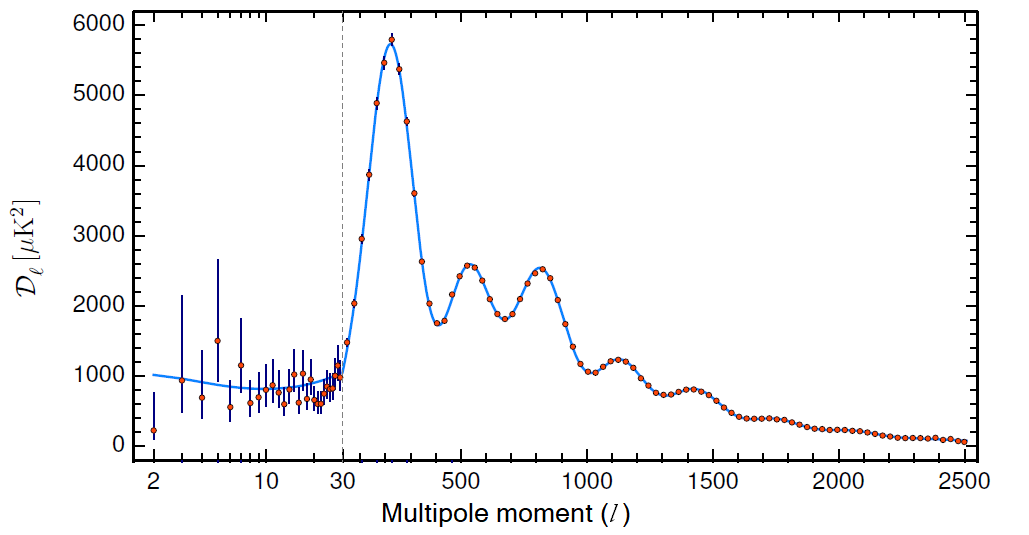}
		\caption{\label{PowerSpectrum}Comparison of the results of the power spectrum corresponding to PLANCK measurement of the CMB fluctuactions (orange dots) and the predictions of the $\Lambda$CDM cosmological model, shown as a blue line. Image from~\cite{Planck:2018vyg}.}
	\end{center}
\end{figure}

High-precision measurements of the CMB anisotropies and the large-scale distribution of galaxies have been used to test the predictions of the Standard Cosmological Model. The angular distribution of the CMB temperature fluctuations can be expressed in terms of the spherical harmonics:
\begin{equation}
	\frac{\Delta T}{T}(\theta,\phi) = \sum_{lm}{a_{lm}Y_{lm}(\theta,\phi)},
\end{equation}
where $l$ and $m$ correspond to a specific angular scale (multipole moment). One of the key quantities obtained from the analysis of the temperature map is the power spectrum $D_l$, obtained by calculating the square of the amplitude of the temperature fluctuations for each $l$:
\begin{equation}
	D_l = \frac{l(l+1)}{2\pi(2l+1)} \sum_{m} |a_{lm}|^2
\end{equation}
The temperature fluctuations in the CMB depend on the density profile of the early Universe. The height of the first peak of the power spectrum is related to the density of all the matter (dark and baryonic), while the second one is related only to the density of baryonic matter. Overall, a modeling of the power spectrum is done parameterizing the Standard Cosmological Model with six independent parameters. These are: the baryon density and dark matter density parameters, the angular size of the sound horizon at recombination, the primordial curvature fluctuation amplitude, the scalar spectral index, and the reionization optical depth. The data collected by Planck is then fitted to that model as shown in Figure~\ref{PowerSpectrum}~\cite{Planck:2018vyg}. The figure presents the power spectrum as a function of angular scale $l$. It can be seen that there is good agreement between the experimental data and the predictions of the $\Lambda$CDM model even at high angular scales.

According to these estimates, the age of the universe is 13.79~$\pm$~0.02~Gyr, and the Hubble constant nowadays, $H_0$, is 67.66~$\pm$~0.42~km/s/Mpc. The present contribution from radiation to the total mass-energy density of the universe is considered to be negligible. The total mass-energy budget of the universe is made up of approximately 69\% dark energy and 31\% of matter, from which 5\% is baryonic~\cite{Planck:2018vyg}. This value is in agreement with the estimated from the comparison of the measured abundances of the light nuclei and the predictions of the Big Bang nucleosynthesis. It also implies that the 26\% of the mass-energy content of the universe is made of non-baryonic dark matter, a contribution 5~times larger than the baryon content.

\section{Dark matter candidates}\label{Section:Intro_DM}
\fancyhead[RO]{\emph{\thesection. \nameref{Section:Intro_DM}}}

As commented, candidates for the dark matter, able to explain all the observations and data exposed in Section~\ref{Section:Intro_StdModel}, have to be searched for beyond the standard model of particle physics. However, the first candidates proposed consisted of normal baryonic matter in form of Massive Compact Halo Objects (MACHOs) and massive neutrinos. MACHOs are large and non-luminous objects, such as brown dwarfs or black holes. They contribute to the total mass budget of the galaxy, but not to the non-baryonic dark matter. Searches for MACHOs in our galaxy at the end of the 20th century allowed to quantify their contribution to the galactic mass, being much lower than that required to explain the observed dynamics. On the other hand, recently, a different type of dark matter candidate has gained attention: the primordial black holes~\cite{Carr:2016drx}. They would be originated in the gravitational collapse of large density fluctuations before the Big Bang Baryogenesis, and therefore they would not be baryonic matter effectively. As they evaporate emitting Hawking radiation, their lifetimes depend on their masses, in such a way that for masses larger than 5$\times$10$^{11}$~kg, they would have lifetimes larger than the current age of the Universe. There are several constraints on their existence~\cite{Carr:2020gox}, but they still remain a potential candidate for dark matter.

Neutrinos seemed to be ideal candidates for the dark matter, because they are known to exist within the standard model of the particle physics, and they interact very weakly with the other particles. However, they were soon discarded as the main component of the dark matter because of different reasons. Neutrino oscillation experiments confirm that  neutrinos are massive, but the absolute scale mass is not yet stablished. The most stringent upper limit of the electron neutrino mass is 0.8~eV (90\% C.L.), obtained by KATRIN collaboration~\cite{Onillon:2023aqz} by determining with accuracy the end-point of the $^3H$ beta decay, while current comological constraints for the sum of the masses of the three neutrino eigenstates is $\lesssim$~0.1~eV~\cite{Workman:2022ynf}. Their low mass is not sufficient to explain the dark matter density of the universe, but more important, they cannot explain the structures formation in the Universe because they are hot dark matter. However, the existence of a fourth neutrino type, called "sterile" because it does not couple to W/Z bosons, is a common ingredient in extensions of the SM. These neutrinos could be cold dark matter or warm dark matter if their mass is in the range from a few~keV to several~MeV. Depending on the model, the properties of these sterile neutrinos make them interesting and viable dark matter candidates if their lifetime is long enough (comparable to the age of the universe) and if their relic density is high enough. 

Axions are hypothetical elementary particles beyond the SM that fulfill all the requirements to be viable cold dark matter candidates. They were proposed to solve the strong CP (charge-parity) problem in quantum chromodynamics (QCD): the conservation of the CP symmetry by the strong interactions cannot be explained within the SM. The Peccei-Quinn mechanism, introduced by Roberto Peccei and Helen Quinn in 1977, offers a solution to this problem by introducing a new symmetry which is spontaneoulsy broken and thus, predicting the existence of the axion as the Goldstone boson associated with this symmetry breaking~\cite{Peccei:1977ur,Kim:2008hd}. Despite being ultra-light, axions are candidates for cold dark matter because they would have been produced  in large quantities through non-thermal mechanisms in the early universe. Other pseudo-scalar particles (generally called Axion-Like Particles, or ALPs) appear in many extensions of the SM and are also good candidates for cold dark matter. There are currently several experiments that search for axions and ALPs mainly through the Primakoff effect~\cite{Sikivie:1983ip}, the conversion of axions into photons in the presence of intense electromagnetic fields. Some of these experiments are looking for axions in the galactic halo (the so-called, haloscopes), as for example the Axion Dark Matter Experiment (ADMX)~\cite{ADMX:2018gho,ADMX:2019uok} and the Center for Axion and Precision Physics Researches (CAPP) Axion Search Experiments~\cite{CAPP:2020utb}. Other experiments look for axions coming from the Sun (known as helioscopes), as the CERN Axion Solar Telescope (CAST)~\cite{CAST:2004gzq,CAST:2017uph}, and the future International Axion Observatory (IAXO)~\cite{IAXO:2012eqj,IAXO:2019mpb}. ALPS can be also searched for by "light shining through a wall" experiments~\cite{Redondo:2010dp}, as done by the ALPS (Any Light Particle Search) collaboration~\cite{Isleif:2022ytq} at DESY (German Electron Synchrotron).

Weakly Interacting Massive Particles (WIMPs) are hypothetical particles that have been considered for decades the most promising candidates for dark matter~\cite{Steigman:1984ac}. In the early universe, WIMPs would have been in thermal equilibrium until their freeze-out. The remaining final relic density of WIMPs results from the mass of the particle (fixing the decoupling time) and the annihilation cross-section. For a particle in the~GeV mass range and weak scale interaction cross-sections, the resulting relic abundance of WIMPs is of the order of the required to explain the missing dark matter. This coincidence is called the "WIMP miracle", and is seen as an important argument supporting WIMPs as robust candidates for the non-baryonic dark matter~\cite{Acharya:2009zt}. In addition, they are predicted in many theoretical models beyond the SM, like supersymmetric (SUSY) extensions of the SM~\cite{Jungman:1995df}. These models introduce a new symmetry between fermions and bosons: every fermion has a bosonic superpartner and viceversa. Most SUSY models include a conserved quantum number, the R-parity. SUSY particles (s-particles), which have odd R-parity, are produced in pairs and the Lightest SUSY Particle (LSP) is stable. In many models, the LSP is the so-called neutralino (a combination of the superpartners of the SM neutral bosons: b-ino, neutral w-ino and neutral higgsinos), which is expected to have a mass in the range from~GeV to hundreds of~TeV.

WIMPs are a wide category of candidates that can stem from different models but share two main properties, they are massive in the range from~GeV to~TeV and have weak interaction cross-sections. Many detection strategies are possible for them. They are described in next section.

\section{Dark matter detection}\label{Section:Intro_Detection}
\fancyhead[RO]{\emph{\thesection. \nameref{Section:Intro_Detection}}}

The strategies for detecting the dark matter are strongly dependent on the candidate properties, being very different for primordial blackholes, for instance, than for particle candidates. The methods to detect axions and ALPs have been overviewed previously, and those for WIMPs are reviewed in this section. They include searches at colliders, indirect detection looking for the products of dark matter particle annihilation, and direct searches of dark matter particles scattering off a detector. All of these detection strategies face similar challenges:
\begin{itemize}
	\item Background events, in particular those that can mimic the dark matter signals, have to be well understood and reduced by all means to improve the sensitivity of the searches. These backgrounds are specific for every detection technique.
	\item The expected dark matter signal strongly depends on the particle dark matter model and on the distribution of those particles at galactic or cluster scales. The experimental constraints derived from any search are then model dependent.
	\item Dark matter candidates can have a wide range of masses and interaction cross-sections depending on the model, which makes it difficult to design experiments that cover all the parameter space.
\end{itemize}
This section reviews the different strategies followed in dark matter searches and summarizes some of the most relevant results up to date.

\subsection{Searches at colliders} \label{Section:Intro_Detection_Colliders}

Colliders produce high-energy particle collisions that can potentially create dark matter particles as WIMPs if they have masses within the energy reach of the collider and relies upon the existence of interactions between the SM particles and the dark matter particles. Collider experiments can search for dark matter particles, but they can also probe the interaction between the SM and dark matter particles by searching for the particles that act as mediators. However, in colliders the stability of the dark matter particles can only be established up to the timescales needed to traverse the detection system~\cite{Goodman:2010ku}.

In order to search for dark matter particles in colliders, the particle model considered is crucial. There are two different approaches: self-consistent models such as SUSY which provide specific features allowing narrowly targeted searches, while simplified or effective models enable more general but less optimal searches. 

Generic dark matter candidates can be searched for in colliders by different methods. Pair production of dark matter particles in the collisions could occur in association with one or more additional SM particles. These mono-X events are selected if they contain a high-momentum object (i.e., a jet, a photon, a vector boson, etc.) in combination with significant missing transverse energy and allow to set constraints on generic dark matter models~\cite{Boveia:2018yeb}. Colliders can also play a crucial role searching for models with new particles mediating the interaction between dark matter and SM particles, by identifying the visible decays of the mediator particles, for instance into pairs of quarks or leptons~\cite{Buchmueller:2014yoa}. This would produce a localized excess of events in the invariant mass spectrum or in specific angular distributions, allowing the presence of dark matter mediators to be inferred.

The Large Hadron Collider (LHC)~\cite{Evans:2008zzb} began operation at CERN in 2008. Two of the LHC experiments (ATLAS~\cite{ATLAS:2008xda} and CMS~\cite{CMS:2008xjf}) have been searching for DM in the proton-proton collisions at high energies (7~TeV during 2010-2012 and 13~TeV during 2015-2018). Although they have not found any significant deviations from the predictions of the Standard Model, their results can be interpreted as constraints on the coupling of dark matter mediators or as limits on the masses of the mediator and dark matter particles~\cite{Trevisani:2018psx}.

Lepton colliders would have many advantages for detecting dark matter particles. They benefit from a lower background and clean environment because the primary collisions are not mediated by the strong nuclear force. In electron-positron colliders, single photon events with missing momentum are a very interesting signal to be searched for as indication of the production of a pair of dark matter particles as $e^- e^+ \rightarrow \gamma \chi \chi$, being $\chi$ the dark matter particle~\cite{Kalinowski:2022unu,Bauer:2017qwy}. Future lepton colliders that can contribute to dark matter searches are the International Linear Collider (ILC)~\cite{Asai:2023dzs}, the Compact Linear Collider (CLIC)~\cite{Roloff:2018dqu}, the Circular Electron-Positron Collider (CEPC)~\cite{CEPCStudyGroup:2018ghi} and Future Circular Colliders (FCC-ee)~\cite{Keintzel:2022ety}.

\subsection{Indirect detection} \label{Section:Intro_Detection_Indirect}

In the indirect dark matter detection approach the signals searched for are the products of the annihilation or decay of the dark matter particles, not the dark matter particles themselves, hence the name~\cite{Gaskins:2016cha}. Although the annihilation is strongly suppressed after the freeze-out, it can occur in regions of high dark matter density, as the galactic center or the galactic halos and subhahlos. Dwarf galaxies orbiting the Milky Way are dark matter dominated regions and they are good targets for indirect detection because their low baryonic content implies very low astrophysical backgrounds. Another interesting target for indirect dark matter searches is the Sun. The galactic halo dark matter particles scattering with the Sun nuclei can lost enough energy to get trapped in the Sun and they can accumulate there~\cite{Nussinov:2009ft}. As the capture and annihilation rates of dark matter particles in the Sun are expected to be in equilibrium (due to the high mass and long-life of the star), the flux of particles resulting from the dark matter annihilation should be constant in time. Being the Sun close to the Earth it is easy to trace back any excess in particle fluxes that could be detected. The same process is expected to happen in the Earth, but at a lower rate.

It is worth to note that the annihilation channels (and therefore the particles of the Standard Model produced) are strongly dependent on the particle model under consideration. The observation of an excess in the fluxes of gamma rays, protons, electrons, positrons, antiprotons, neutrinos, etc, with respect to that expected by the known sources of background can point to the the presence of dark matter annihilation. Indirect dark matter searches can be classified according to the detected particles: gamma rays, neutrinos and charged particles.

\subsubsection{Gamma rays} \label{Section:Intro_Detection_Indirect_Gamma}

Gamma rays can be a clear signature of dark matter annihilation, as they can travel from the source to the detector without being deflected by magnetic fields, which makes easier to identify regions where dark matter annihilates and the gamma ray excess originates~\cite{Bergstrom:1997fj}. Moreover, the gamma rays produced in the dark matter annihilation could have specific energy distributions, such as a peak at a particular energy, which can be used to differentiate dark matter signals from astrophysical backgrounds. These energy distributions depend on the dark matter couplings and mass, i.e. the particle dark matter model. For example, if they annihilate in a quark-antiquark pair, this would produce a particle jet similar to those observed in accelerators, from which a well known gamma spectrum would be released. If they annihilate into two photons, the energy of these photons would be half the mass of the particle and a peak would be observed. This is a very clear signature, not affected by astrophysical backgrounds for WIMP masses above a few GeV.

Depending on the energy of the gamma rays, they can be detected both directly, using satellites, and indirectly, using ground-based Cherenkov telescopes. Fermi Large Area Telescope (Fermi-LAT)~\cite{Fermi-LAT:2009ihh} is a gamma-ray observatory that has been operating since 2008. It is sensitive to energies between 20~MeV and 300~GeV. This experiment detected an excess of gamma rays in the energy range from~2 to 5~GeV coming from the Galactic Center region~\cite{Goodenough:2009gk}. Apart from being a possible dark matter annihilation signal, this gamma-ray excess could be produced by a large number of sources (e.g. pulsars~\cite{Gordon:2013vta}) with such a small angular distance to the Galactic Center that they could not be resolved by Fermi-LAT. Future observatories can provide complementary information to help to understand the nature of this gamma ray excess.

The High Energy Stereoscopic System (HESS)~\cite{HESS:2011zpk}, the Major Atmospheric Gamma-Ray Imaging Cherenkov (MAGIC)~\cite{MAGIC:2009tyk}, and Very Energetic Radiation Imaging Telescope Array System (VERITAS)~\cite{VERITAS:2017tif} are Cherenkov telescopes sensitive to gamma rays with energies above 100~GeV. The Cherenkov Telescope Array (CTA)~\cite{CTAConsortium:2012fwj} is the upcoming next-generation Cherenkov observatory. The largely improved sensitivity of CTA will allow to severely constraint many dark matter models, reaching the thermal annihilation cross-sections.

\subsubsection{Neutrinos} \label{Section:Intro_Detection_Indirect_Neutrinos}

Searching for neutrino signals from the dark matter annihilation is interesting because, as gamma-rays, these particles are neutral and they are not disturbed by magnetic fields but, contrary to gamma-rays, their interaction with the matter is very weak and then, they are not affected by the interstellar medium. The main drawback of this channel is that it requires large detectors and time exposures. Moreover, the angular resolution of neutrino detectors is limited (typically to around 1$^o$ for energies of 100~GeV, while it is about 0.1$^o$ for gamma-rays), which reduces the capability of identifying the source of the signal.

The experiments searching for neutrinos use large amounts of ice or water as detection media, detecting the Cherenkov light emitted by the charged particles produced by the interaction of the neutrinos. IceCube~\cite{IceCube:2012ugg}, located under the ice of the South Pole, detects the Cherenkov light emitted when neutrinos interact with the ice. The Astronomy with a Neutrino Telescope and Abyss Environmental RESearch (ANTARES)~\cite{ANTARES:2016xuh} experiment is an underwater neutrino telescope located in the Mediterranean Sea. Super-Kamiokande~\cite{Super-Kamiokande:2004pou} is a large water Cherenkov detector located in the Kamioka Observatory in Japan. Although it was designed to study the neutrino oscillations, they have also searched for neutrinos produced in the dark matter annihilation at the Sun and the Earth. No significant excess of neutrinos has been detected, but all these experiments have provided constraints on the properties of dark matter particles. The next generation of neutrino telescopes, like IceCube-Gen2~\cite{IceCube-Gen2:2020qha}, KM3NET~2.0~\cite{KM3Net:2016zxf}, and Hyper-Kamiokande~\cite{Hyper-Kamiokande:2016srs}, will have greater sensitivity to this channel of dark matter annihilation.

\subsubsection{Charged cosmic rays} \label{Section:Intro_Detection_Indirect_Cosmic}

Cosmic-ray charged particles, such as electrons, positrons, protons and antiprotons should present distinctive spectral signatures if they are produced by dark matter annihilation, which can provide important information about the properties of dark matter particles. The detection of antimatter particles, in particular, could be a significant indication of the detection of dark matter annihilation products, because antimatter fluxes from astrophysical sources on Earth are relatively rare~\cite{Ibarra:2008qg}. However, charged cosmic rays can be deflected by magnetic fields and interact with the interstellar medium, and therefore their paths are altered making difficult to trace them back to their sources. Additionally, these charged particles can lose energy as they propagate through inverse Compton scattering or synchrotron radiation, which cause their observed energy spectrum to differ from the original one.

Several experiments have been designed to search for charged cosmic rays, many of them located in the space. The Alpha Magnetic Spectrometer (AMS)~\cite{Palmonari:2011zz} is installed on the International Space Station (ISS). Payload for Antimatter Matter Exploration and Light-nuclei Astrophysics (PAMELA) experiment~\cite{PAMELA:2017bna} was installed in a satellite and operated from 2006 to 2016. Fermi-LAT~\cite{Fermi-LAT:2009ihh} has also been used to study cosmic ray electrons and positrons. All these experiments have observed an unexpected increase of the positron fraction (defined as the ratio of positrons to the total number of electrons and positrons in cosmic rays) at energies from~10 GeV to~1 TeV, reaching a maximum at around 300~GeV~\cite{AMS:2019rhg}. This fraction should decrease because the production mechanisms (like nuclear collisions and pion decays) are less efficient for positron than for electrons at high energies. However, when all uncertainties are taken into account, astrophysical backgrounds are able to explain the observed excess, and better modeling of those backgrounds is required to improve the sensitivity of these searches.

\subsection{Direct detection} \label{Section:Intro_Detection_Direct}

Direct dark matter detection experiments are designed to detect the presence of dark matter particles of the Milky Way dark halo through their interactions with ordinary matter in convenient detection systems. The factors that must be taken into account in the design of these experiments, the detection techniques applied, and a summary of the current state of experimental efforts are described in this section.

\subsubsection{Expected dark matter signal and detector requirements} \label{Section:Intro_Detection_Direct_Signature}

\begin{figure}[h!]
	\begin{center}
		\includegraphics[width=0.75\textwidth]{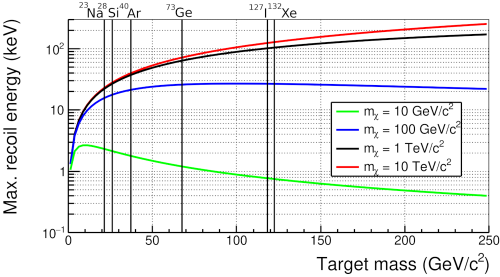}
		\caption{\label{Enr(Mass)}Maximum energy deposited by a WIMP through an elastic scattering for a WIMP velocity of 220~km/s as a function of the target mass for some typical target nuclei used in direct dark matter search experiments and four different WIMP masses.}
	\end{center}
\end{figure}

\begin{figure}[h!]
	\begin{center}
		\includegraphics[width=0.75\textwidth]{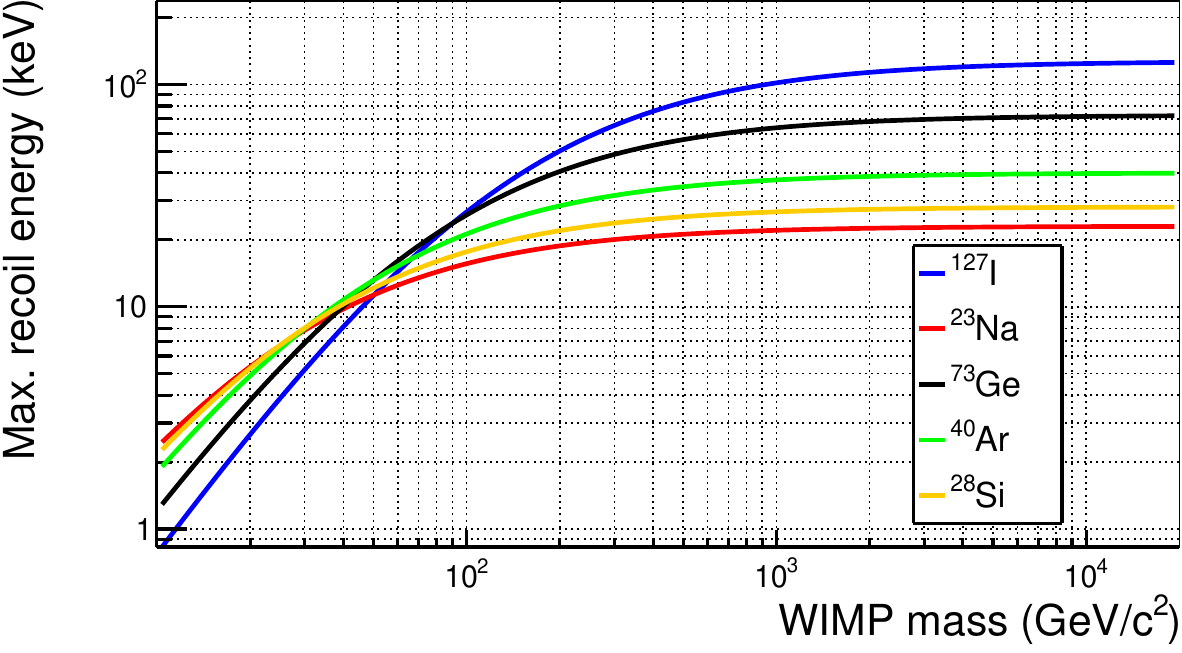}
		\caption{\label{Enr(WIMPMass)}Maximum energy deposited by a WIMP through an elastic scattering for a WIMP velocity of 220~km/s as a function of the WIMP mass for some typical target nuclei used in direct dark matter search experiments.}
	\end{center}
\end{figure}

Most of the experiments aim to detect nuclear recoils produced by a WIMP in convenient detectors, being the most probable interaction channel in most of the WIMP models. In general, all of them look for an elastic scattering between the WIMP and the nucleus, but it is also possible to scatter inelastically, leaving the nucleus in an excited state. This process is not only less probable, but also requires a minimum energy of the particle to excite the nucleus. In the elastic scattering, the nuclear recoil energy $E_{nr}$ depends on the WIMP mass, $m_{\chi}$, the mass of the nucleus, $m_N$, the scattering angle of the nucleus in the laboratory frame, $\theta$, and WIMP velocity in the laboratory frame,  $v$. As WIMPs are not relativistic (as we will see next, they have typical velocities of hundreds of~km/s), the nuclear recoil energy is
\begin{equation}\label{eq:EnrWIMPS}
	E_{nr} = \frac{2\mu_{\chi N}^2}{m_N}v^2 cos^2(\theta),
\end{equation}
where
\begin{equation}\label{eq:muChiN}
	\mu_{\chi N} = \frac{m_{\chi}m_N}{m_{\chi}+m_N}
\end{equation}
is the WIMP-nucleus reduced mass. The maximum value for this nuclear recoil energy is:
\begin{equation}\label{eq:MaxEnrWIMPS}
	E_{nr}^{max} = \frac{2\mu_{\chi N}^2}{m_N}v^2.
\end{equation}
Figure~\ref{Enr(Mass)} shows the maximum energy that a WIMP can deposit through an elastic scattering as a function of the target mass and WIMP masses for a WIMP velocity of 220~km/s. These energies are also shown as a function of the WIMP mass in Figure~\ref{Enr(WIMPMass)}.

These figures show that the maximum recoil energy for low WIMP masses is higher for light nuclei than for massive nuclei, and viceversa. Moreover, it is possible to observe that the energy released by the WIMP is small, below $\sim$100~keV. An important requirement for direct detection experiments is the achievable energy threshold, the minimum energy deposition that can be detected. In order to be able to detect a dark matter particle, there is a minimum velocity required for guaranteeing that the recoil energy is above the energy threshold, $E_{th}$. As it is shown in Figure~\ref{Enr(WIMPMass)} it must be particularly low for low mass WIMPs. The minimum WIMP velocity at which the detectors are sensitive can be written in terms of the experiment energy threshold, $E_{th}$, as:
\begin{equation}\label{eq:MaxEnrWIMPS}
	v_{min} = \sqrt{\frac{E_{th}\cdot m_N}{2\mu_{\chi N}}}
\end{equation}

The mean counting rate expected in a detector due to dark matter interactions can be estimated as:
\begin{equation}\label{eq:R}
	R = N_T \cdot \Phi_{\chi} \cdot \sigma,
\end{equation}
where $N_T$ is the total number of target nuclei, $\Phi_{\chi}$ is the WIMP flux and $\sigma$ is the WIMP-nucleus interaction cross-section. The first one can be obtained as the ratio of the total mass of the detector $M_{det}$ to the mass of the target nuclei $m_N$, and the mean WIMP flux as:
\begin{equation}\label{eq:WIMPFlux}
	\Phi_{\chi} = \frac{\rho_{\chi} \cdot v_{mean}}{m_{\chi}},
\end{equation}
where $\rho_{\chi}$ is the WIMP local density, at the Solar System position, and $v_{mean}$ is the mean velocity of the WIMPs in the laboratory rest frame. These two parameters are strongly dependent on the models of the dark matter halo and the WIMP, and thus also the expected mean counting rate:
\begin{equation}\label{eq:R_2}
	R = \frac{M_{det} \cdot \rho_{\chi} \cdot v_{mean} \cdot \sigma}{m_N \cdot m_{\chi}}.
\end{equation}
If we take into account the WIMP velocity distribution in the halo, conveniently expressed in the Earth reference system, $f(\Vec{v})$, we can write for the differential rate:
\begin{equation}\label{eq:difRate1}
	\frac{dR}{dE_{nr}} = \frac{M_{det} \rho_{\chi}}{m_N m_{\chi}} \int_{v_{min}}^{\infty} \frac{d\sigma(\Vec{v},\Vec{q})}{dE_{nr}} |\Vec{v}| f(\Vec{v}) d^3\Vec{v},
\end{equation}
where $d\sigma/dE_{nr}$ is the differential cross-section. It depends on the WIMP-nucleon velocity, $\Vec{v}$, and on the transferred momentum $\Vec{q}$, and it can be separated into two terms:
\begin{equation}\label{eq:difCS}
	\frac{d\sigma}{dE_{nr}}(\Vec{v},\Vec{q}) = \frac{d\sigma}{dE_{nr}}(\Vec{v},\Vec{q} = 0) \cdot F^2(q),
\end{equation}
where $F(q)$ is the nuclear form factor. In the case of spin-independent interactions $F(q)$ represents the loss of coherence in the scattering at large momentum transfers. For contact interactions in the non-relativistic limit the scattering is isotropic in the center of mass reference system, and therefore the total cross section for zero momentum transfer $\sigma^0$ is:
\begin{equation}\label{eq:difCS}
	\sigma^0 = \int_{0}^{E_{nr}^{max}} \frac{d\sigma}{dE_{nr}}(\Vec{v},\Vec{q} = 0) dE_{nr} = \frac{d\sigma}{dE_{nr}}(\Vec{v},\Vec{q} = 0) E_{nr}^{max},
\end{equation}
and then
\begin{equation}\label{eq:difCS}
	\frac{d\sigma}{dE_{nr}}(\Vec{v},\Vec{q} = 0) = \frac{\sigma^0}{E_{nr}^{max}} = \frac{m_N}{2\mu_{\chi N}^2 v^2} \sigma^0.
\end{equation}
Then, the differential rate can be expressed as:
\begin{equation}\label{eq:difRate2}
	\frac{dR}{dE_{nr}} = \frac{M_{det} \rho_{\chi}}{2 m_{\chi} \mu_{\chi N}^2} \sigma^0 F^2(q) \int_{v_{min}}^{\infty} \frac{f(\Vec{v})}{v} d^3\Vec{v},
\end{equation}
and the integral is known as the "mean inverse speed". The WIMP-nucleus interaction is strongly dependent on the WIMP microscopic particle model. Two main generic types of interaction mechanisms are usually considered, spin-independent (SI) and spin-dependent (SD):
\begin{equation}\label{eq:sigma0}
	\sigma_0 F^2(q) = \sigma_{SI}^0 F_{SI}^2(q) + \sigma_{SD}^0 F_{SD}^2(q).
\end{equation}
SI-interaction cross-section adds coherently for protons and neutrons in the nucleus as:
\begin{equation}\label{eq:sigmaSI1}
	\sigma_{SI}^0 = \left(Z+(A-Z)\left(\frac{f_n}{f_p}\right)\right)^2\frac{\mu^2_{\chi N}}{\mu^2_{\chi n}} \sigma_{SI},
\end{equation}
where $\mu^2_{\chi n}$ and $\sigma_{SI}$ are the WIMP-nucleon reduced mass and WIMP-proton SI cross-section, respectively and $f_n$ and $f_p$ are the WIMP SI-couplings to neutrons and protons, respectively. They are commonly supposed to be equal, and therefore:
\begin{equation}\label{eq:sigmaSI2}
	\sigma_{SI}^0 = A^2 \frac{\mu^2_{\chi N}}{\mu^2_{\chi n}} \sigma_{SI}.
\end{equation}
On the other hand, SD-interaction cross-section does not add coherently the contribution of the nucleons, as it is the result from the coupling to the spin content of the nucleus, $J$:
\begin{equation}\label{eq:sigmaSD}
	\sigma_{SD}^0 = \frac{\mu^2_{\chi N}}{\mu^2_{\chi n}} \sigma_{SD} \frac{4(J+1)}{3J(a_p^2+a_n^2)}\left(\left<S_p\right>a_p+\left<S_n\right>a_n\right)^2,
\end{equation}
where $a_{n,p}$ are the WIMP SD-couplings and $\left<S_{n,p}\right>$ are the expectation values of the spin content of the neutron or proton group in the nucleus. As the SD-interaction depends mainly on the unpaired nucleon, it is supposed to be subdominant if both (SI and SD) are present. Considering only the SI-interaction, the differential rate is:
\begin{equation}\label{eq:difRate3}
	\frac{dR}{dE_{nr}} = \frac{M_{det} \rho_{\chi} A^2}{2 m_{\chi} \mu_{\chi n}^2} \sigma_{SI} F^2(q) \int_{v_{min}}^{\infty} \frac{f(\Vec{v})}{v} d^3\Vec{v}.
\end{equation}
The dark matter local density usually considered is 0.3~GeV/c$^2$/cm$^3$, but as seen in Equation~\ref{eq:difRate3} it can be taken as a scale factor in the counting rate. The most recent estimates, using global fits, of the dark matter local density provide a value of 0.39 $\pm$ 0.03~GeV/c$^2$/cm$^3$~\cite{Catena:2009mf}. The velocity distribution of the particles in the galactic rest frame $f(\Vec{v}_{gal})$ is assumed to follow a Maxwell-Boltzmann distribution truncated at the escape velocity of the Milky Way ($v_{esc} =$~544~km/s, from~\cite{Smith_2007}). This model is known as Standard Halo Model (SHM), and assumes that the halo is an isotropic and isothermal sphere in the galactic frame:
\begin{equation}\label{eq:f(v)}
	f(\Vec{v}_{gal}) d^3\Vec{v}_{gal} = \frac{1}{v_0^3\pi^{3/2}}\exp{\left(-\frac{v_{gal}^2}{v_0^2}\right)} d^3\Vec{v}_{gal},
\end{equation}
where $v_0$, related with the dispersion velocity of the dark matter particles in the halo, is taken as the velocity of the Local Standard of Rest (LSR), which follows the mean motion of the material in our galaxy in the neighborhood of the Sun. Then, assuming spherical symmetry, the velocity distribution as a function of the module of the velocity, $v_{gal}$, is
\begin{equation}\label{eq:f(v)2}
	f(v_{gal}) dv_{gal} = \frac{4v_{gal}^2}{v_0^3\sqrt{\pi}}\exp{\left(-\frac{v_{gal}^2}{v_0^2}\right)} dv_{gal}.
\end{equation}
The velocity included in the differential rate must be defined in the laboratory rest frame, $\Vec{v}$. Since these velocities are non-relativistic, the change of reference frame is just a Galilean boost by the Earth’s velocity in the Galactic frame, $\Vec{v_E}$:
\begin{equation}\label{eq:vEBoost}
	\Vec{v}_{gal} = \Vec{v} + \Vec{v_E},
\end{equation}
Due to that boost, the mean inverse speed is complicated to obtain analytically. A very rough approximation allows to derive the WIMP differential rate in a simple way considering that the Earth is static in the galactic frame ($\Vec{v_E} = 0$), in whose case, the WIMP velocity distribution in the laboratory frame is:
\begin{equation}\label{eq:f(v)3}
	f(v) dv = \frac{4v^2}{v_0^3\sqrt{\pi}}\exp{\left(-\frac{v^2}{v_0^2}\right)} dv,
\end{equation}
which implies that the differential rate for SI-interaction is:
\begin{equation}\label{eq:difRate4}
	\frac{dR}{dE_{nr}} = \frac{2 M_{det} \rho_{\chi} A^2 \sigma_{SI} F^2(q)}{m_{\chi} \mu_{\chi n}^2 v_0^3\sqrt{\pi}} \int_{v_{min}}^{v_{esc}} v \exp{\left(-\frac{v^2}{v_0^2}\right)} dv,
\end{equation}
and then:
\begin{equation}\label{eq:difRate5}
	\frac{dR}{dE_{nr}} = R_0\exp{\left(-\frac{E_{nr}}{E_0}\right)},
\end{equation}
where
\begin{equation}\label{eq:R0}
	R_0 = \frac{M_{det} \rho_{\chi} A^2 \sigma_{SI} F^2(q)}{m_{\chi} \mu_{\chi n}^2 v_0\sqrt{\pi}}
\end{equation}
and
\begin{equation}\label{eq:E0}
	E_0 = \frac{2\mu_{\chi N}^2 v_0^2}{m_N} = \frac{2m_{\chi}^2 m_N v_0^2}{\left(m_{\chi}+m_N\right)^2}.
\end{equation}

\begin{figure}[h!]
	\begin{center}
		\includegraphics[width=0.75\textwidth]{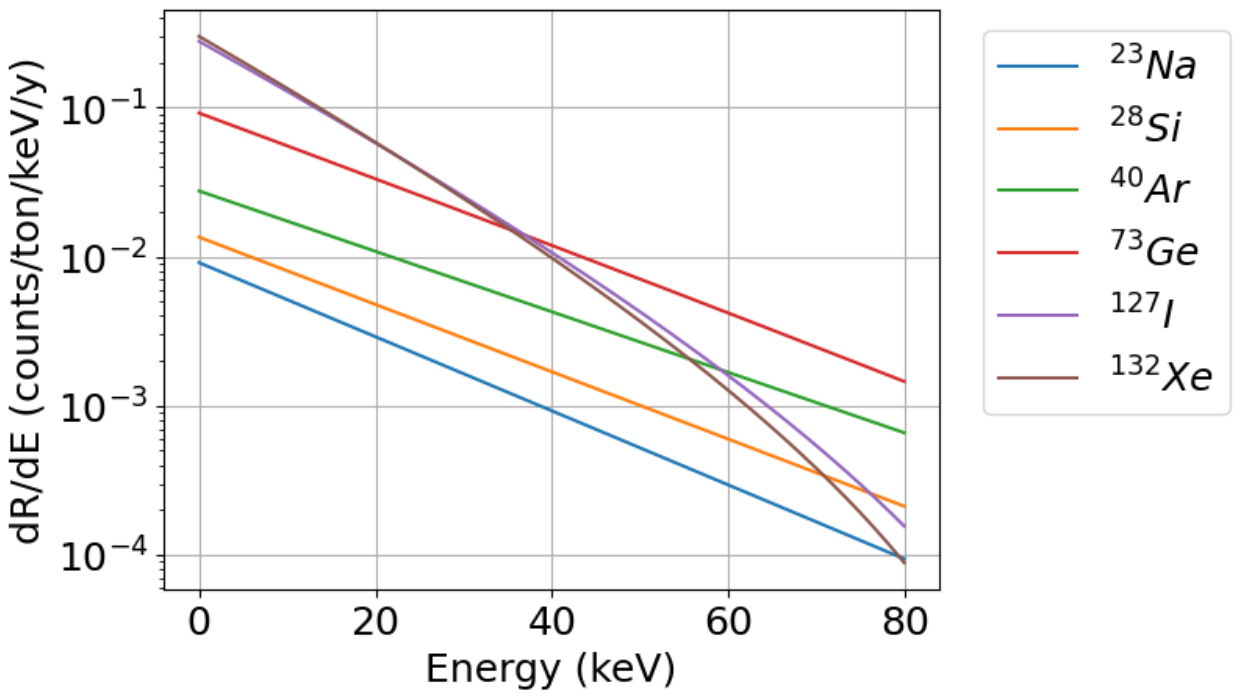}
		\caption{\label{RecoilRate}The differential rate as a function of the nuclear recoil energy (Equation~\ref{eq:difRate5}) for some typical target nuclei used in direct dark matter search experiments considering $m_{\chi}$~=~100~GeV, a SI WIMP-nucleon cross-section of~10$^{-10}$~pb and the dark halo and WIMP-nucleus interaction models explained in text.}
	\end{center}
\end{figure}

This approximation is far from realistic because $v_E$ is of the same order of $v_{gal}$, however it brings an idea of the order of magnitude of the detection rate. For WIMPs with masses, $m_{\chi}$, between 10~GeV and 10~TeV the expected flux in the Earth ranges from $\sim$~10$^7$ to~10$^{10}$ WIMPs/m$^2$/s. Although this flux could seem very high, the WIMP-nucleon cross-sections are very low (below 10$^{-5}$~pb), which implies a very low rate. The approximation of the differential rate obtained in Equation~\ref{eq:difRate5} has been plotted in Figure~\ref{RecoilRate} as a function of the nuclear recoil energy for some typical target nuclei used in direct dark matter search experiments. In this plot, the WIMP mass and SI WIMP-nucleon cross-section considered have been $m_{\chi}$~=~100~GeV and~10$^{-10}$~pb, respectively.

\begin{figure}[h!]
	\begin{subfigure}[b]{0.42\textwidth}
		\includegraphics[width=\textwidth]{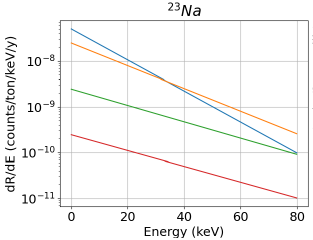}
	\end{subfigure}
	\begin{subfigure}[b]{0.57\textwidth}
		\includegraphics[width=\textwidth]{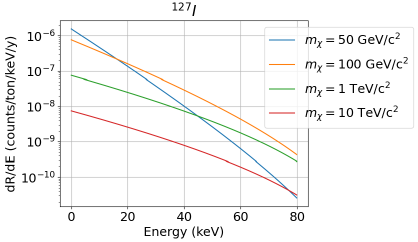}
	\end{subfigure}
	\caption{\label{RecoilRate_WIMPMass}The differential rate as a function of the nuclear recoil energy (Equation~\ref{eq:difRate5}) for $^{23}Na$ (left) and $^{127}I$ (right) for four different WIMP masses considering a SI WIMP-nucleon cross-section of 10$^{-10}$~pb and the dark halo and WIMP-nucleus interaction models explained in text.}
\end{figure}

The spectrum shape is also strongly dependent on the mass of the WIMP. Figure~\ref{RecoilRate_WIMPMass} shows the difference in the spectrum shape for $^{23}Na$ and $^{127}I$ for a SI WIMP-nucleon cross-section of 10$^{-10}$~pb for different WIMP masses. It is possible to observe that (as described in Equation~\ref{eq:E0}) the spectrum shape does not depend on the WIMP mass when $m_{\chi} \gg m_N$, while if $m_{\chi} \ll m_N$, then $m_{\chi}+m_N\sim m_N$ and therefore $E_0$ depends on the $m_{\chi}^2$.

It is possible to observe that the expected rate increases for massive nuclei in the case of SI interacting WIMPs. Depending on the WIMP-nucleon interaction cross-section, the estimated rates can vary from 1~event/kg/day to 1~event/ton/year. These low expected rates require that experiments devoted to dark matter searches operate in extreme low radioactive background conditions, have very large target masses and long exposure times. Moreover, as the differential rate has not distinctive features, but it is approximately exponential even when taking into account more realistic estimates than the presented before within the SHM and the conventional dark matter interaction operators (SI and SD interactions), it is very difficult to guarantee a positive detection of dark matter against the background sources. Searching for active background discrimination techniques is then mandatory to improve the sensitivity.

Understanding the background sources of the experiments, in particular those contributing in the ROI, and developing a robust background modeling is very important for the dark matter analysis. The main sources of background for direct dark matter detection experiments are~\cite{Heusser:1995wd}: environmental radioactivity, internal radioactive contamination from the presence of primordial isotopes in the detector materials, cosmic radiation and cosmogenic activation in the detector materials due to previous exposure to cosmic rays. Ultimately, the background will be limited by the neutrino flux~\cite{EDELWEISS:2020fxc}. The coherent neutrino scattering in the detector target nuclei represents an ultimate background, as it shares most of the characteristic features with those of WIMP scattering. Among the different neutrino sources, the most relevant as background for direct dark matter searches are the Sun, and the atmosphere. Due to the impossibility to shield the detectors from them, this background is very difficult to overcome. In any case, the cross-section of these processes is very small, and the sensitivity to detect these neutrinos has not yet been reached by any experiment.

In order to strongly reduce the cosmic ray induced background, the detectors have to be located underground. The rock overburden is commonly expressed in units of meters of water equivalent (m.w.e.). A depth of several tens of meters of rock is sufficient to make the hadronic component negligible. However, the muon flux is more difficult to attenuate and besides the direct interaction of the muons in the detector, they can produce nuclear reactions in the different detectors components and produce fast neutrons, prompt or delayed, which can imply a relevant background for dark matter searches~\cite{Wang:2001fq}. These neutrons can cause keV-level nuclear recoils by scattering elastically off the target nuclei in the detector~\cite{Mei:2005gm}. To reduce this background, these experiments are carried out in underground laboratories below hundred of meters of rock. In addition to the underground location of the laboratory, the residual muon flux contribution to the background can be further reduced by using active veto detectors to identify and tag the muon reaching the detector or its associated cascade.

Radiogenic neutrons can be produced in ($\alpha$,n) reactions, by alpha particles emitted during radioactive decay (usually from primordial decay chains) in the rocks or building materials of the laboratory~\cite{Mei:2008ir}. Radiogenic neutrons can also be produced through spontaneous fission in isotopes such as $^{238}U$. Passive shielding made of water or materials with high hydrogen content (such as polyethylene or paraffin) are typically used to moderate neutrons and then, reduce strongly the direct contribution to the ROI. To reduce the gamma radiation coming from natural uranium and thorium decay chains, as well as from the decay of isotopes such as $^{40}K$, $^{60}Co$, and $^{137}Cs$ in the surrounding materials, passive shielding made of high-Z materials with low radioactivity (such as lead and copper) is used to enclose the detector. Finally, to reduce the contribution to the background of the airborne radon, the inner part of the detector shielding should be tightly isolated from the laboratory air and can be flushed with clean nitrogen gas or radon-free air.

Internal contamination (present in the detector materials) is the most relevant background for most of the experiments, because it is impossible to shield. It is mandatory to work with the highest radiopure materials in order to achieve good sensitivities. The isotopes that typically contribute to this background are long-lived natural radioisotopes (such as those in the $^{232}Th$, $^{238}U$ chains and $^{40}K$), cosmogenic activation isotopes (such as $^3H$ and $^{39}Ar$), and anthropogenic isotopes (such as $^{60}Co$, $^{85}Kr$ and $^{137}Cs$). To minimize internal radioactivity, all the detector components have to be selected for their low radioactivity using analytical methods such as high purity germanium (HPGe) spectrometry. Isotopes produced by cosmogenic activation due to previous exposure of the detector materials to cosmic radiation before moving underground, can be reduced by minimizing the time spent at the surface and avoiding air transport. Finally, special care must be taken to prevent the accumulation of $^{222}Rn$ and its daughters on surfaces, which can be removed by acid treatment or electropolishing.

In addition to all the previously commented background reduction techniques, data analysis can allow to reject some backgrounds that can be discriminated from the signal searched for. These background rejection techniques include the discrimination of background events using pulse shape information, as for example applying event selection based on the different time responses of the detector for energy depositions corresponding to different particles. The ability to discriminate nuclear recoils from electronic recoils is essential for dark matter searches, since the background is completely dominated by the latter, and the WIMP interactions searched for correspond to nuclear recoil signals. Other possibility is to apply fiducial cuts, when the information on the position of the energy deposition is available. This allows to remove events produced in regions of the detector with higher activity, for instance the surface. Other possibility of discrimination between nuclear recoils and electronic recoils profits from the different energy sharing in the different channels available for the energy conversion (heat, light and charge). This requires a detector designed to measure simultaneously two of these channels and will be further described in Section~\ref{Section:Intro_Detection_Direct_Techinques}.

\subsubsection{Characteristic signatures of the dark matter signal: annual modulation}
\label{Section:Intro_Detection_Direct_AnualMod}

\begin{figure}[h!]
	\begin{center}
		\includegraphics[width=\textwidth]{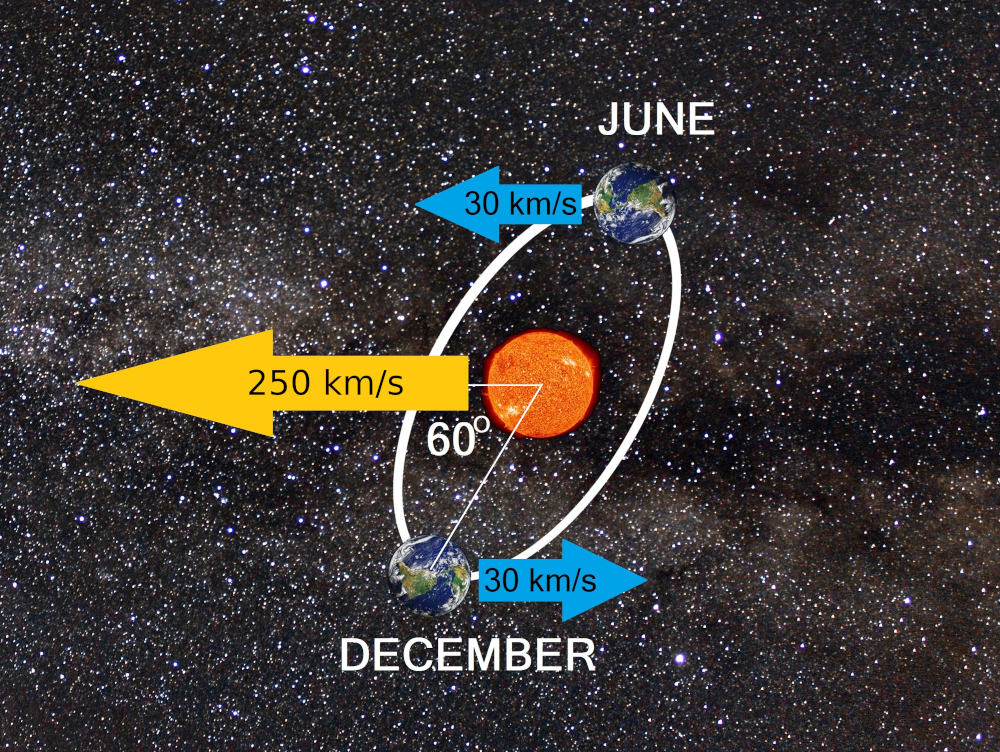}
		\caption{\label{AnnualMod}The orbit of the Earth around the Sun combined with the motion of the Sun around the galactic center produces a predictable annual modulation in the velocity of WIMPs reaching the Earth. Background image from El cielo de Rasal~\cite{ElCieloDeRasalYoutube,ElCieloDeRasalBlog}.}
	\end{center}
\end{figure}

The expected dark matter spectrum has not distinctive features which allow to identify clearly any dark matter contribution over the commented backgrounds. Therefore, it is important to analyze the possible dark matter characteristic features that are not shared by the backgrounds, and then allow the positive identification of the dark matter signal. One of the most clear signatures is the annual modulation in the WIMP detection rate, produced by the change in relative velocity of the dark matter particles with respect to the detector nuclei along the year because of the movement of the Earth around the Sun~\cite{Drukier:1986tm,Freese:1987wu,Freese:2012xd}. The orbit of the Earth around the Sun is almost circular with a period of one year and average velocity of $v_E$~$\sim$~29.8~km/s~\cite{McCabe_2014}. The plane of this orbit is inclined at an angle $\gamma \sim$~60$^o$ respect to the galactic plane. The velocity of the Sun respect to the galactic center is the sum of the local standard of rest velocity, $v_0$, which is $\left(0,238,0\right)$~km/s~\cite{Bland_Hawthorn_2016,Gravity_2021}, and the solar peculiar velocity, which is $\left(11.1,12.2,7.3\right)$~km/s~\cite{Schonrich_2010}. Therefore, in the galactic center rest frame, the velocity of the Sun, $v_S$, has a module of~$\sim$~250~km/s, and then, the projection of the Earth velocity in the direction of motion of the Sun changes with the time of the year following:
\begin{equation}\label{eq:vT}
	v_T = v_S+v_E\cos{\gamma}\cos{\left(\frac{2\pi}{T}(t-t_0)\right)},
\end{equation}
where $T$ is the period of the orbit of the Earth around the Sun (1~year) and $t_0$ corresponds to the 2$^{nd}$ of June, when Sun and Earth velocities are aligned and then the addition of the two velocities is maximal. Figure~\ref{AnnualMod} illustrates this effect.~\cite{Froborg_2020}.

As it was shown in Equation~\ref{eq:difRate4}, the interaction rate depends on the velocity of the WIMPs in the Earth rest frame. The variation of this velocity during one year is small in amplitude, allowing to do a Taylor expansion of the expected dark matter rate in a given energy range ($S_k$) and keep only the first order~\cite{Froborg_2020}:
\begin{equation}
	S_k \approx S_{0k}+S_{mk}\cos{\left(\frac{2\pi}{T}(t-t_0)\right)},
\end{equation}
where $S_{0k}$ is the mean dark matter rate and $S_{mk}$ is the modulation amplitude, both in the energy range $k$. It is worth noting that the WIMPs are in average more energetic near $t_0$, when $v_T$ (Equation~\ref{eq:vT}) is maximal, implying this change in the differential flux of WIMPs that at low recoil energies there is phase reversal and the maximum is expected in December. In summary, the annual modulation of the dark matter detection rate must satisfy four requirements:
\begin{itemize}
	\item The rate of events modulates around the average value, following a cosinoidal behaviour with a period of one year and phase around June 2$^{nd}$.
	\item This modulation is only present in a specific low-energy range where dark matter interactions are expected to occur, according to the kinematical reasons commented in Section~\ref{Section:Intro_Detection_Direct_Signature}.
	\item In a multi-detector setup, the modulation in only present in single-hit events, as the probability of a dark matter particle interacting with multiple detectors is negligible.
	\item The modulation amplitude is a weak effect (from 1 to 10\% of the total dark matter average rate depending on the WIMP mass and energy range).
\end{itemize}
This approximation is only valid for the SHM. If there were anisotropies in the WIMP velocity distribution or the halo was rotating, the phase and the modulation amplitude would be very different from the previously commented scenario. Moreover, if the halo has some kind of substructure, as could be the presence of streams, this approach would not be valid~\cite{Bernabei:2006ya}.

It is important for experiments searching for annual modulation of dark matter to carefully control and understand all potential sources of systematic effects that could mimic the annual modulation signature. This includes carefully accounting for any experimental parameter that may change with time and even modulate annually. For example, it has been observed an annual modulation of the muon flux underground~\cite{OPERA:2018jif}, which could be translated in a modulation in the detection rate produced by the neutrons induced by muons or other muon-related events that could fall in the ROI. The radon concentration in the air of underground laboratories as Canfranc Underground Laboratory (LSC) has been observed to present an annual modulation, which also correlates with other environmental conditions, such as the humidity~\cite{Amare:2022dgr}. Moreover, the detector response can depend also on these conditions (for instance, the temperature can affect the detectors gain). Then, it is mandatory to control all the environmental conditions that can introduce systematic effects in the measurement and try to work in the most stable conditions.

\subsubsection{Detection techniques} \label{Section:Intro_Detection_Direct_Techinques}

Many strategies have been followed to design dark matter detection experiments. All of them are based on a few basic detection techniques largely developed in the nuclear and particle physics. The energy released by a particle in a detection medium produces ionization or excitation in the material, resulting in the production of light and free electric charge that can be conveniently readout. However, most of the energy is converted into heat. 

In ionization detectors, the visible signal is the free charge released in the material when the primary or secondary particles produced following the interaction ionize the medium: electron-holes in a semiconductor or electron-ions in liquid or gas targets. This charge conveniently drifted in an electric field can produce an electric signal in the electrodes, that can be readout as a current or charge signal. In scintillation detectors the visible signal is the light produced in the radiative deexcitation of states excited by the primary or secondary particles. It can either correlate or compete with the ionization process. The wavelength of the light emitted depends on the energy of the excited states, but usually is in the visible or ultraviolet range, allowing for the light detection using Photomultiplier Tubes (PMTs) or Sillicon Photomulitpliers (SiPMs). But most of the energy goes to quantized vibrations in the lattice of the material (phonons). These vibrations effectively increase the temperature of the material, which can be measured using very sensitive phonon or temperature sensors. This technique requires to operate the detector at very low temperatures (10 to 100~mK) to reduce the thermal capacity of the material allowing to produce measurable temperature variations.

The capability of the materials to convert the energy deposited by incident radiation into detectable signal carriers (such as pairs e-h, phonons or scintillating photons) is their signal yield, $Y$, being one of the most relevant parameters determining the achievable threshold and energy resolution. $Y$ is defined as the number of carriers emitted per unit of energy deposited. Then, for a given energy conversion channel, $i$, the number of carriers produced in the material, $S_i$ is
\begin{equation}\label{eq:DetResponse}
	S_i(E) = Y_i(E) \cdot E,
\end{equation}
where $Y_i(E)$ is the signal yield for this particular channel and $E$ is the deposited energy. This yield can depend on the energy, in particular if very wide energy ranges are considered. We will refer to this as non-proportional behaviour. However, many times the signal yield is supposed to be constant for a material and type of interacting particle (proportional behaviour). The sharing of the deposited energy among these three channels depends strongly on the material and the type of interacting particle. Equation~\ref{eq:DetResponse} should be written as:
\begin{equation}\label{eq:DetResponse2}
	S_{ip}(E) = Y_{ip}(E) \cdot E,
\end{equation}
being $p$ the type of interacting particle. This allows to design hybrid detectors able to discriminate the interacting particle by measuring simultaneously two of these channels. 

In the case of scintillation detectors, this dependence of the light yield with the type of particle is related with the saturation of the color centers that highly ionizing particles can produce because of the high density of energy released~\cite{Birks:1964zz}. Particularly relevant for dark matter detectors is the discrimination between nuclear recoil ($nr$) and electron recoils ($er$), as the first are expected to be the signature of a dark matter interaction, while the latter constitute most of the background.

It is important therefore to characterize the detector response using particles with a known energy and that interact with the same process as the searched signal. However, this is not always possible. Although in most of the WIMP models, they are expected to interact with the nuclei of the target, the periodic calibrations of direct dark matter search experiments are usually done with gamma sources, which produce electron recoils. The energy calibrated this way is known as electron equivalent energy ($E_{ee}$, with units of keVee). To obtain the nuclear recoil energy ($E_{nr}$) from the measured electron equivalent energy ($E_{ee}$), it is required to know the convenient scaling factor between $E_{nr}$ and $E_{ee}$. This scaling factor is usually $<$~1 for scintillation and ionization measurements, reason why it is called Quenching Factor ($QF$). It is defined as the ratio of the number of carriers emitted following a nuclear recoil to that emitted following an electronic recoil for the same energy deposition, $E$. There are some models that describe the dependence of the QF with energy for certain detectors, but in general it has to be measured for every target material. Then, for a given signal channel:
\begin{equation}
	QF_i(E) = \frac{S_{inr}(E)}{S_{ier}(E)} = \frac{Y_{inr}(E)}{Y_{ier}(E)}.
\end{equation}
Since detectors are calibrated using electronic recoils, the electron equivalent energy corresponding to a number of carriers measured $S_i(E)$ can be determined, using Equation~\ref{eq:DetResponse2} as:
\begin{equation}\label{eq:Eee}
	E_{ee} = \frac{S_i(E)}{Y_{ier}(E)}.
\end{equation}
Then, the signal of a nuclear recoil releasing an energy $E$ (obtained with the Equation~\ref{eq:DetResponse2}), is associated to an electron equivalent energy given by:
\begin{equation}\label{eq:Eee2}
	E_{ee} = \frac{S_{inr}(E)}{Y_{ier}(E)} = \frac{Y_{inr}(E)}{Y_{ier}(E)} \cdot E = QF_i(E) \cdot E.
\end{equation}
This allows to express the nuclear recoil energy as a function of the electron equivalent energy as
\begin{equation}
	E = \frac{E_{ee}}{QF_i(E)},
\end{equation}
or equivalently,
\begin{equation}\label{eq:QFDefinition}
	QF_i(E) = \frac{E_{ee}}{E}.
\end{equation}
It is important to emphasize that the $QF$ is different for each kind of carrier in each material. Although some theoretical approaches for the calculation of these $QF$ have been proposed~\cite{Birks:1951boa,PhysRev.122.815,LINDHARD,Hitachi:2005ti,Hitachi:2007zz}, a general theory allowing to estimate this factor for different particles, target materials and detection mechanisms is still lacking. The implications of a low QF for a dark matter experiment are clear: the recoil energies expected for dark matter interactions, that are lower than 100~keV (Section~\ref{Section:Intro_Detection_Direct_Signature}), will appear in the gamma-calibrated spectrum at energies $E_{ee}$ lower than QF$\cdot$100~keVee, which makes even more difficult to detect these interactions and increases the importance of achieving low energy thresholds.

\subsubsection{Current experimental status} \label{Section:Intro_Detection_Direct_Experiments}

\begin{figure}[h!]
	\begin{center}
		\includegraphics[width=\textwidth]{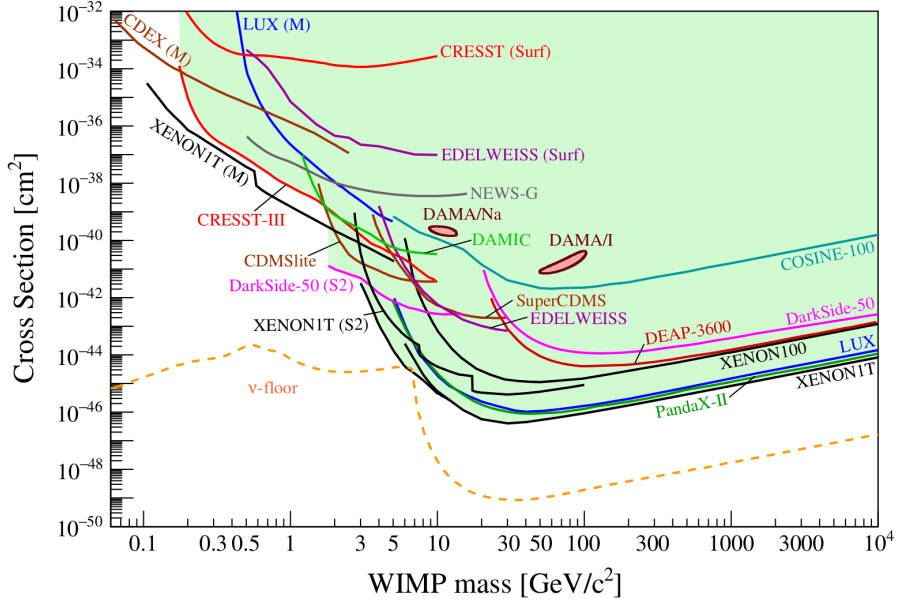}
		\caption{\label{ExclusionPlot}Spin-independent WIMP-nucleon scattering exclusion plot with current limits at 90\% C.L. considering the SHM. Orange islands represent the DAMA/LIBRA positive signal in the [2,6]~keV range, while dashed line represent the neutrino floor~\cite{Billard:2021uyg}.}
	\end{center}
\end{figure}

As it was explained above, disentangling the dark matter signal from residual radioactive backgrounds can be very challenging. Standard searches for SI interacting WIMPs with masses above 10~GeV are dominated by very massive experiments with large atomic number nuclei in the target composition (mainly TPCs of noble elements, see explanation below), while the $m_{\chi}<$~10~GeV region is outside the scope of these experiments, since they are not able to reach the energetic threshold needed to explore it. Most of the experiments are not able to positively identify a dark matter signal, but they can rule out those candidates that would have produced in the detector an interaction rate larger than that measured. This allows to produce exclusion plots in the WIMP-nucleon interaction cross-section as a function of the WIMP mass, with the curves representing the upper limit cross-sections for each WIMP mass that can be excluded by the experiment at a given confidence level (usually 90\%). The excluded region for each experiment strongly depends on the WIMP and galactic halo model considered. Usually, it is assumed only one type of interaction (SI, SD-proton, SD-neutron or SD considering equal coupling to protons and neutrons) and the SHM. Over the years, the exclusion limits have covered a wide range of the WIMP parameter space (m,$\sigma$). Figure~\ref{ExclusionPlot} shows the current status of the exclusion plots (and one positive signal)~\cite{Billard:2021uyg} for the SI-interaction case and the SHM.

DAMA/LIBRA (DArk MAtter/Large sodium Iodide Bulk for RAre processes) experiment~\cite{Bernabei:2020mon}, located at LNGS, uses an array of highly radio-pure NaI(Tl) scintillators, and it is well-known for its claim of observing an annual modulation signal consistent with that expected for dark matter particles distributed in our galaxy following the SHM. Regions of the parameter space compatible with that positive signal are also presented in Figure~\ref{ExclusionPlot} for the SI-interaction in sodium and iodine (labelled as $DAMA/Na$ and $DAMA/I$ in the figure, respectively). This result is still under debate as other experiments have ruled out the region of the parameter space singled out by DAMA in most of the considered scenarios. However, the comparison among experiments is model-dependent, and then, it is not possible a definitive ruling-out of the DAMA/LIBRA result. Some experiments aimed at confirming or refuting DAMA/LIBRA observation in a model-independent way, using a similar setup and detectors made of the same target material: NaI(Tl). (Annual modulation with NAI Scintillators) experiment~\cite{Amare:2021yyu} is taking data at the Canfranc Underground Laboratory (LSC) in Spain, while COSINE-100 (COherent Scattering and Interaction with Nuclei Experiment)~\cite{Adhikari:2017esn} was installed at the Yang Yang Underground Laboratory, in South Korea. In the longer term, there are other experiments that will use NaI(Tl) scintillation crystals for direct dark matter detection, as the SABRE (Sodium Iodide with Active Background Rejection Experiment) project~\cite{SABRE:2018lfp}, aiming at installing twin detectors in Australia and Italy, the COSINUS (Cryogenic Observatory for SIgnatures seen in Next-generation Underground Searches) project~\cite{Angloher:2016ooq}, which is developing cryogenic detectors based on NaI, and the PICOLON (Pure Inorganic Crystal Observatory for LOw-energy Neutr(al)ino) project~\cite{Fushimi:2021mez}, which is working in the development of highly radiopure NaI(Tl) scintillators. NaI scintillator detectors will be covered in detail in Section~\ref{Section:Intro_Scintillators} while other detection techniques are summarized next.

\textbf{Solid-state cryogenic detectors} use crystalline targets operated at very low temperatures, typically in the range from 10~to 20~mK in order to measure the heat channel. Different target materials can be used. However, crystal semiconductors such as germanium and silicon and scintillating crystals such as CaWO$_4$, allow the simultaneous measurement of an additional energy channel, the ionization and the light channels, respectively. This hybrid detection guarantees a high discrimination power against beta/gamma backgrounds. The measurement of the heat signal requires very sensitive thermometers, as can be Transition Edge Sensors (TES) and Neutron Transmutation Doped (NTD) thermistors. TES are devices made of a superconducting material that operate at the transition temperature between the superconducting and normal states. At this transition, the resistivity of the material depends strongly on the temperature. On the other hand, NTD thermistors are semiconductor devices that have been irradiated with neutrons, thus creating impurities that increase the dependence of their resistivities on the temperature. In both cases, the small changes in the temperature produced by the energy deposited by the particles are translated into a measurable change of the resistance of the device, reaching sensitivities of the order of few~$\mu$K. The ionization detection requires that the semiconductor crystal is biased with an electric field to drift the charge carriers  produced by the particle interaction towards the electrodes producing a measurable current. Two experiments using semiconducting targets at very low temperature and measuring heat and ionization are SuperCDMS (Cryogenic Dark Matter Search)~\cite{SuperCDMS:2014cds,SuperCDMS:2017mbc}, located at the SNOLAB in Canada and EDELWEISS (Expérience pour DEtecter Les WIMPs En SIte Souterrain)~\cite{EDELWEISS:2016nzl,EDELWEISS:2017lvq}, located at the Modane Underground Laboratory (LSM) in France. A different approach is that followed by CRESST-III (Cryogenic Rare Event Search with Superconducting Thermometers) experiment~\cite{CRESST:1999ynq,CRESST:2019jnq}, running at the LNGS, which uses scintillating bolometers of CaWO$_4$, and measures the scintillation light channel and the heat channel. This experiment has achieved an energy threshold below 100~eV in five of the ten detectors. This remarkable low threshold combined with the presence of a light isotope in the crystal composition has allowed CRESST-III to achieve the leading limits for SI interacting WIMPs at the lowest mass range explored as of today. This technique is able to provide very low energy threshold, but only for small mass crystalline targets. Moreover, both EDELWEISS and CRESST-III, and other low-threshold experiments are observing unexpected backgrounds below a few hundred of~eVs which is reducing their sensitivities. Further work to understand the backgrounds in this energy range is required and a collective effort known as EXCESS is ongoing~\cite{Fuss:2022fxe}.

\textbf{Noble liquid detectors} use noble elements like xenon or argon in liquid state as target materials (LXe, LAr). These experiments employ two different technologies: single-phase and dual-phase. Single-phase detectors measure the scintillation light emitted by these liquids when a particle releases energy on them. The DEAP-3600 (Dark Matter Experiment with Argon Pulse shape discrimination) experiment~\cite{DEAP-3600:2017ker,DEAP:2019yzn}, located at SNOLAB, in Canada, is a single-phase noble liquid detector that uses 3.3~tons of argon. This element is particularly useful for these kind of detectors as it allows to discriminate the nature of the interacting particle due to the different time profile of the light produced by nuclear recoils and electron recoils. Dual-phase detectors work as Time Projection Chambers (TPCs), measuring both the primary scintillation signal (S1) and the ionization signal. To measure the latter, the electrons produced by the interacting particle are drifted by an electric field towards the gas region (above the liquid). In the gaseous region, the electrons are able to produce a scintillation signal (S2) which is proportional to the initial ionization signal. The light detectors are placed in a grid and can provide a light pattern for this S2 signal which informs the position of the interaction in the plane (X,Y) while the Z position is obtained from the drift time. The S1 signal is used for triggering. Dual-phase detectors allow to discriminate the type of interacting particle through the ratio of S2 to S1 signals. The most-massive dark matter search experiments use this technique. The most recent phases of some of the experiments using LXe are XENONnT~\cite{XENON:2020kmp}, at the LNGS, with an active mass of 5.9~tons, being the upgrade of XENON1T~\cite{Aprile:2012zx,XENON:2018voc}, PandaX-4T (Particle and Astrophysical Xenon Experiments) experiment~\cite{PandaX:2022osq},  at the China Jinping Underground Laboratory (CJPL), with an active mass of 4~tons, the upgrade of PandaX-II~\cite{PandaX-II:2020oim} and LUX-ZEPLIN experiment~\cite{LZ:2022ufs}, located at the Sanford Underground Research Facility (SURF), in the United States and with an active mass of 7~tons, being the fusion of the LUX~\cite{LUX:2016ggv} and ZEPLIN~\cite{ZEPLIN-III:2009htd} experiments. Concerning the use of LAr as a target material, the DarkSide-20k experiment~\cite{DarkSide-20k:2017zyg}, at LNGS, the next phase of DarkSide-50 experiment~\cite{DarkSide-50:2022qzh}, is in construction at LNGS and will use an active mass of 23~tons.

\textbf{Metastable superheated liquids} are liquids kept above their boiling point in such a metastable stable that the interaction of a particle is able to trigger the phase transition producing a gas bubble. The operation conditions can be fine-tuned in order to be sensitive to nuclear recoil energy depositions but not to electron recoils. These bubbles can be optically detected and counted, thus providing a measurement of the number of particle interactions. Moreover, the formation of these bubbles is detected through the change of the sonic pressure in piezoelectric materials. The combination of this information allows to discriminate between different particle interactions. The energy threshold at which the phase transition is triggered can be adjusted with the temperature and the pressure at which the liquid is maintained. The main search currently using this technique is PICO experiment~\cite{PICO:2019vsc,Garcia-Viltres:2021swf} (the fusion of PICASSO~\cite{PICASSO:2012ngj} and COUPP~\cite{COUPP:2012jrk} experiments), which uses C$_3$F$_8$ in a gel matrix. Thanks to the use of $^{19}F$ in the target, PICO is leading the sensitivity for WIMP candidates with spin-dependent coupling to protons.

\textbf{Silicon charge-coupled devices (CCDs)} are silicon-based detectors that consist of a grid of pixels, each acting as an individual detector. In these detectors, particle interactions with silicon atoms create a number of electron-hole pairs proportional to the energy deposited. The possibility of measuring both the energy and position with high accuracy, allows to distinguish WIMPs from background events, as the firsts are only expected to happen in a single pixel. DArk Matter In CCDs (DAMIC) experiment ~\cite{DAMIC:2011khz} is using this technology, with its first phase at SNOLAB. The next phase (DAMIC-M)~\cite{Castello-Mor:2020jhd} will be located in Modane Underground Laboratory (LSM), in France. They expect to reduce its background by a factor of~50 respect to the first phase and the experiment will consist of 50~CCDs with more than 36~million pixels each, with a total mass of the order of the~kg. It will be sensitive to DM particles interacting with electrons with masses down to 1~MeV/c$^2$. In the longer term, the Oscura experiment (DAMIC+SENSEI)~\cite{Oscura:2022vmi} will be able to confirm or rule out the dark matter-electron scattering for masses down to 500~keV.

\section{NaI(Tl) scintillators in dark matter detection}\label{Section:Intro_Scintillators}
\fancyhead[RO]{\emph{\thesection. \nameref{Section:Intro_Scintillators}}}

In this section, the properties of the NaI(Tl) scintillator crystals are overviewed. They are used in the ANAIS-112 experiment and both, NaI and NaI(Tl) are under consideration as targets for the ANAIS+ project. This makes this material particularly relevant for the goals of this work. The most common light detectors used for the scintillation readout are the PMTs, whose properties are also overviewed. Finally, the DAMA/LIBRA experiment and the remarkable measurement on the annual modulation will be presented, as well as the status and results of other experiments aiming at confirming or refuting that result. The ANAIS-112 experiment will be explained in more detail in Chapter~\ref{Chapter:ANAIS}. 

\subsection{NaI(Tl) crystals} \label{Section:Intro_Scintillators_NaI(Tl)}

\begin{figure}[h!]
	\begin{center}
		\includegraphics[width=\textwidth]{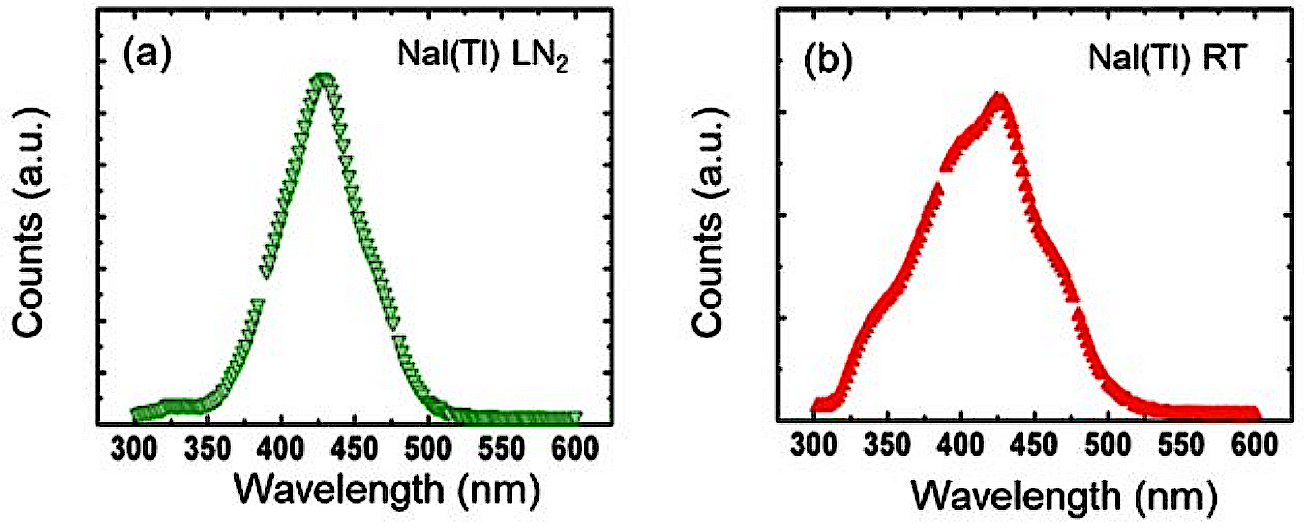}
		\caption{\label{NaIEmission}Emission spectrum of NaI(Tl) crystal both at liquid nitrogen (LN$_2$) temperature (left) and at room temperature (right). Image from~\cite{Sibczynski:2012hr}.}
	\end{center}
\end{figure}

NaI(Tl) crystals have been used since the 1940s due to their remarkable scintillation properties. They are known for its excellent light yield of about 40~photons/keV. The main scintillation component has a decay time of 230~ns with an emission spectrum peaked at 420~nm~\cite{Birks:1964zz,Robertson1961,Schweitzer1983,Eby1954,Sibczynski:2012hr}, shown in Figure~\ref{NaIEmission}. Slower decay time components of 1.5~$\mu$s and $\sim$~0.15~s have been measured~\cite{Cuesta:2013vpa}. Other possible components could be present at much longer timescales, but there is little information about them~\cite{Emigh1954}. Crystals can be machined into a wide variety of sizes and shapes, but they are fragile and tend to be damaged under mechanical or thermal shocks. Moreover, NaI(Tl) is hygroscopic and must be stored in an air-tight container to prevent deterioration.

\begin{figure}[h!]
	\begin{center}
		\includegraphics[width=0.75\textwidth]{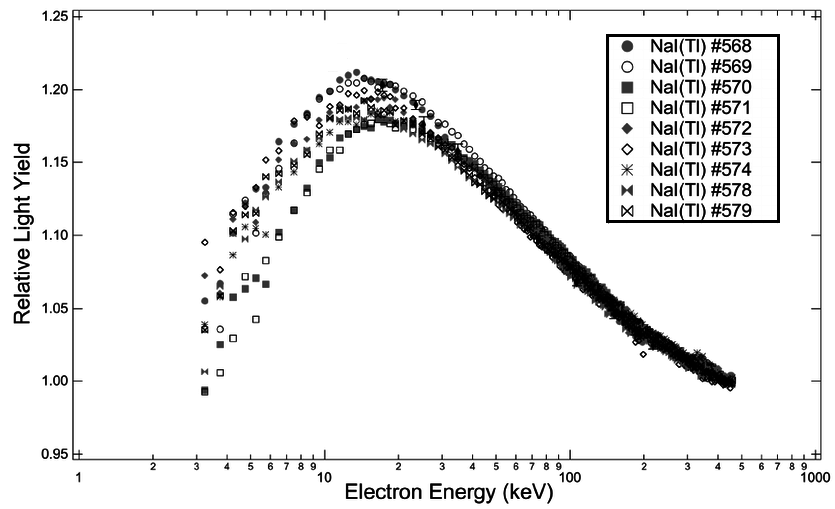}
		\caption{\label{NaI(Tl)NonProportionality}Light yield of NaI(Tl) under electron bombardment normalized at 450~keV for 10~different crystals. Image from~\cite{Hull:2009}.}
	\end{center}
\end{figure}

NaI(Tl) is a convenient target for DM searches due to several compelling reasons. It is sensitive both to low-mass WIMPs through the interaction with $Na$ and also to massive WIMPs through the interaction with $I$. Moreover, both elements are 100\% sensitive to SD interactions. Its high light yield propitiates a low energy threshold, and their scintillation time is long enough to discriminate easily fast emissions as Cherenkov in PMTs or even electronic noise, which allows to reduce the background level.  However, this also implies a disadvantage, as it makes difficult to trigger events with low number of photons. Moreover, non-linearities in its response have been reported~\cite{PhysRev.122.815,Rooney:1997,Moses:2001vx,Choong:2008,Hull:2009,Payne:2009,Khodyuk:2010ydw,Payne:2011}. Figure~\ref{NaI(Tl)NonProportionality}, for example, shows the light yield of NaI(Tl) normalized at 450~keV for 10~crystals as reported in~\cite{Hull:2009}. Variations from 10 to 15\% are observed  from $\sim$~3~keV to $\sim$~20~keV. This implies that the possible systematics related with the detector calibration must be taken into account and that the calibration must be made close to the Region Of Interest (ROI) of the experiment.

The scintillation process of NaI(Tl) is described with great detail in~\cite{Birks:1964zz} using the band theory. The introduction of a dopant in inorganic scintillators (thalium in the case of NaI(Tl)) generates states in the gap between the valence and the conduction bands. These states act as luminescent centers. When a particle deposits energy in the crystal, the electrons can be excited and move from the valence to the conduction band. If they reach a luminescent center they will emit a photon after a time distributed following an exponential having as time constant the lifetime of the corresponding state. They can also reach a quenching center and release the energy into phonons, or they can reach a metastable center, being the electron trapped for typically a time much longer than the characteristic scintillation. These trapped electrons can eventually be promoted to a radiatively decaying state, emitting then a retarded photon. These traps can be associated to crystal impurities or defects in the crystalline structure.

The concentration of thallium in NaI(Tl) crystals is a critical parameter for its light yield. If it is too low it will not efficiently convert the absorbed energy into visible photons, while if too high the self-absorption of the emitted photons increases, which also reduces the effective light yield. Therefore, there is a concentration that maximizes the light yield, which at room temperature is approximately 0.1\% of the molar mass.

\begin{figure}[h!]
	\begin{center}
		\includegraphics[width=\textwidth]{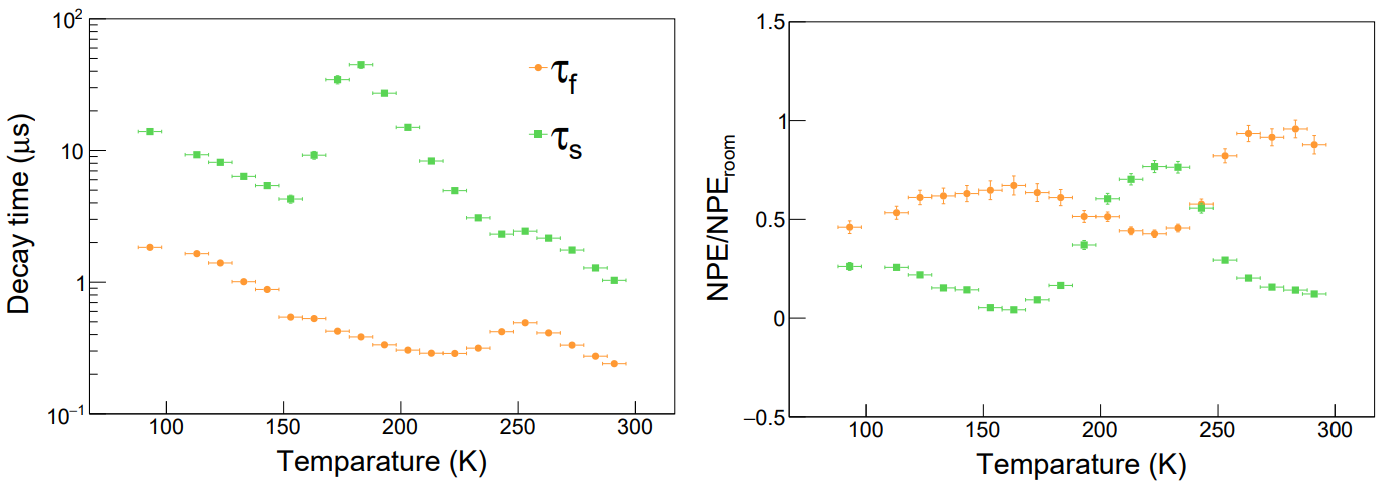}
		\caption{\label{NaI(Tl)Scint}Scintillation times (left) and light yields normalized to that of the fast component at room temperature (right) of the two scintillation components of NaI(Tl) crystal. Image from~\cite{Lee:2021jfx}.}
	\end{center}
\end{figure}

It is worth noting that the crystal properties change with the temperature, and therefore also the response for different thalium concentrations. For the usual concentrations, it has been observed a small increase of the light yield while decreasing the temperature, followed by a stronger decrease at temperatures below 240~K~\cite{Lee:2021aoi,Lee:2021jfx}. The scintillation time constants are also affected by the temperature. For instance, in~\cite{Lee:2021jfx} the scintillation at different temperatures is fit to two exponential components (slow and fast). Figure~\ref{NaI(Tl)Scint} shows the light yields (normalized at that of the fast component at room temperature) and scintillation times of the two components of NaI(Tl) crystal presented in~\cite{Lee:2021jfx}.

The response of NaI(Tl) crystal to nuclear recoils must be known to compare the measured spectrum with theoretical predictions for WIMP interactions. Therefore, as it has been explained in Section~\ref{Section:Intro_Detection_Direct_Techinques} the response has to be corrected by the QF.

\begin{figure}[ht]
	\begin{center}
		\includegraphics[width=\textwidth]{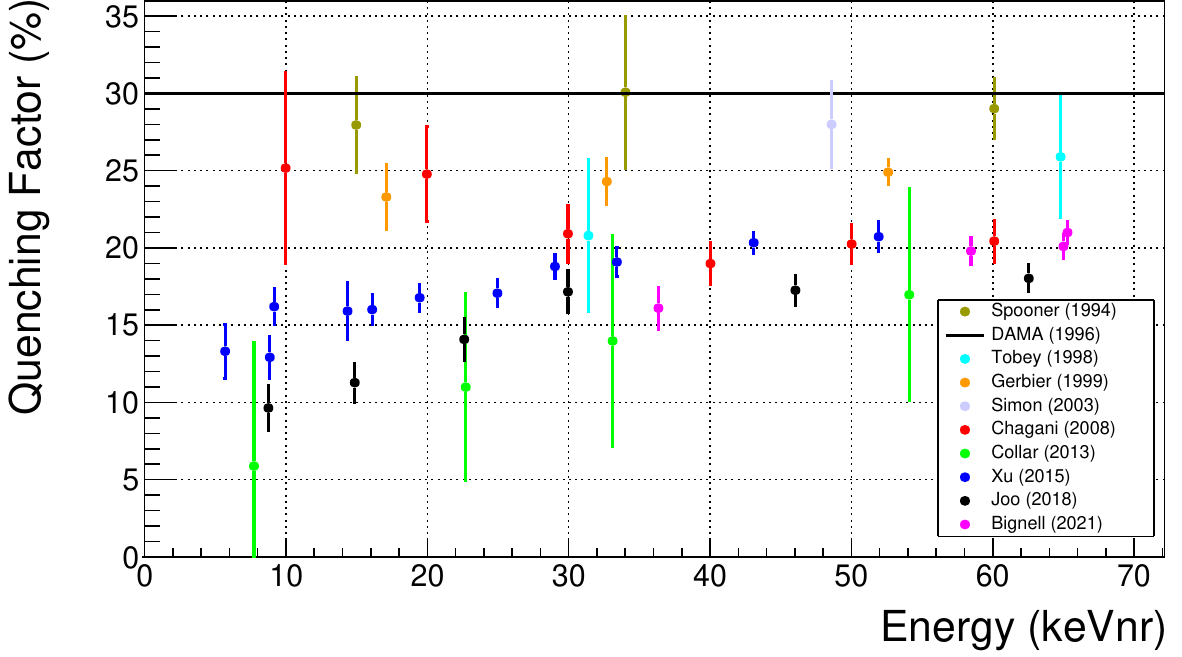}
		\caption{\label{NaQF_measurements}Quenching factor measurements for sodium nuclei in NaI(Tl) crystals~\cite{Spooner,Bernabei:1996vj,Tovey:1998ex,Gerbier:1998dm,Simon:2002cw,Chagani,Collar,Xu,Joo:2018hom,Bignell_2021}.}
	\end{center}
\end{figure}

\begin{figure}[ht]
	\begin{center}
		\includegraphics[width=\textwidth]{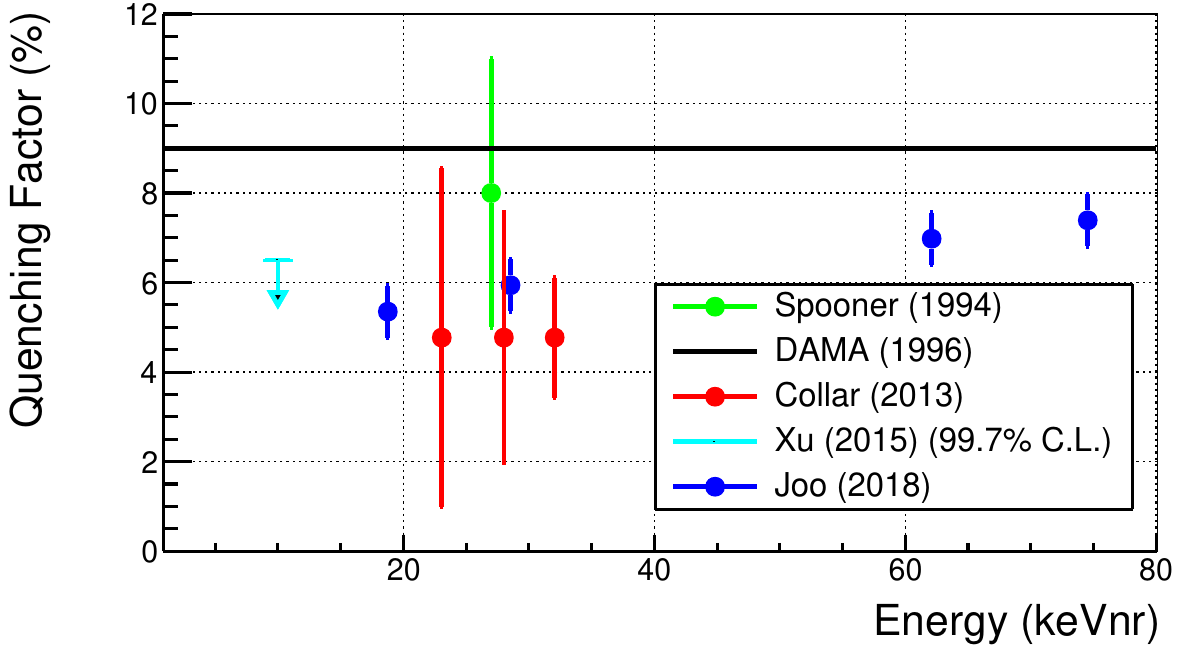}
		\caption{\label{IQF_measurements}Quenching factor measurements for iodine nuclei in NaI(Tl) crystals~\cite{Spooner,Bernabei:1996vj,Collar,Xu,Joo:2018hom}. Cyan arrow is the upper limit at 99.7\% C.L. obtained in~\cite{Xu}.}
	\end{center}
\end{figure}

Some experiments have been carried out since the 90s to measure the sodium an iodine QFs in NaI(Tl). The principle followed in most of the measurements is to induce nuclear recoils of a known energy using quasi-monoenergetic neutrons elastically scattered in the crystal at a known angle. This is done by detecting the scattered neutron in an array of detectors surrounding the crystal. The energy deposited in the crystal is calibrated using electronic recoils, and then, the QF can be calculated as the ratio of the measured to the deposited energy as in Equation~\ref{eq:QFDefinition}. The results of some sodium QF measurements are shown in Figure~\ref{NaQF_measurements}, while in Figure~\ref{IQF_measurements} the iodine QF measurements are shown. In Section~\ref{Section:QF_NaIQFOverview} we will review the experimental methods followed in each measurement.

Most recent measurements of the sodium QF agree in a decrease of this factor at low nuclear recoil energies, but there is some disagreement in this energy dependence among the different measurements. It can be due to different systematics affecting each experiment, since the experimental set-ups and the data analysis are different, but it is also possible that its value is different for crystals with different characteristics. Concerning the iodine QF, as the nucleus is much heavier than sodium, the energy transferred in elastic scattering by a neutron of the same energy is much lower and therefore the recoil spectrum is concentrated at lower energies. This results in experimental difficulties to determine the iodine QF in NaI(Tl) and explains the lack of results. In some cases only upper limits are obtained.

\begin{figure}[ht]
	\begin{center}
		\includegraphics[width=\textwidth]{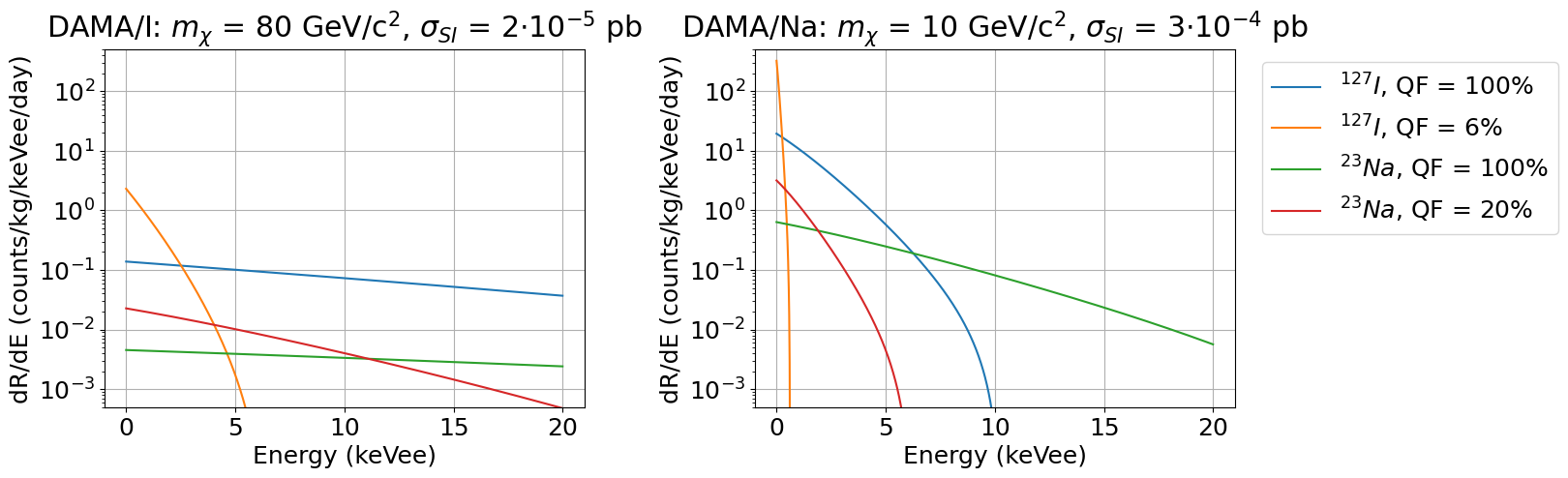}
		\caption{\label{RecoilSpectrum_Eee}Nuclear recoil spectrum produced by WIMPs as a function of the electron equivalent energy for the $Na$ and $I$ nuclei considering QFs of 100\% and typical values found in bibliography in NaI(Tl) detectors (20\% and 6\%, respectively). It has been plotted considering the DAMA/LIBRA positive signals in the [2,6]~keV range shown in Figure~\ref{ExclusionPlot}: DAMA/I (left plot, $m_{\chi}=80$~GeV/c$^2$ and $\sigma_{SI}=2\cdot10^{-5}$~pb) and DAMA/Na (right plot, $m_{\chi}=10$~GeV/c$^2$ and $\sigma_{SI}=3\cdot10^{-4}$~pb).}
	\end{center}
\end{figure}

As it has already been emphasized, a small quenching factor (QF) has significant implications in the search for dark matter. Figure~\ref{RecoilSpectrum_Eee} shows the nuclear recoil spectrum produced by WIMPs as a function of the electron equivalent energy for the $Na$ and $I$ nuclei for QFs of 100\% and typical values found in the bibliography in NaI(Tl) detectors (20\% and 6\%, respectively). It has been plotted considering the DAMA/LIBRA positive signals in the [2,6]~keV range shown in Figure~\ref{ExclusionPlot}: DAMA/I ($m_{\chi}=80$~GeV/c$^2$ and $\sigma_{SI}=2\cdot10^{-5}$~pb) and DAMA/Na ($m_{\chi}=10$~GeV/c$^2$ and $\sigma_{SI}=3\cdot10^{-4}$~pb). It is clear that the electron equivalent energy where the WIMP signal is expected is much below the energy deposited in the form of a nuclear recoil, and that the parameter space region that is reached by the experiments that use NaI(Tl) crystals can be significantly altered by different values of the sodium and iodine QFs. Therefore, a precise measurement of the QF in NaI(Tl) crystals is necessary to accurately compare results from NaI(Tl) experiments among themselves and with those from other targets.

To measure the QF of the NaI(Tl) crystals, a collaborative effort between members of COSINE-100, COHERENT, and ANAIS-112 started in 2018. The crystal dependence of the QF and the systematics affecting the measurements were evaluated by measuring different crystals in the same set-up at the Triangle Universities Nuclear Laboratory (TUNL) and applying the same analysis. These measurements and the analysis developed were an important part of my job during my PhD, and they will be discussed in Chapter~\ref{Chapter:QF}.

\subsection{PMTs for scintillation detection} \label{Section:Intro_Scintillators_PMTs}

\begin{figure}[h!]
	\begin{center}
		\includegraphics[width=\textwidth]{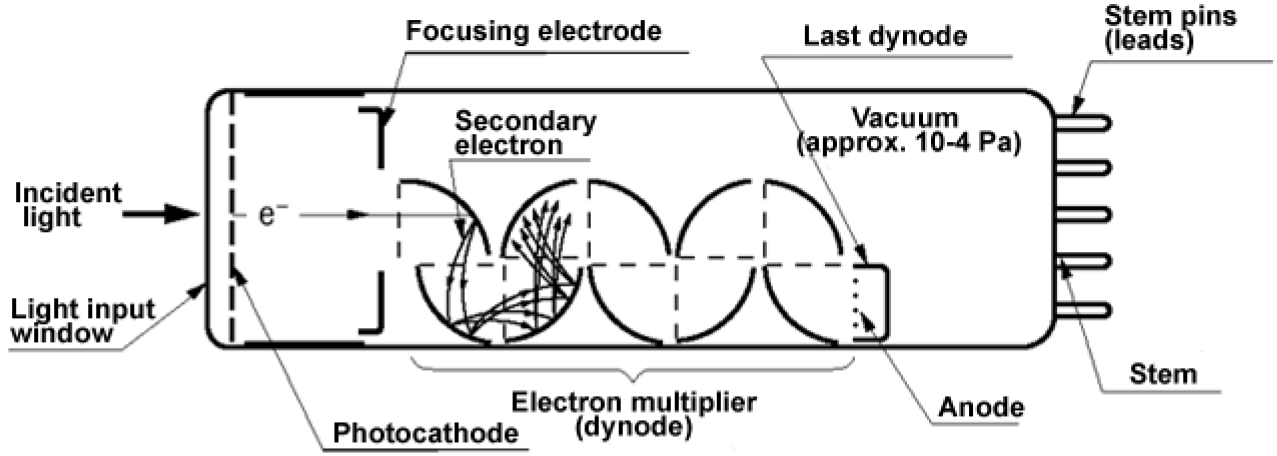}
		\caption{\label{SchemePMT}Photomultiplier tube (PMT) scheme. Image from~\cite{Hamamatsu}.}
	\end{center}
\end{figure}

The scintillation light is commonly detected using Photomultiplier tubes (PMTs), which are highly sensitive devices for detecting light in the ultraviolet, visible, and near-infrared ranges. A PMT consists of a tube with vacuum inside and with an optical window made of quartz or borosilicate, where the photosensitive material, called photocathode, is deposited in a very thin layer, a series of electron multipliers (also known as dynodes) and an anode~\cite{Leo:1987kd,Knoll:2000fj}. When a photon reaches the photocathode, it causes the emission of a so-called photoelectron with a probability given by the quantum efficiency. Photoelectrons are then accelerated by a high voltage applied to the device, which focus them onto the first dynode. In each dynode the incident electrons convert the kinetic energy acquired by the acceleration in the electric field into a number of secondary electrons. This process is repeated through the dynode string, creating an electron cascade. Finally, the electrons emitted by the last dynode are collected at the anode. An scheme of a typical PMT is shown in Figure~\ref{SchemePMT}~\cite{Hamamatsu}.

To ensure optimal operation, a PMT must be shielded from external light and protected from electromagnetic interference, and must be equipped with a suitable signal processing chain. The choice of the photocathode material, dynode configuration, and other design factors can affect the sensitivity, resolution, and overall performance of the PMT. The most relevant parameters of a PMT are:

\begin{itemize}
	\item \textbf{Quantum efficiency}: Probability of generating a photoelectron from an incident photon. With typical values from 10\% to 50\%, it depends on the specific photocathode and the wavelength of the incident light.
	\item \textbf{Gain:} It is a measure of the amplification of the electrical signal produced by the device, typically expressed as the number of collected electrons for each photoelectron produced. It can range from around 10$^3$ to over 10$^6$, depending on the PMT and the operating conditions. It is strongly dependent on the HV applied, so it is important to maintain a stable supply in order to ensure its stability.
	\item \textbf{Temporal response:} The rise (fall) time of the photoelectron signal is the time it takes for the current to change from 10\% to 90\% (from 90\% to 10\%) of its maximum amplitude. This response is related with the dispersion in the arrival times of the electrons to the anode, and it is strongly dependent on the focusing capacity of the system. 
	\item \textbf{Transit time:} is the time interval between the arrival of light at the photocathode and the appearance of the resulting electrical pulse at the anode.
	\item \textbf{Dark  current:}  It is the rate of dark events produced without light excitation of the photocathode. It is mainly related with thermal excitations of electrons in the photocathode, that result in random emission of photoelectrons that will be amplified by the dynode chain. However, there can be contributions from any other process able to release an electron from the photocathode. This explains that for instance the PMT dark current is lower in low radioactive background environment.
	\item \textbf{Single electron response (SER):} It is the output signal produced by a single photoelectron.
\end{itemize}

These devices have been for long time the most efficient light detectors. They are highly sensitive and can detect very low levels of light, at single photon level. They also have a fast response in time, allowing them to measure short duration light pulses. They present a good linear response and a wide dynamic range, something useful for experiments involving the detection of both weak and strong light signals. Another advantage of PMTs is their long lifetime, they can last for tens of years, making them a reliable choice for long-term experiments. All of these reasons make them a valuable tool for a wide range of applications.

However, they also have some disadvantages that may make them less suitable for scintillation detection in dark matter search experiments. They are generally large and quite massive, and they consist of many different material components. All of this is a disadvantage in experiments that require low background, because the larger the mass and number of components, the more difficult is to achieve a high radiopurity. In addition, usually they consist of transparent media, as quartz or borosilicate, where energetic charged particles can produce Cherenkov light, being a source of background light worrisome for some applications where low light signals are searched for.

Moreover, direct light emission from them has been observed, for example by Double Chooz collaboration~\cite{DoubleChooz:2016ibm}. It is a reactor neutrino experiment that uses 390 low-background PMTs with diameters of 25~cm. During the commissioning of this experiment, a relevant background found was produced by the emission of light inside the optical volumes of the PMTs. A study and characterization of these events allowed to identify their origin: the transparent epoxy resin covering the electric components of the PMT base emits light due to the combined effect of high voltages and heat. Such kind of light emission would interfere with the scintillation detection resulting in backgrounds difficult to cope with. This will be overviewed more in detail in next chapters.

\subsection{Annual modulation search with NaI(Tl) crystals} \label{Section:Intro_Scintillators_DAMA}

As it has been explained in Section~\ref{Section:Intro_Detection_Direct_Experiments}, the DAMA/LIBRA experiment (located in the LNGS, in Italy), uses highly radiopure NaI(Tl) crystal detectors to search for WIMPs. The setup consists of 25~parallelepipedic crystals produced by Saint-Gobain company~\cite{SaintGobain}, of 9.7~kg mass each in a 5$\times$5~matrix configuration. These crystals are coupled to two PMTs on each side through 10~cm Suprasil-B light guides and surrounded by low-radioactive OFHC (Oxygen Free High Conductivity) freshly electrolyzed copper shields. The detectors are enclosed in an anti-radon box made of OFHC copper which is continuously flushed with high-purity nitrogen gas. Moreover, it has a passive shield made of copper, low-radioactive lead, cadmium foil, polyethylene and concrete.

\begin{figure}[h!]
	\begin{center}
		\includegraphics[width=\textwidth]{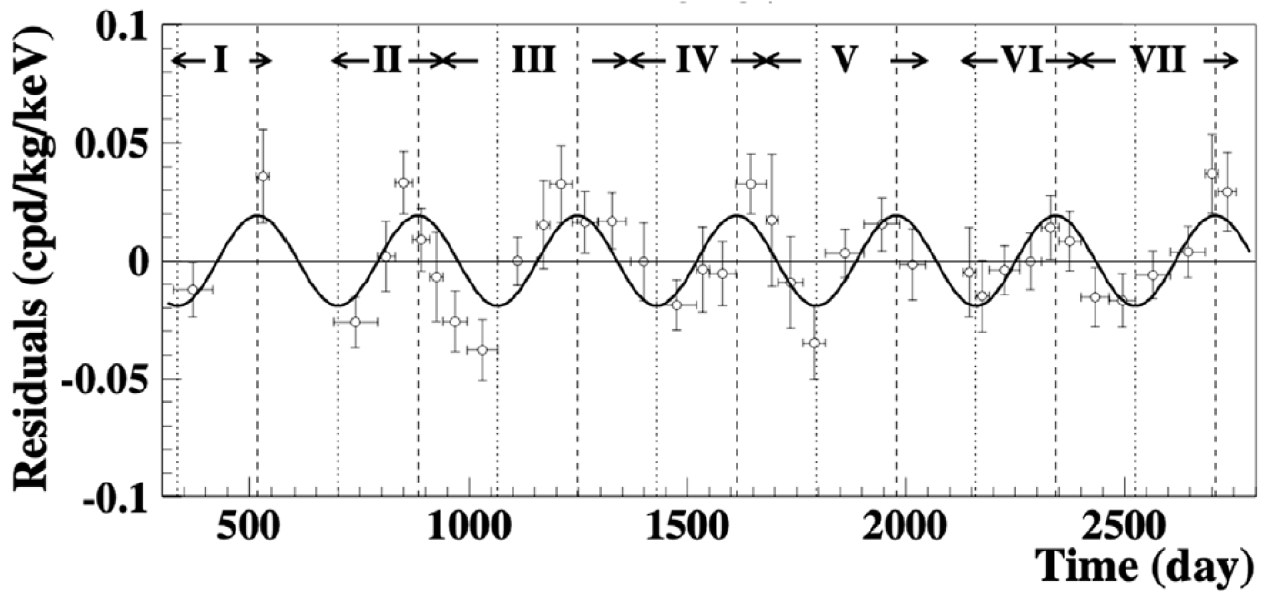}
		\caption{\label{Modulation1}Experimental residual rate of the single-hit scintillation events measured in the [2,6]~keVee energy region by DAMA/NaI. Image from~\cite{Bernabei:2020mon}.}
	\end{center}
\end{figure}

\begin{figure}[h!]
	\begin{center}
		\includegraphics[width=\textwidth]{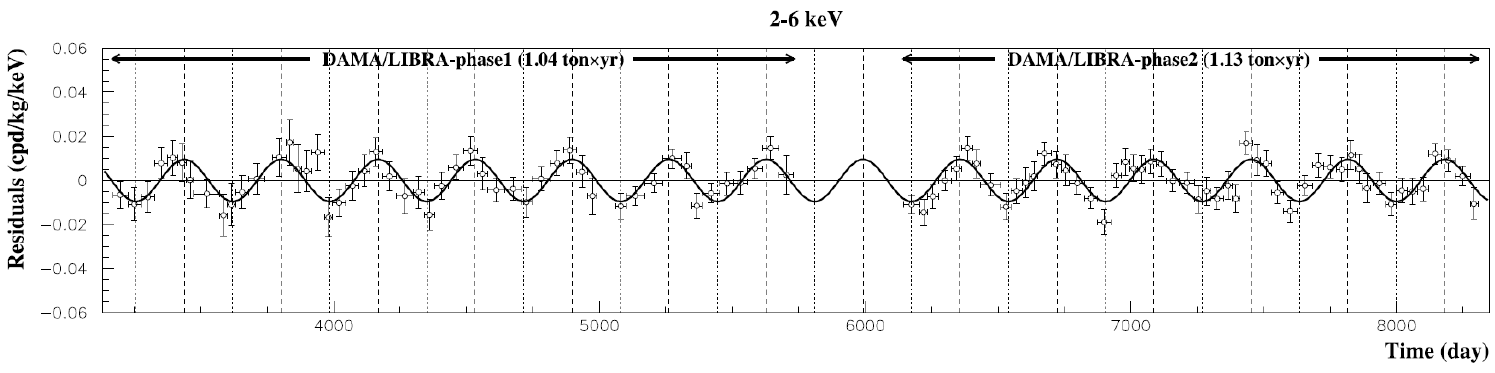}
		\caption{\label{Modulation2}Experimental residual rate of the single-hit scintillation events measured in the [2,6]~keVee energy region by DAMA/LIBRA. Image from~\cite{Bernabei:2020mon}.}
	\end{center}
\end{figure}

\begin{figure}[h!]
	\begin{center}
		\includegraphics[width=\textwidth]{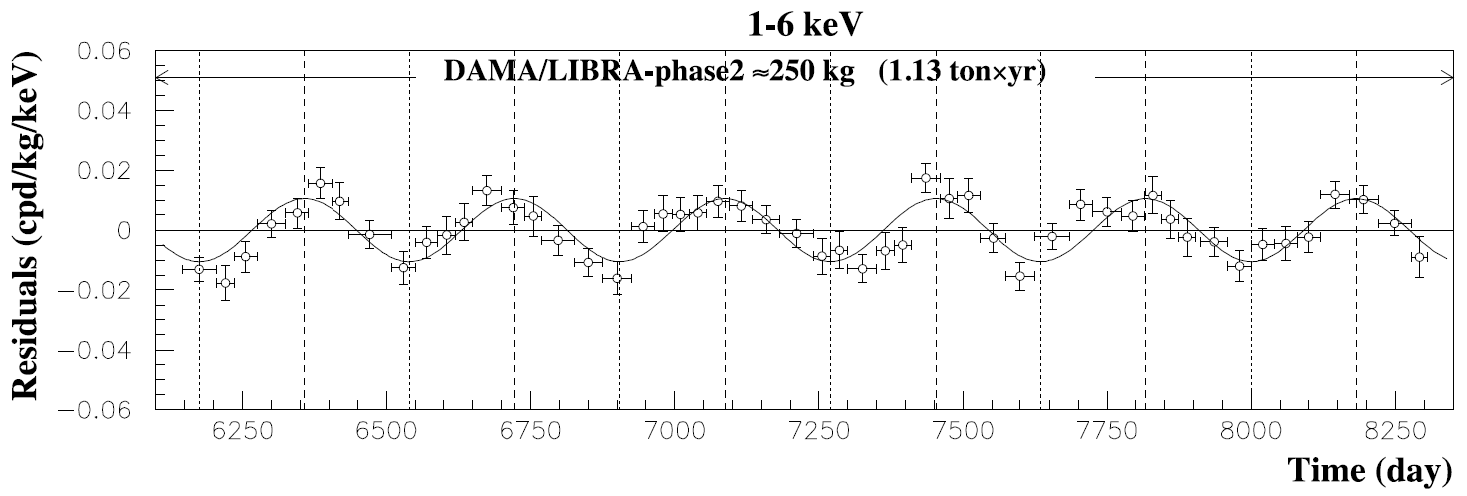}
		\caption{\label{Modulation3}Experimental residual rate of the single-hit scintillation events measured in the [1,6]~keVee energy region by DAMA/LIBRA-phase2. Image from~\cite{Bernabei:2020mon}.}
	\end{center}
\end{figure}

DAMA has been reporting from the end of the 1990's a positive result of an annual modulation in their data, compatible with the signal expected for WIMPs in the Standard Halo Model~\cite{Bernabei:1996vj,Bernabei:1998fta,Bernabei:2003za,Bernabei:2004lli,Bernabei:2013xsa,Bernabei:2018yyw,Bernabei:2020mon}. Figure~\ref{Modulation1} shows the residual rate of the single-hit scintillation events measured in the [2,6]~keVee energy region in the first set-up of DAMA, known as "DAMA/NaI" (1995 - 2002)~\cite{Bernabei:2003za}. Figure~\ref{Modulation2} and~\ref{Modulation3} show this rate in the [2,6]~keVee and [1,6]~keVee energy ranges, respectively, in DAMA/LIBRA (2003 - 2010)~\cite{Bernabei:2013xsa} and DAMA/LIBRA-phase2 (2011 - present)~\cite{Bernabei:2018yyw}. These residual rates have been calculated by subtracting the yearly averaged rate. The black curves in the figures correspond to the function 
\begin{equation}\label{eq:modDAMA}
	S(t) = S_m \cos{\left(\frac{2\pi}{T}(t-t_0)\right)},
\end{equation}
fitted to the data with free $S_m$ and the period $T$ and the phase $t_0$ fixed at 1~year and the 2$^{nd}$ of June, respectively. The combined fit of all data to Equation~\ref{eq:modDAMA} provides as result modulation amplitudes $S_m$ = 10.3 $\pm$ 0.8~c/keVee/ton/day at 12.9$\sigma$ C.L. in the [2,6]~keVee range and 10.6 $\pm$ 1.1~c/keVee/ton/day at 9.6$\sigma$ C.L. in the [1,6]~keVee range.

The DAMA collaboration concludes that this effect cannot be explained by any known systematic effects or backgrounds and that it is likely to be a signal of dark matter~\cite{DAMA:2008jlt,DAMA:2010gpn,Bernabei:2013xsa,Bernabei:2018jrt}. The signal in the [2-6]~keVee can be interpreted in terms of a standard WIMP
(see Figure~\ref{ExclusionPlot} for SI). However, as it can be appreciated also in that figure, there is large number of experiments that have excluded the space region singled out by DAMA. This has produced a very wide debate about whether the signal comes from dark matter or just some unaccounted for systematic effect.

As the interpretation of this signal in the context of dark matter is highly dependent on the specific model being used, a model-independent confirmation is essential in order to fully understand the implications of the DAMA/LIBRA results. Currently, the COSINE-100 and ANAIS-112 experiments are using the same target material and detection technique as DAMA/NaI and DAMA/LIBRA in an attempt to clarify this puzzle. The ANAIS-112 experimental setup and latest results will be extensively covered in Chapter~\ref{Chapter:ANAIS}.

Concerning the COSINE-100 experiment, it is located at the Y2L in South Korea~\cite{COSINE-100:2019lgn}. It uses a total mass of 106~kg of low-background NaI(Tl) crystals produced by Alpha Spectra Inc.~\cite{AlphaSpectra} in a 4$\times$2~matrix configuration. They are hermetically housed in 1.5~mm-thick OFC (Oxygen-Free Copper) tubes and attached with quartz windows to two Hamamatsu PMTs. These encapsulations have a calibration window with either a reduced copper thickness of 0.5~mm or a 0.13~mm-thick Mylar film (depending on the detector) to allow for low energy calibrations. The matrix is housed in an acrylic box, which is immersed in a wall reflecting tank of 2200~L filled with a liquid scintillator, which acts as a veto for the $^{40}K$ emissions from the crystals. This scintillation is also detected with Hamamatsu PMTs. The tank is surrounded by a OFC box, lead and by panels made of plastic scintillator for muon veto. It started the data taking on the 30th September, 2016.

The analysis of the annual modulation from 3~years of data was presented in~\cite{COSINE-100:2021zqh}. It corresponds to a total exposure of 173~kg$\times$yr, for an effective mass of 61.3~kg.  The best fit for the modulation amplitude of the single-hit events when the period and phase are fixed to the one expected for the Standard Halo Model are $S_m$ = 5.1 $\pm$ 4.7~c/keVee/ton/day for the [2,6]~keVee range and 6.7 $\pm$ 4.2~c/keVee/ton/day for the [1,6]~keVee range, consistent with both the null hypothesis and the DAMA/LIBRA’s best fit value at $\sim$1$\sigma$ C.L. Therefore, more data are needed to confirm or refute the DAMA/LIBRA signal. The experiment was decommissioned in early 2023 and now they are setting up COSINE-200~\cite{COSINE:2020egt}, with a larger mass of more radiopure NaI(Tl) crystals.

\begin{figure}[h!]
	\begin{center}
		\includegraphics[width=0.61\textwidth]{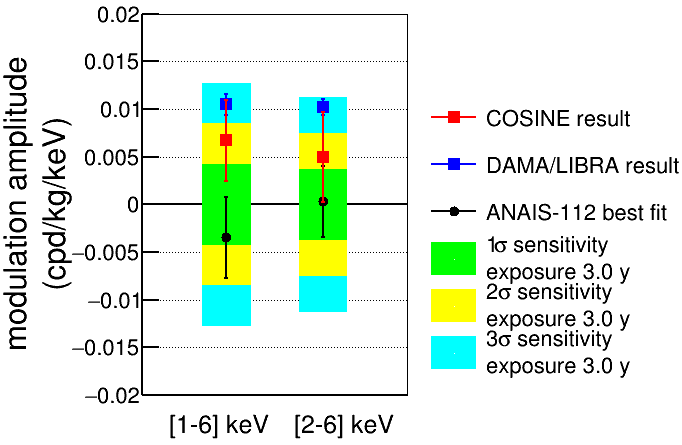}
		\caption{\label{ANAIS_res}Comparison between results on annual modulation from DAMA/LIBRA~\cite{Bernabei:2020mon} (blue), COSINE-100 three-year exposure~\cite{COSINE-100:2021zqh} (red) and ANAIS-112 three-year exposure~\cite{Amare:2021yyu} (black). Estimated sensitivity of ANAIS-112 is shown at different C.L. as the colored bands.}
	\end{center}
\end{figure}

The current status of the annual modulation analysis is presented in Figure~\ref{ANAIS_res}, using the first three years of ANAIS-112 data and COSINE-100 data. ANAIS-112 has reached an effective exposure of 313.95 kg$\times$yr, and has obtained for the best fit in the [1–6] keV ([2–6] keV) energy region a modulation amplitude of -3.4 $\pm$ 4.2~c/keVee/ton/day (0.3 $\pm$ 3.7~c/keVee/ton/day). Therefore, it supports the absence of modulation at 3.3 (2.6)$\sigma$, for a sensitivity of 2.5 (2.7)$\sigma$.

%% file: ANAIS-112-New2.tex
\chapter{ANAIS-112 experiment}\label{Chapter:ANAIS}

\fancyhead[LE]{\emph{Chapter \thechapter. \nameref{Chapter:ANAIS}}}

The ANAIS-112 (Annual modulation with NaI Scintillators) experiment~\cite{Amare:2018sxx} sear-ches for dark matter annual modulation with ultra-pure NaI(Tl) scintillators at the Canfranc Underground Laboratory (LSC) in Spain. It consists of nine 12.5 kg NaI(Tl) modules in a 3$\times$3 matrix configuration (with a total mass of 112.5~kg) and it started taking data in "dark matter search" mode on the 3$^{rd}$ of August 2017.

The ANAIS-112 experimental setup is described in Section~\ref{Section:ANAIS_Setup}. The performance of the ANAIS-112 modules is summarized in Section~\ref{Section:ANAIS_Module}. The procedure designed for the analysis of the data collected by the experiment is described in Section~\ref{Section:ANAIS_Analysis}, and the method to calibrate the response of the detectors to low-energy depositions in Section~\ref{Section:ANAIS_Calibration}. To achieve a low energy threshold and low background in the experiment, it is important to reject the non bulk-scintillation events, which dominate the experiment’s rate below 10 keV. To do that, an event filtering process has to be applied to the raw data. The procedure followed for filtering those events is explained in Section~\ref{Section:ANAIS_Filter}, as well as the efficiencies of the selections applied. Finally, the background spectrum was modeled as presented in Section~\ref{Section:ANAIS_Background} using the GEANT4 package~\cite{GEANT4:2002zbu}, a simulation tool that allows for detailed tracking of particles and their interactions with matter. Because the measured background presents an excess with respect to the estimates of this background model, which has been reduced but not completely eliminated by applying machine learning techniques, other strategies trying to understand this excess and improve both the threshold and the background in the ROI are being followed (see section~\ref{Section:ANAIS_Improves}).

\section{Experimental setup} \label{Section:ANAIS_Setup}
\fancyhead[RO]{\emph{\thesection. \nameref{Section:ANAIS_Setup}}}

\begin{figure}[h!]
	\begin{center}
		\includegraphics[width=0.75\textwidth]{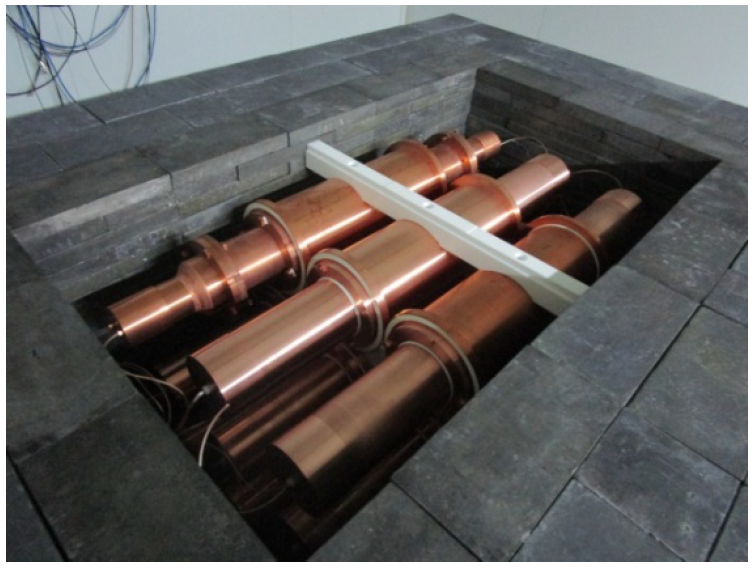}
		\caption{\label{ModulesANAIS}Picture of the ANAIS-112 modules inside the shielding.}
	\end{center}
\end{figure}

The ANAIS-112 experiment consists of nine modules in a 3$\times$3~matrix, each weighing 12.5~kg, thus giving a total mass of 112.5~kg. They are made of highly pure NaI(Tl) crystals, which were manufactured by Alpha Spectra in Colorado, US~\cite{AlphaSpectra}, and have a cylindrical shape with a length of 11.75" and a diameter of 4.75", coated with teflon film. The two sides of the crystals are optically coupled to two 3~mm-thick layer of silicone and 1~cm-thick synthetic quartz windows. These windows are coupled to PMTs with optical gel. The crystals are encapsulated in 1.5~mm-thick Oxygen Free Electronic (OFE) copper, which has a Mylar window with a diameter of 1~cm in the lateral face, allowing for low-energy calibration with external sources. Figure~\ref{ModulesANAIS} shows a picture of the modules in the ANAIS-112 setup.

\begin{figure}[h!]
	\begin{center}
		\includegraphics[width=0.75\textwidth]{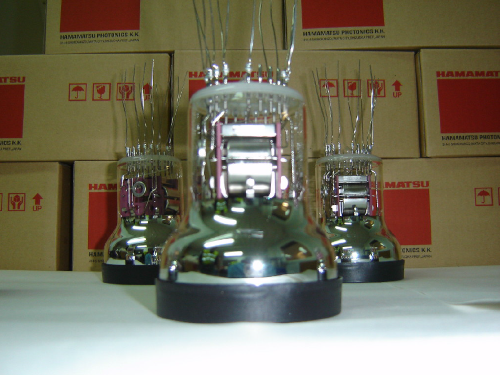}
		\caption{\label{PMTsANAIS}Hamamatsu R12669SEL2 photomultipliers.}
	\end{center}
\end{figure}

The ANAIS-112 photomultipliers are Hamamatsu R12669SEL2. They were selected for their appropriate characteristics. This model has a bialkali photocathode, which provides a high quantum efficiency (higher that 33\%), with a peak response at 420~nm. Additionally, its dark current is below 500~Hz, and it has ten dynode stages for enhanced signal amplification. At the nominal voltage value, the model exhibits a gain factor of approximately 10$^6$. Moreover, they are similar to the PMTs used in the DAMA/LIBRA-phase2 experiment~\cite{Bernabei:2020mon,Bernabei:2012zzb}, and they were also later selected by the KIMS/COSINE collaboration for use in their experiments~\cite{Adhikari:2017esn,Kim:2014toa}. An image of these PMTs is shown in Figure~\ref{PMTsANAIS}.

\begin{figure}[h!]
	\begin{center}
		\includegraphics[width=0.75\textwidth]{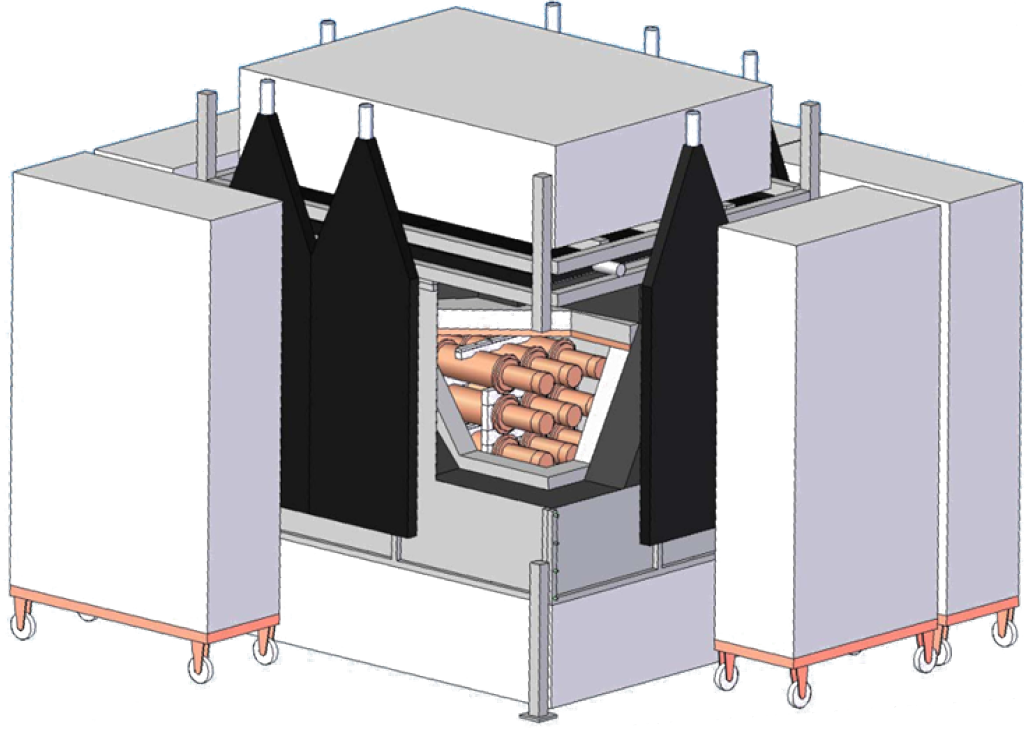}
		\caption{\label{SetupANAIS}Artistic view of the ANAIS-112 experimental setup.}
	\end{center}
\end{figure}

The ANAIS-112 shielding consists of 10~cm of archaeological lead, 20~cm of low activity lead, an anti-radon box flushed with radon-free nitrogen gas, an active muon veto system with 16 plastic scintillators covering the top and sides, and a 40~cm neutron moderator consisting of water tanks and polyethylene blocks. The experiment is housed in hall~B of LSC under 2450~m.w.e. rock overburden. The complete shield of the ANAIS-112 experiment is sketched in Figure~\ref{SetupANAIS}.

\subsection{Electronics and DAQ system} \label{Section:ANAIS_DAQ}

Custom-made preamplifiers are used for the initial amplification of each PMT signal in the experiment. They are located close to the PMTs to minimize the cable length and optimize signal-to-noise ratio, while the rest of the electronics is in an air-conditioned cabinet where the temperature is controlled. The PMT high voltage supply is provided by a CAEN~SY2527 Universal Multichannel Power Supply, which allows continuous monitoring of individual voltages.

\begin{figure}[h!]
	\begin{center}
		\includegraphics[width=\textwidth]{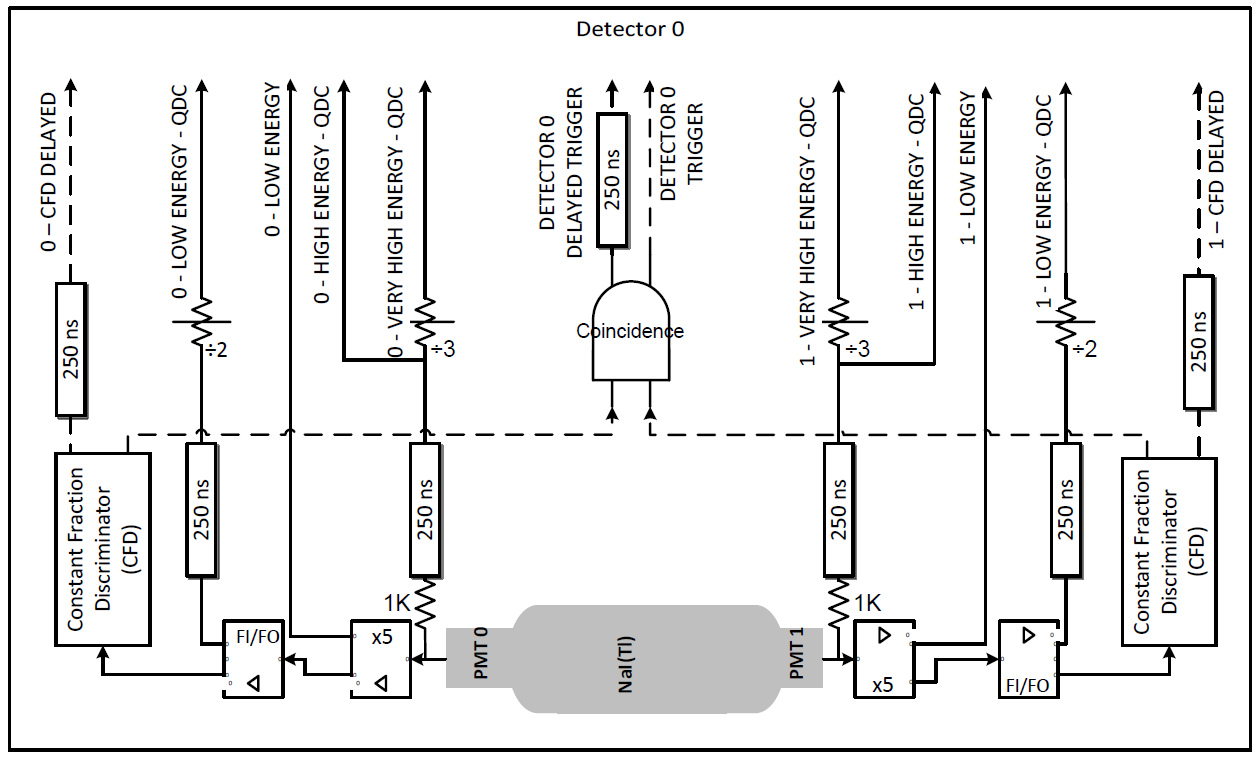}
		\caption{\label{AnaisDAQ2}Scheme of the electronic signal processing chain and trigger of a single ANAIS-112 module. Image from~\cite{Amare:2018sxx}.}
	\end{center}
\end{figure}

The signal of each PMT is processed separately, as it is presented in Figure~\ref{AnaisDAQ2}. It is divided into a trigger signal and low energy~(LE) and high energy~(HE) signals. The first two are preamplified, while the last one is acquired without any amplification. The PMT trigger is carried out by a CAEN~N843 Constant Fraction Discriminator~(CFD) with a threshold below the level of a single photoelectron. Then, the trigger of a given module is produced by the coincidence of the two PMT trigger signals in a 200~ns window. While the HE signals are attenuated and sent to CAEN~V792 Charge to Digital Converter modules with a 1~$\mu$s integration window and 12-bit resolution, the LE signals are directed to MATACQ-CAEN~V1729A digitizers. These digitizers record 2520~samples and have the option of sampling at 1 or 2~GHz. In ANAIS it is configured at 2~GHz, giving a time acquisition window of 1260~ns. They have a dynamic range of 1~V with a resolution of 14~bits and a dead time per event of the order of 4~ms.

\begin{figure}[h!]
	\begin{center}
		\includegraphics[width=\textwidth]{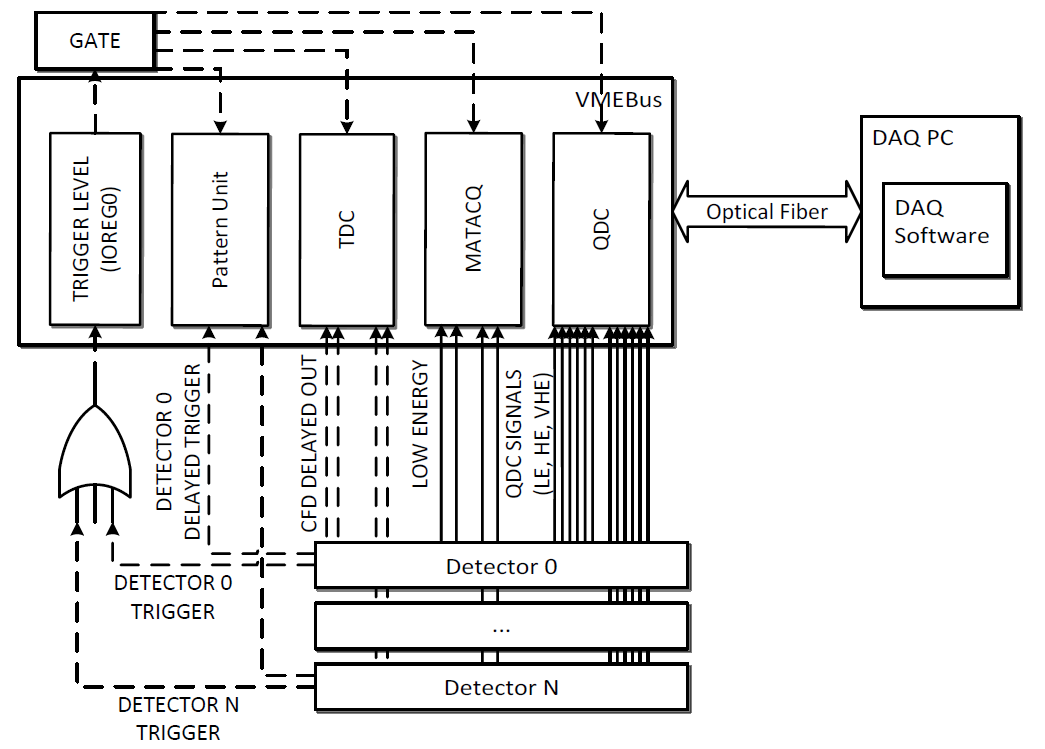}
		\caption{\label{AnaisDAQ3}Scheme of the main acquisition trigger. Image from~\cite{Amare:2018sxx}.}
	\end{center}
\end{figure}

The main acquisition trigger (Figure~\ref{AnaisDAQ3}) combines the individual triggers of the nine modules using a logical OR operation. This global trigger signal is sent to a CAEN~V977~I/O Register~(IOREG0) that opens the gate for all VME boards and generates the VME interrupt request that triggers the acquisition software. The register changes to BUSY state until it is reset via VME command by the DAQ program, allowing a measurement of the acquisition dead-time via a latched scaler. Moreover, a Pattern Unit module records which detectors have triggered, allowing the DAQ system to acquire only the relevant data, which reduces the dead time and saves space in the disk.

The DAQ software stores in a ROOT file~\cite{Brun:1997pa} the timestamp of the event, the triggering pattern (identification of modules that triggered), and the signals of the two PMTs of the modules that triggered (HE signal from the QCDs and the waveforms of both PMTs LE signals from the MATACQ digitizers). This output ROOT file is later analyzed offline to extract the relevant information of the event, as explained in Section~\ref{Section:ANAIS_Analysis}.

\subsection{Muon veto system} \label{Section:ANAIS_Setup_Muon}

The muon flux at LSC~Hall~A has been measured as~\cite{Trzaska:2019kuk}
\begin{equation*}
	\Phi = (5.26 \pm 0.21)\times10^{-3}m^{-2}s^{-1},
\end{equation*}
which is four orders of magnitude lower than the surface muon flux, but significant enough to have an impact on ANAIS-112 detectors.

Although direct muon interactions in NaI(Tl) detectors are expected to contribute at very high energy and therefore well above the region of interest (ROI) for dark matter modulation analysis, muons can also interact in other components of the experimental setup and produce different kind of events that could fall in the ROI of ANAIS-112 and endanger the fulfillment of the experiment's goals. Muons can for instance produce neutrons in the ANAIS-112 shielding, but also they can interact in the glass of the PMTs, quartz windows and other components and produce there Cherenkov light or other weak scintillation. Moreover, following the large energy deposition from muons in the crystals and because of the long phosphorescence time of 0.15~s measured in NaI(Tl)~\cite{TesisClara}, photons arriving in the tail of a muon pulse are able to produce multiple triggers in ANAIS DAQ~\cite{Amare:2018sxx}.

\begin{figure}[h!]
	\begin{center}
		\includegraphics[width=\textwidth]{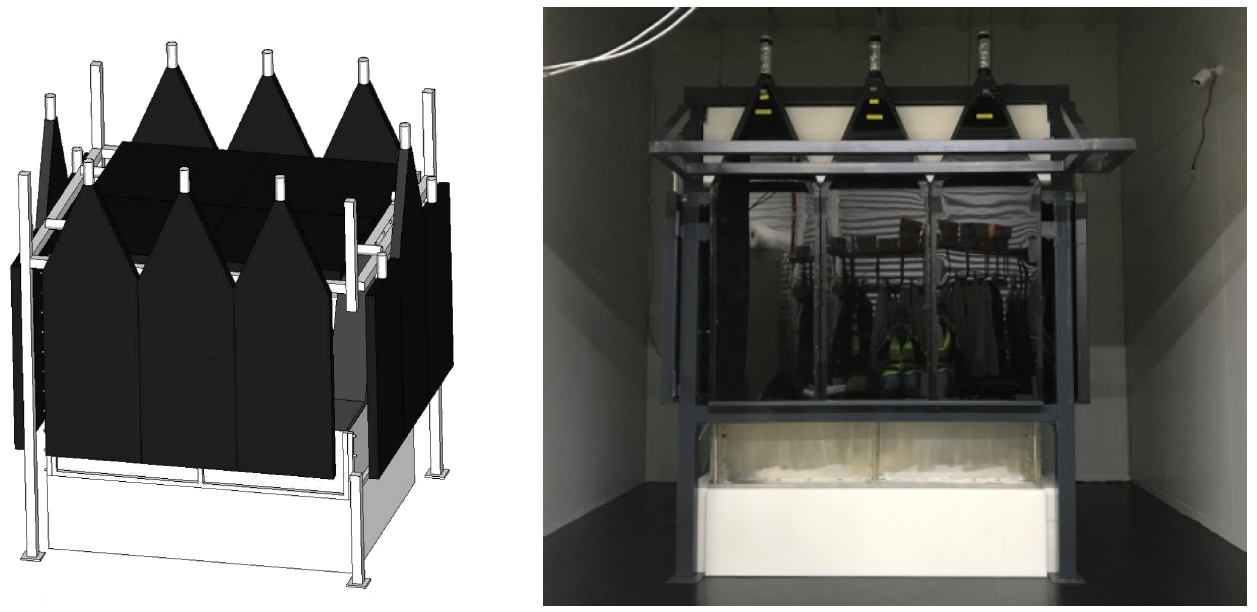}
		\caption{\label{MuonVetoSystem}Artistic view (left) and picture (right) of the ANAIS-112 veto system.}
	\end{center}
\end{figure}

That is why the veto scintillator system in the ANAIS-112 experiment plays a crucial role in data selection and annual modulation analysis, as it allows to relate muon events in the plastic scintillators with energy depositions in the NaI(Tl) crystals. It consists of sixteen 5~cm-thick plastic scintillators arranged on the top and four sides of the anti-radon box. Half of them were previously used in the IGEX neutrinoless double-beta decay experiment~\cite{IGEX:2002bce}, while the other half were specifically purchased to Scionix~\cite{Scionix} for use in the ANAIS-112 setup. The positions of the scintillators on the different faces of the setup can be seen in Figure~\ref{MuonVetoSystem}.

The system follows a tagging strategy, aiming at measuring the residual muon flux in the LSC, but it was designed to study the correlations between muon hits in the plastic scintillators and events in the crystals. This allows to reject offline those events in the NaI(Tl) crystals occurring after the passage of a muon through the ANAIS-112 set-up. The trigger rate of the veto system is high (between 5 and 6 Hz), and it is mostly dominated by background events in the plastic scintillators. Nevertheless, the strong particle discrimination capabilities of these scintillators allow us to remove background events using offline pulse shape analysis (PSA) techniques.

The muon veto system has a dedicated DAQ independent from that of the ANAIS-112 modules. However, both DAQ systems, although running in parallel in different computers, are synchronized while data taking. The muon veto front-end electronics processes the signals from the 16~plastic scintillators, which are integrated in two different time windows (corresponding to the total pulse and to the tail of the pulse). The veto DAQ software stores the time of each trigger, the pattern of triggering, the total pulse area and tail pulse area for each event. The selection parameters and operating conditions of the plastic scintillators were carefully chosen to maximize the acceptance of real muon events and the rejection of background events.

\subsection{Blank module} \label{Section:ANAIS_Setup_Blank}

\begin{figure}[h!]
	\begin{center}
		\includegraphics[width=0.75\textwidth]{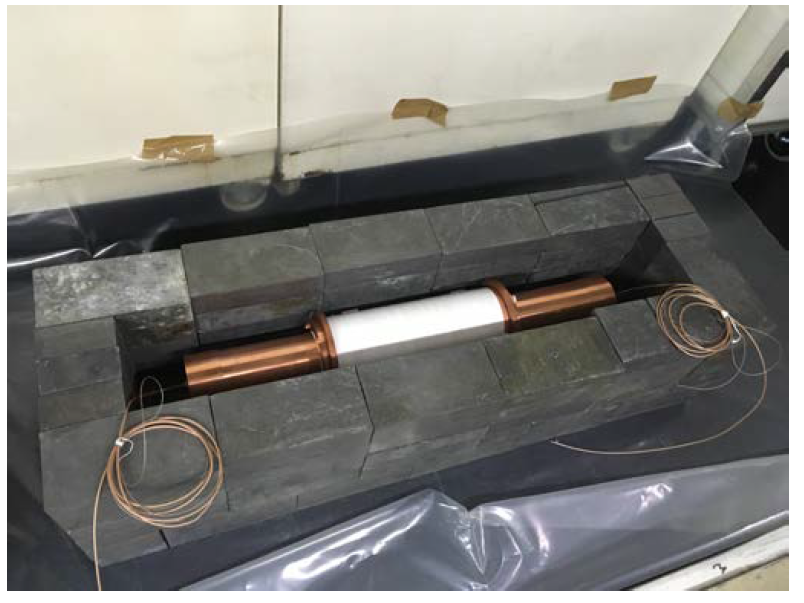}
		\caption{\label{Blank}Picture of the setup of the ANAIS Blank module.}
	\end{center}
\end{figure}

In order to study non-bulk scintillation events that contribute to the background in the ANAIS-112 experiment, a Blank module (whose picture can be seen in Figure~\ref{Blank}) was installed inside the ANAIS hut at LSC~Hall~B in August~2018.

This module is similar to the nine modules of ANAIS-112, but it does not contain a NaI(Tl) crystal. The two PMTs, are coupled to two quartz optical windows inside a copper housing, lined with Teflon film as diffusor material. The inner volume is not tightly closed, and then, shares properties on humidity and composition with the surrounding atmosphere. The module is contained within a lead shielding which can be flushed with gas. Radon free air, for instance, has been used for most of the operation period, but at the beginning, normal laboratory air was inside and around the blank module (see Section~\ref{Section:SIM_Res_Blank}). The PMT signals are processed similarly to that of the ANAIS-112 modules, by using the same preamplifiers and electronic chain, but independent high voltage supply. The data acquisition system for the Blank module has been integrated into the ANAIS-112 system and it is referred to as detector D9. The data analysis for D9 follows the same procedures as for the other modules. 

The Blank module was designed and commissioned for better understanding events at low energy attributable to non-bulk NaI(Tl) scintillation and the determination of the efficiency of the rejection of such events by the ANAIS-112 filtering protocols. In Chapter~\ref{Chapter:OptSim}, an optical simulation of the Blank module will be used to understand the origin of some of the events triggering that module, as well as the expected contribution from both, $^{40}K$ contamination in the PMTs borosilicate and $^{222}Rn$ in the laboratory air, to the trigger rate of the Blank module.

\section{Module performance} \label{Section:ANAIS_Module}
\fancyhead[RO]{\emph{\thesection. \nameref{Section:ANAIS_Module}}}

In this section, the main characteristics of the detector response to energy depositions are overviewed, as they will be required for the optical simulation of the detector in Chapter~\ref{Chapter:OptSim}.

The detector response to any light signal is obtained by the convolution of the Single Electron Response (SER) and the light emission profile. This SER is determined by accumulating signals that can be attributed to only one electron produced in the photocathode. They allow to build the distributions of some of the SER relevant parameters (area, amplitude and width) and they can be averaged to produce the mean SER pulse. As the SER is strongly dependent on the operational conditions (temperature and voltage), it is very important to control them in the ANAIS-112 experiment.

\begin{figure}[h!]
	\begin{center}
		\includegraphics[width=\textwidth]{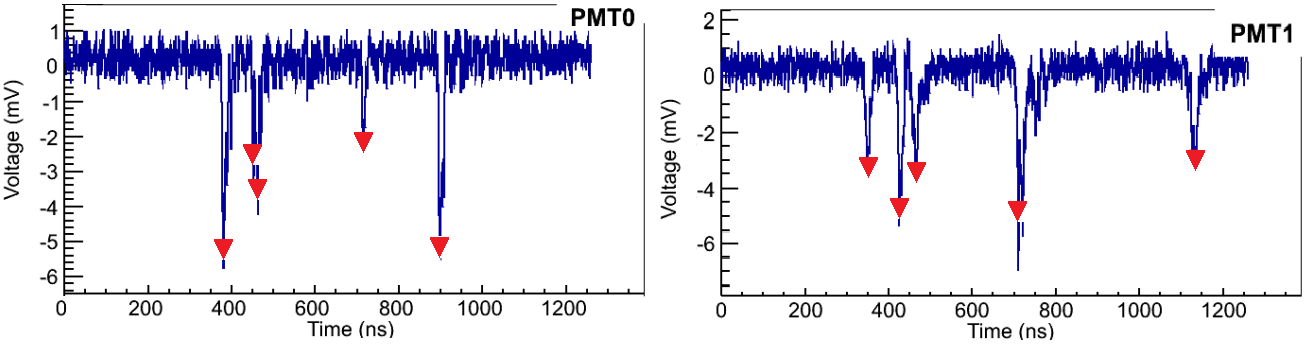}
		\caption{\label{PeakFindingAlgo}Example of acquired waveforms for the two PMTs of the same module for a 0.9 keV event. The red triangles mark the peaks found by the algorithm.}
	\end{center}
\end{figure}

\begin{figure}[h!]
	\begin{center}
		\includegraphics[width=\textwidth]{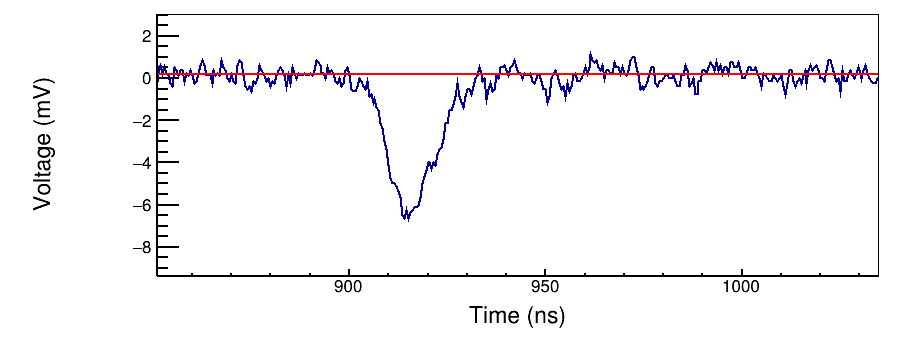}
		\caption{\label{pheANAIS}Example of a single photoelectron. Red line represents the level of the baseline.}
	\end{center}
\end{figure}

To calculate it, an unbiased phe population is needed. It is obtained by applying a peak identification algorithm, which is based on a low pass filter and detecting a sign change in the derivative waveform. This algorithm can find peaks in the waveform down to the baseline noise. The identification of the single phes by this algorithm is correct at the low light regime, being its efficiency limited by the phe overlap. This is the reason why this analysis is done using very low-energy background events, in which the number of photoelectrons is expected to be low, and in addition, only the last peak in the tail of the pulse is considered to guarantee that the probability of overlapping is very low. Once the peaks have been identified, the waveform is integrated in a narrow window around the maximum of the last peak $\left[-30,+60\right]$~ns, to obtain the corresponding area distribution. An example of the acquired waveforms for the two PMTs of the same module for a 0.9~keV event and the peaks found by the algorithm is shown in Figure~\ref{PeakFindingAlgo}, while Figure~\ref{pheANAIS} shows the waveform of a single photoelectron. It is possible to observe that it is approximately gaussian, with an average standard deviation, $\sigma$, of 6~ns. It is very similar for all the ANAIS-112 PMTs. 

\begin{figure}[h!]
	\begin{center}
		\includegraphics[width=0.8\textwidth]{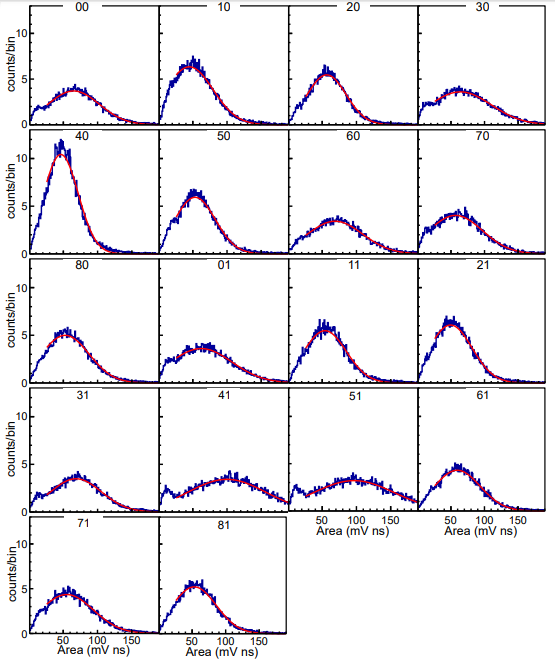}
		\caption{\label{DistributionsSER}Example of SER area distributions for all the PMTs of the experiment and their fits to Gaussian functions.}
	\end{center}
\end{figure}

\begin{figure}[h!]
	\begin{subfigure}[b]{0.49\textwidth}
		\includegraphics[width=\textwidth]{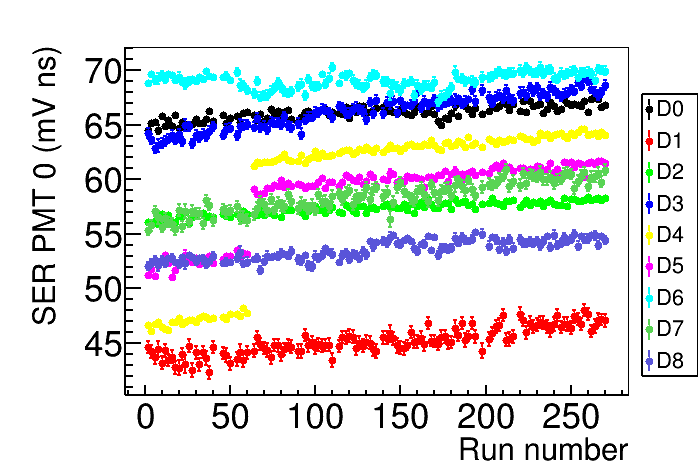}
	\end{subfigure}
	\begin{subfigure}[b]{0.49\textwidth}
		\includegraphics[width=\textwidth]{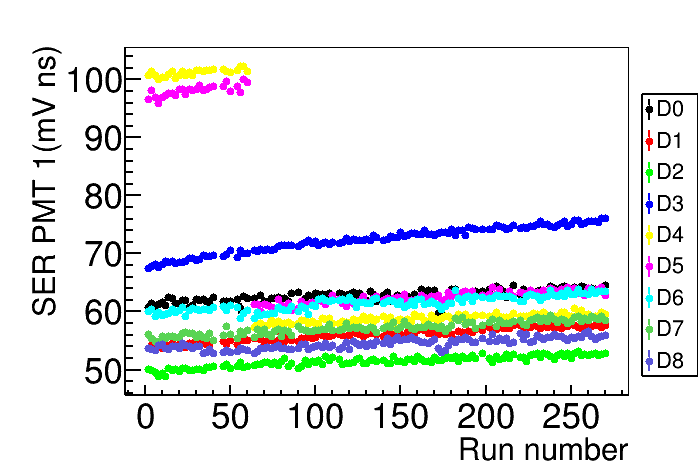}
	\end{subfigure}
	\caption{\label{SEREvolution}Evolution in time of the mean area of the SER for the PMTs-0 (left plot) and PMTs-1 (right plot) of each module of the ANAIS-112 experiment. It can be observed how the HV of PMTs in D4 and D5 was modified after the first year of data taking. The PMTs gain shows some drift along the ANAIS-112 operation.}
\end{figure}

Figure~\ref{DistributionsSER} shows the SER area distributions for all the ANAIS-112 PMTs corresponding to a single background run (about two weeks of exposure). These distributions are obtained in each run to monitor and analyze the stability of the detectors. They are partially biased at low energies because of baseline noise and threshold effects in the application of the peak finding algorithm, reason why they are fitted to Gaussian functions leaving free the amplitude, mean and standard deviation. These two last variables are continuously monitored to analyze the stability conditions of the experiment, as it is shown in Figure~\ref{SEREvolution}.

Another parameter defined for each PMT individually is the number of phes per unit of deposited energy in the NaI(Tl) crystal, the so-called Light Collection (LC). It has to be calculated for a well known energy deposition (in ANAIS-112 is used for this purpose the 22.6~keV peak from $^{109}Cd$ calibration runs), as the ratio between the waveform area and the SER mean area for each PMT. The LC for a given module is later calculated as the addition of the LC obtained for the two PMTs. It can be understood as the Light Yield (LY) of the crystal (number of scintillation photons produced per energy unit) multiplied by the Light Collection Efficiency ($LCE$) (ratio of detected photoelectrons to emitted photons). This efficiency depends on the ratio of photons reaching the photocathode to those emitted by the crystal ($\eta$), the ratio of phe produced in a PMT to the total number of photons reaching the photocathode, known as Quantum Efficiency ($QE$), and the probability of the produced phes to reach the first dynode and start an avalanche ($\epsilon$)~\cite{PMTsHandbook}. Therefore, the dependence between the LC and LY can be factorized:
\begin{equation}\label{LC_PMTs_ANAIS}
	LC = LY \cdot LCE = LY \cdot \eta \cdot QE \cdot \epsilon
\end{equation}
Mean LC values for the first three years of data taking are summarized in Table~\ref{table:moduleCharacteristics} for the 18~PMTs, together with their QE (as given by the manufacturer). The averaged QE of all the PMTs is 40.6\% with a standard deviation of 2.0\%. The LC for the nine modules is very high, with and average of 14.44~$\pm$~0.01~phes/keV and a standard deviation of 0.91~phes/keV. This high LC impacts positively the resolution, the energy threshold and the efficiencies of the filtering protocols.

\begin{table}[h!]
	\centering
	\begin{tabular}{|c|c|c|}
		\hline
		PMT & QE (\%) & LC (phe/keV) \\
		\hline
		D0 - PMT0 & 38.2 & 7.06$\pm$0.01\\
		D0 - PMT1 & 37.2 & 7.42$\pm$0.01\\
		D1 - PMT0 & 39.7 & 7.76$\pm$0.01\\
		D1 - PMT1 & 39.7 & 6.89$\pm$0.01\\
		D2 - PMT0 & 39.2 & 6.59$\pm$0.01\\
		D2 - PMT1 & 42.6 & 7.68$\pm$0.02\\
		D3 - PMT0 & 37.3 & 7.11$\pm$0.01\\
		D3 - PMT1 & 39.4 & 7.23$\pm$0.01\\
		D4 - PMT0 & 40.1 & 7.01$\pm$0.03\\
		D4 - PMT1 & 41.8 & 7.33$\pm$0.03\\
		D5 - PMT0 & 43.6 & 7.54$\pm$0.04\\
		D5 - PMT1 & 43.9 & 7.26$\pm$0.06\\
		D6 - PMT0 & 40.4 & 6.38$\pm$0.01\\
		D6 - PMT1 & 38.9 & 6.36$\pm$0.01\\
		D7 - PMT0 & 41.9 & 7.24$\pm$0.01\\
		D7 - PMT1 & 41.5 & 7.32$\pm$0.01\\
		D8 - PMT0 & 41.6 & 7.99$\pm$0.01\\
		D8 - PMT1 & 43.4 & 7.84$\pm$0.01\\
		\hline
		Weighted mean & 40.6 & 7.22$\pm$0.01\\
		\hline
	\end{tabular} \\
	\caption{Quantum Efficiency and Light Collection for the 18 PMTs used in ANAIS-112 experiment. QE are the nominal values given by the manufacturer and the LC is that obtained for the three first years of data.}
	\label{table:moduleCharacteristics}
\end{table}

\section{Data analysis} \label{Section:ANAIS_Analysis}
\fancyhead[RO]{\emph{\thesection. \nameref{Section:ANAIS_Analysis}}}

The data is analyzed using the ROOT software framework~\cite{Brun:1997pa}. As explained in Section~\ref{Section:ANAIS_DAQ}, the DAQ system stores the data in ROOT files for offline analysis. These files are organized into two types of measurement runs: background and calibration. In addition, during background runs, the Veto DAQ system stores the information in separated files.

The background and calibration data files contain the information directly read from the ANAIS-112 modules: PMT waveforms (LE signals), QDC values (HE signals), the timestamp of the event and the triggering pattern. In the analysis of these files, different parameters are determined for each acquired PMT waveform:
\begin{itemize}
	\item Mean value and standard deviation of the baseline, calculated using the samples of the waveform in the pretrigger region.
	\item Pulse onset, defined as the point where the signal exceeds a threshold level.
	\item Puse area, obtained by integrating the waveform from the pulse onset and subtracting the contribution from the baseline.
	\item Number and position of peaks identified by the peak-finding algorithm.
\end{itemize}

Other parameters are determined for each module, by combining the information from the two corresponding PMT waveforms, for instance pulse shape parameters as $p_1$, $\mu$, etc. Some of them will be commented later (see Section~\ref{Section:ANAIS_Filter}) because they are important in the filtering protocols. Moreover, each event gets the value of the time elapsed after the last previous trigger in the muon veto by comparing the corresponding timestamps.

Later, on a different level of analysis, the energy associated to each event is calculated by applying the results of the calibration method explained in Section~\ref{Section:ANAIS_Calibration} using the sum of the pulse areas of the two PMTs as an energy estimator.

\section{Energy calibration} \label{Section:ANAIS_Calibration}
\fancyhead[RO]{\emph{\thesection. \nameref{Section:ANAIS_Calibration}}}

The response of the ANAIS-112 detectors is calibrated using photons (gammas and x-rays), which produce electronic recoils. However, WIMPs are expected to elastically scatter off the target nuclei, producing a nuclear recoil with the kinetic energy lost by the WIMP (see Section~\ref{Section:Intro_Detection_Direct}). Then, the detector response to nuclear recoils has to be well known to interpret the energy deposited spectra in terms of WIMPs. In this section both energy calibration procedures will be presented and discussed.

\subsection{Electron equivalent energy calibration} \label{Section:ANAIS_Calibration_Eee}

\begin{figure}[h!]
	\begin{center}
		\includegraphics[width=0.75\textwidth]{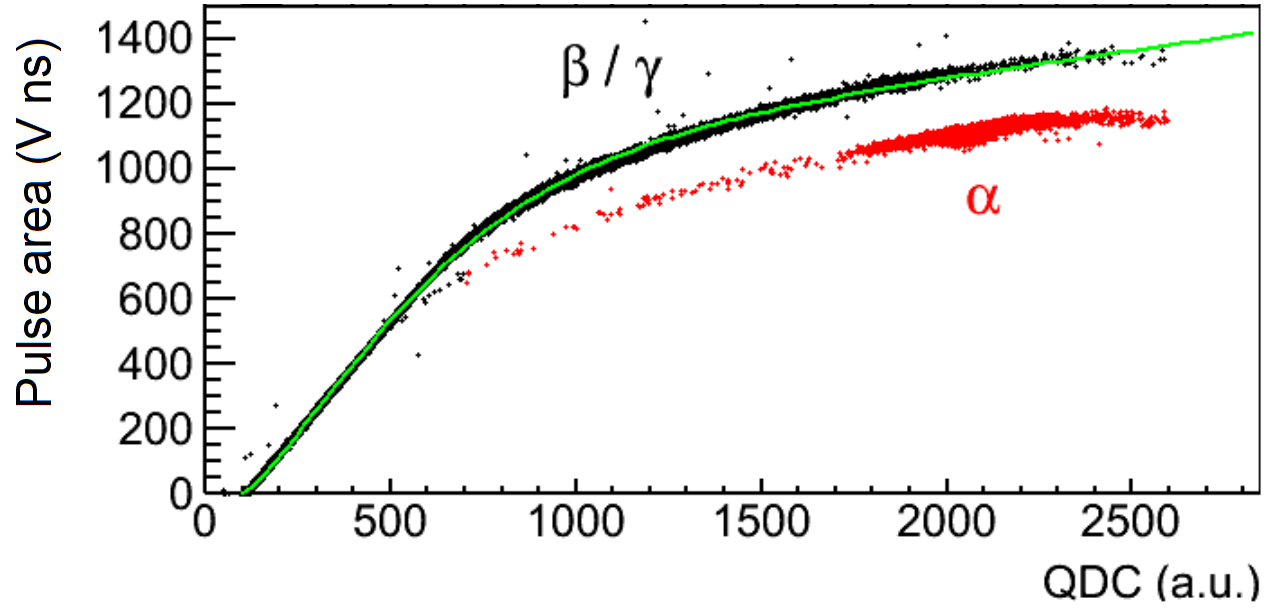}
		\caption{\label{HE-Linear}Sum of the pulse area of the two digitized signals for one of the ANAIS-112 detectors during the two first weeks of data taking as a function of the QDC readout. Green line represents the modified logistic function used to linearize the signal. Red (black) dots are identified as the $\alpha$ ($\beta$ / $\gamma$) population due to the different ratio of pulse area to QDC. Image from~\cite{Amare:2018sxx}.}
	\end{center}
\end{figure}

Due to the slightly non linear response of NaI(Tl), the electron equivalent energy response of ANAIS-112 detectors is calibrated in two energy ranges: low-energy (LE) and high-energy (HE). The pulses digitized by the MATACQ are saturated for events with energies above 500~keV. On the other hand, the HE signal acquired from QCDs has a 12-bit resolution (see Section~\ref{Section:ANAIS_DAQ}), reason why they are not used as energy estimators in the HE range. Instead, their information is used to linearize the response from the area of the digitized pulse using a modified logistic function, as it is shown in Figure~\ref{HE-Linear}. This linearized variable is used as estimator of the HE (above 50~keV) assigned to each event. This double readout system at HE allows to discriminate $\alpha$ events (shown in red in Figure~\ref{HE-Linear}) from $\beta$ / $\gamma$ (shown in black). The reason is that $\alpha$ events are faster and therefore for the same QDC value its saturated pulse integral is smaller.

\begin{figure}[h!]
	\begin{center}
		\includegraphics[width=0.75\textwidth]{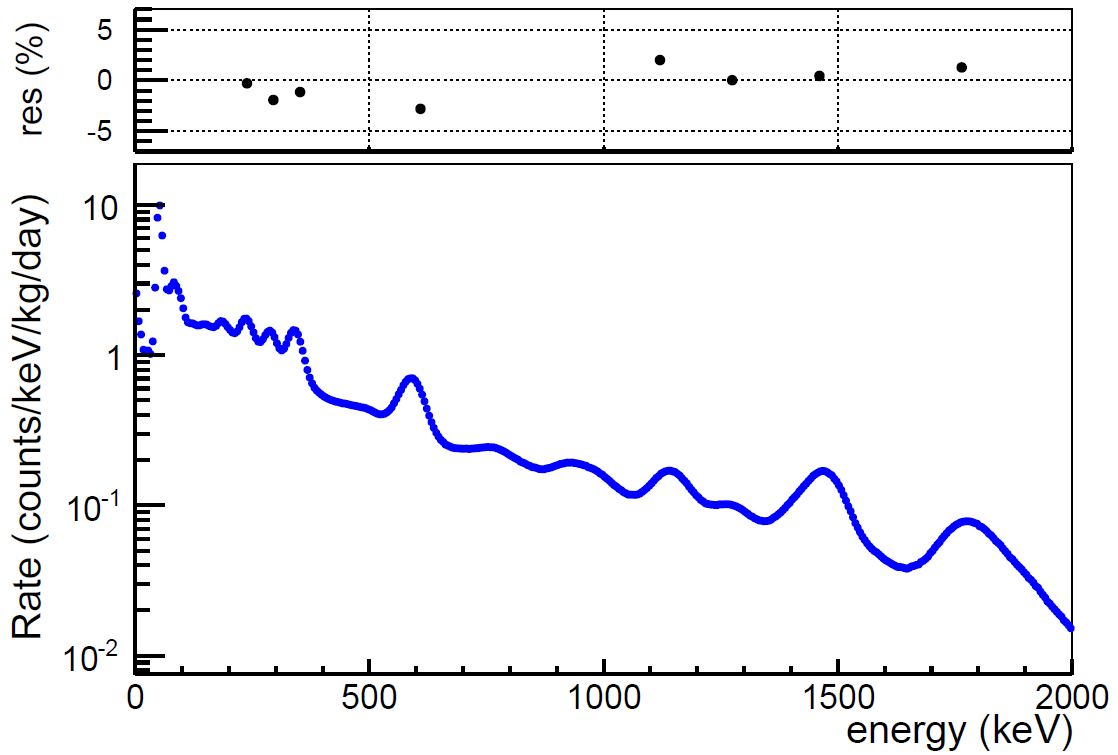}
		\caption{\label{HighEnergyCalANAIS}Total high energy anticoincidence spectrum (i.e., discarding coincidence events among different modules) measured along the three years of ANAIS-112 operation. The top panel shows the residuals for the positions of the main peaks identified in the background. Image from~\cite{CoarasaCasas:2021euy}.}
	\end{center}
\end{figure}

After this linearization, the HE region is calibrated using several easily identifiable peaks in the background data. The number of peaks used for calibration varies between 6 and 8, depending on their presence in the background spectrum of each detector corresponding to approximately two weeks. Among them are: 238.6~keV ($^{212}Pb$), 295.2~keV ($^{214}Pb$), 338.3~keV ($^{228}Ac$) together with 351.9~keV ($^{214}Pb$), 583.2~keV ($^{208}Tl$) combined with 609.3~keV ($^{214}Bi$), 1120.3~keV ($^{214}Bi$), 1274.5~keV ($^{22}Na$), 1460.8~keV ($^{40}K$) and 1764.5~keV ($^{214}Bi$). Each peak is fitted to a Gaussian function plus a second-degree polynomial, and the calibration is performed using a linear regression between the nominal energies of the peaks and the Gaussian means. The energy calibrated background spectrum for single-hit events for the first three years of data of ANAIS-112 is shown in Figure~\ref{HighEnergyCalANAIS}, with residuals for the main peaks positions below 4\%~\cite{CoarasaCasas:2021euy}.

\begin{figure}[h!]
	\begin{center}
		\includegraphics[width=\textwidth]{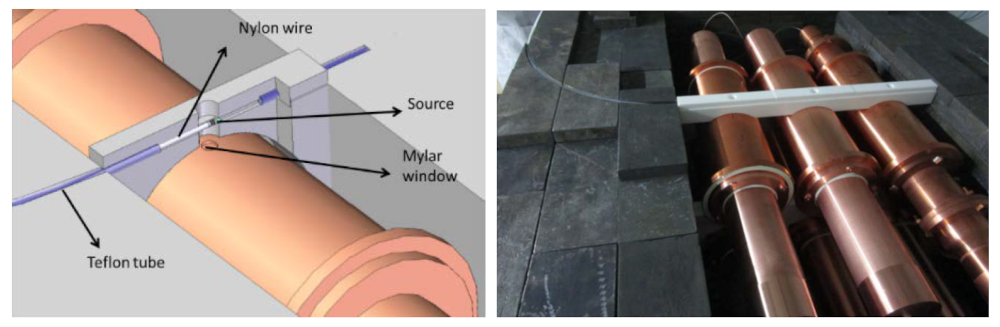}
		\caption{\label{CalibrationSytemANAIS}Schematic view (left) and picture (right) of the ANAIS-112 calibration system.}
	\end{center}
\end{figure}

For the calibration of the LE range, a measurement using external $^{109}Cd$ sources is done every two weeks. The nine modules are calibrated simultaneously using a multi-source system, which allows to reduce the time required for calibration. It consists of two flexible wires where the sources are fixed, and it ensures radon-free operation by moving the wires along a sealed tube inside the shielding. The sources can be easily positioned in front of the Mylar window built in the ANAIS modules. The design and implementation of this system is shown in Figure~\ref{CalibrationSytemANAIS}.

\begin{figure}[h!]
	\begin{center}
		\includegraphics[width=0.75\textwidth]{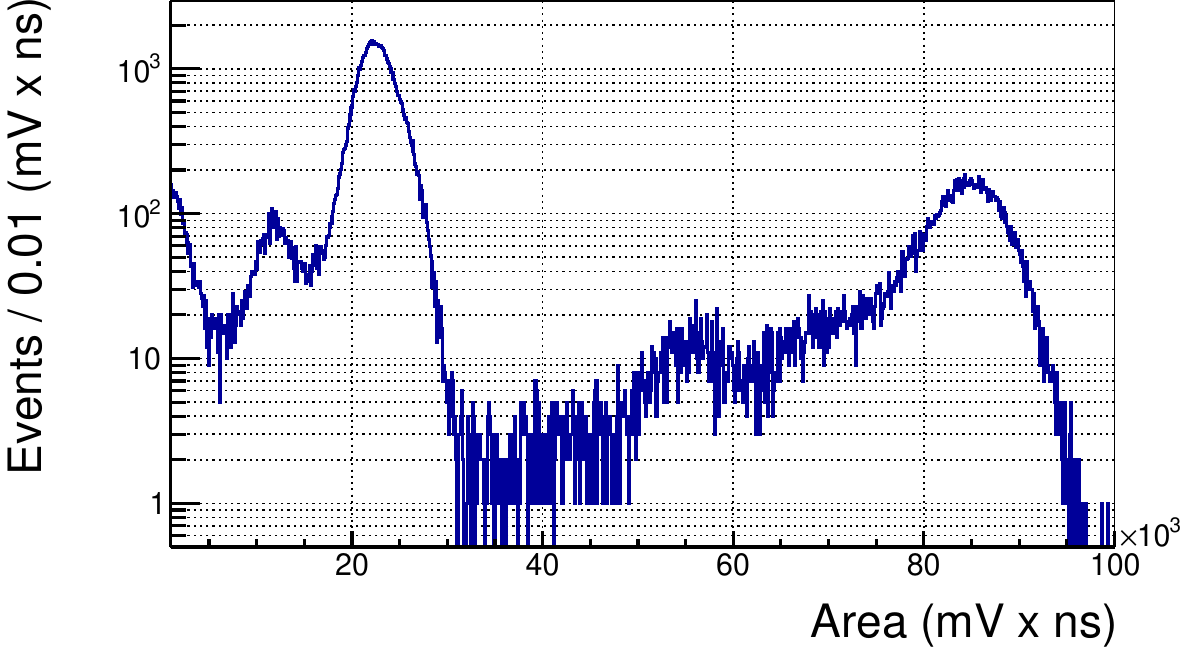}
		\caption{\label{CalibrationANAIS}Pulse area spectrum for the D0 module in one of the periodical calibration runs using $^{109}Cd$ external sources. Similar spectra are obtained for the nine modules.}
	\end{center}
\end{figure}

$^{109}Cd$ decays by electron capture to $^{109}Ag$, emitting a gamma ray with an energy of 88.0~keV in 3.6\% of the decays. In addition, K$_{\alpha}$ and K$_{\beta}$ $Ag$ x-rays are emitted with average energies of 22.1~and 25.0~keV and intensities of 69.8\% and 13.8\%, respectively. The resolution of the ANAIS-112 detectors is not good enough to distinguish between so similar energies, resulting in a broader peak, sum of both contributions with an averaged energy of 22.6~keV. The spectrum also exhibits a peak attributed to the x-rays of the bromine atoms, present in the heat-shrink plastic covering the $^{109}Cd$ sources. This results in the emission of the bromine K-shell x-rays produced after the photoelectric absorption of $^{109}Cd$ photons by the bromine. These bromine x-rays have an averaged energy of 11.9~keV. These lines can be identified in the calibration spectra of Figure~\ref{CalibrationANAIS}.

\begin{figure}[h!]
	\begin{center}
		\includegraphics[width=0.75\textwidth]{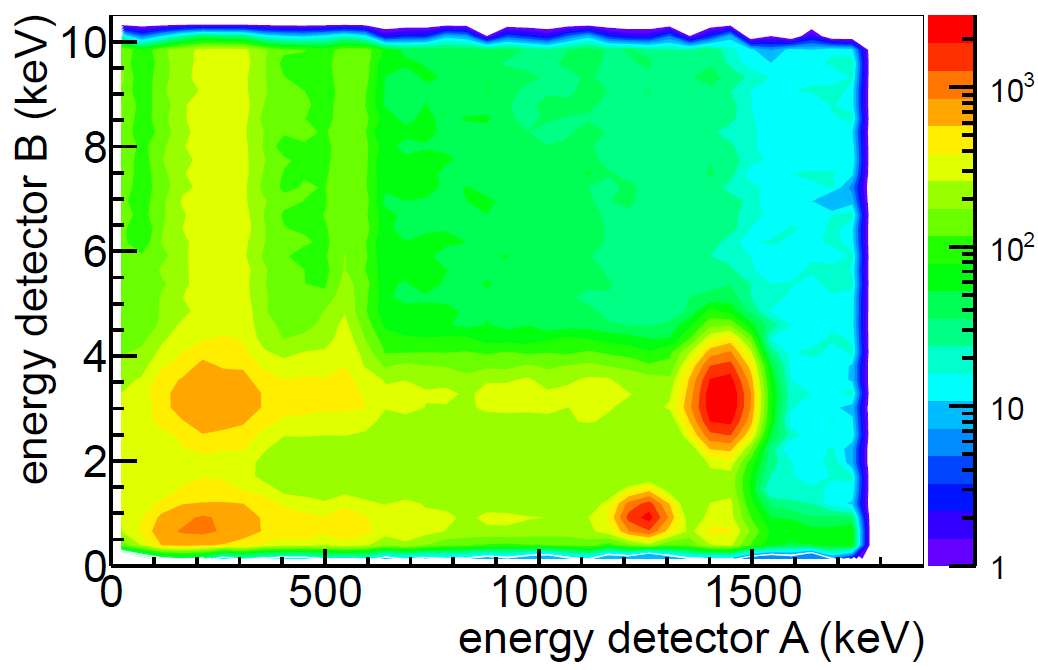}
		\caption{\label{LowEnergyCalANAIS}Scatter plot of the energies recorded in two detectors triggered in coincidence when one of the detectors had an energy deposition below 10~keV over the course of the first three years of ANAIS-112 exposure. Energy depositions corresponding to 1274.5~keV (1460.8~keV) and 0.9~keV (3.2~keV) in detectors~A and~B, respectively, are clearly observed, corresponding to the full absorption in detector~A of the high-energy gamma emitted by $^{22}Na$ ($^{40}K$) after the EC decay occurring in detector~B. Image from~\cite{CoarasaCasas:2021euy}.}
	\end{center}
\end{figure}

In order to accurately calibrate the low energy response, the 11.9 and 22.6~keV lines from the $^{109}Cd$ calibration are used, besides two background lines that help to increase the accuracy of the calibration in the ROI. These lines are those produced by the $^{40}K$ and $^{22}Na$ contamination in the crystal. These isotopes decay via electron capture, process which leaves the final atom with a hole in the internal shells. Moreover, in both cases, the final state of the EC decay is an excited state of the daughter nuclei. Then, the EC process is followed by the atomic de-excitation, which releases in the case of the K-shell EC an energy of 0.9~keV for $^{22}Na$ and 3.2~keV for $^{40}K$, and later, depending on the lifetime of the excited nuclear state, the emission of gamma photons with energies of 1274.5~keV and 1460.8~keV, for $^{22}Na$ and $^{40}K$, respectively. In the case of these isotopes the lifetime of those states is very short, and all the emissions can be considered simultaneous. The binding energy of the atomic shell is fully absorbed within the crystal where the decay occurs, while the gamma photons can escape and interact in another detector, resulting in a coincident event. Figure~\ref{LowEnergyCalANAIS} shows the low energy events recorded in one detector when another detector has been triggered in coincidence, for the first three years of ANAIS-112 exposure.

\begin{figure}[h!]
	\begin{center}
		\includegraphics[width=0.75\textwidth]{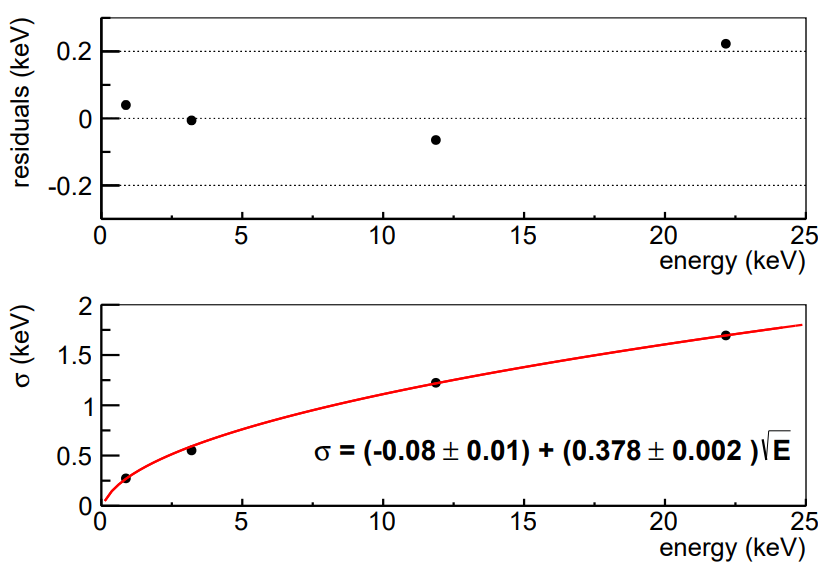}
		\caption{\label{CalResANAIS}Top plot: residuals of the low energy calibration. Bottom plot: energy resolution of the same energy lines as a function of energy. Red line corresponds to the fit to a $\alpha+\beta\sqrt{E}$ function. Image from~\cite{Amare:2018sxx}.}
	\end{center}
\end{figure}

The low rate of the 0.9~and 3.2~keV peaks and the low efficiency for the detection of the coincidence to select them accurately requires the accumulation of background data over a period of approximately 1.5~months to allow using them for calibration. The median of the resulting pulse area distributions is used in the calibration, and a linear fit is then applied to the expected energies of the four selected peaks against their areas for each detector. The residuals of this fit (defined as the difference between the reconstructed mean position of the peak minus the nominal energy) are shown in Figure~\ref{CalResANAIS}, together with the resolutions obtained for the four peaks and the fit to a $\alpha+\beta\sqrt{E}$ function~\cite{Amare:2018sxx}. It can be seen that the calibration deviations are below 0.2~keV in the whole range and below 0.04~keV in the ROI.

\subsection{Nuclear recoil energy calibration} \label{Section:ANAIS_Calibration_Enr}

The parameter space region that is reached by the ANAIS experiment can be significantly altered by different values of the sodium and iodine QFs (see Section~\ref{Section:Intro_Scintillators_NaI(Tl)}). Therefore a measurement of the QF for NaI(Tl) crystals similar to those used in ANAIS-112 is required to interpret the data in terms of a dark matter signal. The response to neutrons of different small cylindrical NaI(Tl) crystals was characterized at TUNL laboratories. One of the crystals measured was produced in the same ingot that some of the ANAIS-112 detectors. The analysis of these measurements will be considered in depth in Chapter~\ref{Chapter:QF}.

\begin{figure}[h!]
	\begin{center}
		\includegraphics[width=\textwidth]{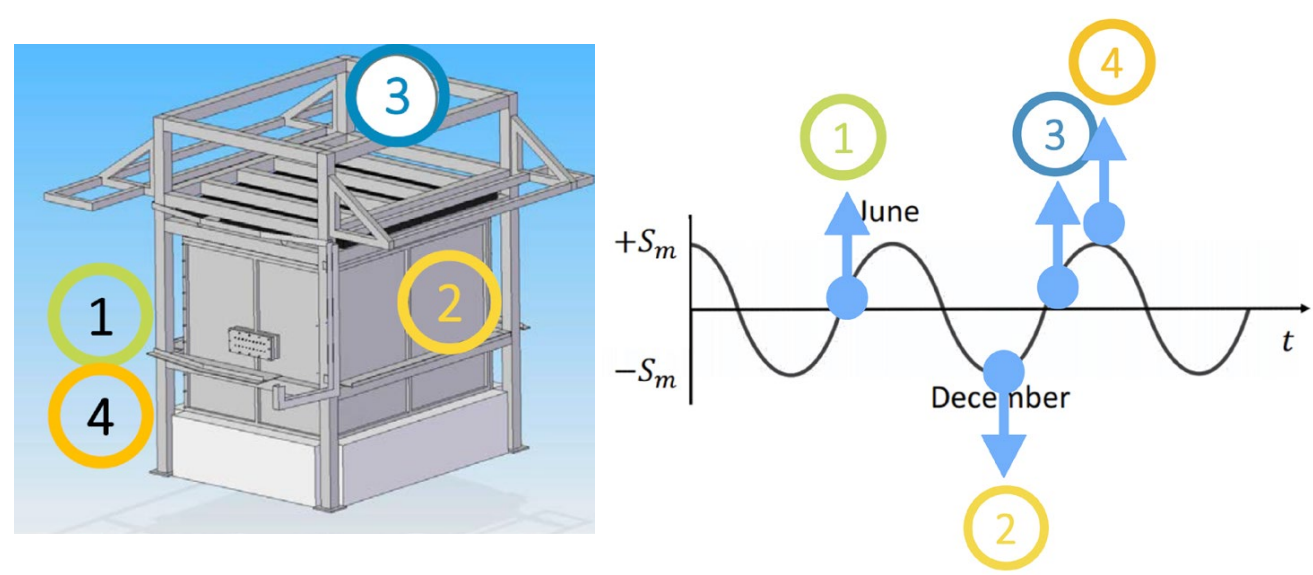}
		\caption{\label{Cf252Calibrations}Position of the $^{252}Cf$ neutron source (left) and time of the year (right) of the four neutron calibration runs carried out in the ANAIS-112 experiment.}
	\end{center}
\end{figure}

Apart from those QF measurements, four calibration runs have been carried out since April 2021 using a $^{252}Cf$ neutron source. This isotope has a half-life of 2.6~years and decays by spontaneous fission with a probability of 3\%, emitting neutrons. As it is shown in Figure~\ref{Cf252Calibrations}, these calibrations have been done at different moments of the year placing the source at different positions in the ANAIS-112 setup.

\begin{figure}[h!]
	\begin{center}
		\includegraphics[width=0.75\textwidth]{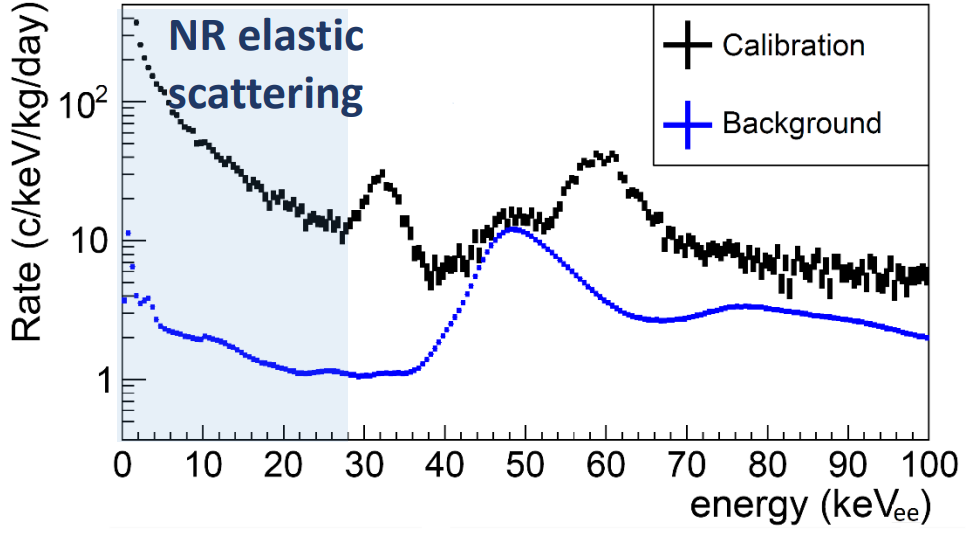}
		\caption{\label{NRScatteringANAIS}Energy spectra comparison of a $^{252}Cf$ calibration (black) and the ANAIS-112 background (blue). It is possible to observe the contribution of the elastic scattering at energies below 30~keV.}
	\end{center}
\end{figure}

Figure~\ref{NRScatteringANAIS} shows a comparison between the spectrum obtained in one of these calibrations and the background. It is possible to observe the contribution of the elastic nuclear recoil spectrum at energies below 30~keV. The neutrons can experience different interactions within the NaI(Tl) crystals, elastic and inelastic, with sodium and iodine nuclei. Although they are explained in Appendix~\ref{Chapter:Anexo_NeutronNaI}  we comment next briefly on two of the features observed in Figure~\ref{NRScatteringANAIS}. The capture of the neutron by $^{127}I$ results in the production of $^{128}I$, which decays by EC in 6.9\% of the cases, resulting in the emission of Te x-rays, at around 31~keV for K-shell EC, which is observed in the spectrum. Moreover, the inelastic scattering on $^{127}I$ produces the nuclear excitation to the first energy state, which corresponds to an energy of 57.6~keV. When the nucleus returns to the ground state, it releases this energy, as it is observed in the spectrum.

The analysis of the data taken during these measurements (still in progress), can be an important cross-check for the results presented in Chapter~\ref{Chapter:QF}. Moreover, they can be also used to analyze the effect of the crystal size in the detector response to nuclear recoils, being the multiple scattering one of the most relevant differences between both measurements. Precise simulations of these measurements are ongoing. Introducing in the simulation the QFs and comparing simulation and measurement, limits on the QF values for both, Na and I recoils, can be derived.

\section{Filtering protocols and efficiencies} \label{Section:ANAIS_Filter}
\fancyhead[RO]{\emph{\thesection. \nameref{Section:ANAIS_Filter}}}

The filtering protocols applied in the ANAIS-112 experiment aim to select bulk scintillation events in the NaI(Tl) crystals compatible with a DM signal rejecting other background events having different origins. They are explained in this section, as well as the way their efficiencies are calculated. 

\begin{figure}[h!]
	\begin{center}
		\includegraphics[width=0.75\textwidth]{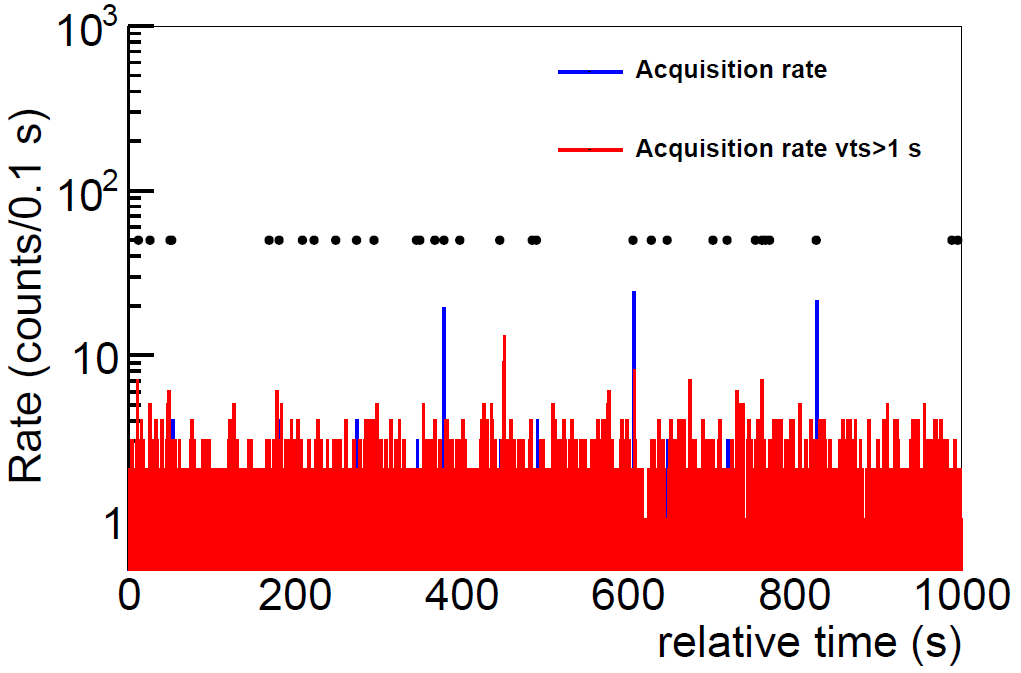}
		\caption{\label{AcqRateANAIS}Trigger rate of ANAIS-112 in a period of 1000~s. Blue line shows the total trigger rate, while red line represents the rate after the rejection of events in a period of 1~s after the veto trigger. Black dots indicate events in the veto scintillators. Image from~\cite{Amare:2018sxx}.}
	\end{center}
\end{figure}

First, WIMPs are expected to interact with ordinary matter through processes with very low cross-sections, which means that the probability of a DM particle interacting in two modules or in coincidence with a background event is negligible. Therefore, coincidences among different modules can be attributed to backgrounds and then, rejected. As it was explained in Section~\ref{Section:Intro_Scintillators_NaI(Tl)}, NaI(Tl) crystals have a scintillation component with a very slow time constant ($\sim$~150~ms) which can be observed after high-energy depositions, as those produced by muons~\cite{Cuesta:2013vpa}. As the dead time per event is of the order of 4~ms, this slow scintillation is able to produce a strong increase in the module trigger rate, because the scintillation pulse tail can trigger several times the acquisition. The ANAIS veto system (explained in Section~\ref{Section:ANAIS_Setup_Muon}) allows to reject this kind of events produced by muons. In Figure~\ref{AcqRateANAIS}, the ANAIS-112 total trigger rate over a period of 1000~s with a 0.1~s time binning is represented by the blue line, while the timing of veto scintillator triggers is marked by black dots. By rejecting events arriving within one second of a veto scintillation trigger (red line shows the trigger rate after application of this selection), most of these high rate periods are eliminated. Although most of the triggers in the veto scintillator system do not produce any rate increase in the ANAIS detectors, this selection criterion is useful to reject other possible single-hit and low energy events that muons could also produce, as neutron emission in the shielding materials. The introduction of this selection criterion in the ANAIS data analysis chain implies a reduction in the live time of about 3\%.

\begin{figure}[h!]
	\begin{center}
		\includegraphics[width=\textwidth]{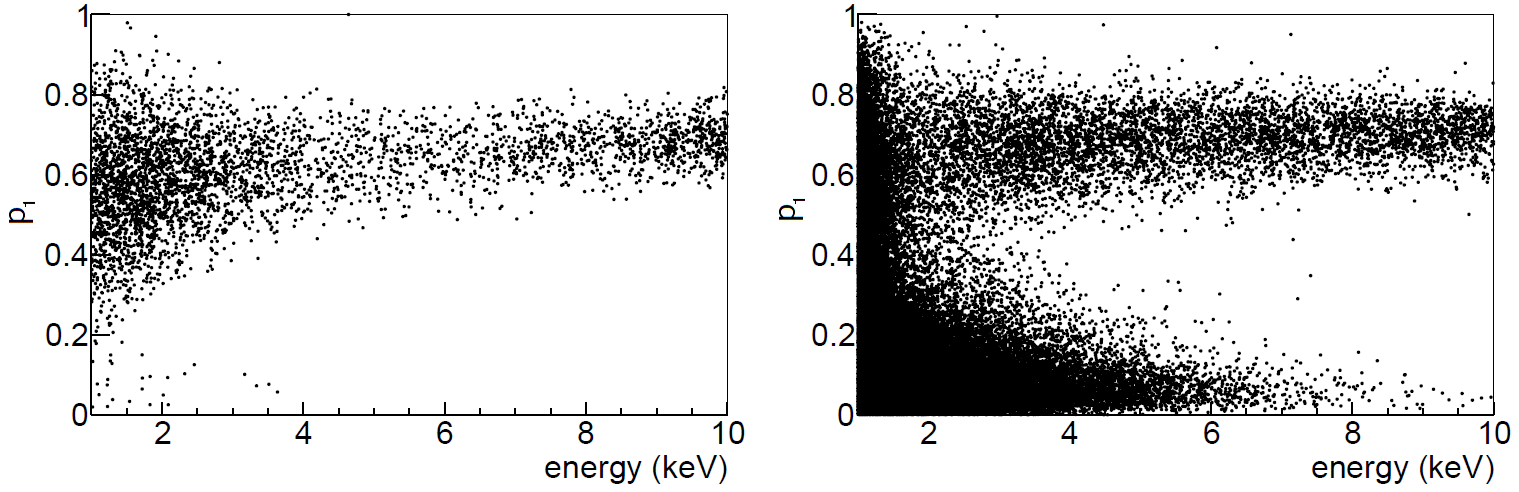}
		\caption{\label{p1ANAIS}$p_1$ distribution as a function of the energy for detector D1 for a $^{109}Cd$ calibration run (left plot) and a background run (right plot). Image from~\cite{CoarasaCasas:2021euy}.}
	\end{center}
\end{figure}

The emission of Cherenkov light in the borosilicate of the PMTs can be produced by the radioactive contamination in the PMTs themselves or in the materials close to them~\cite{Knoll:2000fj,Amare:2014eea}. In particular, the $^{40}K$ present in the PMTs borosilicate~\cite{Amare:2018ndh} decays $\beta^-$ with a Q-value of 1311~keV with a probability of 89\%. A large part of the electrons are above the threshold energy for Cherenkov production, which is 168~keV. Any fast scintillation event, as it is Cherenkov emission, would produce a signal pulse with the PMT SER temporal behaviour. These events can be distinguished easily by pulse-shape analysis (PSA) from bulk scintillation events in the NaI(Tl) crystal, which show the typical scintillation time constant of 230~ns. Other scintillation emissions in materials as the quartz or silicone could be more difficult to disentangle, depending on the corresponding characteristic scintillation time constants. The PSA parameter used to discriminate between events with different time behaviour is $p_1$, defined as~\cite{DAMA:2008bis}:
\begin{equation}
	p_1 = \frac{\sum_{t = 100s}^{t = 600s} V(t)}{\sum_{t = 0}^{t = 600}V(t)},
\end{equation}
where $V(t)$ is the pulse amplitude of the sum of the waveforms from the two PMTs of the same module at time $t$ after the pulse onset. Figure~\ref{p1ANAIS} shows the distribution of this parameter for detector D1 as a function of the energy for a $^{109}Cd$ calibration run and a background run. It can be observed in the calibration that the $p_1$ parameter for bulk scintillation events in NaI(Tl) is approximately 0.65. Below 3~keV the dispersion in the values of $p_1$ increases strongly as the number of phes is lower. Most events in the background run have $p_1$ values lower than 0.4, and then, they cannot be considered NaI(Tl) bulk scintillation events.

\begin{figure}[h!]
	\begin{center}
		\includegraphics[width=\textwidth]{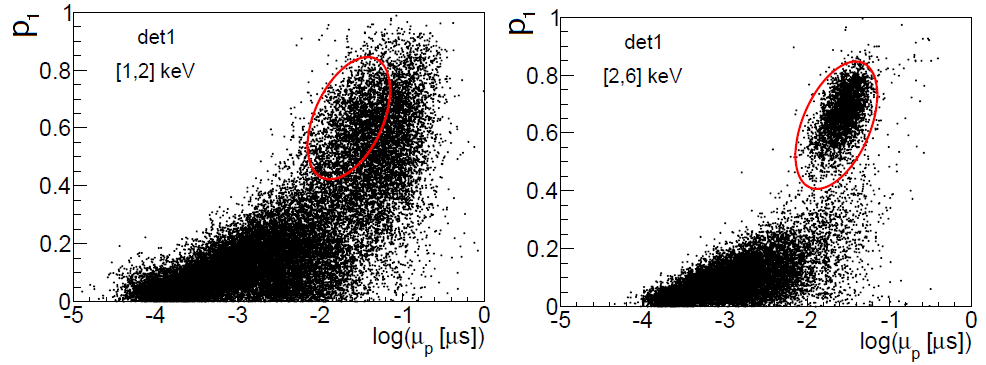}
		\caption{\label{p1MuANAIS}$p_1$:$log(\mu)$ scattering plot for background events in the detector D1 for two energy ranges: [1,2]~keV (left) and [2,6]~keV (right). The red line represents the $PSV$ cut defined in the ANAIS-112 experiment. Image from~\cite{CoarasaCasas:2021euy}.}
	\end{center}
\end{figure}

It is difficult to distinguish between both populations at low energies using only the $p_1$ parameter because they overlap. Moreover, $p_1$ parameter is not useful for identifying fake low energy events that may be produced by long time phosphorescence in the crystal or by pulse tails. To address this issue, a new parameter based on the time associated to the individual peaks found in the low energy events and that can be correlated with the arrival times of the individual phes has been defined:
\begin{equation}
	\mu = \frac{\sum_i A_i t_i}{\sum_i A_i},
\end{equation}
where $A_i$ and $t_i$ are the amplitude and time of the $i^{th}$ peak identified in the waveform, respectively. Figure~\ref{p1MuANAIS} shows the $p_1$:$log(\mu)$ scattering plot for background events in D1 for two energy ranges ([1,2] and [2,6]~keV). As both variables are correlated, the selection of events is done using both of them and assuming that $\Vec{x} = \left(p_1,\log\left(\mu\right)\right)$ follows a 2D Gaussian distribution with mean $\Vec{k}$ and covariance matrix $V$ (both of them are calculated using the 0.9~keV bulk scintillation population from $^{22}Na$ decay). The resulting Pulse Shape Variable ($PSV$) is defined as:
\begin{equation}
	PSV = \left(\Vec{x}-\Vec{k}\right)^T V^{-1}\left(\Vec{x}-\Vec{k}\right).
\end{equation}
An event is considered as bulk NaI(Tl) scintillation if its $PSV$ is smaller than a conveniently chosen $PSV_{cut}$. The $PSV_{cut}$ chosen in the ANAIS-112 analysis is shown in Figure~\ref{p1MuANAIS} as a red line: events inside the ellipse correspond to bulk NaI(Tl) scintillation while those found outside are rejected. Figure~\ref{p1MuANAIS} shows clearly that the PSV parameter for bulk scintillation events has a strong dependence on the event energy. The $PSV_{cut}$ value is chosen by requiring a 77.7\% of efficiency in the acceptance of events in the 1-2 keV energy region for $^{40}K$ and $^{22}Na$ events tagged by the detection of a HE gamma in coincidence.

\begin{figure}[h!]
	\begin{subfigure}[b]{0.49\textwidth}
		\includegraphics[width=\textwidth]{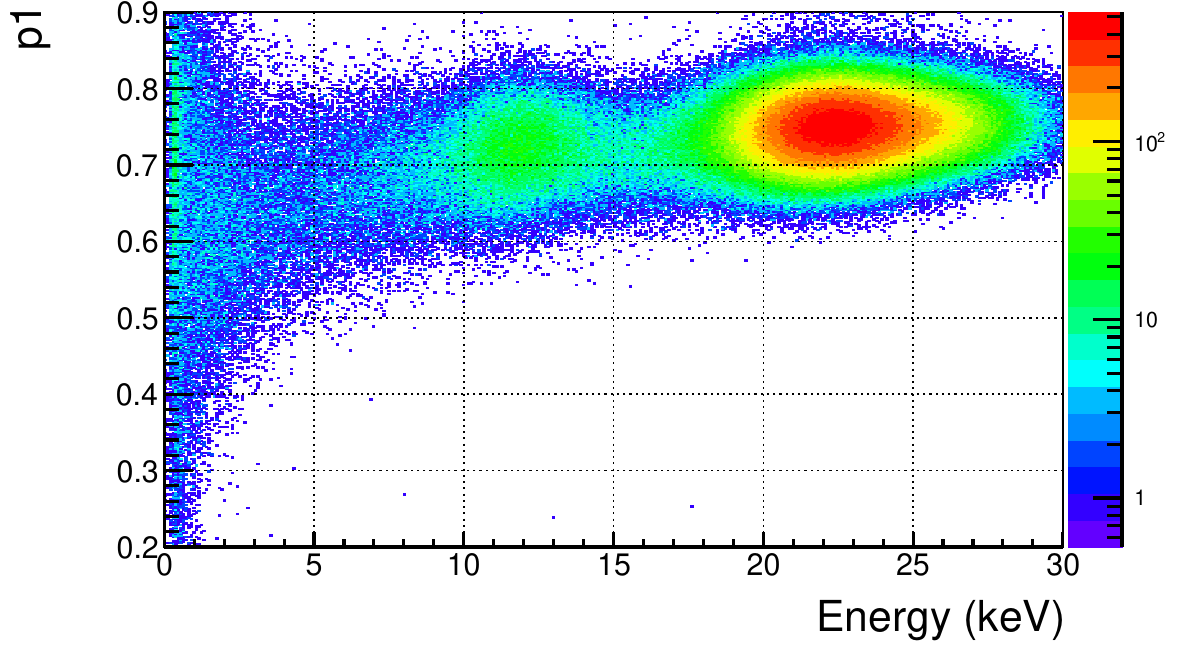}
	\end{subfigure}
	\begin{subfigure}[b]{0.49\textwidth}
		\includegraphics[width=\textwidth]{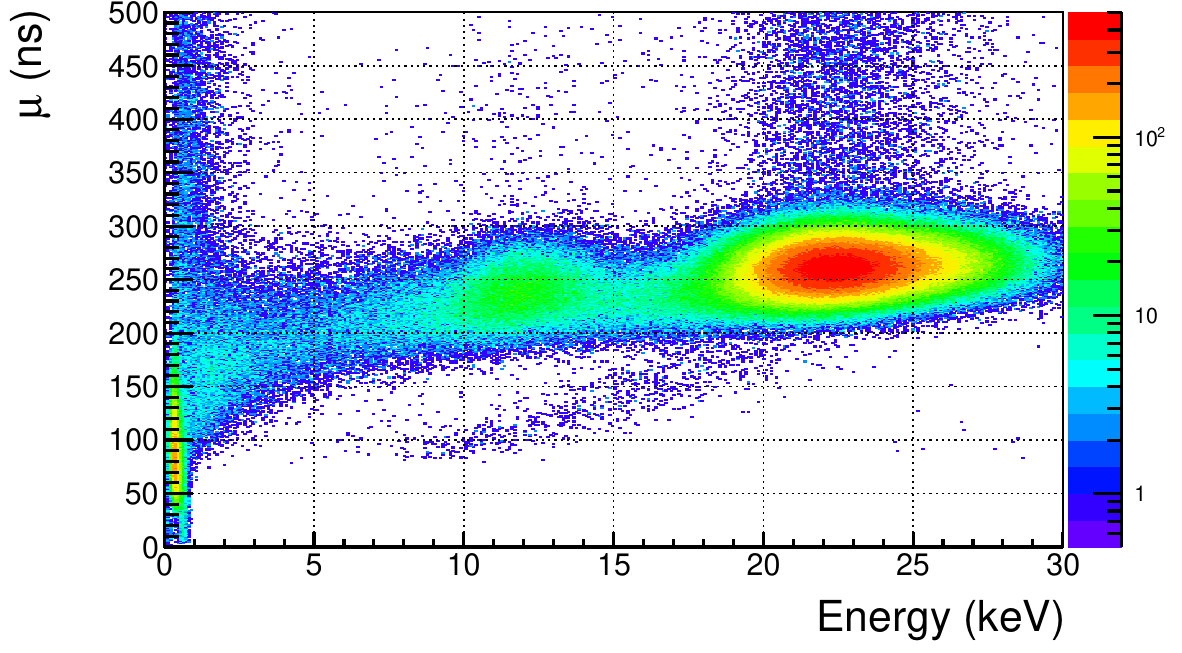}
	\end{subfigure}
	\caption{\label{muP1(E)_Exp}Scatter plots of $p_1$ (left) and $\mu$ (right) variables as a function of the energy for the experimental data corresponding to a calibration run of the D0.}
\end{figure}

\begin{figure}[h!]
	\begin{subfigure}[b]{0.49\textwidth}
		\includegraphics[width=\textwidth]{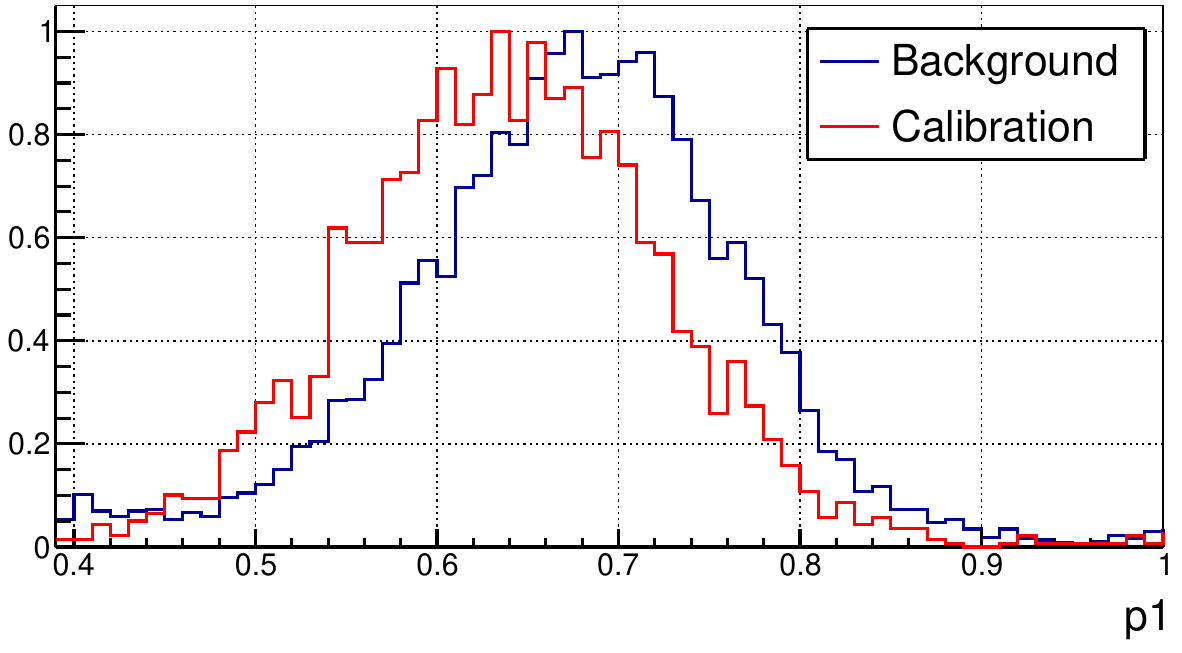}
	\end{subfigure}
	\begin{subfigure}[b]{0.49\textwidth}
		\includegraphics[width=\textwidth]{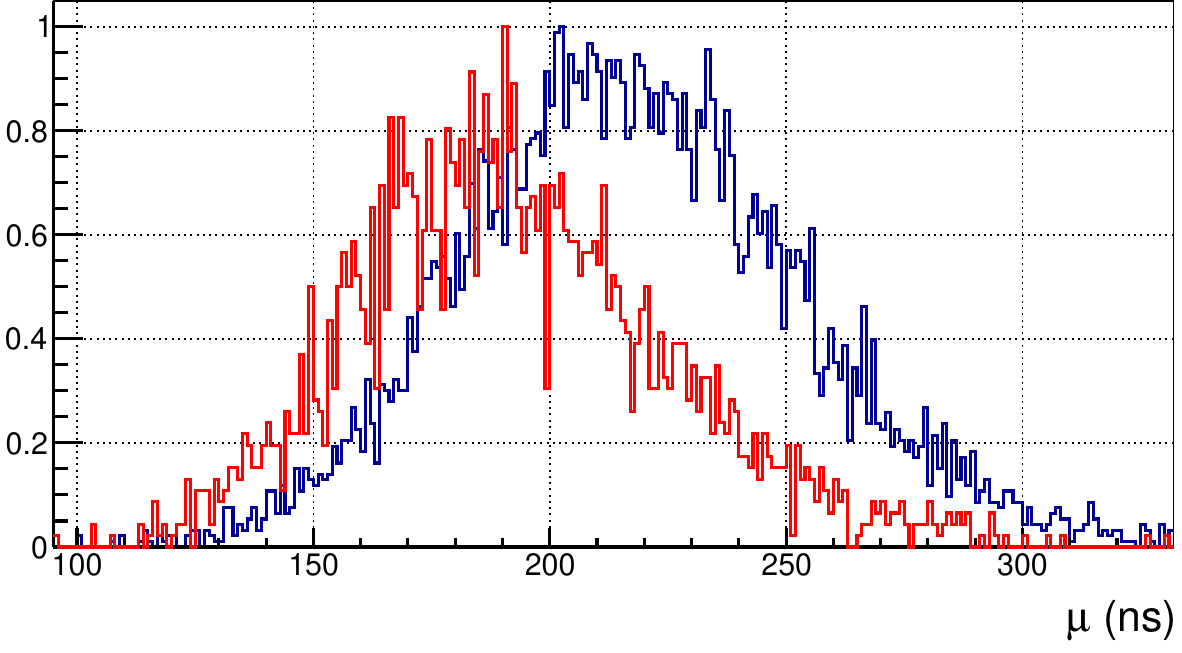}
	\end{subfigure}
	\caption{\label{P1MuExp_3keV}Comparison of the $p_1$ (left) and $\mu$ (right) distributions for events with energies in the range from 3~to 4~keV for calibration runs (red line) and background runs (blue line). Histograms are normalized to have a maximum equal to 1.}
\end{figure}

It is worth noting that both $p_1$ and $\mu$ variables depend on the energy, decreasing at low energies, as Figure~\ref{muP1(E)_Exp} shows for a calibration run of the D0. Moreover, there is a clear difference in the pulse shape variables observed between calibration and background measurements at low energies. This difference is observed in Figure~\ref{P1MuExp_3keV} for events between 3~and 4~keV. The reason behind this different pulse shape is not yet understood, although it could be related with the position within the crystal where the scintillation is produced. In calibration events the scintillation is produced near the surface and in a centered position, while for background the distribution is quite homogeneous in the crystal volume. The energy dependence of the pulse shape variables and their difference between bulk (background) and surface ($^{109}Cd$ calibration) events can be related to systematical effects in the light generation and propagation, and therefore they can be analyzed using an optical simulation, as that presented in Chapter~\ref{Chapter:OptSim}.

In the energy range below 2~keV, a distinct population of background events is observed characterized by an asymmetry between the light collected by the two PMTs of the same module. These asymmetric events have also been detected by the COSINE collaboration and may be connected to the light emission observed in PMTs by other experiments~\cite{Akimov:2015cta,Li:2015qhq}. They can be removed by using the number of peaks identified in each PMT ($n_0$ and $n_1$).

\begin{figure}[h!]
	\begin{center}
		\includegraphics[width=\textwidth]{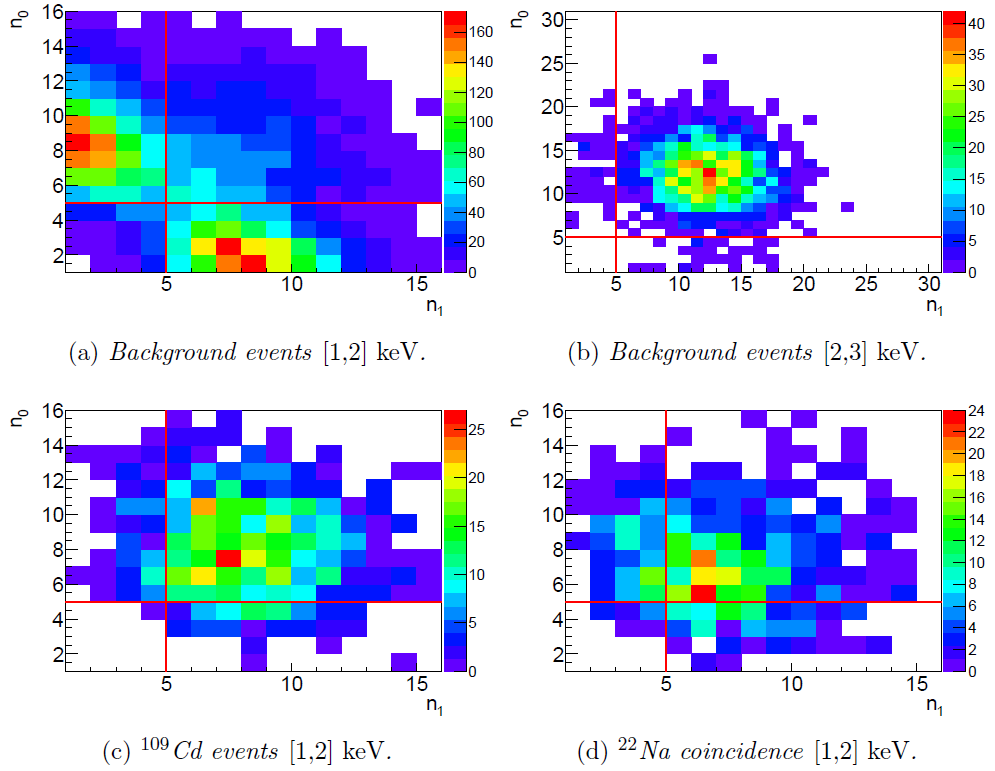}
		\caption{\label{AsymCutANAIS}Distribution in the ($n_0,n_1$) plane (number of peaks detected by the algorithm for the two PMT signals from the same module) for: background events with energy between 1 and 2~keV (a), background events with energy between 2~and 3~keV (b), $^{109}Cd$ calibration events with energy between 1 and 2~keV (c), and bulk scintillation events from the decay of $^{22}Na$ corresponding to an averaged energy of 0.9~keV that have been selected by coincidence with a 1274.5~keV deposition in another crystal (d). Image from~\cite{CoarasaCasas:2021euy}.}
	\end{center}
\end{figure}

In Figure~\ref{AsymCutANAIS}, the distribution of $n_0$ and $n_1$ values for background events in the [1-2]~keV energy range is shown. It can be observed that most of the events are highly asymmetric, with one or two peaks in one PMT and more than five in the other. However, the probability of observing a scintillation event with 1~or 2~photoelectrons in one PMT for an energy deposition of 1~keV  is approximately 3\%, as the LC measured for ANAIS modules is of about 7~phe/keV/PMT (see Section~\ref{Section:ANAIS_Module}). These events are only visible in background runs below 2~keV and are not present neither in the $^{109}Cd$ calibration runs nor in the 0.9~keV bulk scintillation population from $^{22}Na$. This suggests that these events are produced by light emission at or near the PMTs.

As the $PSV$ parameter is not able to differentiate these events from bulk NaI(Tl) scintillation, a cut based on the number of peaks detected by the peak-finding algorithm at each PMT is implemented. Specifically, a minimum of five peaks identified at each PMT signal is required to select an event (as shown by the red lines in Figure~\ref{AsymCutANAIS}).

The total efficiency of the filtering protocol for the selection of bulk NaI(Tl) scintillation events is the result of combining the trigger efficiency, the $PSV$ cut efficiency and that corresponding to the asymmetry cut. Next, the procedure followed to obtain each one of them is explained.

\begin{figure}[h!]
	\begin{center}
		\includegraphics[width=\textwidth]{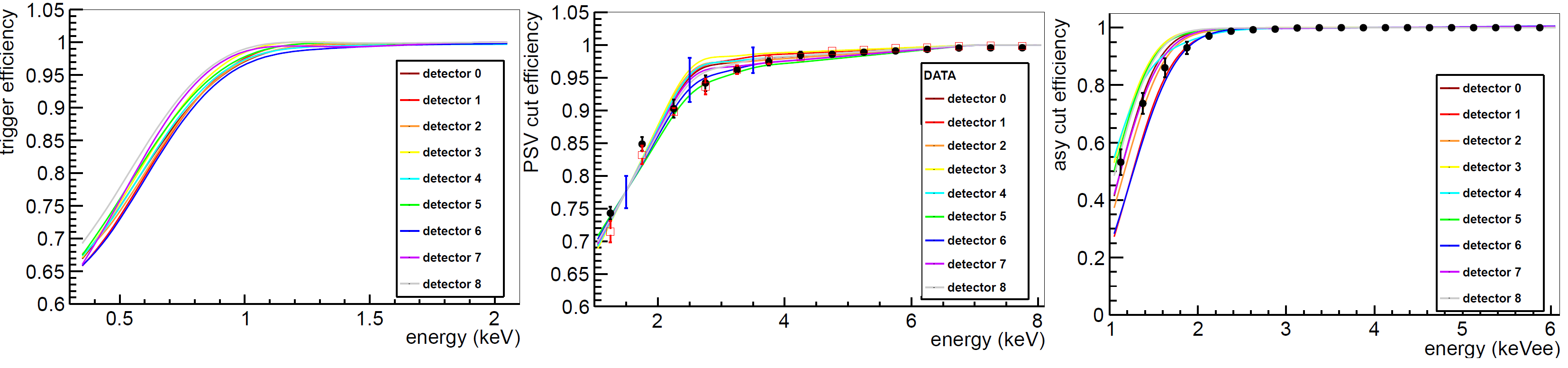}
		\caption{\label{TriggerEfficiency}Efficiencies as a function of energy for each detector and each selection criterion: trigger (left), $PSV$ cut (middle) and asymmetry cut (right). The trigger efficiency has been calculated using the MC technique, the $PSV$ efficiency is calculated both with $^{22}Na$/$^{40}K$ events (lines) and with MC (black dots for scintillation time of 230~ns and red dots for 205~ns). Blue lines of this plot represent the average of the statistical uncertainties for these efficiencies. The asymmetry cut efficiency is calculated using the $^{109}Cd$ calibration events accumulated along one year. Figures from~\cite{Amare:2018sxx}.}
	\end{center}
\end{figure}

To determine the trigger efficiency for each detector, a Monte Carlo technique is used, simulating 2$\times$10$^4$ pulses for the two PMTs with an energy randomly sampled between 0 and 10~keV. The number of phes detected in each PMT is chosen following a Poisson distribution with a mean equal to $E\cdot LC$, where $LC$ is that measured for the specific PMT. Photoelectrons are modeled as Gaussian peaks with a standard deviation of 6~ns, while their amplitudes are sampled from the SER amplitude distribution of the corresponding PMT. The arrival time of each phe is sampled from an exponential PDF with a time constant of 230~ns. The trigger efficiency as a function of energy is then calculated as the ratio of the number of events in which the first phes above the trigger threshold in both PMTs of the same module fall in a 200~ns window to the total number of simulated events generated for that energy. The results (presented in Figure~\ref{TriggerEfficiency}) show that the corresponding trigger efficiency is 97.3~$\pm$~0.8\% at 1~keV.

For the calculation of the efficiency of the $PSV$ selection, $\Vec{k}$ and $V$ are calculated in the energy range from 1 to 2~keV, and the $PSV$ cut is fixed to 3 (red ellipse in Figure~\ref{p1MuANAIS}) in order to have a 77.7\% efficiency at that energy. This calculation is done for every detector using events from $^{22}Na$ and $^{40}K$ tagged by the coincidence with a HE gamma. As these populations are polluted with fast events, a value of $p_1$ larger than 0.4 is required before calculating $\Vec{k}$ and $V$ to reject those events. Then, the efficiency is calculated in 1~keV bins for energies below 4~keV as the number of $^{22}Na$/$^{40}K$ events in the ellipsoid divided by the total number of $^{22}Na$/$^{40}K$ events having $p_1 >$~0.4. Results are shown in Figure~\ref{TriggerEfficiency}, and they are very similar for all the ANAIS-112 modules, with efficiencies increasing from 77.7\% at 1.5 keV up to 95\% at 2.5 keV and 98\% at 3.5 keV. These efficiencies were checked by using the MC technique explained above, which moreover, allowed to include the effect of the difference in the scintillation time between electronic (230~ns) and nuclear recoils (205~ns). Slight differences in the corresponding $PSV_{cut}$ efficiencies were derived (see Figure~\ref{TriggerEfficiency}~\cite{Amare:2018sxx}).

Finally, the efficiency of the asymmetry cut is calculated using all the $^{109}Cd$ calibration events accumulated along one year to have enough events in the energy region from 1~to 2~keV. It is calculated as the ratio of events surviving the selection to the total number of events, as a function of energy. Results (presented also in Figure~\ref{TriggerEfficiency}~\cite{Amare:2018sxx}) show some spread between the different detectors, being 43\% the mean value of the efficiency at 1~keV and 10\% the standard deviation. It is worth noting the strong decrease of this efficiency below 2~keV, which shows that the asymmetric events are currently the most important background contribution that limits the improvement in energy threshold in the ANAIS-112 experiment.

\section{Background model} \label{Section:ANAIS_Background}
\fancyhead[RO]{\emph{\thesection. \nameref{Section:ANAIS_Background}}}

As it was explained in Section~\ref{Section:Intro_Detection_Direct}, it is crucial for dark matter experiments to have a deep understanding of the background. In the present section, the background sources are presented as well as the model for the estimate of the different contributions to the measurement. The background model is fully explained in~\cite{Amare:2018ndh}.

The reliability of the models depends on three key factors: a precise identification of the sources of background, a careful calculation of how much impact these sources have on the experiment using specific Monte Carlo simulations, and testing of the model’s predictions against the experimental data. In the ANAIS-112 experiment, the sources of background can be divided into external and internal, within the NaI(Tl) crystal. The internal contributions to the background are the dominant ones.

Concerning the external contributions, the activity of four primordial radioactive isotopes has been measured for the most relevant components of the setup. These are: $^{40}K$, $^{232}Th$, $^{238}U$ and $^{226}Ra$. They have been quantified using a HPGe spectrometer at LSC. Table~\ref{table:ActivityExternal} summarizes the measured activities (or derived upper limits) for all the components except the PMTs, which are shown in Table~\ref{table:ActivityPMTs}.

\begin{table}[h!]
	\centering
	\begin{tabular}{|c|c|c|c|c|c|}
		\cline{2-6}
		\multicolumn{1}{c}{} & \multicolumn{5}{|c|}{Activities (mBq/kg)} \\
		\hline
		Component & $^{40}K$ & $^{232}Th$ & $^{238}U$ & $^{226}Ra$ & Other \\
		\hline
		Copper & $<$~4.9 & $<$~1.8 & $<$~62 & $<$~0.9 & $^{60}Co$: $<$~0.4\\
		Quartz & $<$~12 & $<$~2.2 & $<$~100 & $<$~1.9 & \\
		Silicone & $<$~181 & $<$~34 &  & 51$\pm$7 & \\
		Archaeological lead & & $<$~0.3 & $<$~0.2 &  & $^{210}Pb$: $<$~20 \\
		\hline
	\end{tabular} \\
	\caption{Measured activities (or derived upper limits at 95\% C.L.) for the most relevant external components of the ANAIS-112 modules and setup, except the PMTs. Values obtained from~\cite{Amare:2018ndh}.}
	\label{table:ActivityExternal}
\end{table}

\begin{table}[h!]
	\centering
	\begin{tabular}{|c|c|c|c|c|}
		\cline{2-5}
		\multicolumn{1}{c}{} & \multicolumn{4}{|c|}{Activities (mBq/PMT)} \\
		\hline
		Detector & $^{40}K$ & $^{232}Th$ & $^{238}U$ & $^{226}Ra$ \\
		\hline
		D0 - PMT 0 & 97~$\pm$~19 & 20~$\pm$~2 & 128~$\pm$~38 & 84~$\pm$~3 \\
		D0 - PMT 1 & 133~$\pm$~13 & 20~$\pm$~2 & 150~$\pm$~34 & 88~$\pm$~3 \\
		D1 - PMT 0 & 105~$\pm$~15 & 18~$\pm$~2 & 159~$\pm$~29 & 79~$\pm$~3 \\
		D1 - PMT 1 & 105~$\pm$~21 & 22~$\pm$~2 & 259~$\pm$~59 & 59~$\pm$~3 \\
		D2 - PMT 0 & 155~$\pm$~36 & 20~$\pm$~3 & 144~$\pm$~33 & 89~$\pm$~5 \\
		D2 - PMT 1 & 136~$\pm$~26 & 18~$\pm$~2 & 187~$\pm$~58 & 59~$\pm$~3 \\
		D3 - PMT 0 & 108~$\pm$~29 & 21~$\pm$~3 & 161~$\pm$~58 & 79~$\pm$~5 \\
		D3 - PMT 1 & 95~$\pm$~24 & 22~$\pm$~2 & 145~$\pm$~29 & 88~$\pm$~4 \\
		D4 - PMT 0 & 98~$\pm$~24 & 21~$\pm$~2 & 162~$\pm$~31 & 87~$\pm$~4 \\
		D4 - PMT 1 & 137~$\pm$~19 & 26~$\pm$~2 & 241~$\pm$~46 & 64~$\pm$~2 \\
		D5 - PMT 0 & 90~$\pm$~15 & 21~$\pm$~1 & 244~$\pm$~49 & 60~$\pm$~2 \\
		D5 - PMT 1 & 128~$\pm$~16 & 21~$\pm$~1 & 198~$\pm$~39 & 65~$\pm$~2 \\
		D6 - PMT 0 & 83~$\pm$~26 & 23~$\pm$~2 & 238~$\pm$~70 & 53~$\pm$~3 \\
		D6 - PMT 1 & 139~$\pm$~21 & 24~$\pm$~2 & 228~$\pm$~52 & 67~$\pm$~3 \\
		D7 - PMT 0 & 104~$\pm$~25 & 19~$\pm$~2 & 300~$\pm$~70 & 59~$\pm$~3 \\
		D7 - PMT 1 & 103~$\pm$~19 & 26~$\pm$~2 & 243~$\pm$~57 & 63~$\pm$~3 \\
		D8 - PMT 0 & 127~$\pm$~19 & 23~$\pm$~1 & 207~$\pm$~47 & 63~$\pm$~2 \\
		D8 - PMT 1 & 124~$\pm$~18 & 21~$\pm$~2 & 199~$\pm$~44 & 61~$\pm$~2 \\
		\hline
		Weighted mean & 114.9~$\pm$~4.6 & 21.6~$\pm$~0.4 & 180.2~$\pm$~9.8 & 66.7~$\pm$~0.6 \\
		\hline
	\end{tabular} \\
	\caption{Measured activities for all the PMTs of the experiment. Weighted means are also presented. Values obtained from~\cite{Amare:2018ndh}.}
	\label{table:ActivityPMTs}
\end{table}

The primary background contributions are those from NaI(Tl) crystals themselves, and they have been quantified for every ANAIS-112 detector. As explained before, $^{22}Na$ and $^{40}K$ activities were quantified using the coincidences between different modules, considering the efficiency of this process, which has been calculated using a MC simulation. The activity of $^{210}Pb$, $^{232}Th$ and $^{238}U$ was determined through the measurement of the alpha rate using pulse shape analysis and delayed coincidences ($Bi$/$Po$ sequences). All of them are presented in Table~\ref{table:ActivityNaI}.

Cosmogenic isotopes produced in the crystals when they were exposed to cosmic radiation were also examined. Among them, the most relevant were several $I$ and $Te$ isotopes with half-lives from tens to hundreds of days, but more worrisome were others, having longer half-life as $^{109}Cd$, $^{113}Sn$ and $^{22}Na$ (463~days, 115~days and 2.6~years, respectively). The two first isotopes are potentially harmful, as they generate energy depositions in the ROI following L-shell EC. Finally, an additional background source was observed in the very low energy region, which fits with a tritium activity of 0.20~mBq/kg for the D0 and D1 and 0.09~mBq/kg for the others. The initial activities of all of these cosmogenically produced isotopes in every ANAIS-112 module when going underground are presented in Table~\ref{table:ActivityNaI2}. As the detectors D0 and D1 were installed underground years before this full analysis was carried out, the activity of the $^{109}Cd$ and $^{113}Sn$ isotopes was too low to observe their contribution.

\begin{table}[h!]
	\centering
	\begin{tabular}{|c|c|c|c|c|}
		\cline{2-5}
		\multicolumn{1}{c}{} & \multicolumn{4}{|c|}{Activities (mBq/kg)} \\
		\hline
		Module & $^{40}K$ & $^{232}Th$ & $^{238}U$ & $^{210}Pb$ \\
		\hline
		D0 & 1.33~$\pm$~0.04 & (4.0~$\pm$~1.0)$\times$10$^{-3}$ & (10.0~$\pm$~2.0)$\times$10$^{-3}$ & 3.15~$\pm$~0.10 \\
		D1 & 1.21~$\pm$~0.04 & & & 3.15~$\pm$~0.10 \\
		D2 & 1.07~$\pm$~0.03 & (0.7~$\pm$~0.1)$\times$10$^{-3}$ & (2.7~$\pm$~0.2)$\times$10$^{-3}$ & 0.70~$\pm$~0.10 \\
		D3 & 0.07~$\pm$~0.03 & & & 1.80~$\pm$~0.10 \\
		D4 & 0.54~$\pm$~0.04 & & & 1.80~$\pm$~0.10 \\
		D5 & 1.11~$\pm$~0.02 & & & 0.78~$\pm$~0.01 \\
		D6 & 0.95~$\pm$~0.03 & (1.3~$\pm$~0.1)$\times$10$^{-3}$ & & 0.81~$\pm$~0.01 \\
		D7 & 0.96~$\pm$~0.03 & (1.0~$\pm$~0.1)$\times$10$^{-3}$ & & 0.80~$\pm$~0.01 \\
		D8 & 0.76~$\pm$~0.02 & (0.4~$\pm$~0.1)$\times$10$^{-3}$ & & 0.74~$\pm$~0.01 \\
		\hline
	\end{tabular} \\
	\caption{Measured activities for the non-cosmogenic radioactive isotopes in ANAIS-112 crystals. Values obtained from~\cite{Amare:2018ndh}.}
	\label{table:ActivityNaI}
\end{table}

\begin{table}[h!]
	\centering
	\begin{tabular}{|c|c|c|c|}
		\cline{2-4}
		\multicolumn{1}{c}{} & \multicolumn{3}{|c|}{Activities (kg$^{-1}$day$^{-1}$)} \\
		\hline
		Module & $^{22}Na$ & $^{109}Cd$ & $^{113}Sn$ \\
		\hline
		D0 & 155.0~$\pm$~11.0 &   &   \\
		D1 & 168.0~$\pm$~11.0 &   &   \\
		D2 & 43.9~$\pm$~6.0 & 11.7~$\pm$~2.4 & 18.7~$\pm$~4.4 \\
		D3 & 68.6~$\pm$~4.6 & 6.5~$\pm$~2.1 & 11.5~$\pm$~4.2 \\
		D4 & 61.8~$\pm$~3.1 & 10.0~$\pm$~2.7 & 20.5~$\pm$~6.2 \\
		D5 & 43.7~$\pm$~2.3 & 9.2~$\pm$~2.3 & 19.2~$\pm$~4.6 \\
		D6 & 53.8~$\pm$~2.7 & 8.9~$\pm$~1.6 & 14.8~$\pm$~2.9 \\
		D7 & 55.6~$\pm$~2.7 & 7.9~$\pm$~1.4 & 13.4~$\pm$~2.5 \\
		D8 & 56.4~$\pm$~2.8 & 7.2~$\pm$~1.4 & 12.2~$\pm$~2.5 \\
		\hline
	\end{tabular} \\
	\caption{Obtained initial activities for the cosmogenic radioactive isotopes in ANAIS-112 crystals. Values obtained from~\cite{Amare:2018ndh}.}
	\label{table:ActivityNaI2}
\end{table}

The GEANT4 package~\cite{GEANT4:2002zbu} was used to simulate all background sources. It is a software toolkit developed at CERN for the simulation of the particle passage through matter. Its main applications are in high-energy, nuclear, and accelerator physics, but it is useful for modeling the interaction of particles/radiation in any physical system, allowing to simulate a wide range of physical processes across a broad range of energies (from~eV to~TeV).

\begin{figure}[h!]
	\begin{center}
		\includegraphics[width=0.75\textwidth]{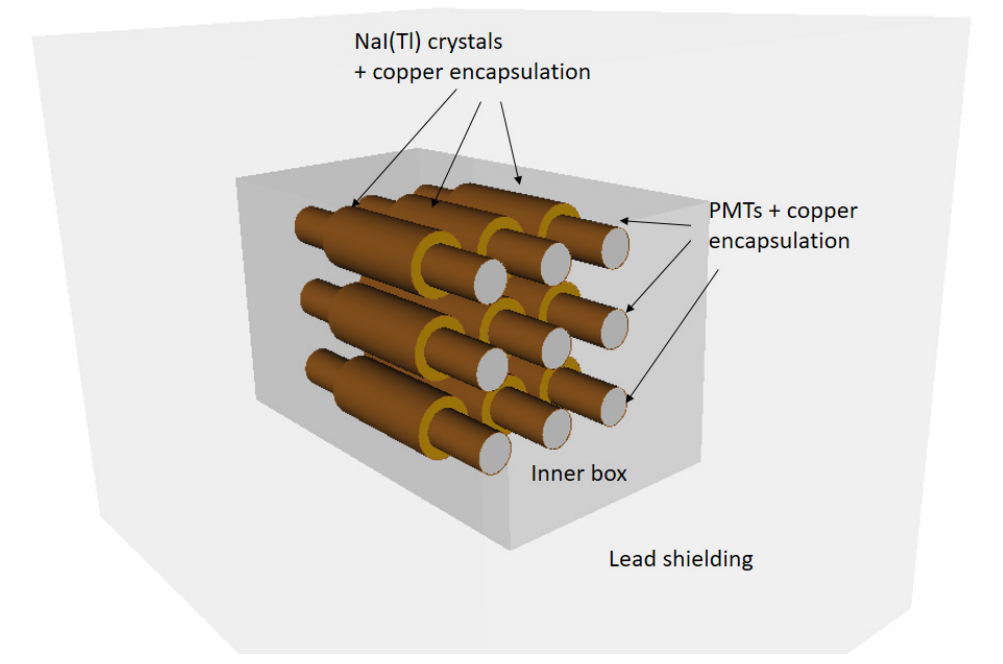}
		\caption{\label{GeoSimANAIS}Geometry of the ANAIS–112 setup implemented in the GEANT4 simulations. Image from~\cite{Amare:2018ndh}.}
	\end{center}
\end{figure}

\begin{figure}[h!]
	\begin{center}
		\includegraphics[width=0.75\textwidth]{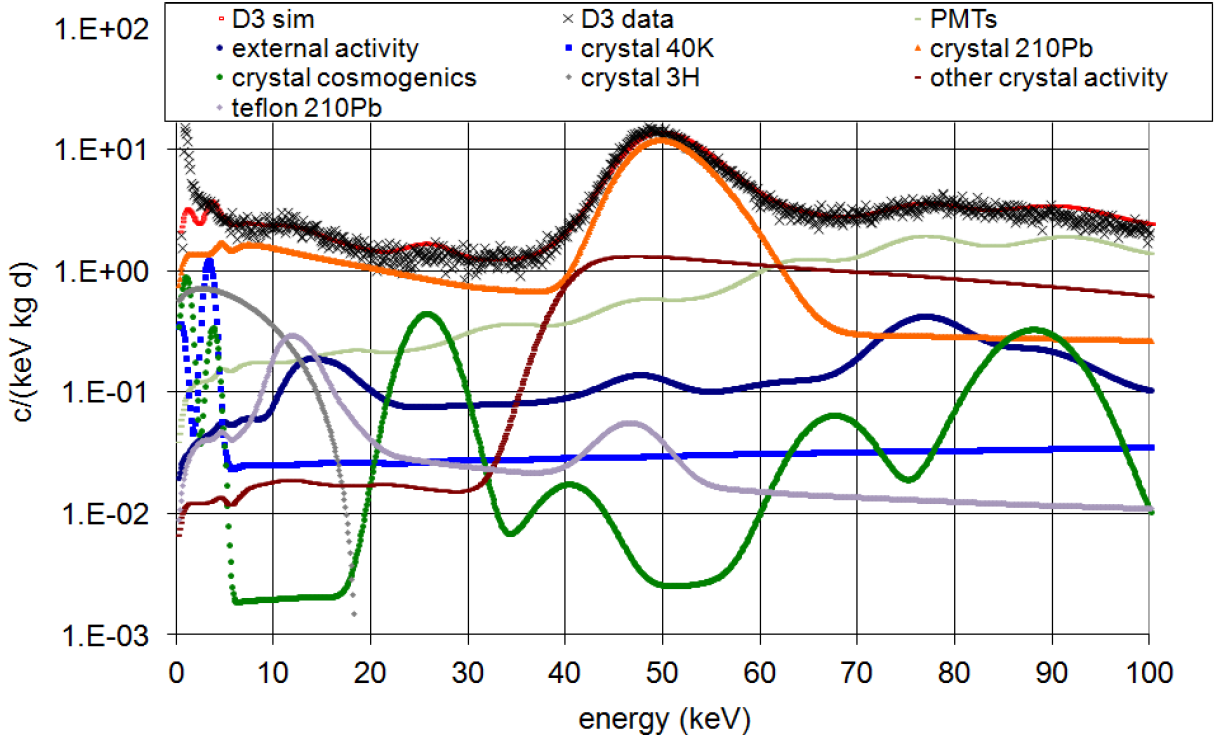}
		\caption{\label{LEbkgANAIS}Low energy spectra from each background source, the sum of all of them and the measured background. Data corresponds to single-hits in D3 detector for the first year of data taking. Image from~\cite{Amare:2018ndh}.}
	\end{center}
\end{figure}

\begin{figure}[h!]
	\begin{center}
		\includegraphics[width=0.75\textwidth]{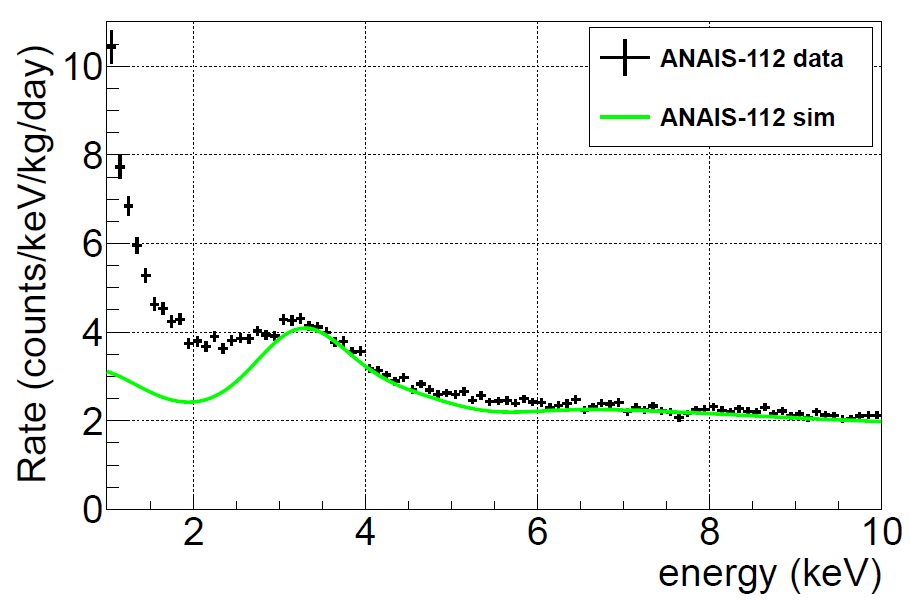}
		\caption{\label{LEBkgComparison}Comparison between the modeled and experimental data of the low energy spectrum for single-hit events in all detectors after filtering and efficiency correction. Image from~\cite{Amare:2018ndh}.}
	\end{center}
\end{figure}

The simulation of the ANAIS-112 experiment included the lead shielding, detectors, NaI(Tl) crystals, teflon wrapping, copper encapsulation, mylar windows, silicone pads, quartz windows, and PMTs. The complete geometry is shown in Figure~\ref{GeoSimANAIS}. The energy spectra of the ANAIS-112 detectors were constructed by summing the spectra of all simulated contributions. The spectrum for the detector D3, as example, is shown in Figure~\ref{LEbkgANAIS}. Figure~\ref{LEBkgComparison} shows a comparison between the simulated and experimental data of the low energy spectrum for single-hit events in all detectors after filtering and efficiency correction.

The predictions of the model for single hit events at energies above 100~keV agree with the measured background within 5.6\% across all the detectors. When considering events from 200~to 2000~keV that involve energy being deposited in more than one detector, the model matches the measurements within 2.9\%. At low energies, a good agreement between model and measurements is found above 3~keV but there are inconsistencies at lower energies. In the 1~to 2~keV energy range, the model deviated by 54\%. However, in the 2~to 6~keV range, the deviation was only 10.7\%. This event excess may be caused by light events originated in other detector components or radioactive background contributions that have not been accounted for in the model. In particular, the possible contribution of the light emission at the PMTs will be analysed in Section~\ref{Section:SIM_Res_cherenkov}. Understanding the origin of this excess would be important to improve the sensitivity of the experiment and better evaluate the prospects of future projects under development aiming at a reduced threshold energy. 

\section{Strategies to improve the sensitivity} \label{Section:ANAIS_Improves}
\fancyhead[RO]{\emph{\thesection. \nameref{Section:ANAIS_Improves}}}

\subsection{Optical simulation of ANAIS-112} \label{Section:ANAIS_Improves_Sim}

As shown in previous sections, there are different performance features in the ANAIS-112 data requiring further understanding. Some of them could be related with the light production, propagation and absorption because, for instance, of spatial inhomogeneities in the crystal properties, or the contribution of different scintillation mechanisms. A very interesting way to account for these effects is to use MC techniques that reproduce both the particles interaction processes and the light production and transport mechanisms. The GEANT4 package~\cite{GEANT4:2002zbu} allows to simulate an ANAIS-112 module including all these effects.

This simulation could allow to improve the understanding of the detector response to different energy depositions in different positions of the crystals, but also in other detector components, enabling the development of more efficient filtering protocols. Among those performance features requiring understanding we can analyze with such an optical simulation for instance, the Cherenkov emission in the PMTs, quartz and other transparent media in the module, the possible different pulse shape for superficial/bulk events, and separating the different contributions to the experimentally observed energy resolution. Additionally, a dataset of pulses from both bulk NaI(Tl) scintillation and pulses with different origin can be generated. These datasets could be used to train an event identification neural network that could be incorporated into the filtering process of the experiment.

A first implementation of such an optical simulation of one of the ANAIS-112 modules has been developed and it is explained in Chapter~\ref{Chapter:OptSim}. First results suggest this simulation could have many interesting applications in the future.

\subsection{Replacing PMTs by Silicon Photomultipliers (SiPMs)} \label{Section:ANAIS_Improves_SiPMs}

As it was explained in Chapter~\ref{Chapter:Intro}, PMTs have been for long time the most efficient light detectors. They are highly sensitive in the wavelength range of the NaI(Tl) scintillation peaks, have a fast response time, linear response and wide dynamic range. However, they can be a relevant source of background, which can be a disadvantage in experiments that require low background, as ANAIS-112. In addition, the Cherenkov light produced within their transparent components can contribute to the background of the experiment. Moreover, they can emit light through various mechanisms, as observed by Double Chooz collaboration~\cite{DoubleChooz:2016ibm} (as it was explained in Section~\ref{Section:Intro_Scintillators_PMTs}). Light produced at or near the PMTs could explain some of the strongly asymmetric events observed in ANAIS-112. Because of this, removing the PMTs without losing LCE will be important to progress towards lower energy threshold. 

Recently, there has been a surge in the use of Silicon PhotoMultipliers (SiPMs) in physics experiments~\cite{Buzhan:2003ur,Golovin:2004jt,Renker:2006ay,Danilov:2008fi,Simon:2018xzl}. They are solid-state single-photon sensitive devices based on Single-Photon Avalanche Diodes (SPAD) implemented on a common silicon substrate and operated in Geiger mode (see Chapter~\ref{Chapter:SiPM_Intro}). They can be as efficient as the PMTs in the detection of scintillation~\cite{Kovalchuk:2005cp,Laurenti:2008zz}, but are much less massive and can be easily made of very radio-pure materials.

SiPMs are already established as a versatile and effective technology for photon detection in experimental physics, and their use in the ANAIS experiment can help to improve its sensitivity. Their radiopurity and low mass implies lower background contribution than PMTs. Moreover, other possible light emission effects from PMTs as those observed in~\cite{DoubleChooz:2016ibm} associated to the HV applied to the PMTs, would not be present in SiPMs which operate at low voltage. However, these devices present also some noise effects as optical crosstalk and high dark count rate, which can be an inconvenient for their application in the ANAIS experiment.

With the objective of a possible future upgrade of the ANAIS experiment using these devices, and to analyze the possible sensitivity improvement, a new research line within ANAIS started, focusing first in the characterization of SiPMs from different providers and later on the light readout of NaI and NaI(Tl) crystals. For such an application, the SiPM dark current has to be reduced by operating the system at low temperatures and this requires an special prototype design. In Chapter~\ref{Chapter:SiPM_Intro}, an overview of the SiPM characteristics is presented. In Chapter~\ref{Chapter:SiPMStar2} they were applied to the light readout of both NaI and NaI(Tl) crystals at different temperatures at LNGS. Finally, in Chapter~\ref{Chapter:SiPMZgz} the design and first steps of the characterization of the first NaI(Tl)+SiPM detector prototype are shown.

%% file: ASim2.tex
\chapter{Optical simulations for ANAIS-112} \label{Chapter:OptSim}

\fancyhead[LE]{\emph{Chapter \thechapter. \nameref{Chapter:OptSim}}}

The background events rate measured in the ROI by the ANAIS-112 experiment exceeds the estimated rate using a background model specifically developed for this experiment. Such background excess can be due to events having a different origin than those produced by the NaI(Tl) scintillation that pass the event filtering, but also, to some radioactive background non well accounted for in the ANAIS-112 background model. Moreover, differences in the pulse shape parameters have been observed between calibration and background measurements, something that has not been understood yet. As optical processes play a very fundamental role in the building of the detector signal, an optical simulation becomes necessary for a deep understanding of the pulse shape characteristics corresponding either to bulk scintillation produced by energy depositions in the NaI(Tl) crystal or to other light signals with different origin.

For this purpose, a MC simulation using GEANT4 package has been developed with the geometry of one ANAIS module. This simulation reproduces the optical processes (all of them modeled in the unified optical model of GEANT4~\cite{Geant4App}), from the light emission after an energy deposition to the collection of the photons in the PMT-photocathode, considering the propagation, reflection, diffusion and absorption processes. Moreover, this simulation generates the detector output signal taking into account the PMT response. In this process, the information about the LC and the SER of ANAIS PMTs is required. In this chapter we will describe the optical simulation developed and some applications. On the one hand, we will analyse the effect of the digitizer parameters on the energy calibration and resolution. We will also analyse the detection of Cherenkov emission produced by the $^{40}K$ contamination in the borosilicate of the PMTs and by the $^{222}Rn$ in the air of the laboratory. These events have to be disentangled from bulk scintillation ones, and we can use the simulation to estimate the efficiencies both for triggering and filtering of these events.

Geometry, optical properties of the materials, and optical processes included in the simulation are described in Section~\ref{Section:SIM_Const}. The validation of the simulation and the comparison between simulated and measured pulses for $^{109}Cd$ calibration events and other populations are reported in Section~\ref{Section:SIM_Val}. In Section~\ref{Section:SIM_Res} the main results obtained from the simulation are presented. Finally, in Section~\ref{Section:SIM_Conclusions} some conclusions are drawn, emphasizing that this work is still in progress, and therefore some future developments have to be implemented before drawing more robust conclusions or moving towards more ambitious applications.

\section{Construction of the simulation} \label{Section:SIM_Const}
\fancyhead[RO]{\emph{\thesection. \nameref{Section:SIM_Const}}}

Simulations of the ANAIS-112 setup, as well as of previous prototypes, had been carried out previously~\cite{Cebrian:2012yda,Amare:2016rbf,Amare:2018ndh}. These previous simulations produced as output the energy depositions in the NaI(Tl) sensitive volume from the different particles produced in the decay of the radioactive isotopes contaminating the detector components. In the simulations presented in this section the output of the simulation is the number of photons reaching the photocathode of each of the two PMTs in one ANAIS-112 module, which later is converted into an electrical pulse. This development, summarized in the following sections, has been carried out in the frame of two final degree projects in 2021 (Marta Villalba~\cite{MartaVillalba}, and Diego Alcón~\cite{DiegoAlcon}), and a master thesis in 2023 (Víctor Pérez~\cite{VictorPerez}), all of them under my supervision.

\subsection{Description of the simulated setup} \label{Section:SIM_Const_Geometry}

\begin{figure}[h!]
	\begin{center}
		\includegraphics[width=\textwidth]{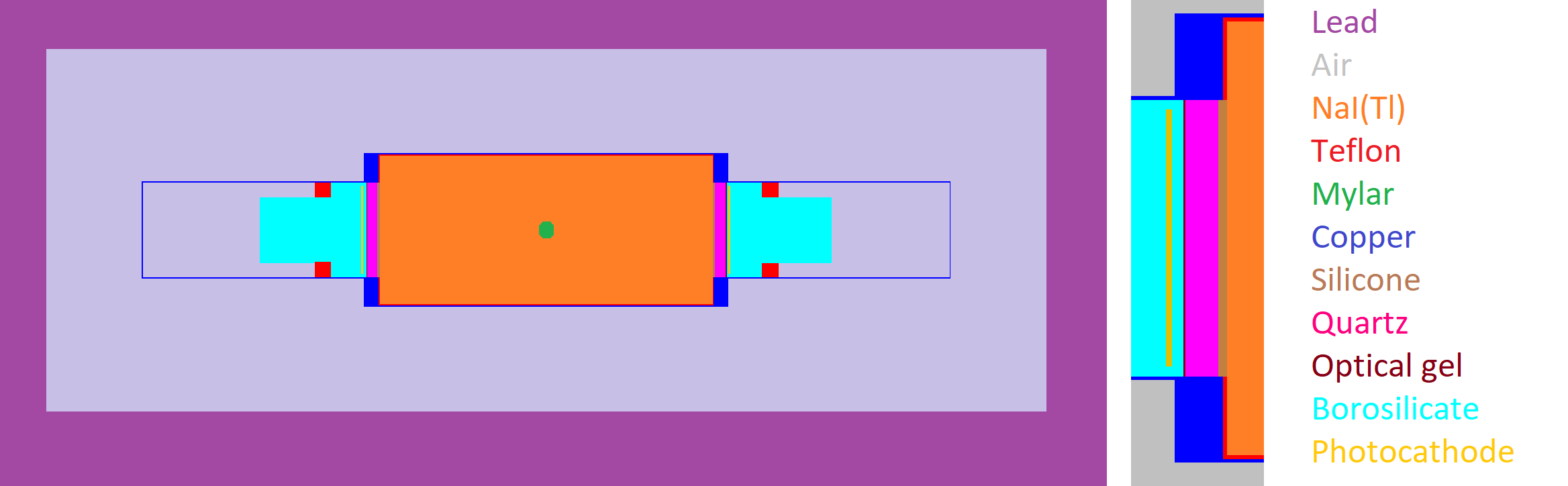}
		\caption{\label{OptSimGeo}Geometry of the ANAIS-112 module defined in the simulation. The left picture shows the complete geometry, while in the right picture a close view of the PMT-NaI(Tl) optical coupling region is shown.}
	\end{center}
\end{figure}

In this simulation, a single module of ANAIS (described in Section~\ref{Section:ANAIS_Setup}) has been considered. The simulation code is an improved update from the code developed in~\cite{TesisPatricia}. The module is inside a 20~cm-thick lead shielding, whose dimensions are 70$\times$70$\times$130~cm$^3$. The module consists of a cylindrical NaI(Tl) crystal (11.75" length and 4.75" diameter) coated by a 0.5~mm-thick teflon layer and coupled to two quartz windows (1~cm-thick) at both sides with a 3~mm thick silicone pad as refractive index coupling medium. This volume is housed in 1.5~mm-thick OFE copper and tightly closed to prevent humidity from damaging the highly hygroscopic NaI(Tl) crystal. By mistake, in the simulation the copper housing was considered 1~mm thick. No relevant effect on the conclusions drawn is expected. A Mylar window (20~$\mu$m-thick) was built on one lateral of the copper housing to allow for low-energy gamma calibration with external sources (see Section~\ref{Section:ANAIS_Calibration}). Two High QE PMTs are coupled to the quartz windows with optical gel (0.1~mm-thick). The PMTs are also housed inside copper caps. The inner part of the lead cavity, as simulated, can be seen in Figure~\ref{OptSimGeo}.

\begin{figure}[h!]
	\begin{center}
		\includegraphics[width=\textwidth]{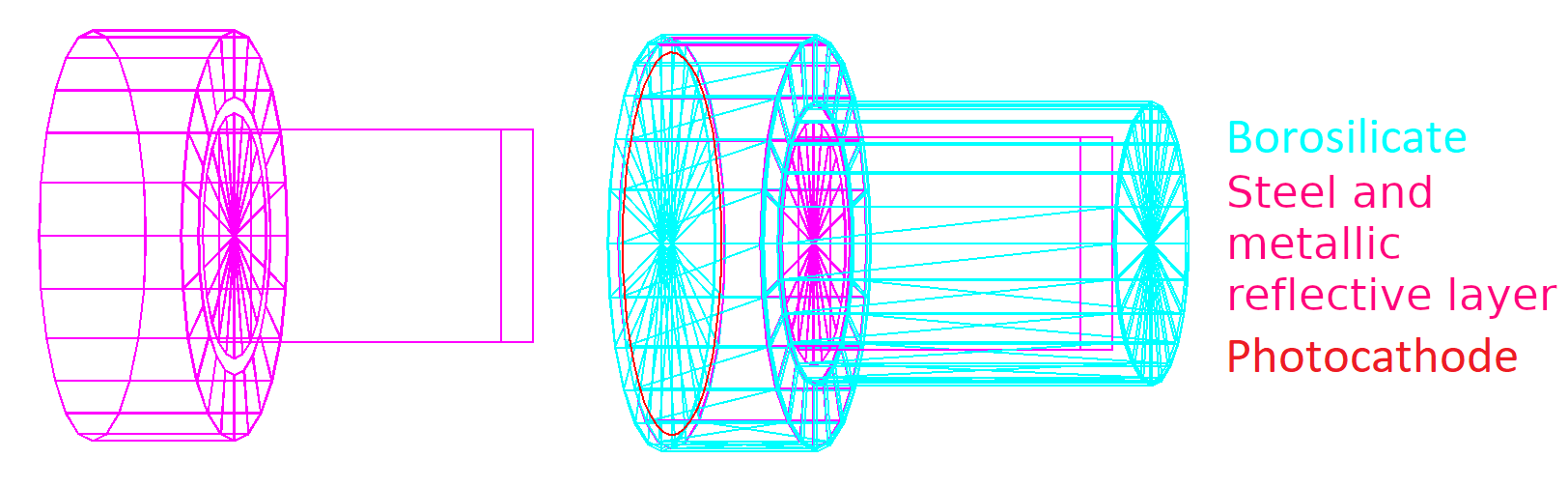}
		\caption{\label{OptSimGeoPMT}Modelled geometry of a PMT. The left picture shows all the metallic part of the PMT structure, which includes the reflective layer covering partially the inner "head" surface and the dynodes structure in the "body". The right picture includes also the borosilicate and the photocathode.}
	\end{center}
\end{figure}

The PMTs have been modelled in the simulation following the design of the R12669SEL2 PMTs used in the ANAIS-112 modules (see Section~\ref{Section:ANAIS_Setup}). A picture of these PMTs was shown in Figure~\ref{PMTsANAIS}. They have been modelled as two cylinders with different radius made of 0.5~mm-thick borosilicate with vacuum inside. The two cylinders are the "head" (including the photocathode and PMT region before the first dynode) and the "body" (containing the structure of dynodes). The dimensions of these cylinders were obtained by a direct measurement. The radius of the head is 38.1~mm (the same as that of all the volumes that conform the optical window) while the radius of the body is 26~mm. Their lengths are 29.4~and 63.6~mm, respectively. On the other hand, the dimensions of the internal components of the PMTs were aproximated, as they could not be directly measured and there is not information on this issue in the HAMAMATSU technical description of the PMT. The optical window of the PMT head consists of 2~mm-thick borosilicate. In the internal surface of the PMT head there are two different thin-layer depositions: the 2~nm-thick Bialkali photocathode (Sb-Rb-Cs)~\cite{Motta:2004yx} with 35~mm radius, covering almost the whole optical window, and a 0.1~mm-thick metallic reflector covering the rest of the internal surface but a small region in the head-body frontier. The PMT body is connected to the head and contains the dynode structure, which has been modelled as a compact parallelepiped of steel with dimensions 23$\times$39$\times$54~mm$^3$. A schematic picture of the geometry of the PMTs can be seen in Figure~\ref{OptSimGeoPMT}. Finally, two pieces of teflon are placed around the PMTs to fix them in their positions. These pieces have been defined as rings with a length of 15~mm, an internal radius equal to the radius of the "body" and as the external radius, that of the copper housing of the PMTs. The space between the copper housing and the PMTs is filled with air. The PMT modeling can be improved in the future, as for example defining a more complex dynode structure or including some components of the PMT wiring and electric socket.

\begin{figure}[h!]
	\begin{center}
		\includegraphics[width=0.75\textwidth]{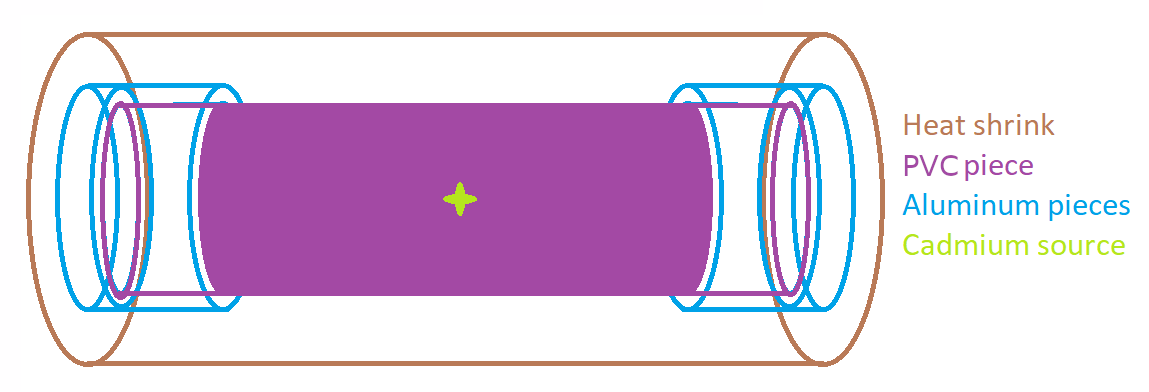}
		\caption{\label{OptSimGeoMacarron}Geometry simulated for the encapsulation of the $^{109}Cd$ source.}
	\end{center}
\end{figure}

In order to simulate the ANAIS-112 calibration with $^{109}Cd$ external sources, it is required to define the geometry of the piece where the radioactive isotope is encapsulated (see Section~\ref{Section:ANAIS_Calibration} for more information). The geometry of this piece is shown in Figure~\ref{OptSimGeoMacarron}. The piece is made of polyvinyl chloride (PVC) and consists of a cylinder with a radius of 1.5~mm and a length of 5~mm, being the radioactive source considered point-like and placed at the center. This PVC piece is connected to a nylon wire with two aluminum cylinders (1.5~mm length, 1.6~mm radius), one at each side, partially covering the PVC piece. PVC and aluminum cylinders are covered by a heat-shrink tube, which is made of PVC with a high bromine content, having 6~mm length and 0.3~mm thickness. We will refer to the whole piece as $^{109}Cd$ source in the following.

Once defined the different volumes, with the corresponding dimensions, material composition and relevant properties, active and non-active volumes have to be established. Active volumes in the geometry are those in which the simulation stores the relevant information after every interaction (hit), such as the energy deposition, the time and position of the hit and the identification of the track and the parent particle. In the simulations presented in this chapter, we have considered as active volumes:
\begin{itemize}
	\item NaI(Tl) crystal (NaI)
	\item All the teflon volumes (Tef)
	\item Optical windows: Silicone-pads, quartz windows and optical gel (Win)
	\item Borosilicate of the PMTs (Boro)
	\item Photocathode (PK)
\end{itemize}
While the non-active volumes are:
\begin{itemize}
	\item Lead shielding
	\item Air inside and outside the lead shielding
	\item $^{109}Cd$ source
	\item Copper housing
\end{itemize}
In addition to the physical processes required for the simulation of the passage of radiation and particles through our system, in our simulation two mechanisms for the conversion of the energy deposited into light are activated: Cherenkov and scintillation light emissions, enabling also optical processes for the propagation, absorption and re-emission of the produced photons. This requires to define the optical properties of the relevant materials, which are summarized next. 

\subsection{Optical properties of the relevant materials} \label{Section:SIM_Const_Prop}

The properties required for the optical simulation are the refractive index and optical absorption length for dielectric materials, and the reflectivity of the surface for metals. These properties depend on the wavelength range of interest, which in our case is determined by the scintillation of the NaI(Tl) and the Cherenkov emission.

\begin{figure}[h!]
	\begin{center}
		\includegraphics[width=0.75\textwidth]{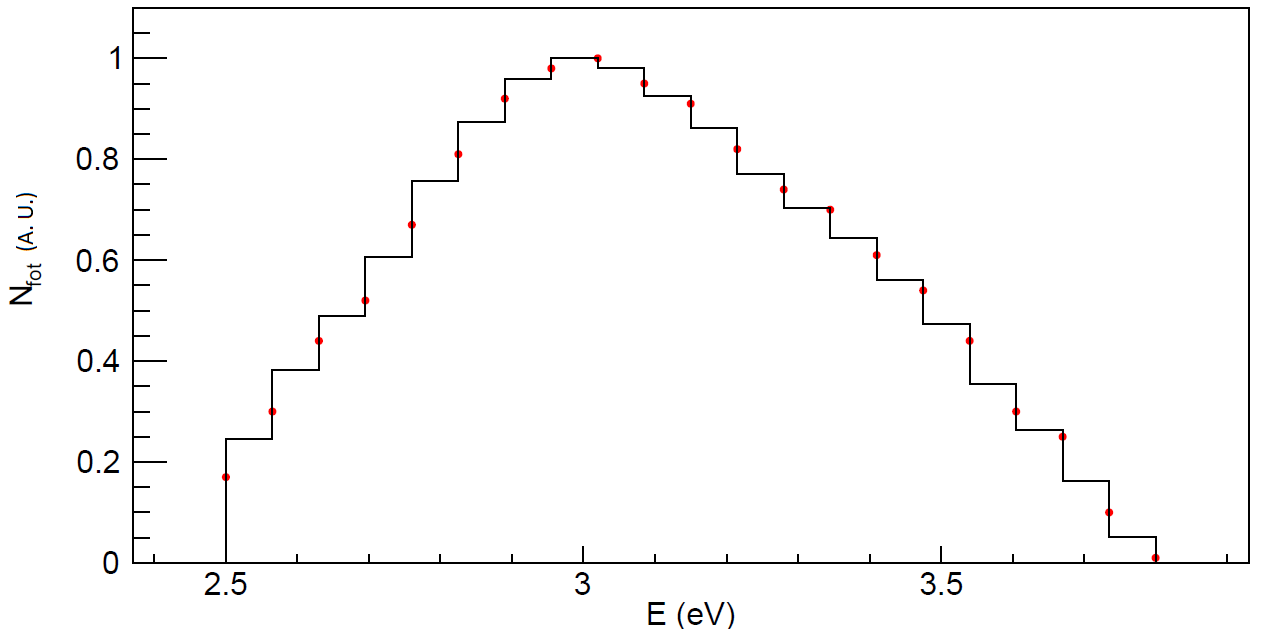}
		\caption{\label{EmisionSpectraSim}Probability distribution function of the emission spectrum of the NaI(Tl) crystal (black line). The red points represent the discrete values of the emission provided as input (taken from~\cite{Knoll:2000fj}). Image from~\cite{MartaVillalba}.}
	\end{center}
\end{figure}

The characteristics of the NaI(Tl) scintillation have been introduced as in~\cite{Knoll:2000fj}: a main scintillation time constant of 230~ns, a light yield of 44~photons/keV and the emission spectrum that peaks at 420~nm ($\sim$~3~eV). For the latter, GEANT4 requires as input a series of discrete values (red points in Figure~\ref{EmisionSpectraSim}). The scintillation wavelength will be sampled from the emission spectrum as shown by the continuous black line in Figure~\ref{EmisionSpectraSim}, i.e., GEANT4 does not interpolate between two neighbor input points of the spectrum, but takes a constant value equal to the mean of both.

\begin{figure}[h!]
	\begin{subfigure}[b]{0.49\textwidth}
		\includegraphics[width=\textwidth]{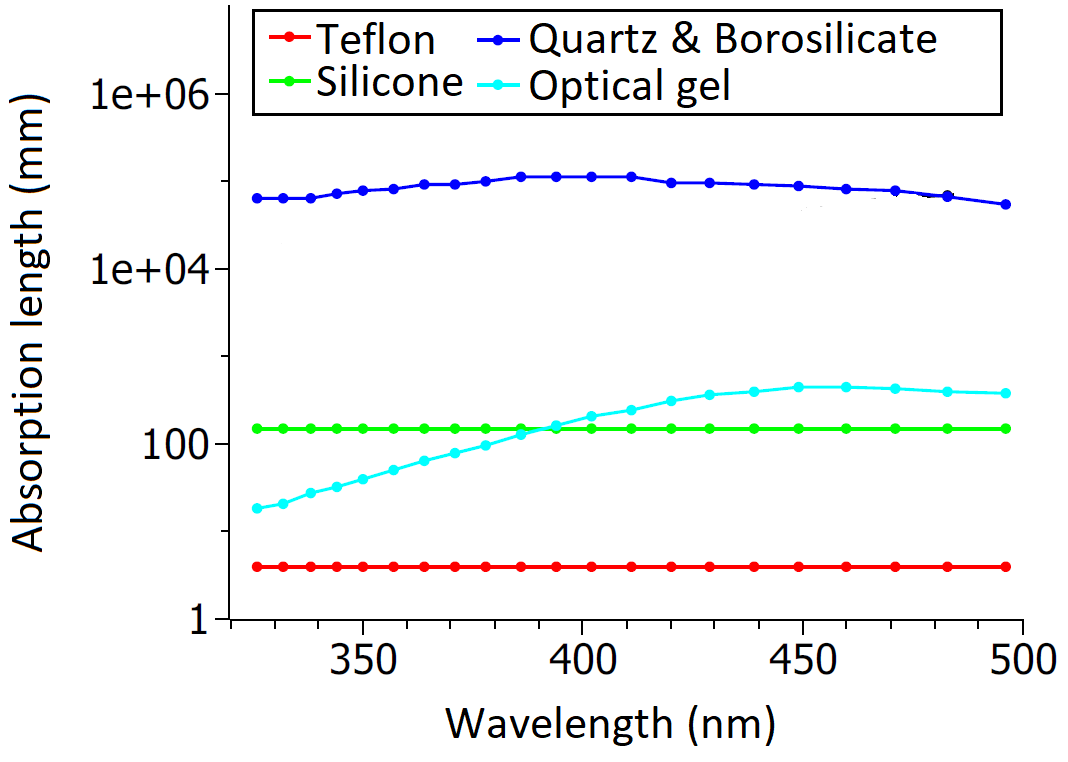}
	\end{subfigure}
	\begin{subfigure}[b]{0.49\textwidth}
		\includegraphics[width=\textwidth]{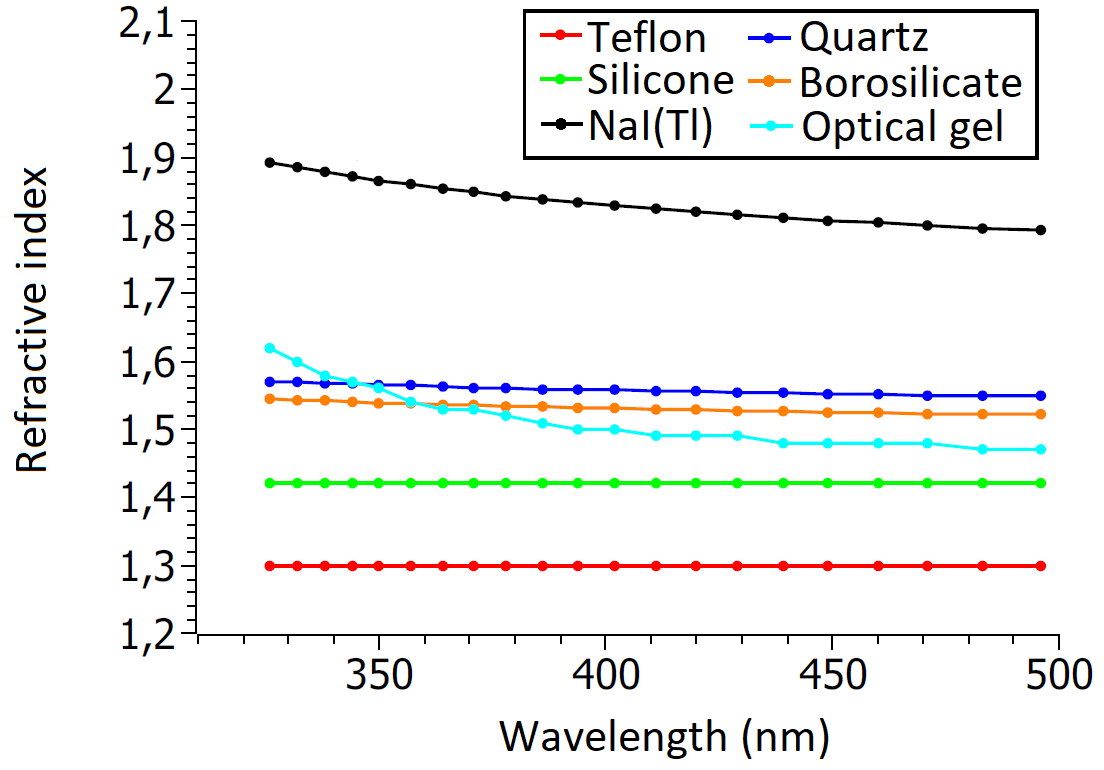}
	\end{subfigure}
	\caption{\label{OpticalPropSim}Absorption length (left) and refractive index (right) as a function of the photon wavelength for the dielectric materials~\cite{OpticalGel,Haerus,Matmatch,SiPad,Li1976RefractiveIO,Malitson:65}. Images from~\cite{MartaVillalba}.}
\end{figure}

The range of optical photon energies considered in the simulation is from 2.5~to 3.8~eV, which corresponds to wavelengths between 326~and 496~nm. This range covers the emission of the NaI(Tl) crystal~\cite{Knoll:2000fj}. Figure~\ref{OpticalPropSim} shows the absorption length and the refractive index considered for the different materials~\cite{OpticalGel,Haerus,Matmatch,SiPad,Li1976RefractiveIO,Malitson:65}. For NaI(Tl), the absorption length has been fixed at a constant value of 100~m, similar to that of quartz and borosilicate, because it has not been possible to find such information for NaI(Tl). The effect of this choice in the simulation results will be analyzed in Section~\ref{Section:SIM_Val}.

After an energy deposition in the NaI(Tl), light is produced by sampling a Poisson distribution with mean $E\cdot LY$, with $E$ the energy deposited and $LY$ the light yield. Then, the energy of the emitted photons is sampled from the energy distribution shown in Figure~\ref{EmisionSpectraSim}, and the photons direction is sampled from an isotropic distribution. Every optical photon propagates until it is absorbed, scattered or reaches a surface.

The "optical surfaces" are found between two different media, and they define the probabilities of the photons to be reflected, refracted or absorbed, as well as the angle respect to the surface at which they will continue traveling. There are different types of surfaces that can be defined in GEANT4 between two dielectric materials, but we have used only two of them: \textit{Polished} and \textit{GroundBackPainted}.

In a \textit{Polished} surface, the probability of a photon being reflected, $R$, is calculated according to the Fresnel equation~\cite{Dietz-Laursonn:2016tpy}:
\begin{equation}\label{eq:reflectivity}
	R = \frac{\left(n_2\cos{\theta_i}-n_1\sqrt{1-\frac{n_1^2}{n_2^2}\sin^2{\theta_i}}\right)^2+\left(n_1\cos{\theta_i}-\sqrt{n_2^2-n_1^2\sin^2{\theta_i}}\right)^2}{\left(n_2\cos{\theta_i}+n_1\sqrt{1-\frac{n_1^2}{n_2^2}\sin^2{\theta_i}}\right)^2+\left(n_1\cos{\theta_i}+\sqrt{n_2^2-n_1^2\sin^2{\theta_i}}\right)^2},
\end{equation}
where $\theta_i$ is the incidence angle with respect to the normal of the surface and $n_1$ and $n_2$ are the refractive indices for the initial and final medium at the wavelength of the simulated photon, respectively. If the incidence angle is larger than the critical angle $\theta_i > \theta_c = \arcsin{(n_2/n_1)}$, then there is total reflection ($R = 1$). If the photons are reflected, the angle at which they will continue traveling with respect to the normal of the surface, $\theta_{refl}$, is calculated from the law of reflection:
\begin{equation}\label{eq:reflection}
	\theta_{refl} = \theta_i,
\end{equation}
while if the photons are not reflected, then they are transmitted (refracted) with a probability $T = 1-R$. In such case, the angle of the refracted photons with respect to the normal of the surface, $\theta_r$, will be obtained from the Snell's law:
\begin{equation}\label{eq:refraction}
	n_2\sin{\theta_r} = n_1\sin{\theta_i}.
\end{equation}

\begin{figure}[h!]
	\begin{center}
		\includegraphics[width=\textwidth]{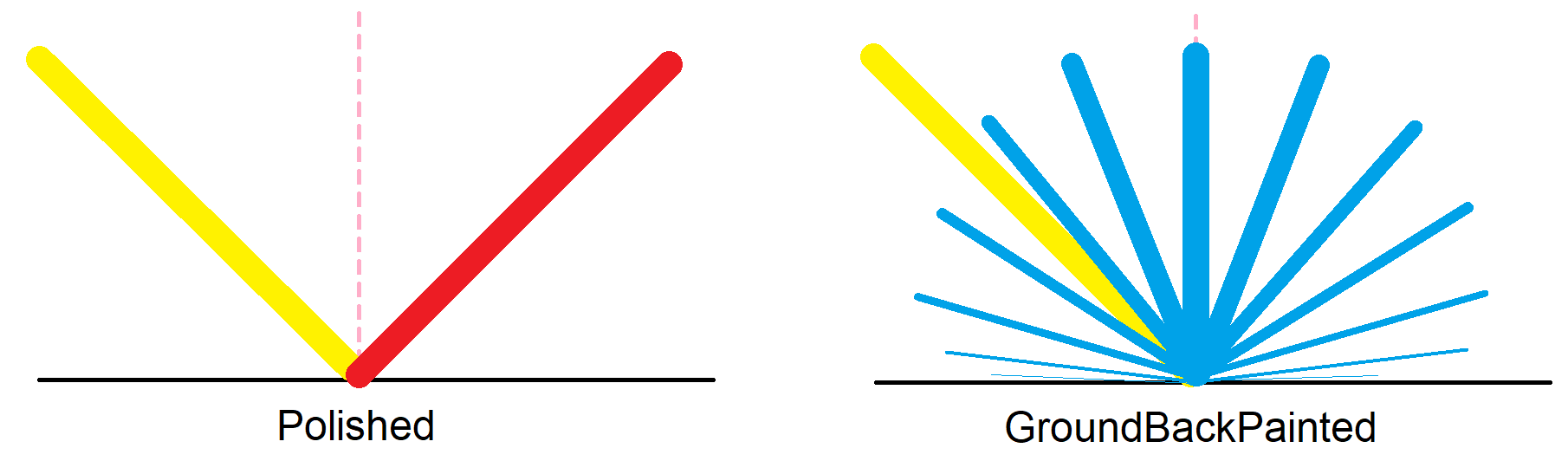}
		\caption{\label{Reflection}Difference between the reflection in a \textit{Polished} surface (left plot) and the reflection in a \textit{GroundBackPainted} surface (right plot) produced by the same incident light (yellow line). The thickness of the blue line in the \textit{GroundBackPainted} surface represents the reflection probability given by the Equation~\ref{eq:LambertRefl}.}
	\end{center}
\end{figure}

On the other hand, \textit{GroundBackPainted} surfaces are used to simulate rough (non-specular) surfaces following the Lambert's cosine law. In this case, the reflection angle, $\theta_{refl}$, is obtained from a probability distribution given by:
\begin{equation}\label{eq:LambertRefl}
	P(\theta_{refl}) d\theta_{refl} = \frac{1}{2} \cos{\theta_{refl}} d\theta_{refl}.
\end{equation}
Then, if the normal of the surface is parallel to the Z-axis, the direction of the reflected photon is given by:
\begin{equation}\label{eq:LambertRefl}
	\vec{u}_{ph} = \left(\sin{\theta_{refl}}\sin{\phi},\sin{\theta_{refl}}\cos{\phi},\cos{\theta_{refl}}\right),
\end{equation}
where $\phi$ is the angle formed by the direction of the reflected photon with the X axis, and it is randomly sampled. In \textit{GroundBackPainted} surfaces, the reflectivity is not obtained from the Fresnel equation, but introduced by hand~\cite{Dietz-Laursonn:2016tpy}. The difference in the reflectivity properties of the two defined surfaces are shown in Figure~\ref{Reflection}.

In our simulation, all the surfaces between dielectrics have been considered as \textit{Polished} except the one between the NaI(Tl) crystal and the teflon, that has been defined as \textit{GroundBackPainted}, as it is diffusive. This surface is very relevant in the transport of optical photons. The reflectivity of this surface was set to 99\% for the preliminary test shown in this Chapter. However, as we do not know the real value, the effect of modifying the value of this parameter in the results should be analyzed in future work.

Concerning the surfaces between dielectric and metallic media, all the surfaces have been considered as \textit{Polished} and, unlike what happened in the surface between two dielectrics, in this case no transmission is allowed and photons which are not reflected are directly absorbed. In these cases, the reflectivity must be also defined. In this simulation there are several dielectric-metal surfaces, all of them involving either the copper housing or the reflective metal layer and steel dynode structure inside the PMTs. For all of them a reflectivity of 100\% has been assumed in this work. This reflectivity should not be a relevant parameter when analyzing the scintillation light produced in the NaI(Tl) crystal and collected by the PMTs, however it could be more important for the Cherenkov light propagation through the system. Because of this, the effect of modifying this parameter should be also analyzed in the future.

\subsection{Pulse generation} \label{Section:SIM_Const_PulseGen}

The only parameter that is known for the photocathode of the ANAIS PMTs is the $QE$, defined in Equation~\ref{LC_PMTs_ANAIS} as the ratio of phe produced in the PMT to the total number of photons reaching the photocathode. This means, that $1-QE$ should be the amount of photons that are either transmitted or absorbed without producing phe, because we have taken $R = 0$ for the photocathode surfaces. The $QE$ considered in the simulation was the average value for the 18~PMTs of the ANAIS-112 experiment, which was obtained in Section~\ref{Section:ANAIS_Module} as 40.6\% (see Table~\ref{table:moduleCharacteristics}). Although the photocathode is semi-transparent, it was defined as opaque ($T = 0$) for the first tests of the simulation. As commented before, this factor is not important for studying the NaI(Tl) scintillation. However, it can be relevant in simulations of the light produced in the PMTs, as for example the Cherenkov light after a $^{40}K$ decay. In this simulation, the opacity of the photocathode reduces the propagation of Cherenkov light from one PMT to the other, thus reducing the triggering probability. The effect of the transparency of the photocathode in that case will be analyzed in Section~\ref{Section:SIM_Res_cherenkov}.

A data structure is stored in a ROOT file for each simulated event. It contains information of the energy deposited in the active volumes and the time associated  to the phe produced in the photocathode with respect to the time of generation of the primary particle, separately for the two PMTs: $t0_n$ ($t1_n$) for the n$^{th}$ phe in the PMT0 (PMT1) of the module. These phe are randomly sampled with probability QE among all the photons reaching the corresponding photocathode.

\begin{figure}[h!]
	\begin{center}
		\includegraphics[width=\textwidth]{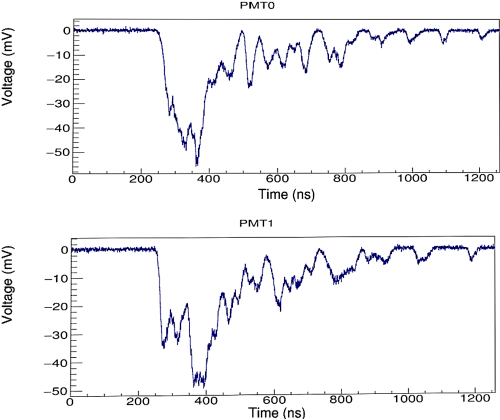}
		\caption{\label{PulsosCd109_2e9}Pulses generated (one for each PMT) with a sampling rate of 2~GHz for an event that has deposited 22.1~keV in the crystal during a simulation of the calibration with the $^{109}Cd$ source.}
	\end{center}
\end{figure}

\begin{figure}[h!]
	\begin{center}
		\includegraphics[width=\textwidth]{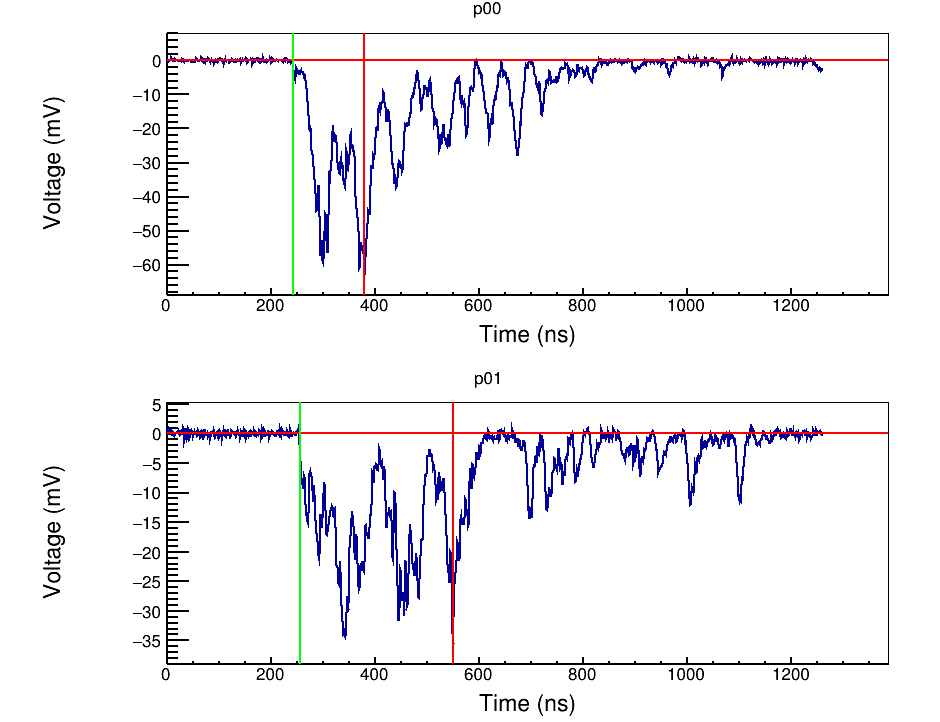}
		\caption{\label{PulsosExperimentales}Example of pulses acquired in the ANAIS-112 experiment. Horizontal red lines are the baselines, vertical red lines indicate the maximum of the pulse and green lines represent the pulse onsets.}
	\end{center}
\end{figure}

A second level of processing allows to produce the PMT output signal. Combining the information of the SER obtained experimentally (see Figure~\ref{DistributionsSER}) with the number and arrival times of simulated photoelectrons, a realistic output pulse can be generated. As it was explained in Section~\ref{Section:ANAIS_DAQ}, the digitizer used in the ANAIS-112 experiment is a MATACQ-32, which records 2520 samples and has the option of sampling at 1 or 2~GHz. In ANAIS it is configured at 2~GHz, resulting in a time digitization window $\Delta t_{acq}$ of 1260~ns. Then, for each simulated event, a waveform is generated for each PMT following these steps:
\begin{enumerate}
	\item The waveform baseline $V_{bsl}(t)$ is sampled using a gaussian distribution with mean zero and standard deviation taken as the typical root mean squared from ANAIS pulses baselines (0.35~mV). This implies considering a white noise contribution in the signal.
	\item For each photoelectron (the n$^{th}$ of the considered event), the relative position of the photoelectron arrival in the waveform frame ($T_n$) has to be obtained from $t_n$, the time associated to the production of that photoelectron with respect to the time of generation of the primary particle. To do that, a pretrigger time ($T_0 = 0.2\cdot \Delta t_{acq}$) is defined to fix the pulse onset in the waveform reference. This is the time assigned to the first detected photoelectron and the time reference for the others. Then
	\begin{equation}
		T_n = t_n - t_1 + T_0,
	\end{equation}
	where $t_1$ is the absolute time of the first detected photoelectron in any of the PMTs of the module.
	\item The pulse area of each photoelectron $a$ is obtained from a gaussian PDF whose mean, $\mu_{SER}$, and standard deviation, $\sigma_{SER}$ are typical of the SER of the ANAIS detectors shown in Section~\ref{Section:ANAIS_Module}. The values used are $\mu_{SER}$ = 64~mV~ns and $\sigma_{SER}$ = 32~mV~ns.
	\item The time-width of each photoelectron $t_W$ is obtained from another gaussian PDF whose mean and standard deviation are those obtained in ANAIS ($\mu$ = 6~ns, $\sigma$ = 0.5~ns).
	\item The shape of the n$^{th}$ photoelectron is defined as a gaussian function where $T_n$, $a$ and $t_W$ are included as:
	\begin{equation}
		phe_n(t) = \frac{a}{t_W\sqrt{2\pi}} \exp\left(-\frac{(t-T_n)^2}{2t_W^2}\right).
	\end{equation}
	\item This is done in a loop for all the photoelectrons producing signal in the same PMT, adding consecutively their signals.
	\begin{equation}
		V(t) = V_{bsl}(t) + \sum_n phe_n(t)
	\end{equation}
\end{enumerate}
In the Figure~\ref{PulsosCd109_2e9}, pulses generated (one for each PMT) corresponding to a simulation of the $^{109}Cd$ calibration are shown for an event that deposited 22.1~keV in the crystal. Moreover, in Figure~\ref{PulsosExperimentales}, two acquired pulses in the ANAIS-112 experiment are shown. It is possible to observe that they are very similar to the simulated ones.

This pulse generation method allows for the application of the analysis of the ANAIS experiment (which was explained in Section~\ref{Section:ANAIS_Analysis}) to the simulated events, enabling the determination of relevant event parameters. These are: the pulse shape variables $p_1$ and $\mu$, the pulse onset at each PMT ($t_{00}$ and $t_{01}$) and the number of peaks found by the peak-finding algorithm in each pulse ($nPeak_0$ and $nPeak_1$) and its sum ($Peaks$). The area of each pulse is calculated as the integral of the signal of each PMT in the whole acquisition window:
\begin{equation}\label{eq:areaOptSim}
	area_i = t_R\sum_{t = T_0}^{t = \Delta t_{acq}} V_i(t),
\end{equation}
where $t_R$ is the time resolution of the digitizer. Additionally, the sum of both areas is calculated. Moreover, the time of the first photoelectron produced for each event is saved in a variable named \textit{tPulse} and the sum of the energy depositions in the NaI(Tl) volume is stored in a variable called $E_{NaI}$. Finally, the number of photoelectrons in each PMT ($n_0$ and $n_1$) and its sum ($Phes$) are also saved. To analyze the Cherenkov emission, a boolean variable called \textit{Chere} is also defined, storing the volumes where this emission is produced. After the application of this analysis to the simulated pulses, a new ROOT file is saved including all of these variables.

In order to simulate the behavior of the detector accurately, it is necessary to take into account the DAQ dead time. In the ANAIS DAQ there is a period of about 4~ms after the end of the acquisition window during which the arriving photoelectrons cannot generate new signals. If there are additional photoelectrons arriving after this time, the process is repeated, with two more pulses being generated using the same procedure, and then a new event will result, saving in this case the time difference between multiple events generated. This process is repeated recursively until all of the photoelectrons from a given event have been accounted for. As the scintillation time of the crystal is set to 230~ns, in general for low or mid energy events we do not expect any photon after the 4~ms dead time. However, this consideration of the dead time can be important for events that involve long-duration energy depositions, such as those caused by neutrons or long decay chains, for example, the one initiated by the $^{222}Rn$. Moreover, as it was explained in Section~\ref{Section:ANAIS_Filter}, we have measured scintillation times of the order of 200~ms after a muon interaction, which is traduced in a multiple triggering after that single energy deposit~\cite{Cuesta:2013vpa}. Analyzing the effect of the long phosphorescence in NaI(Tl) signals, would be an interesting follow-up of this work.

\section{Code validation} \label{Section:SIM_Val}
\fancyhead[RO]{\emph{\thesection. \nameref{Section:SIM_Val}}}

For the validation of the simulation and the results presented in the next section, three radioactive isotopes were simulated in different positions of the setup for different purposes. 10$^7$ decays of $^{109}Cd$ were simulated in the source position (Figure~\ref{OptSimGeoMacarron}) to reproduce the low energy calibration of the experiment, and 10$^4$ decays of $^{22}Na$ and $^{40}K$ homogeneously distributed in the crystal to reproduce two of the most relevant background contributions in the region of interest of ANAIS-112 (see Section~\ref{Section:ANAIS_Background}). These three isotopes' decays are fundamental for the ROI energy calibration protocol of ANAIS-112 experiment (see Section~\ref{Section:ANAIS_Calibration}), as they produce the four energies of reference: 0.9~keV ($^{22}Na$), 3.2~keV ($^{40}K$), 11.8~keV and 22.6~keV ($^{109}Cd$).

The first validation of the simulation is related to the optical absorption length of the NaI(Tl), which is not known. This is a very important parameter that can strongly affect the results of the simulation, as the photons can travel of the order of several meters inside the module before reaching the photocathodes. To analyze its effect on the LC, nine simulations of the $^{40}K$ decay were run fixing the light yield of the crystal to the nominal value of 40~photons/keV given by AlphaSpectra~\cite{AlphaSpectra}. In each of these simulations, the value of the absorption length was changed from 0.1~m to 10$^4$~m, and the LC was obtained for each 3.2~keV energy deposition as the number of collected photoelectrons in both PMTs in a time window of 1~$\mu$s divided by the deposited energy. Figure~\ref{LC(Labs,Ly)} shows the LC obtained for each of these simulations as a function of the absorption length. It is possible to observe that the LC saturates for lengths larger than 100~m. This was the value used in all the following simulations, because for higher values, no improvement in LC is achieved, and for lower values we cannot reproduce the ANAIS measured LC, in average 14.44~$\pm$~0.01 phe/keV (see Section~\ref{Section:ANAIS_Module}). In fact, in order to reproduce such LC, we decided to increase the NaI(Tl) light yield up to 44~photons/keV. With this configuration, the LC recovered from the simulation at 3.2~keV is 14.62~$\pm$~0.08~phe/keV.

\begin{figure}[h!]
	\begin{center}
		\includegraphics[width=0.75\textwidth]{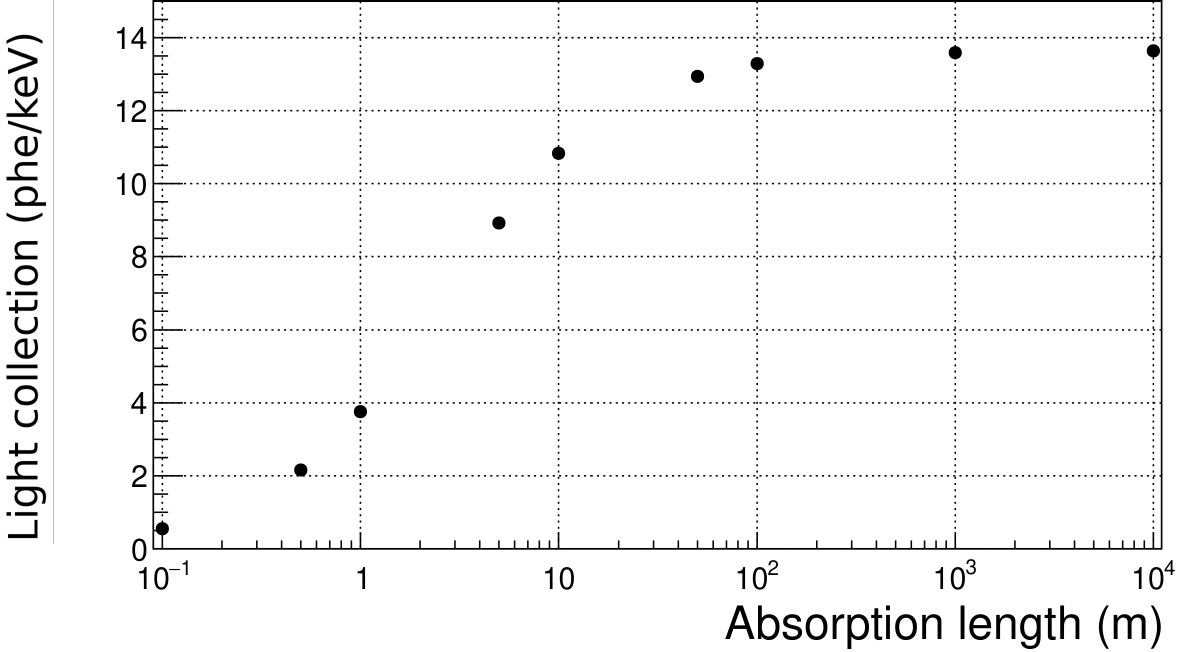}
		\caption{\label{LC(Labs,Ly)}Results of the LC as a function of the NaI(Tl) optical absorption length for a light yield of 40~photons/keV.}
	\end{center}
\end{figure}

\begin{figure}[h!]
	\begin{subfigure}[b]{0.49\textwidth}
		\includegraphics[width=\textwidth]{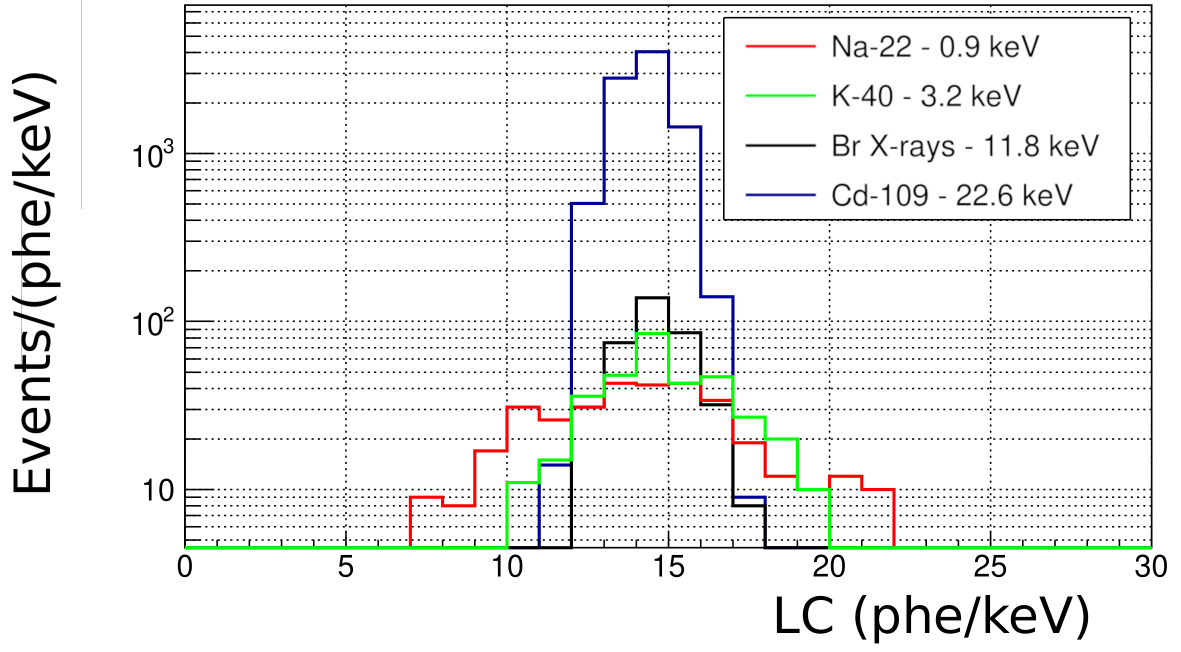}
	\end{subfigure}
	\begin{subfigure}[b]{0.49\textwidth}
		\includegraphics[width=\textwidth]{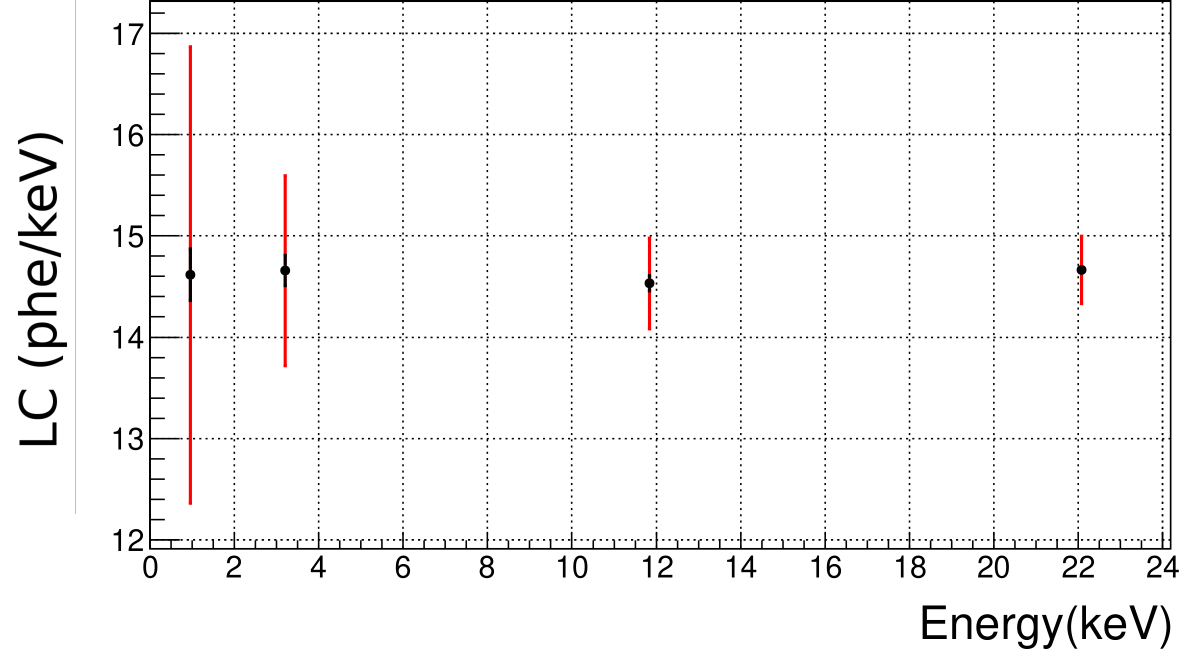}
	\end{subfigure}
	\caption{\label{LC(E)}Left plot: distribution of the LC for the four lines used in the low energy calibration of the experiment. Right plot: mean values of the LC for the same four energies. Black lines are statistical errors while the standard deviations are plotted as the red lines.}
\end{figure}

The results for the LC for the four lines used in the low energy calibration of the experiment using a light yield of 44~photons/keV are shown in Figure~\ref{LC(E)}. It can be observed that LC values obtained are fully compatible within statistical errors (black lines in right panel). This is interesting because of the different distribution of $^{22}Na$ and $^{40}K$ isotopes with respect to the energy depositions from $^{109}Cd$ events. Energy depositions from the emissions of the two first isotopes are distributed homogeneously in the crystal bulk, while those from $^{109}Cd$ emissions happened preferentially near the crystal surface at the crystal center. This result implies that light propagation between the interaction point and the PMTs is not affecting significantly the LC. However, in the experimental measurements the LC depends on the deposited energy due to the energy dependence of the light yield. Future work with the simulation should include this dependence.

\begin{figure}[h!]
	\begin{center}
		\includegraphics[width=0.75\textwidth]{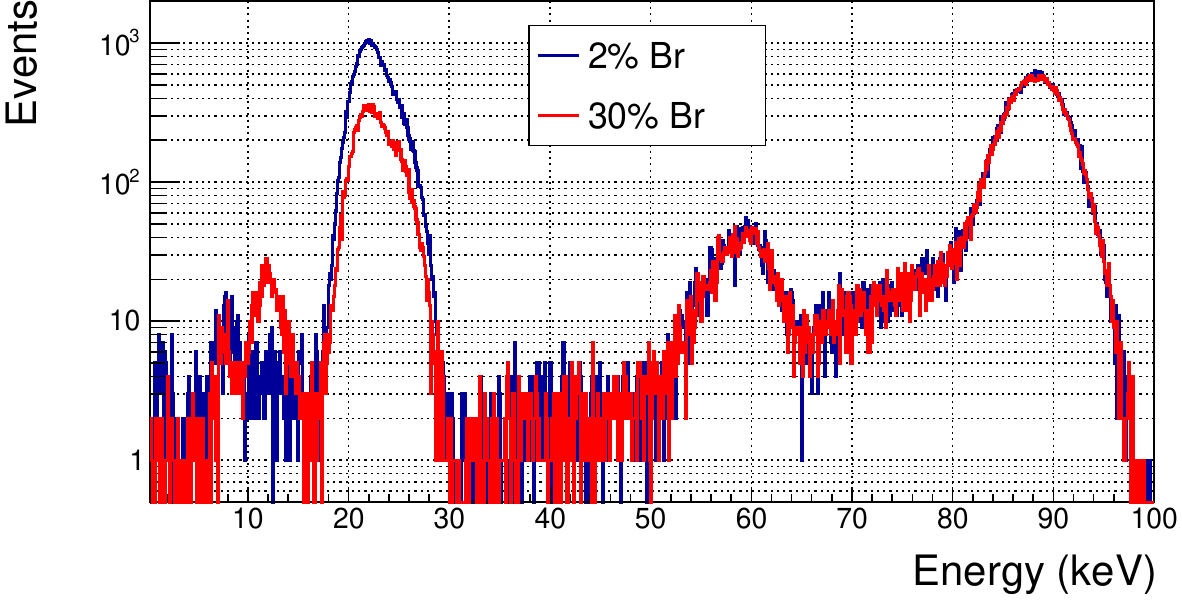}
		\caption{\label{BrContent}Comparison between simulated spectra for a Br mass content of 2\% (blue line) and 30\% (red line).}
	\end{center}
\end{figure}

It is interesting to check if our simulation is able to reproduce the measured spectrum of $^{109}Cd$. The mass content of bromine in the heat-shrink tube is unknown, but it determines the detection ratio of the 22.6 and 11.8~keV peaks, in such a way that the higher is the amount of bromine, the lower is the number of 22.6~keV events arriving to the crystal and the higher the number of 11.8~keV events. Figure~\ref{BrContent} illustrates this effect by comparing the simulated spectra for a bromine mass content of 2\% and 30\%. Trying to determine the bromine content able to reproduce the experimental measurement, some simulations were run with a different mass content of bromine. Then, the ratio between the number of events of the 22.6~keV line to those of the 11.8~keV line was compared with the experimental value, considering the interval between 9 and 15~keV for the 11.8~keV line and between 16 and 30~keV for the 22.6~keV line. The value of mass content of bromine that makes this ratio closer to the one obtained experimentally is 30\%.

\begin{figure}[h!]
	\begin{center}
		\includegraphics[width=0.75\textwidth]{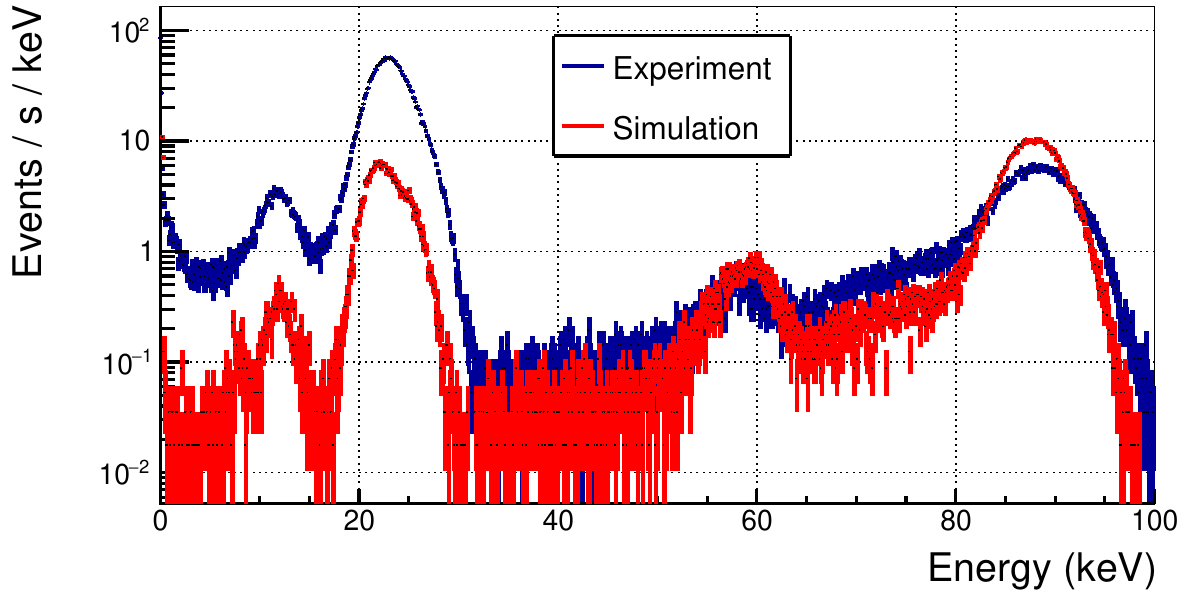}
		\caption{\label{ExpSimOpt}Comparison between simulated (red line) and experimental (blue line) spectra for the $^{109}Cd$ calibration for a 30\% $Br$ content in the heat-shrink tube for D3.}
	\end{center}
\end{figure}

Figure~\ref{ExpSimOpt} compares the experimental and simulated spectra applying the same energy calibration. The simulated spectrum has been scaled with the time exposure, considering the nominal activity of the $^{109}Cd$ source (1.5~$\mu$Ci) and its age at the time of the measurement.

In these spectra is possible to identify the 22.6 and 88.1~keV peaks emitted in the $^{109}Cd$ decay and the x-ray emitted by the bromine at 11.8~keV. The peak at about 60~keV is produced by the iodine K-shell x-ray escape after the absorption of the 88.1~keV photon. It is possible to observe a good agreement between the event ratio of 11.8~to 22.6~keV peaks and the spectral shapes both for simulation and experimental measurement. However, the intensities of the lines do not agree. The reason is that the source encapsulation modeling is too simple, and the total number of x-rays and gammas reaching the crystal is strongly dependent on the thickness and the materials considered in the housing of the source and its position respect to the Mylar window. More effort to develop a realistic geometry for the source is now ongoing. In any case, for the goal of this work, this modelling is sufficient, because we will use these simulations to obtain events in the ROI produced by the $^{109}Cd$ emissions and characterize the 11.8~keV / 22.6~keV peak properties.

It is worth noting that no convolution was applied to the simulated spectra in order to take energy resolution into account. In spite of that, the agreement between simulated and experimental data as regards the widths of the 11.8~and 22.6~keV peaks is remarkable, which supports that the light production and transport, and the width of the SER distribution are the most relevant contributions to the resolution of the detector. This will be analyzed in more detail in the next section.

\section{Simulation results} \label{Section:SIM_Res}
\fancyhead[RO]{\emph{\thesection. \nameref{Section:SIM_Res}}}

The optical simulation has allowed us to study the impact of several factors on the detector performance, as the effect of the digitizer time window on energy calibration and resolution or the systematics observed in the shapes of the pulses for different populations, and to estimate the trigger and event filtering efficiencies in the ROI. In addition, it allows us to estimate the detection rate of the Cherenkov events produced by the contamination of $^{40}K$ in the borosilicate of the PMTs and their mixing with scintillation events, and the effect that the $^{222}Rn$ and $^{40}K$ have on the acquisition rate of the Blank module. All these analysis are still in progress, but preliminary results will be presented in this section.

\subsection{Acquisition window effects on energy calibration and resolution} \label{Section:SIM_Res_SR}

We are evaluating the possibility of modifying the ANAIS acquisition window. The reason is that part of the scintillation pulse is not digitized in the current acquisition window of 1260~ns. The MATACQ digitizer allows to switch that time to 2520~ns, but it also implies a change in the sampling rate from 2~GS/s to 1~GS/s (as the number of recorded samples remains constant to 2520) and therefore a deterioration of the time resolution. Several parameters that could be affected by the change in sampling rate as $p_1$ and $\mu$ were analyzed. Results obtained from the simulations were identical for both sampling rates. In this section, I will present the results obtained for the energy calibration and resolution.

\begin{figure}[h!]
	\begin{center}
		\includegraphics[width=\textwidth]{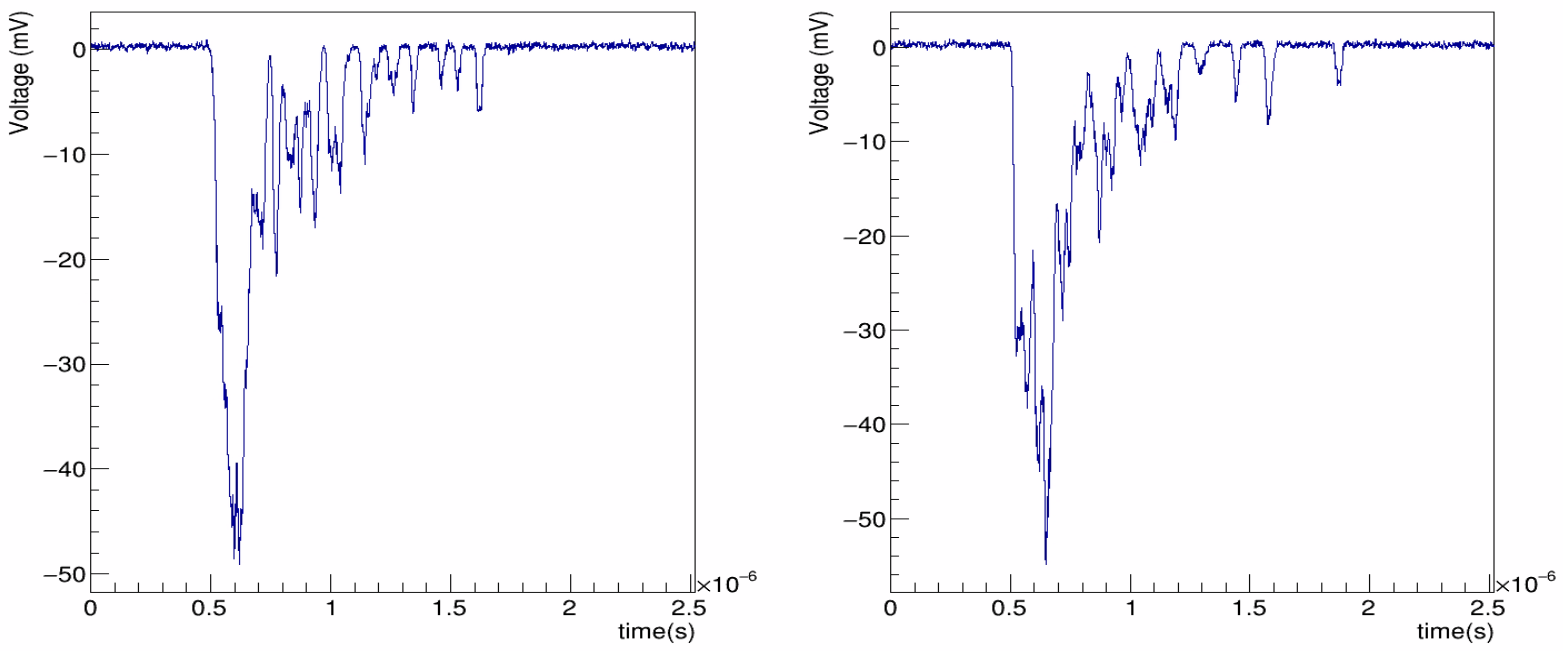}
		\caption{\label{PulsosCd109_1e9}Pulses generated (one for each PMT) with a sampling rate of 1~GHz for an event that has deposited 22.1~keV in the crystal from the simulated $^{109}Cd$ calibration.}
	\end{center}
\end{figure}

For this purpose, the same $^{109}Cd$, $^{22}Na$ and $^{40}K$ simulations shown in previous section were used, generating pulses according to the two different digitization windows and sampling rates. As an example, Figure~\ref{PulsosCd109_1e9} shows the two generated pulses using a sampling rate of 1~GHz for an event that deposited 22.1~keV in the crystal from the simulated $^{109}Cd$ calibration.

\begin{figure}[h!]
	\begin{center}
		\includegraphics[width=0.75\textwidth]{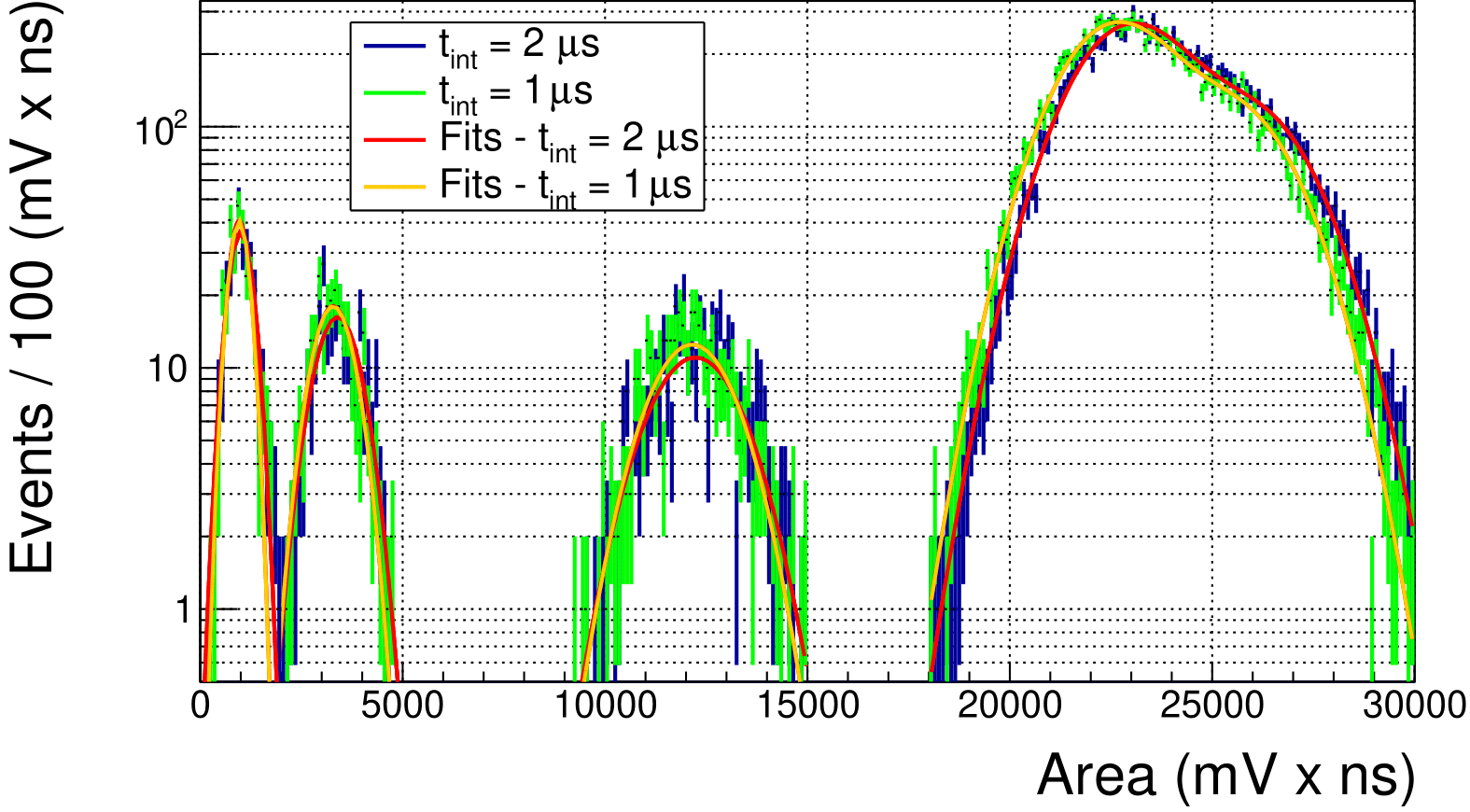}
		\caption{\label{calibration_Fits}Pulse area distributions of the four calibration peaks for an integration time of 1~$\mu$s (green) and 2~$\mu$s (blue), and the corresponding gaussian fits: orange for 1~$\mu$s and red for 2~$\mu$s.}
	\end{center}
\end{figure}

The area of the pulses (Equation~\ref{eq:areaOptSim}) was computed for integration times of 1~$\mu$s (for the 2~GHz sampled pulses) and 2~$\mu$s (1~GHz). Figure~\ref{calibration_Fits} shows the pulse area histograms obtained for both sampling rates. Each peak corresponds to a given energy selected by the $E_{NaI}$ variable. They were fitted to gaussians leaving the amplitude, mean and standard deviation parameters free. Fits are also shown in Figure~\ref{calibration_Fits}. The $^{109}Cd$ x-rays were fitted to two gaussians, centered at 22.1~and 25.0~keV. In Table~\ref{tabla:calibrationFits} the results of the fits are presented.

\begin{table}[h]
	\centering
	\begin{tabular}{|c|c|c|c|c|c|c|}
		\cline{2-7}
		\multicolumn{1}{c|}{} & \multicolumn{2}{|c|}{$\mu$ (mV~$\cdot$~ns)} & \multicolumn{2}{|c|}{$\sigma$ (mV~$\cdot$~ns)} & \multicolumn{2}{|c|}{$\chi^2$} \\
		\hline
		Energy(keV) & 1~$\mu$s & 2~$\mu$s & 1~$\mu$s & 2~$\mu$s & 1~$\mu$s & 2~$\mu$s \\
		\hline
		0.9 & 950$\pm$17 & 995$\pm$19 & 254$\pm$12 & 304$\pm$20 & 1.02 & 0.81  \\
		3.2 & 3276$\pm$36 & 3398$\pm$40 & 524$\pm$33 & 548$\pm$31 & 0.63 & 1.30 \\
		11.8 & 12130$\pm$64 & 12250$\pm$67 & 1049$\pm$64 & 1072$\pm$62 & 0.90 & 0.97 \\
		22.1 & 22602$\pm$26 & 22947$\pm$26 & 1382$\pm$15 & 1393$\pm$15 & \multirow{2}{*}{0.95} & \multirow{2}{*}{0.70} \\
		25.0 & 25675$\pm$46 & 25698$\pm$40 & 1426$\pm$19 & 1434$\pm$20 &  &  \\
		\hline
	\end{tabular} \\
	\caption{Results of the fit of the simulated pulse area distributions to gaussians for the four peaks used in the ANAIS-112 LE calibration. The corresponding reduced $\chi^2$ is also reported.}
	\label{tabla:calibrationFits}
\end{table}

\begin{figure}[h!]
	\begin{center}
		\includegraphics[width=0.75\textwidth]{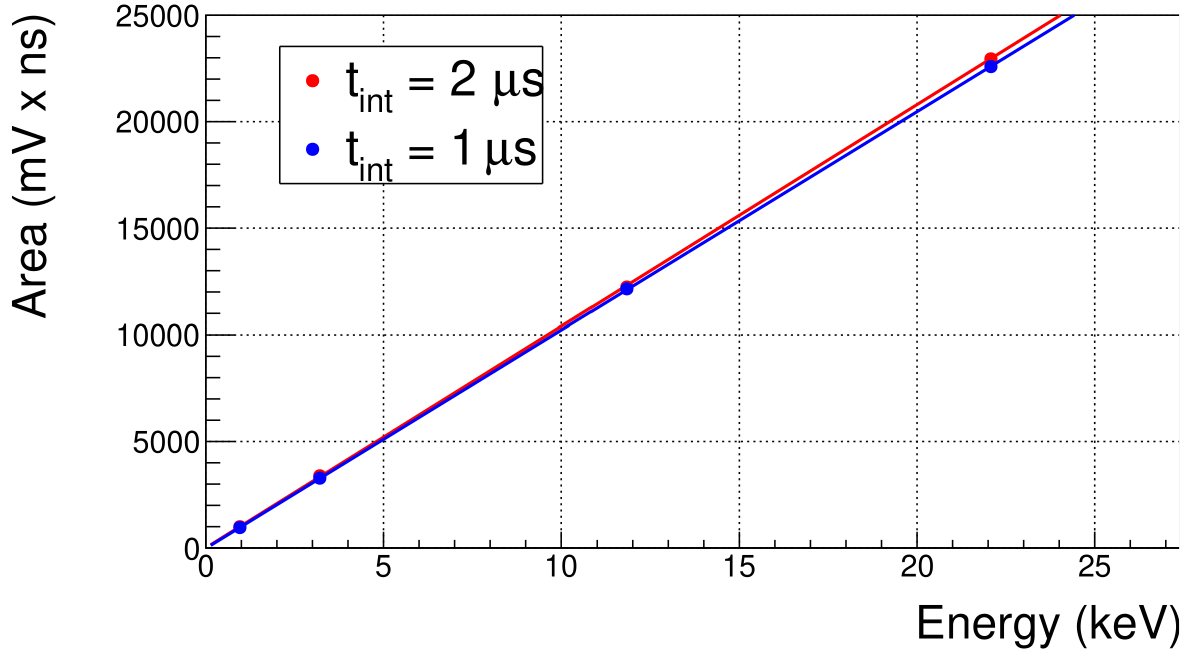}
		\caption{\label{CalibrationOptSim}Linear calibration between the mean area of the pulses corresponding to the four peaks used in the ANAIS LE calibration for both integration windows (blue circles for 1~$\mu$s and red circles for 2~$\mu$s).}
	\end{center}
\end{figure}

The mean values were fitted to a linear function of the energy as $area = c_0+c_1\cdot E$. Fits are shown in Figure~\ref{CalibrationOptSim}, and their resulting fit parameters in Table~\ref{tabla:Calibration}. As expected, pulse areas for the longer integration time are slightly larger, because some photons can arrive beyond the first $\mu$s. In fact, considering the scintillation time of 230~ns, the pulse area should increase in 1.25\% when increasing the integration time from 1~to 2~$\mu$s. On the other hand, it is very interesting to observe that although a proportional relation is expected in both cases, for a time integration of 1~$\mu$s (the same as in ANAIS-112 experiment) the independent term of the calibration is not compatible with zero (see Table~\ref{tabla:Calibration}). This is something that is also observed in the calibration of the experimental data. Although it has not been understood yet, it could be related with the integration window chosen in ANAIS-112 experiment and/or the time resolution. Experimental testing of this effect can be faced in the future. 

\begin{table}[h]
	\centering
	\begin{tabular}{|c|c|c|}
		\hline
		$t_{int}$ ($\mu s$) & $c_0$ (mV$\cdot$ns) & $c_1$ (mV$\cdot$ns/keV) \\
		\hline
		1 & -29~$\pm$~17 & 1026~$\pm$~1 \\
		2 & 9~$\pm$~18 & 1040~$\pm$~1 \\
		\hline
	\end{tabular} \\
	\caption{Results for the parameters of the linear relation between pulse area and energy (energy calibration).}
	\label{tabla:Calibration}
\end{table}

\begin{figure}[h!]
	\begin{center}
		\includegraphics[width=0.75\textwidth]{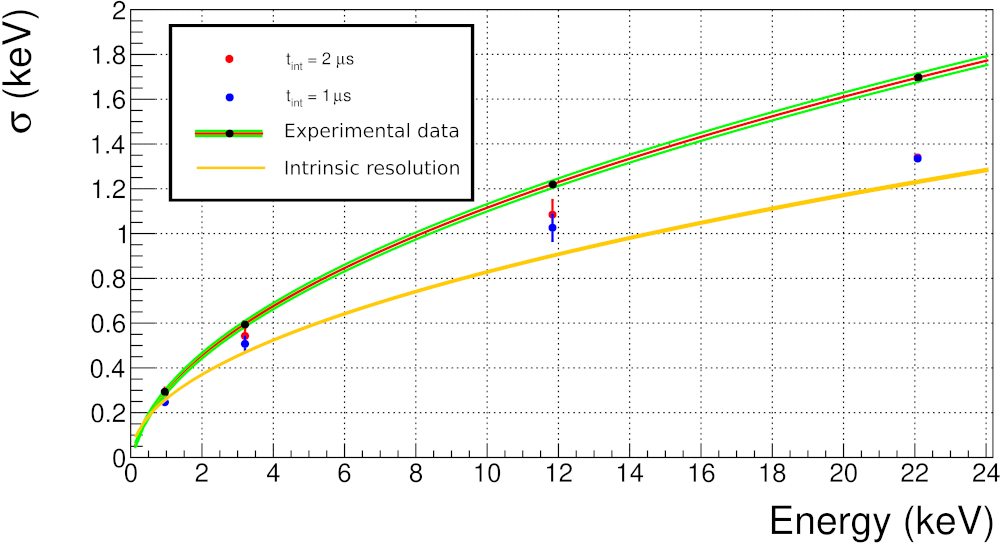}
		\caption{\label{ResolutionOptSim}Standard deviation of the peaks as a function of the energy for both integration windows. Values from ANAIS-112 data (see Section~\ref{Section:ANAIS_Calibration}) are also presented (red line) including the uncertainties of the fit function (green shadow), as well as the intrinsic resolution derived from the LC (orange line). See the text for more details.}
	\end{center}
\end{figure}

The standard deviation of the peaks is shown in the Figure~\ref{ResolutionOptSim}. This plot also displays the resolution obtained in the experimental measurements including the uncertainties (as presented in~\cite{Amare:2018sxx}), and the intrinsic resolution corresponding to the LC obtained from the simulation, that assuming poissonian behavior, it can be estimated as $\sigma = a\cdot \sqrt{E}$, where $a = LC^{-1/2}$. Thus, the value of $a$ is 0.262~$\pm$~0.001~keV$^{1/2}$. This is the same value that in ANAIS-112 experimental data, as the LC is the same. The results obtained are very similar for both digitization time windows and compatible with each other. However, a slightly worse resolution is systematically observed for the integration time of 2~$\mu$s, which could be related to the lower temporal resolution. However, this analysis shows that the change in the sampling rate in the ANAIS-112 experiment could be carried out without a serious effect in the energy resolution. From now on, all the analysis will be performed simulating the pulses with a sampling rate of 2~GHz and an digitization window of 1260~ns.

\begin{figure}[h!]
	\begin{center}
		\includegraphics[width=0.75\textwidth]{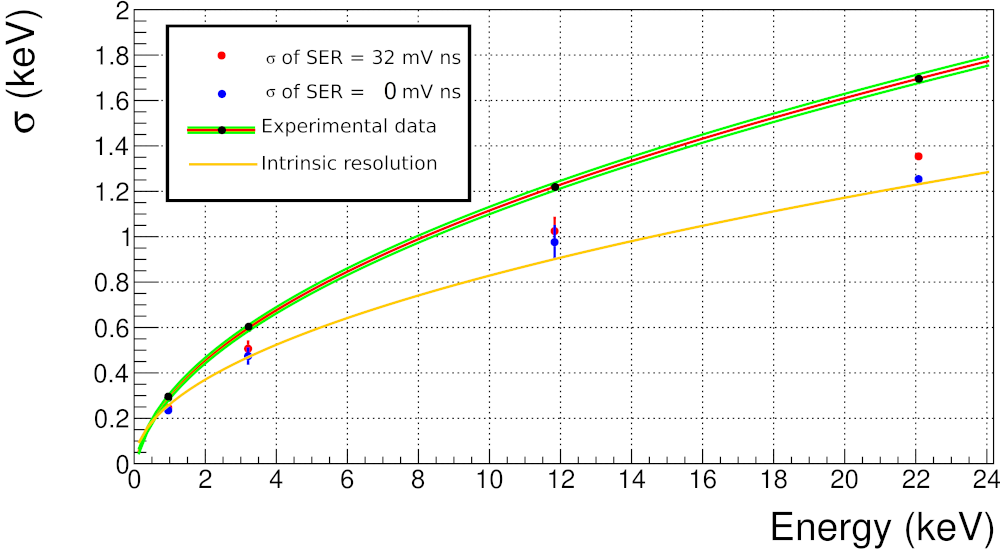}
		\caption{\label{ResolutionSER}Comparison of the standard deviation obtained for each peak simulating the SER distribution as explained in Section~\ref{Section:SIM_Const_PulseGen} ($\sigma_{SER}$~=~32~mV$\cdot$ns, red circles) or applying the same area to all the photoelectrons ($\sigma_{SER}$~=~0~mV$\cdot$ns, blue circles). The values from experimental data are also presented including the uncertainties of the fit function (see Section~\ref{Section:ANAIS_Calibration}).}
	\end{center}
\end{figure}

It is worth noting that the energy resolution derived from simulation is higher than the poissonian contribution, directly related to the fluctuation in the number of collected photoelectrons. This could be explained by the effect of the SER, which is also sampled from a distribution with some fluctuation. To analyze this effect, two different SER distributions have been compared: the distribution derived from ANAIS data (see Section~\ref{Section:SIM_Const_PulseGen} and Section~\ref{Section:ANAIS_Module}) with $\sigma_{SER}$~=~32~mV$\cdot$ns and a distribution with $\sigma_{SER}$~=~0~mV$\cdot$ns, with all the photoelectrons having the same area. The results are shown in Figure~\ref{ResolutionSER}. The resolution obtained without the SER effect is compatible with the intrinsic resolution. On the one hand, it means that the $\sigma_{SER}$ has a relevant effect in the resolution, and on the other hand, that there are some other effects contributing to the energy resolution in the experimental data that have not been taken into account in this simulation, as inhomogeneities in the scintillation properties of the crystal implying a dependence of the light yield on the crystal position, for instance. These effects can be analyzed in future work with this simulation.

\subsection{Pulse shape analysis} \label{Section:SIM_Res_PSA}

\begin{figure}[h!]
	\begin{subfigure}[b]{0.49\textwidth}
		\includegraphics[width=\textwidth]{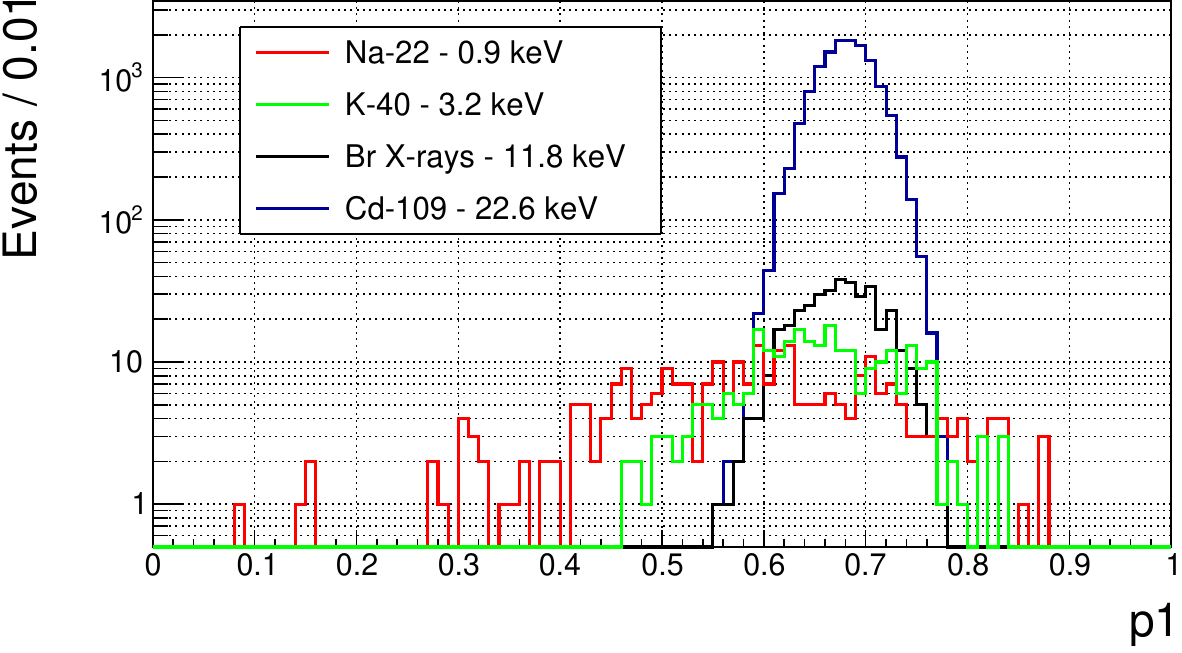}
	\end{subfigure}
	\begin{subfigure}[b]{0.49\textwidth}
		\includegraphics[width=\textwidth]{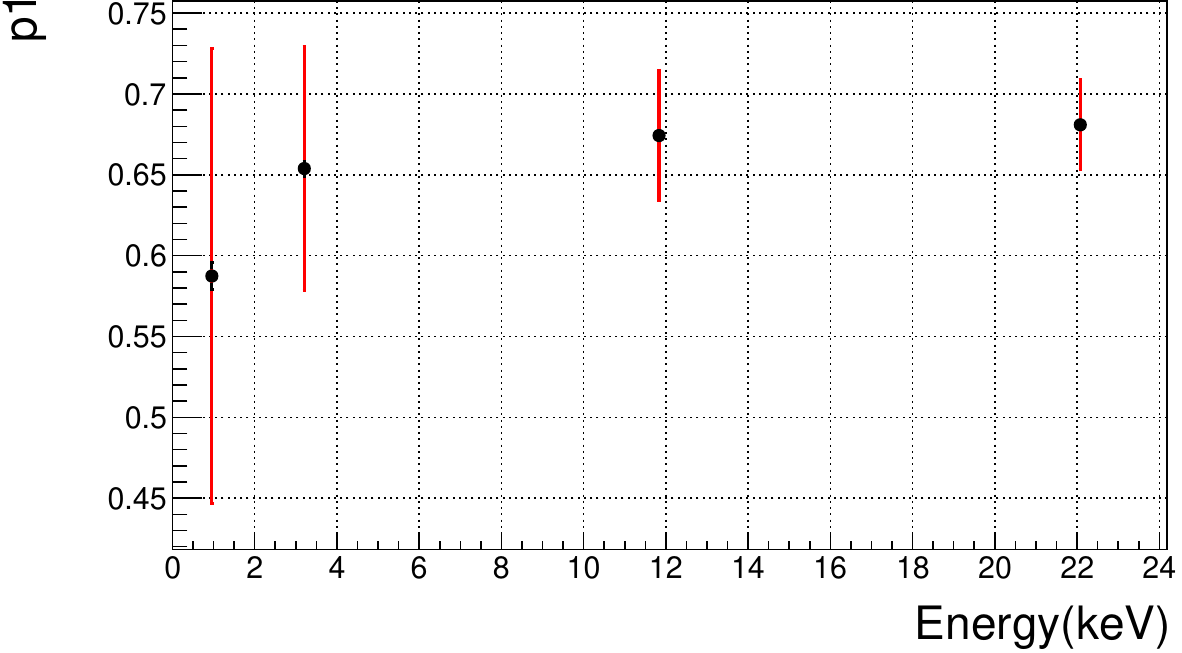}
	\end{subfigure}
	\caption{\label{p1OptSim}Left plot: distributions of the $p_1$ variable for the four peaks used in the low energy calibration of the experiment. Right plot: means and statistical errors (black) of these distributions. The standard deviations are plotted as the red lines.}
\end{figure}

\begin{figure}[h!]
	\begin{subfigure}[b]{0.49\textwidth}
		\includegraphics[width=\textwidth]{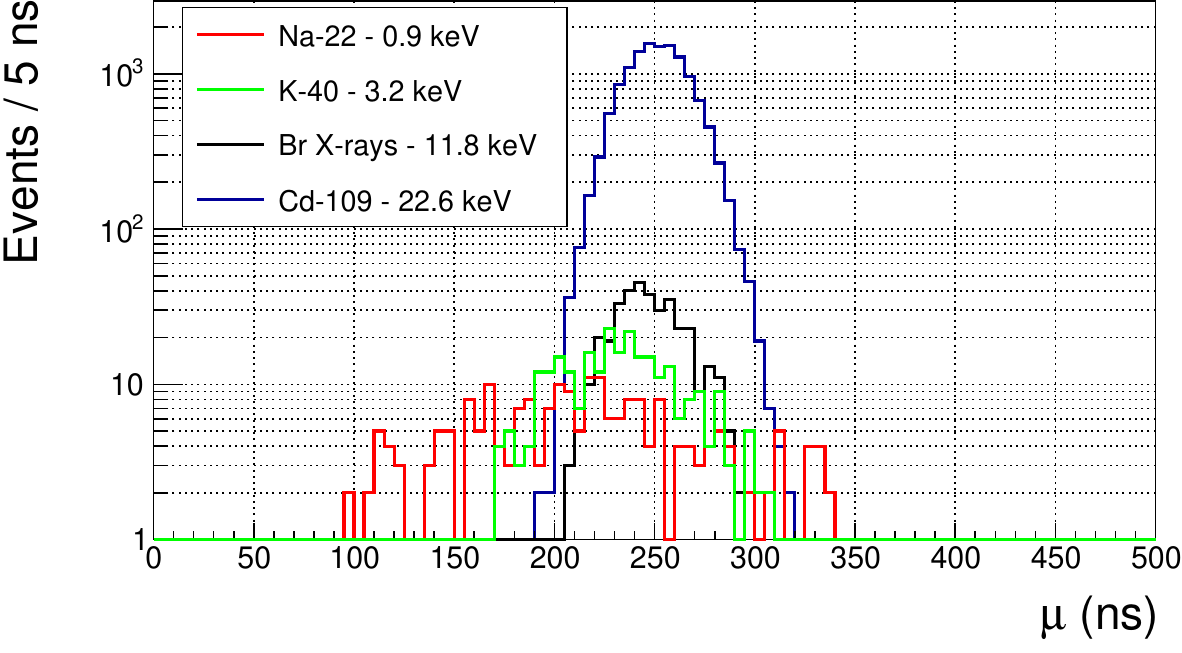}
	\end{subfigure}
	\begin{subfigure}[b]{0.49\textwidth}
		\includegraphics[width=\textwidth]{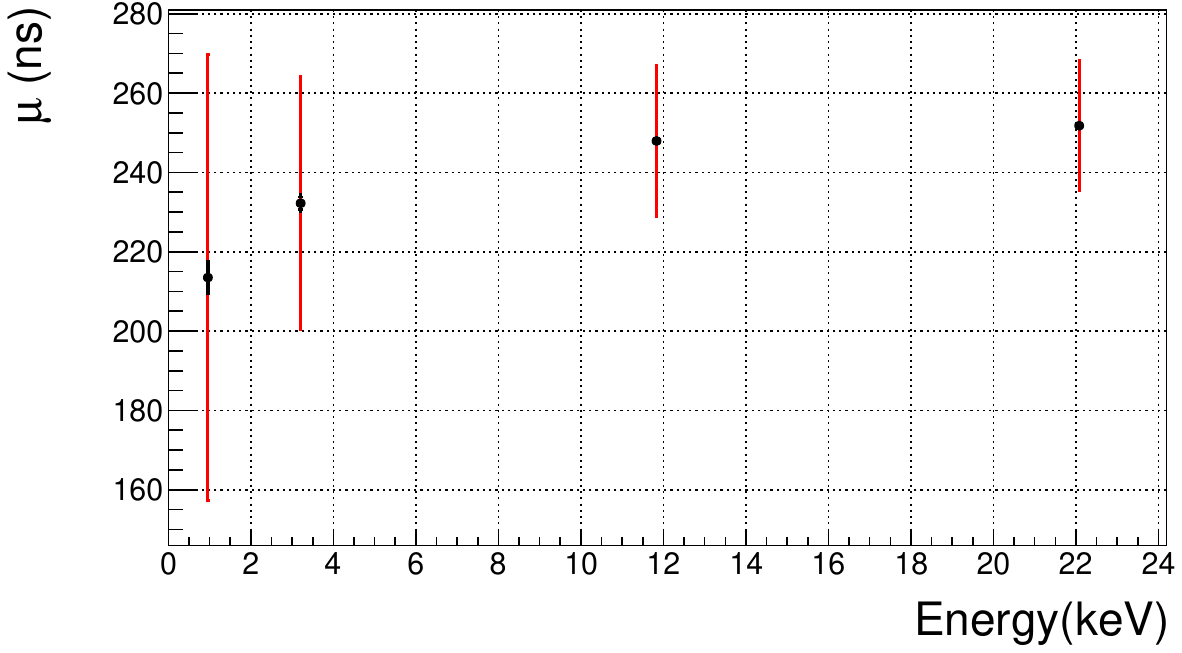}
	\end{subfigure}
	\caption{\label{muOptSim}Left plot: distributions of the $\mu$ variable for the four peaks used in the low energy calibration of the experiment. Right plot: means and statistical errors (black) of these distributions. The standard deviations are plotted as the red lines.}
\end{figure}

The two pulse shape variables used in ANAIS-112 for the event filtering ($p_1$ and $\mu$, see Section~\ref{Section:ANAIS_Filter}) have been analyzed using the simulation for the three different populations ($^{109}Cd$, $^{22}Na$ and $^{40}K$) and compared with the experimental data. The $p_1$ distributions for the four peaks used in the low energy calibration of the experiment have been obtained by selecting the energy depositions using the $E_{NaI}$ variable, as explained in Section~\ref{Section:SIM_Res_SR}, and they are shown in Figure~\ref{p1OptSim}, together with the means and standard deviations of these distributions. The same procedure has been followed for the $\mu$ variable, shown in Figure~\ref{muOptSim}.

There is a clear increase in the dispersion of both variables at lower energies, which is due to the decrease of the number of photoelectrons per event. Something remarkable is that, although only one scintillation component has been considered in the simulation and it does not depend on the deposited energy, both $p_1$ and $\mu$ mean values decrease at low energies. These two effects have been observed in the experimental data, as it was shown in Figure~\ref{muP1(E)_Exp}, where these variables were plotted as a function of the deposited energy for a $^{109}Cd$ calibration run. However, in the case of the experimental data we cannot discard there is a different scintillation time contributing to the light emission for different energy depositions. In the simulation, on the other hand, all the pulse shapes should be identical, and then, understanding why systematically the pulses show to be faster at low energies in the simulation is relevant for the ANAIS-112 analysis. These results of the simulation allow to conclude that because these parameters are used in the filtering process, both selection criteria and corresponding efficiencies have to be estimated with events of similar energy depositions.

\begin{figure}[h!]
	\begin{subfigure}[b]{0.49\textwidth}
		\includegraphics[width=\textwidth]{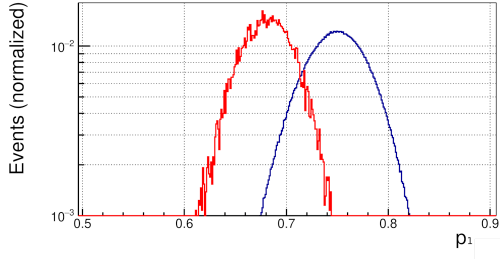}
	\end{subfigure}
	\begin{subfigure}[b]{0.49\textwidth}
		\includegraphics[width=\textwidth]{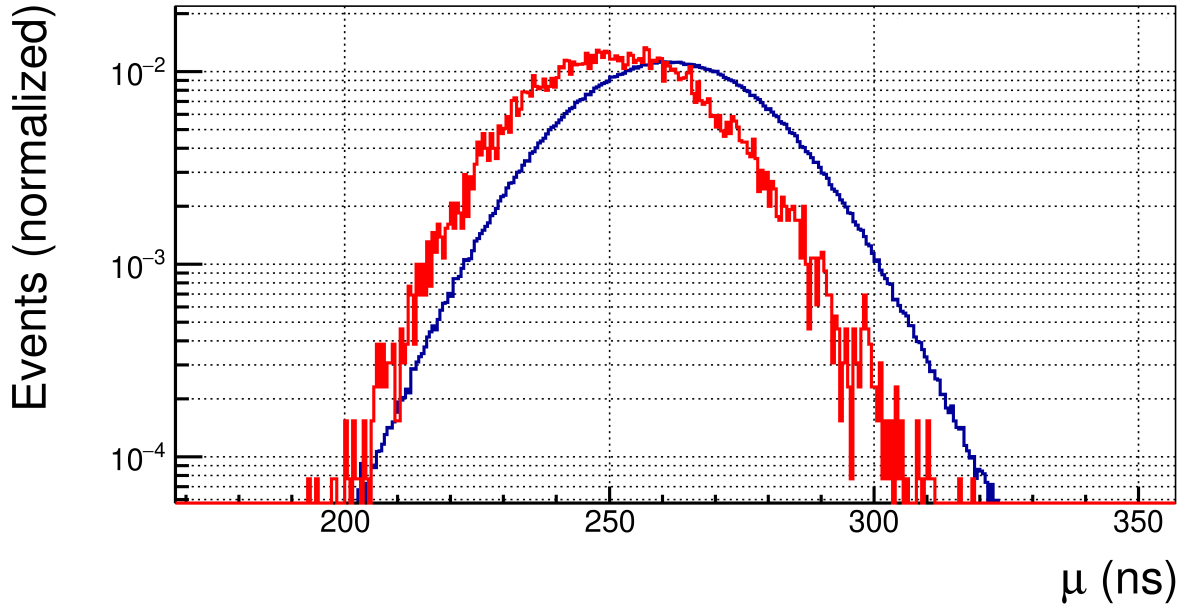}
	\end{subfigure}
	\caption{\label{p1MuExpSim}Comparison of the $p_1$ (left) and $\mu$ (right) distributions for the 22.6~keV peak between simulation (red line) and experimental (blue line) data. Histograms are normalized to have a total area of 1.}
\end{figure}

Another interesting comparison is between the mean values of the pulse shape variables for experimental and simulated data. For the experimental data a calibration run has been used, selecting events in a range of energy depositions from 16~keV to 30~keV corresponding to the 22.6~keV peak. This peak has been selected because it has energy enough to have low dispersion in the distributions. The results for the $p_1$ and $\mu$ distributions are shown in Figure~\ref{p1MuExpSim}, and the means of both variables are shown in Table~\ref{table:p1mu_ExpSim}. Both variables are clearly larger in the experimental data, which could be explained by the contribution of another scintillation component of the NaI(Tl) crystal that has not been taken into account in the simulation. The two slow scintillation components observed in NaI(Tl) (see Section~\ref{Section:ANAIS_Setup}) have time decays of 1.5~$\mu$s and 0.15~ms~\cite{Cuesta:2013vpa}.

As it was explained in Section~\ref{Section:ANAIS_Filter}, an important difference in the pulse shape variables distribution is observed between calibration and background measurements at low energies, as Figure~\ref{P1MuExp_3keV} showed for events between 3~and 4~keV. The mean values of these distributions are compared in Table~\ref{table:p1mu_ExpSim}. This difference is not understood, but it could be related to the different spatial distribution of the energy depositions corresponding to both populations. The simulation does not show any difference between the pulse shapes corresponding to $^{109}Cd$ energy depositions, occurring in the surface, and those of $^{40}K$ and $^{22}Na$ events, discarding the light propagation as responsible of any of the observed differences. However, as the simulation is not able to reproduce the observed pulse shapes, further work is required. The introduction of both, additional scintillation components and spatial effects in the light production, both in the crystal bulk and in the crystal surface could be a very important improvement in future work with the simulation.

\begin{table}[h]
	\centering
	\begin{tabular}{|c|c|c|c|c|c|}
		\cline{3-6}
		\multicolumn{2}{c|}{} & \multicolumn{2}{|c|}{Mean of $\mu$ (ns)} & \multicolumn{2}{|c|}{Mean of $p_1$} \\
		\hline
		Run & E (keV) & Sim. & Exp. & Sim. & Exp. \\
		\hline
		Cal. & 22.6 & 251$\pm$16 & 262$\pm$20 & 0.681$\pm$0.028 & 0.750$\pm$0.037 \\
		Cal. & [3,4] & 236$\pm$16 & 192$\pm$4 & 0.654$\pm$0.027 & 0.640$\pm$0.001 \\
		Bkg. & [3,4] & 232$\pm$2 & 217$\pm$4 & 0.654$\pm$0.005 & 0.675$\pm$0.001 \\
		\hline
	\end{tabular} \\
	\caption{Results of the mean of the pulse shape variables for the 22.6~keV peak and for events in the energy range between 3~and 4~keV. They are compared for experimental and simulation data and for background and calibration measurements. For the simulated background, the events of the 3.2~keV peak from the $^{40}K$ decays have been used.}
	\label{table:p1mu_ExpSim}
\end{table}

\subsection{Efficiency calculation} \label{Section:SIM_Res_Efficiency}

The simulation can be very useful to check the efficiencies of the ANAIS-112 trigger and the filtering protocols applied in the data analysis, enabling also studies in order to improve them. The filtering protocols of ANAIS were explained in Section~\ref{Section:ANAIS_Filter}, as well as the corresponding efficiencies at energies at or near the ROI and the methods used for their experimental estimate. In this section, the value of these efficiencies at 1~keV will be compared with the results from the simulation for the 0.9~keV peak.

\begin{figure}[h!]
	\begin{subfigure}[b]{0.49\textwidth}
		\includegraphics[width=\textwidth]{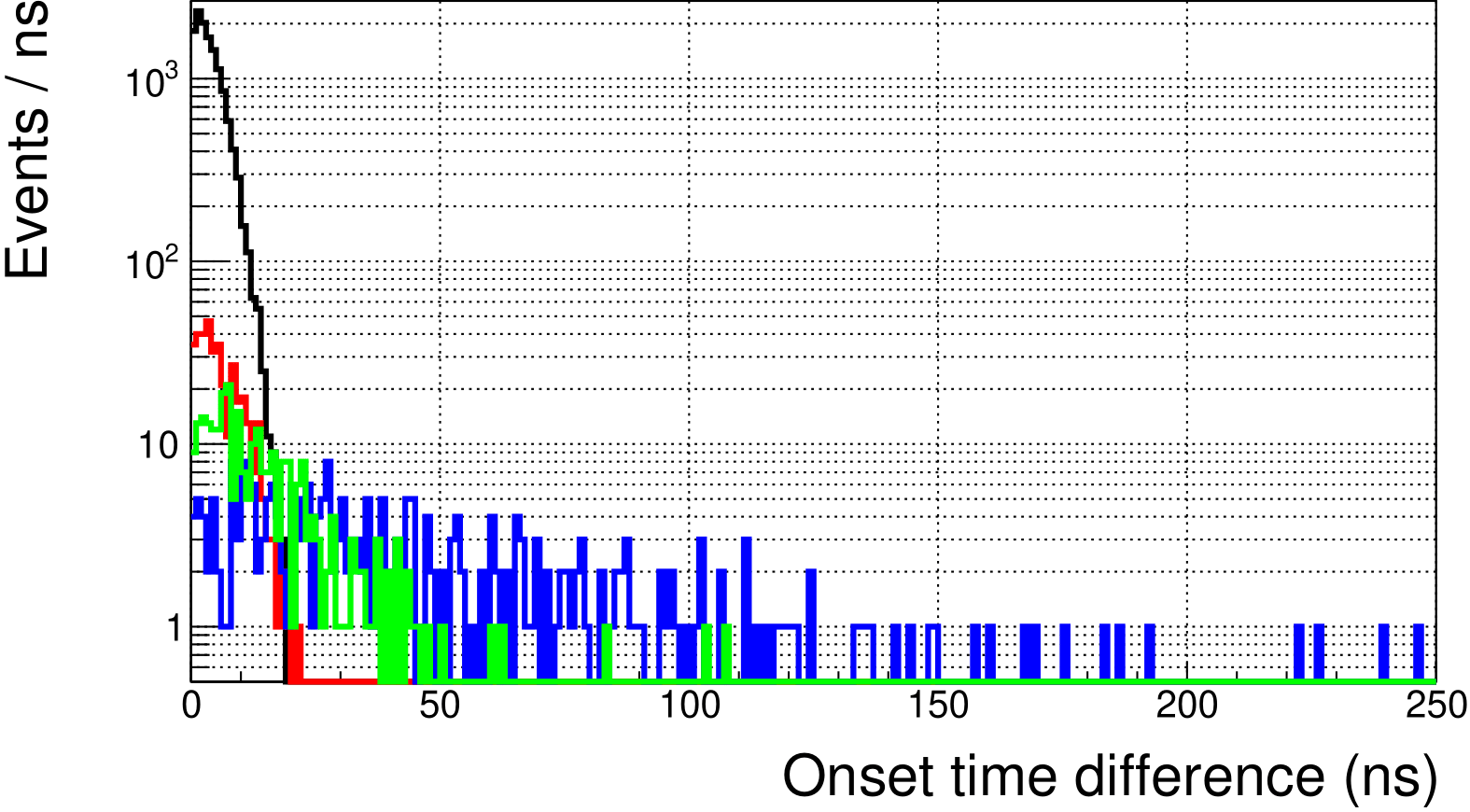}
	\end{subfigure}
	\begin{subfigure}[b]{0.49\textwidth}
		\includegraphics[width=\textwidth]{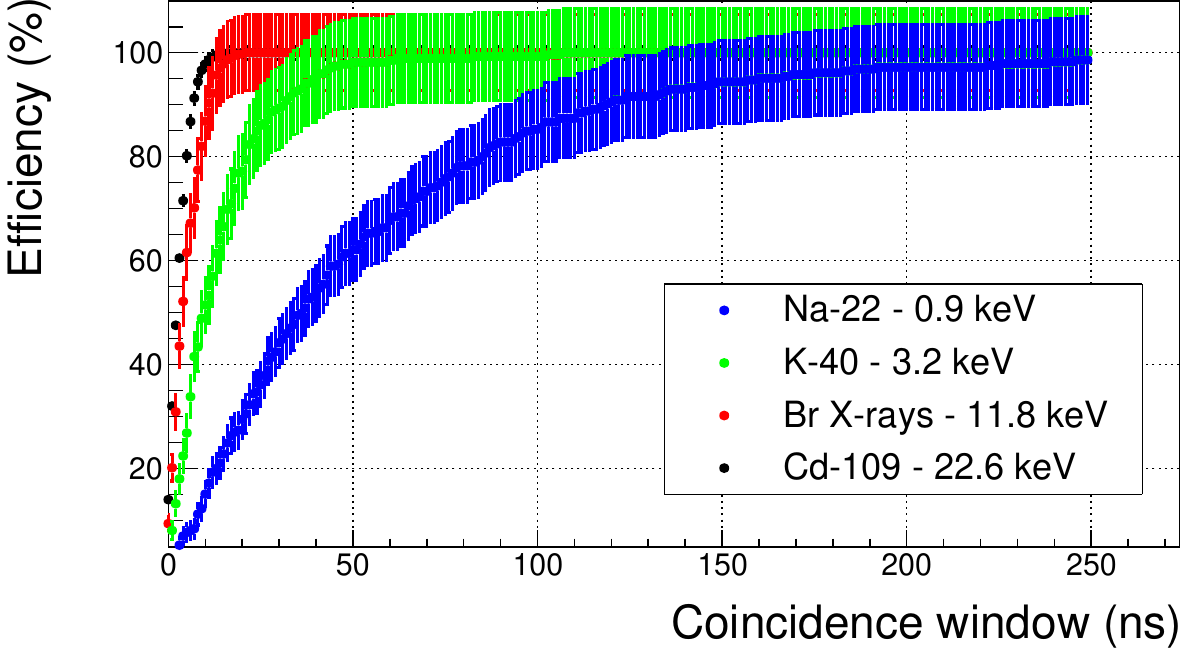}
	\end{subfigure}
	\caption{\label{timeDifEff}Left plot: Distribution of the onset time difference for the four peaks used in the low energy calibration of the experiment. Right plot: trigger efficiencies as a function of the coincidence window applied in the trigger protocol.}
\end{figure}

First, the trigger efficiency has been analyzed  as a function of the coincidence window used to define the trigger. In the ANAIS experiment, the trigger is setup as the logical AND between the two PMTs trigger signals (requiring one photoelectron each) in a 200~ns time window. The module of the difference between the pulse onset time in both PMTs, providing the difference between the arrival of the first photoelectrons to both PMTs, is shown for the four peaks in the ROI in Figure~\ref{timeDifEff}, as well as the corresponding trigger efficiencies as a function of the coincidence window. These efficiencies are 100\% for a time window of 200~ns for all the peaks except the 0.9~keV of the $^{22}Na$, which is 96~$\pm$~11\% (1\% of them were not coincident and 3\% were coincident in a time window larger than 200~ns). The reason for the large uncertainty of this result is the low number of events obtained in the simulation. This trigger efficiency was obtained as 97.3~$\pm$~0.8\% for events of 1~keV (see Section~\ref{Section:ANAIS_Filter}) which is in agreement with the value obtained for the simulation presented in this Chapter.

\begin{figure}[h!]
	\begin{subfigure}[b]{0.49\textwidth}
		\includegraphics[width=\textwidth]{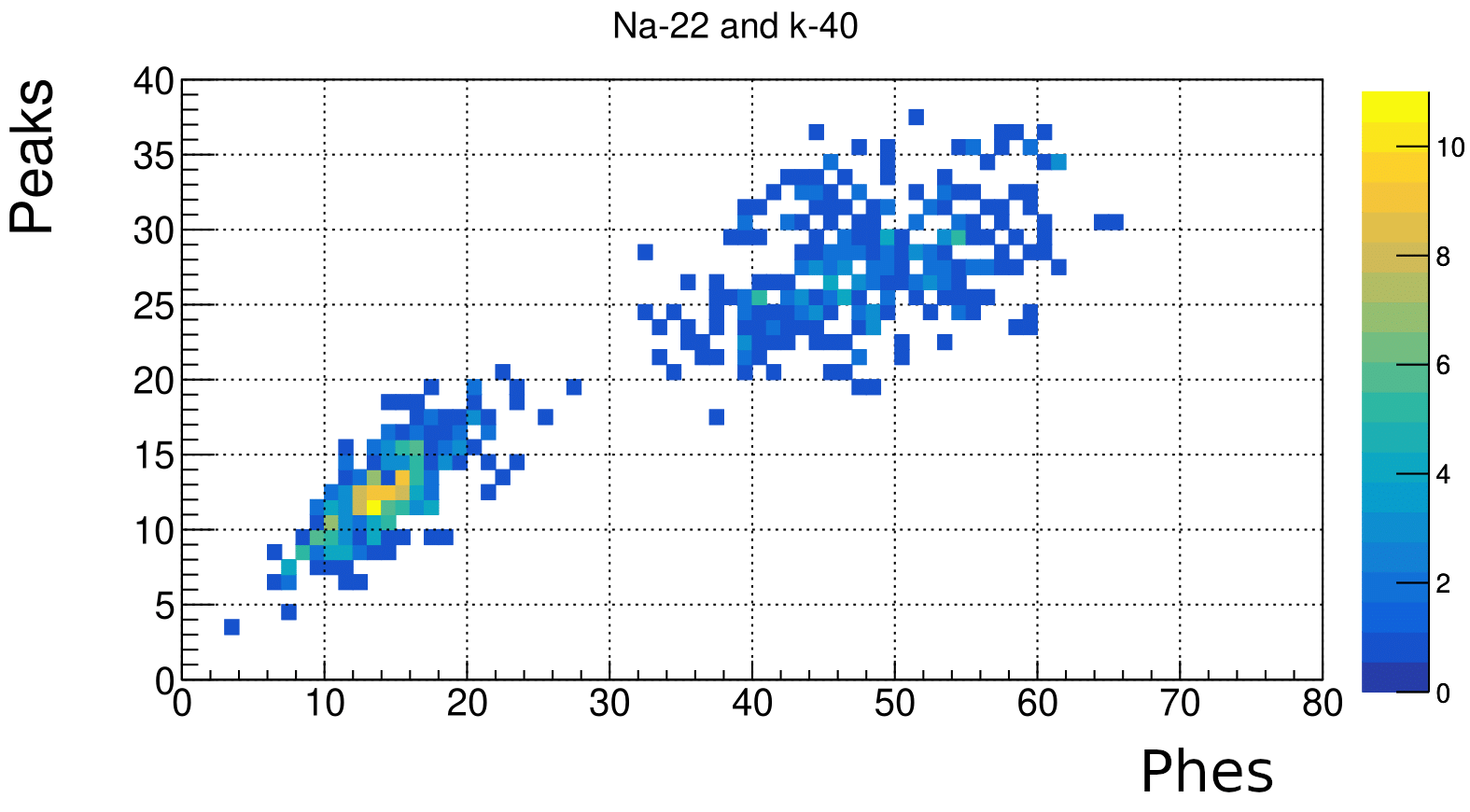}
	\end{subfigure}
	\begin{subfigure}[b]{0.49\textwidth}
		\includegraphics[width=\textwidth]{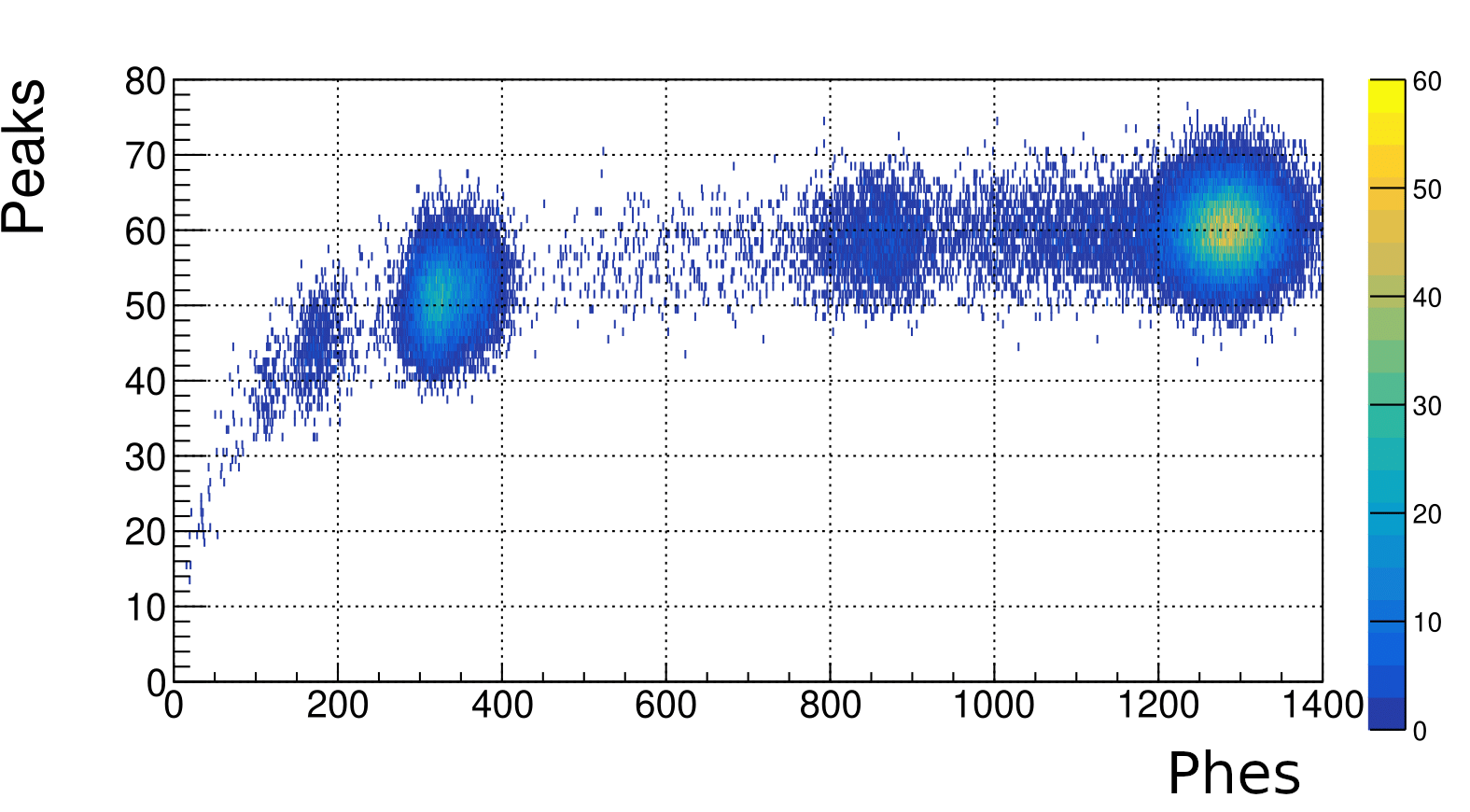}
	\end{subfigure}
	\caption{\label{peaks_vs_photons}Scatter plot of the number of peaks found by the peak-finding algorithm versus the number of photoelectrons produced in the photocathodes for the three simulated populations. On the left, 0.9~keV of the $^{22}Na$ ad 3.2~keV of the $^{40}K$. On the right: all the events recorded in the simulation with the $^{109}Cd$ source.}
\end{figure}

With respect to the bulk scintillation events selection, in ANAIS-112 two criteria are applied: the PSV and the asymmetry cut (see Section~\ref{Section:ANAIS_Filter}). The same peak-finding algorithm applied to ANAIS-112 pulses has been applied to the simulated pulses. Figure~\ref{peaks_vs_photons} shows the scatter plot of the number of peaks found by the peak-finding algorithm versus the number of photoelectrons produced in the photocathodes for the three simulated populations. It is possible to observe that the number of peaks saturates at a value of around~60, but it becomes appreciably lower than the number of photoelectrons even for 3.2~keV energy depositions due to the high photoelectron overlap probability.

\begin{figure}[h!]
	\begin{subfigure}[b]{0.49\textwidth}
		\includegraphics[width=\textwidth]{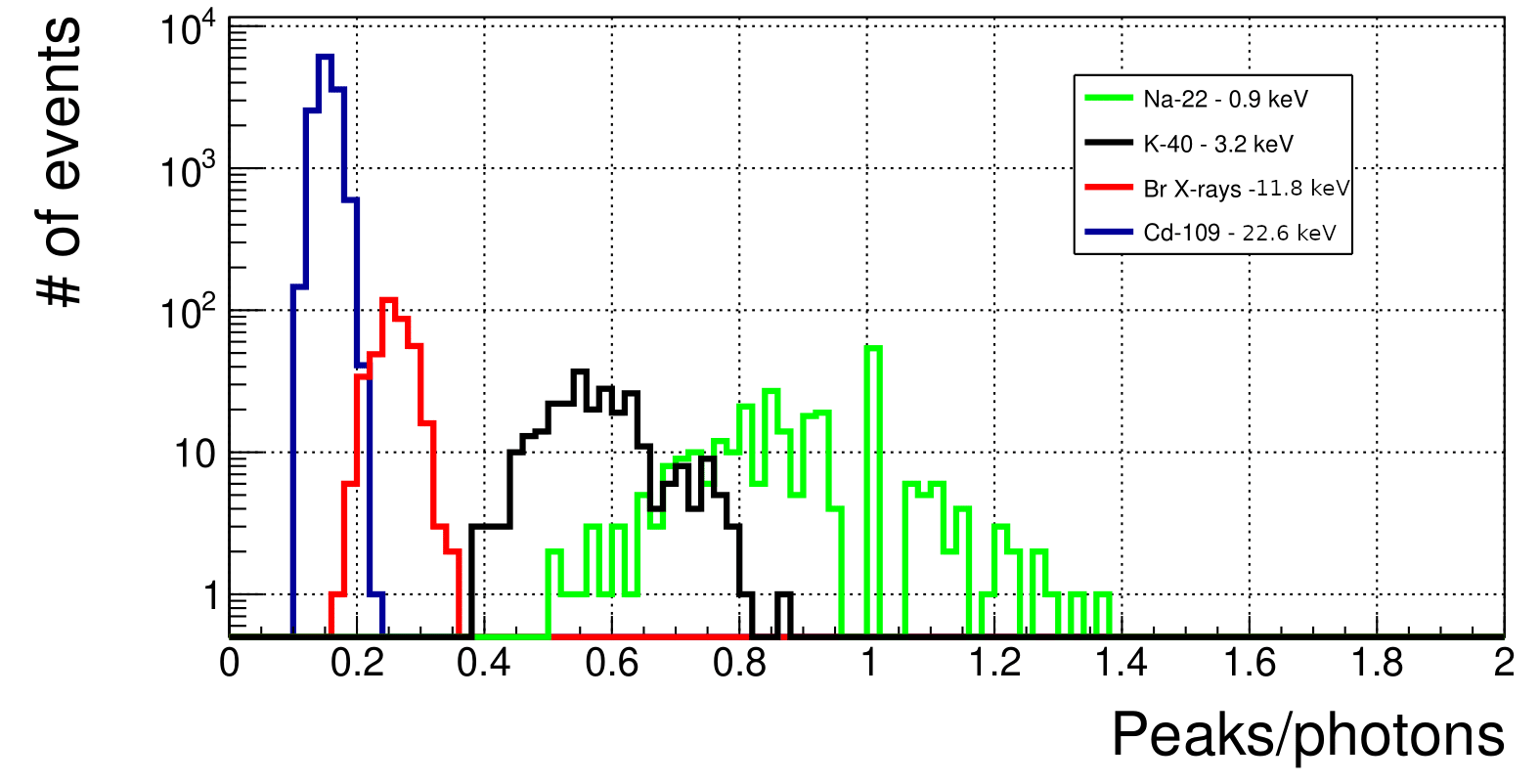}
	\end{subfigure}
	\begin{subfigure}[b]{0.49\textwidth}
		\includegraphics[width=\textwidth]{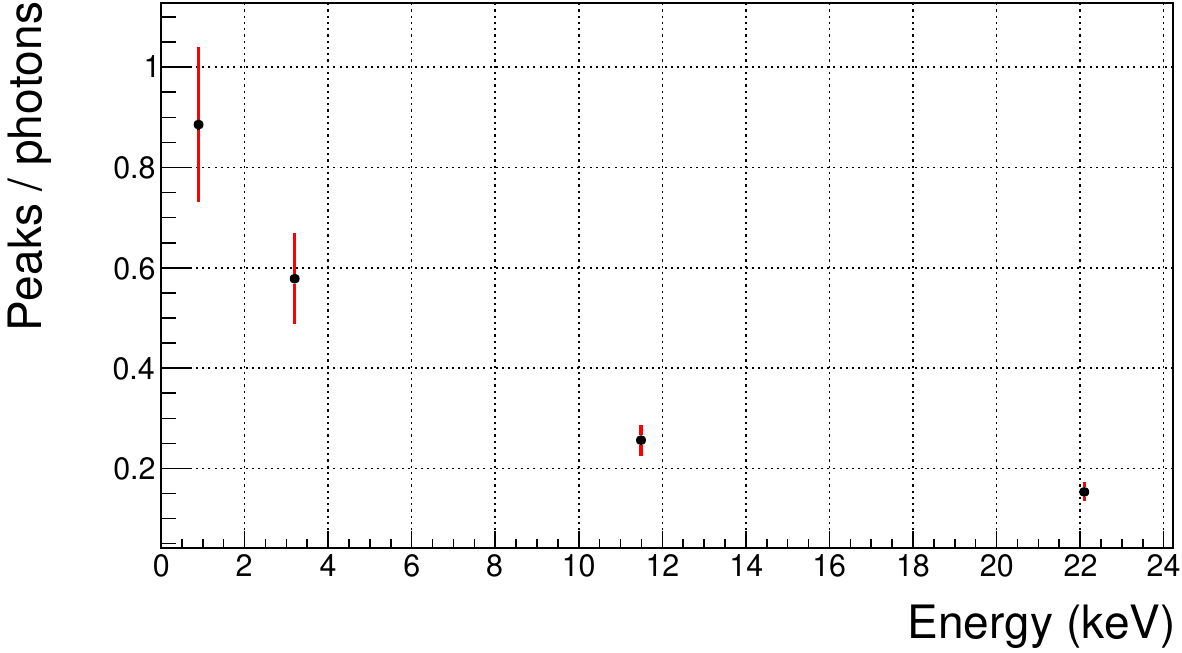}
	\end{subfigure}
	\caption{\label{peaks/photons}Left plot: distribution of the ratio of peaks identified to the total number of photoelectrons. Right plot: means of these distributions. Standard deviations of the distributions are also shown in this plot as the red lines.}
\end{figure}

The distribution of the ratio between the number of peaks found and the number of photoelectrons is represented in Figure~\ref{peaks/photons}, as well as the mean values. This ratio for pulses having a very low number of photoelectrons inside the time acquisition window can be understood as the efficiency of the algorithm. For the 0.9~keV peak, which is below the minimum of the ROI of the experiment, it has been obtained a value of 89\% with a standard deviation of 15\%. It confirms the robustness of this algorithm in the identification of individual photoelectrons, which is very important in order to guarantee the correct performance of the filtering of the asymmetry cut, based on the number of peaks identified in the pulse.

\begin{figure}[h!]
	\begin{center}
		\includegraphics[width=0.75\textwidth]{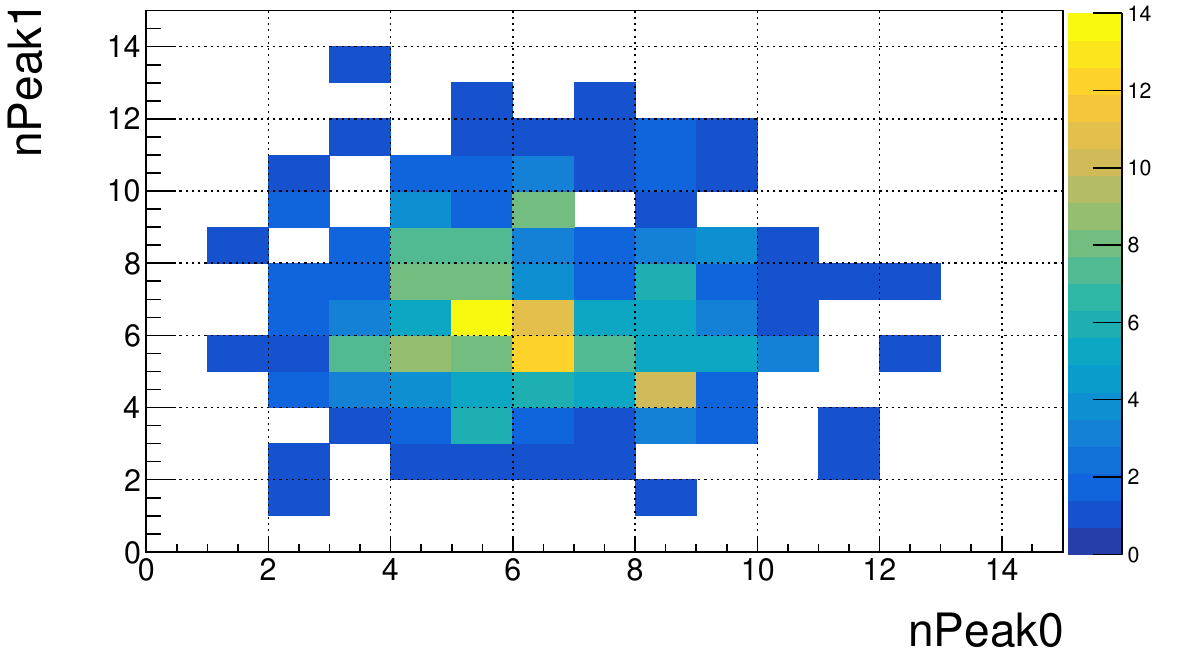}
		\caption{\label{Na22Peaks}Scatter plot of the number of peaks identified in each PMT for the simulated events of the 0.9~keV peak that have passed the trigger condition.}
	\end{center}
\end{figure}

Figure~\ref{Na22Peaks} shows the scatter plot of the number of peaks identified in each PMT for the simulated events of the 0.9~keV peak that have passed the trigger condition. The efficiency derived from the simulated events from $^{22}Na$ for the selection criterion applied in ANAIS-112 (at least 5 peaks identified in each PMT signal) is 55~$\pm$~5\%, while in ANAIS-112 data the average efficiency for all the modules is 43\% with a standard deviation of 10\% depending on the module (see Figure~\ref{TriggerEfficiency}) for calibration events with energies of 1~keV. It is worth noting that, although they energy of the events used for that comparison is slightly different, both values are in agreement.

\begin{figure}[h!]
	\begin{subfigure}[b]{0.49\textwidth}
		\includegraphics[width=\textwidth]{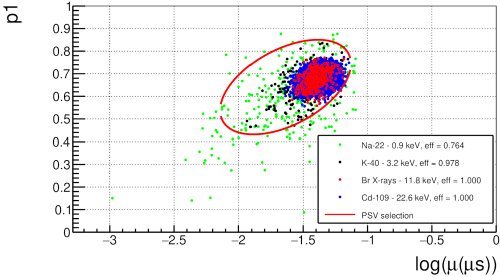}
	\end{subfigure}
	\begin{subfigure}[b]{0.49\textwidth}
		\includegraphics[width=\textwidth]{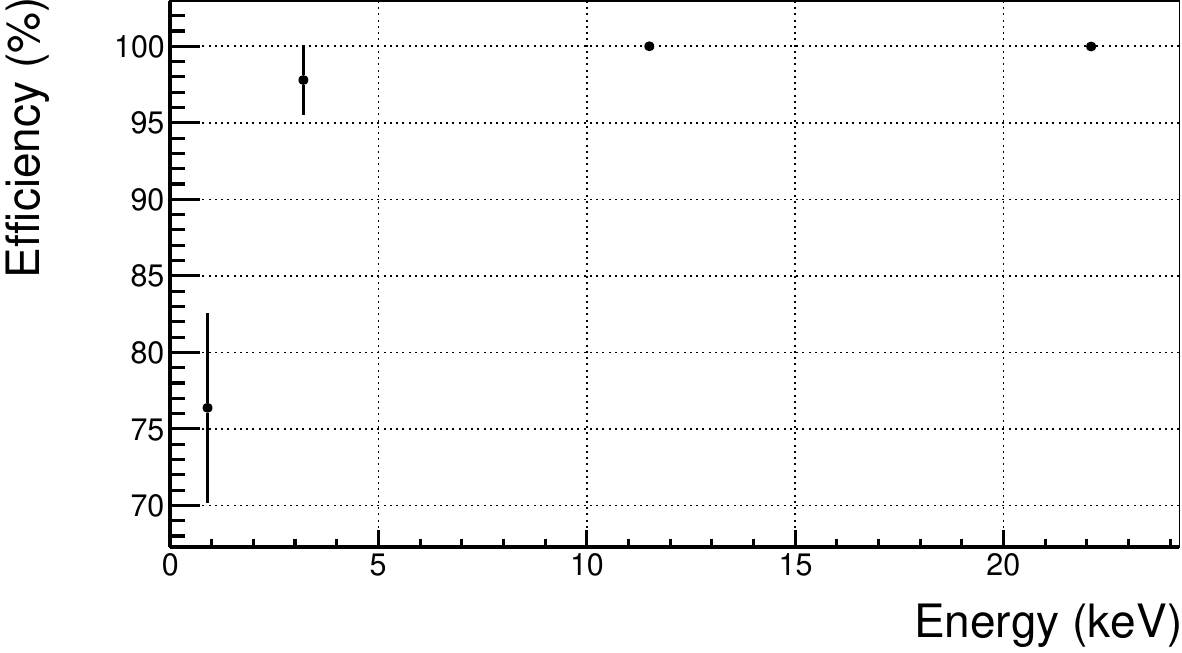}
	\end{subfigure}
	\caption{\label{PSVcut}Left plot: scatter plot in the $p_1$:$\log\left(\mu\right)$ plane of the four simulated energy peaks. The PSV selection applied in the ANAIS-112 experiment is plotted as the red line. Right plot: PSV selection efficiency for each of these energy peaks.}
\end{figure}

The second selection criterion is based on the PSV parameter. Because the simulated events do not show the same pulse shape than the experimental ones, the following analysis is just an exercise to show the potential of the simulations, when refined to reproduce conveniently all the relevant signal features. Figure~\ref{PSVcut} shows the scatter plot for ($p_1$:$\log\left(\mu\right)$) pulse shape variables for the four simulated peaks, keeping only those events fulfilling the trigger condition. Corresponding acceptance efficiencies are also shown for the PSV cut (red line) used in the ANAIS experiment.

\begin{figure}[h!]
	\begin{center}
		\includegraphics[width=0.75\textwidth]{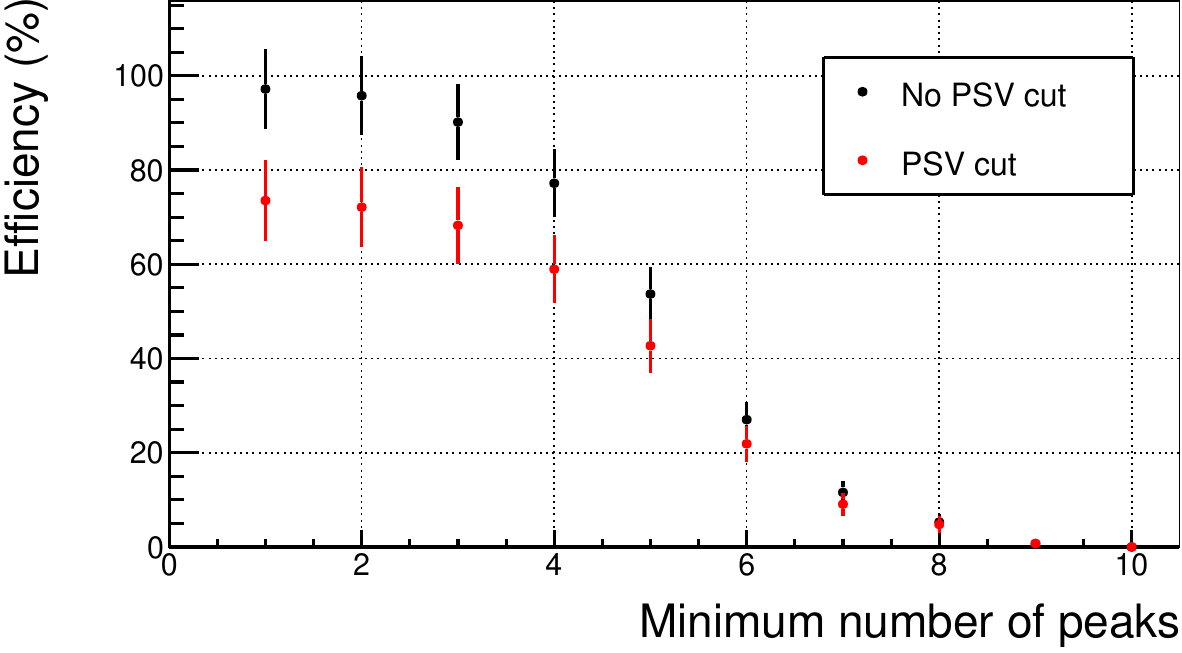}
		\caption{\label{efficiencyNa22}Selection efficiency as a function of the minimum required number of peaks in each PMT for the simulated events of the 0.9~keV peak that have passed the trigger condition, both with and without the application of the PSV selection.}
	\end{center}
\end{figure}

In the following, we focus our interest on the 0.9~keV peak, which is in fact below the ROI, just at the experiment’s energy threshold. For this peak, both selection criteria have low efficiencies. Figure~\ref{efficiencyNa22} shows the acceptance efficiency as a function of the minimum number of peaks required as selection criterion, with and without application of the PSV selection.

Thus, the current selection efficiency at 0.9~keV calculated from this simulation is 55~$\pm$~5\% applying only the asymmetry cut and 43~$\pm$~5\% considering also the PSV selection. This values in the experimental data for events with energies of 1~keV are 43\% and 30\%, with standard deviation of 10\% and 7\%, respectively. A decrease of the minimum number of peaks required from 5 to 3 would increase the efficiency from 43\% to 70\%. However, it would also imply the reduction of the rejection of the asymmetric events observed in the ANAIS-112 experiment, as it was explained in Section~\ref{Section:ANAIS_Filter}. It gives an idea of the importance of the identification of this background contribution, and the high increase in the sensitivity of the experiment if this kind of events could be rejected from the data. If their origin is in the PMTs or in the optical windows, then alternatives to these light detectors, as the SiPMs, have to be explored (see Chapter~\ref{Chapter:SiPMZgz}).

In order to improve this efficiency estimate, in particular that of the PSV cut, further work is required to include the slower scintillation components for instance, and to model the differences observed between $^{109}Cd$ calibration events and bulk scintillation events. 

\subsection{Study of Cherenkov emission in the PMTs} \label{Section:SIM_Res_cherenkov}

Apart from scintillation in the NaI(Tl) crystal, this simulation also allows to reproduce and analyze other light production mechanisms as it is the Cherenkov emission in transparent media. This emission occurs when particles travel within a medium with a velocity higher than that of the light in this medium. Therefore, a minimum energy is required for a particle to produce Cherenkov light, the so-called Cherenkov energy threshold, which depends on the mass in rest, $m$, and the refractive index of the medium, $n$, as:
\begin{equation}
	E_{th} = mc^2\left(\left(1-1/n^2\right)^{-1/2}-1\right).
\end{equation}
It is expected to happen in the borosilicate of the PMTs, in the quartz windows, in the silicone pads and in the optical gel, where the Cherenkov energy thresholds for electrons are 161, 158, 209 and 139~keV, respectively (see Section~\ref{Section:SIM_Const_Prop} for information on the refractive index considered for these media). The activities of the four primordial radioactive isotopes present in these materials ($^{238}U$, $^{232}Th$, $^{226}Ra$ and $^{40}K$) have been either measured or constrained by determining upper limits (as it was presented in Section~\ref{Section:ANAIS_Background}), obtaining that the most important contribution comes from the PMTs. From all of these isotopes, $^{40}K$ is expected to be concentrated in the borosilicate of the PMT due to its affinity with that material, while the other three could also be present in the dynodes, wiring and other internal components.

This isotope decays $\beta^-$ with a branching ratio of 89\% and a Q-value of 1311~keV, which means that a large fraction of the electrons produced will have an energy above the Cherenkov energy threshold. The other 11\% of the decays are EC processes to an excited state, emitting a gamma ray of 1461~keV. This gamma ray can deposit energy through photoelectric or Compton effect in the borosilicate of the PMT or in the optical windows transferring its energy to the electrons of the material, which in turn can produce Cherenkov light. Moreover, the decay of $^{40}K$ is much more simple than the decay chains of the $^{238}U$ and $^{232}Th$. All these factors made it the most interesting isotope to use in the first analysis of the detection of the Cherenkov light in the ANAIS-112 experiment using this simulation. A continuation of this work will be the analysis of the effect that natural chains can have in the ANAIS data depending on their distribution in the PMT.

Cherenkov events should be very fast, easy to remove by the asymmetry and PSV cuts. However, in case of the decay of $^{40}K$ by EC, it will emit a 1461~keV gamma ray which, as commented before, can produce the Cherenkov light in the PMT or the optical windows following a Compton effect in those volumes, while the scattered photon can deposit energy in the NaI(Tl) crystal, producing scintillation. These events would have contribution of both Cherenkov and scintillation light and therefore they could be more difficult to remove and be contaminating ANAIS data. The simulation presented in this section was designed to bring more information on this possibility. Then, 10$^6$ decays of $^{40}K$ were simulated distributed homogeneously in all the borosilicate volumes (head and body) of the two PMTs of the module.

\begin{figure}[h!]
	\begin{center}
		\includegraphics[width=0.75\textwidth]{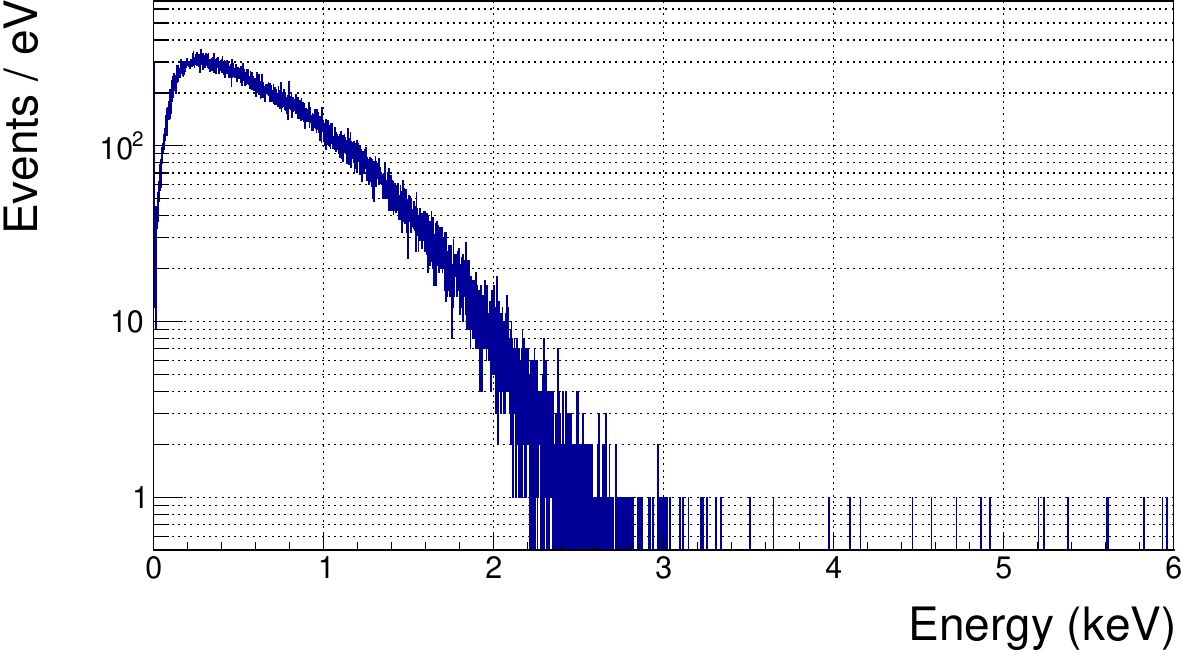}
		\caption{\label{Energycherenkov}Energy calibrated spectrum of the Cherenkov events that pass the trigger condition in the simulation of $^{40}K$ decays in the PMTs. Spectrum is only shown in the ROI of the experiment.}
	\end{center}
\end{figure}

From all simulated decays, 28.69~$\pm$~0.06\% of them produced light events fulfilling the trigger condition. The average of $^{40}K$ content in the PMTs used in the experiment is 114.9~$\pm$~4.6~mBq/PMT and therefore for each module is 229.8~$\pm$~9.2~mBq (see Section~\ref{Section:ANAIS_Background}). This results in a contribution to the trigger rate of 65~$\pm$~3~mHz per module, or equivalently, 593~$\pm$~24~mHz for the whole experiment, which is more than 10\% of the trigger rate of the experiment. Figure~\ref{Energycherenkov} shows the energy equivalent to the pulse areas of the events which pass the trigger condition. Conversion from pulse areas to energy is done using the calibration obtained in Section~\ref{Section:SIM_Res_SR}, specifically in Table~\ref{tabla:Calibration}. The percentage of decays producing events within the ROI of ANAIS is 5.27~$\pm$~0.02\%.

\begin{figure}[h!]
	\begin{center}
		\includegraphics[width=\textwidth]{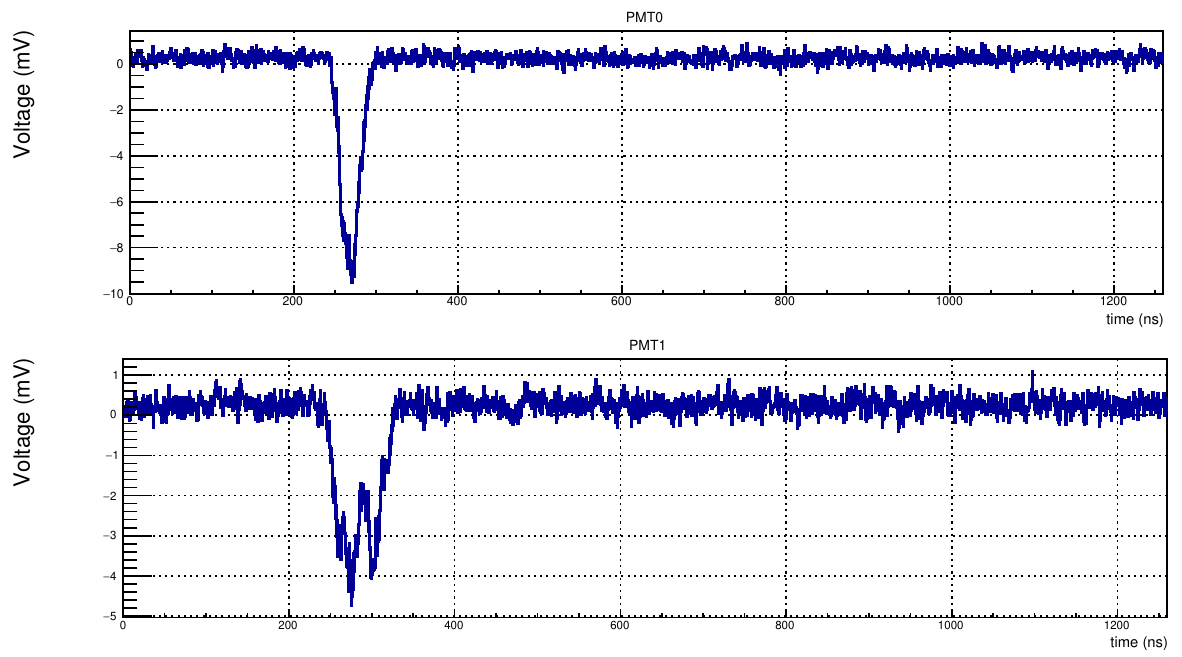}
		\caption{\label{cherenkovPulse}Cherenkov pulse that pass the trigger condition. In this pulse, four photoelectrons were produced in each photocathode, all together in the PMT0 and separated in time in the PMT1.}
	\end{center}
\end{figure}

\begin{figure}[h!]
	\begin{center}
		\includegraphics[width=0.75\textwidth]{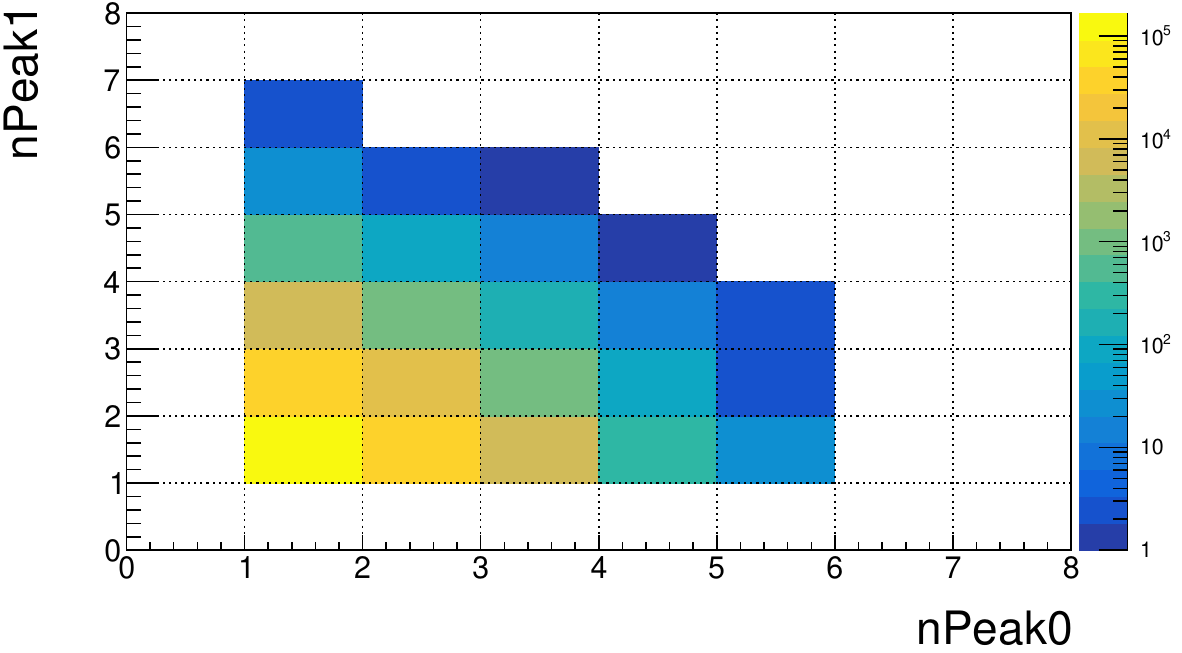}
		\caption{\label{peakscherenkov}Scatter plot of the number of peaks identified in each PMT for the events that have passed the trigger condition and have an equivalent energy below 6~keV in the simulation of $^{40}K$ decays in the PMTs.}
	\end{center}
\end{figure}

Figure~\ref{cherenkovPulse} shows a Cherenkov event passing the trigger condition. In that event, four photoelectrons were produced in each photocathode, all together in the PMT0 and separated in time in the PMT1. This example shows on the one hand the short characteristic time of these events compared with those from scintillation, and on the other hand the small number of peaks that the peak-finding algorithm can identify. Both properties combined allow to reject most of these events.
In Figure~\ref{peakscherenkov}, we can observe the number of peaks identified in each pulse by the peak-finding algorithm for the simulated events resulting from the decays of $^{40}K$ in the PMTs which would produce a valid trigger in the experiment and would have equivalent energies below 6~keV. All of these events are rejected by requiring at least five peaks in each PMT signal, criterion followed in ANAIS. It is possible to observe that the \textit{nPeak0:nPeak1} distribution is very symmetric, which indicates that Cherenkov events should not be the reason of the asymmetrical events observed in the ANAIS-112 experiment.

\begin{figure}[h!]
	\begin{center}
		\includegraphics[width=0.75\textwidth]{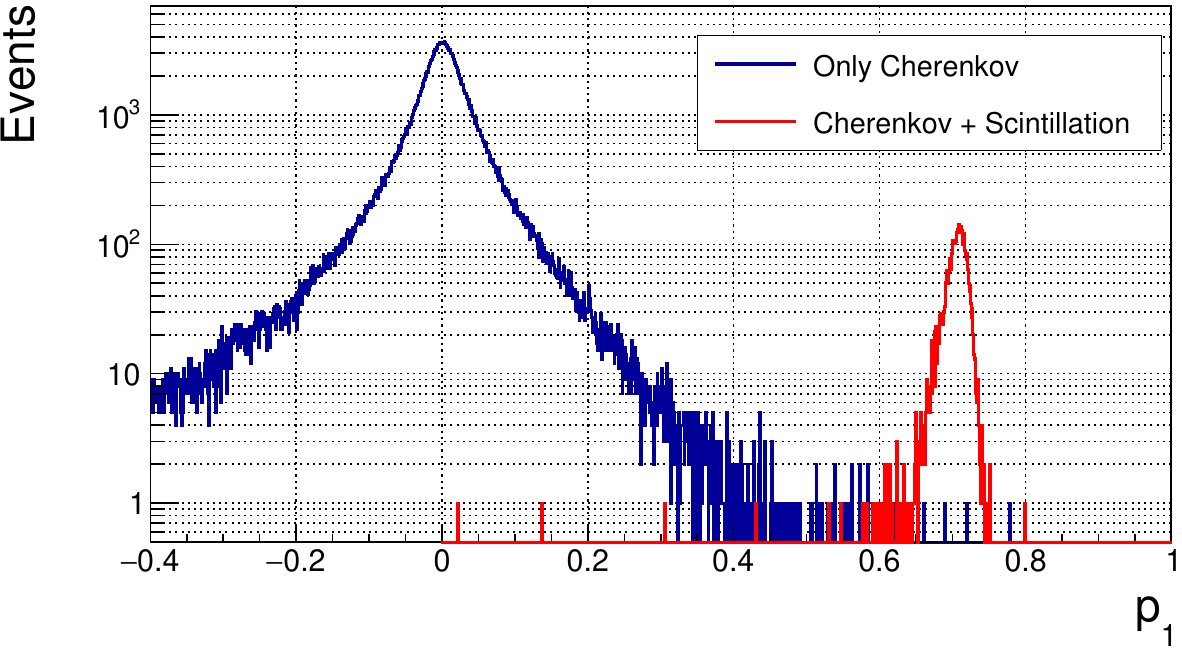}
		\caption{\label{p1_K40PMT}$p_1$ distributions for pure Cherenkov events (blue line) and for events with Cherenkov mixed with Scintillation light (red line) in the simulation of $^{40}K$ decays in the PMTs.}
	\end{center}
\end{figure}

\begin{figure}[h!]
	\begin{subfigure}[b]{0.49\textwidth}
		\includegraphics[width=\textwidth]{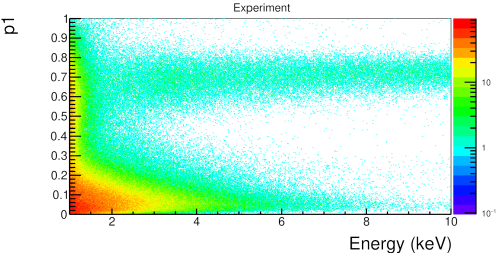}
	\end{subfigure}
	\begin{subfigure}[b]{0.49\textwidth}
		\includegraphics[width=\textwidth]{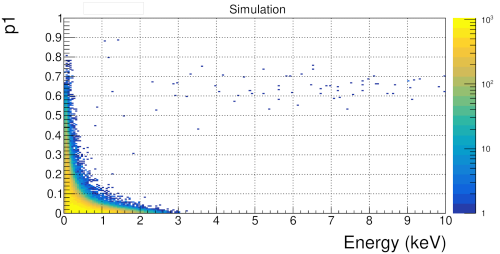}
	\end{subfigure}
	\caption{\label{p1Ene_Cherenkov}Scatter plot in the $p_1$:\textit{Energy} plane of experimental data events (left plot) and simulation of $^{40}K$ in the borosilicate of the PMTs (right plot).}
\end{figure}

From all the triggered events, some of them will have only Cherenkov light, other with only scintillation light and other with a mixture of both. In Figure~\ref{p1_K40PMT}, which compares the $p_1$ distributions for pure Cherenkov and events with Cherenkov mixed with Scintillation light, it is observed that the mixed population has a $p_1$ value similar to that expected for NaI(Tl) scintillation. Moreover, in Figure~\ref{p1Ene_Cherenkov}, which presents the $p_1$:\textit{Energy} scatter plots for the background in experimental data and this simulation, it is possible to observe in both a similar tendency of the low $p_1$ population.

\begin{figure}[h!]
	\begin{center}
		\includegraphics[width=0.75\textwidth]{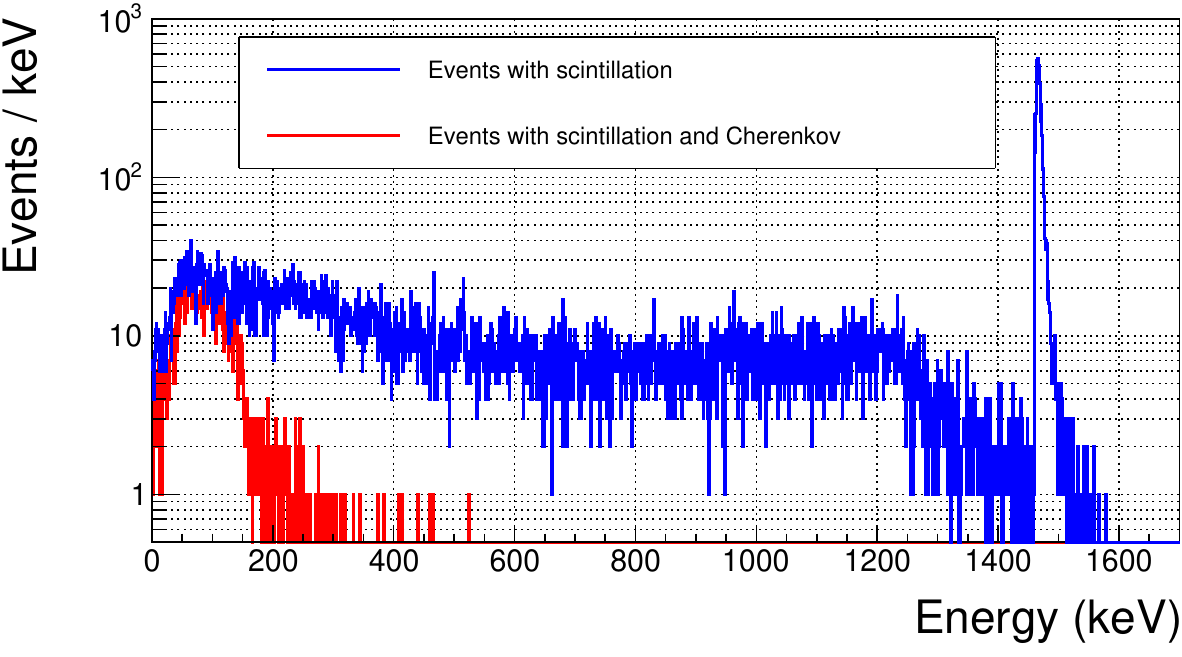}
		\caption{\label{Energy_K40PMTs}Spectrum of the energy deposited in the crystal in the simulation of $^{40}K$ decays in the PMTs for all the events (blue line) and only for those consisting of both, scintillation and Cherenkov light (red line).}
	\end{center}
\end{figure}

The total number of events that have deposited energy in the crystal and pass the asymmetry cut is 1.94~$\pm$~0.01\% of the $^{40}K$ decays, which implies a contribution to the background of 31~$\pm$~1~c/kg/day. Among them, those events which include Cherenkov light from borosilicate or optical windows are 0.21~$\pm$~0.01\% of the decays. Figure~\ref{Energy_K40PMTs} shows the total spectrum of the energy deposited in the crystal in this simulation compared with that of those events consisting of both, scintillation and Cherenkov light. In the first one it is possible to identify the 1461~keV photopeak and the Compton spectrum, while the second one is concentrated at low energies, as it implies at least one Compton scattering in a transparent volume. The contribution of all the triggered events in the ROI ([1,6]~keV) is 0.003~$\pm$~0.001\% of the simulated events, corresponding to a rate of 0.049~$\pm$~0.009~c/kg/day. Table~\ref{tabla:CherenkovRates} shows the percentages relative to the number of $^{40}K$ decays in the PMTs borosilicate of events fulfilling the ANAIS-112 trigger criterion, falling in the ROI, and passing the filtering procedures for two different values of the photocathode transparency.  

The photocathode was simulated as opaque, but the number of events passing the trigger condition is expected to increase with its transparency. Therefore, to analyse the effect that this parameter has on the results, the same simulation was run but with a photocathode transparency of 10\% and simulating 10$^4$ events. The percentage of events passing each selection criterion is shown in Table~\ref{tabla:CherenkovRates}. It can be observed an small but non-negligible effect of this parameter on the trigger rate of Cherenkov events, while in pure scintillation events, results are compatible within one standard deviation, as expected. It seems clear that the photocathode modelling in the optical simulation is important for this kind of events, and it should be improved, moving to more realistic models, in future works.

\begin{table}[h]
	\centering
	\begin{tabular}{|c|c|c|}
		\cline{2-3}
		\multicolumn{1}{c|}{} & \multicolumn{2}{|c|}{Events (\%)} \\
		\hline
		Selection & $T = 0\%$ & $T = 10\%$ \\
		\hline
		Trig. (All)	 								& 28.69~$\pm$~0.06 	& 32.25~$\pm$~0.65 \\
		Trig. (Pure Cherenkov)						& 26.75~$\pm$~0.06 	& 30.46~$\pm$~0.63 \\
		Trig. (Pure Scintillation)		 			& 1.73~$\pm$~0.01 	& 1.61~$\pm$~0.13 \\
		Trig. (Cherenkov + Scintillation) 			& 0.21~$\pm$~0.01 	& 0.18~$\pm$~0.04 \\
		Trig. (All) + in ROI					 	& 5.27~$\pm$~0.02   & 7.04~$\pm$~0.27 \\
		Trig. (All) + the Assymetry cut + in ROI	& 0.003~$\pm$~0.001 & $<$~0.01 \\
		\hline
	\end{tabular} \\
	\caption{Percentages relative to the number of $^{40}K$ decays in the PMTs borosilicate of events passing each selection criterion for a photocathode transparency of 0\% or 10\%.}
	\label{tabla:CherenkovRates}
\end{table}

\subsection{Contribution of $^{222}Rn$ and $^{40}K$ decays to the Blank trigger rate} \label{Section:SIM_Res_Blank}

As it was explained in Section~\ref{Section:ANAIS_Setup_Blank}, to study non-bulk scintillation events originated in the PMTs, a Blank module was installed in the Hall B of the LSC, very close to the ANAIS experimental setup. This module is similar to the nine modules but without NaI(Tl) crystal. It is enclosed in a lead shielding that can be flushed with Radon-Free Air (RFA) or nitrogen gas.

\begin{figure}[h!]
	\begin{center}
		\includegraphics[width=0.75\textwidth]{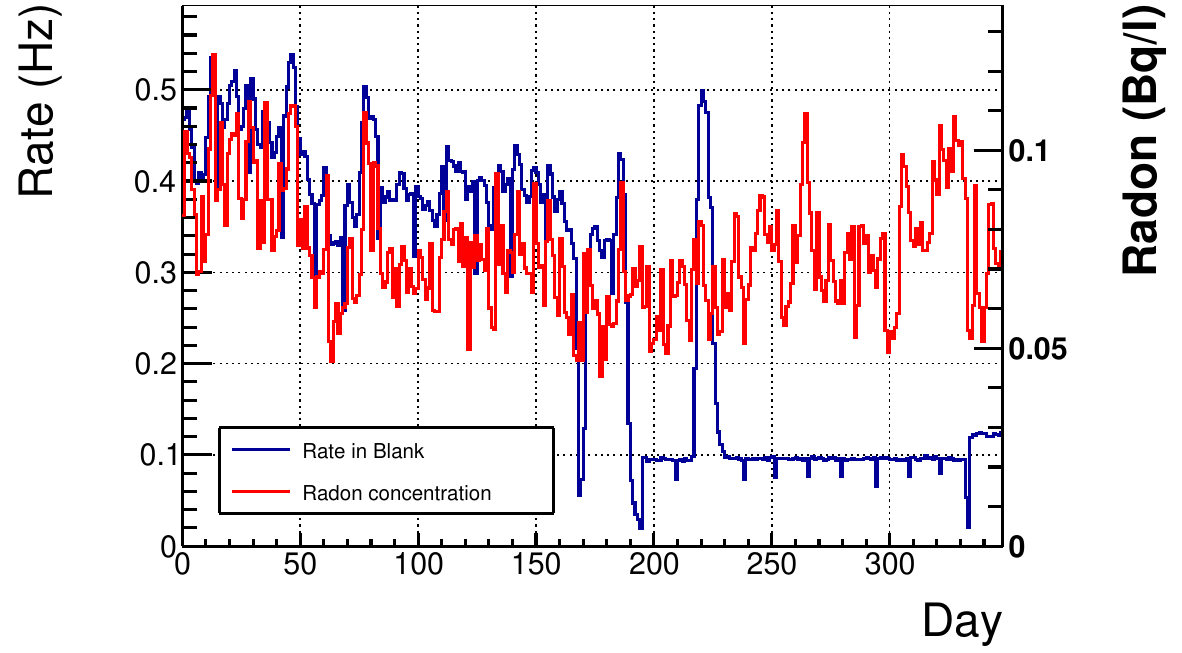}
		\caption{\label{RateBlank}Averaged trigger rate in the Blank module for each day of data taking (blue line) and averaged $^{222}Rn$ concentration measured by the AlphaGUARD for the laboratory air on the same day (red line).}
	\end{center}
\end{figure}

The first 160~days of data taking of this module, the RFA was not connected to the lead shielding, and therefore there was normal air around and within the Blank module. This implies that near the module there was $^{222}Rn$, whose content was being continuously monitored by a dedicated $Rn$-detector AlphaGUARD. Later on, the shielding was connected to the RFA line, and an important reduction was observed in the trigger rate, as it is shown in Figure~\ref{RateBlank}. The averaged trigger rate in the first 160~days of data taking was 412.3~$\pm$~0.2~mHz, while after that time (considering for the calculation the days from 250~to 350), the rate decreased down to 98.4~$\pm$~0.1~mHz. The concentration of $^{222}Rn$ measured by the Alpha-GUARD in the same period of time is also shown in Figure~\ref{RateBlank}.

\begin{figure}[h!]
	\begin{center}
		\includegraphics[width=0.75\textwidth]{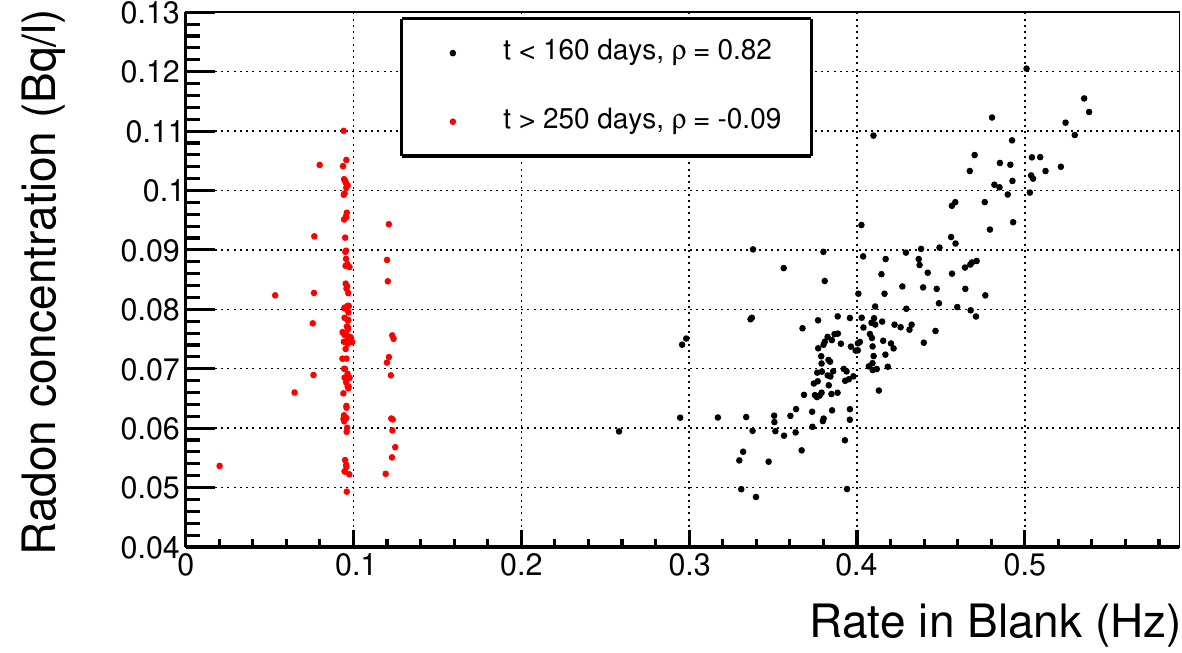}
		\caption{\label{corBlankRadon}Scatter plot of the concentration of $^{222}Rn$ as a function of the trigger rate in the Blank module obtained for each day in two different periods of time: t~$<$~160~days (black dots) and t~$>$~250~days (red dots).}
	\end{center}
\end{figure}

Figure~\ref{corBlankRadon} presents the scatter plot of the concentration of $^{222}Rn$ as a function of the trigger rate in the Blank module obtained for each day in the two different periods of time (t~$<$~160~days and t~$>$~250~days). A clear correlation is observed during the first period ($\rho = 0.82$) in contrast with that of the second period ($\rho = -0.09$), which indicates the important effect that the $^{222}Rn$ has on the trigger rate of this module.

For this purpose, a simulation of the module without crystal was run for 10$^6$ decays of $^{222}Rn$ homogeneously distributed in all the volumes containing air, which are: the space between the module and the lead walls (75.49~l), the volume inside the module (3.41~l) and the two volumes between the PMTs and the copper housing (0.54~l each one). This gives a total air volume of 79.98~l. The averaged content of $^{222}Rn$ measured by the AlphaGUARD in the Hall~B during the first period of data taking of the Blank module was 78.27~Bq/m$^3$, which is equivalent to 6.26~Bq in the simulated volume. Then, the 10$^6$ decays simulated are equivalent to 1.6$\times$10$^5$~s, i.e., 1.85~days.

\begin{figure}[h!]
	\begin{center}
		\includegraphics[width=0.5\textwidth]{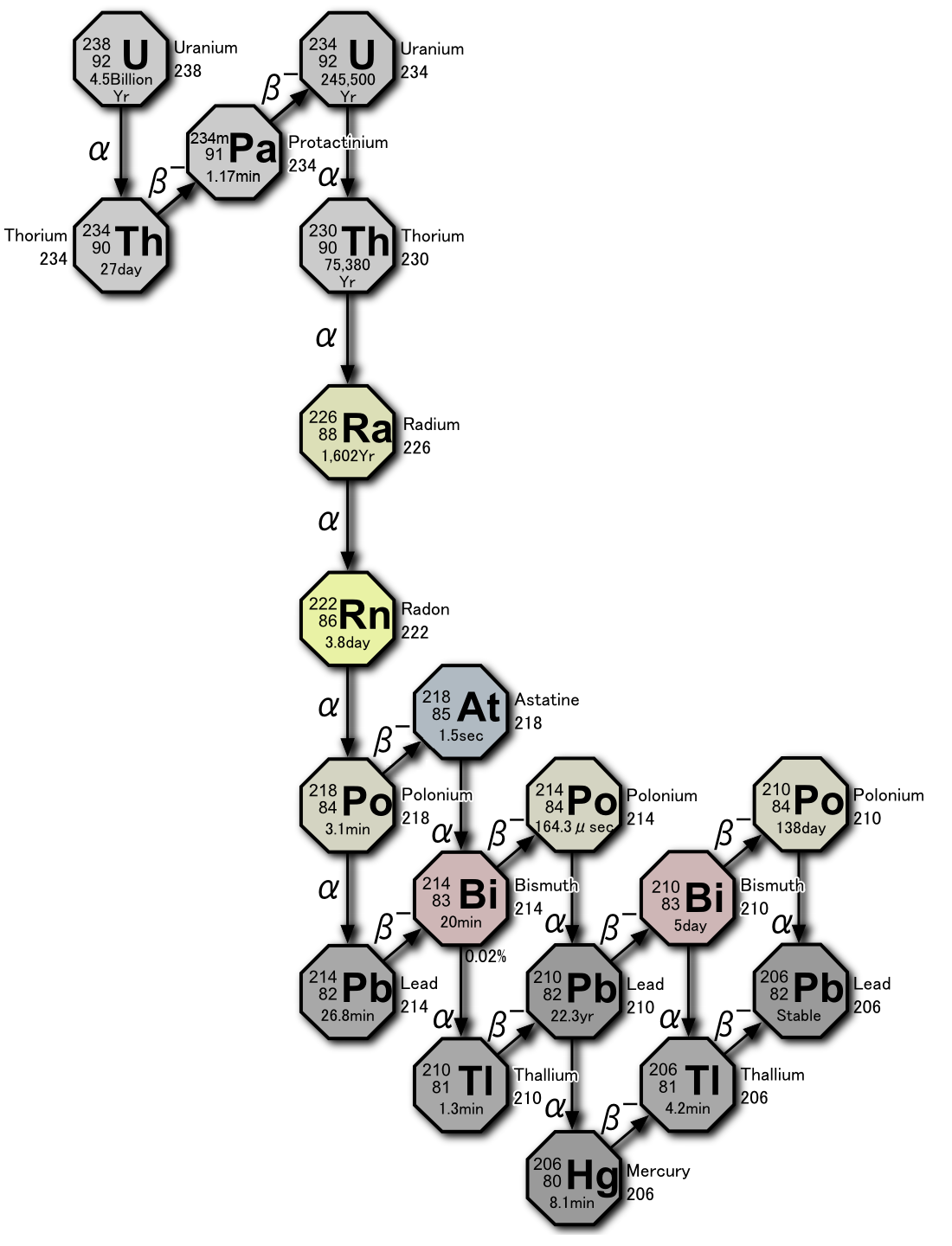}
		\caption{\label{Rn222Chain}Decay chain of the $^{238}U$. Image from~\cite{wikimedia}.}
	\end{center}
\end{figure}

The GEANT4 simulation allows to reproduce the decay chain of the $^{238}U$ (shown in Figure~\ref{Rn222Chain}) from the $^{222}Rn$ until the $^{210}Pb$, which is a long-lived isotope (22.3~years of half-life) and acts as a bottleneck in the chain. All the isotopes between $^{222}Rn$ and this isotope will decay in few hours, but the $^{210}Pb$ and its daughters will decay years after that, until the equilibrium in the chain is reached, and that is why they have not been simulated. Among them, only $^{214}Pb$ and $^{214}Bi$ will produce Cherenkov light because they decay $\beta^-$ with Q-values of 1024~and 3272~keV, respectively. However, as commented in the previous sections, all the emitted gammas could produce Compton interactions able to generate Cherenkov emission in the transparent media of the module.

\begin{figure}[h!]
	\begin{center}
		\includegraphics[width=0.75\textwidth]{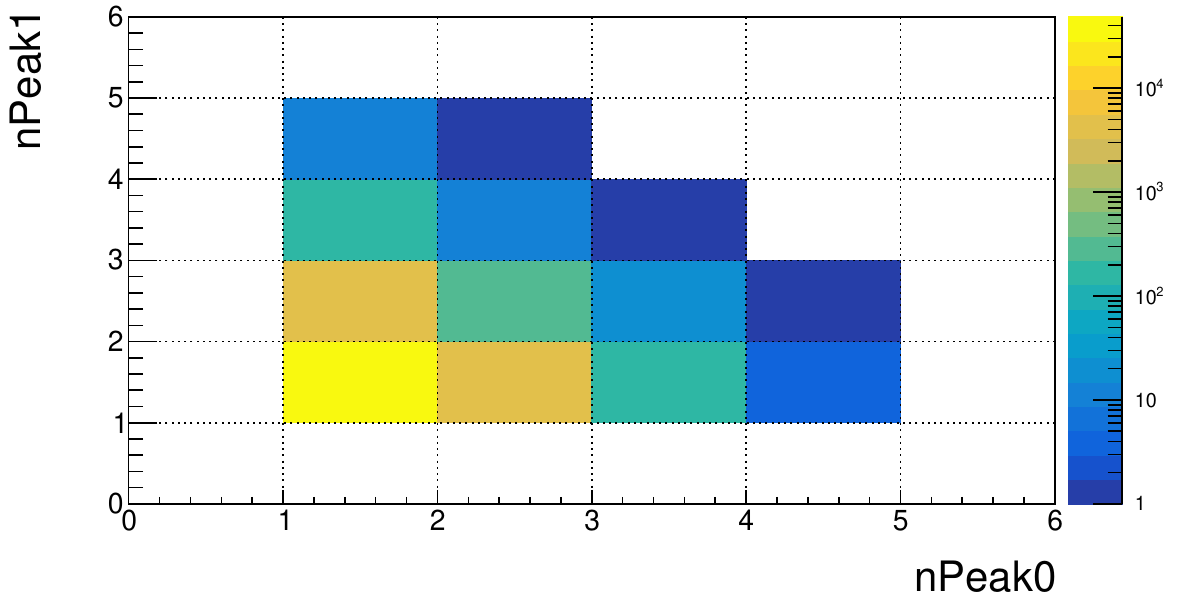}
		\caption{\label{peaks_Rn222}Scatter plot of the number of peaks identified in each PMT for the events that have passed the trigger condition in the simulation of the $^{222}Rn$ decays in the air surrounding the Blank module.}
	\end{center}
\end{figure}

The total number of the events that pass the trigger condition for the 10$^6$ simulated $^{222}Rn$ decays were 42725~$\pm$~207, giving as a result a contribution to the trigger rate of 267.5~$\pm$~1.3~mHz, while the difference of the trigger rate between the two periods (with and without $^{222}Rn$) obtained experimentally, was 314.0~$\pm$~0.2~mHz. The comparison of both values allows to conclude that Cherenkov events from $^{222}Rn$ decays in the transparent media of the module are able to explain reasonably the order of magnitude of its trigger rate increase. Although they deviate beyond the uncertainty margins, the reason may be due to the supposition of a constant concentration of radon (which is far from reality) or to the simplification of the geometry, being the simulation of the photocathode as opaque the most limiting effect. Figure~\ref{peaks_Rn222} shows the scatter plot of the number of peaks identified in each PMT for the simulated events. Although improvements in the modeling of the system, in particular of the PMT, could modify some of these conclusions, it seems that asymmetric events observed in ANAIS should be associated to some light emission process different than Cherenkov.

To analyze the contribution of the $^{40}K$ to the rate in the Blank module after its connection to the RFA line, 10$^6$ decays of this isotope have been simulated homogeneously in the borosilicate of the PMTs, exactly as it was done in Section~\ref{Section:SIM_Res_cherenkov}. From all the simulated events, 16.36~$\pm$~0.04\% passed the trigger condition, which is much less than those obtained in Section~\ref{Section:SIM_Res_cherenkov} when there was a crystal between the PMTs and there was not scintillation light, which was 26.75~$\pm$~0.06\%. The reason for the higher rate in a module with crystal is that it acts as an optical guide, allowing the Cherenkov light produced in one PMT to reach the other by reducing the total reflections, which increases the triggering coincidence. This could also happen for other scintillation occurring in the PMTs, explaining that anomalous event rate in ANAIS-112 modules is higher than in the Blank module. In any case, the simulation indicates that the contribution of $^{40}K$ events to the trigger rate is 37.6~$\pm$~0.1~mHz, approximately 38\% of the rate in the Blank module after the connection to the RFA line, which is a very important contribution. In the future, including in the simulation the $^{238}U$ and $^{232}Th$ contaminations in the PMTs borosilicate, we expect to be able to explain most of the Blank module trigger rate with Cherenkov events. The different rate contributions obtained from the simulation of the Blank module are summarized in Table~\ref{tabla:BlankRates}.

\begin{table}[h]
	\centering
	\begin{tabular}{|c|c|}
		\cline{2-2}
		\multicolumn{1}{c|}{} & \multicolumn{1}{|c|}{Rate (mHz)} \\
		\hline
		Before RFA connection (experimental) & 412.3~$\pm$~0.2 \\
		After RFA connection (experimental) & 98.4~$\pm$~0.1 \\
		Difference (experimental) & 314.0~$\pm$~0.2 \\
		$^{222}Rn$ contribution (simulation) & 267.5~$\pm$~1.3 \\
		$^{40}K$ contribution (simulation) & 37.6~$\pm$~0.1 \\
		\hline
	\end{tabular} \\
	\caption{Experimental rates in the Blank module before and after its connection to the RFA and the rate contributions of $^{222}Rn$ and $^{40}K$ isotopes obtained in the simulation. $^{222}Rn$ would contribute only before RFA connection while the contribution of $^{40}K$ would be present in both periods.}
	\label{tabla:BlankRates}
\end{table}

\section{Conclusions and future work} \label{Section:SIM_Conclusions}
\fancyhead[RO]{\emph{\thesection. \nameref{Section:SIM_Conclusions}}}

The background event excess observed in the ANAIS-112 experimental data with respect to that estimated by the background model, the origin of the asymmetric events at very low energy, and the differences in pulse shape observed between $^{109}Cd$ calibration events and bulk scintillation events have been three relevant motivations for the development of the optical simulation presented in this chapter.

This simulation is based on previous simulations of the ANAIS experiment using GEANT4 package, but it has included the optical processes taking part in the detection mechanism (as light emission, propagation, diffusion and absorption processes) as they play a fundamental role in the detector response. By including the information of the SER of the PMTs of the ANAIS-112 experiment, the detector output signals are also obtained, allowing to apply the same analysis as in the experiment and obtaining relevant variables as the pulse area, $p_1$, $\mu$, and the number of peaks identified in each PMT pulse.

This new simulation also includes a more precise geometry of the PMTs and the housing of the $^{109}Cd$ calibration source with respect to the previous simulations. However, some improvements remain for future work on this direction, as the definition of a more complex geometry of the PMTs (for example including the dynodes structure or the electric components and wiring and even the PMT socket) or the calibration piece modeling. For the first tests presented in this Chapter, some simplifications have been done assuming some properties of the detector that are unknown, as for example the teflon and copper reflectivities or the photocathode transparency. The effects of these parameters on the signal output should be analyzed in the future to establish a more realistic modelling of the ANAIS-112 module.

The simulation of the event populations used in the ANAIS-112 experiment for the low energy calibration ($^{22}Na$ and $^{40}K$ decays homogeneously distributed in the crystal and $^{109}Cd$ decays in the calibration piece), has allowed to do some preliminary analysis, as the effect of the sampling rate on the energy calibration and detector resolution, the systematics observed in the pulse shape variables, calculation of trigger and selection efficiencies and estimation of detection rate of Cherenkov events produced by the $^{40}K$ present in the borosilicate of the PMTs.

There have been two systematical effects coincident between simulation and experimental data. First, the independent term of the energy calibration is not compatible with zero when the integration window is 1~$\mu$s with a 2~GHz sampling rate, somethingdifferent when an integration window of 2~$\mu$s and 1~GHz sampling rate is considered. It can imply that this non-proportional effect can be related with the integration window or the sampling rate. The other effect is the decrease of the values of the $p_1$ and $\mu$ parameters at the lowest energies. However, as the slow scintillation components of the NaI(Tl) crystal have not been yet introduced in the simulation, the values obtained for these variables have not been properly reproduced. A more realistic modelling of the NaI(Tl) scintillation should be considered in future works.

Concerning the detector resolution, it has been found that it would be not affected if, in the future, the experiment changes the sampling rate from 2~to 1~GHz. Moreover, it has been observed that $\sigma_{SER}$ has a relevant effect on it, and that there are some mechanisms that can be impoverishing the resolution and that have not been included in the simulation, as inhomogeneities in scintillation properties of the crystal implying a dependence of the light yield on the crystal position, for instance. These effects can be analyzed in future work with this simulation.

The values obtained for the trigger and asymmetry selection efficiencies are compatible with those corresponding to ANAIS data (presented in Section~\ref{Section:ANAIS_Filter}). All the Cherenkov events not mixed with scintillation are rejected with the asymmetry cut. In fact, if the minimum number of peaks required in the selection changes from~5 to~3, rejection of Cherenkov events is still done at 99.9\%. With this change, bulk scintillation events acceptance efficiency would increase at 0.9~keV from 43\% to 70\% (from simulation). However, it would also imply a much lower rejection of the asymmetric events present in the ANAIS-112 data. This gives an idea on how much the sensitivity of the experiment would increase if those asymmetric events were not present. As the analysis of the Cherenkov events produced by the $^{40}K$ and $^{222}Rn$ decays show that they have symmetrical light sharing between both PMTs, Cherenkov emission cannot be responsible of those asymmetrical events. That is why, if their origin is some other scintillation mechanisms in the PMTs or in the optical windows, then alternatives to these light detectors, as the SiPMs, have to be explored (see Chapter~\ref{Chapter:SiPMZgz}).

In the medium term, future work with this simulation will focus on including the inhomogeneities, the defects in the light propagation, the energy dependence of the light yield of the detector, the slow scintillation components of the NaI(Tl) crystal, a better description of the PMT geometry and a better modeling of the optical properties of the detector components (as the photocathode transparency and the reflectivity of the teflon and copper). This should allow to reproduce the pulse shape variables obtained in the experimental data, the LC dependence on the energy and the global energy resolution of the detector when an specific module is simulated. In the long term, all the electronic chain can be modelled and introduced in the signal-building (including preamplifier gain, bandwidth, signal attenuation in cables, etc). This simulation can be upgraded to include the nine modules of the ANAIS-112 experiment, and each particular module can be simulated by incorporating the SER and QE of its PMTs.

Ultimately, the objective is to improve the selection of the NaI(Tl) scintillation from other types of events. It can be done by generating a dataset of pulses for various event sources and training an event identification neural network with them, which could be integrated into the filtering process of the ANAIS-112 experiment. It will also allow to improve the design of future detection systems and to confirm the calculations of efficiency of filtering processes, finally improving our understanding of the events whose origin is still unclear.

%% file: QF2.tex
\chapter{NaI(Tl) Quenching Factor measurement}\label{Chapter:QF}

\fancyhead[LE]{\emph{Chapter \thechapter. \nameref{Chapter:QF}}}

The response of NaI(Tl) detectors to energy deposited by nuclear recoils has to be well understood to calibrate the ROI for dark matter searches. In general, the detector response is periodically determined using gamma sources that produce electron recoils in most of the dark matter experiments, in particular in ANAIS we use external sources ($^{109}Cd$) and emissions from internal radioactive contaminations ($^{22}Na$ and $^{40}K$) as explained in Chapter~\ref{Chapter:ANAIS}. Then, to convert the response to electron recoils into the response to nuclear recoils, a precise knowledge of the relative scintillation produced by nuclear recoils of sodium and iodine nuclei with respect to electron recoils depositing the same energy (QF) is required.

Different dedicated experiments have been carried out since the 90s to measure the QF for sodium and iodine nuclei in NaI(Tl). They will be reviewed and commented in Section~\ref{Section:QF_NaIQFOverview}. Then, a new measurement at Triangle Universities Nuclear Laboratory (TUNL) will be presented. This measurement took place in August and October of 2018 in a collaborative effort between members of COSINE-100, COHERENT and ANAIS-112 experiments. In this measurement, quasi-monoenergetic neutrons are directed to the NaI(Tl) crystal where they can scatter producing a nuclear recoil, and the scattered neutron be detected by one of the surrounding neutron detectors, which allows to determine the energy deposited in the NaI(Tl). The light produced by the nuclear recoil energy deposition in the crystal is independently measured.

To analyze the possible crystal-dependence of the QF, five crystals were measured in the same setup. All the five crystals were produced by Alpha Spectra company using the same growth process but different powder qualities. One of them (Crystal~5) was produced in the same growing batch than some of the ANAIS-112 crystals. The measurement procedure, the objectives and the experimental setup are descibed in Section~\ref{Section:QF_TUNL}, as well as the data acquisition system and the run plan followed. The event analysis is presented in Section~\ref{Section:QF_Analysis_WaveformRec}. Section~\ref{Section:QF_Analysis_NeutronBeam} describes the method followed to calculate the energy of the neutron beam.

A GEANT4 simulation of the experiment (explained in Section~\ref{Section:QF_GEANT4}) was developed and applied to obtain the distributions of the energy depositions in the crystal by the nuclear recoils in the neutron runs, and by the gamma emissions from radioactive sources in the calibration runs. Moreover, it has allowed to estimate systematic uncertainties related, for instance, with the inaccuracies in the knowledge of the detector positions, and it has provided information related with the event selection. The analysis of the data, event identification and energy calibration of the detectors are explained in Section~\ref{Section:QF_Analysis}. The sodium QF values are calculated by fitting simulated nuclear recoil energy distributions and electron equivalent energy calibrated spectra of the crystal signal. The iodine recoils could not be separated from the background. Therefore, they were analyzed by following a different method, profiting from the inelastic scattering signal on $^{127}I$. All the results are presented in Section~\ref{Section:QF_Results}.

\section{NaI(Tl) QF measurements overview}\label{Section:QF_NaIQFOverview}
\fancyhead[RO]{\emph{\thesection. \nameref{Section:QF_NaIQFOverview}}}

As explained in Section~\ref{Section:Intro_Detection_Direct_Techinques}, WIMPs are expected to interact with the target nuclei. As the periodic calibrations of dark matter search experiments are usually done with gamma sources, which produce electron recoils, and the light yield in scintillators is strongly quenched for nuclear recoils, the corresponding scintillation QF have to be measured, i.e. QF for sodium and iodine recoils in NaI(Tl), in the case of ANAIS. For this purpose, different experiments have been carried out since the 90s, as it was briefly overviewed in Section~\ref{Section:Intro_Scintillators_NaI(Tl)}. Figures~\ref{NaQF_measurements} and~\ref{IQF_measurements} show the results of these measurements for the QF for sodium and iodine nuclei, respectively. The different methods followed by these experiments are reviewed next.

The nuclear recoils are typically induced by monoenergetic neutrons, produced in reactions like $^7Li$(p,n)$^7Be$ and $D-D$ fusion. Both, protons and deuterium ions have to be accelerated and beamed onto a target containing $^7Li$ or deuterium, respectively. In these cases, the neutrons are monoenergetic if the protons/deuterium ions are also monoenergetic. These neutrons are collimated and directed to the NaI(Tl) crystal, where they can elastically scatter the sodium and iodine nuclei producing nuclear recoils. Crystals used have commonly small size (16~cm$^3$ or less) to reduce multiple scattering as much as possible, and their scintillation is measured with PMTs. After scattering, neutrons can reach one of the neutron detectors (in the following called backing detectors or BD) that surround the crystal and are placed at different scattering angles. They are commonly made of a liquid scintillator, which can discriminate neutron and gamma events by the pulse shape, due to the different scintillation times for each particle. These BD allow to identify the scattered neutron and via the corresponding scattering angle, to obtain the energy transferred to the nucleus. The energy transferred by the neutron to the target nucleus through an elastic scattering depends on the neutron initial energy $E_n$, the mass of the neutron $M_n$, the mass of the target nucleus $M_T$ and the neutron scattering angle in the laboratory frame $\theta$, following the Equation~\ref{Equation_Enr}~\cite{Arneodo:2000vc}.
\begin{equation}\label{Equation_Enr}
	E_{nr} = 2E_n\frac{M_n^2}{(M_n+M_T)^2}\left(\frac{M_T}{M_n}+sin^2\theta-cos\theta\sqrt{\left(\frac{M_T}{M_n}\right)^2-sin^2\theta}\right)
\end{equation}
Most recent measurements of the sodium QF agree in a decrease of this factor at low nuclear recoil energies, below 30~keVnr, but there is some disagreement in the energy dependence among the different measurements (see Figure~\ref{NaQF_measurements}). Possible systematics affecting these measurements are analyzed next, as well as mitigation strategies applied in our experiment's design and analysis procedures in order to reduce their effect in the QF estimates: 

\begin{enumerate}
	\item Multiple scattering in the crystal, its housing, and other components of the experimental setup introduce a background to the single-scattered neutrons. This contribution is always present, but can be reduced by using a small size crystal and also by applying a time-of-flight (TOF) selection, since multiple scattered neutrons that reach the BDs are slower than single scattered ones. Crystals either cylindrical or cubic have been used with dimensions ranging from 2 to 5~cm. In the case of cylindrical crystals, shapes having the same length than diameter are used. In our measurement, cylindrical crystals with dimensions between 1.5~cm and 2.5~cm were used (see Section~\ref{Section:QF_TUNL_Setup_Crystals}).
	
	\item Channeling of ions in crystals has been observed for incidence directions matching the principal axes of the crystalline structure~\cite{Bozorgnia:2010xy}. This effect would bias the QF determination, because different scintillation would be observed for different orientations of the crystal with respect to the neutron beam. Only Collar~\cite{Collar} searched for the channelling signal without finding any hint of that effect. To reduce the effects of this possible systematic contribution, we rotated the crystal 30$^o$ every eight hours of exposure.
	
	\item The detector gain (i.e. the factor to convert a given deposited energy into a measurable signal) can vary over time due to PMT gain drifts, changes in the PMT-crystal coupling, changes in temperature of the system, etc. It can imply an incorrect energy calibration and loss of resolution in the measurements. These effects can be reduced either by calibrating with external sources with high periodicity, or by applying a time-dependent gain correction profiting from neutron induced lines available along the measurements. For instance, in Ref.~\cite{Chagani}, authors calibrated the crystal response every three hours using $^{57}Co$ external sources, while in~\cite{Xu} and~\cite{Bignell_2021}, they used the $^{127}I$ inelastic peak for correcting the response along the whole measurements. In this work, possible gain drifts were corrected using $^{127}I$ inelastic peak (Section~\ref{Section:QF_Analysis_NaIcal_GainCorrection}), but external sources were periodically used for energy calibration. 
	
	\item All the previous experiments required the coincidence between the NaI(Tl) crystal and the BD signals for triggering the acquisition. This introduces a threshold in the energy deposited in the crystal, whose effect on the measurement should be corrected with the trigger efficiency. This trigger efficiency can be obtained as the fraction of pulses of known energy that produces a trigger. First QF measurements~\cite{Spooner,Tovey:1998ex,Gerbier:1998dm,Simon:2002cw,Chagani} did not correct the spectra by the trigger efficiency, and therefore this systematic affects the QF values obtained: QFs at low energy are higher than those in other measurements because the measured energy is overestimated. It is very important to design specific protocols for determining all of these efficiencies, as it was done in the other measurements, where different techniques have been followed, as a calibration measurement~\cite{Xu,Joo:2018hom,Bignell_2021} or a calculation of the probability of triggering~\cite{Collar}. In our measurement, such a threshold effect was avoided by triggering in the BDs (Section~\ref{Section:QF_TUNL_Setup_BDs}), instead of requiring a coincidence between the NaI(Tl) crystal and the BDs. The integration window in the crystal waveform was fixed taking into account the TOF of the neutron between both detectors (Section~\ref{Section:QF_Analysis_WaveformRec_NaI}).
	
	\item Although the energy distribution of the neutrons before being scattered in the crystal is determined by the energy applied to the accelerator and the reaction used to produce them, in general, it is expected a reduction of the neutron mean energy at the crystal position and an increase of the energy distribution width due to different factors, as the deposition of energy by the proton or deuteron in the target before producing the neutron and the scattering of the neutron in the collimator and shields. Different approaches have been followed to obtain the neutron energies. In~\cite{Spooner,Tovey:1998ex,Gerbier:1998dm,Chagani} neutrons were supposed to be mono-energetic with energy fixed by the specifications of the neutron gun. In this case, the nominal energy should include the energy loss by the proton or the deuteron before producing the neutron to avoid the introduction of relevant systematical errors. In~\cite{Simon:2002cw}, the neutron energies were calculated using TOF between a Beam Pulse Monitor (BPM) and a neutron detector. In this procedure, the measured TOF distribution of the neutrons has to be deconvolved with the time response of the detector. In~\cite{Xu,Bignell_2021}, the neutron production process and possible energy loss was simulated, and in~\cite{Collar,Joo:2018hom} the neutron energy was directly measured with $^3He$ detectors. In the measurement presented in this Chapter, the neutron initial energy was calculated with dedicated TOF measurements between a BPM and a neutron detector placed on-beam. The neutron production was simulated, obtaining a TOF distribution that was convolved with the time response of the detector, thus getting a PDF for each neutron energy distribution. The experimental data was later fitted to that PDF thus obtaining the neutron energy distributions (see Section~\ref{Section:QF_Analysis_NeutronBeam}).
	
	\item The energy deposited by each scattered neutron in the NaI(Tl) crystal is obtained from Equation~\ref{Equation_Enr}, but if the source is not completely monochromatic, for each scattering angle there will be a deposited energy distribution dependent on the initial neutron energy distribution. Moreover, the finite size of the NaI(Tl) crystal and BDs implies that every BD covers a range of scattering angles. There have been two approaches to obtain the nuclear recoil energies corresponding to a triggered BD, which are used in the estimate of the QF. In~\cite{Spooner,Tovey:1998ex,Gerbier:1998dm,Chagani}, the Equation~\ref{Equation_Enr} was used to derive the mean energy of the nuclear recoils using the scattering angles corresponding to the center of each BD with respect to the center of the NaI(Tl) crystal. Then, the QF was calculated by dividing the mean energy of the measured coincident nuclear recoils signals in the NaI(Tl), calibrated in electron equivalent energy, by the mean nuclear recoil energy derived from Equation~\ref{Equation_Enr} as commented. This approach is affected by systematics related with the initial neutron energy distribution previously commented and the size of both BDs and NaI(Tl) crystal, and also by the non-proportionality in the light response of the NaI(Tl) and energy-dependent resolution. This approach cannot explain, for instance, the asymmetric nuclear recoil distributions that are typically observed in these measurements. For the second approach, a MC simulation reproducing the experimental setup was developed~\cite{Bernabei:1996vj,Simon:2002cw,Collar,Xu,Joo:2018hom,Bignell_2021}. In that case, the QF was obtained as a free parameter in the fitting of the simulated nuclear recoil energy and the measured electron equivalent energy distribution converted into nuclear recoil energy by a floating QF. This approach allows to include in the simulated nuclear recoil distributions all the previously commented possible systematics: energy distribution of the incident neutrons, finite size of the BDs and NaI(Tl) crystal, non-proportional light response, etc. In our measurement, 18~BDs were placed surrounding the crystal, and a MC simulation (Section~\ref{Section:QF_GEANT4}) was developed to obtain the nuclear recoil energy distributions for each triggered BD. The initial energy of the neutrons has been simulated as the one determined in Section~\ref{Section:QF_Analysis_NeutronBeam}. Before fitting, the nuclear recoil energy distributions derived from the simulation were gaussian convolved to take into account the detector resolution. The free parameters of the fit were the QF, as a scaling factor, and the resolution of the detector  (Section~\ref{Section:QF_Results}). The same procedure has been followed applying an energy dependent and an energy independent resolution function obtaining incompatible QF values. Therefore, the differences in the QF values obtained have been considered as a systematic uncertainty. Finally, two different electron equivalent energy calibrations (propotional and non-proportional) have been applied to the crystal energy spectrum, obtaining very different energy depencences of the QF. This result is very relevant and it will be further commented in next paragraph.
	
	\item The light yield of the NaI(Tl) crystal is nonproportional at energies below    $\sim$~50~keV~\cite{PhysRev.122.815,Rooney:1997,Moses:2001vx,Choong:2008,Hull:2009,Payne:2009,Khodyuk:2010ydw,Payne:2011}. Therefore, the selection of the reference energies and the functional form assumed for the energy calibration will affect strongly the results at the lowest energies. Most of the experiments applied a proportional calibration, as in~\cite{Tovey:1998ex,Gerbier:1998dm,Joo:2018hom} using the 59.5~keV line from $^{241}Am$, in~\cite{Xu} using the 57.6~keV from the neutron inelastic scattering in $^{127}I$ and in~\cite{Chagani} with the 122.1~keV line from $^{57}Co$. Beyond assuming proportionality, these measurements are using only one energy as reference, which in addition is far from the ROI. Then, this approach can result in miscalibration, and the incorrect determination of the QF. Non-proportionality can be taken into account in different ways: a non-linear calibration (as done in~\cite{Collar,Bignell_2021}) or a linearization using reference lines within or very near to the ROI to minimize the impact of the related systematics (as done in~\cite{Spooner,Simon:2002cw}). In this work, two different calibration procedures have been considered in the ROI to analyze related systematics in the QF estimates: A proportional and a non-proportional linear calibration (Section~\ref{Section:QF_Analysis_NaIcal_EnergyCal}).
\end{enumerate}

The QF measurements made by DAMA~\cite{Bernabei:1996vj} are different from the others and therefore require a separated explanation. They were performed at the ENEA-Frascati Laboratory with the same kind of crystals as the DAMA experiment. They applied a Compton calibration using a $^{137}$Cs source and a neutron calibration with a $^{252}Cf$ source, which was shielded by a suitable thick moderator and placed at 35~cm from the crystal. Neutron multiple scatterings were rejected analysing the pulse characteristics recorded during 3250~ns. The low-energy spectrum collected with the $^{252}Cf$ source was background subtracted and fitted to a function
\begin{equation}
	Y(E_{ee}) = \alpha_{Na}G_{Na}\left(\frac{E_{ee}}{QF_{Na}}\right)+\alpha_{I}G_{I}\left(\frac{E_{ee}}{QF_{I}}\right),
\end{equation}
being
\begin{equation}\label{eq:QFDAMA}
	G_x(E_R) = exp(a_{1x}E_R^3+a_{2x}E_R^2+a_{3x}E_R),
\end{equation}
where $x$ stands for $Na$ and $I$, and the corresponding $E_R = E_{ee}/QF_x$. The $a_{ix}$ values were estimated through a Monte Carlo method following~\cite{Fushimi:1993nq}, and $\alpha_x$ and $QF_x$ were the free parameters of the fit. It is possible to observe that this procedure to calculate the QF is very different from the rest of the measurements. First, the neutron source is not mono-energetic and the scattering angle is not measured by a BD allowing the identification of the scattered neutrons. Both things imply a wide dispersion of the nuclear recoil energy distribution which is modelled following a MC method to a very general exponential shape (Equation~\ref{eq:QFDAMA}). Moreover, it supposes that the QF of both the iodine and sodium nuclei are constant. Finally, the lack of information presented in the article makes difficult the analysis of the systematics affecting the measurement. 

\section{QF measurements at TUNL}\label{Section:QF_TUNL}
\fancyhead[RO]{\emph{\thesection. \nameref{Section:QF_TUNL}}}

The measurements presented here were carried out in collaboration between members of COHERENT, COSINE (Yale University) and ANAIS (Zaragoza University) teams. They were performed in August and October of 2018 at the Advanced Neutron Calibration Facility, at the Triangle Universities Nuclear Laboratory (TUNL) Tandem accelerator facility, located in Durham, North Carolina, US.

Neutrons were generated using the $^7Li$(p,n) reaction. The neutrons that elastically scattered off the NaI(Tl) crystal were detected in a BD placed at a known scattering angle, allowing the determination of the nuclear recoil energy. These measurements have been designed to reduce as much as possible the systematic effects that could have affected the previous measurements (see Section~\ref{Section:QF_NaIQFOverview}), and to match the range of energies interesting for dark matter searches as ANAIS-112 or COSINE-100 and CE$\nu$NS experiments as COHERENT. In the case of ANAIS-112, the ROI in electron equivalent energies is below 6~keV, which would correspond to 30~keV for sodium nuclei and 60~keV for iodine nuclei for typical QF's ($\sim$~20$\%$ and $\sim$~8$\%$, respectively). The facility allowed the BD to be placed in convenient positions to access nuclear recoil energies for sodium nuclei below 100~keV and for iodine nuclei below 20~keV (see Section~\ref{Section:QF_TUNL_Setup_BDs}). The complete experimental setup and data acquisition system are described next.

\subsection{Neutron beam}\label{Section:QF_TUNL_Setup_NeutronProduction}

\begin{figure}[h!]
	\begin{center}
		\includegraphics[width=\textwidth]{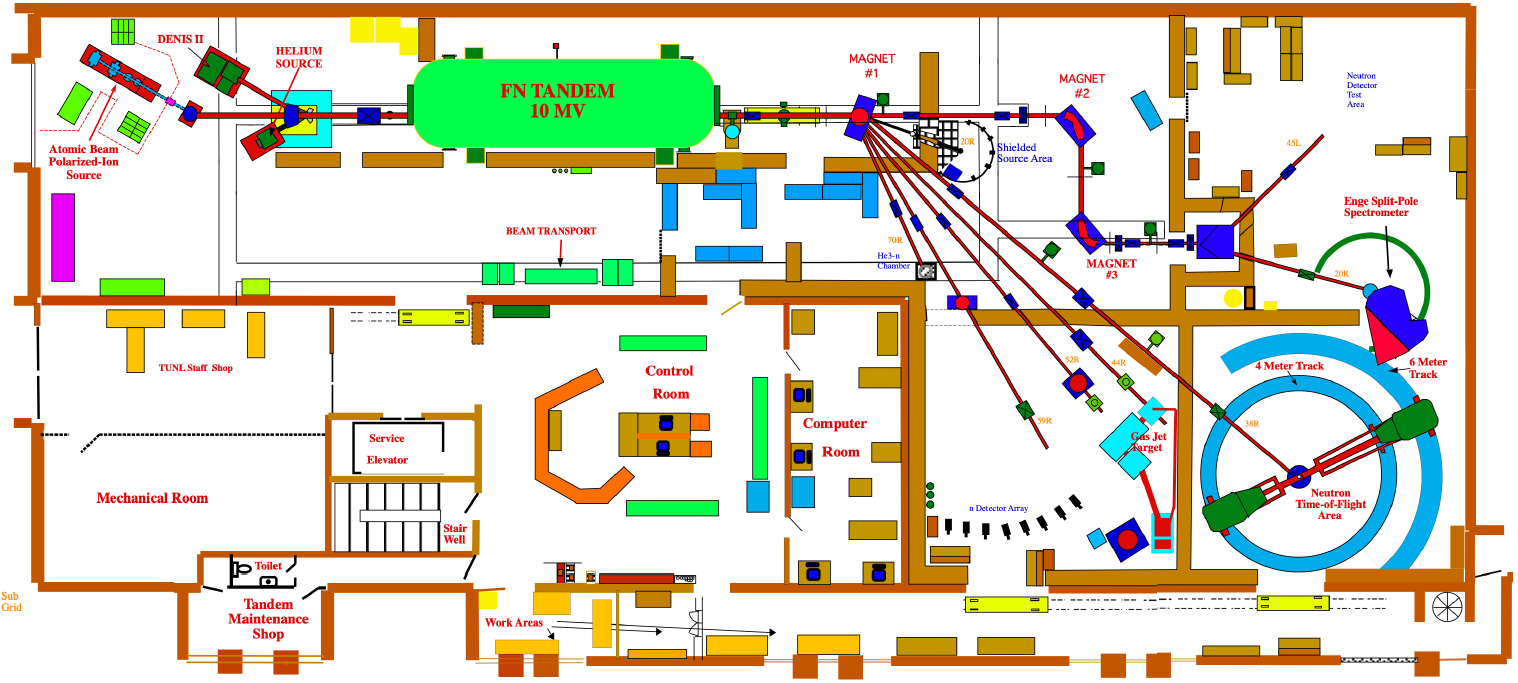}
		\caption{\label{accelerator_scheme}Triangle Universities Nuclear Laboratory (TUNL) Tandem accelerator facility~\cite{TandemAccelerator}. The measurements were done in the Time-of-flight area (on the bottom-right section).}
	\end{center}
\end{figure}

In the Tandem accelerator facility at TUNL~\cite{TandemAccelerator}, protons were produced using the Direct Extraction Negative Ion Source (DENIS)~\cite{Moehs:2005ca}, and were accelerated using a 10~MV Tandem van de Graaff accelerator. The facility provides pulsed proton beams with a frequency of 1.25~MHz (for August runs) or 2.5~MHz (for October runs), corresponding to periodicities of 800~ns and 400~ns, respectively, and with a typical width of $\sim$~2~ns. The scheme of the accelerator facility is shown in Figure~\ref{accelerator_scheme}~\cite{TandemAccelerator}.

In the experiment, protons were accelerated to reach a kinetic energy of 2.7~MeV. The proton beam can be deflected using magnets and driven towards the experimental room (the Time-of-flight area shown in Figure~\ref{accelerator_scheme}) where the NaI(Tl) crystals and the BDs were placed. A Beam Pickoff Monitor (BPM), consisting of an induction coil where the proton beam induces a current, was placed around the proton line to control the beam pulsing and provided relevant information for the TOF analysis (see Section~\ref{Section:QF_TUNL_DAQ}).

\begin{figure}[h!]
	\begin{center}
		\includegraphics[width=\textwidth]{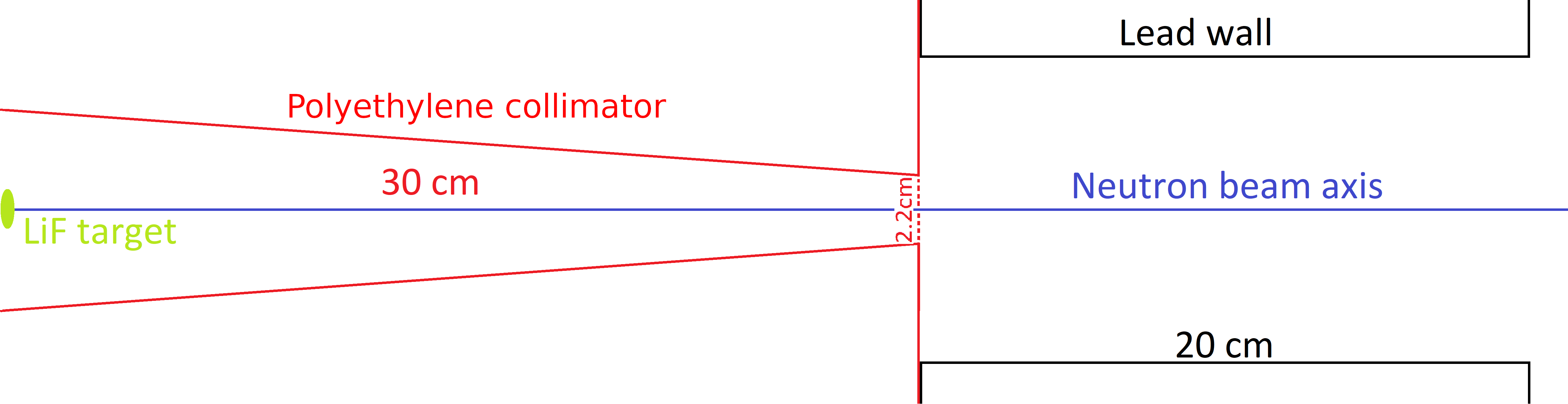}
		\caption{\label{colimator}Collimator and lead wall scheme while the QF measurement. The lead wall was not present while the measurements of the beam profile.}
	\end{center}
\end{figure}

Protons hit a lithium fluoride (LiF) target and produce neutrons with an energy of the order of 1~MeV through the $^7Li$(p,n) reaction. The LiF target was thermally evaporated onto a 0.1~mm-thick tantalum backing to guarantee that protons were stopped and were not contaminating the resulting neutron beam with products from other side reactions. Between the LiF target and the sample under study (the NaI(Tl) crystal) there is a polyethylene collimator with an aperture of 22~mm diameter at 30~cm from the target followed by a 20~cm lead wall. The lead helps to reduce the beam-produced gammas that could reach the detectors. Both the collimator and the lead wall are sketched in Figure~\ref{colimator}.

\begin{figure}[h!]
	\begin{center}
		\includegraphics[width=0.75\textwidth]{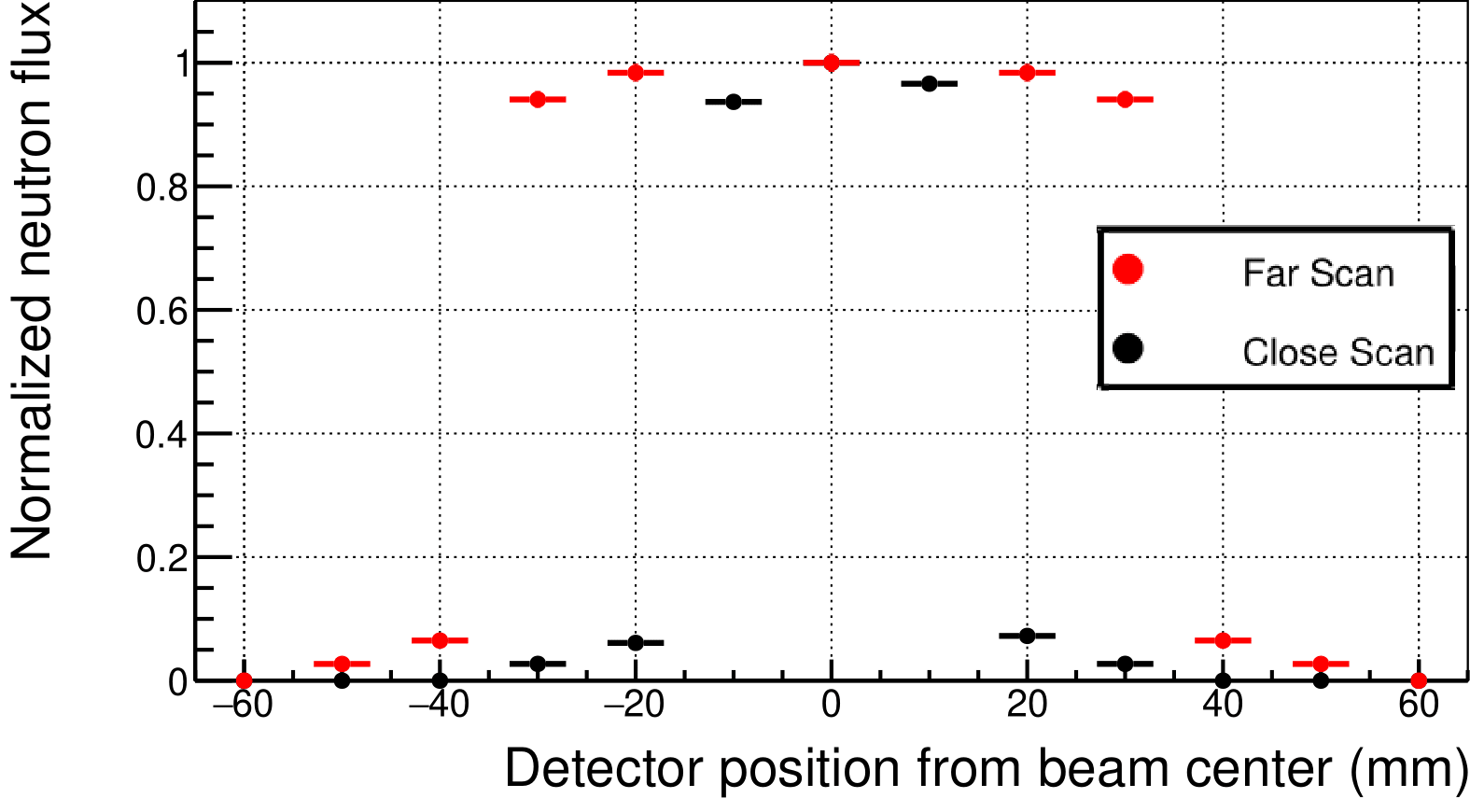}
		\caption{\label{beam_scan}Beam scan measurements. The neutron rate normalized to the maximum is plotted as a function of the distance to the beam axis. The maximum of both histograms is at the beam axis.}
	\end{center}
\end{figure}

Before the QF measurements campaign, a dedicated measurement was carried out in the same setup to characterize the beam shape. A plastic scintillator (0.5$\times$0.5$\times$2~cm$^3$) allowed to scan the profile of the beam with a resolution of 1~cm at two different distances from the target (47.4~and 98.6~cm). The measurements were taken without the lead wall to allow for a closer scan. The results of these measurements, which are presented in Figure~\ref{beam_scan}, show a constant flux around the beam axis, within a characteristic radius, while beyond the flux decreases abruptly. The half-width of the beam derived from this measurement is 1.5~$\pm$~0.5~cm for the close scan and 3.5~$\pm$~0.5~cm for the far scan, being the uncertainty dominated by the spatial resolution of the scan. Therefore, considering a point-like neutron source, it resulted in a beam angle of 1.9~$\pm$~0.6$^o$ for the close measurement and 2.0~$\pm$~0.3$^o$ for the far measurement, both compatible with the value obtained with the distance between the LiF target and the end of the collimator (30~cm) and the radius of the collimator (1.1~cm), which is 2.1$^o$.

\subsection{NaI(Tl) crystals}\label{Section:QF_TUNL_Setup_Crystals}

\begin{figure}[h!]
	\begin{subfigure}[b]{0.49\textwidth}
		\includegraphics[width=\textwidth]{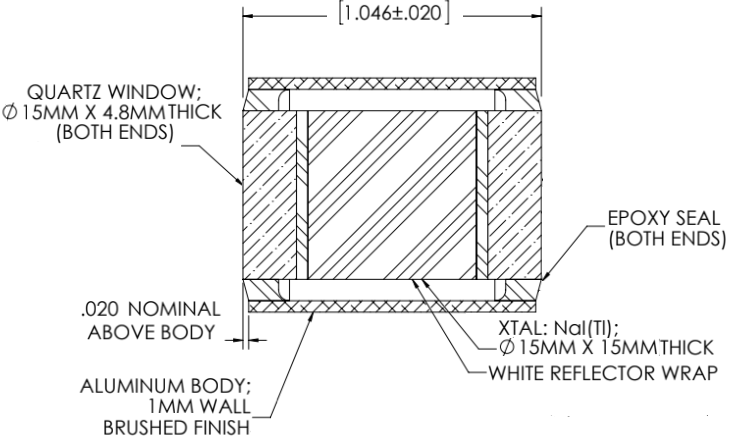}
	\end{subfigure}
	\begin{subfigure}[b]{0.49\textwidth}
		\includegraphics[width=\textwidth]{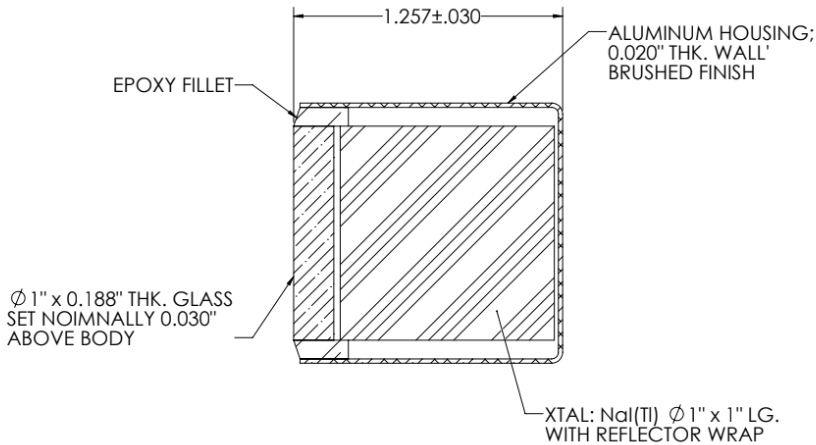}
	\end{subfigure}
	\caption{\label{NaI_housing}Housing specifications for Zaragoza (left plot) and Yale (right plot) crystals.}
\end{figure}

Five cylindrical crystals of NaI(Tl) grown at Alpha Spectra (AS)~\cite{AlphaSpectra} were measured in similar conditions. They were grown using the same method but with different powder quality, and one of them (the number~5) was selected from the same ingot as some of the ANAIS-112 crystals. The properties of the five crystals are shown in Table~\ref{tabla:crystalSpec}. All crystals were encapsulated by the manufacturer in an aluminum housing (with a width of 0.5~mm for Yale crystals and 1~mm for those of Zaragoza), with 4.8~mm-thick optical windows (one glass window for Yale crystals and two quartz windows for those of Zaragoza). Figure~\ref{NaI_housing} shows the housing design for Zaragoza (left) and Yale (right) crystals. 

\begin{table}[h!]
	\centering
	\begin{tabular}{|c|c|c|c|c|c|c|}
		\hline
		Crystal~$\#$ & Run & Group & Powder quality & Length and $\phi$ (mm) \\
		\hline
		1 & August & Yale & WS-I & 25\\
		2 & August & Yale & WS-II & 25\\
		3 & October & Yale & WS-III & 25\\
		4 & October & Zaragoza & Std. & 15\\
		5 & October & Zaragoza & WS-III & 15\\
		\hline
	\end{tabular} \\
	\caption{Specifications of the 5 measured crystals. All of them were grown at Alpha Spectra Inc., Colorado, US~\cite{AlphaSpectra}. WS powder quality refers to WIMPScint.}
	\label{tabla:crystalSpec}
\end{table}

Zaragoza crystals were designed with two optical windows, but in these measurements only one PMT was coupled to the crystal. Therefore, we covered one of the windows with a diffusive material (teflon), inside a copper housing. All the crystals were optically coupled to the same PMT (HAMAMATSU H11934-200-10 ultra-bialkali~\cite{PMTNaIQFManual}, whose main characteristics are summarized in Table~\ref{tabla:pmtNaI} and Figure~\ref{pmtNaISpec}) using the same optical gel (Eljen Technology EJ-550 Optical Grade Silicone Grease~\cite{EJ-550:Eljen}), which has a refractive index of~1.46~and whose optical transmission in a 0.1~mm-thick layer at 300~nm is $>$~90$\%$ and at 400~nm is $>$~99$\%$.

\begin{figure}[h!]
	\begin{center}
		\includegraphics[width=\textwidth]{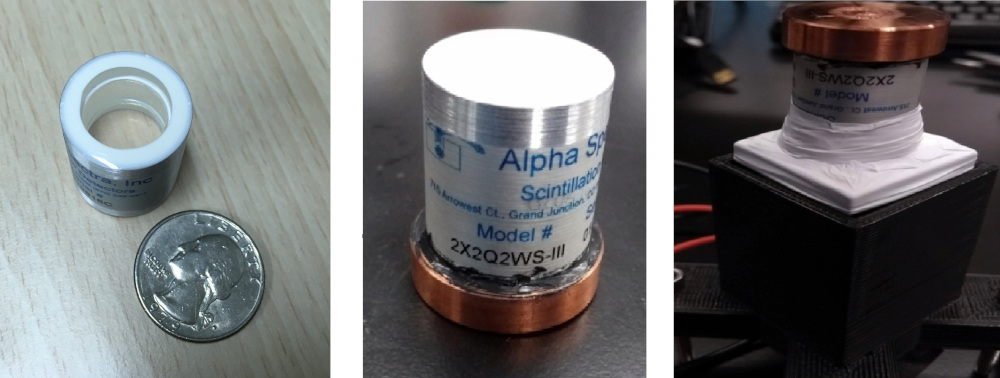}
		\caption{\label{crystal+pmt}From left to right: crystal~5 before and after covering one of the optical windows, and PMT+crystal system before covering with the isolating tape and placing in the beam for measurement.}
	\end{center}
\end{figure}

\begin{figure}[h!]
	\begin{center}
		\includegraphics[width=0.5\textwidth]{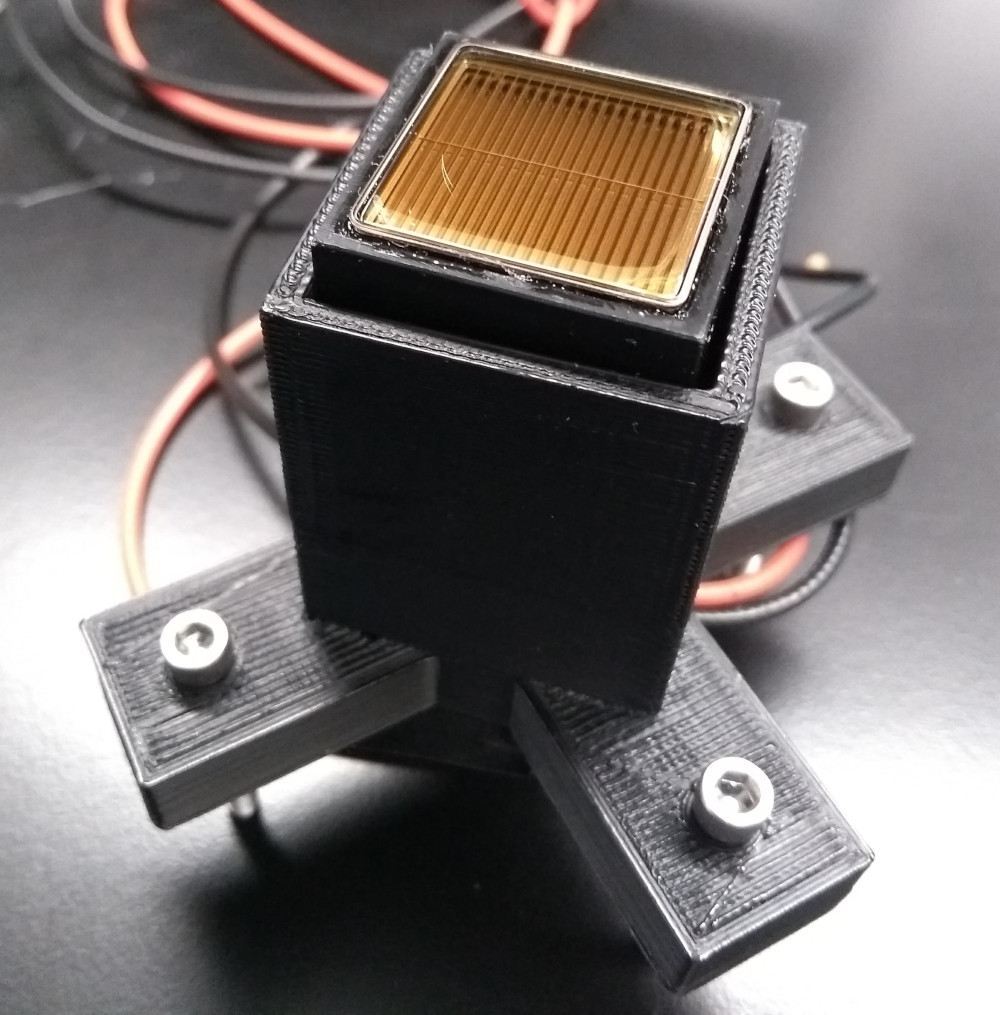}
		\caption{\label{pmtQF}PMT model HAMAMATSU H11934-200-10 ultra-bialkali~\cite{PMTNaIQFManual}, used for the QF measurements of the five NaI(Tl) crystals. Its main characteristics are summarized in Table~\ref{tabla:pmtNaI} and Figure~\ref{pmtNaISpec}.}
	\end{center}
\end{figure}

In both cases, the crystal and the PMT were covered by teflon and PVC isolating tapes to fix the structure, avoiding displacement and preventing external light from interfering in the NaI(Tl) scintillation detection. Pictures of one ANAIS crystal, the PMT used and the coupling of both are shown in Figures~\ref{crystal+pmt} and~\ref{pmtQF}. For the energy calibration of the NaI(Tl), a $^{133}Ba$ source was placed at 1~cm from the crystal. The crystals were placed at a distance from the LiF target enough to ensure they were completely covered by the neutron beam: 112~cm in August measurements  (angle subtended by the crystals of 0.7$^o$) and 66~cm in October (subtended angles of 1.1$^o$ for crystal 3 and 0.7$^o$ for crystals 4 and 5).

\subsection{Backing Detectors}\label{Section:QF_TUNL_Setup_BDs}

\begin{figure}[h!]
	\begin{center}
		\includegraphics[width=0.5\textwidth]{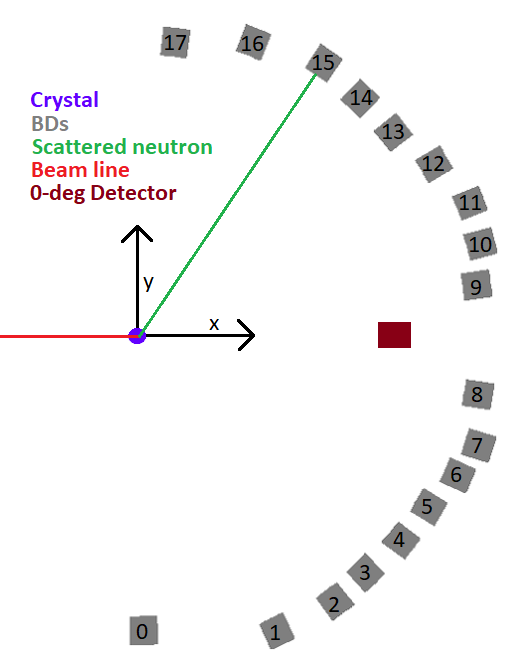}
		\caption{\label{Set-Up_Scheme}BD distribution around the NaI(Tl) crystal. BD are numbered according to their positions, as in Tables~\ref{tabla:BDpos_Ago} and \ref{tabla:BDpos_Oct}.}
	\end{center}
\end{figure}

\begin{figure}[h!]
	\begin{center}
		\includegraphics[width=0.5\textwidth]{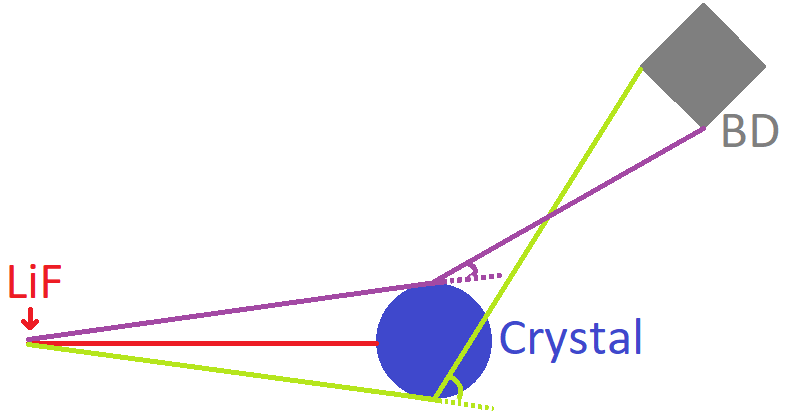}
		\caption{\label{ScatteringAngles}Example of the maximum and minimum scattering angles defining the acceptance range for a BD, shown as green and purple lines, respectively. Detectors are not to scale.}
	\end{center}
\end{figure}

\begin{figure}[h!]
	\begin{center}
		\includegraphics[width=0.75\textwidth]{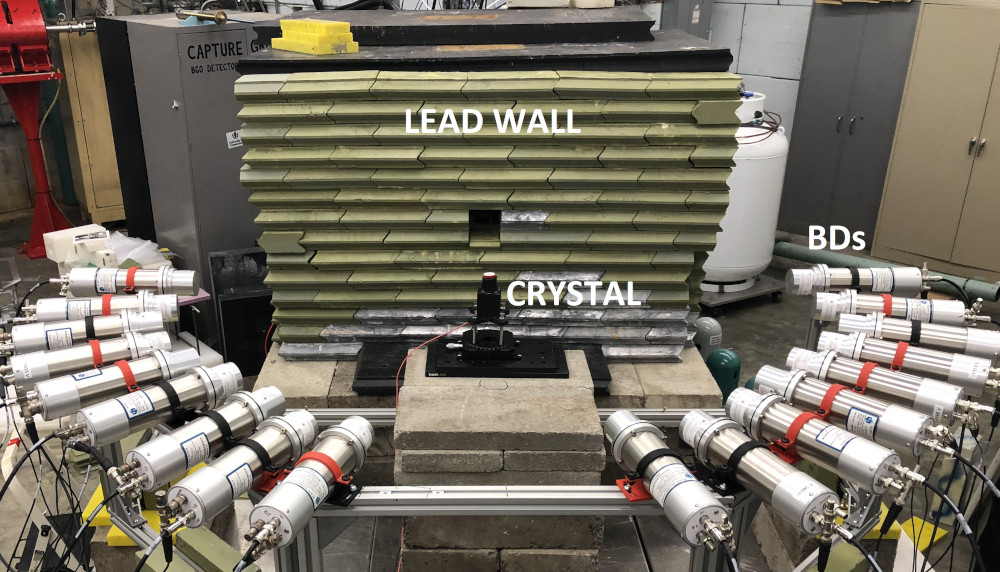}
		\caption{\label{bds_pic}Picture of the experimental setup. The lead wall, the crystal and the BDs can be observed.}
	\end{center}
\end{figure}

After scattering in the NaI(Tl) crystal, the neutrons can reach one of the eighteen BDs, which cover a wide range of scattering angles (from $\sim$~15$^o$ to $\sim$~90$^o$) and are located at distances ranging from 60~to 70~cm from the NaI(Tl) crystal. Their distribution around the crystal is shown in Figure~\ref{Set-Up_Scheme}. Due to the finite dimensions of the NaI(Tl) and BDs, every BD can detect neutrons for a range of scattering angles. The minimum and maximum angles defining this range are illustrated in Figure~\ref{ScatteringAngles}. Positions, acceptance scattering angles range and total solid scattering angle covered for each BD are presented in Tables~\ref{tabla:BDpos_Ago} and~\ref{tabla:BDpos_Oct}, for August and October runs, respectively. A picture of the complete experimental setup is shown in Figure~\ref{bds_pic}. The nuclear recoil energies accessible with this BD configuration for 1~MeV neutrons are [3,83]~keV for sodium and [0.5,16]~keV for iodine (using the Equation~\ref{Equation_Enr} with the angles corresponding to the central position of each BD).

\begin{figure}[h!]
	\begin{center}
		\includegraphics[width=0.75\textwidth]{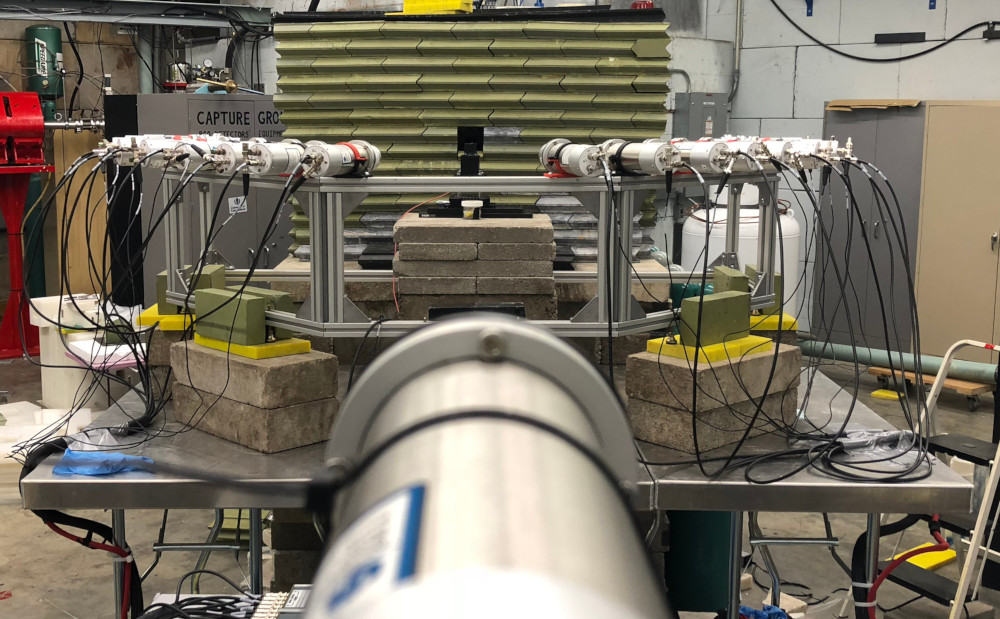}
		\caption{\label{deg0_pic}Picture of the experimental setup taken from the 0-deg detector, aligned with the neutron beam.}
	\end{center}
\end{figure}

Additionally, another liquid scintillator detector (see Figure~\ref{deg0_pic}) similar to those used as BDs was located aligned with the beam to perform dedicated TOF runs to measure the neutron beam energy. We will refer to this detector as "0-deg".

\begin{table}[h!]
	\centering
	\begin{tabular}{|c|c|c|c|c|c|c|}
		\hline
		BD & X(cm) & Y(cm) & $\theta_{min}$ (deg) & $\theta_{max}$ (deg) & $\Omega$ (deg$^2$) \\
		\hline
		0 & 0.00$\pm$2.05 & -62.60$\pm$0.13 & 83.08$\pm$0.03 & 92.33$\pm$0.03 & 6.75$\pm$0.02 \\
		1 & 13.17$\pm$0.79 & -62.60$\pm$0.13 & 75.85$\pm$0.01 & 80.39$\pm$0.01 & 6.40$\pm$0.02 \\
		2 & 23.31$\pm$0.45 & -62.60$\pm$0.13 & 67.40$\pm$0.01 & 71.75$\pm$0.01 & 5.87$\pm$0.02 \\
		3 & 32.77$\pm$0.31 & -58.90$\pm$0.13 & 58.76$\pm$0.01 & 63.07$\pm$0.01 & 5.77$\pm$0.02 \\
		4 & 40.09$\pm$0.24 & -52.05$\pm$0.13 & 50.18$\pm$0.01 & 54.61$\pm$0.01 & 6.07$\pm$0.02 \\
		5 & 45.68$\pm$0.19 & -44.45$\pm$0.13 & 41.93$\pm$0.01 & 46.50$\pm$0.01 & 6.45$\pm$0.02 \\
		6 & 54.75$\pm$0.15 & -35.90$\pm$0.13 & 31.03$\pm$0.01 & 35.48$\pm$0.01 & 6.11$\pm$0.02 \\
		7 & 61.89$\pm$0.13 & -30.60$\pm$0.13 & 24.20$\pm$0.01 & 28.42$\pm$0.01 & 5.50$\pm$0.02 \\
		8 & 64.68$\pm$0.11 & -22.15$\pm$0.13 & 16.78$\pm$0.01 & 21.03$\pm$0.01 & 5.60$\pm$0.02 \\
		9 & 63.61$\pm$0.11 & 17.50$\pm$0.13 & 13.18$\pm$0.01 & 17.59$\pm$0.01 & 6.02$\pm$0.02 \\
		10 & 62.87$\pm$0.12 & 28.60$\pm$0.13 & 22.35$\pm$0.01 & 26.57$\pm$0.01 & 5.49$\pm$0.02 \\
		11 & 55.46$\pm$0.15 & 36.70$\pm$0.13 & 31.31$\pm$0.01 & 35.68$\pm$0.01 & 5.92$\pm$0.02 \\
		12 & 47.44$\pm$0.18 & 43.40$\pm$0.13 & 40.19$\pm$0.01 & 44.71$\pm$0.01 & 6.34$\pm$0.02 \\
		13 & 43.88$\pm$0.21 & 49.70$\pm$0.13 & 46.36$\pm$0.01 & 50.75$\pm$0.01 & 5.95$\pm$0.02 \\
		14 & 37.32$\pm$0.27 & 57.20$\pm$0.13 & 54.75$\pm$0.01 & 59.00$\pm$0.01 & 5.62$\pm$0.02 \\
		15 & 27.42$\pm$0.39 & 62.60$\pm$0.13 & 64.22$\pm$0.01 & 68.48$\pm$0.01 & 5.61$\pm$0.02 \\
		16 & 15.96$\pm$0.65 & 62.70$\pm$0.13 & 73.47$\pm$0.01 & 77.96$\pm$0.01 & 6.26$\pm$0.02 \\
		17 & 2.97$\pm$3.48 & 62.90$\pm$0.13 & 84.99$\pm$0.03 & 89.61$\pm$0.03 & 6.61$\pm$0.02 \\
		\hline
	\end{tabular} \\
	\caption{Positions (X and Y) with the corresponding uncertainties at 1~$\sigma$ of the center of the face of every BD looking to the crystal, $\theta_{min}$ and $\theta_{max}$ defining the acceptance scattering angles range and acceptance solid scattering angle ($\Omega$) for August measurements. The angle uncertainties take into account the position uncertainties. The x axis is aligned with the neutron beam and the y axis is perpendicular in the horizontal plane (see Figure~\ref{Set-Up_Scheme}).}
	\label{tabla:BDpos_Ago}
\end{table}

\begin{table}[h!]
	\centering
	\begin{tabular}{|c|c|c|c|c|c|c|}
		\hline
		BD & X(cm) & Y(cm) & $\theta_{min}$ (deg) & $\theta_{max}$ (deg) & $\Omega$ (deg$^2$) \\
		\hline
		0 & 1.09$\pm$4.29 & -59.56$\pm$0.13 & 86.51$\pm$0.03 & 91.39$\pm$0.03 & 7.38$\pm$0.02 \\
		1 & 28.42$\pm$0.36 & -60.64$\pm$0.13 & 62.72$\pm$0.01 & 67.07$\pm$0.01 & 5.84$\pm$0.02 \\
		2 & 40.51$\pm$0.24 & -54.49$\pm$0.13 & 51.23$\pm$0.01 & 55.51$\pm$0.01 & 5.68$\pm$0.02 \\
		3 & 46.74$\pm$0.20 & -48.67$\pm$0.13 & 44.00$\pm$0.01 & 48.31$\pm$0.01 & 5.75$\pm$0.02 \\
		4 & 54.17$\pm$0.16 & -41.99$\pm$0.13 & 35.66$\pm$0.01 & 39.90$\pm$0.01 & 5.58$\pm$0.02 \\
		5 & 59.90$\pm$0.14 & -35.12$\pm$0.13 & 28.29$\pm$0.01 & 32.48$\pm$0.01 & 5.43$\pm$0.02 \\
		6 & 66.06$\pm$0.36 & -28.73$\pm$0.13 & 21.48$\pm$0.01 & 25.52$\pm$0.01 & 5.05$\pm$0.01 \\
		7 & 70.46$\pm$0.11 & -22.50$\pm$0.13 & 15.74$\pm$0.01 & 19.68$\pm$0.01 & 4.79$\pm$0.01 \\
		8 & 70.19$\pm$0.10 & -11.95$\pm$0.13 & 7.62$\pm$0.01 & 11.71$\pm$0.01 & 5.17$\pm$0.01 \\
		9 & 69.87$\pm$0.10 & 10.37$\pm$0.13 & 6.38$\pm$0.01 & 10.50$\pm$0.01 & 5.25$\pm$0.01 \\
		10 & 70.75$\pm$0.11 & 18.94$\pm$0.13 & 13.00$\pm$0.01 & 16.97$\pm$0.01 & 4.88$\pm$0.01 \\
		11 & 68.63$\pm$0.12 & 27.25$\pm$0.13 & 19.69$\pm$0.01 & 23.63$\pm$0.01 & 4.80$\pm$0.01 \\
		12 & 61.03$\pm$0.14 & 35.24$\pm$0.13 & 27.94$\pm$0.01 & 32.07$\pm$0.01 & 5.28$\pm$0.01 \\
		13 & 52.47$\pm$0.16 & 41.67$\pm$0.13 & 36.29$\pm$0.01 & 40.63$\pm$0.01 & 5.84$\pm$0.02 \\
		14 & 45.68$\pm$0.20 & 48.70$\pm$0.13 & 44.66$\pm$0.01 & 49.01$\pm$0.01 & 5.88$\pm$0.02 \\
		15 & 38.16$\pm$0.26 & 56.04$\pm$0.13 & 53.60$\pm$0.01 & 57.89$\pm$0.01 & 5.70$\pm$0.02 \\
		16 & 23.36$\pm$0.43 & 60.27$\pm$0.13 & 66.56$\pm$0.01 & 71.06$\pm$0.01 & 6.27$\pm$0.02 \\
		17 & 7.36$\pm$1.34 & 59.95$\pm$0.13 & 80.59$\pm$0.03 & 85.41$\pm$0.03 & 7.18$\pm$0.02 \\
		\hline
	\end{tabular} \\
	\caption{Same as in Table~\ref{tabla:BDpos_Ago} but for October measurements.}
	\label{tabla:BDpos_Oct}
\end{table}

These BDs consist of EJ-309, a liquid scintillator based on the solvent xylene. They were produced by Eljen Technology~\cite{EJ-309:Eljen} with a cylindrical shape with same diameter and length (50.8~mm) and encapsulated in a 0.425~mm-thick aluminium housing (specifications are presented at Appendix~\ref{Anexo_Specifications}, in Table~\ref{tabla:LSSpec} and Figure~\ref{LSSpec}). They were optically coupled to a HAMAMATSU PMT R7724 (NEG-HV)~\cite{PMT_LS:Hamamatsu}, as it is shown in Figure~\ref{BDSpec}. Characteristics of this PMT are presented in Table~\ref{tabla:PMT_BD_Spec} and Figure~\ref{PMT_BD_Spec}.

EJ-309 was selected because of its excellent particle discrimination capabilities. The average scintillation time of the EJ-309 liquid scintillator for electronic recoils (produced by gammas or electrons) is usually half of that for nuclear recoils (produced by neutrons). This results in a different pulse shape in the BD signal for neutrons and gammas/electrons, thus separating the neutron events from background events. For the energy calibration of these detectors, a $^{137}Cs$ source was placed close to the BDs 8 and 9, on the beam line at 60~cm from the crystal.

\subsection{Measurements schedule}\label{Section:QF_TUNL_Schedule}

The time intervals of measurement for each crystal are shown in Table~\ref{tabla:dataAcq}. The run scheme was similar in August and October, with small differences between them. For all the crystals, different acquisition batches were taken in intervals of approximately eight hours: NaI(Tl) energy calibration with a $^{133}Ba$ source, BDs energy calibration using a $^{137}Cs$ source, beam-off background measurement both in crystal and BDs, and the beam-on measurement. In every new batch of measurements, the crystal was rotated 30$^o$ with respect to the vertical axis to reduce possible systematics related to channeling (explained in Section~\ref{Section:QF_NaIQFOverview}). The acquisition time of the calibrations was 3 minutes, the background measurements were 5 minutes long and the rest of the time was dedicated to the beam-on measurements. Both the number of rotations and the duration of the beam-on data taking depended on the crystal, as it is presented in Table~\ref{tabla:timeAcq}.

\begin{table}[h!]
	\centering
	\begin{tabular}{|c|c|c|}
		\hline
		Crystal~$\#$ & From & To \\
		\hline
		1 & 8/19 at 9PM & 8/21 at 11AM \\
		2 & 8/21 at 12PM & 8/22 at 9PM \\
		3 & 10/20 at 1AM & 10/21 at 12PM \\
		4 & 10/21 at 3PM & 10/23 at 12PM \\
		5 & 10/23 at 2PM & 10/24 at 10AM \\
		\hline
	\end{tabular} \\
	\caption{Time intervals of measurement of each crystal.}
	\label{tabla:dataAcq}
\end{table}

\begin{table}[h!]
	\centering
	\begin{tabular}{|c|c|c|c|c|c|c|c|}
		\cline{3-8}
		\multicolumn{2}{c}{} & \multicolumn{6}{|c|}{Acquisition times (hours)} \\
		\hline
		Cr.~$\#$ & $\#$~batches & 0$^o$ & 30$^o$ & 60$^o$ & 90$^o$ & 120$^o$ & Total \\
		\hline
		1 & 4 & 10.02 & 7.33 & 7.58 & 9.32 & 0 & 34.25 \\
		2 & 4 & 9.08 & 7.72 & 8.62 & 4.75 & 0 & 30.17 \\
		3 & 4 & 8.17 & 7.90 & 7.53 & 8.30 & 0 & 31.90 \\
		4 & 5 & 8.35 & 8.18 & 8.33 & 9.62 & 8.70 & 43.18 \\
		5 & 2 & 8.58 & 9.02 & 0 & 0 & 0 & 17.60 \\
		\hline
	\end{tabular} \\
	\caption{Number of batches (corresponding to different orientation of the crystal) for each crystal and corresponding acquisition times (in hours) for the beam-on measurements (for each orientation and in total).}
	\label{tabla:timeAcq}
\end{table}

\subsection{Data acquisition system}\label{Section:QF_TUNL_DAQ}

\begin{figure}[h!]
	\begin{center}
		\includegraphics[width=0.5\textwidth]{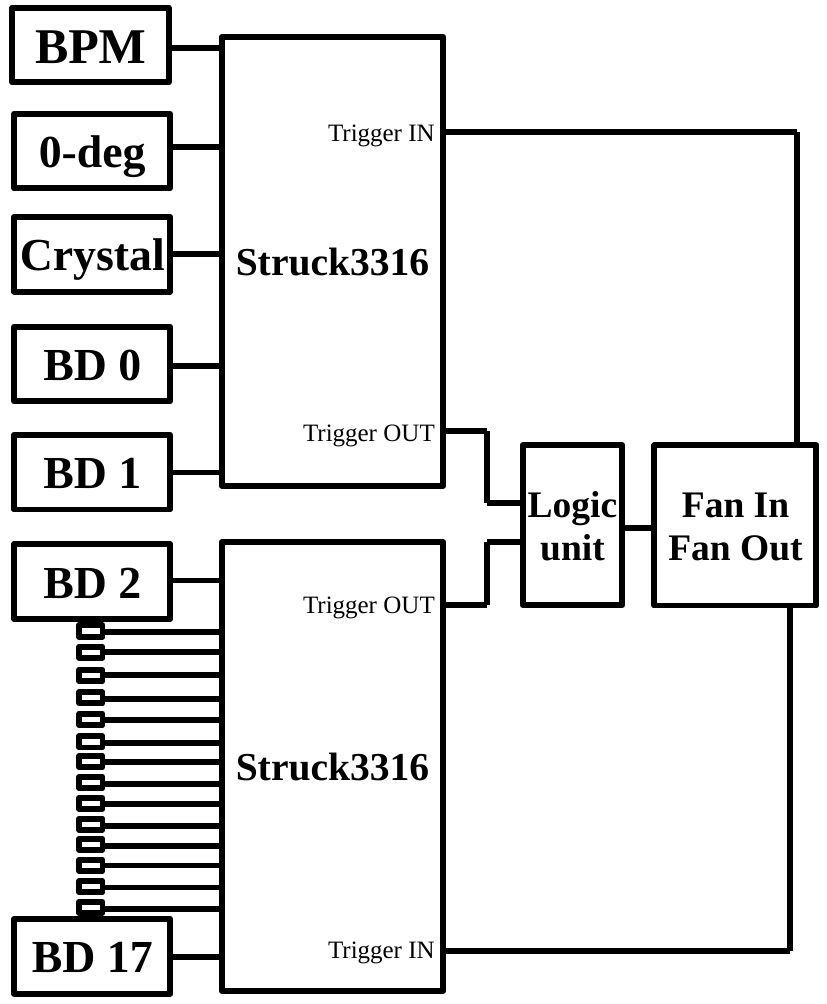}
		\caption{\label{ElectronicChain}Scheme of the electronic acquisition system.}
	\end{center}
\end{figure}

Two 14-bit, 250~MHz, 16-channels Struck 3316 digitizers~\cite{StruckDAQ} were used to acquire the data. The electronic acquisition system is shown in Figure~\ref{ElectronicChain}. The signals of all the detectors were connected to the digitizers. Each digitizer acted as a discriminator with a configuration that depended on the measurement. In the beam-on measurements, trigger was done by any of the BDs. In the background measurements, any of the detectors except the BPM could trigger. In the energy calibrations, the trigger was done by the corresponding detector which was measured: the crystal or any of the liquid scintillators for $^{133}Ba$ and $^{137}Cs$ measurements, respectively. Finally, in the TOF measurements the 0-deg detector was the only one which triggered. The output trigger signal from both digitizers was sent to a logical unit in OR mode, which generated a global trigger signal. This trigger signal, duplicated in a Fan-In Fan-Out module, was used as external trigger of both digitizers (see Figure~\ref{ElectronicChain}).

\begin{figure}[h!]
	\begin{center}
		\includegraphics[width=\textwidth]{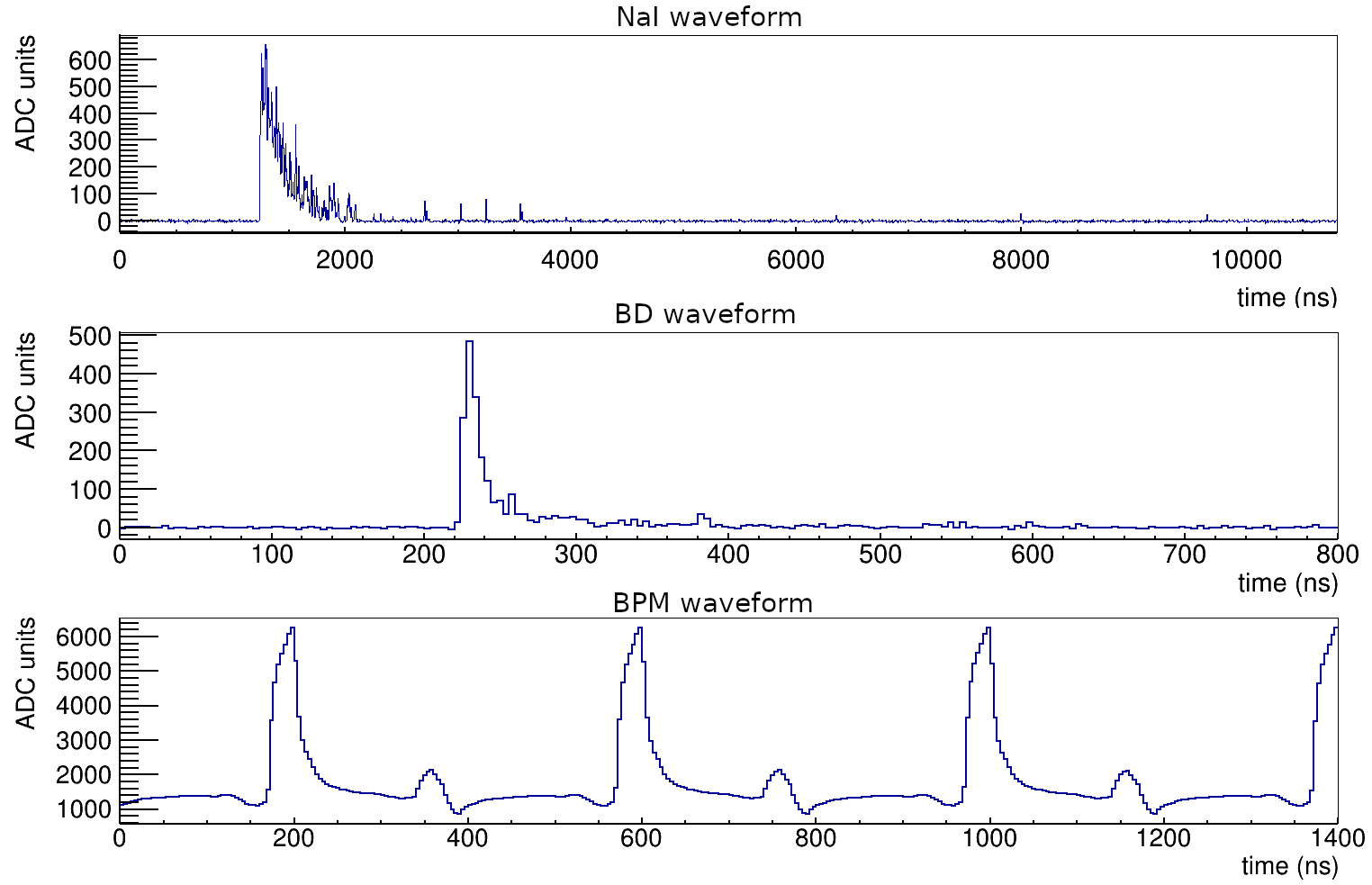}
		\caption{\label{waveforms}Example of the acquired waveforms. From top to bottom: NaI(Tl) detector, BD and BPM. In this event, corresponding to a beam-on measurement, the BD triggered the data acquisition. From the time difference between the BD pulse onset and the previous proton pulse in the BPM signal, information on the TOF of the particle, allowing the identification of the neutrons, can be drawn, as it will be shown in Section~\ref{Section:QF_Analysis_WaveformRec_BDs}.}
	\end{center}
\end{figure}

For each event corresponding to one of those global triggers, the timestamp variable (with a resolution of 1~ns) and 21~signals from the digitizers were acquired: the BPM (explained in Section~\ref{Section:QF_TUNL_Setup_NeutronProduction}), the NaI(Tl) crystal, the 18 BDs and the 0-deg signals. Different digitization windows are used for each signal: 1400~ns for the BPM, 10800~ns for the NaI(Tl) crystal, and 800~ns for the liquid scintillators (BDs and 0-deg detector). Pre-trigger was defined as 10$\%$ of the waveform for crystal calibrations and as 25$\%$ of the waveform in any other case. An example of the recorded waveforms is presented in Figure~\ref{waveforms}.

\section{Event analysis}\label{Section:QF_Analysis_WaveformRec}

\fancyhead[RO]{\emph{\thesection. \nameref{Section:QF_Analysis_WaveformRec}}}

Raw data files from the acquisition software were converted into ROOT files~\cite{Brun:1997pa} using a code developed by members of TUNL. These files were then processed in a second level analysis, which defined one "event" for each trigger, containing the information of all the waveforms recorded. Then, the BD and crystal waveforms were analyzed differently, as detailed in the following.

\subsection{Crystal waveforms}\label{Section:QF_Analysis_WaveformRec_NaI}

The first step in the NaI(Tl) waveform analysis was to obtain the baseline level and the corresponding root mean square (RMS). It is done by averaging the first and last 200~ns of the waveform. However, single photoelectrons randomly distributed due to the PMT dark current can contribute to the waveform and affect the baseline calculation. This is the reason to develop and apply a baseline quality cut. If the difference between both baseline levels was lower than a value (defined as three times the smallest root mean square), the level used for the baseline subtraction was the average of both values. However, if the baseline level calculated in the first 200~ns is more than three RMS above the baseline level calculated in the last 200~ns, the event is rejected, while if it is below that level, the event is kept and the baseline used is that corresponding to the first 200~ns.

After subtracting the baseline level from the waveform, an algorithm which looks for a pulse after the first 200~ns is applied. We set the analysis threshold at 5~RMS from the baseline. This criterion, could imply that threshold effects appeared as consequence of different noise levels during the measurements. We checked that the RMS of the baseline was stable along the data taking. The waveform position where the pulse is above the threshold will be stored as \textit{t0NaI}. Finally, the integral in a 2~$\mu$s window of the pulse, was stored in the variable \textit{fixedIntegral} and taken as the estimator of the energy deposited in the NaI(Tl). This variable has units of ADC because it was obtained as the sum of the values of the amplitude of the waveform:
\begin{equation}
	fixedIntegral = \sum ADC(t).
\end{equation}
This variable can be converted into the pulse area with physical dimensions (mV$\cdot$ns) just by multiplying by the temporal bin size of the digitizer (in our case 4~ns) and the voltage resolution (2000~mV/2$^{14}$). To avoid threshold effects, the starting time of this integral is fixed instead of using the \textit{t0NaI} variable, and it depends on the measurement. In calibration measurements, where the hardware trigger is done by the NaI(Tl), the pulse area is calculated in a fixed time window: from 1.5~to 3.5~$\mu$s. In the beam-on measurements, the time difference between the pulse onset in the crystal and BD waveforms was determined by the TOF of the scattered particle. As the trigger was done by the BDs and the neutrons are slower than the gammas, the integration window to determine the event energy is dependent on the type of particle. Therefore, pulses in the NaI(Tl) correlated with neutrons triggering the BDs should appear before in the waveform trace than those correlated with gammas. The average of the \textit{t0NaI} values was obtained after a neutron selection in BDs, and it was used in Section~\ref{Section:QF_Analysis_Selection_NeutronsBDs} as the start time of the integration window in the NaI(Tl) waveform for the quenching analysis.

\subsection{BD waveforms}\label{Section:QF_Analysis_WaveformRec_BDs}

For each one of the BD waveforms a similar analysis was developed: baseline level and RMS were calculated as the mean and standard deviation of the waveform in its first 160~ns. If the maximum of the waveform is more than 5 RMS above the baseline level, this BD was considered to have a pulse, and the number of this BD is saved as the triggered BD. The first time bin of the waveform which is 5 RMS above the baseline level is defined as the start time of the pulse ($t_o$). Applying the same analysis to all of the BD waveforms in the same event, a variable called 'multiplicity' was defined as the number of BD with signal above the threshold. Events with multiplicity higher than 1 ($\sim$~1$\%$ of the events) were rejected from the analysis as the direction of the scattered neutron cannot be determined.

\begin{figure}[h!]
	\begin{center}
		\includegraphics[width=0.75\textwidth]{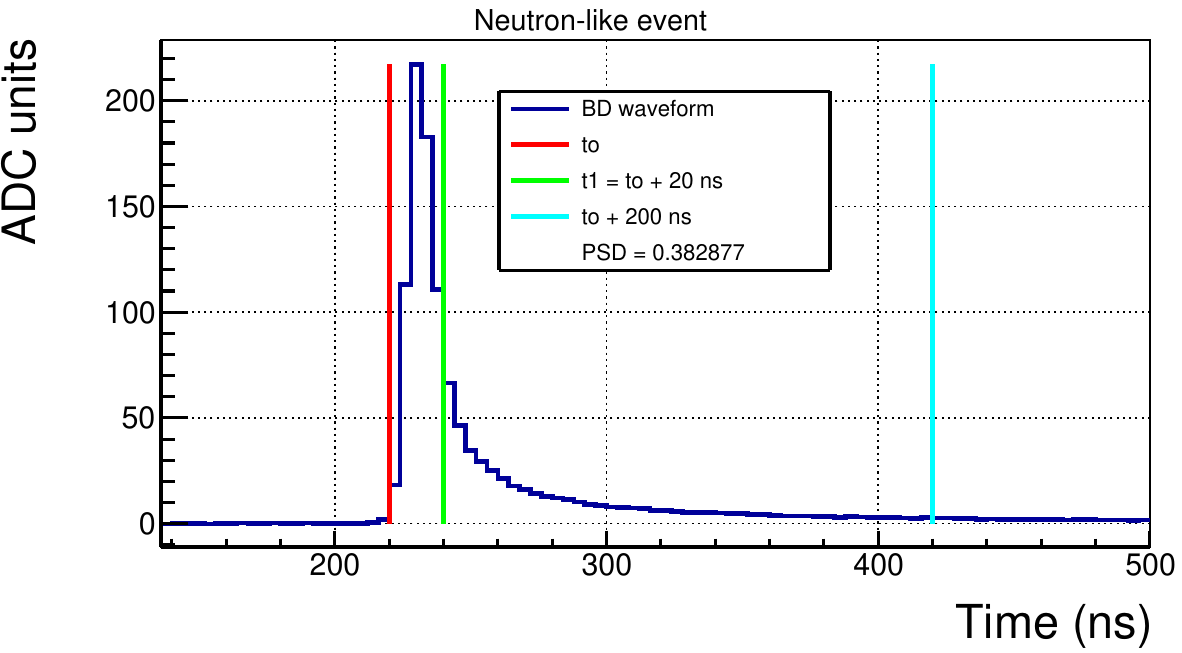}
		\caption{\label{PulsePSD_Neutron}BD waveform for a neutron event, with the integral intervals used for \textit{PSD} calculation. Edges of the integration windows are represented with lines: Red for $t_o$, green for $t_1$ (in this case defined as $t_o$+20~ns), and blue for $t_o$+200~ns.}
	\end{center}
\end{figure}

\begin{figure}[h!]
	\begin{center}
		\includegraphics[width=0.75\textwidth]{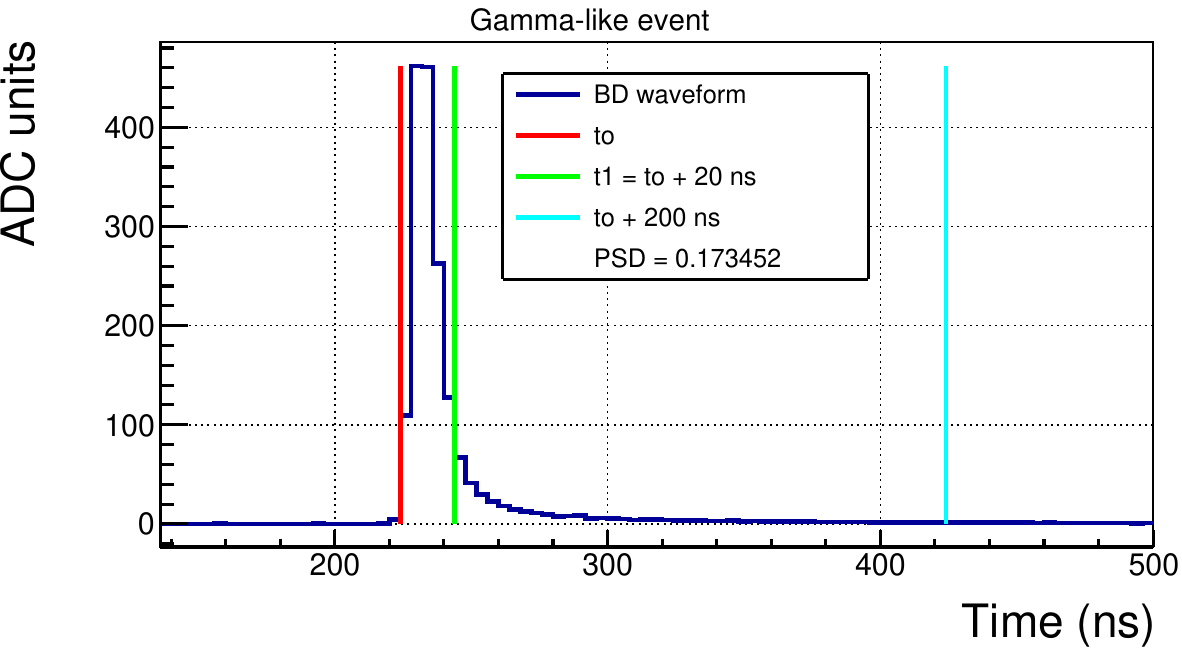}
		\caption{\label{PulsePSD_Gamma}BD waveform for a gamma event, with the integral intervals used for \textit{PSD} calculation. Edges of the integration windows are represented with lines: Red for $t_o$, green for $t_1$ (in this case defined as $t_o$+20~ns), and blue for $t_o$+200~ns.}
	\end{center}
\end{figure}

As it has been explained, neutrons can be easily differentiated from gammas or electrons in liquid scintillators by their different scintillation time. The discrimination is commonly done through a pulse shape variable defined for each BD waveform as the ratio of the pulse tail integral to the total pulse integral in a 200~ns time window:
\begin{equation}
	PSD = \frac{\sum_{t=t_1}^{t=t_o+200~ns}V_{BD}(t)}{\sum_{t=t_o}^{t=t_o+200~ns}V_{BD}(t)},
\end{equation}
where $V_{BD}(t)$ is the value of the waveform at a given time, $t_o$ is the initial time of the pulse and $t_1$ defines the starting of the pulse tail, which is chosen to maximize the discrimination capability. Figures~\ref{PulsePSD_Neutron} and~\ref{PulsePSD_Gamma} show the BD waveforms for a neutron and a gamma event, respectively. Different $t_1$ values were used and the discrimination power of the generated \textit{PSD} values were analyzed (Section~\ref{Section:QF_Analysis_Selection_NeutronsBDs}).

\begin{figure}[h!]
	\begin{center}
		\includegraphics[width=0.75\textwidth]{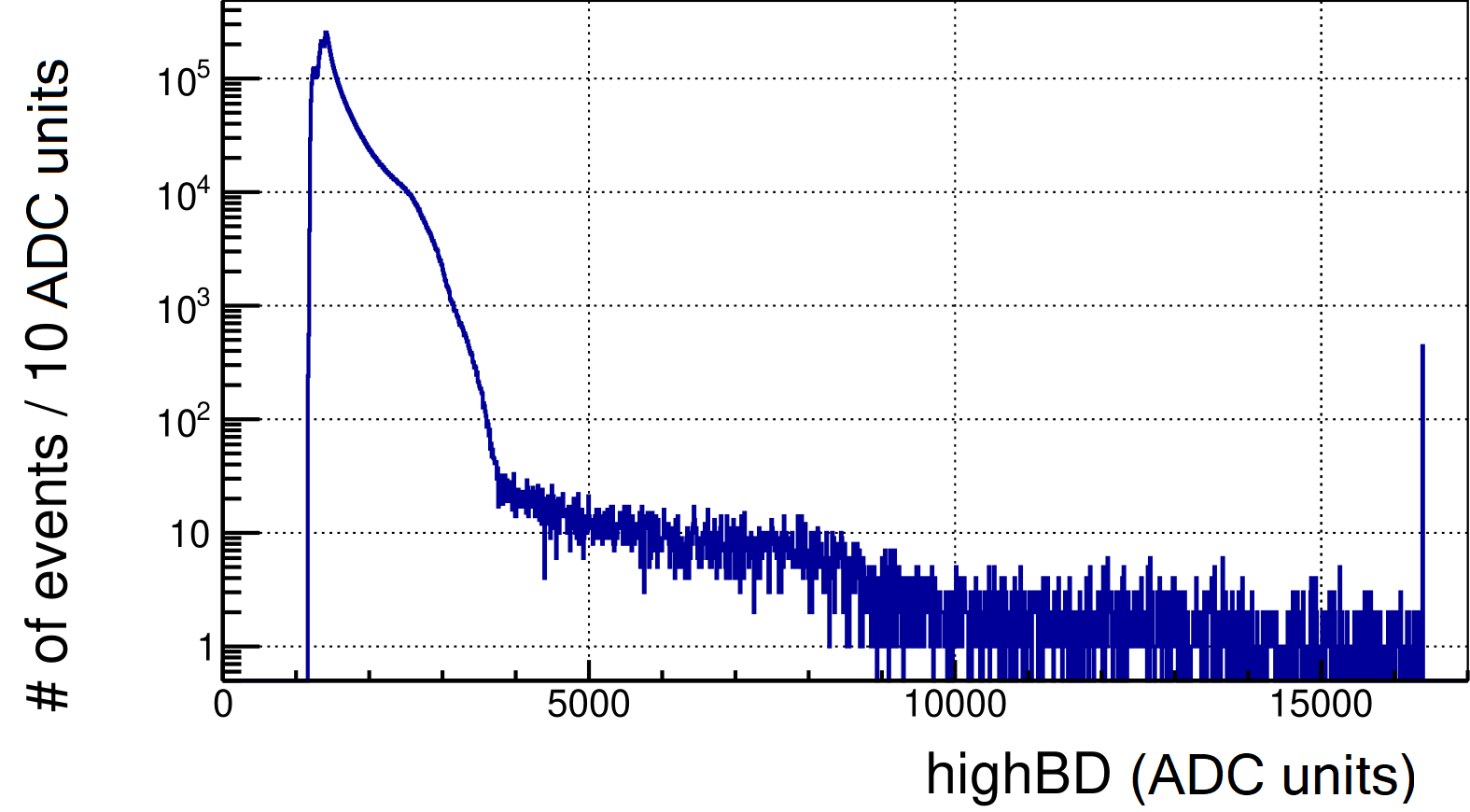}
		\caption{\label{highBD}Maximum of the pulse in BD waveforms, saturation level corresponds to 16380~ADC~units.}
	\end{center}
\end{figure}

Saturated waveforms in the BD would produce wrong \textit{PSD} values, and therefore, they have to be rejected. This is done using the value of the pulse maximum, defined as \textit{highBD}. Figure~\ref{highBD} shows the histogram of these pulse amplitudes for one of the BD detectors. The saturation level was found to be 16380~ADC~units, and the percentage of saturated events was 0.005\%.

\begin{figure}[h!]
	\begin{center}
		\includegraphics[width=0.75\textwidth]{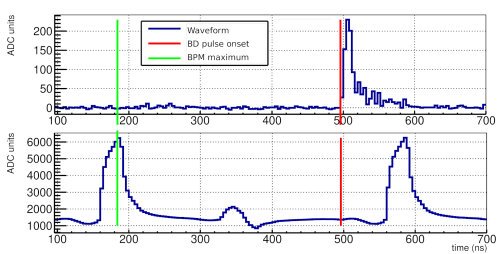}
		\caption{\label{PulsetSPB}Pulse in a BD waveform (top plot) and BPM waveform (bottom plot). Red line represents the initial time of the pulse ($t_o$), while green line represents the last maximum of the BPM waveform previous to the pulse in the BD. The value of the \textit{timeSincePrevBPM} is the difference between both times.}
	\end{center}
\end{figure}

Another variable built using the BD and BPM waveforms is \textit{timeSincePrevBPM}, which is defined as the difference between $t_o$ and the last maximum of the BPM waveform before $t_o$ (see Figure~\ref{PulsetSPB}). This variable can be understood as the TOF of the neutron from the LiF target to the triggered BD plus an offset related with the TOF of the protons from BPM to the target and the signal processing. Both \textit{PSD} and \textit{timeSincePrevBPM} variables will be used to discriminate between gammas/electrons and neutrons in BDs.

\section{Neutron beam energy determination} \label{Section:QF_Analysis_NeutronBeam}
\fancyhead[RO]{\emph{\thesection. \nameref{Section:QF_Analysis_NeutronBeam}}}

\begin{figure}[h!]
	\begin{center}
		\includegraphics[width=0.75\textwidth]{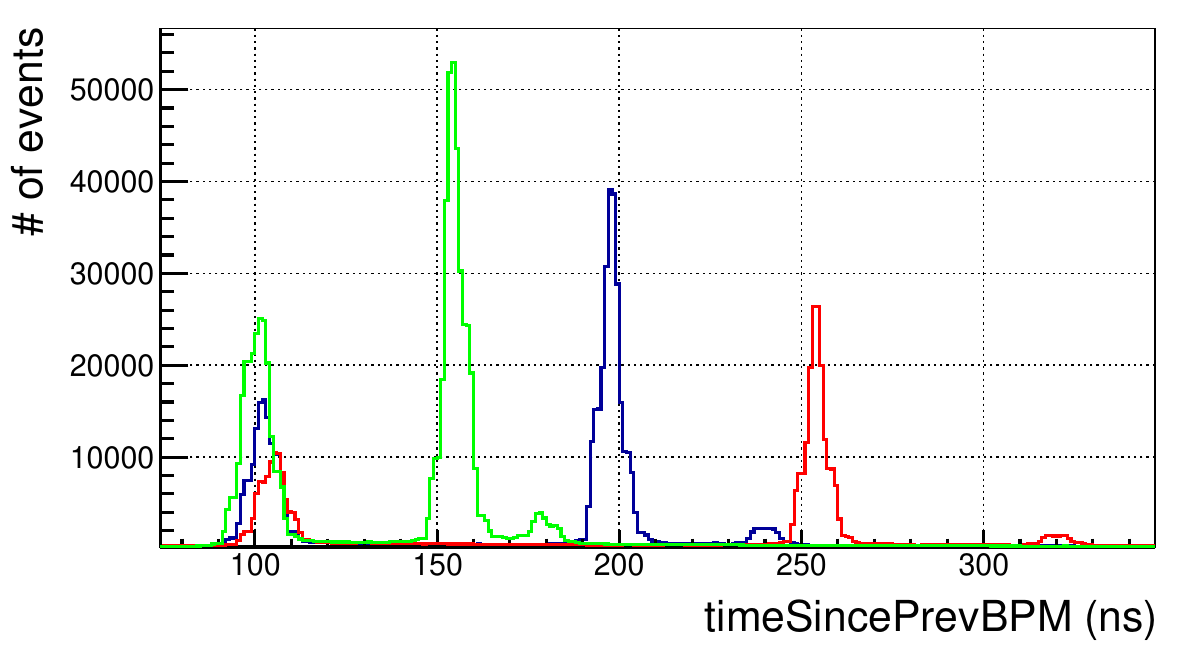}
		\caption{\label{tSPBdeg0}Distributions of \textit{timeSincePrevBPM} variable for the three positions of the 0-deg detector during TOF measurements of the October run: green for close, blue for middle and red for far position (see text for more details). Photons produced in the LiF target are identified as the short TOF population, while two neutron populations are observed at longer TOF: the main one, corresponding to neutrons of $\sim$~1~MeV and the other to $\sim$~500~keV.}
	\end{center}
\end{figure}

The neutron beam energy distributions for both runs will be a relevant input in the simulations developed to obtain the expected nuclear recoil energy distribution (Section~\ref{Section:QF_GEANT4}). They were calculated through the TOF of the neutrons between the LiF target and the 0-deg detector. For this purpose, dedicated TOF measurements were done both in August and October runs, placing the 0-deg detector at three different distances from the LiF target. These distances were 296.3~cm, 343.6~cm and 394.3~cm in August run and 74.5~cm, 133.2~cm and 210.8~cm in October run. The \textit{timeSincePrevBPM} variable (which is directly the TOF plus an offset) is calculated as for the BDs (see Section~\ref{Section:QF_Analysis_WaveformRec_BDs}). The distributions of this variable for the three positions of the 0-deg detector are presented in Figure~\ref{tSPBdeg0}.

The LiF target under proton irradiation is expected to produce photons, and they are identified in the \textit{timeSincePrevBPM} histogram shown in Figure~\ref{tSPBdeg0} as the short TOF population. The distribution of the \textit{timeSincePrevBPM} for photons allows to estimate the offset $\delta_t$ in this variable:
\begin{equation}
	\delta_t = t_{\gamma} - TOF_{\gamma},
\end{equation}
where $t_{\gamma}$ is the most probable value of the \textit{timeSincePrevBPM} for photons and $TOF_{\gamma} = D/c$, being $D$ the distance of the 0-deg detector to the LiF target. Therefore, the TOF for neutrons ($TOF_n$) is estimated as 
\begin{equation}\label{eq:TOF_offset}
	TOF_n = t_n - \delta_t,
\end{equation}
being $t_n$ the most probable value for the \textit{timeSincePrevBPM} variable for neutrons. $TOF_n$ has been calculated for the three different distances and results are shown in Tables~\ref{tabla:TOFAgo} and~\ref{tabla:TOFOct}, together with the values of $TOF_{\gamma}$ and $\delta_t$. It is possible to observe that, as expected, the value of the offset is the same for the three measurements of each run.

\begin{table}[h!]
	\centering
	\begin{tabular}{|c|c|c|c|c|c|}
		\hline
		Distance (m) & $t_{\gamma}$ (ns) & $t_n$ (ns) & $TOF_n{\gamma}$ (ns) & $TOF_n$ (ns) & $\delta_t$ (ns)\\
		\hline
		2.963 & 17 & 230 & 10 & 222 & 7 \\
		\hline
		3.436 & 19 & 265 & 11 & 259 & 8 \\
		\hline
		3.943 & 21 & 302 & 13 & 293 & 8 \\
		\hline
	\end{tabular} \\
	\caption{Distance between 0-deg detector and LiF target, the corresponding most probable values of the \textit{timeSincePrevBPM} variable for gammas and neutrons ($t_{\gamma,n}$), estimated TOF of gammas and neutrons ($TOF_{\gamma,n}$) and time offset $\delta_t$, for August run.}
	\label{tabla:TOFAgo}
\end{table}

\begin{table}[h!]
	\centering
	\begin{tabular}{|c|c|c|c|c|c|}
		\hline
		Distance (m) & $t_{\gamma}$ (ns) & $t_n$ (ns) & $TOF_{\gamma}$ (ns) & $TOF_n$ (ns) & $\delta_t$ (ns) \\
		\hline
		0.745 & 101 & 154 & 2 & 54 & 99 \\
		\hline
		1.332 & 102 & 197 & 4 & 100 & 98 \\
		\hline
		2.108 & 105 & 254 & 7 & 155 & 98 \\
		\hline
	\end{tabular} \\
	\caption{Distance between 0-deg detector and LiF target and corresponding most probable values of the \textit{timeSincePrevBPM} variable for gammas and neutrons ($t_{\gamma,n}$), estimated TOF of gammas and neutrons ($TOF_{\gamma,n}$) and time offset $\delta_t$, for October run.}
	\label{tabla:TOFOct}
\end{table}

\begin{figure}[h!]
	\begin{subfigure}[b]{0.49\textwidth}
		\includegraphics[width=\textwidth]{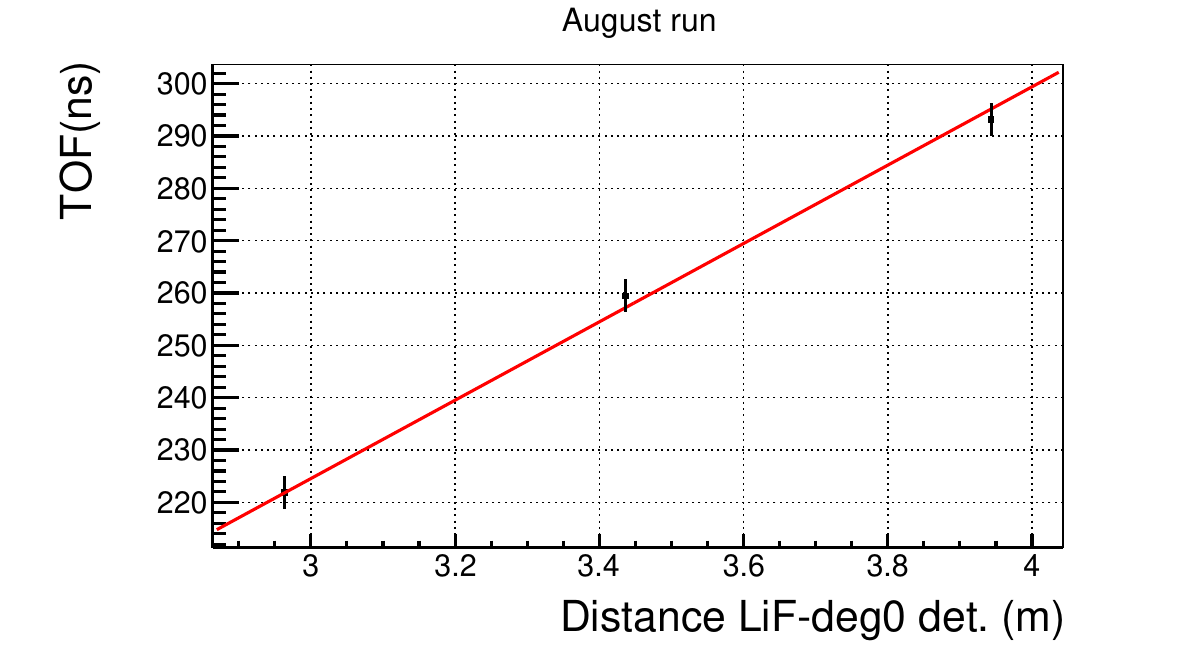}
	\end{subfigure}
	\begin{subfigure}[b]{0.49\textwidth}
		\includegraphics[width=\textwidth]{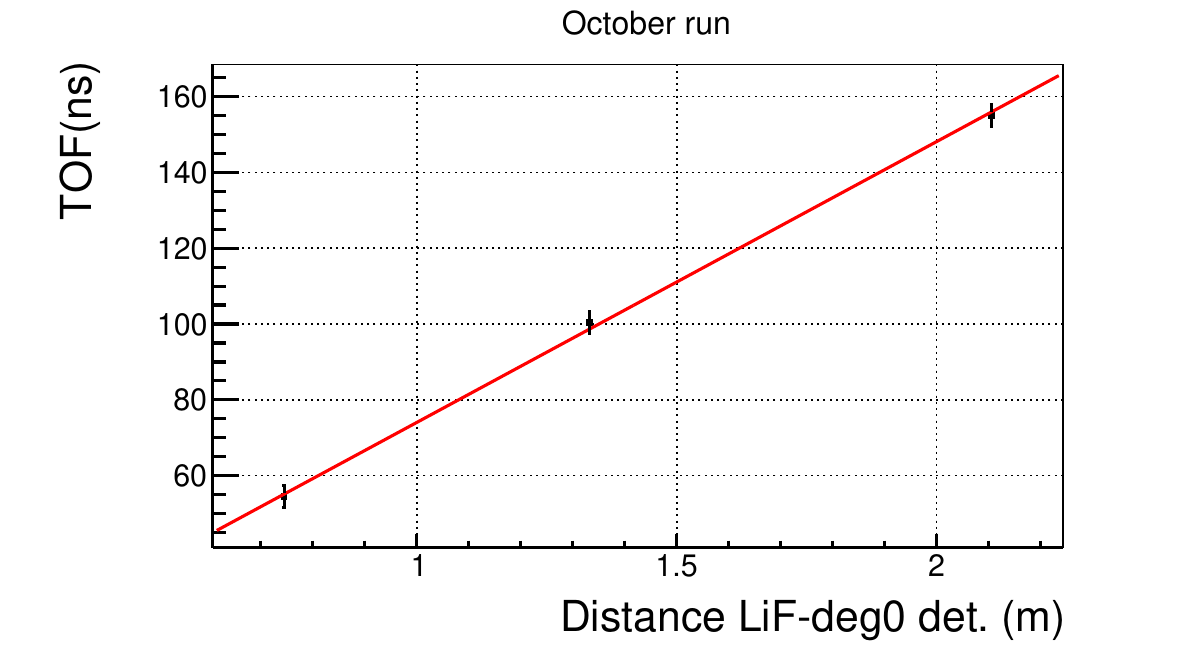}
	\end{subfigure}
	\caption{\label{TOFmeas}Linear fit of data from Tables~\ref{tabla:TOFAgo} and~\ref{tabla:TOFOct} to the Equation~\ref{eq:D_Vs_TOF}. Left plot corresponds to the August run, resulting in a mean neutron energy of 933~$\pm$~12~keV and a $\chi^2/nDF=3.36$, and right plot corresponds to the October run, resulting in a mean neutron energy of 953~$\pm$~28~keV and a $\chi^2/nDF=1.68$.}
\end{figure}

As the energy of the neutrons is about 1~MeV, they are not relativistic and their kinetic energy can be calculated as:
\begin{equation}\label{eq:En_Vs_TOF}
	E_n = \frac{mD^2}{2{TOF_n}^2},
\end{equation}
where $m$ is the neutron mass. The neutron energy was calculated with a linear fit between $TOF_n$ and $D$ following the equation
\begin{equation}\label{eq:D_Vs_TOF}
	TOF_n = \sqrt{\frac{m}{2E_n}} D.
\end{equation}
The same analysis was done for both runs, and the fits are shown in Figure~\ref{TOFmeas}. As a first result, the mean neutron energy of August run was 933~$\pm$~12~keV, while for October run was 953~$\pm$~28~keV.

It was clear that this method had important limitations. It was not possible to calculate the neutron energy distribution using this method because, although in principle the corresponding neutron energy can be calculated for each event using its TOF and the Equation~\ref{eq:En_Vs_TOF}, the time resolution limits strongly the energy resolution. As an example, the resolution of 1~ns in the TOF of the events of the close measurement in October run implies a resolution in energies of the order of 10~keV. Moreover, the asymmetrical shape of the TOF distribution could introduce some systematics in the calculation of the neutron energies. Some contributions which widened the TOF distribution were considered:

\begin{itemize}
	\item The finite size of the 0-deg detector introduces a dispersion in the measured TOF, as the neutron can interact in any point along the detector axis, and then, the distance travelled from the target has a dispersion.
	\item As the time width of the pulsed proton beam is around 2~ns, the neutrons can arrive at the detector at different times even if they have the same energy.
	\item The TOF distribution must be convolved with the detector time response (assumed gaussian).
\end{itemize}

All of these contributions can be included in an global time response, whose convolution with the $TOF_n$ distribution purely associated with the neutron energy, will produce the measured $TOF_n$ distribution.

Sam Hedges, who was a member of the collaboration taking part in the measurements and analysis presented in this Chapter, developed a method to obtain the distribution of the neutron energies from the measured $TOF_n$ distribution. As I did not participate in this particular analysis, this procedure is here only summarized for completeness. A fully explanation of this method can be found in his PhD thesis~\cite{ThesisSam}. It consists of the next steps:
\begin{enumerate}
	\item Simulate the 0-deg detector and 1~MeV gammas generated at the LiF target position using Monte Carlo N-Particle Transport Code (MCNP)~\cite{POZZI2003550} and fit the measured $TOF_{\gamma}$ distribution to that obtained from the simulation after having convolved the latter with a gaussian whose FWHM is a free parameter representing the time response of the detector.
	\item Simulate protons with different energies inciding onto the LiF target and calculate, using SRIM (Stopping and Range of Ions in Matter) from~\cite{ZIEGLER20101818}, the energy loss in the LiF target. For each initial proton energy, a broad distribution of proton energies is obtained. These distributions of proton energies can be converted into neutron energies distributions using the information on the Nuclear Data Tables~\cite{LISKIEN197557} for the neutron production cross-sections and neutron energies produced for the $^7Li$(p,n)$^7Be$ reaction.
	\item Simulate the transport of the neutrons produced at the LiF target position for each initial proton energy and arrival to the 0-deg detector in the three positions, using MCNP. The neutron energy is sampled according to the previously obtained energy distribution and the beam divergence parameters are taken into account (see Section~\ref{Section:QF_TUNL_Setup_NeutronProduction}). From each initial proton energy the $TOF_n$ distribution can be obtained and later, be convolved with the time response obtained from the fit of $TOF_{\gamma}$ (step 1), generating a PDF for the $TOF_n$ dependent on the different initial proton energies. The measured $TOF_n$ distribution is fitted to this PDF to determine the incident proton energy. With the latter, the neutron energy distribution can be calculated following step~2.
\end{enumerate}

\begin{figure}[h!]
	\begin{center}
		\includegraphics[width=0.75\textwidth]{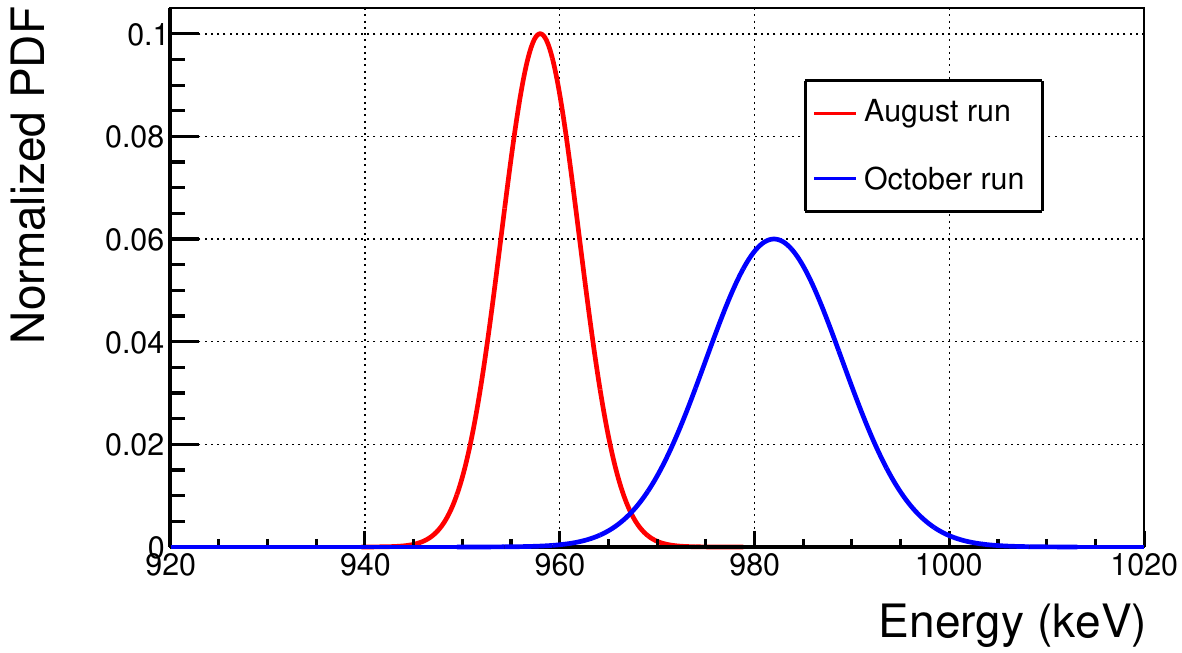}
		\caption{\label{EnDistr}Neutron energy distributions of each run using the second method. They are presented normalized in area, as they will be used as PDF in the simulation.}
	\end{center}
\end{figure}

The time response, proton energies and mean and standard deviation of the neutron energies obtained from these fits are presented in Table~\ref{tabla:NeutronEnergies}. Both analysis of the neutron energies result in higher mean neutron energy for the October run than for the August run. Moreover, the energies obtained with the first method were 25~and 29~keV lower than those obtained with the second one. It is probably due to the effect of the time response of the detector, which increased the measured $TOF_n$, thus reducing the neutron energy obtained with the first method. The neutron energy distributions obtained from the second method are presented in Figure~\ref{EnDistr}. They will be used as PDFs in the simulations carried out for the QF analysis (see Section~\ref{Section:QF_GEANT4}).

\begin{table}[h!]
	\centering
	\begin{tabular}{|c|c|c|c|c|}
		\hline
		Run & Time resp. (ns) & E$_P$ (keV) & Mean E$_n$  (keV) & Std. dev. E$_n$(keV) \\
		\hline
		August & 3.40~$\pm$~0.06 & 2670.9$^{+1.5}_{-3.1}$ & 958~$\pm$~5 & 4~$\pm$~3 \\
		\hline
		October & 1.21~$\pm$~0.03 & 2696.8$^{+0.3}_{-0.8}$ & 982~$\pm$~7 & 7~$\pm$~5 \\
		\hline
	\end{tabular} \\
	\caption{Time responses (FWHM of the gaussian), proton energies and mean and standard deviation of the neutron energy distributions derived for the two runs (see text for more detail on the procedure followed).}
	\label{tabla:NeutronEnergies}
\end{table}

\section{Monte Carlo simulation with GEANT4}\label{Section:QF_GEANT4}

\fancyhead[RO]{\emph{\thesection. \nameref{Section:QF_GEANT4}}}

As it was explained before (see Section~\ref{Section:QF_NaIQFOverview}), to reduce systematic effects related to the calculation of the nuclear recoil energies, a MC simulation was developed using the GEANT4 package~\cite{GEANT4:2002zbu}. This simulation has been also a fundamental tool for obtaining the energy depositions in the crystals during their corresponding calibrations, and also provided relevant information on the TOF, which allowed to improve the event identification in the BDs. In this section, a detailed description of the simulation implementation is presented, as well as the procedures followed to validate it and the results obtained for the nuclear recoil energy distributions and the energy calibrations.

\subsection{Neutron beam definition} \label{Section:QF_GEANT4_Beam}

\begin{figure}[h!]
	\begin{center}
		\includegraphics[width=\textwidth]{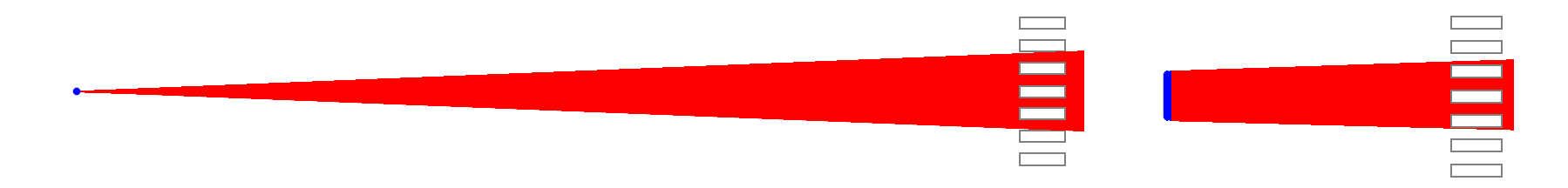}
		\caption{\label{beam_scan_geo}Geometries constructed in the GEANT4 simulation of the beam scan in the far position. On the left, point-like neutron source in the LiF target position. On the right, disc shape source in the collimator position. The neutron source is presented in blue color while red color represents the neutron beam generated in each case. The neutron detectors are the grey rectangles.}
	\end{center}
\end{figure}

\begin{figure}[h!]
	\begin{subfigure}[b]{0.49\textwidth}
		\includegraphics[width=\textwidth]{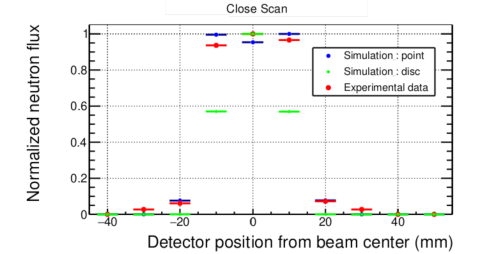}
	\end{subfigure}
	\begin{subfigure}[b]{0.49\textwidth}
		\includegraphics[width=\textwidth]{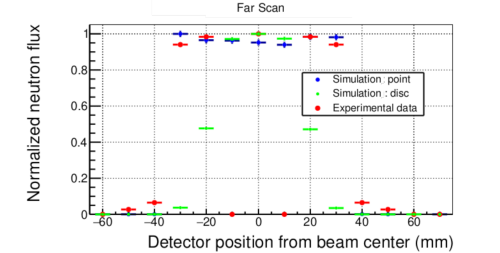}
	\end{subfigure}
	\caption{\label{beam_scan_res}Comparison between experimental and simulated data of the number of neutrons detected in the close (left plot) and far (right plot) positions in the beam scan. Red circles are the experimental data, blue circles correspond to the neutron source simulated in the LiF position, while for green circles the neutron source was a disk in the collimator position. All the data have been scaled to their maximum value for a suitable comparison.}
\end{figure}

The beam energy is described by the PDF obtained in Section~\ref{Section:QF_Analysis_NeutronBeam}, with the energy distributions presented in Figure~\ref{EnDistr}. For the angular distribution, two different models were considered: a point-like source in the LiF target position with 2.4$^o$ beam angle and a disc shape source with 1.1~cm of radius in the collimator position with the same beam angle. In both cases, the beam angle is defined as the maximum angle respect to the beam-axis at which the neutrons can emerge from the source. The plastic scintillators used in the experimental measurement of the beam shape (see Section~\ref{Section:QF_TUNL_Setup_NeutronProduction}) were also included in the simulation, as it is shown in Figure~\ref{beam_scan_geo}, and the neutron rate resulting from the simulation is shown in Figure~\ref{beam_scan_res} compared with the experimental measurements.

Figure~\ref{beam_scan_res} shows clearly that, as expected, the point-like source in the LiF target position with a beam angle of 2.4$^o$ reproduces correctly the measured beam shape. This was the neutron source modelling chosen for the following simulations. We chose a beam angle larger than the estimated in Section~\ref{Section:QF_TUNL_Setup_NeutronProduction}, but within one standard deviation of that result, which is not supposed to produce any effect on the nuclear recoil energy distributions because the NaI(Tl) crystals are completely covered by the neutron beam (the angular size of the NaI(Tl) crystal seen from the LiF target position is 1.1$^o$ for the crystal~3, and smaller for the other crystals, see Section~\ref{Section:QF_TUNL_Setup_Crystals}).

\subsection{Geometry and output variables} \label{Section:QF_GEANT4_Geometry}

\begin{figure}[]
	\begin{center}
		\includegraphics[width=\textwidth]{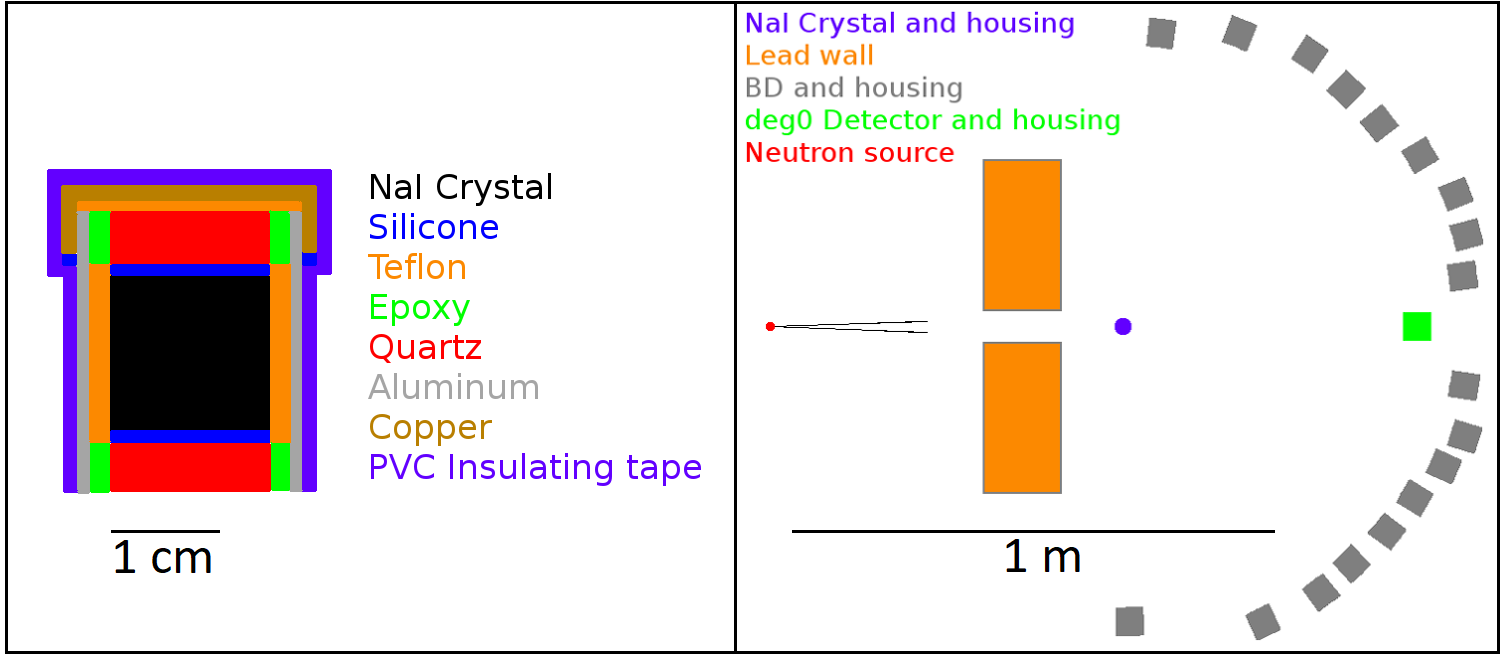}
		\caption{\label{geant_geo}Left panel: complete geometry of the crystal and its housing defined in the simulation. Right panel: View of the GEANT4 simulation geometry which includes the NaI(Tl) crystal from Zaragoza and its encapsulation, the BDs, the lead wall and the 0-deg detector in the close position.}
	\end{center}
\end{figure}

A simplified geometry was implemented, including only the main volumes (see Figure~\ref{geant_geo}):

\begin{itemize}
	\item NaI(Tl) crystal (NaI)
	\item 18 BDs (BD)
	\item Housing of the crystal (Cap)
	\item Housing of the BDs (Al)
	\item Lead wall (Pb)
	\item Air of the laboratory (Air)
\end{itemize}

In the TOF simulation, the 0-deg detector was also included in addition to the 18 BDs. Dimensions of the crystal and the components of its housing were obtained from specifications of AS (see Section~\ref{Section:QF_TUNL_Setup_Crystals}). The PVC isolating tape was included in the geometry of the housing of the crystal assuming 0.5~mm thickness, although the precise value was not known. Simulated dimensions and positions of the BDs and its housing correspond to measurements shown in Section~\ref{Section:QF_TUNL_Setup_BDs}. 

All the volumes were defined as active (i.e., the simulation keeps track of the deposited energy on them) independently on the simulation purpose. There are three kinds of output variables for each volume: energy deposition, multiplicities (number of energy depositions in a volume) and the time of the first energy deposition (taking as reference the neutron production time). For the case of the energy depositions in the crystal volume, the energy transferred to sodium or iodine recoils was also stored in separate variables (\textit{ENa} and \textit{EI}, respectively). The number of the BD hit is saved in a variable called \textit{BD}. In the following we will also use the term \textit{channel} for the BD, both in simulation and measurements.

\subsection{Validation: TOF measurement} \label{Section:QF_GEANT4_Validation}

\begin{figure}[h!]
	\begin{center}
		\includegraphics[width=0.75\textwidth]{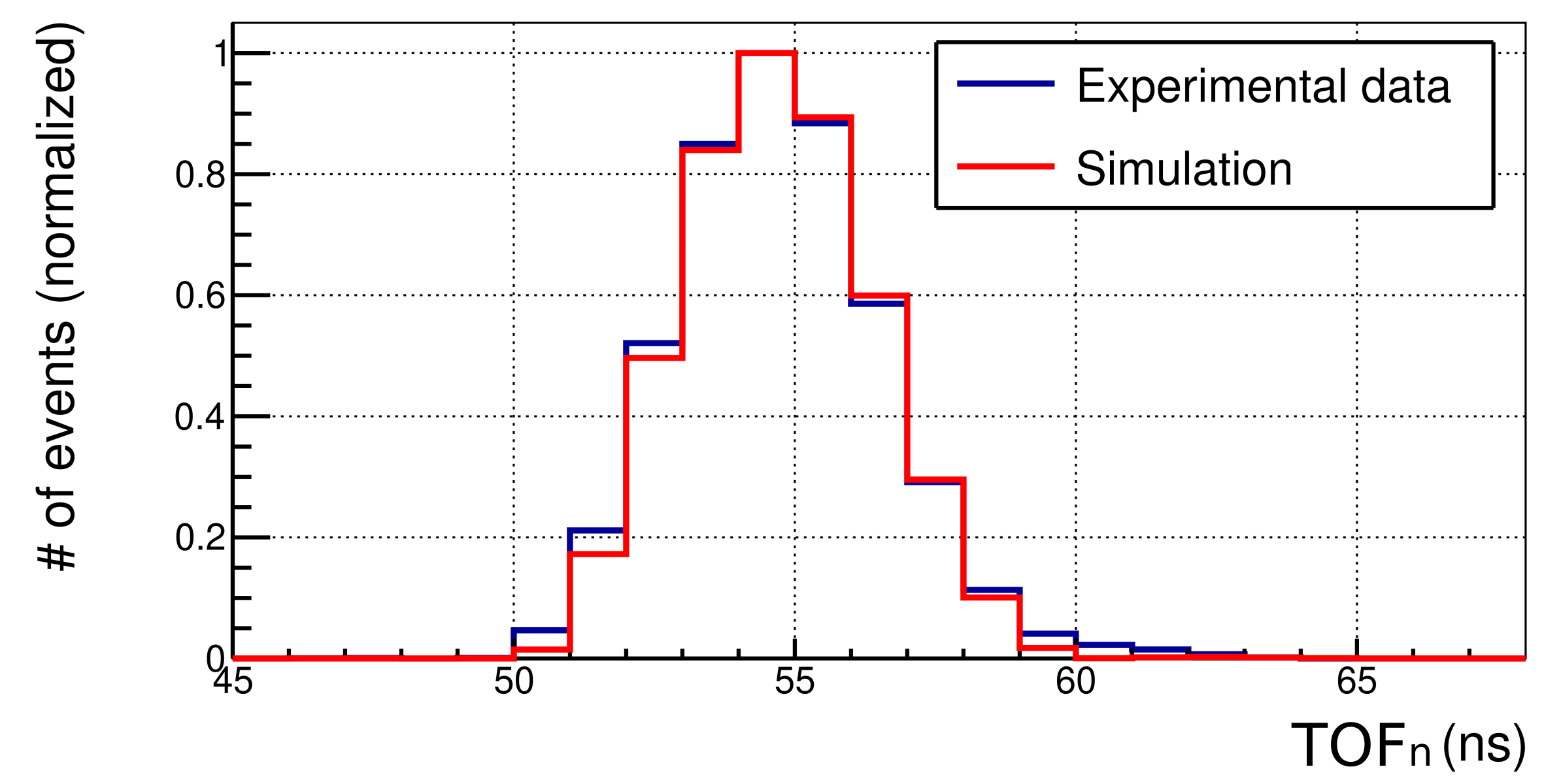}
		\caption{\label{TOF_SimVsExp}$TOF_n$ distribution comparison between simulation and experimental data. Both distributions were normalized to the maximum value.}
	\end{center}
\end{figure}

The TOF measurements with the 0-deg detector were used to validate the simulation. The 0-deg detector was placed in the close position and the time of the energy deposition by a neutron in this detector was saved. As it was explained in Section~\ref{Section:QF_Analysis_NeutronBeam}, in order to compare simulations and measurements the simulated $TOF_n$ distribution has to be convolved with the detector time response, assumed gaussian with FWHM of 1.21~$\pm$~0.03~ns for the October run (see Table~\ref{tabla:NeutronEnergies}). Moreover, the offset $\delta_t$ in the \textit{timeSincePrevBPM} variable (defined in the Equation~\ref{eq:TOF_offset} and obtained in Section~\ref{Section:QF_Analysis_NeutronBeam}) was subtracted to do a proper comparison. The result for the close measurement in the October run is shown in Figure~\ref{TOF_SimVsExp}.

Simulation and measurement are in clear agreement, and therefore, we can conclude that the neutron energy distribution reconstructed following the procedure explained in Section~\ref{Section:QF_Analysis_NeutronBeam} and shown in Figure~\ref{EnDistr} reproduces satisfactorily the TOF measurement, something essential to proceed to carry on simulations with different goals, as shown next.

\subsection{Simulation results}\label{Section:QF_GEANT4_Results}

\subsubsection{Nuclear recoil energies} \label{Section:QF_GEANT4_Results_EnrDistr}

\begin{figure}[h!]
	\begin{center}
		\includegraphics[width=\textwidth]{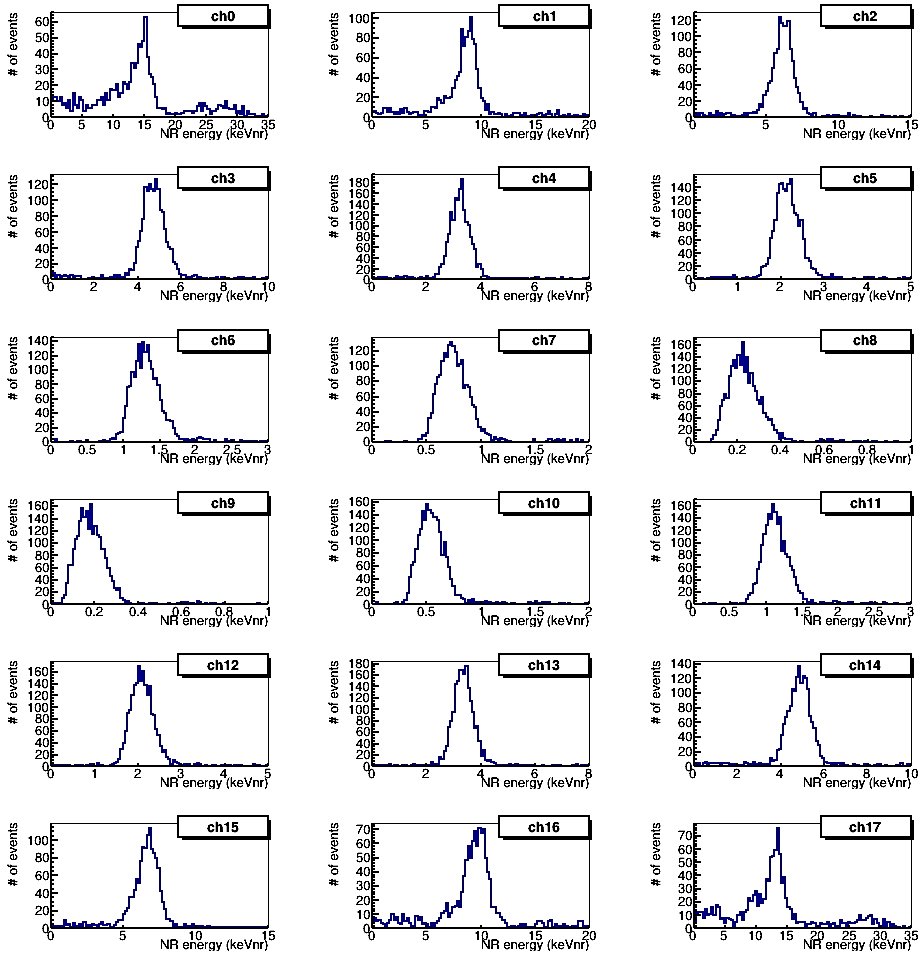}
		\caption{\label{enr_yodo}Nuclear recoil energy distributions for the iodine nuclei from $10^9$~simulated neutrons. They are shown for each channel. BD positions correspond to the August run.}
	\end{center}
\end{figure}

\begin{figure}[h!]
	\begin{center}
		\includegraphics[width=\textwidth]{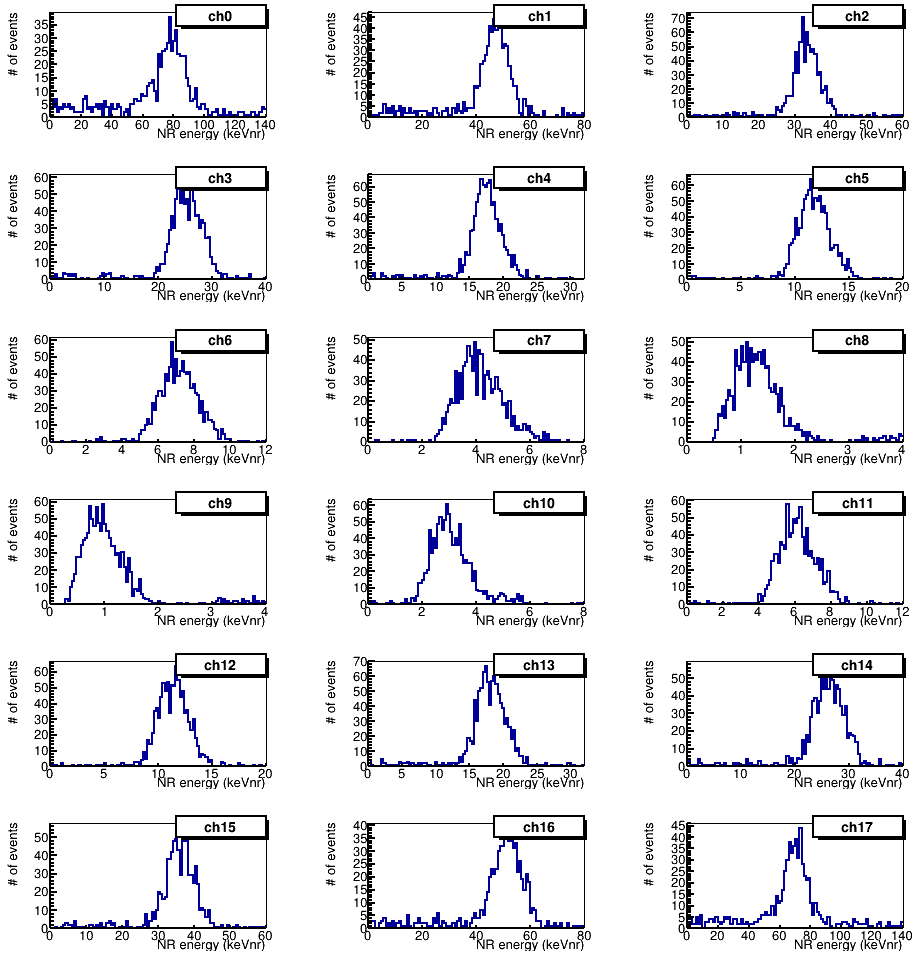}
		\caption{\label{enr_sodio}Nuclear recoil energy distributions for the sodium nuclei from $10^9$~simulated neutrons. They are presented for each channel. BD positions correspond to the August run.}
	\end{center}
\end{figure}

The nuclear recoil energy distributions in the crystal for each triggered BD derived from the simulation are presented in Figures~\ref{enr_yodo} and~\ref{enr_sodio} for iodine and sodium nuclei recoils, respectively, for the BD positions corresponding to August measurements.

Three different geometrical configurations were simulated: the August configuration, the October configuration with the Yale crystal (number~3) and the October configuration with the Zaragoza crystals (crystal differences can be seen in Section~\ref{Section:QF_TUNL_Setup_Crystals}). Moreover, to evaluate the systematic uncertainties in the nuclear recoil energies (and therefore, in the QF) due to the BD position uncertainties (shown in Section~\ref{Section:QF_TUNL_Setup_BDs}), two more simulations were run for each configuration, placing the BDs in the corresponding maximum and minimum scattering angles compatible with the position uncertainties. The systematic uncertainties were calculated with the following procedure: if (x,y) is the position of the center of the face looking to the crystal of a given BD and (dx,dy) the uncertainties at 1~$\sigma$, the corresponding scattering angle is $\theta = \arctan(|y|/x)$ (following the same reference system of Section~\ref{Section:QF_TUNL_Setup_BDs}). The position corresponding to $\theta_{minimum}$ is (x+dx,y-dy), and the position corresponding to $\theta_{maximum}$ is (x-dx,y+dy). The systematic uncertainties in the nuclear recoil energies are then derived as the difference between the mean nuclear recoil energies obtained in each configuration and the mean nuclear recoil energies corresponding to the BDs in the maximum and minimum scattering angle positions for that configuration. They are presented in Table~\ref{tabla:EnrNaValues} for sodium nuclei and in Table~\ref{tabla:EnrIValues} for iodine nuclei for the three simulated configurations.

\begin{table}[h]
	\centering
	\begin{tabular}{|c|c|c|c|}
		\cline{2-4}
		\multicolumn{1}{c}{} & \multicolumn{3}{|c|}{E$_{nr}$ (keV)} \\
		\hline
		BD~$\#$ & Crystals 1 and 2 & Crystal 3 & Crystals 4 and 5 \\
		\hline
		0 & 79.70$^{+1.13}_{-0.93}$ & 81.05$^{+1.84}_{-1.32}$ & 81.16$^{+1.80}_{-1.60}$ \\
		1 & 64.44$^{+0.69}_{-0.92}$ & 48.01$^{+0.69}_{-0.68}$ & 48.41$^{+0.70}_{-0.68}$ \\
		2 & 53.11$^{+0.45}_{-0.62}$ & 33.57$^{+0.47}_{-0.97}$ & 34.43$^{+0.88}_{-0.17}$ \\
		3 & 42.17$^{+0.44}_{-0.43}$ & 25.75$^{+0.46}_{-0.75}$ & 25.59$^{+0.20}_{-0.52}$ \\
		4 & 32.26$^{+0.25}_{-0.36}$ & 17.93$^{+0.55}_{-0.72}$ & 18.03$^{+0.17}_{-0.52}$ \\
		5 & 23.28$^{+0.18}_{-0.05}$ & 11.81$^{+0.31}_{-0.36}$ & 11.78$^{+0.61}_{-0.40}$ \\
		6 & 13.69$^{+0.17}_{-0.17}$ & 7.18$^{+0.30}_{-0.32}$ & 7.10$^{+0.46}_{-0.24}$ \\
		7 & 8.58$^{+0.11}_{-0.04}$ & 4.33$^{+0.24}_{-0.09}$ & 4.25$^{+0.22}_{-0.23}$ \\
		8 & 4.67$^{+0.12}_{-0.09}$ & 1.51$^{+0.17}_{-0.02}$ & 1.24$^{+0.13}_{-0.11}$ \\
		9 & 3.19$^{+0.14}_{-0.10}$ & 1.24$^{+0.13}_{-0.03}$ & 0.97$^{+0.13}_{-0.10}$ \\
		10 & 7.53$^{+0.16}_{-0.17}$ & 3.31$^{+0.25}_{-0.14}$ & 3.21$^{+0.11}_{-0.05}$ \\
		11 & 13.94$^{+0.17}_{-0.32}$ & 6.39$^{+0.27}_{-0.04}$ & 6.27$^{+0.22}_{-0.23}$ \\
		12 & 21.68$^{+0.10}_{-0.25}$ & 11.71$^{+0.43}_{-0.31}$ & 11.97$^{+0.47}_{-0.71}$ \\
		13 & 27.68$^{+0.36}_{-0.75}$ & 18.29$^{+0.44}_{-0.44}$ & 18.24$^{+0.42}_{-0.40}$ \\
		14 & 37.51$^{+0.47}_{-0.21}$ & 26.84$^{+0.48}_{-0.85}$ & 27.01$^{+0.92}_{-0.84}$ \\
		15 & 49.11$^{+0.63}_{-0.44}$ & 36.57$^{+0.50}_{-0.66}$ & 36.95$^{+0.64}_{-0.51}$ \\
		16 & 61.53$^{+0.83}_{-1.07}$ & 53.04$^{+0.81}_{-0.81}$ & 53.12$^{+0.59}_{-1.13}$ \\
		17 & 77.16$^{+1.05}_{-1.16}$ & 71.55$^{+0.81}_{-0.83}$ & 71.61$^{+1.04}_{-1.22}$ \\
		\hline
	\end{tabular} \\
	\caption{Mean sodium recoil energies for the three simulated configurations: August BD positions (crystals~1 and~2), October BD positions with Yale crystal (crystal~3) and October BD positions with Zaragoza crystals (crystals~4 and~5). Systematic errors associated to the uncertainties in the BD positions are shown.}
	\label{tabla:EnrNaValues}
\end{table}

\begin{table}[h]
	\centering
	\begin{tabular}{|c|c|c|c|}
		\cline{2-4}
		\multicolumn{1}{c}{} & \multicolumn{3}{|c|}{E$_{nr}$ (keV)} \\
		\hline
		BD~$\#$ & Crystals 1 and 2 & Crystal 3 & Crystals 4 and 5 \\
		\hline
		0 & 14.18$^{+0.14}_{-0.06}$ & 14.24$^{+0.18}_{-0.11}$ & 14.22$^{+0.38}_{-0.17}$ \\
		1 & 10.90$^{+0.07}_{-0.22}$ & 8.73$^{+0.10}_{-0.08}$ & 8.71$^{+0.27}_{-0.02}$ \\
		2 & 6.03$^{+0.16}_{-0.35}$ & 6.02$^{+0.13}_{-0.04}$ & 5.96$^{+0.08}_{-0.03}$ \\
		3 & 4.96$^{+0.11}_{-0.18}$ & 4.66$^{+0.13}_{-0.10}$ & 4.65$^{+0.05}_{-0.04}$ \\
		4 & 3.42$^{+0.19}_{-0.10}$ & 3.22$^{+0.08}_{-0.08}$ & 3.19$^{+0.07}_{-0.08}$ \\
		5 & 3.08$^{+0.06}_{-0.12}$ & 2.13$^{+0.08}_{-0.06}$ & 2.10$^{+0.07}_{-0.04}$ \\
		6 & 2.41$^{+0.02}_{-0.02}$ & 1.33$^{+0.04}_{-0.05}$ & 1.33$^{+0.02}_{-0.07}$ \\
		7 & 1.55$^{+0.01}_{-0.01}$ & 0.78$^{+0.02}_{-0.03}$ & 0.79$^{+0.03}_{-0.04}$ \\
		8 & 0.80$^{+0.01}_{-0.01}$ & 0.25$^{+0.02}_{-0.02}$ & 0.25$^{+0.02}_{-0.02}$ \\
		9 & 0.55$^{+0.02}_{-0.02}$ & 0.20$^{+0.01}_{-0.01}$ & 0.20$^{+0.01}_{-0.01}$ \\
		10 & 1.35$^{+0.04}_{-0.04}$ & 0.58$^{+0.02}_{-0.04}$ & 0.60$^{+0.01}_{-0.03}$ \\
		11 & 2.44$^{+0.03}_{-0.03}$ & 1.15$^{+0.04}_{-0.06}$ & 1.17$^{+0.02}_{-0.04}$ \\
		12 & 3.46$^{+0.01}_{-0.08}$ & 2.08$^{+0.05}_{-0.08}$ & 2.07$^{+0.07}_{-0.07}$ \\
		13 & 4.35$^{+0.13}_{-0.16}$ & 3.31$^{+0.10}_{-0.08}$ & 3.28$^{+0.06}_{-0.07}$ \\
		14 & 6.20$^{+0.01}_{-0.02}$ & 4.77$^{+0.13}_{-0.11}$ & 4.72$^{+0.08}_{-0.03}$ \\
		15 & 5.91$^{+0.26}_{-0.05}$ & 6.45$^{+0.19}_{-0.08}$ & 6.35$^{+0.10}_{-0.07}$ \\
		16 & 10.50$^{+0.12}_{-0.02}$ & 9.43$^{+0.23}_{-0.05}$ & 9.34$^{+0.09}_{-0.03}$ \\
		17 & 13.99$^{+0.08}_{-0.25}$ & 13.39$^{+0.09}_{-0.09}$ & 13.51$^{+0.16}_{-0.35}$ \\
		\hline
	\end{tabular} \\
	\caption{Mean iodine recoil energies for the three simulated configurations: August BD positions (crystals~1 and~2), October BD positions with Yale crystal (crystal~3) and October BD positions with Zaragoza crystals (crystals~4 and~5). Systematic errors associated to the uncertainties in the BD positions are shown.}
	\label{tabla:EnrIValues}
\end{table}

\subsubsection{Crystal calibrations}\label{Section:QF_GEANT4_Results_Calibrations}

\begin{figure}[h]
	\begin{center}
		\includegraphics[width=0.75\textwidth]{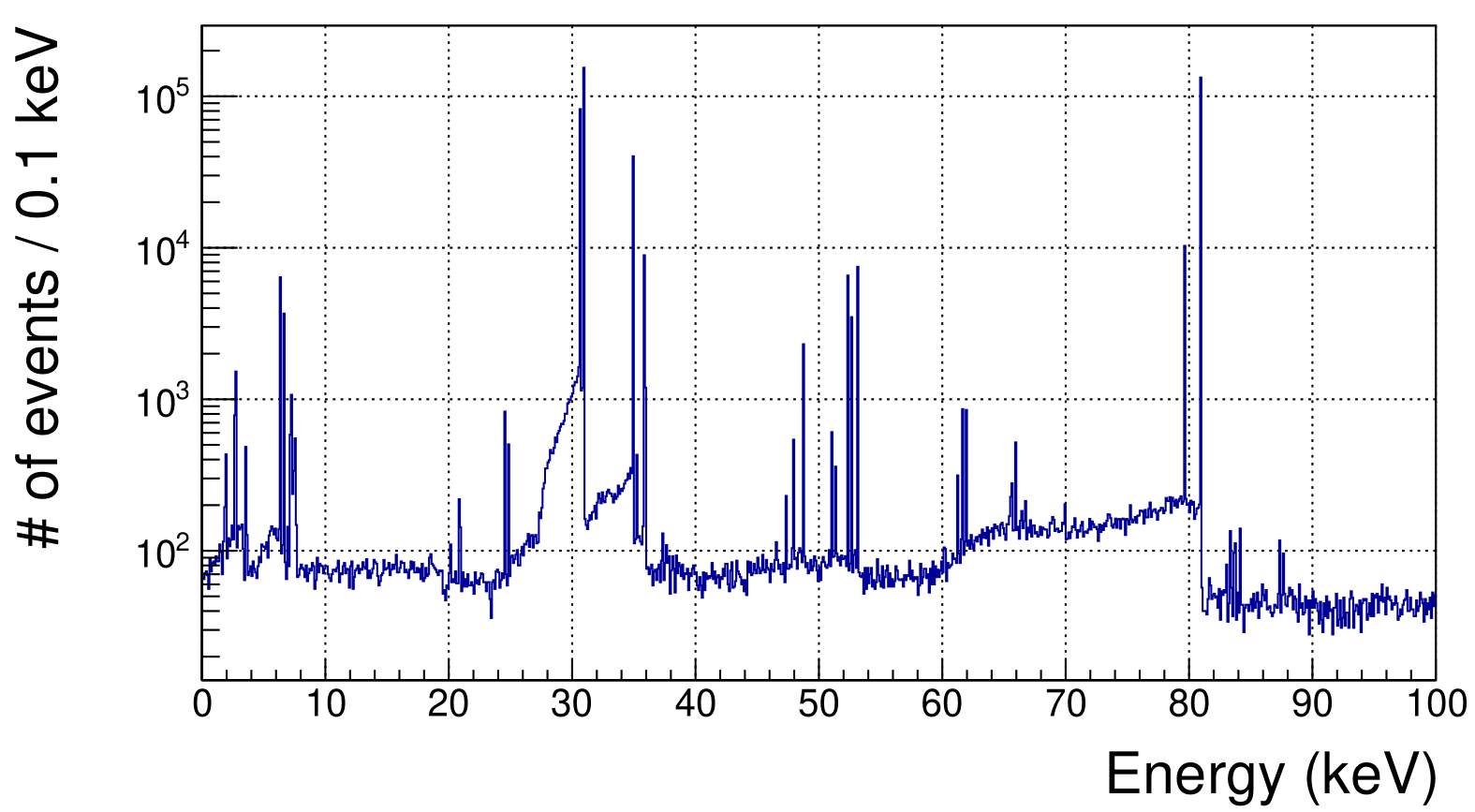}
		\caption{\label{CalibrationSimBa133}Simulated spectra of the NaI(Tl) crystal energy calibration with the $^{133}Ba$ source before convolving with the detector energy resolution.}
	\end{center}
\end{figure}

\begin{figure}[h]
	\begin{center}
		\includegraphics[width=0.75\textwidth]{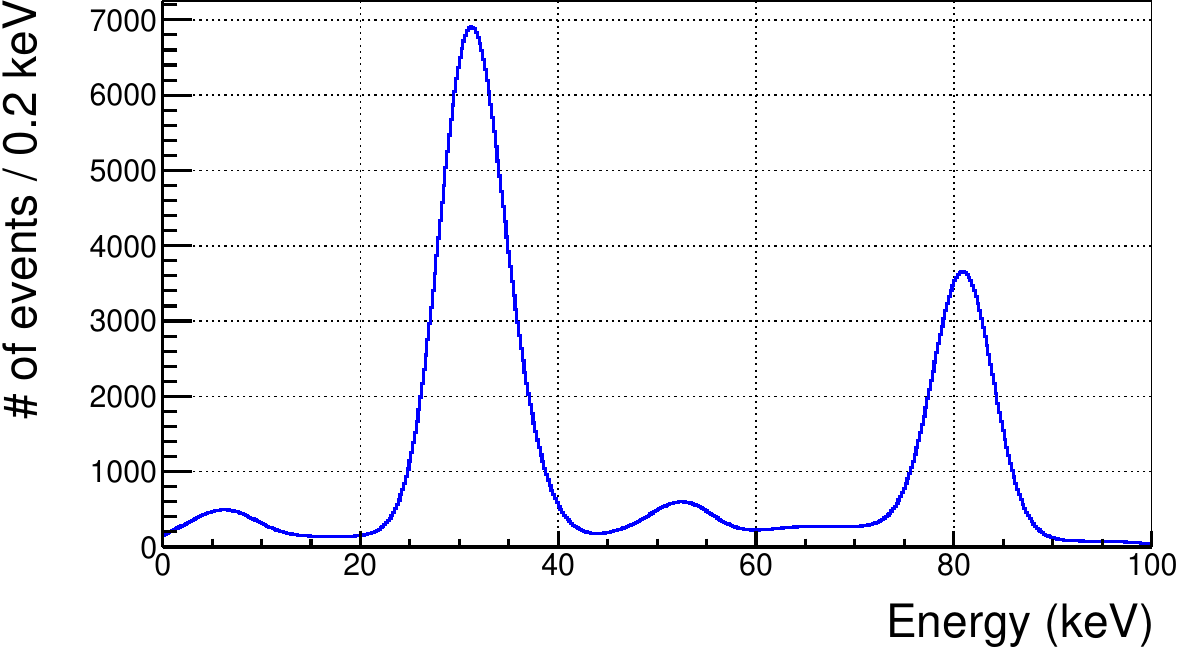}
		\caption{\label{Ba133_simConv_spectum}$^{133}Ba$ energy spectrum simulated for crystal~5 geometry and convolved with a Gaussian of $\sigma$ = 3~keV, independent on the energy.}
	\end{center}
\end{figure}

One of the objectives of the simulation was to obtain the spectra of energy depositions during the crystal calibrations with $^{133}Ba$, in order to improve the calibration in energy. The source was defined as point-like and non-encapsulated and placed at the same position for all the experimental configurations, as indicated in Section~\ref{Section:QF_TUNL}. The obtained spectrum is shown in Figure~\ref{CalibrationSimBa133}. To take into account the detector resolution, this spectrum was convolved with a Gaussian of $\sigma$ = 3~keV, independent on the energy. The corresponding spectrum is displayed in Figure~\ref{Ba133_simConv_spectum}.

This crystal calibration simulation will be used in Section~\ref{Section:QF_Analysis_NaIcal_EnergyCal} to obtain the averaged energy depositions of the observed lines. The main emissions from $^{133}Ba$ source are shown in Table~\ref{tabla:Ba133_Xray} and Table~\ref{tabla:Ba133_gamma}, for x-rays and gammas, respectively, obtained from~\cite{TabRad}. $^{133}Ba$ isotope has a half-life of 10.5~years and decays to $^{133}Cs$ through electron capture. The x-ray emissions of 30.63~keV and higher energies are identified in the simulated spectra, but it is not the case of the L-shell x-rays at around 4~keV, that will be absorbed in the crystal housing. The peak with energy below 10~keV observed in the simulated spectrum was identified as the iodine K-shell x-ray escape after the photoelectric absorption in the NaI(Tl) crystal of the $\approx$~35~keV x-rays, and according to the simulation results, has a mean energy of 6.6~keV.

\begin{table}[h]
	\centering
	\begin{tabular}{|c|c|c|}
		\hline
		Energy (keV) & BR($\%$) & Assignement \\
		\hline
		4.29 & 6.0 & Cs L$_{\alpha1}$ \\
		4.62 & 3.8 & Cs L$_{\beta1}$ \\
		4.93 & 1.2 & Cs L$_{\beta2}$ \\
		30.63 & 34.9 & Cs K$_{\alpha2}$ \\
		30.97 & 64.5 & Cs K$_{\alpha1}$ \\
		34.92 & 6.0 & Cs K$_{\beta3}$ \\
		34.99 & 11.6 & Cs K$_{\beta1}$ \\
		35.82 & 3.6 & Cs K$_{\beta2}$ \\
		\hline
	\end{tabular} \\
	\caption{X-ray emissions from the $^{133}Ba$ source with a branching ratio (BR) higher than 1$\%$~\cite{TabRad}.}
	\label{tabla:Ba133_Xray}
\end{table}

\begin{table}[h]
	\centering
	\begin{tabular}{|c|c|c|}
		\hline
		Energy (keV) & BR($\%$) \\
		\hline
		53.2 & 2.2 \\
		79.6 & 2.6 \\
		81.0 & 34.1 \\
		276.4 & 7.2 \\
		302.9 & 18.3 \\
		356.0 & 62.1 \\
		383.9 & 8.9 \\
		\hline
	\end{tabular} \\
	\caption{Gamma emissions from the $^{133}Ba$ source with a branching ratio (BR) higher than 1$\%$~\cite{TabRad}.}
	\label{tabla:Ba133_gamma}
\end{table}

\begin{figure}[]
	\begin{center}
		\includegraphics[width=0.75\textwidth]{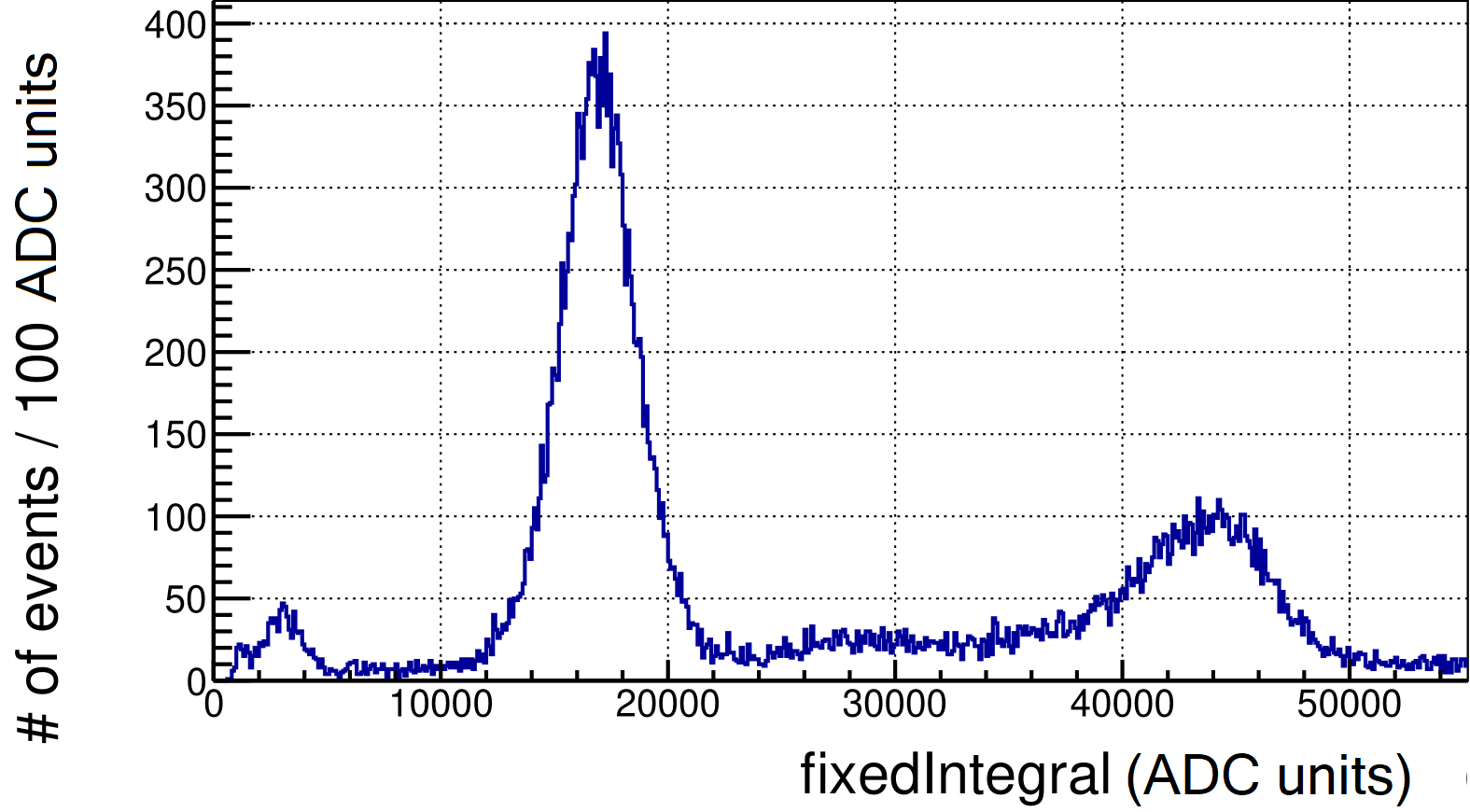}
		\caption{\label{Ba133_exp_spectum}$^{133}Ba$ calibration spectrum measured for crystal~5.}
	\end{center}
\end{figure}

Figure~\ref{Ba133_exp_spectum} shows the experimental measurement for the calibration of crystal~5 with $^{133}Ba$ in terms of the \textit{fixedIntegral} variable. The comparison between the simulation shown in Figure~\ref{Ba133_simConv_spectum} and the measurement allows to identify and associate mean energies to the three structures observed in the measurement: 6.6, 30.9~and 35.1~keV. 

\section{Data Analysis}\label{Section:QF_Analysis}

\fancyhead[RO]{\emph{\thesection. \nameref{Section:QF_Analysis}}}

To analyze possible systematics derived from the analysis protocols, two different analysis codes were developed. In this section it is described the analysis developed by the ANAIS group at the University of Zaragoza. Information about the analysis developed by other members of the collaboration can be found in~\cite{ThesisSam}. Compatible results for the QF were derived from both analysis procedures.

In this section, first, the neutron selection procedure developed is presented, and then, the calibration protocol of the NaI(Tl) data, consisting first in the correction of possible gain drifts using the $^{127}I$ inelastic peak observed in the beam-on measurements and later, on the electron equivalent energy calibration of the ROI using the measurements with $^{133}Ba$.

\subsection{Neutrons selection} \label{Section:QF_Analysis_Selection_NeutronsBDs}

\begin{figure}[h!]
	\begin{subfigure}[b]{0.49\textwidth}
		\includegraphics[width=\textwidth]{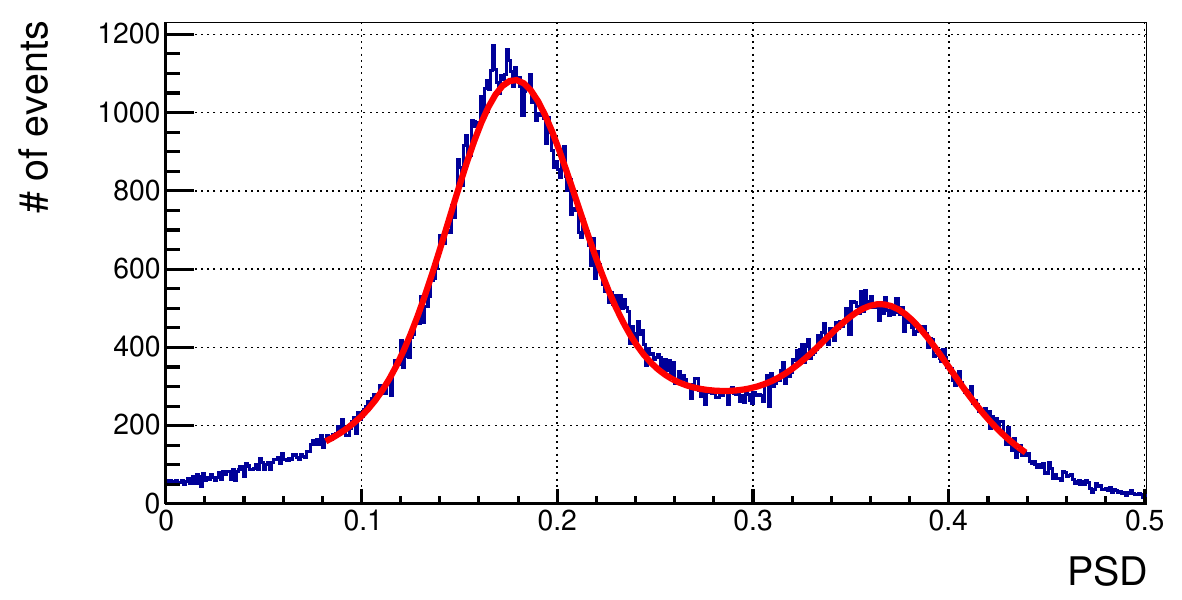}
	\end{subfigure}
	\begin{subfigure}[b]{0.49\textwidth}
		\includegraphics[width=\textwidth]{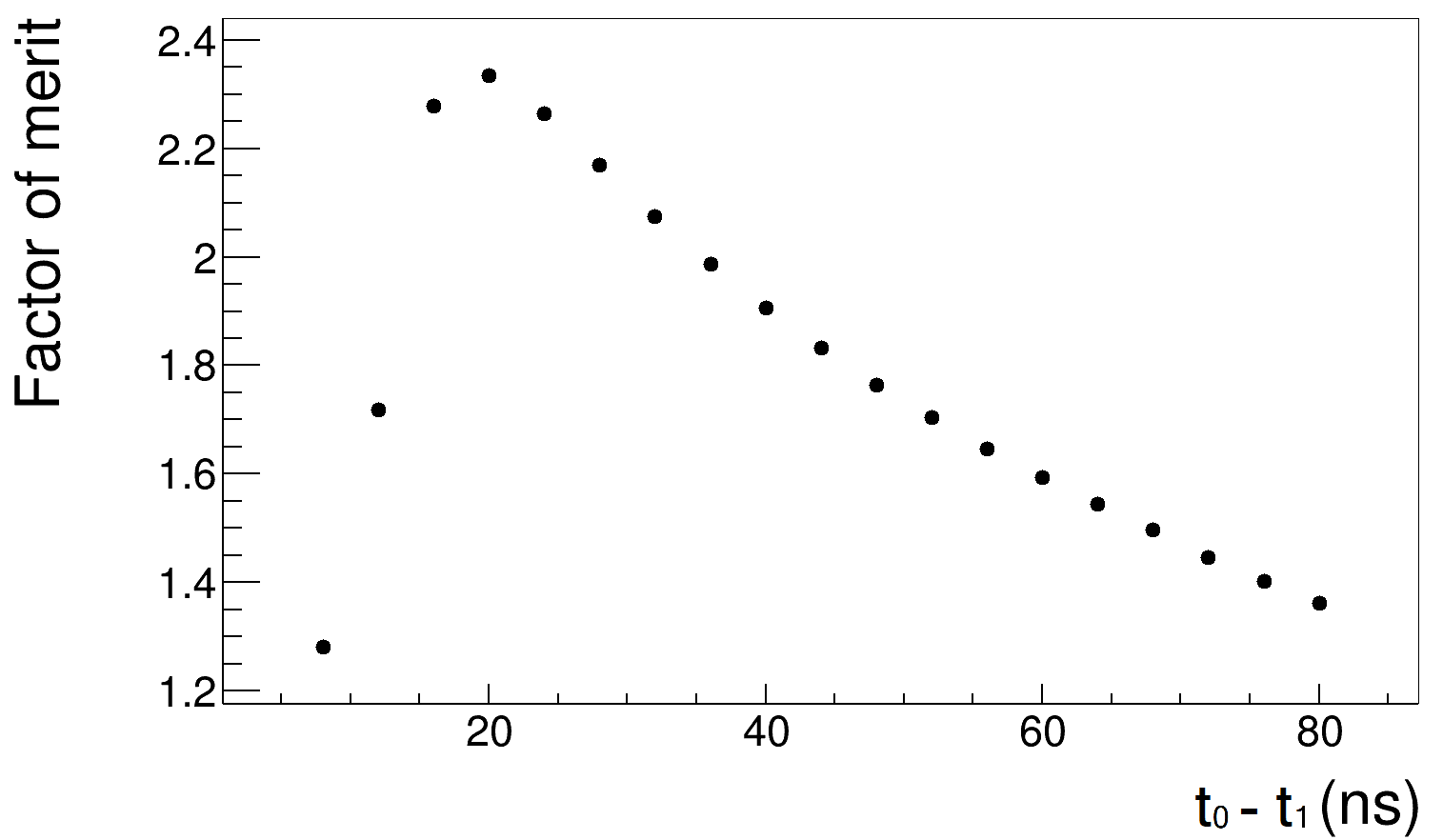}
	\end{subfigure}
	\caption{\label{factor_merit}\textit{PSD} histogram for $10^5$~events of any triggered BD using $t_1$ = $t_0$+20~ns being clear the discrimination between neutrons (higher \textit{PSD} value) and gammas (left plot), and factor of merit for different $t_1$ values (right plot).}
\end{figure}

One of the most important issues in the QF analysis is the development of a robust protocol for the identification of the neutrons that reach a BD after scattering in the NaI(Tl) crystal against the different backgrounds present in the data. Some of those backgrounds were well-identified and therefore could be rejected. In this section we revise the different events selection protocols applied to discriminate neutron events from background. As it was briefly explained before, neutrons can be identified in a BD through the pulse shape analysis. In order to obtain the best discrimination between neutron interactions and electron/gamma events, the \textit{PSD} parameter was defined in Section~\ref{Section:QF_Analysis_WaveformRec_BDs}. The start time of the pulse tail integration $t_1$ was not fixed in the definition, and it had to be optimized. To do that, the \textit{PSD} distributions were obtained for $10^5$~events of any triggered BD, and the two populations obtained (corresponding to neutrons and electron/gamma events) were fitted to gaussians (see left plot in Figure~\ref{factor_merit}). It was done for $t_1$ values ranging from 8 to 80~ns, and a factor of merit, $p$, was defined as:
\begin{equation}
	p = \frac{\mu_N-\mu_{\gamma}}{\sigma_N+\sigma_{\gamma}},
\end{equation}
where $\mu_N$ and $\mu_{\gamma}$ are the mean values of the \textit{PSD} variable of neutron and gamma distributions, and $\sigma_N$ and $\sigma_{\gamma}$ their standard deviations obtained from the fit. The parameter $p$ helps to select the most discriminating value of $t_1$. As it is observed in the right plot of the Figure~\ref{factor_merit}, the discrimination improves while reducing $t_1$ from 80~ns down to 20~ns. This is the value chosen for all the following analysis.

\begin{figure}[h!]
	\begin{center}
		\includegraphics[width=0.75\textwidth]{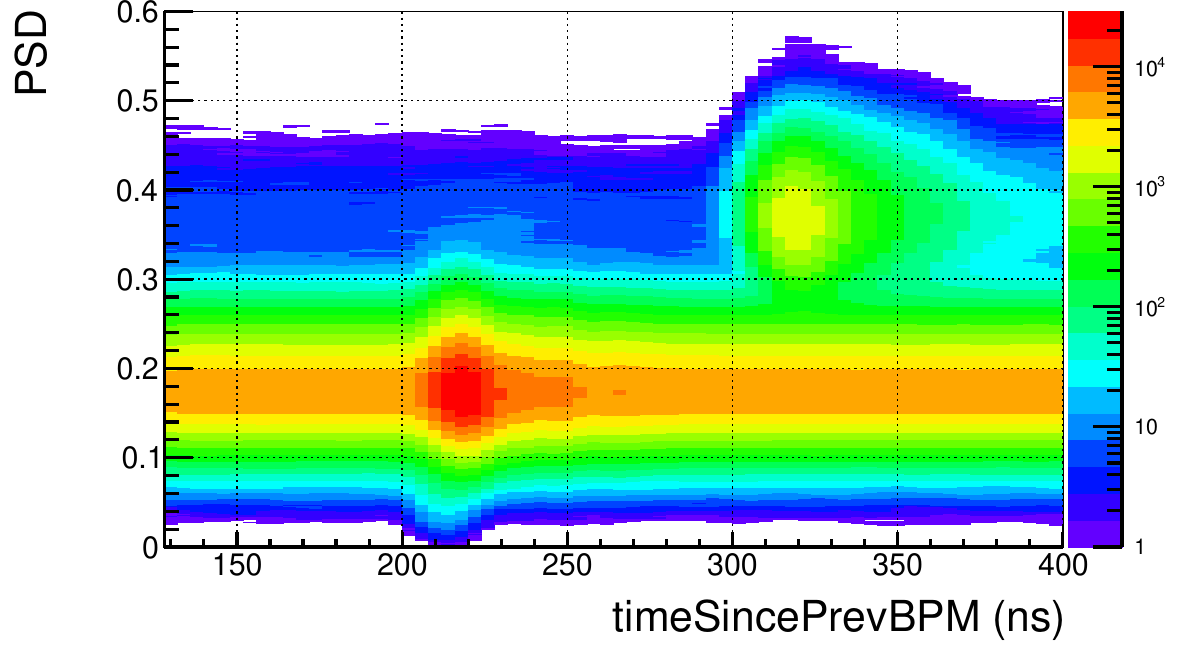}
		\caption{\label{psd_tspb}\textit{PSD} vs \textit{timeSincePrevBPM} corresponding to all the BD waveforms for the measurement of crystal~5. Neutrons correlated with the beam are clearly identified (\textit{PSD}$>$0.3 and \textit{timeSincePrevBPM}$>$300~ns).}
	\end{center}
\end{figure}

\begin{figure}[h!]
	\begin{center}
		\includegraphics[width=0.75\textwidth]{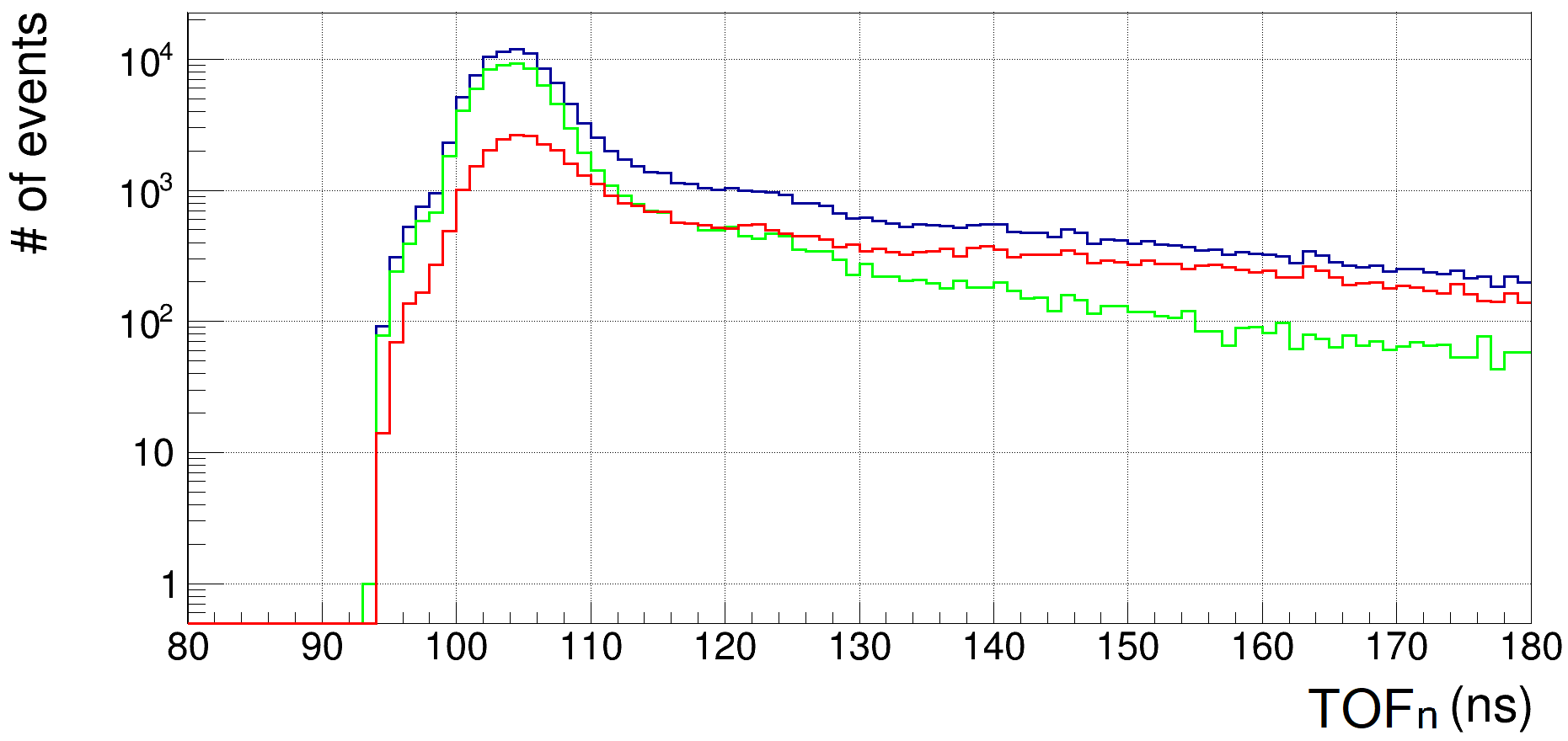}
		\caption{\label{t0bdn}Time of arrival of the neutrons to the BDs from simulation for single scattered neutrons (green), multiple scattered neutrons (red) and total (blue).}
	\end{center}
\end{figure}

The \textit{PSD} variable is important in the neutron selection, but discrimination power can be further improved adding a cut in TOF, through the \textit{timeSincePrevBPM} variable. Three different populations can be identified in Figure~\ref{psd_tspb}, representing the \textit{PSD} parameter vs \textit{timeSincePrevBPM}: non-beam correlated gammas/electrons (flat \textit{timeSincePrevBPM} distribution with \textit{PSD} around 0.18), and two beam-correlated populations: short TOF events (\textit{PSD}~$\sim$~0.18 and \textit{timeSincePrevBPM}~$\sim$~220~ns) are gammas generated in the LiF target by protons, and long TOF events (\textit{PSD}~$\sim$~0.35 and \textit{timeSincePrevBPM}~$\sim$~320~ns) are neutrons. As result of this analysis, the neutron selection criterion chosen was: 0.3$<$\textit{PSD}$<$0.5. Once applied this selection and analyzed the \textit{timeSincePrevBPM} distribution, a tail at long time values of the TOF distribution is observed (see Figure~\ref{psd_tspb}). It could be produced by neutron multiple scattering, as it would reduce the energy of the neutron reaching the BD and then, increase the TOF. This hypothesis was checked using simulation data. Figure~\ref{t0bdn} shows an histogram of the time of arrival of neutrons to the BD obtained from the simulation, using as time origin the neutron generation time.

It was found that the contribution of multiple scattering becomes dominant for $TOF_n$ above 25~ns after the most probable value. Therefore, multiple scattered neutrons contaminating the selected data can be reduced by taking an upper cut in \textit{timeSincePrevBPM}. By selecting those events having a value of this variable below the most probable value +25~ns, the percentage of single scattered neutrons increases from 68\% to 80\% according to the simulations. Additionally, in the $TOF_n$ distribution obtained from the simulation (see Figure~\ref{t0bdn}) is observed that there is a minimum $TOF_n$, which is $\sim$~10~ns below the most probable value. Before this time, events are not expected to be produced by beam neutrons. As the most probable value of \textit{timeSincePrevBPM} for neutrons in the experimental data is $\sim$~315~ns, the criterion chosen is 304~ns$\leq$\textit{timeSincePrevBPM}$\leq$340~ns. The same selection was applied for all the channels, and the single scattering acceptance efficiency calculated with the simulation was 80\%.

\begin{figure}[h!]
	\begin{center}
		\includegraphics[width=\textwidth]{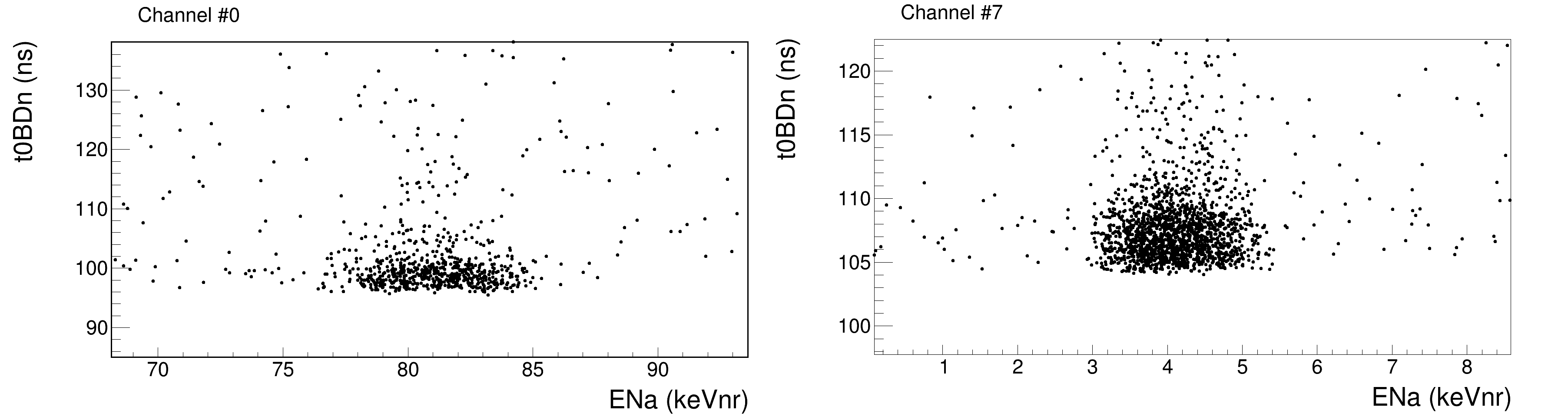}
		\caption{\label{t0bdn_vs_ena}Time of the first energy deposition by neutrons in BDs vs the sodium nuclei recoil energy produced by the same neutron, for channels $\#$~0 and $\#$~7. No correlation between both variables is observed.}
	\end{center}
\end{figure}

The event selection using TOF is an indirect neutron energy selection. The dependency of the sodium nuclei recoil energies with the TOF was analyzed using the simulation data. In Figure~\ref{t0bdn_vs_ena}, which represents the time of the first energy deposit in the BDs versus the nuclear recoil energies in the NaI(Tl) crystal, it is not observed any correlation between both variables, which implies that a cut in TOF would not affect the shape of the NR energy distributions.

There are other methods to improve the neutron identification, for instance, using the energy released by the neutron in the BD. Nuclear recoil energy spectra in the BD can be simulated in order to find optimal values of minimum and maximum deposited energy in the BD to discriminate neutron events from other backgrounds, but the limited knowledge of QF for the liquid scintillators introduces a very relevant systematic contribution when comparing the information drawn from the simulation with the experimental measurements. For this reason, although these distributions were studied and analysed, no selection based on the energy measured at the BD was applied.

\begin{figure}[h!]
	\begin{center}
		\includegraphics[width=0.75\textwidth]{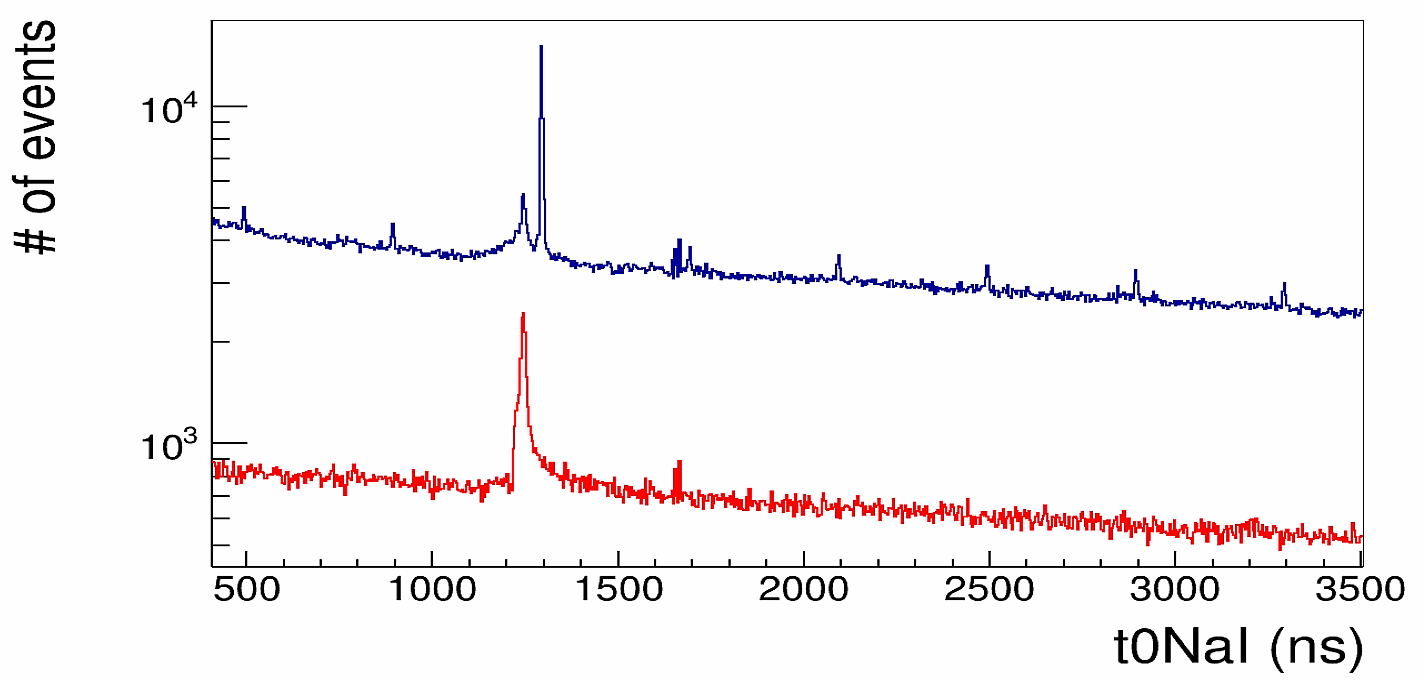}
		\caption{\label{t0NaI2}Comparison of the distribution of the \textit{t0NaI} variable without (blue line) and with (red line) neutron event selection in BDs for crystal~5.}
	\end{center}
\end{figure}

\begin{figure}[h!]
	\begin{center}
		\includegraphics[width=0.75\textwidth]{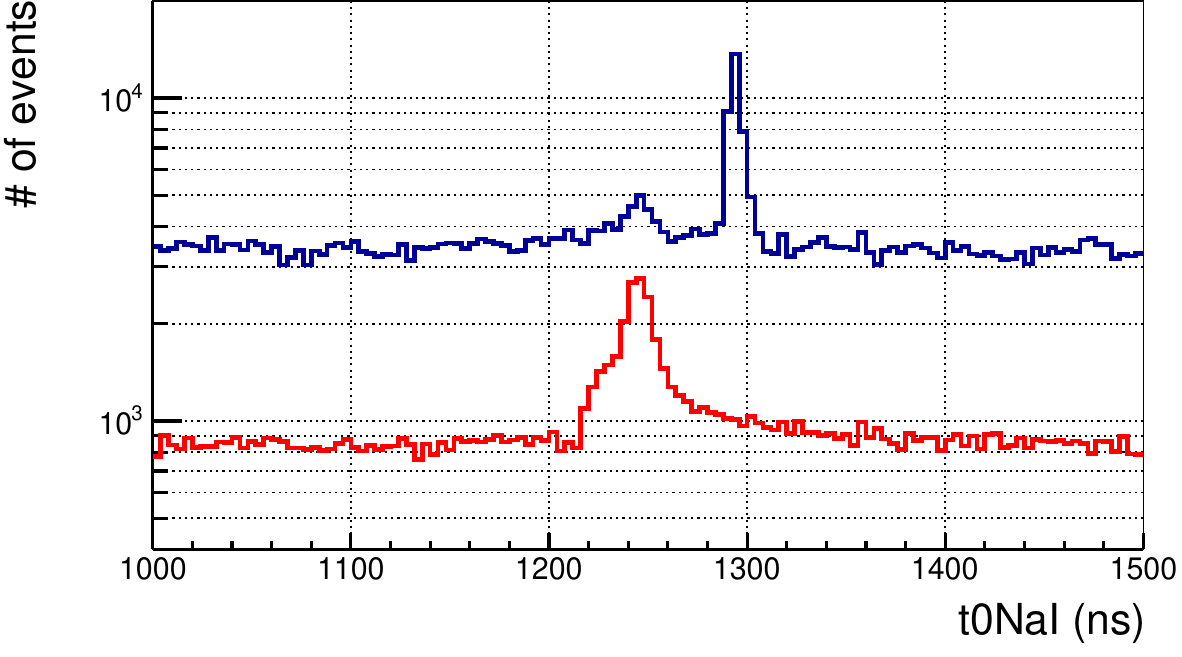}
		\caption{\label{t0NaI3}Comparison of the distribution of the \textit{t0NaI} variable without (blue line) and with (red line) neutron event selection in BDs for crystal~5.}
	\end{center}
\end{figure}

Once we have identified that the BD trigger was produced by a neutron, the next step is to  determine the onset in the NaI(Tl) waveform that corresponds to the nuclear recoil associated to that neutron. This is done using \textit{t0NaI} (variable defined in Section~\ref{Section:QF_Analysis_WaveformRec_NaI}) for neutron events in the BDs. Figure~\ref{t0NaI2} shows the \textit{t0NaI} distribution before and after applying the neutron event selection, while Figure~\ref{t0NaI3} shows a closer view of the distribution of this variable after a neutron event selection in BDs for crystal~5. Non beam-correlated events had a contribution at any time. Beam-correlated events show the same periodicity of the beam (400~ns). Rate is fully dominated by gamma/electron events. After removing them, only one clear peak in the \textit{t0NaI} distribution remains.

As the trigger was done in one of the BDs, this detector fixed the trigger time, and thus, crystal events are observed back in time. That is the reason why events that were triggered in BDs by beam-correlated gammas appear in the crystal signal between 50 and 100~ns after the beam-correlated neutrons in Figure~\ref{t0NaI2}: this time is the TOF difference between gammas and neutrons from the crystal to the BD. This analysis revealed that the events with neutrons in both crystal and BD appeared at about 1200~ns from the beginning of the waveform. In order to avoid threshold effects affecting the \textit{t0NaI} variable, we decided to integrate the crystal signal for every BD trigger using a fixed time interval: from 1200~ns to 3200~ns (\textit{fixedIntegral} variable, defined in Section~\ref{Section:QF_Analysis_WaveformRec_NaI}).

\begin{figure}[h!]
	\begin{subfigure}[b]{0.55\textwidth}
		\includegraphics[width=\textwidth]{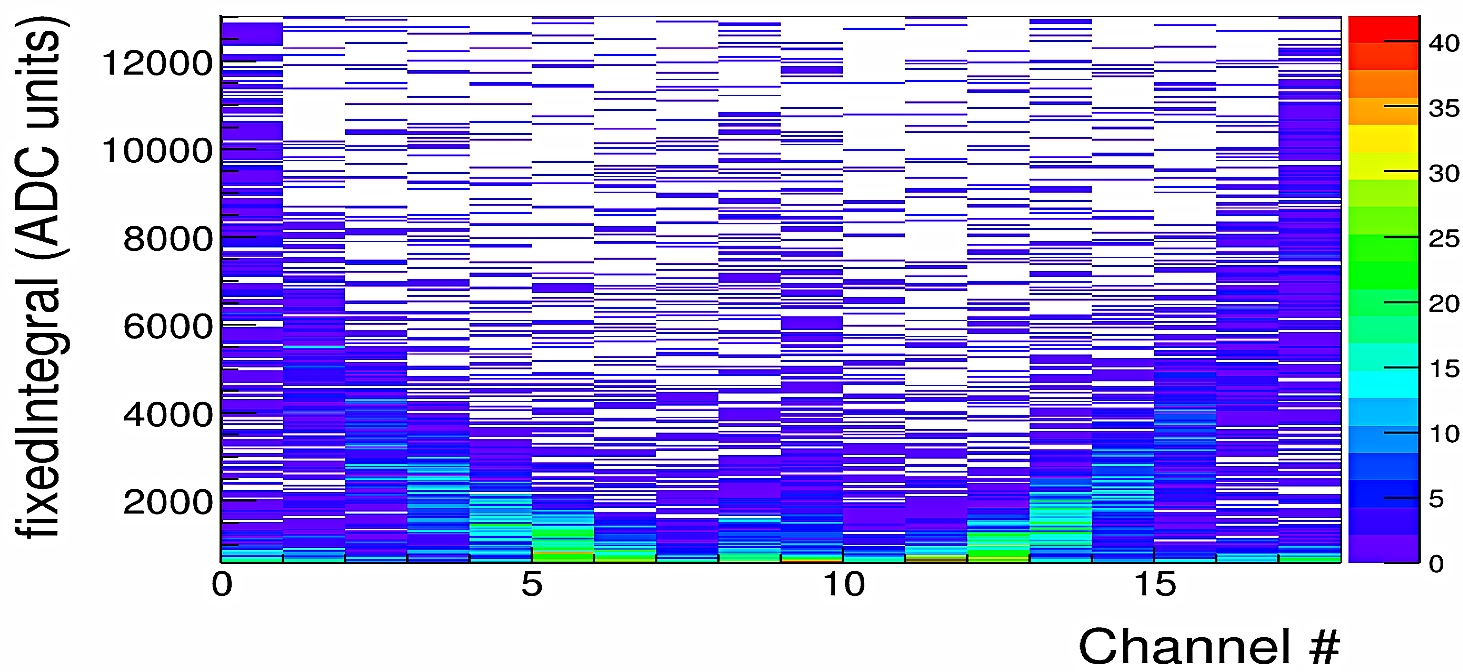}
	\end{subfigure}
	\begin{subfigure}[b]{0.44\textwidth}
		\includegraphics[width=\textwidth]{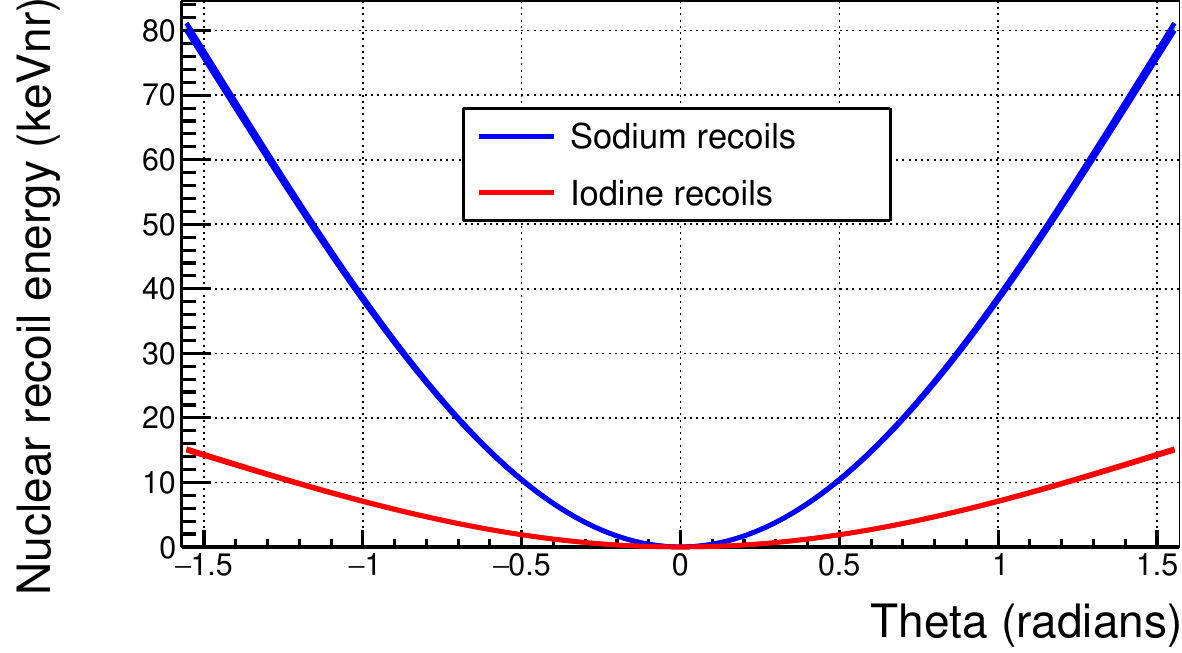}
	\end{subfigure}
	\caption{\label{integralNaI_channel}Left plot: \textit{fixedIntegral} distribution for crystal~5 as a function of the channel number. Right plot: nuclear recoil energies as a function of the scattering angle for sodium (blue) and iodine (red) recoils, obtained from Equation~\ref{Equation_Enr}.}
\end{figure}

After applying this selection procedure, sodium nuclei recoils could be observed in the \textit{fixedIntegral} spectrum in the highest scattering angle channels, which correspond to the highest nuclear recoil energies (see Figure~\ref{integralNaI_channel}). At the small scattering angle channels (between $\#$~7 and $\#$~10), recoils cannot be distinguished from the background. In the case of iodine recoils even at high scattering angle channels they could not be identified among other backgrounds.

With respect to the NaI(Tl) signal, a relevant background is the presence of individual photoelectrons from the dark current and other fast events with origin in the PMT. In order to discriminate those background events from bulk scintillation events, specific Pulse Shape Analysis techniques can be developed. However, as it was explained in Section~\ref{Section:QF_NaIQFOverview}, they would introduce a threshold effect in the selection of crystal events, which is something that we want to avoid. Then, we do not apply any PSA for the crystal events selection.

\subsection{NaI(Tl) Calibration}\label{Section:QF_Analysis_NaIcal}

\subsubsection{Stability study and gain drift correction}\label{Section:QF_Analysis_NaIcal_GainCorrection}

\begin{figure}[h!]
	\begin{center}
		\includegraphics[width=0.75\textwidth]{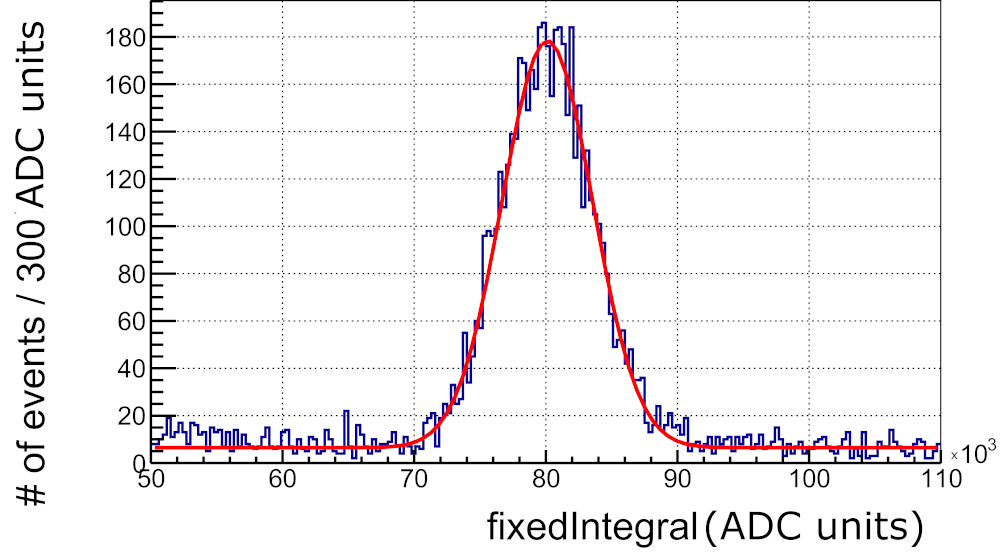}
		\caption{\label{57keVpeak}57.6~keV peak fit for crystal~1 in 1~hour measurement.}
	\end{center}
\end{figure}

\begin{figure}[h!]
	\begin{subfigure}[b]{0.49\textwidth}
		\includegraphics[width=\textwidth]{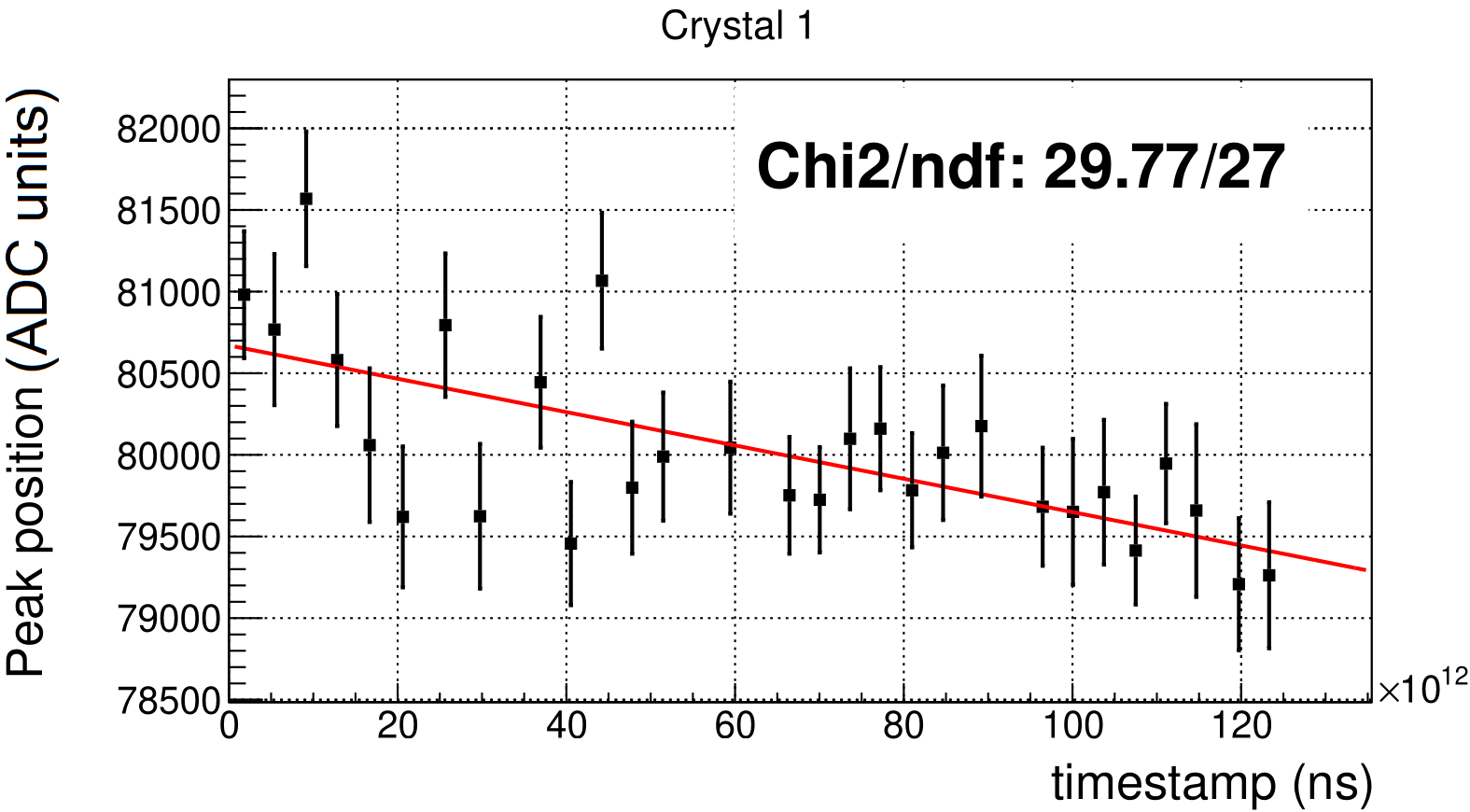}
	\end{subfigure}
	\begin{subfigure}[b]{0.49\textwidth}
		\includegraphics[width=\textwidth]{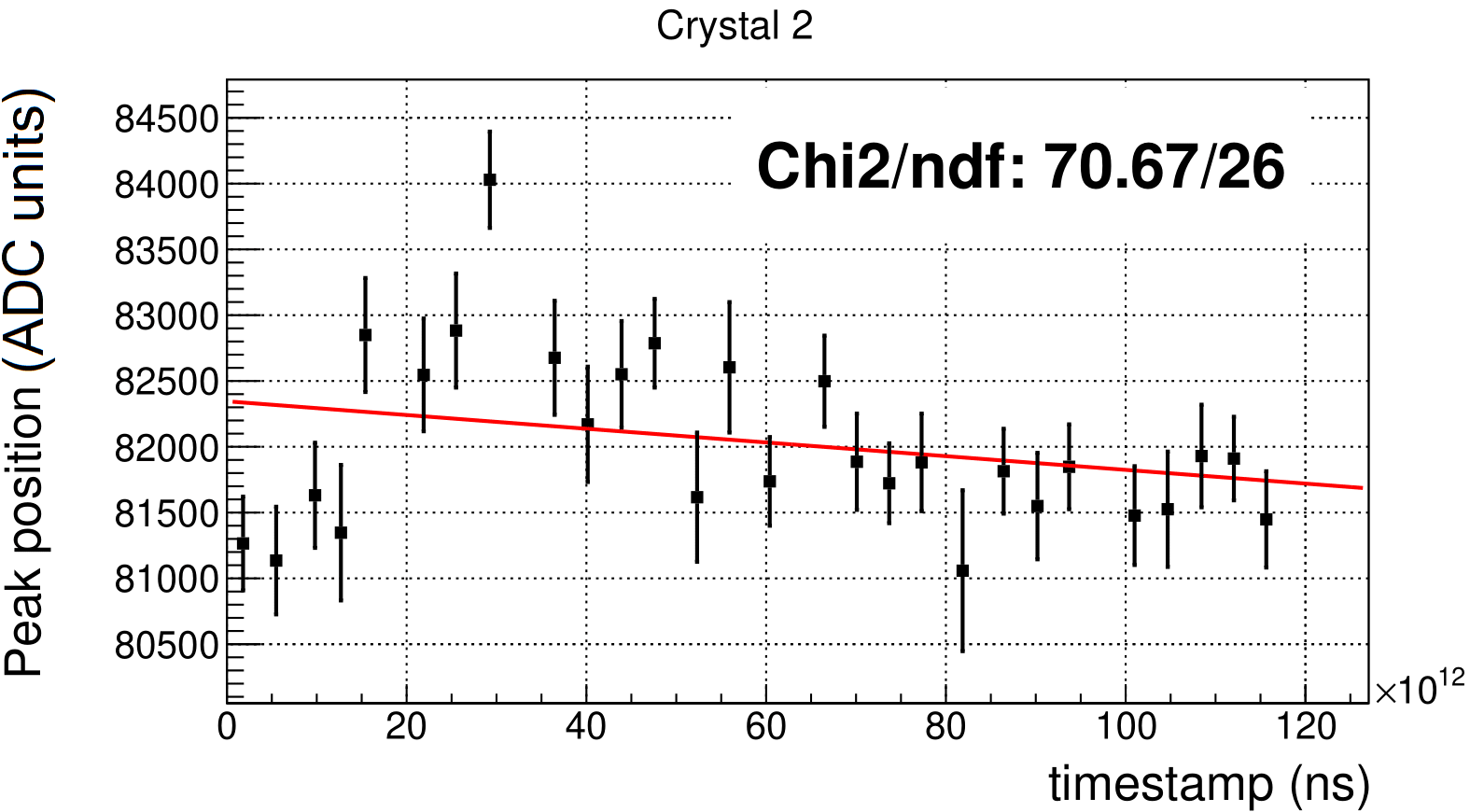}
	\end{subfigure}
	\begin{subfigure}[b]{0.49\textwidth}
		\includegraphics[width=\textwidth]{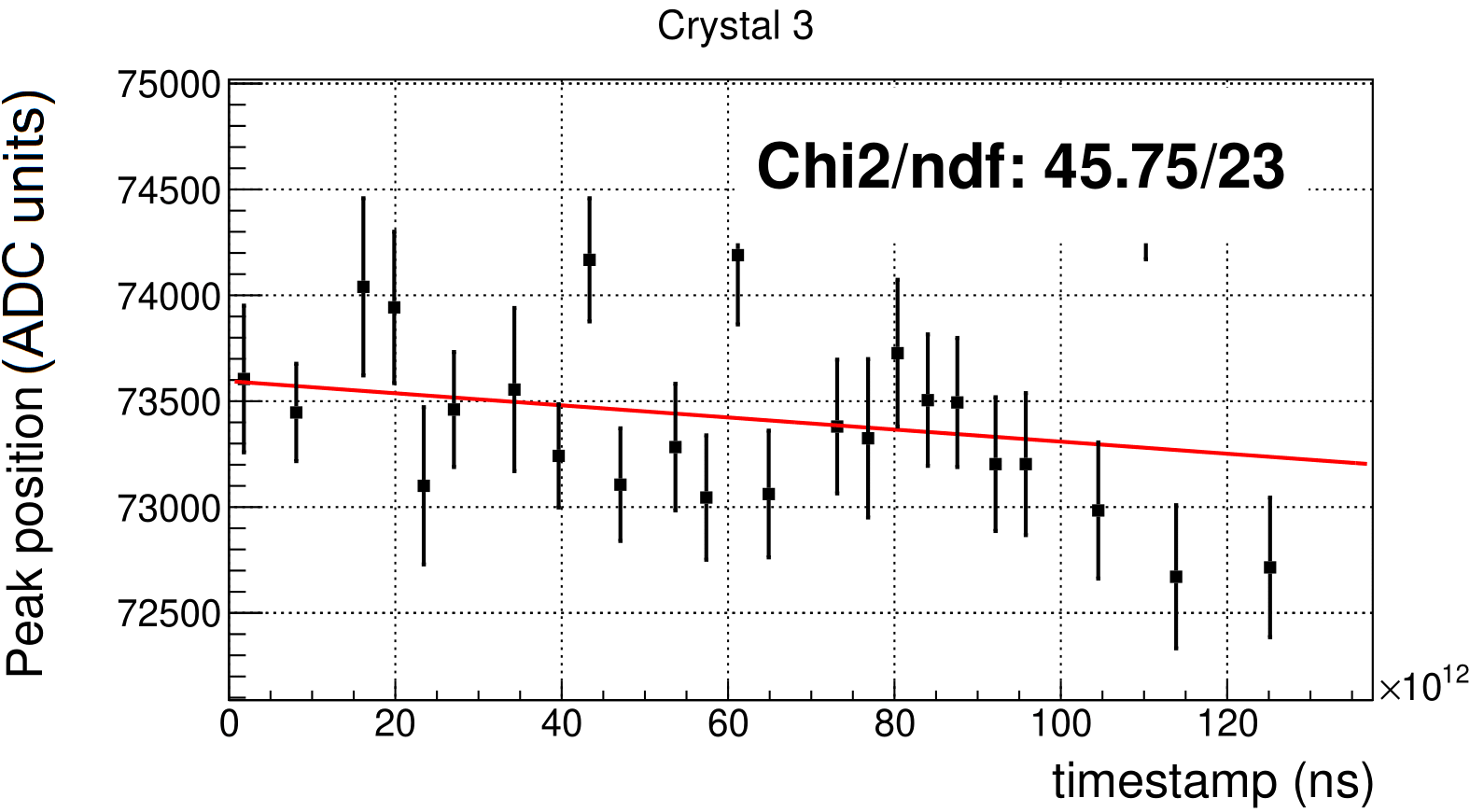}
	\end{subfigure}
	\begin{subfigure}[b]{0.49\textwidth}
		\includegraphics[width=\textwidth]{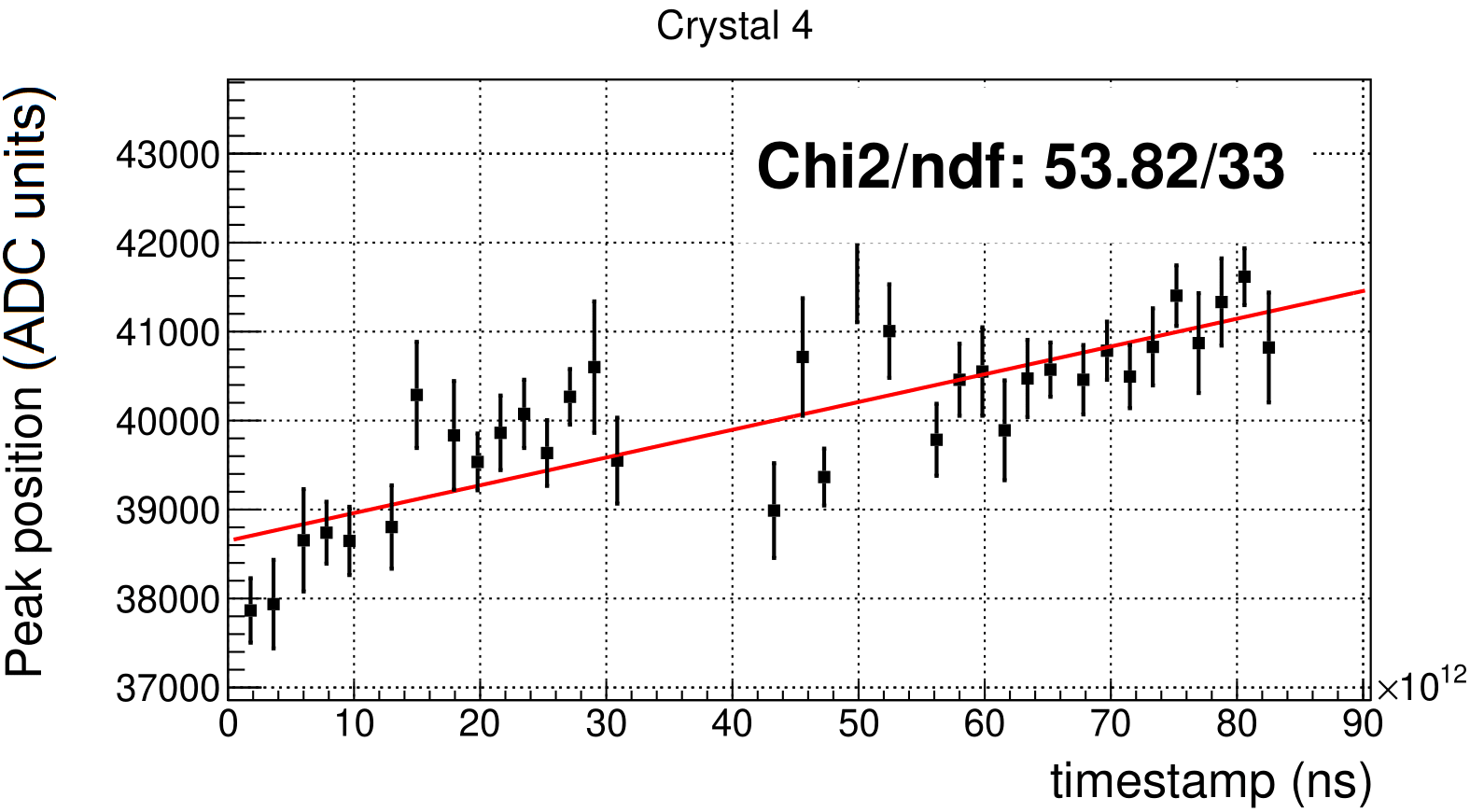}
	\end{subfigure}
	\begin{subfigure}[b]{0.49\textwidth}
		\includegraphics[width=\textwidth]{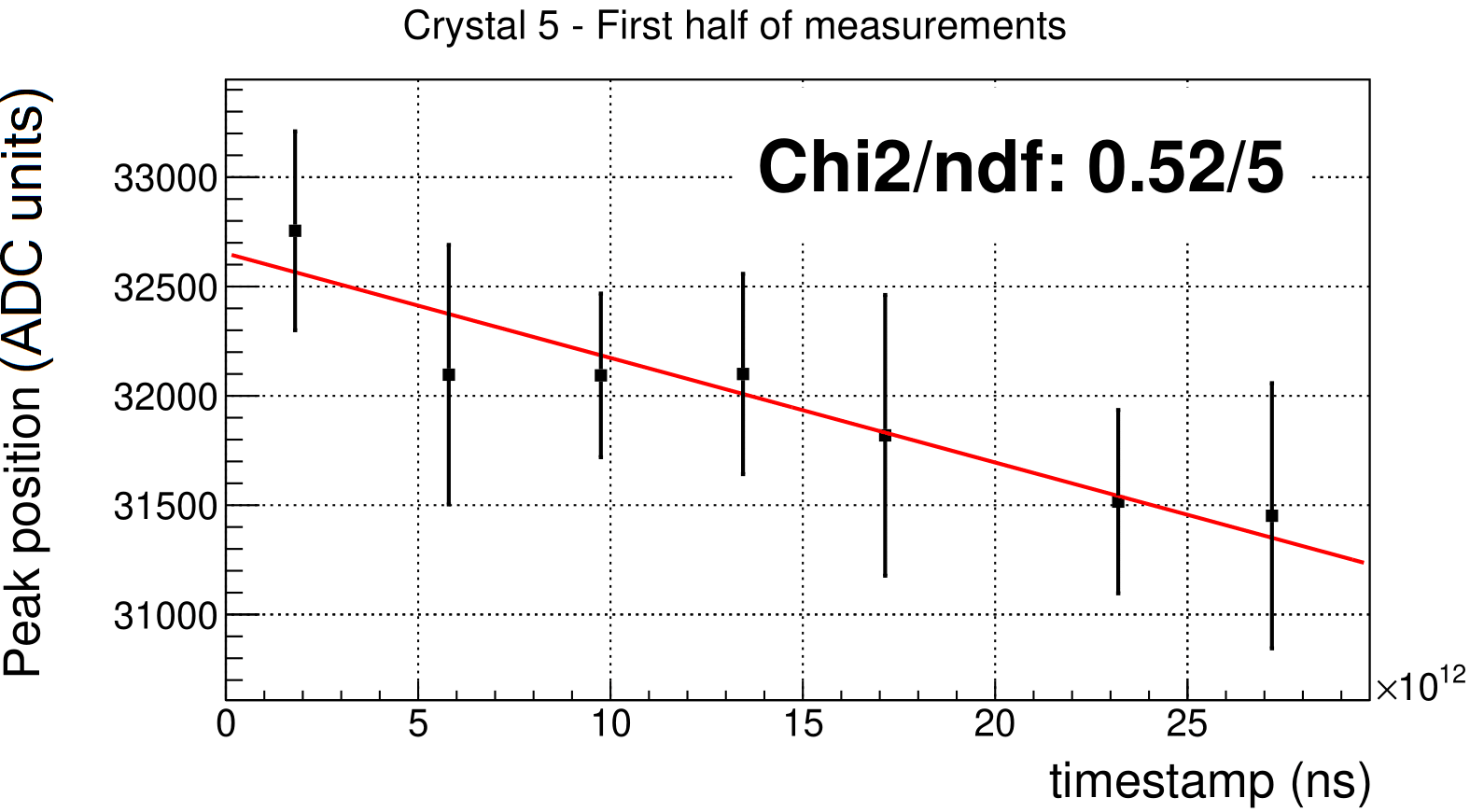}
	\end{subfigure}
	\begin{subfigure}[b]{0.49\textwidth}
		\includegraphics[width=\textwidth]{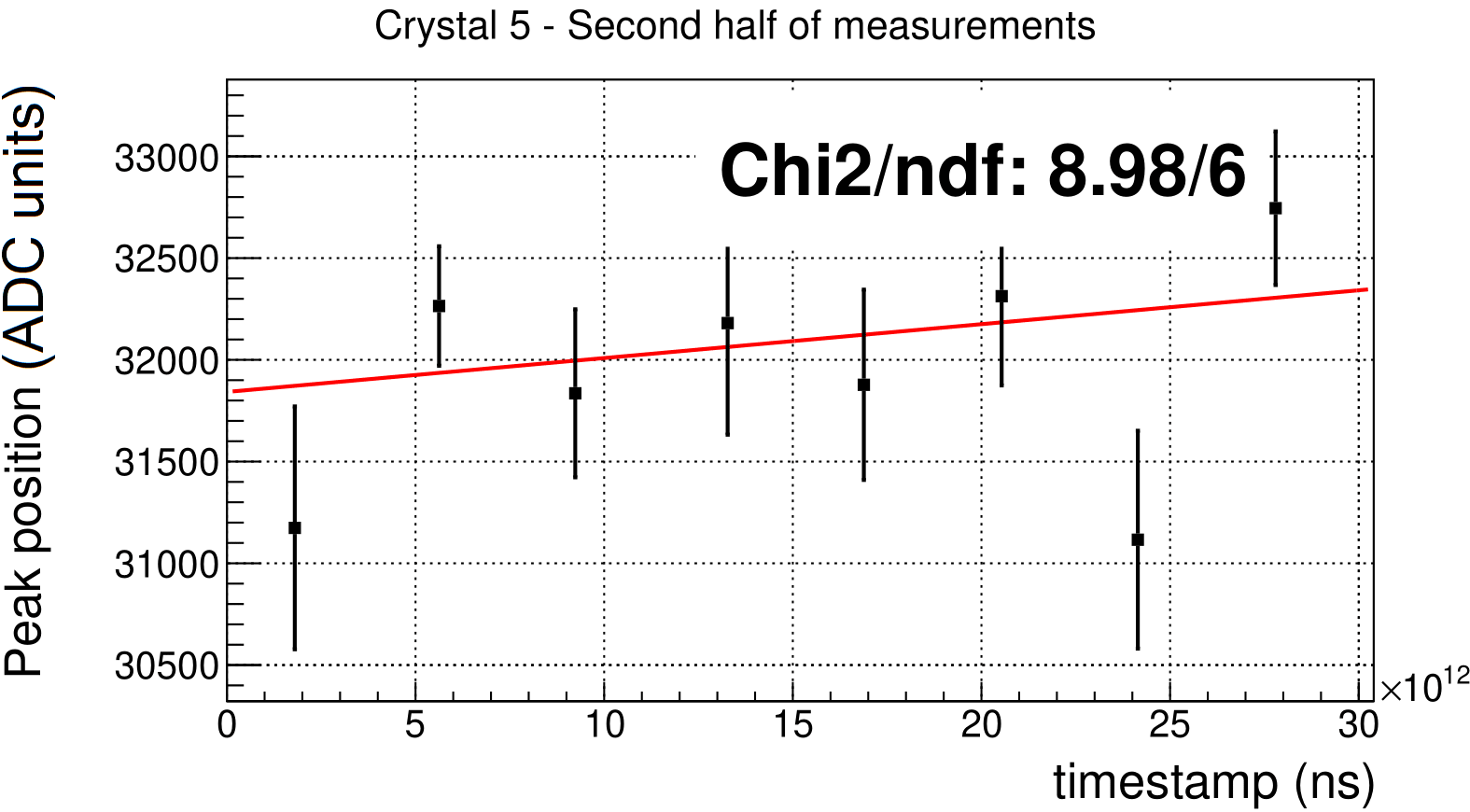}
	\end{subfigure}
	\caption{\label{StabCorrection}Fits of the pulse integral mean of the 57.6~keV peak to a linear dependence with time following Equation~\ref{eq:Int_vs_time} for each crystal. Results are presented in Table~\ref{tabla:StabCorrection}.}
\end{figure}

Before calibrating in energy the crystal signal, it is important to analyse possible drifts in the response of the detector. It can be done by studying the behaviour of a line with known energy that appears during all the beam-on runs, as it is the case for the gamma emitted following the inelastic scattering of a neutron with $^{127}I$ nuclei (see Appendix~\ref{Chapter:Anexo_NeutronNaI} for more information). The first excited state of $^{127}I$ has an energy of 57.6~keV above the ground state and can be produced by neutron inelastic scattering, being a good option to monitor and correct, if necessary, any gain drift, being intense enough and close to the ROI. Therefore, the 57.6~keV peak was analyzed every hour after applying the neutron selection procedure explained in Section~\ref{Section:QF_Analysis_Selection_NeutronsBDs} to the BDs signals. Figure~\ref{57keVpeak} shows the distribution of the corresponding \textit{fixedIntegral }variable, that was fitted to a gaussian function summed to a constant background. The mean from each gaussian fit as a function of the time for all the crystals is shown in Figure~\ref{StabCorrection}. A drift in \textit{fixedIntegral} was clearly observed in crystals~1 and~4, while in~2, 3 and 5 there is a high dispersion but not a clear drift. The case of the crystal~5 is more complicated because of the large uncertainties. Data from crystal~5 was divided in two different periods, as it was observed a different behaviour after a calibration run, in the middle of the beam-on measurements. For all the crystals, a linear dependence of the pulse integral with time was used to model this drift, following
\begin{equation}\label{eq:Int_vs_time}
	fixedIntegral = p_0 + p_1 \cdot timestamp.
\end{equation}
The fit parameters for each crystal are shown in Table~\ref{tabla:StabCorrection}, as well as the mean of the pulse integral for each run. To apply a time-dependent correction of the gain, the \textit{fixedIntegral} variable of each event was divided by the value obtained following the Equation~\ref{eq:Int_vs_time} for the time of the event, and multiplied by the mean value of this variable for the $^{127}I$ inelastic peak for the corresponding crystal. This correction was applied to all the data of the beam-on measurements. It was also applied to the $^{133}Ba$ calibration measurements by extrapolating the detector behaviour at the time when they were acquired.

\begin{table}[h!]
	\centering
	\begin{tabular}{|c|c|c|c|}
		\hline
		Crystal $\#$ & $p_0$ (ADC u.) & $p_1$ & Mean integral (ADC u./hour) \\
		\hline
		1 & 80671~$\pm$~152 & -37~$\pm$~7 & 79949~$\pm$~162 \\
		2 & 82345~$\pm$~150 & -19~$\pm$~8 & 81850~$\pm$~154 \\
		3 & 73595~$\pm$~126 & -10~$\pm$~7 & 72864~$\pm$~183 \\
		4 & 38649~$\pm$~140 & 112~$\pm$~10 & 39629~$\pm$~179 \\
		5 (First half) & 32652~$\pm$~349 & -172~$\pm$~80 & 31622~$\pm$~346 (*) \\
		5 (Second half) & 31842~$\pm$~292 & 60~$\pm$~63 & 31622~$\pm$~346 (*) \\
		\hline
	\end{tabular} \\
	\caption{Parameters obtained from the linear fit between the mean value of the \textit{fixedIntegral} variable for the 57.6~keV peak and the timestamp for each crystal (Equation~\ref{eq:Int_vs_time}). (*)Mean obtained with the complete crystal~5 data.}
	\label{tabla:StabCorrection}
\end{table}

\begin{figure}[h!]
	\begin{center}
		\includegraphics[width=0.75\textwidth]{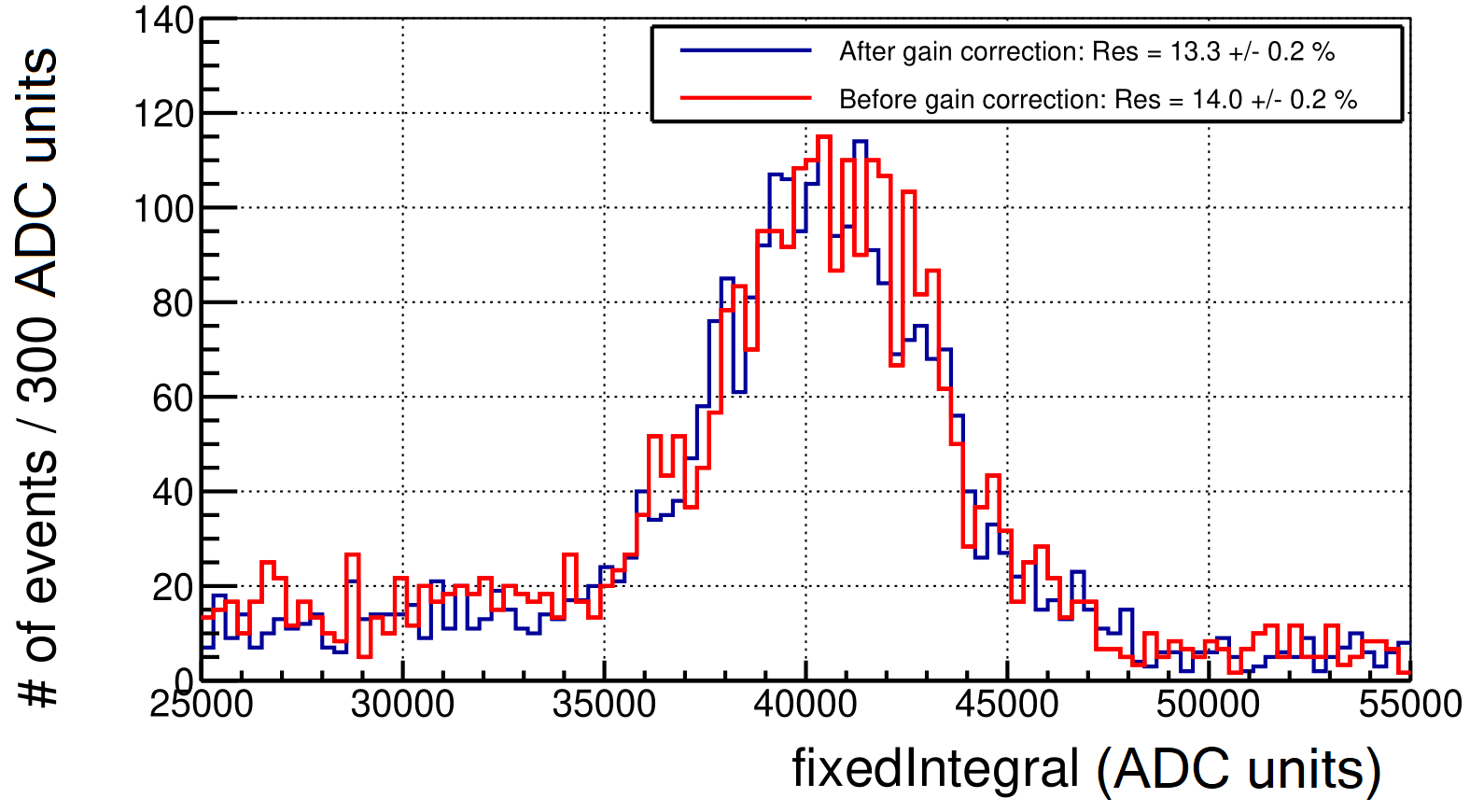}
		\caption{\label{57keVpeakCorrection}57.6~keV peak for crystal~4 before (red line) and after (blue line) the gain correction. The resolution of both distributions (calculated as the standard deviation of the gaussian divided by the mean, in percentage) is also shown.}
	\end{center}
\end{figure}

To evaluate if this method effectively corrects for the gain drift, we analysed the energy resolution of the 57.6~keV peak, which should improve. Figure~\ref{57keVpeakCorrection} shows the peak before and after applying the drift correction method in the crystal~4. The resolution of both distributions was calculated as the standard deviation of the gaussian divided by the mean, in percentage, and it was found to be 14.0~$\pm$~0.2$\%$ before and 13.3~$\pm$~0.2$\%$ after the correction, which is a significant improvement. For the other crystals, this correction is only producing slight improvement in the resolution.

\subsubsection{Energy calibration}\label{Section:QF_Analysis_NaIcal_EnergyCal}

In the calculation of the QF, the electron equivalent energy calibration of the energy depositions in the NaI(Tl) is one of the most relevant points. The light yield of the NaI(Tl) is non-proportional with the energy, with variations of a few percent up to $\sim$~20~keV~\cite{PhysRev.122.815,Rooney:1997,Moses:2001vx,Choong:2008,Hull:2009,Payne:2009,Khodyuk:2010ydw,Payne:2011}. Because of this non-proportional response, we have considered two different approaches for the calibration in electron equivalent energy of the NaI(Tl) crystals which allow to identify and take into account the effect of the possible non-proportional and non-linear dependencies with the energy of the light produced:

\begin{enumerate}
	\item Assume a proportional response calibrating with only one known energy deposition, the 57.6~keV inelastic peak. This is the most common approach followed by the previous sodium QF measurements (see Section~\ref{Section:QF_NaIQFOverview}).
	\item Apply a linear calibration in the ROI using energy depositions produced by the interaction of the x and gamma rays emitted by an external $^{133}Ba$ source (listed in Tables~\ref{tabla:Ba133_Xray} and~\ref{tabla:Ba133_gamma}).
\end{enumerate}

The comparison of the results obtained for the QF with both calibrations will be relevant for the understanding of some possible systematics contributing to the disagreement between different measurements found in the literature (see Figure~\ref{NaQF_measurements}).

In the first method, the gain corrected spectra of the 57.6~keV peak was fitted to a gaussian summed to a constant background. The mean value of the gaussian divided by the energy of the peak allows obtaining a proportional relation between the \textit{fixedIntegral} variable, $A$, and the energy, $E$, as $E = A/c$, with $c$ the calibration parameter. Results of this parameter for each crystal are shown in Table~\ref{tabla:57Cal}. It is important to emphasize that the 57.6~keV peak is above our energy range of interest and assuming proportionality in a well-known non-proportional scintillator is expected to introduce relevant systematic effects in the determination of the QF.

\begin{table}[h!]
	\centering
	\begin{tabular}{|c|c|}
		\hline
		Crystal~$\#$ & $c$ (ADC units/keV) \\
		\hline
		1 & 1388~$\pm$~3 \\
		2 & 1421~$\pm$~3 \\
		3 & 1265~$\pm$~3 \\
		4 & 688~$\pm$~3 \\
		5 & 549~$\pm$~6 \\
		\hline
	\end{tabular} \\
	\caption{Calibration parameter obtained in the proportional calibration of each crystal.}
	\label{tabla:57Cal}
\end{table}

In the second method, the energy depositions in the NaI(Tl) crystal inside the ROI (below 35~keV) during the $^{133}Ba$ calibrations (spectrum shown in Section~\ref{Section:QF_GEANT4_Results_Calibrations}) were considered. The procedure to calibrate the crystals was first to define a linear energy calibration function as
\begin{equation}\label{eq:NaICal}
	E = c_1\cdot fixedIntegral + c_0.
\end{equation}
Then, a PDF is defined as a flat distribution (background contribution) summed to three gaussians, whose means are the energies obtained from the simulation in Section~\ref{Section:QF_GEANT4_Results_Calibrations}: 6.6, 30.9~and 35.1~keV, and the standard deviations considered are 
\begin{equation}\label{eq:NaIRes}
	\sigma = \sqrt{a+b\cdot E}.
\end{equation}
Finally, the experimental data is fitted to this PDF, scaling the conversion between ADC units and energy with the Equation~\ref{eq:NaICal}. The free parameters of the fit are $a$, $b$, those from the calibration function ($c_o$ and $c_1$), the areas of the three gaussians and the level of the constant background.

\begin{figure}[h!]
	\begin{center}
		\includegraphics[width=\textwidth]{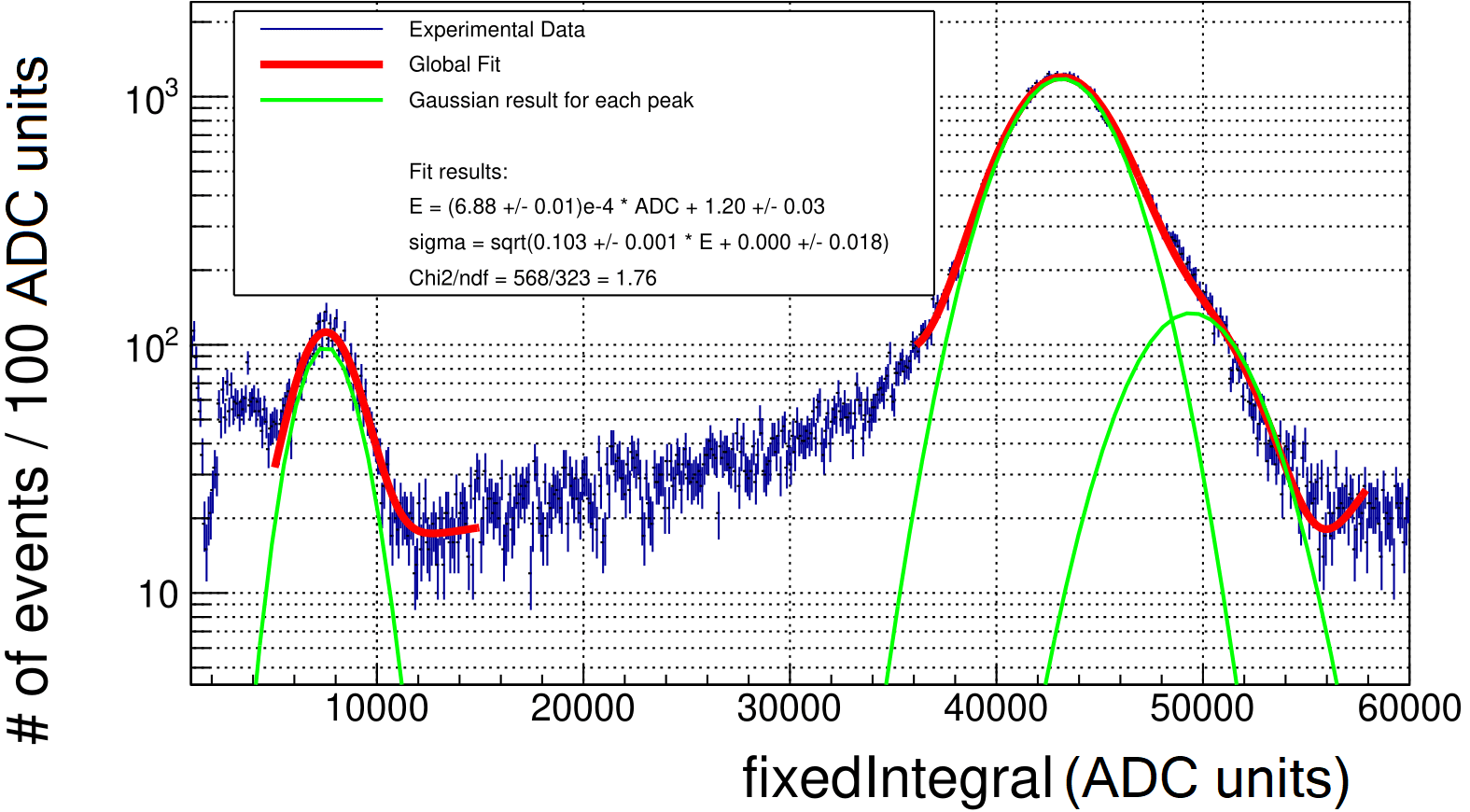}
		\caption{\label{Ba133_fitSeparated}$^{133}Ba$ spectrum fit for crystal 1 (see text for more details).}
	\end{center}
\end{figure}

In the experimental spectrum, there may be background contributions between the peaks that are not accounted for in the simulations. Therefore, it was decided to apply the fit only in a region close to the peaks: from $\sim$~4~keV to $\sim$~10~keV and from $\sim$~25~keV to $\sim$~40~keV. As an example, the fit for the crystal~1 is shown in Figure~\ref{Ba133_fitSeparated}. The results of this calibration are presented in Table~\ref{tabla:Ba133Cal}. It is possible to observe that for all the crystals, the value of the parameter $a$ is compatible with zero.

\begin{table}[h!]
	\centering
	\begin{tabular}{|c|c|c|c|c|}
		\hline
		Cr.~$\#$ & $a$ (keV$^2$) & $b$ (keV) & $c_0$ (keV) & $c_1$ (eV/ADC)\\
		\hline
		1 & 0.000~$\pm$~0.018 & 0.103~$\pm$~0.001 & 1.195~$\pm$~0.032 & 0.688~$\pm$~0.001  \\
		2 & 0.000~$\pm$~0.064 & 0.084~$\pm$~0.001 & 0.825~$\pm$~0.031 & 0.673~$\pm$~0.001 \\
		3 & 0.111~$\pm$~0.084 & 0.095~$\pm$~0.003 & 0.819~$\pm$~0.034 & 0.754~$\pm$~0.001 \\
		4 & 0.000~$\pm$~0.634 & 0.176~$\pm$~0.006 & 1.254~$\pm$~0.082 & 1.366~$\pm$~0.005 \\
		5 & 0.000~$\pm$~0.501 & 0.275~$\pm$~0.017 & 1.470~$\pm$~0.080 & 1.674~$\pm$~0.006 \\
		\hline
	\end{tabular} \\
	\caption{Calibration and resolution parameters obtained from the fitting of the measured $^{133}Ba$ spectra for each crystal.}
	\label{tabla:Ba133Cal}
\end{table}

\begin{figure}[h!]
	\begin{subfigure}[b]{0.49\textwidth}
		\includegraphics[width=\textwidth]{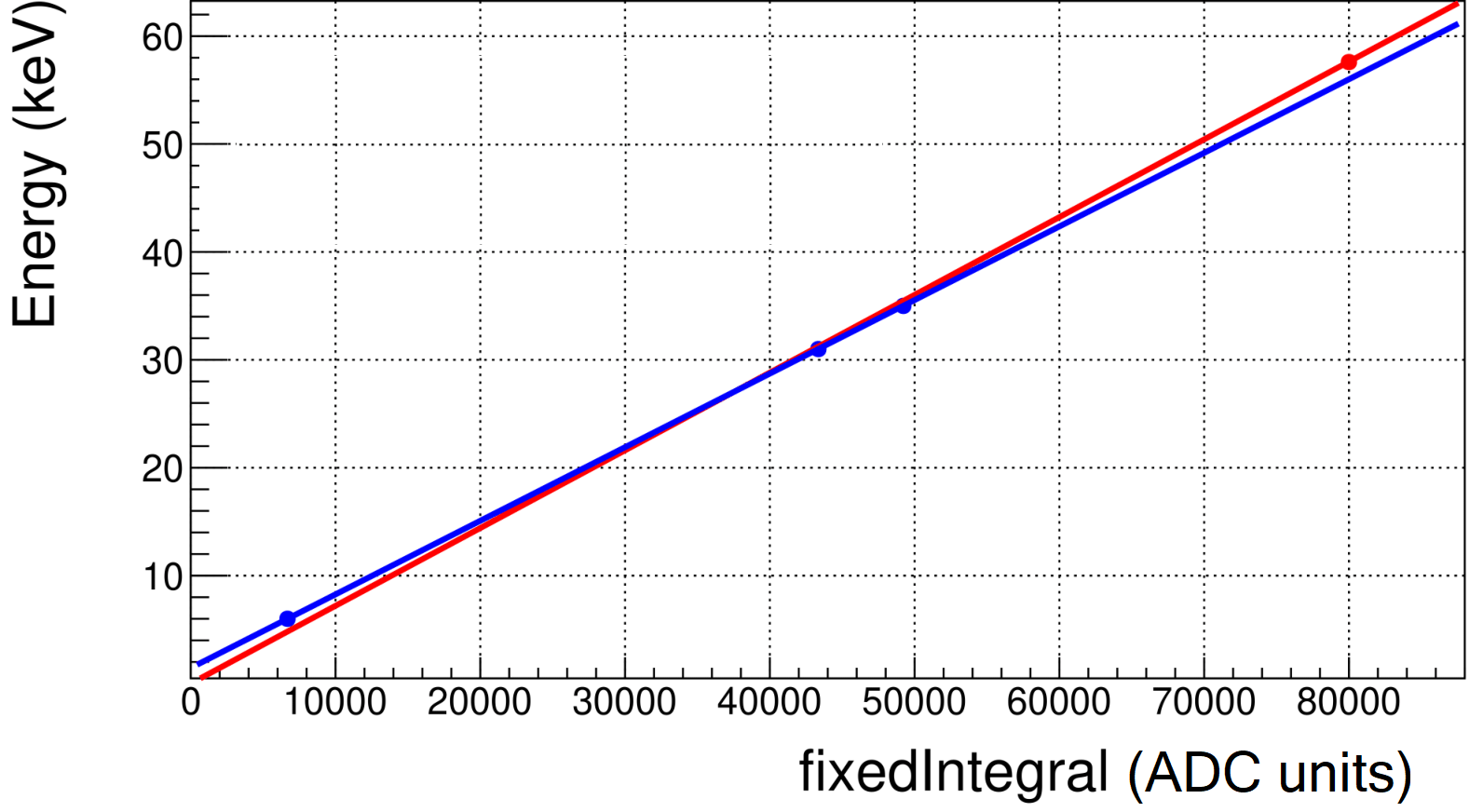}
	\end{subfigure}
	\begin{subfigure}[b]{0.49\textwidth}
		\includegraphics[width=\textwidth]{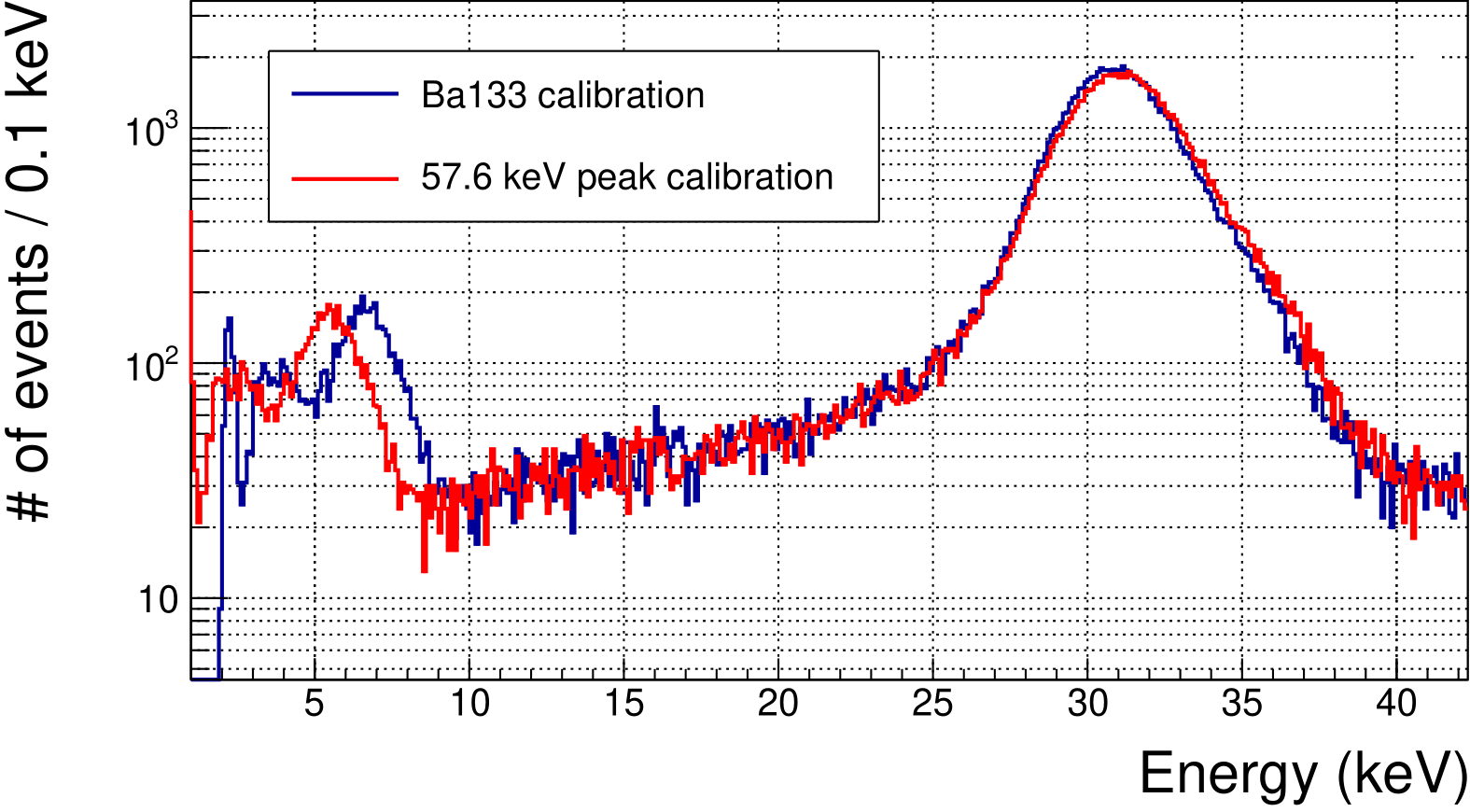}
	\end{subfigure}
	\caption{\label{CalComparison}Comparison of the two energy calibrations for the crystal~1. In the left plot, the two dependencies Energy - ADC units are shown, while in the right plot they were applied to the measured $^{133}Ba$ spectrum to convert the ADC units into electron equivalent energies.}
\end{figure}

Both calibration functions are compared in the left plot of the Figure~\ref{CalComparison} for crystal~1. To evaluate the goodness of each energy calibration to reproduce the deposited energies, they were applied to the measured $^{133}Ba$ spectrum and compared in the ROI, as it is shown in the right plot of the Figure~\ref{CalComparison}. The residual of the lower energy peak (6.6~keV) for that crystal is -1.16~keV for the first calibration and 0.02~keV for the second one. Similar results are obtained for all the crystals. As it can be observed, the difference between the proportional and the non-proportional calibration is more relevant, as expected, at lower energies. Therefore, the selection of the energy calibration procedure will affect strongly the QF results, as it will be shown in next section.

\section{Spectrum fitting and QF calculation} \label{Section:QF_Results}
\fancyhead[RO]{\emph{\thesection. \nameref{Section:QF_Results}}}

In this section, the procedures followed to calculate the QF values for the sodium and iodine recoils are explained.

\subsection{QF of the sodium nuclei}\label{Section:QF_QFcal_Na}

The gain corrected and energy calibrated experimental spectra associated with neutrons in the BDs for each channel and crystal were fitted to the simulated nuclear recoil energy distributions shown in Figure~\ref{enr_sodio}. 

\begin{figure}[h!]
	\begin{center}
		\includegraphics[width=0.75\textwidth]{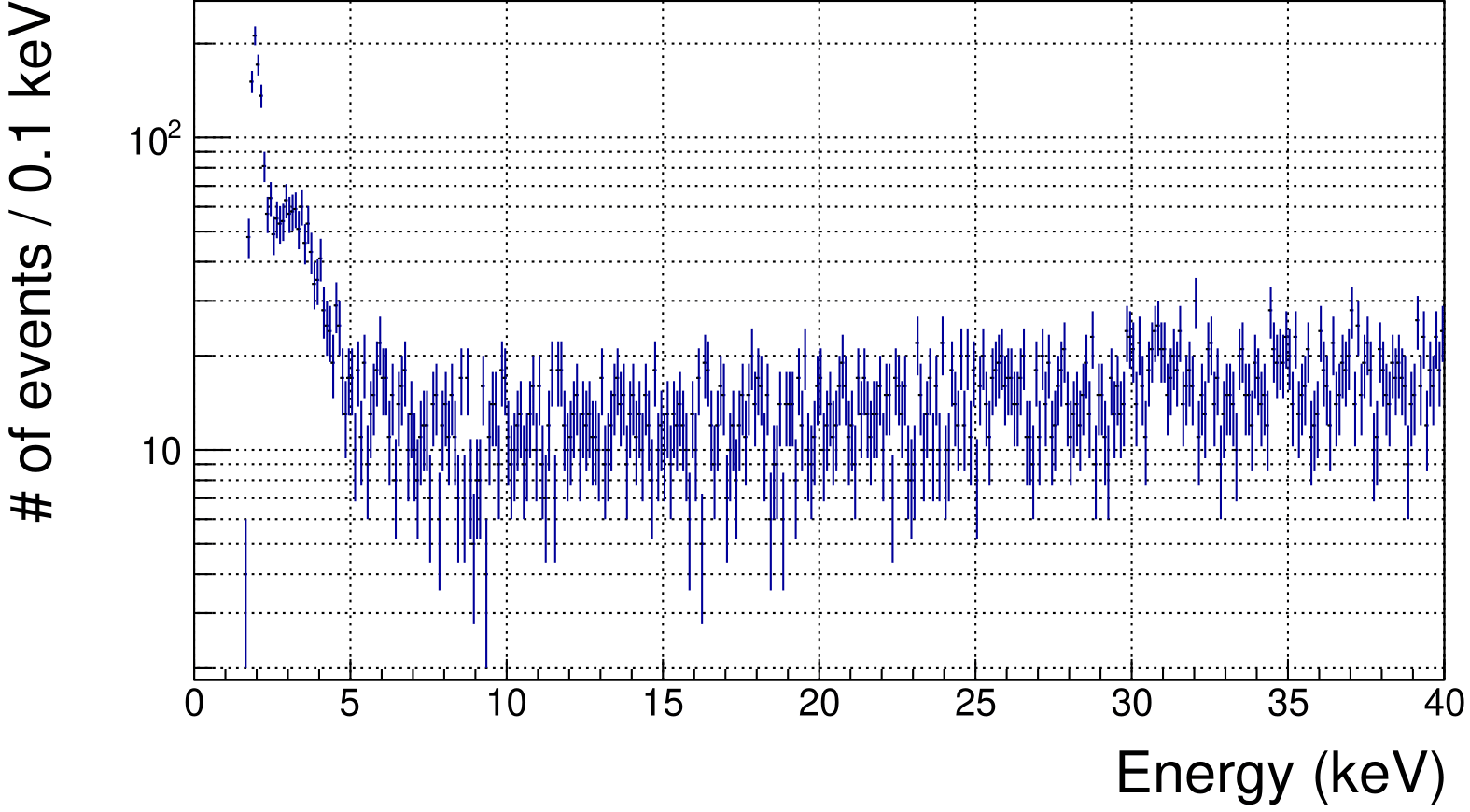}
		\caption{\label{BgndSpectrum}Background spectrum of crystal~1 obtained during beam-on measurements selecting events not related to neutrons without any selection in channel.}
	\end{center}
\end{figure}

The sources of the experimental background were difficult to identify and then, they were not simulated. However, the background contribution has to be considered in the fit. Background measurements in dedicated runs had low statistics in the ROI and did not include beam related events which are expected to be the most relevant background contribution. We considered a better option to use for the background spectra those events that do not fulfill the neutron selection criteria explained in Section~\ref{Section:QF_Analysis_Selection_NeutronsBDs} (0.3~$<$~\textit{PSD}~$<$~0.5 and 304~ns~$<$~\textit{timeSincePrevBPM}~$<$~340~ns). These background spectra were gain corrected and converted into electron equivalent energy. As an example, the background spectrum for crystal~1 with the calibration method~2 is shown in Figure~\ref{BgndSpectrum} without any selection in channel. No clear differences were observed in this spectrum among different channels.

The spectrum of the iodine recoils contributes significantly to the experimental data in some of the channels, and therefore it was also included in the fit using the data from the simulation. Nevertheless, the peak is truncated by the detection threshold, and no information on the iodine QF can be easily derived from these spectra, therefore a constant QF for iodine has been adopted in the fit. To account for the energy resolution, a gaussian PDF was also used to convolve the signal of the sodium recoils corresponding to each channel, and two different modelings were applied for the standard deviation: constant (energy independent, $\sigma = \sigma_0$) and proportional to the square root of the electron equivalent energy ($\sigma = a\sqrt{E_{ee}}$), as it would correspond to a poissonian resolution. 

The procedure for each fit is the following: First, the region above the sodium recoils peak for each channel (between 30 and 40~keV) is fitted to the background PDF ($S_{bkg}$) with a scale factor, $N_{bkg}$, as free parameter. Then, the simulated sodium recoil spectrum is converted into electron equivalent energy with a constant QF as free parameter, different for each channel, and convolved with a gaussian with the standard deviation modelled as commented above. The resulting PDF is called $S_{Na}$. The PDF for iodine recoils ($S_I$) is built following a similar procedure but with constant QF = 5\% and standard deviation = 1~keV. It was checked that the fitting procedure was not sensitive to slight variations of these values, but the systematic contribution of the change in these parameters to the final QF results was also analyzed, as it is explained next. Finally, a PDF was constructed as a linear combination of sodium ($S_{Na}$) and iodine ($S_I$) spectra and the background spectrum ($S_{bkg}$) as:
\begin{equation}\label{eq:PDF_QFfit}
	PDF = N_{Na}S_{Na} + N_I S_I + N_{bkg}S_{bkg}.
\end{equation}
The experimental spectrum for each crystal and channel was fitted to this PDF with $N_{bkg}$ parameter fixed to the value determined in the previous fitting. The free parameters were: the scale factors of the sodium and iodine signals ($N_{Na}$ and $N_{I}$), the sodium QF and the parameters corresponding to the resolution modeling ($\sigma_0$ or $a$, in each case).

\begin{figure}[h!]
	\begin{center}
		\includegraphics[width=0.75\textwidth]{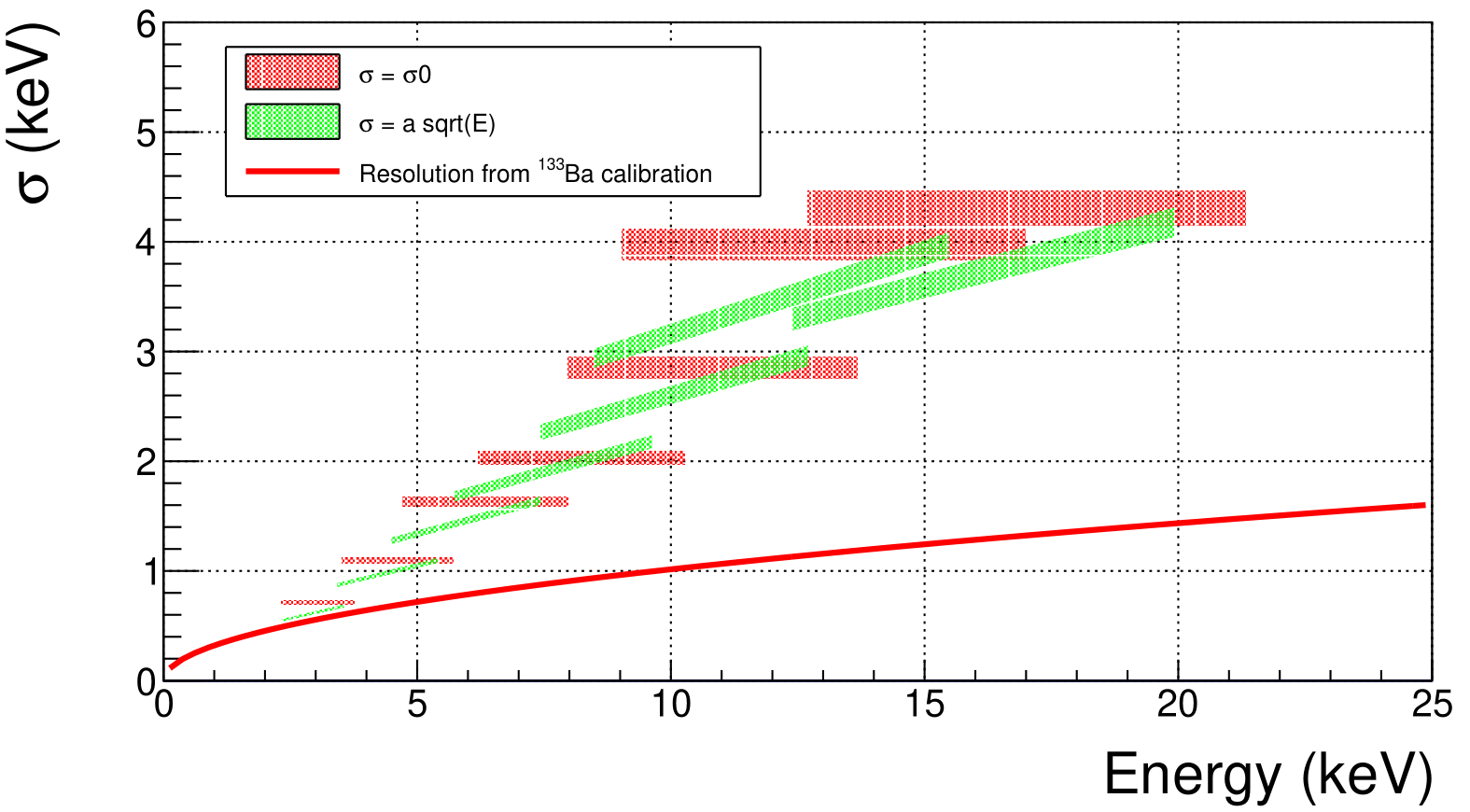}
		\caption{\label{ResolutionResults}Results of the energy resolution for the channels~0 to~6 of the crystal~1, with the corresponding uncertainties represented by the shadows (red for constant and green for energy dependent resolution). The range plotted for each channel corresponds to $\left[E_{mean}-\sigma(E_{mean}),E_{mean}+\sigma(E_{mean})\right]$being E the electron equivalent energy. The resolution obtained from the calibration with $^{133}Ba$ is shown as a red line.}
	\end{center}
\end{figure}

These fits were done for channels from 0 to 6, in a range from 2.5 to 40~keV. The p-values of these fits, which are shown in Table~\ref{tabla:pVal_resolutions}, do not show a clear preference for one resolution modelling over the other. The results for the energy resolutions obtained are plotted together as a function of the energy in Figure~\ref{ResolutionResults} for the crystal~1. The shadowed regions represent the results, with the corresponding uncertainties, plotted for each channel in the range $\left[E_{mean}-\sigma(E_{mean}),E_{mean}+\sigma(E_{mean})\right]$. The energy resolution obtained from the calibration with $^{133}Ba$ is also shown.

\begin{table}[h!]
	\centering
	\begin{tabular}{|c|c|c|}
		\hline
		BD~$\#$ & p-values ($\sigma = \sigma_0$) & p-values ($\sigma = a \sqrt{E}$) \\
		\hline
		0 & 0.72 & 0.62 \\
		1 & 0.11 & 0.31 \\
		2 & 0.01 & 0.03 \\
		3 & 0.41 & 0.24 \\
		4 & 0.08 & 0.38 \\
		5 & 0.66 & 0.99 \\
		6 & 0.23 & 0.79 \\
		11 & 0.03 & 0.71 \\
		12 & 0.01 & 0.81 \\
		13 & 0.18 & 0.86 \\
		14 & 0.59 & 0.18 \\
		15 & 0.32 & 0.22 \\
		16 & 0.34 & 0.19 \\
		17 & 0.49 & 0.39 \\
		\hline
	\end{tabular} \\
	\caption{p-values of the fits of the NaI(Tl) experimental spectra associated to neutrons in the BDs for the crystal~1 to the PDF shown in Equation~\ref{eq:PDF_QFfit}. They are shown for the two models applied for the energy resolution.}
	\label{tabla:pVal_resolutions}
\end{table}

This plot shows clearly that we obtain an energy resolution for nuclear recoils worse than that obtained for electronic recoils independently on the resolution function applied in the fit. It is worth to note that the resolution functions obtained from the fits of each channel do not converge in a common function, and therefore it was not possible to apply the same resolution for all the fits of the same crystal. As there is no preference for the use of any of these two resolution functions, fits were done using both of them leaving free the resolution parameters $\sigma_0$ and $a$. The differences between the QF obtained were therefore considered as a systematical uncertainty. Further work is required to understand the origin of the worse energy resolution for nuclear recoils. The $\sigma$ obtained for the 57.6~keV peak (bulk scintillation) after applying the gain correction (see Figure~\ref{57keVpeakCorrection}) goes from 7 to 9~keV depending on the crystal, which is much larger than that obtained from the $^{133}Ba$ calibration (surface scintillation). It indicates that the resolution could be affected by the spatial dependence of the scintillation properties of the material.

A procedure was followed systematically in all the channels to fix the range of energies considered in the fit. First, the experimental data was fitted to the PDF from Equation~\ref{eq:PDF_QFfit} from 2.5~to 40~keV applying the constant resolution, thus obtaining preliminary $QF_p$ and $\sigma_{0p}$ values. After this preliminary fit, and using the mean value of the sodium recoil energy distribution obtained from the simulation for each channel, $E_{nr}$, the upper energy considered in the fit will be $E_{nr}\cdot QF_p + 5\sigma_{0p}$. However, the spectrum at low energies could not be correctly reproduced by our model, and the $\chi^2$ and the fit results were very dependent on the minimum energy of the range considered in the fit. The procedure followed to choose this minimum energy was to perform multiple fits in a loop, changing in each iteration the minimum energy of the fit range. The minimum energy was then selected as the lowest value for which the reduced $\chi^2$ changed less than 10$\%$ for at least three iterations.

\begin{figure}[]
	\begin{center}
		\begin{subfigure}[b]{0.32\textwidth}
			\includegraphics[width=\textwidth]{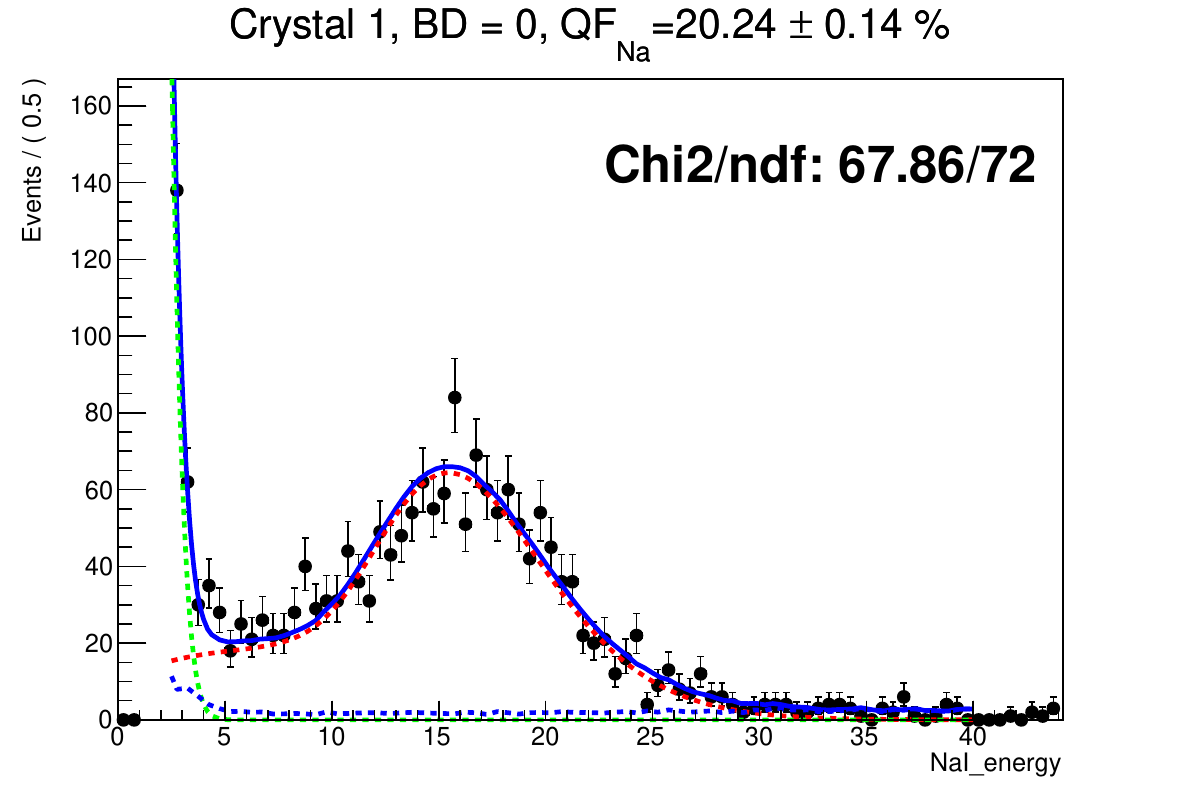}
		\end{subfigure}
		\begin{subfigure}[b]{0.32\textwidth}
			\includegraphics[width=\textwidth]{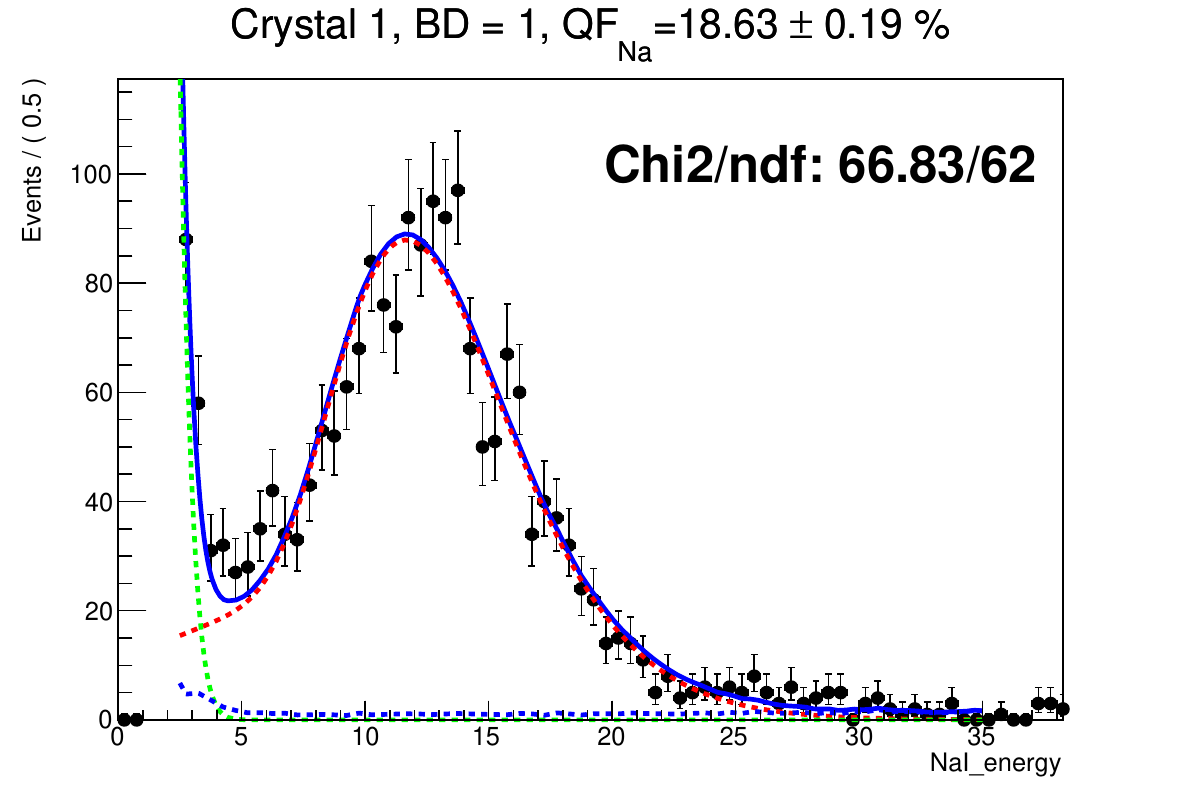}
		\end{subfigure}
		\begin{subfigure}[b]{0.32\textwidth}
			\includegraphics[width=\textwidth]{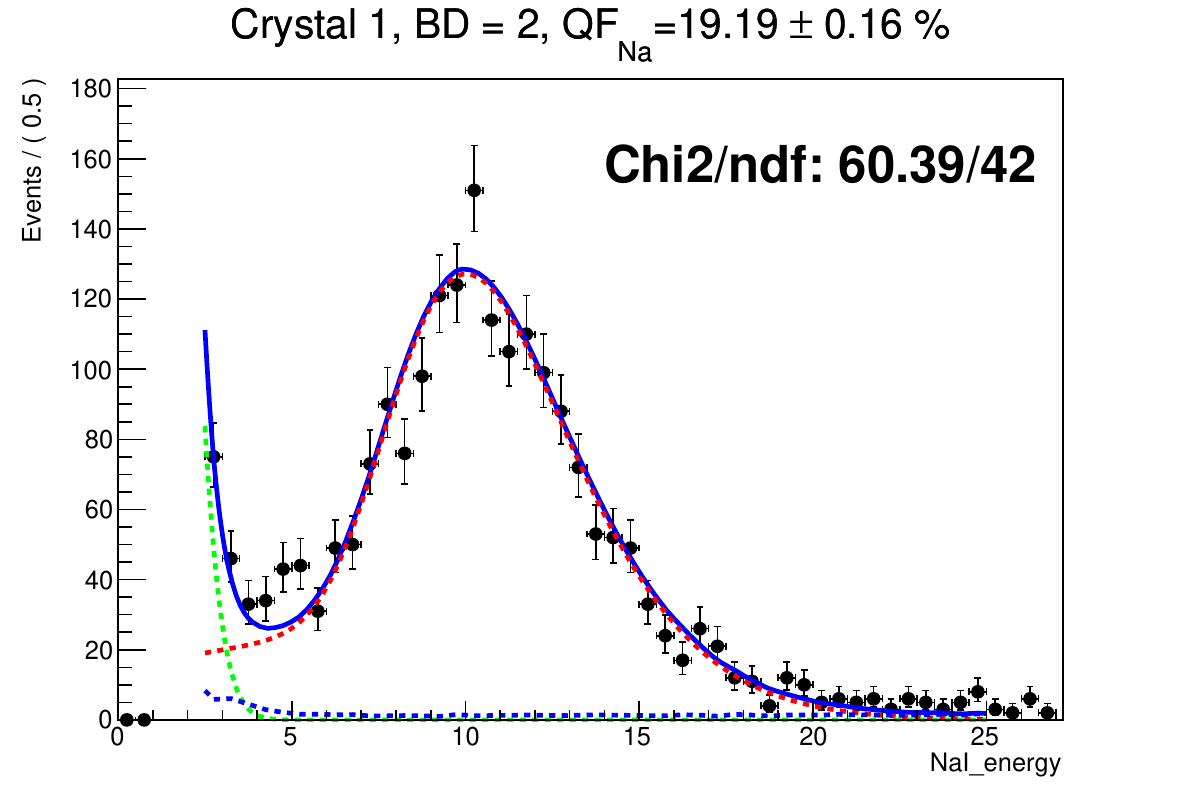}
		\end{subfigure}
		\begin{subfigure}[b]{0.32\textwidth}
			\includegraphics[width=\textwidth]{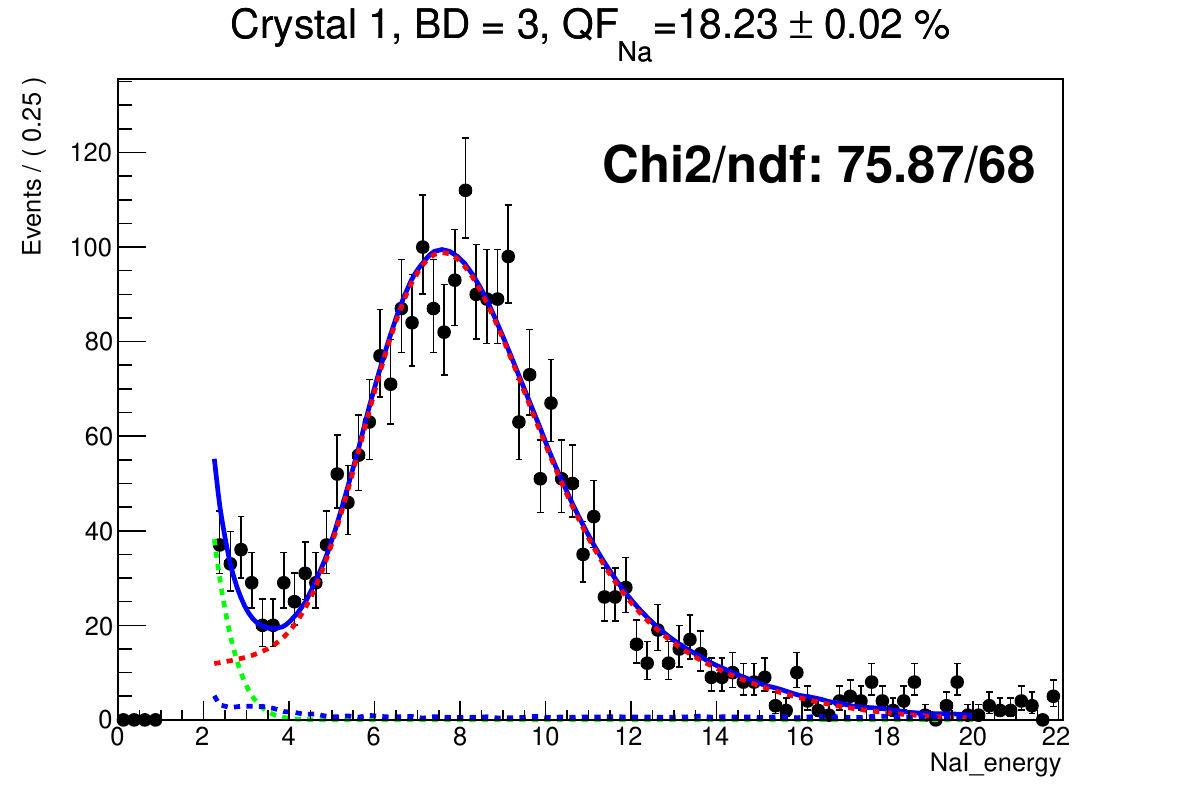}
		\end{subfigure}
		\begin{subfigure}[b]{0.32\textwidth}
			\includegraphics[width=\textwidth]{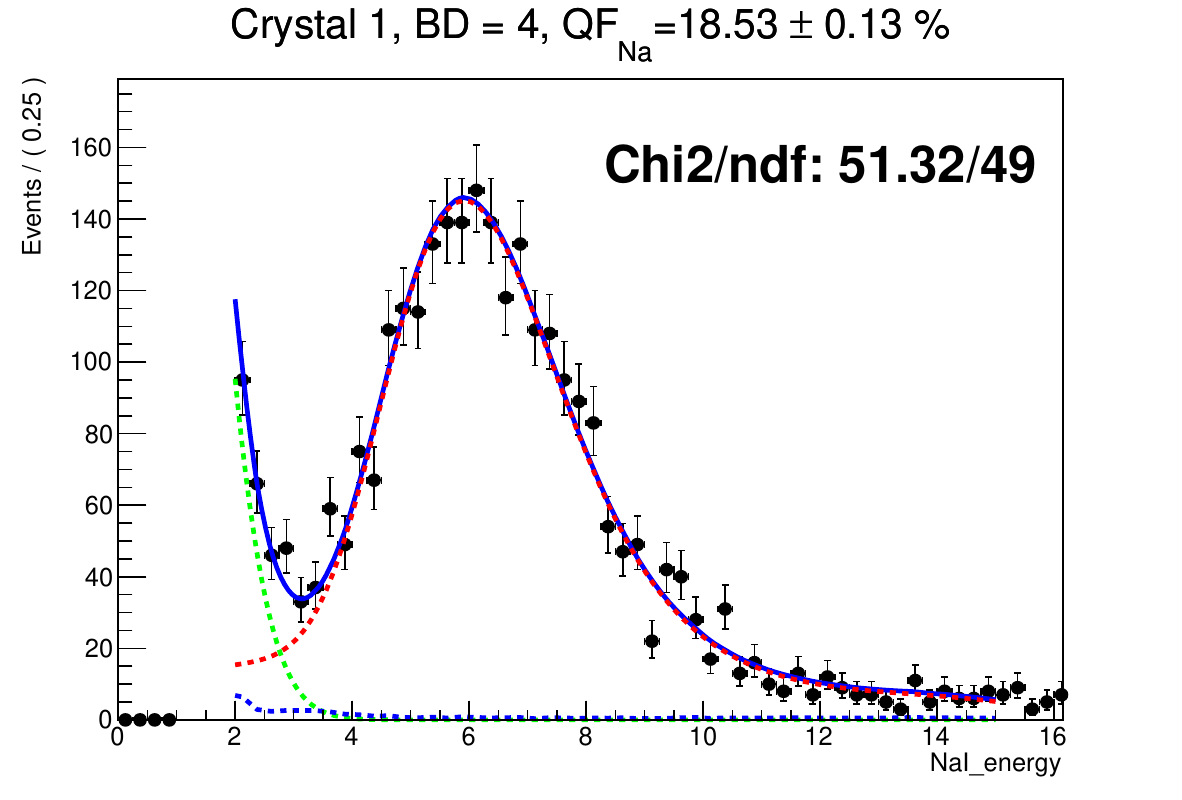}
		\end{subfigure}
		\begin{subfigure}[b]{0.32\textwidth}
			\includegraphics[width=\textwidth]{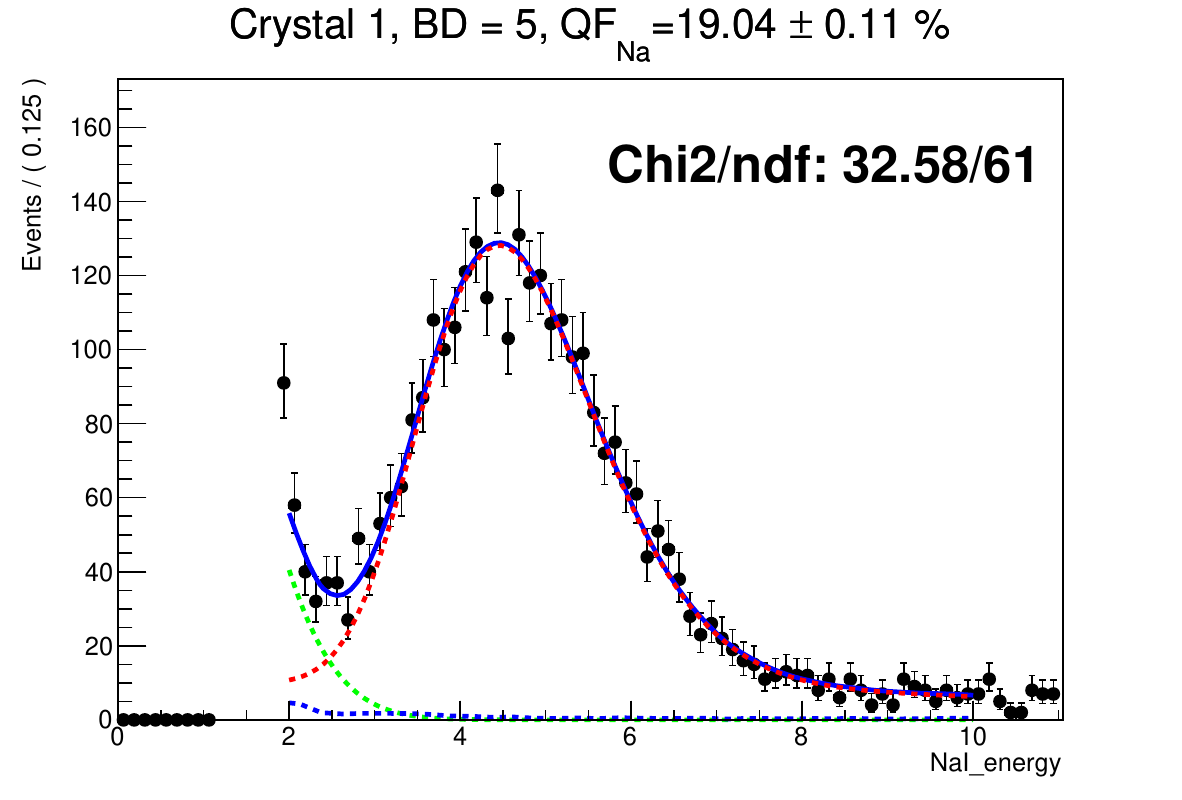}
		\end{subfigure}
		\begin{subfigure}[b]{0.32\textwidth}
			\includegraphics[width=\textwidth]{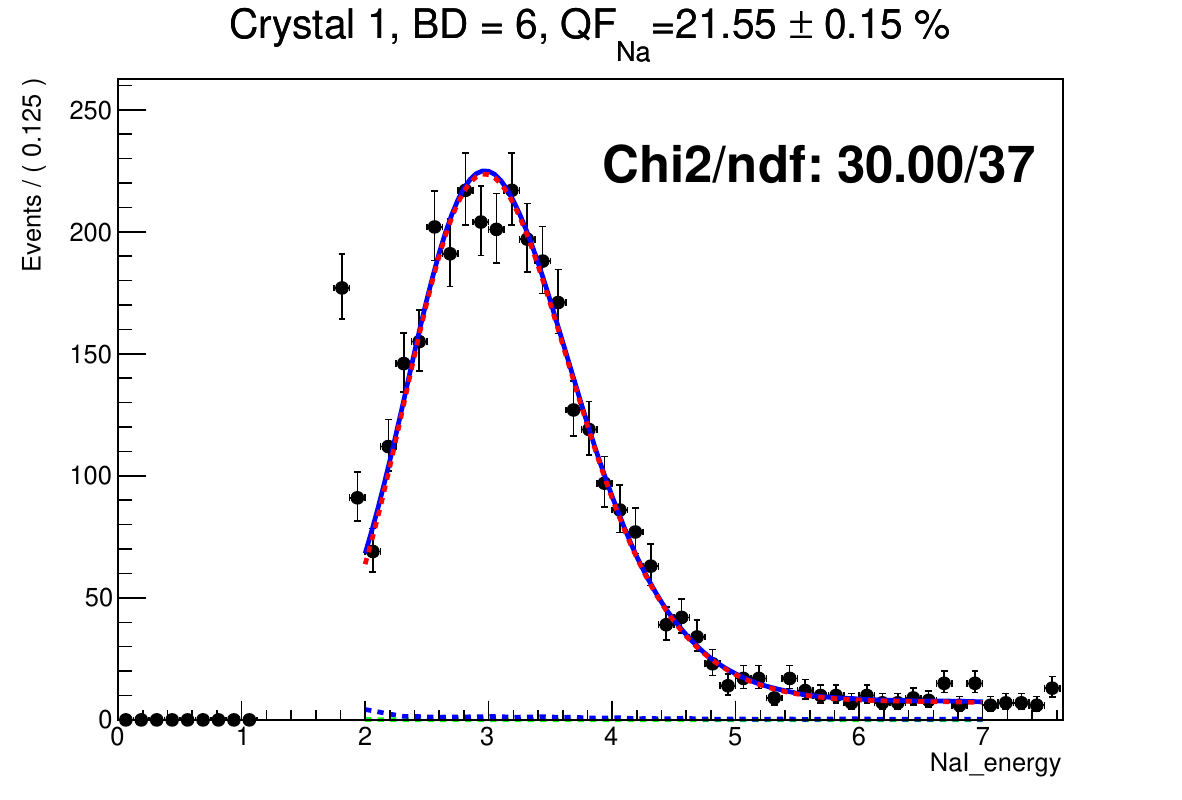}
		\end{subfigure}
		\begin{subfigure}[b]{0.32\textwidth}
			\includegraphics[width=\textwidth]{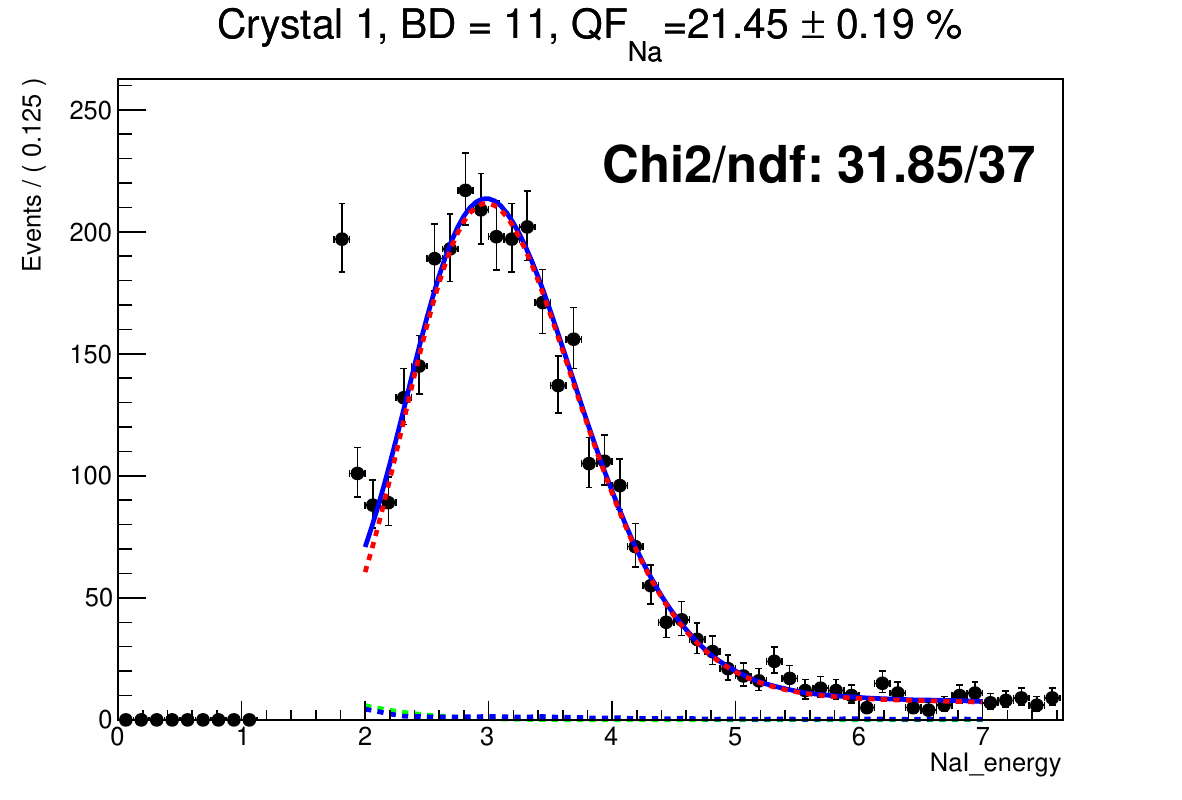}
		\end{subfigure}
		\begin{subfigure}[b]{0.32\textwidth}
			\includegraphics[width=\textwidth]{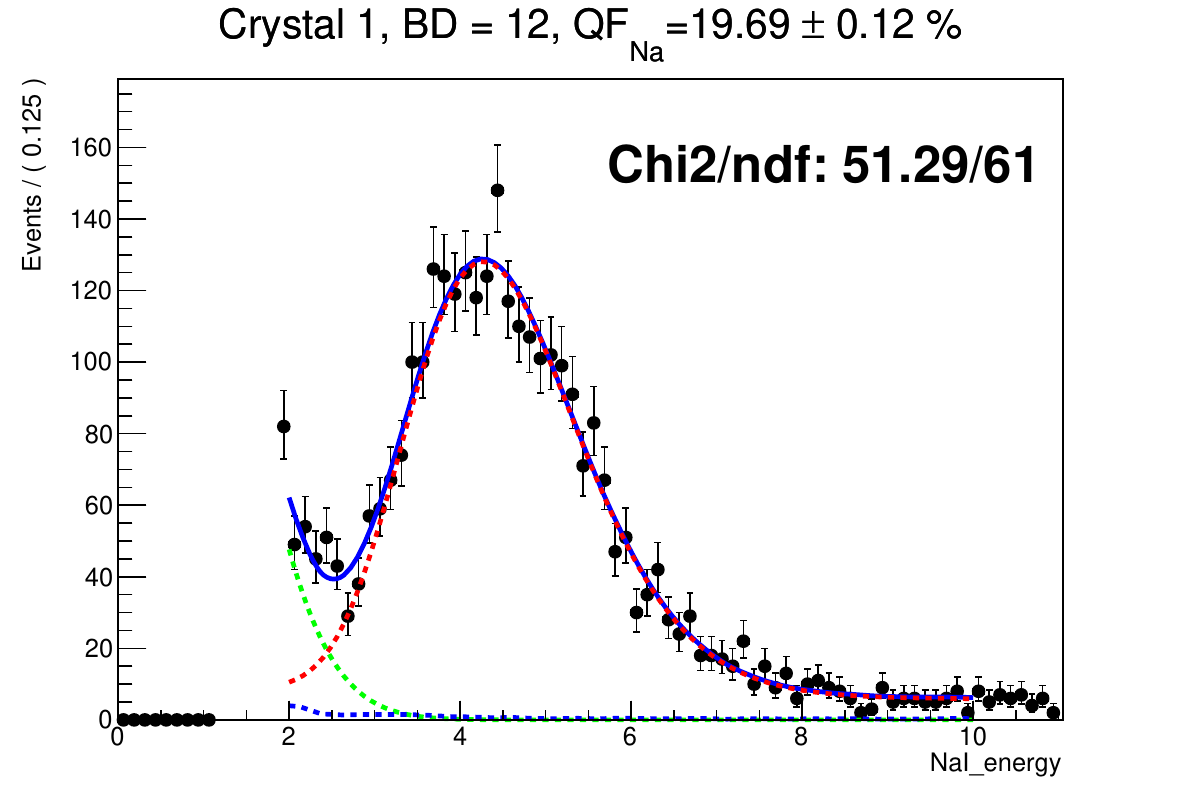}
		\end{subfigure}
		\begin{subfigure}[b]{0.32\textwidth}
			\includegraphics[width=\textwidth]{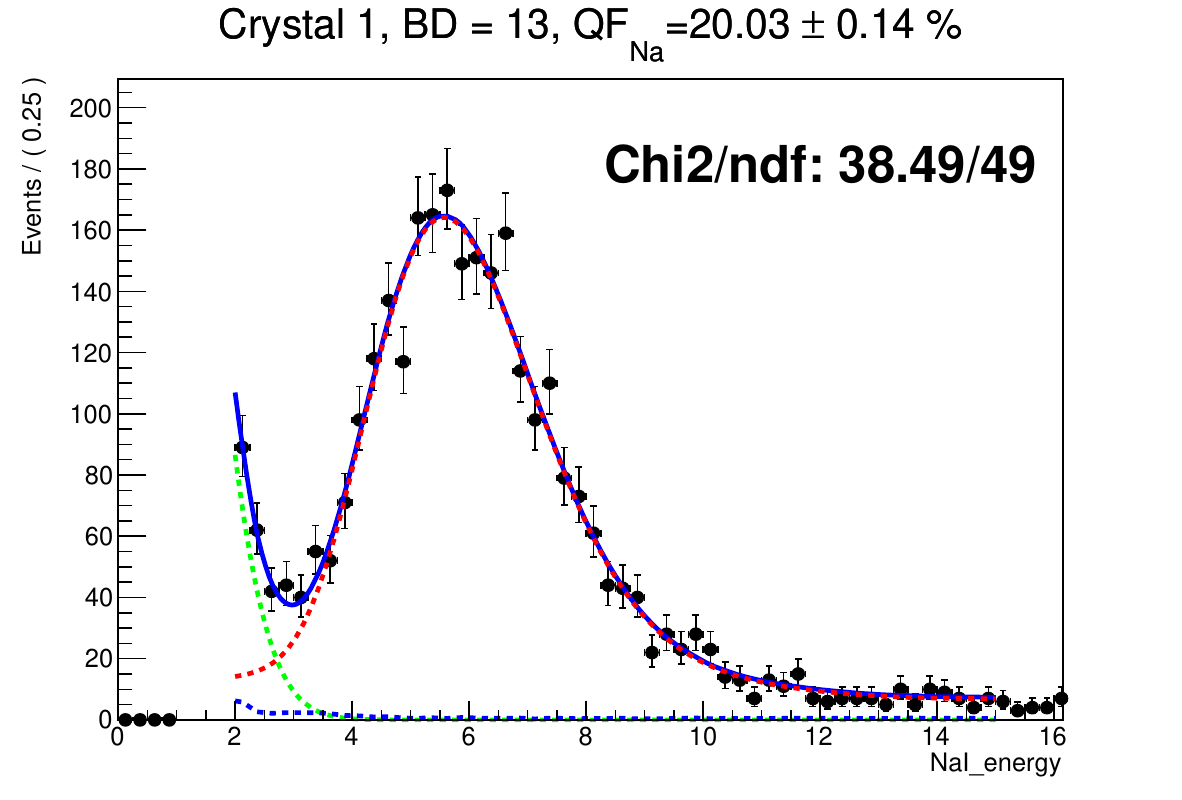}
		\end{subfigure}
		\begin{subfigure}[b]{0.32\textwidth}
			\includegraphics[width=\textwidth]{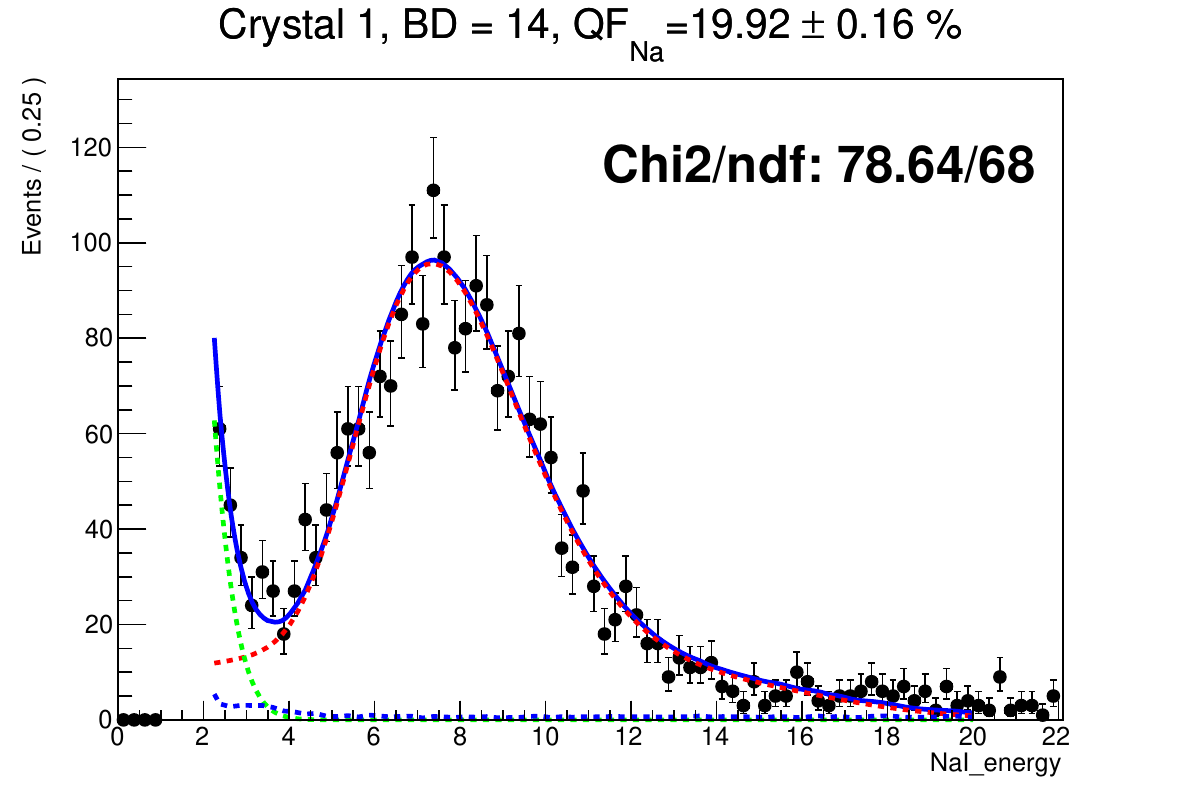}
		\end{subfigure}
		\begin{subfigure}[b]{0.32\textwidth}
			\includegraphics[width=\textwidth]{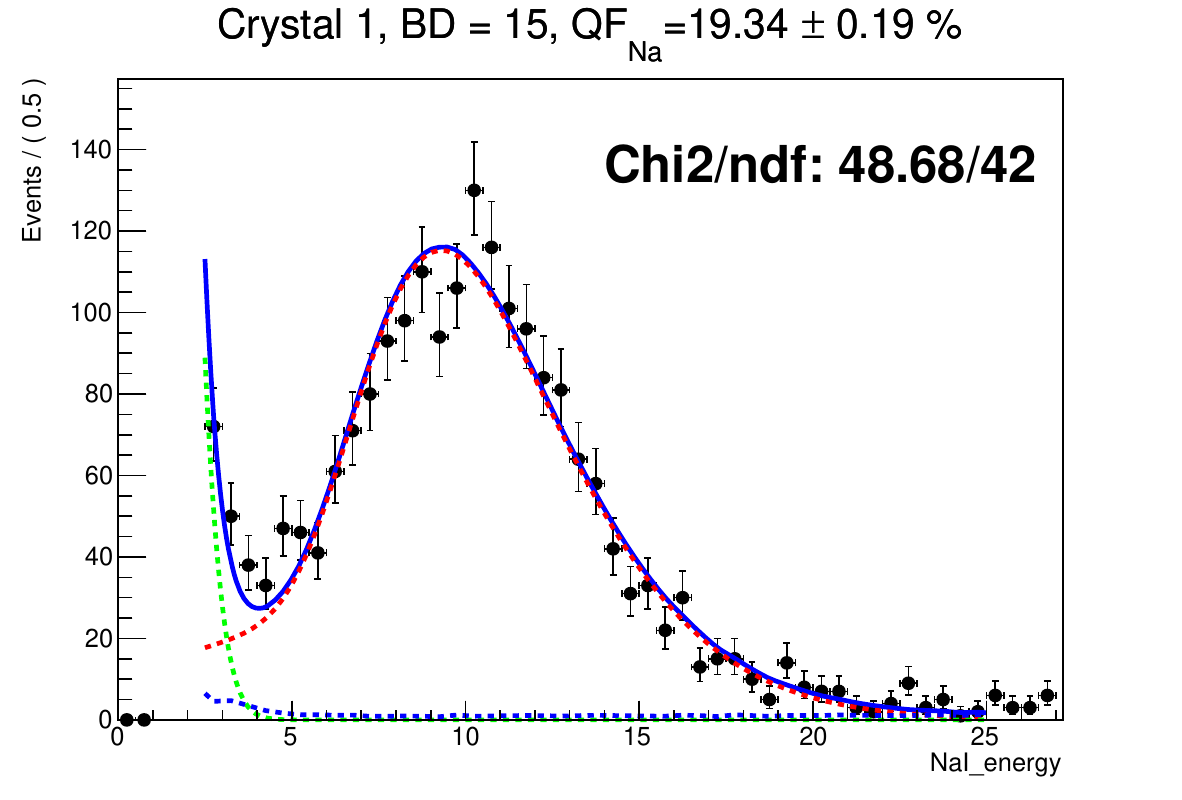}
		\end{subfigure}
		\begin{subfigure}[b]{0.32\textwidth}
			\includegraphics[width=\textwidth]{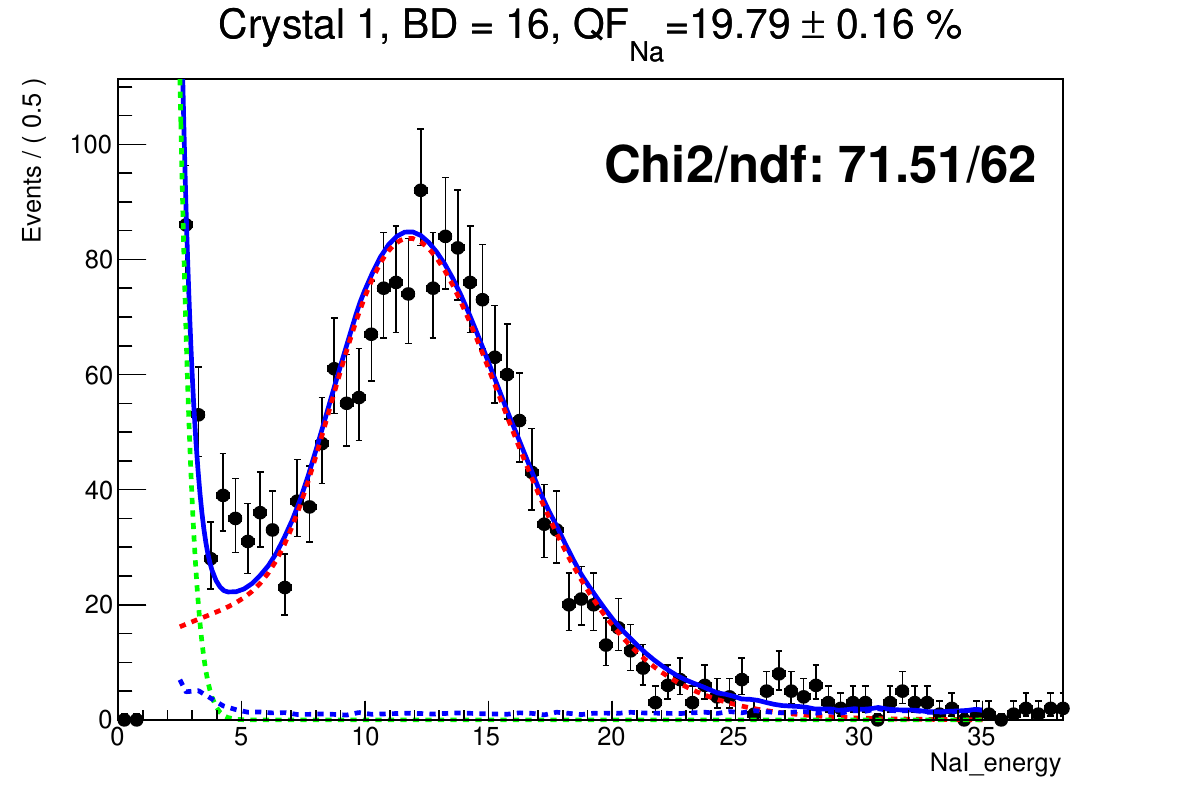}
		\end{subfigure}
		\begin{subfigure}[b]{0.32\textwidth}
			\includegraphics[width=\textwidth]{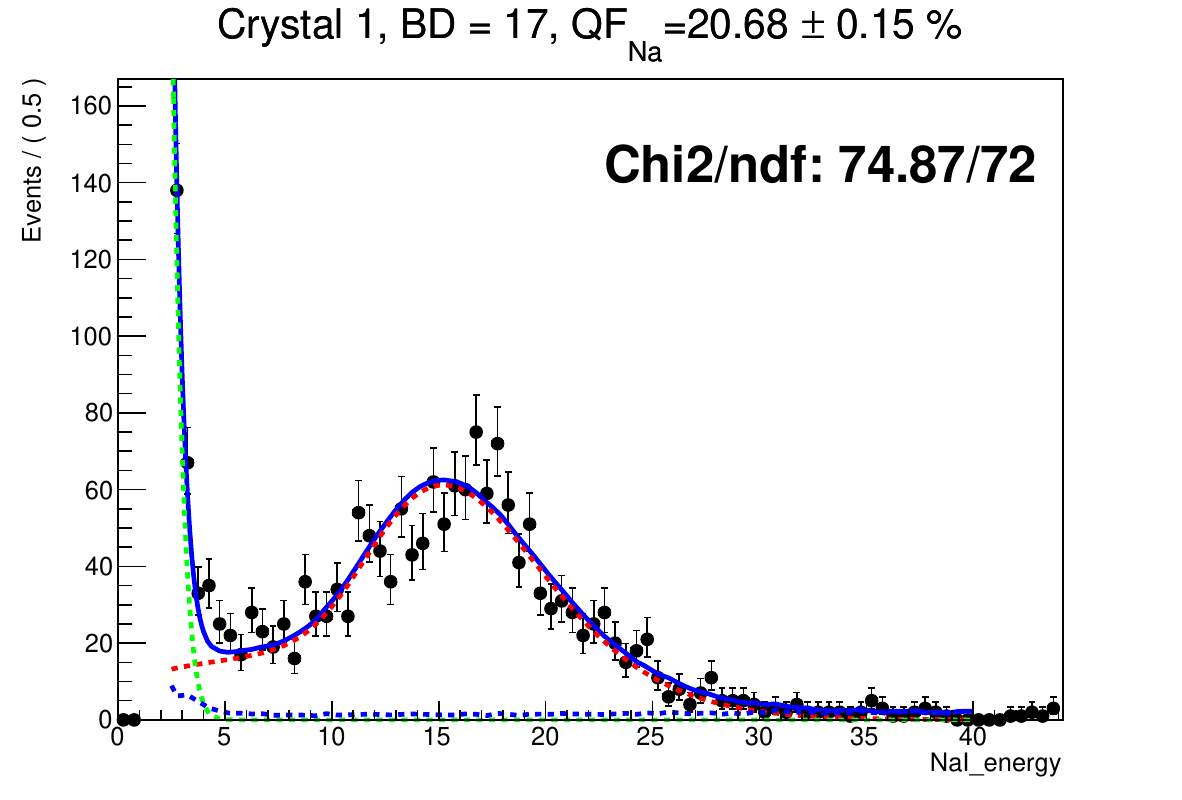}
		\end{subfigure}
	\end{center}
	\caption{\label{QfFit_Cr1}Fits of the experimental data to simulated distributions for crystal~1 for all the channels allowing the identification of the sodium nuclear recoil distribution: from $\#$~0 to $\#$~6 and from $\#$~11 to $\#$~17, corresponding to scattering angles from 92$^o$ to 31$^o$ and sodium nuclear recoil energies from 80~keV to 14~keV. Dashed lines represent the three contributions (blue for the background, red for sodium recoils and green for iodine recoils), while continuous line is the sum of the three PDFs.}
\end{figure}

\begin{figure}[]
	\begin{center}
		\begin{subfigure}[b]{0.34\textwidth}
			\includegraphics[width=\textwidth]{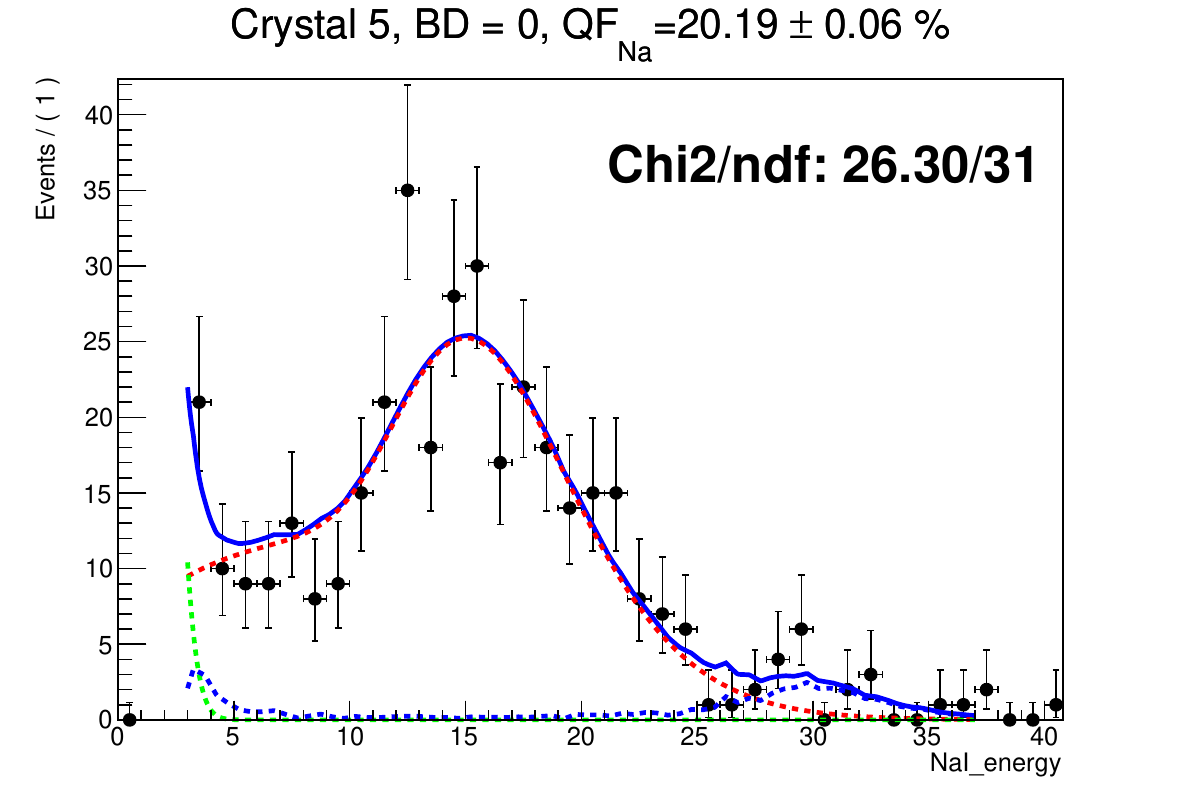}
		\end{subfigure}
		\begin{subfigure}[b]{0.34\textwidth}
			\includegraphics[width=\textwidth]{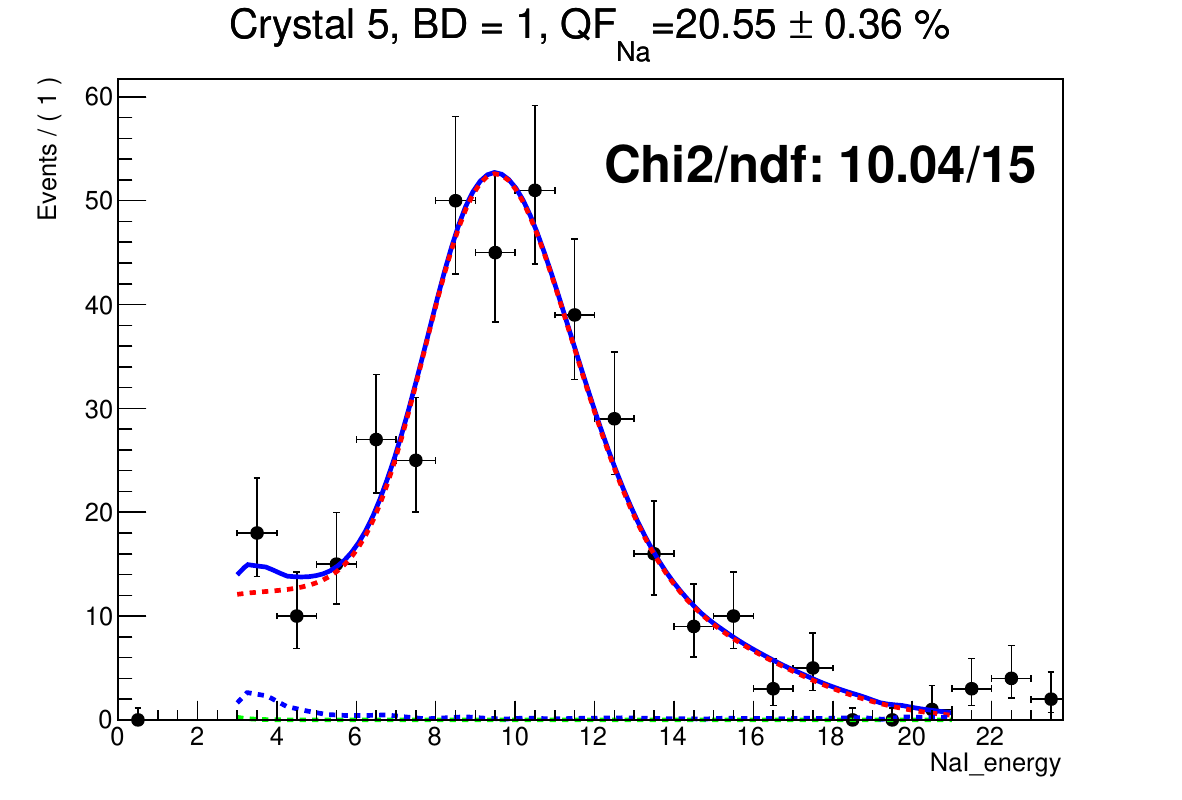}
		\end{subfigure}
		\begin{subfigure}[b]{0.34\textwidth}
			\includegraphics[width=\textwidth]{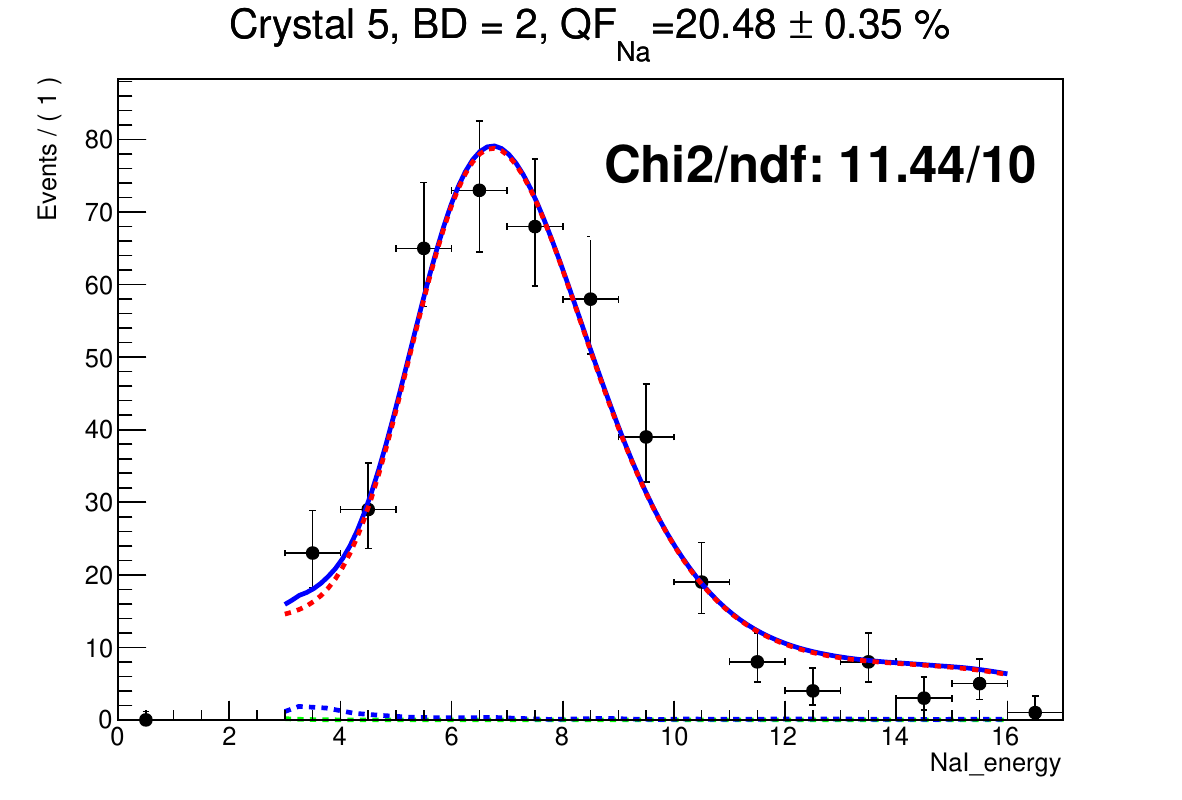}
		\end{subfigure}
		\begin{subfigure}[b]{0.34\textwidth}
			\includegraphics[width=\textwidth]{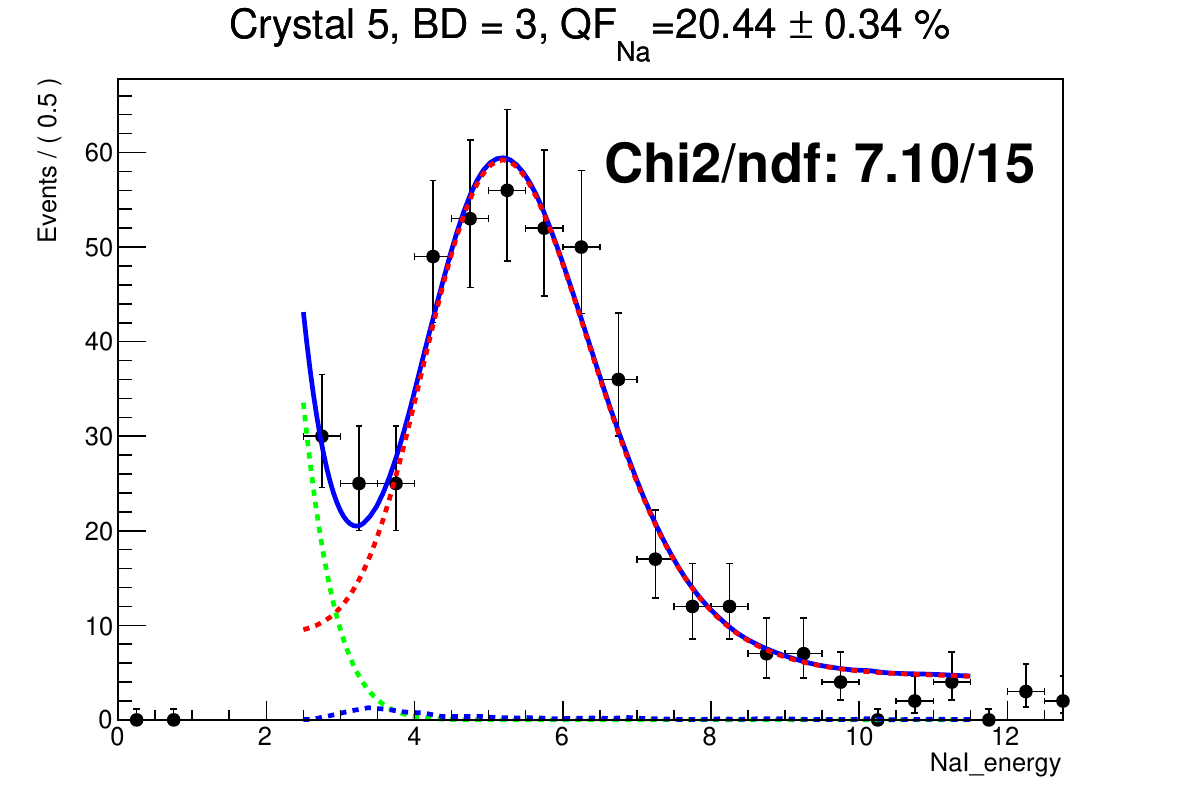}
		\end{subfigure}
		\begin{subfigure}[b]{0.34\textwidth}
			\includegraphics[width=\textwidth]{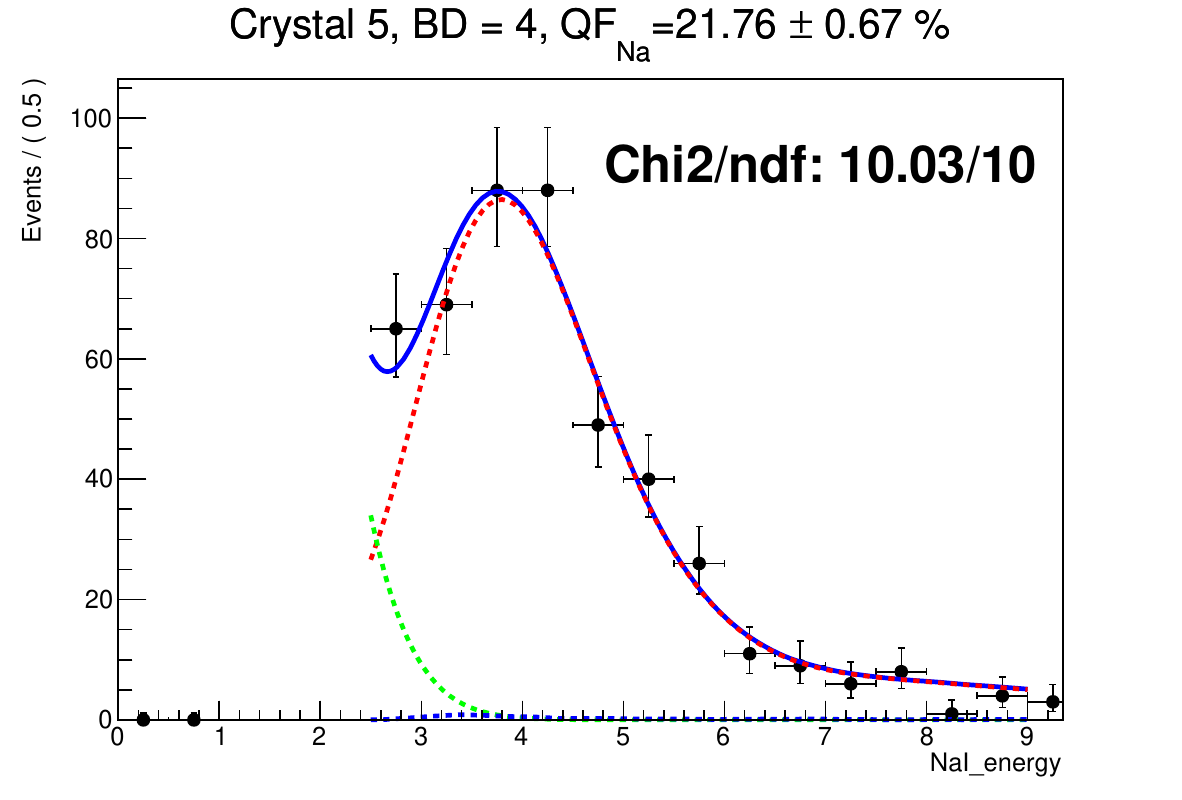}
		\end{subfigure}
		\begin{subfigure}[b]{0.34\textwidth}
			\includegraphics[width=\textwidth]{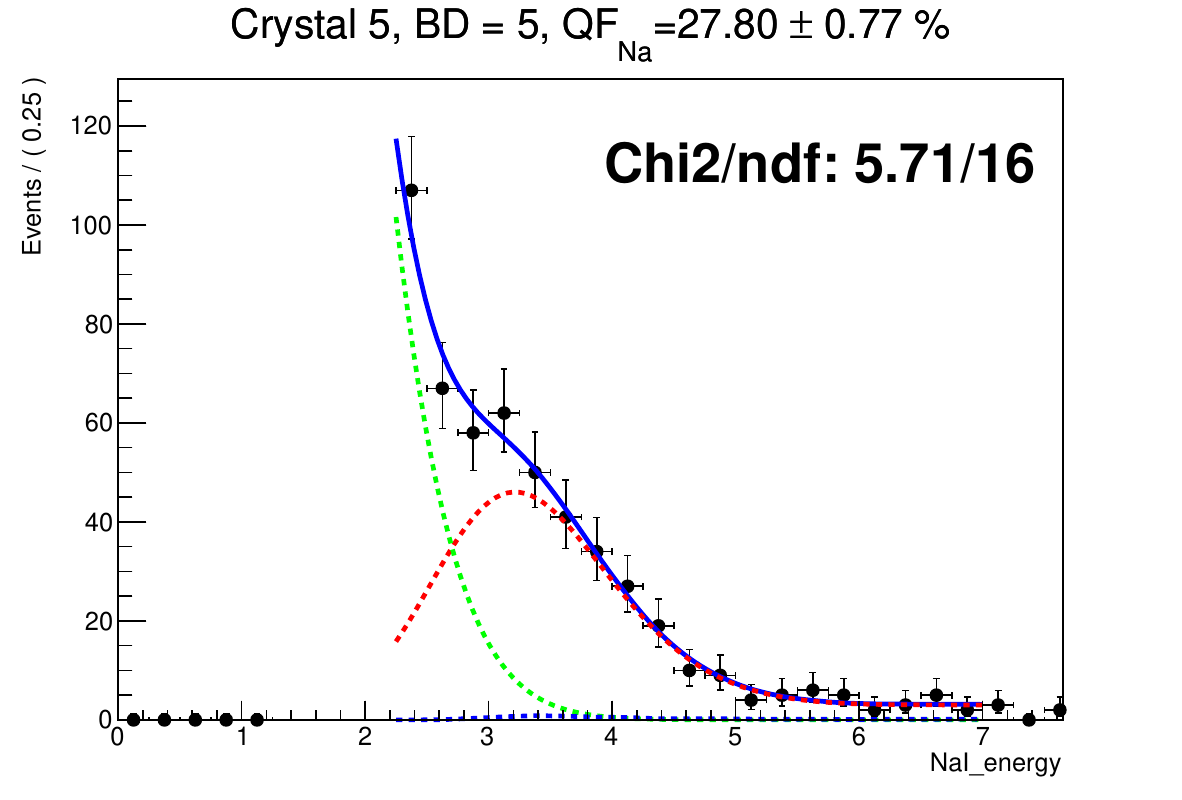}
		\end{subfigure}
		\begin{subfigure}[b]{0.34\textwidth}
			\includegraphics[width=\textwidth]{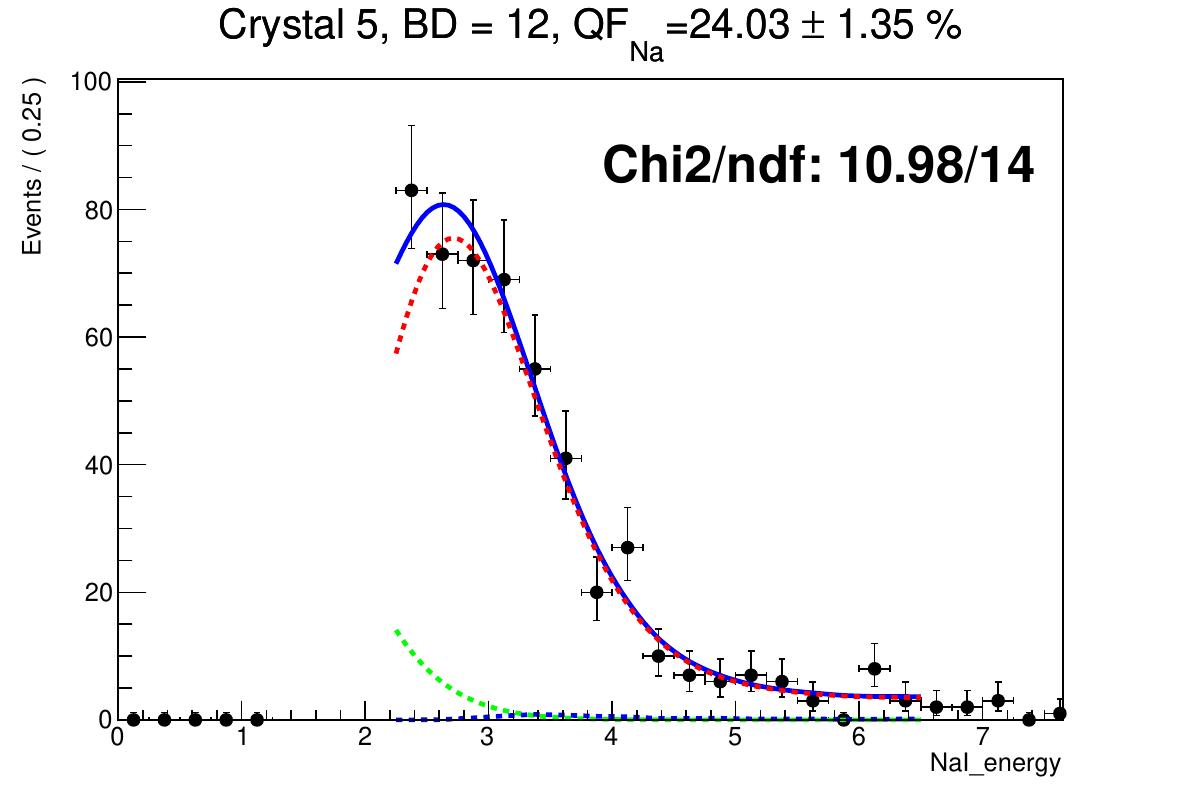}
		\end{subfigure}
		\begin{subfigure}[b]{0.34\textwidth}
			\includegraphics[width=\textwidth]{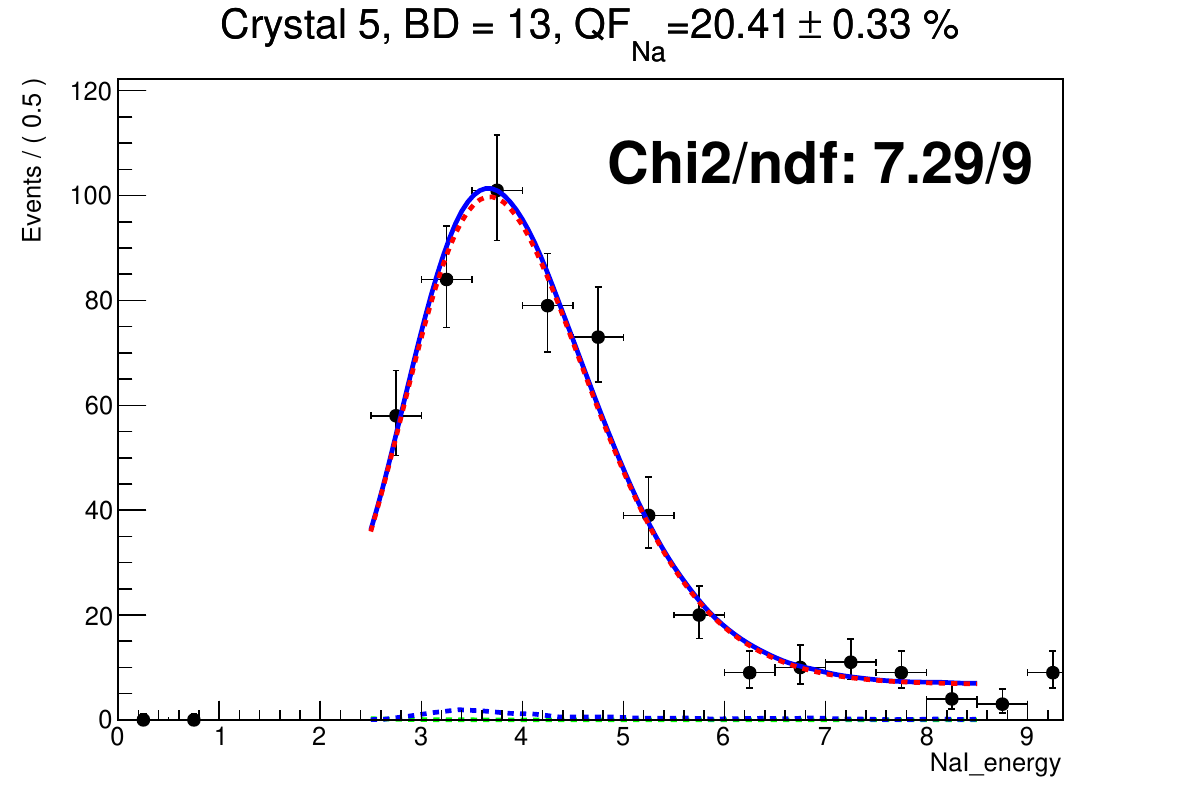}
		\end{subfigure}
		\begin{subfigure}[b]{0.34\textwidth}
			\includegraphics[width=\textwidth]{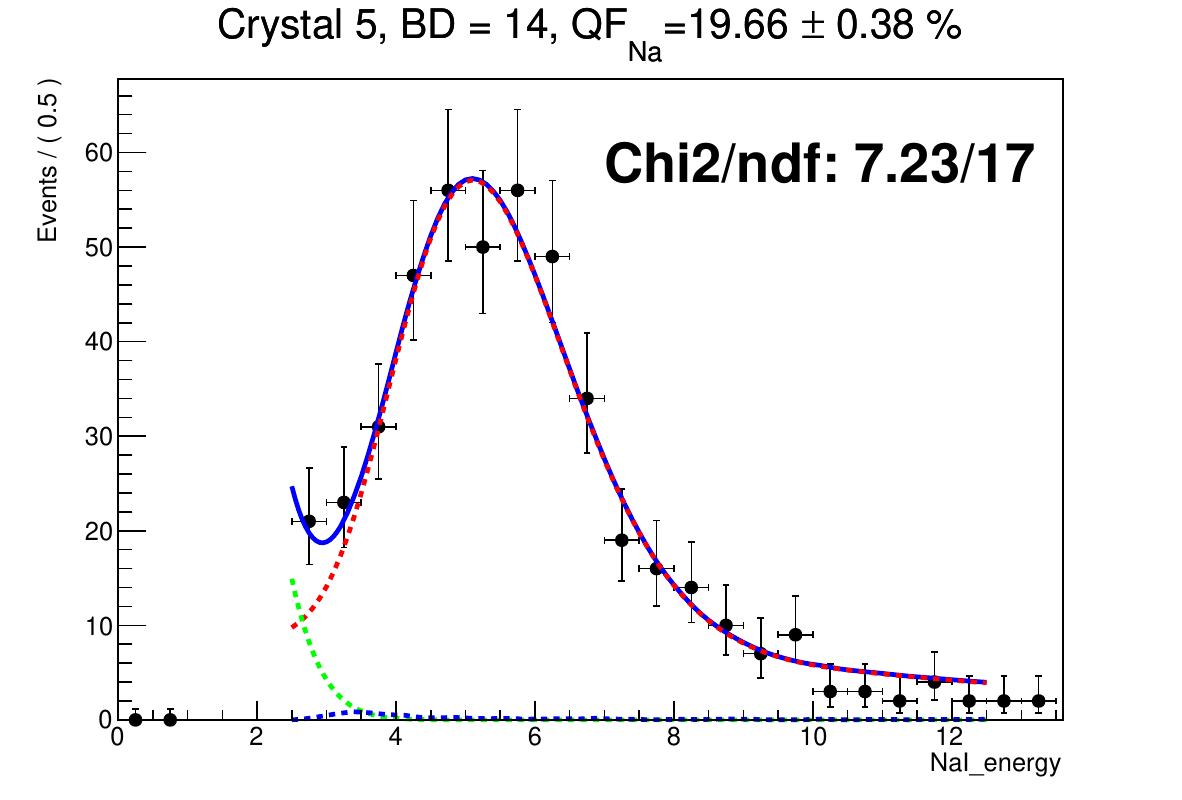}
		\end{subfigure}
		\begin{subfigure}[b]{0.34\textwidth}
			\includegraphics[width=\textwidth]{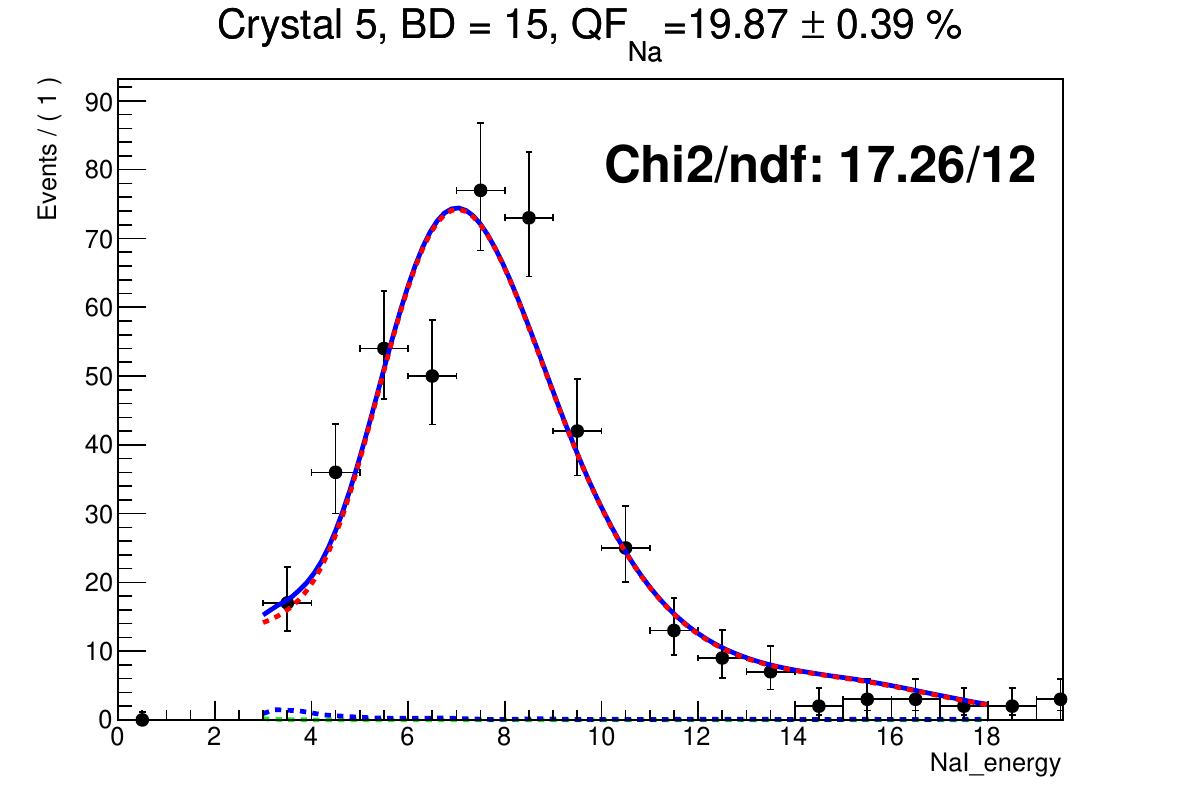}
		\end{subfigure}
		\begin{subfigure}[b]{0.34\textwidth}
			\includegraphics[width=\textwidth]{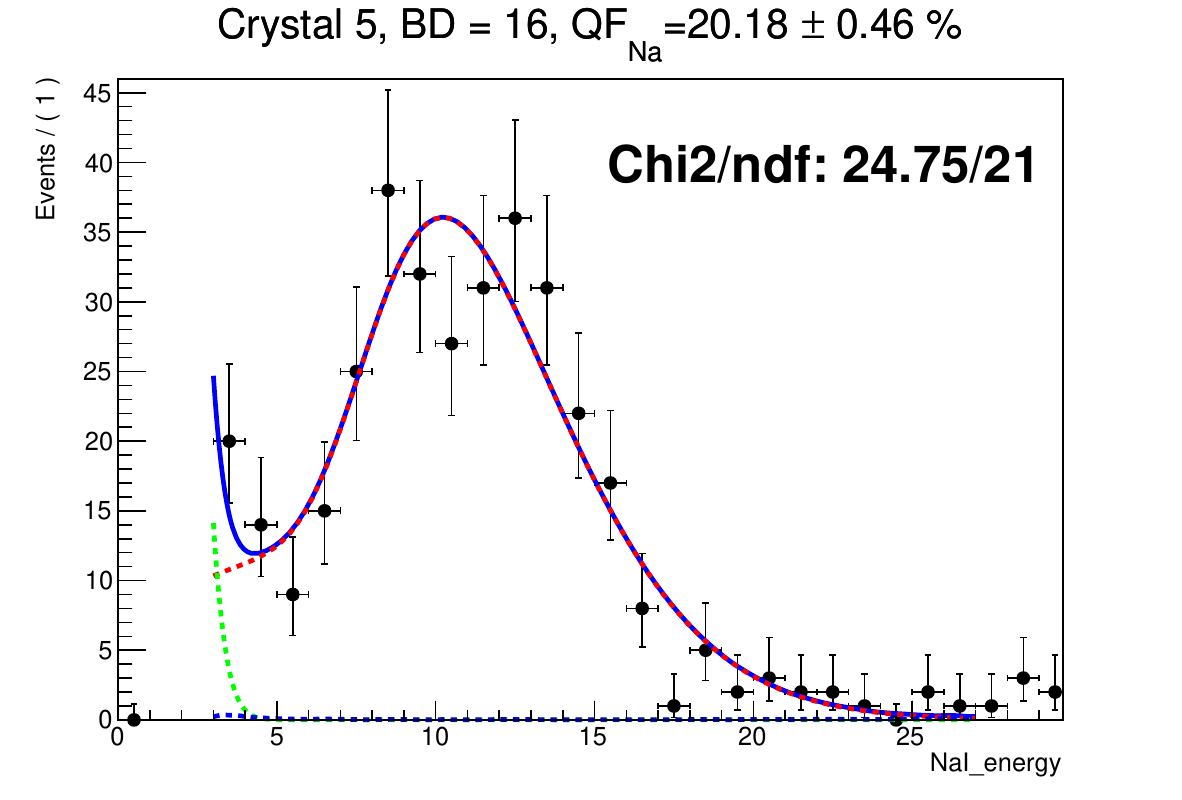}
		\end{subfigure}
		\begin{subfigure}[b]{0.34\textwidth}
			\includegraphics[width=\textwidth]{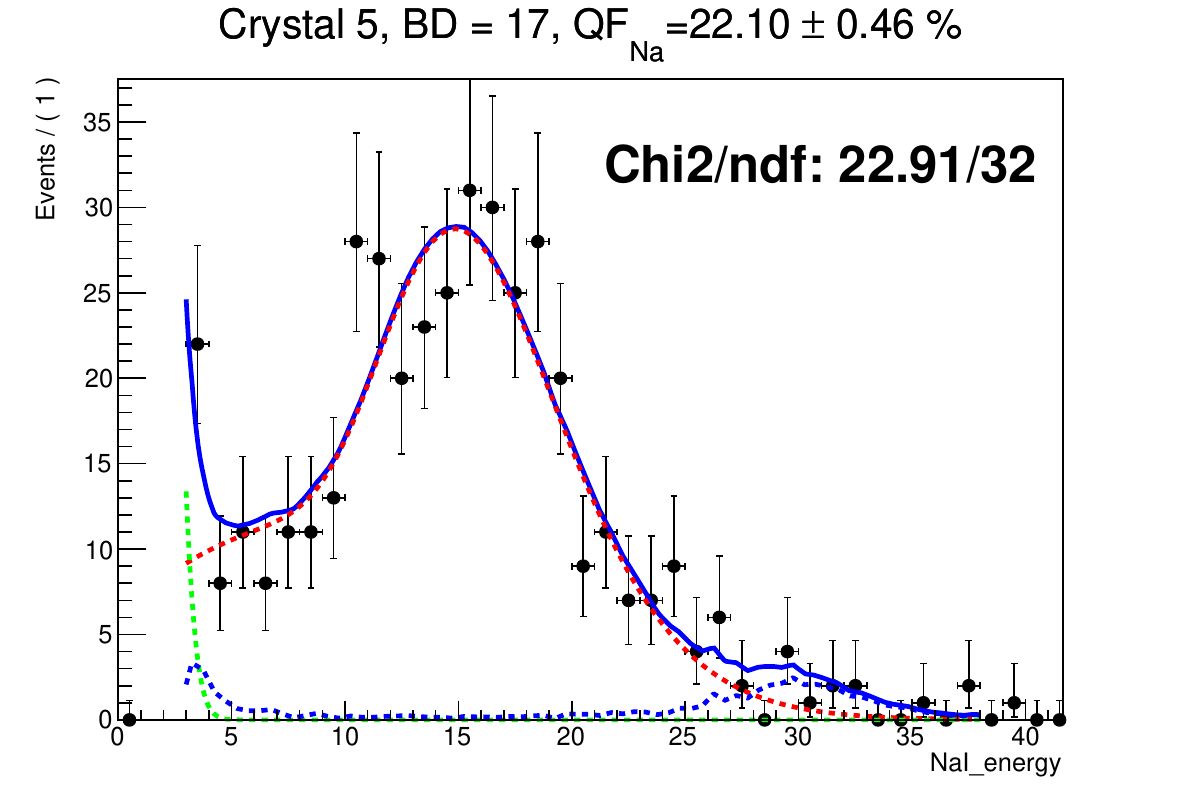}
		\end{subfigure}
	\end{center}
	\caption{\label{QfFit_Cr5}Fits of the experimental data to simulated distributions for crystal~5 for all the channels allowing the identification of the sodium nuclear recoil distribution: from $\#$~0 to $\#$~5 and from $\#$~12 to $\#$~17, corresponding to scattering angles from 91$^o$ to 28$^o$ and sodium nuclear recoil energies from 81 to 12~keV. Dashed lines represent the three contributions (blue for the background, red for sodium recoils and green for iodine recoils), while continuous line is the sum of the three PDFs.}
\end{figure}

For each crystal, this procedure was followed for all the channels where the sodium recoil peak could be disentangled from the background and the iodine recoil spectrum, applying both resolution functions. As examples of the fits, Figure~\ref{QfFit_Cr1} presents those for crystal~1, while Figure~\ref{QfFit_Cr5} presents those for crystal~5, which was taken from the same ingot of the ANAIS-112 crystals. In both figures, the resolution function applied in the fit is that dependent on the square root of the energy.

\begin{figure}[h!]
	\begin{subfigure}[b]{0.49\textwidth}
		\includegraphics[width=\textwidth]{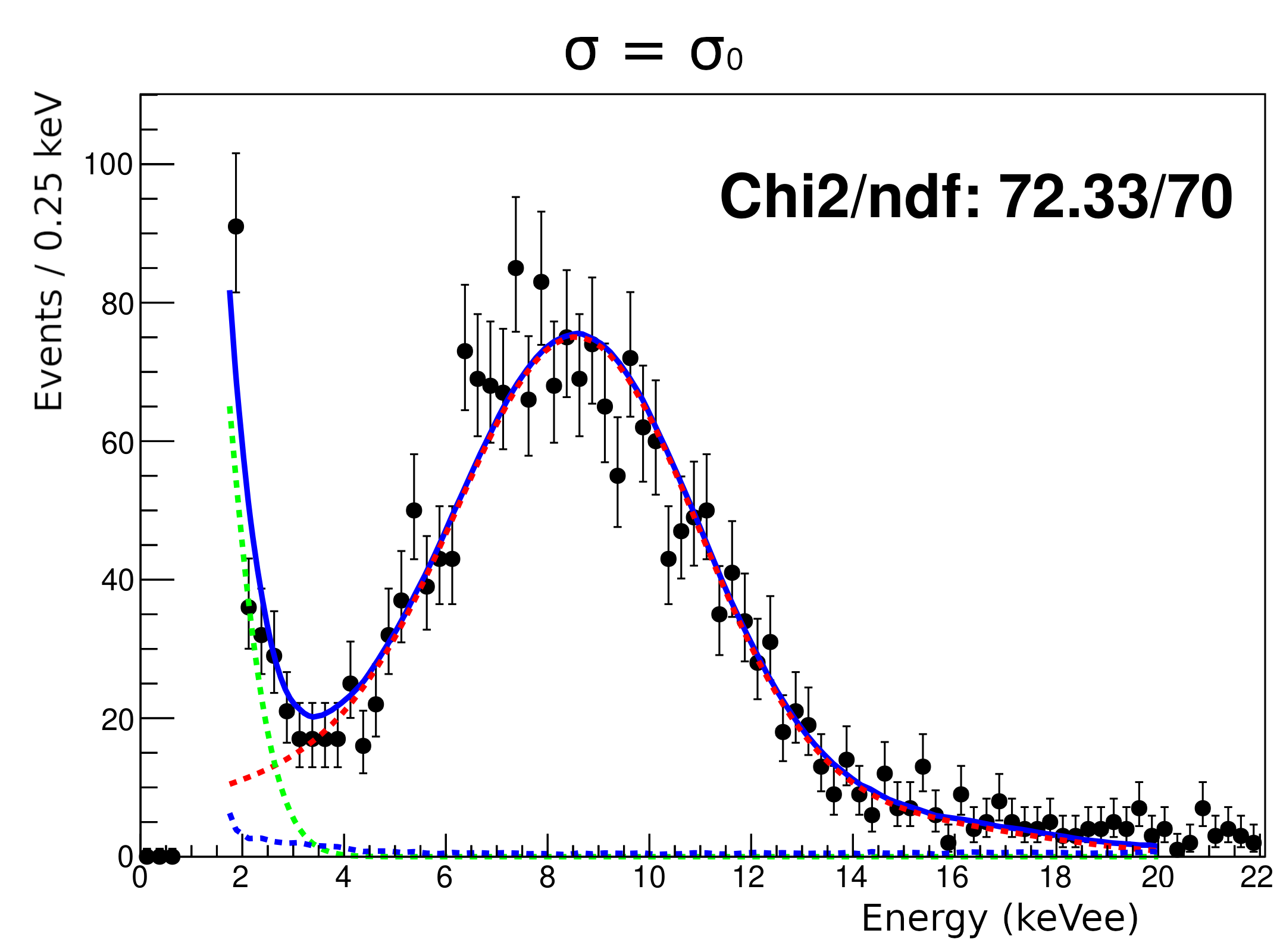}
	\end{subfigure}
	\begin{subfigure}[b]{0.49\textwidth}
		\includegraphics[width=\textwidth]{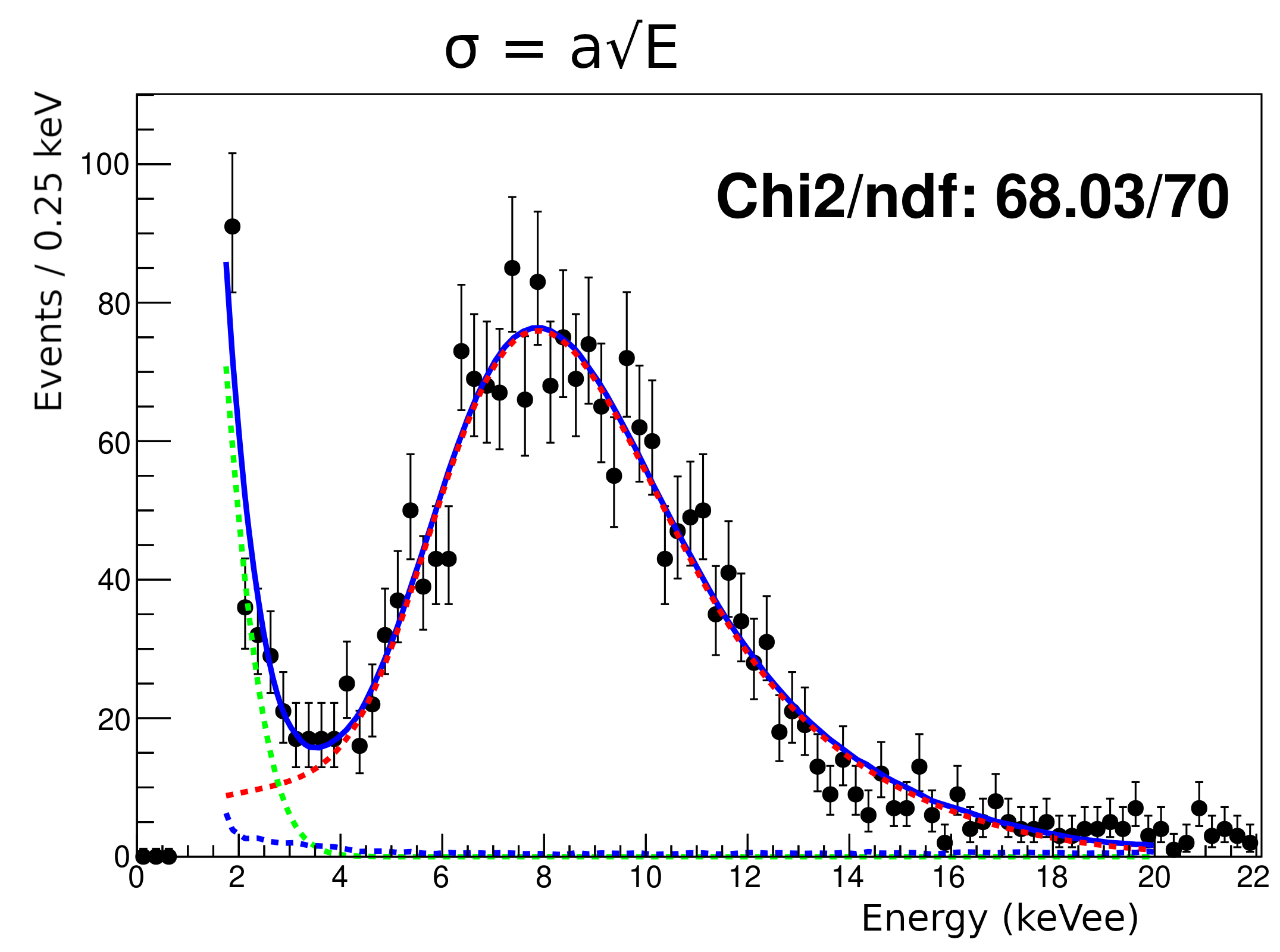}
	\end{subfigure}
	\caption{\label{QF_Fits_EnergyRes}Fits of the experimental data to simulated distributions for crystal~2 for channel~3 using each resolution function.}
\end{figure}

\begin{figure}[h!]
	\begin{center}
		\includegraphics[width=0.75\textwidth]{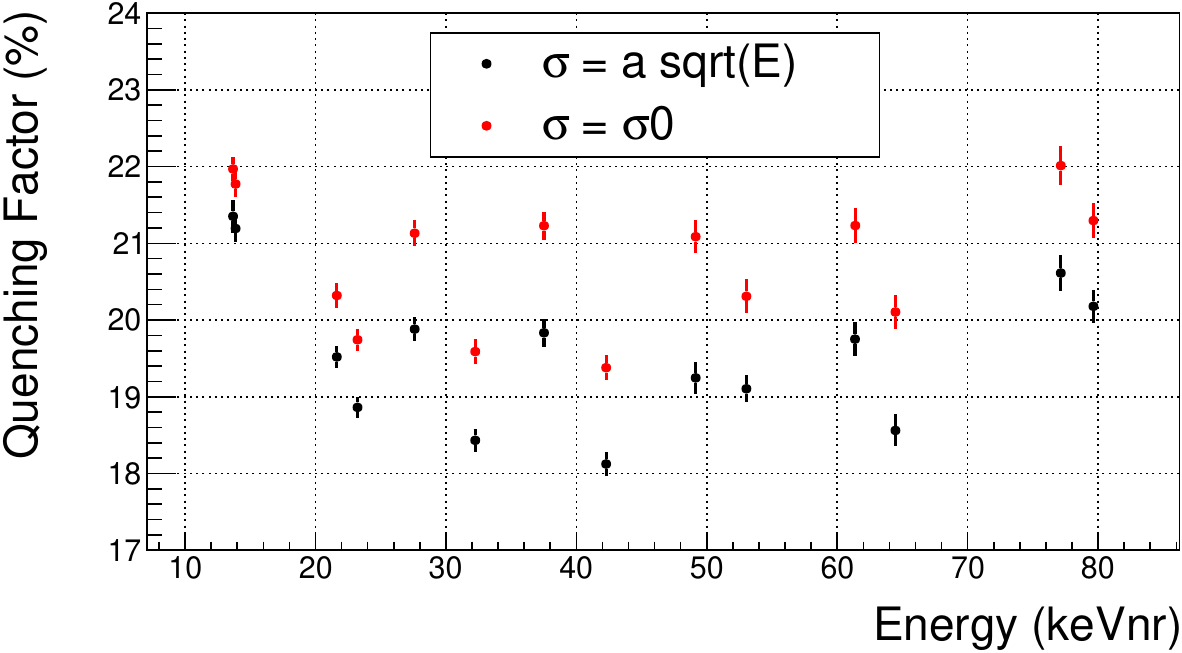}
		\caption{\label{QF_vs_RES}Sodium QF results for the fits of the crystal~1 applying both resolution models: red dots for $\sigma = \sigma_0$ and black dots for $\sigma = a\sqrt{E}$.}
	\end{center}
\end{figure}

Very similar fits are obtained with the constant energy resolution, as it is compared in Figure~\ref{QF_Fits_EnergyRes} for the channel~3 of the crystal~2. However, the QF results obtained using each resolution function are not compatible to each other within their statistical errors, as it can be observed in Figure~\ref{QF_vs_RES}. In fact, the QF results are systematically lower when the resolution independent on the energy is considered. Therefore, the QF values were calculated as the average of both, and their semi-differences were taken as the associated systematical error. For a given channel, the statistical error of the averaged QF was obtained by propagating the individual errors from both fits.

Apart from the systematical error associated with the resolution function, we have analyzed other systematics contributing to the QF calculation: the uncertainty in the positions of the components of the experiment (source, crystal and BDs), the uncertainty in the electron equivalent energy calibration and the value selected for the QF of the iodine and the resolution considered, $\sigma_I$, for the iodine recoils. For the estimate of the first contribution, simulations with the neutron source, NaI(Tl) crystal and BDs displaced for their nominal positions within their corresponding uncertainties (as explained in Section~\ref{Section:QF_GEANT4_Results_EnrDistr}) were carried out. Thus, the results of these simulations for each crystal and channel were used for a fitting similar to the previously explained, and corresponding QF values were obtained. For the second contribution, two fits were done calibrating the spectra with energy calibration functions obtained by pushing the calibration parameters within their standard deviations at 1~sigma. For the third contribution, fits were done fixing the QF of the iodine to 1$\%$ and to 9$\%$ and the resolution applied to the iodine recoil energy distribution to 0.8~keV and to 1.2~keV. The corresponding systematical uncertainties of each contribution were calculated as the difference between the QF values obtained in these fits and the one obtained in the original situation, for each channel. All of these fits were done applying both resolution functions and fixing the resolution parameters to that obtained in the original fits. 

The systematical errors associated to the selection of the QF and the resolution of the iodine recoils were from one to two orders of magnitude lower than the statistical uncertainty, and therefore they were not considered in the error propagation. The uncertainties of the other two contributions were found to be compatible for both resolution functions applied, and therefore the maximum of them was considered in the error calculation for each channel. They were also computed as symmetrical by considering the total uncertainties of the contribution as the maximum between the upper and lower errors. Finally, the three systematical uncertainties were combined with the statistical contribution to obtain the total error in each calculated QF value. As an example, Table~\ref{tabla:QF_errors} shows each contribution for the fits of the crystal~1 applying the second calibration method (similar values are obtained for these contributions using the first calibration method). That table also shows the combination of the contribution of the $QF_I$ and the $\sigma_I$ to the error (although they were not considered in the error propagation) just to show the order of magnitude of those errors.

\begin{table}[h!]
	\centering
	\begin{tabular}{|c|c|c|c|c|c|c|}
		\cline{2-7}
		\multicolumn{1}{c|}{} & \multicolumn{6}{|c|}{Uncertainties (\%)} \\
		\cline{2-7}
		\multicolumn{1}{c|}{} & \multirow{2}{*}{Statistical} & \multicolumn{4}{|c|}{Systematical} & \multirow{2}{*}{Total} \\
		\cline{1-1} \cline{3-6}
		BD~$\#$ &  & Calibration & Positions & Resolution & $QF_I + \sigma_I$ &  \\
		\hline
		0 & 0.13 & 0.07 & 0.25 & 0.57 & 0.02 & 0.64 \\
		1 & 0.11 & 0.07 & 0.40 & 0.78 & 0.01 & 0.89 \\
		2 & 0.12 & 0.10 & 0.22 & 0.61 & 0.01 & 0.67 \\
		3 & 0.08 & 0.13 & 0.24 & 0.62 & 0.01 & 0.68 \\
		4 & 0.10 & 0.15 & 0.24 & 0.59 & 0.03 & 0.66 \\
		5 & 0.09 & 0.19 & 0.23 & 0.45 & 0.01 & 0.55 \\
		6 & 0.11 & 0.28 & 0.39 & 0.35 & 0.01 & 0.60 \\
		11 & 0.13 & 0.27 & 0.42 & 0.28 & 0.01 & 0.59 \\
		12 & 0.10 & 0.19 & 0.41 & 0.41 & 0.02 & 0.62 \\
		13 & 0.11 & 0.17 & 0.37 & 0.63 & 0.02 & 0.76 \\
		14 & 0.10 & 0.10 & 0.36 & 0.70 & 0.02 & 0.80 \\
		15 & 0.14 & 0.14 & 0.25 & 0.92 & 0.03 & 0.97 \\
		16 & 0.14 & 0.09 & 0.40 & 0.77 & 0.03 & 0.88 \\
		17 & 0.14 & 0.08 & 0.19 & 0.70 & 0.02 & 0.74 \\
		\hline
	\end{tabular} \\
	\caption{Error contributions for the fits of the crystal~1 for the first calibration method. The combination of the contribution of the $QF_I$ and the $\sigma_I$ to the error is also presented to show their order of magnitude, although they were not considered in the error propagation. Similar uncertainties were obtained when applying the second calibration method.}
	\label{tabla:QF_errors}
\end{table}

\begin{figure}[h!]
	\begin{center}
		\includegraphics[width=\textwidth]{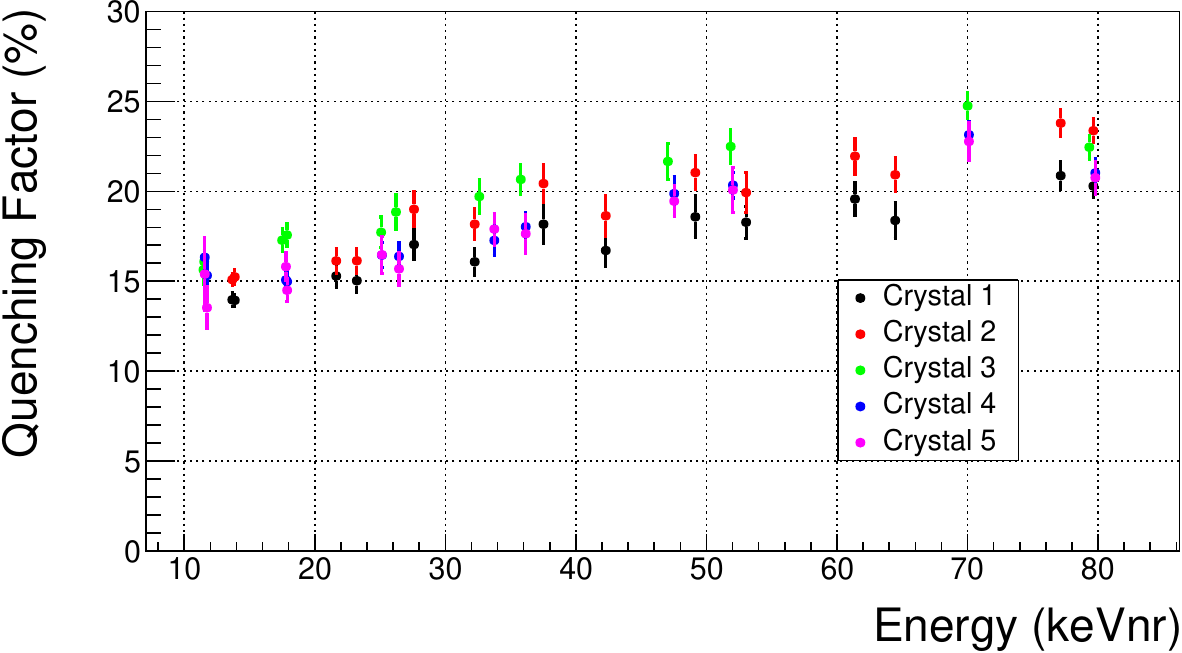}
		\caption{\label{QF_57}Sodium QF results for the 5 crystals using the first calibration method. Uncertainties are the combination of systematical and statistical, as shown in Table~\ref{tabla:QF_errors}.}
	\end{center}
\end{figure}

\begin{figure}[h!]
	\begin{center}
		\includegraphics[width=\textwidth]{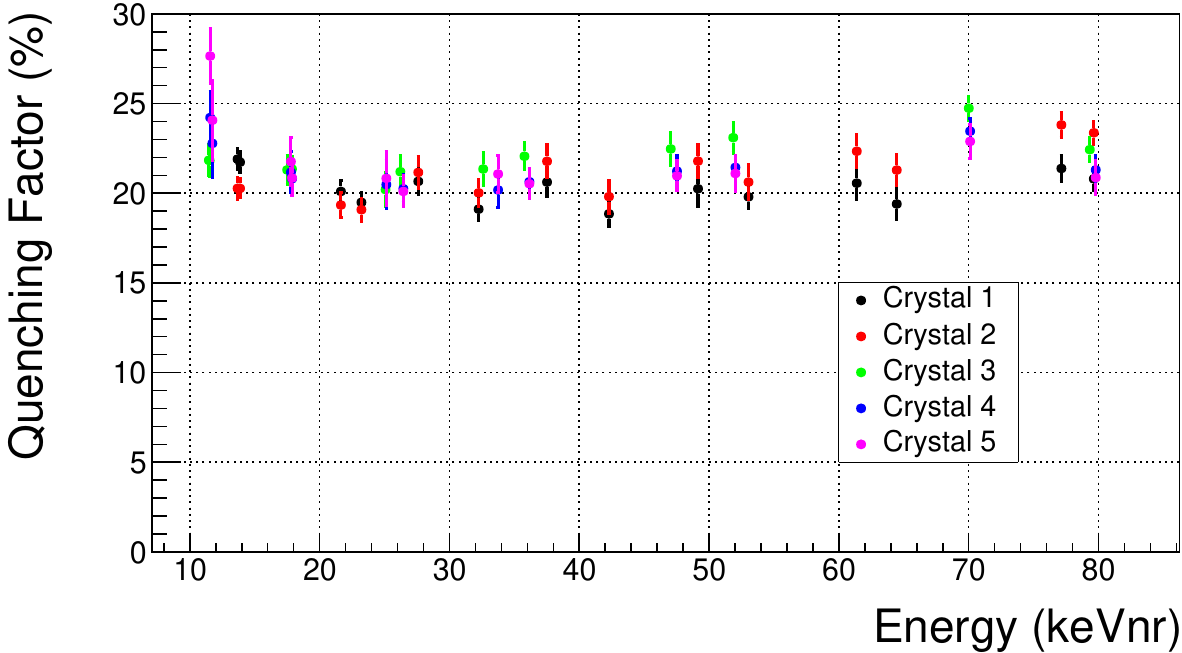}
		\caption{\label{QF_Ba133}Sodium QF results for the 5 crystals using the second calibration method. Uncertainties are the combination of systematical and statistical, as shown in Table~\ref{tabla:QF_errors} (for the first calibration method).}
	\end{center}
\end{figure}

\begin{figure}[h!]
	\begin{center}
		\includegraphics[width=\textwidth]{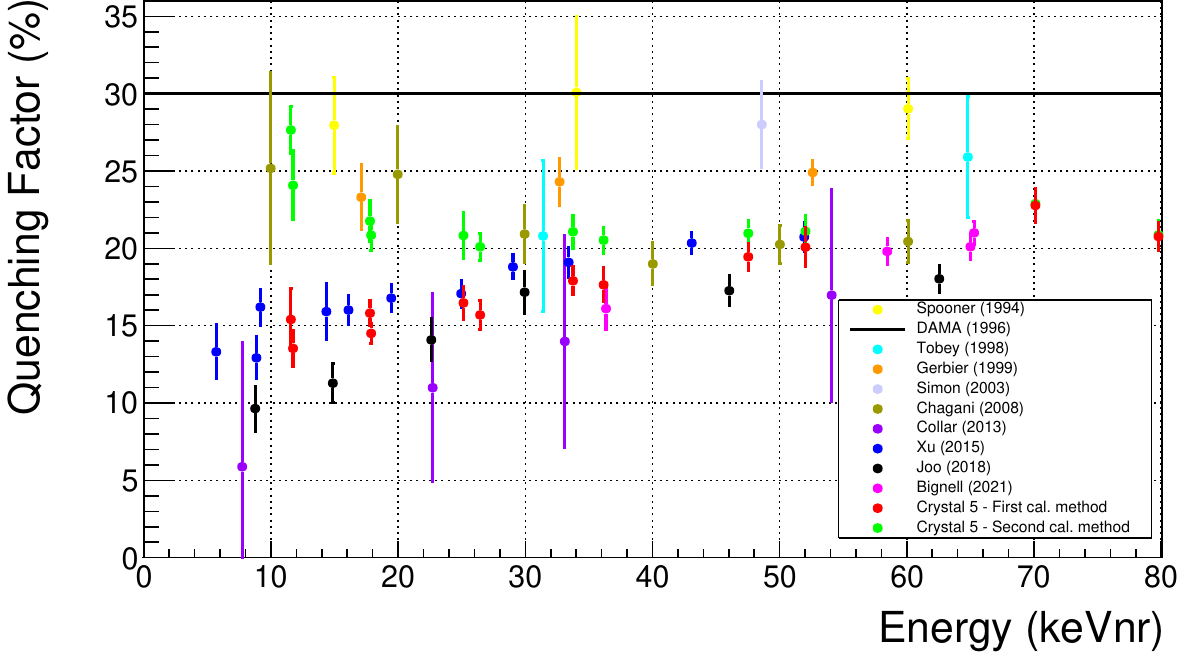}
		\caption{\label{QFComparison}Sodium QF results for the crystal~5 applying both calibrations and the results presented by other experiments~\cite{Spooner,Bernabei:1996vj,Tovey:1998ex,Gerbier:1998dm,Simon:2002cw,Chagani,Collar,Xu,Joo:2018hom,Bignell_2021}.}
	\end{center}
\end{figure}

QF results are shown in Figures~\ref{QF_57} and \ref{QF_Ba133}, using calibration methods 1 and 2 (see Section~\ref{Section:QF_Analysis_NaIcal_EnergyCal}), respectively. They are also presented in Table~\ref{tabla:QFvaluesAgo},~\ref{tabla:QFvaluesCr3} and~\ref{tabla:QFvaluesOct}. A comparison between the results of this analysis for the crystal~5 and the results from previous measurements is shown in Figure~\ref{QFComparison}.

\begin{table}[h]
	\centering
	\begin{tabular}{|c|c|c|c|c|c|}
		\cline{3-6}
		\multicolumn{2}{c}{} & \multicolumn{4}{|c|}{Quenching Factor ($\%$)} \\
		\cline{3-6}
		\multicolumn{2}{c}{} & \multicolumn{2}{|c|}{Calibration method 1} & \multicolumn{2}{|c|}{Calibration method 2}\\
		\hline
		Ch. & Mean $E_{nr}$~(keV) & Crystal 1 & Crystal 2 & Crystal 1 & Crystal 2 \\
		\hline
		0 & 79.64 & 20.29$\pm$0.65 & 23.37$\pm$0.68 & 20.80$\pm$0.63 & 23.37$\pm$0.65 \\
		1 & 64.46 & 18.38$\pm$1.00 & 20.92$\pm$0.94 & 19.40$\pm$0.87 & 21.29$\pm$0.88 \\
		2 & 53.04 & 18.28$\pm$0.88 & 19.93$\pm$1.10 & 19.80$\pm$0.66 & 20.62$\pm$0.96 \\
		3 & 42.29 & 16.71$\pm$0.91 & 18.64$\pm$1.13 & 18.85$\pm$0.68 & 19.81$\pm$0.91 \\
		4 & 32.25 & 16.08$\pm$0.76 & 18.17$\pm$0.85 & 19.11$\pm$0.65 & 20.02$\pm$0.75 \\
		5 & 23.22 & 15.03$\pm$0.63 & 16.14$\pm$0.70 & 19.49$\pm$0.54 & 19.08$\pm$0.64 \\
		6 & 13.68 & 13.97$\pm$0.38 & 15.09$\pm$0.33 & 21.90$\pm$0.59 & 20.27$\pm$0.59 \\
		11 & 13.87 & 13.94$\pm$0.32 & 15.24$\pm$0.39 & 21.73$\pm$0.58 & 20.27$\pm$0.52 \\
		12 & 21.64 & 15.29$\pm$0.63 & 16.13$\pm$0.68 & 20.10$\pm$0.61 & 19.34$\pm$0.69 \\
		13 & 27.6 & 17.04$\pm$0.83 & 19.00$\pm$0.96 & 20.66$\pm$0.75 & 21.16$\pm$0.87 \\
		14 & 37.52 & 18.17$\pm$1.05 & 20.43$\pm$1.04 & 20.62$\pm$0.79 & 21.78$\pm$0.92 \\
		15 & 49.15 & 18.58$\pm$1.17 & 21.04$\pm$0.96 & 20.25$\pm$0.96 & 21.79$\pm$0.92 \\
		16 & 61.38 & 19.57$\pm$0.91 & 21.95$\pm$1.00 & 20.56$\pm$0.88 & 22.34$\pm$0.92 \\
		17 & 77.14 & 20.87$\pm$0.77 & 23.79$\pm$0.74 & 21.38$\pm$0.74 & 23.81$\pm$0.71 \\
		\hline
	\end{tabular} \\
	\caption{Sodium QF values obtained for crystals 1 and 2 using both NaI(Tl) energy calibrations.}
	\label{tabla:QFvaluesAgo}
\end{table}

\begin{table}[h]
	\centering
	\begin{tabular}{|c|c|c|c|}
		\cline{3-4}
		\multicolumn{2}{c}{} & \multicolumn{2}{|c|}{Quenching Factor ($\%$)} \\
		\hline
		Ch. & Mean $E_{nr}$~(keV) & Calibration method 1 & Calibration method 2 \\
		\hline
		0 & 79.33 & 22.45$\pm$0.67 & 22.43$\pm$0.65 \\
		1 & 47.05 & 21.66$\pm$1.00 & 22.47$\pm$0.93 \\
		2 & 32.60 & 19.71$\pm$0.94 & 21.35$\pm$0.92 \\
		3 & 25.08 & 17.72$\pm$0.85 & 20.20$\pm$0.93 \\
		4 & 17.50 & 17.28$\pm$0.65 & 21.30$\pm$0.81 \\
		5 & 11.53 & 16.09$\pm$0.61 & 21.82$\pm$0.90 \\
		12 & 11.47 & 15.64$\pm$0.82 & 21.84$\pm$0.87 \\
		13 & 17.87 & 17.57$\pm$0.62 & 21.35$\pm$0.72 \\
		14 & 26.22 & 18.85$\pm$0.97 & 21.20$\pm$0.91 \\
		15 & 35.78 & 20.66$\pm$0.84 & 22.06$\pm$0.76 \\
		16 & 51.86 & 22.49$\pm$0.96 & 23.10$\pm$0.86 \\
		17 & 70.01 & 24.75$\pm$0.72 & 24.74$\pm$0.67 \\
		\hline
	\end{tabular} \\
	\caption{Sodium QF values obtained for crystal 3 using both NaI(Tl) energy calibrations.}
	\label{tabla:QFvaluesCr3}
\end{table}

\begin{table}[h]
	\centering
	\begin{tabular}{|c|c|c|c|c|c|}
		\cline{3-6}
		\multicolumn{2}{c}{} & \multicolumn{4}{|c|}{Quenching Factor ($\%$)} \\
		\cline{3-6}
		\multicolumn{2}{c}{} & \multicolumn{2}{|c|}{Calibration method 1} & \multicolumn{2}{|c|}{Calibration method 2}\\
		\hline
		Ch. & Mean $E_{nr}$~(keV) & Crystal 4 & Crystal 5 & Crystal 4 & Crystal 5 \\
		\hline
		0 & 79.79 & 21.02$\pm$0.78 & 20.76$\pm$0.91 & 21.30$\pm$0.80 & 20.87$\pm$0.94 \\
		1 & 47.53 & 19.87$\pm$0.94 & 19.45$\pm$0.87 & 21.23$\pm$0.89 & 20.97$\pm$0.84 \\
		2 & 33.75 & 17.27$\pm$0.86 & 17.90$\pm$0.89 & 20.18$\pm$0.98 & 21.07$\pm$1.05 \\
		3 & 25.14 & 16.45$\pm$0.71 & 16.47$\pm$1.04 & 20.48$\pm$1.32 & 20.83$\pm$1.50 \\
		4 & 17.79 & 15.07$\pm$0.62 & 15.81$\pm$0.78 & 21.17$\pm$1.15 & 21.76$\pm$1.34 \\
		5 & 11.57 & 16.33$\pm$0.78 & 15.40$\pm$2.01 & 24.22$\pm$1.43 & 27.65$\pm$1.54 \\
		12 & 11.74 & 15.32$\pm$1.06 & 13.53$\pm$1.18 & 22.78$\pm$1.91 & 24.07$\pm$2.22 \\
		13 & 17.88 & 14.98$\pm$0.63 & 14.50$\pm$0.64 & 20.81$\pm$0.75 & 20.85$\pm$0.98 \\
		14 & 26.45 & 16.38$\pm$0.75 & 15.69$\pm$0.92 & 20.25$\pm$0.79 & 20.10$\pm$0.86 \\
		15 & 36.16 & 18.03$\pm$0.81 & 17.64$\pm$1.09 & 20.64$\pm$0.69 & 20.53$\pm$0.84 \\
		16 & 52.03 & 20.34$\pm$0.71 & 20.07$\pm$1.24 & 21.44$\pm$0.61 & 21.10$\pm$1.01 \\
		17 & 70.11 & 23.14$\pm$0.75 & 22.77$\pm$1.07 & 23.46$\pm$0.72 & 22.89$\pm$0.94 \\
		\hline
	\end{tabular} \\
	\caption{Sodium QF values obtained for crystals 4 and 5 using both NaI(Tl) energy calibrations.}
	\label{tabla:QFvaluesOct}
\end{table}

When the response of the detector is considered proportional in energies up to 57.6~keV (first calibration method), it is observed an apparent decrease in QF at low energies, similar to that reported by other experiments (as seen in Figure~\ref{QF_57}). On the other hand, there is no clear dependence with energy of the QF when a non-proportional but linear energy calibration in the ROI of the crystal signal is applied. In fact, both results are not compatible among them at energies below 50~keV, approximately.

Some hint of a QF increase at low energies is observed in Figure~\ref{QF_Ba133}, but those points correspond to channels in which the sodium recoil energy spectrum was partially mixed with the background. Therefore, an additional systematic may be affecting the results, as it is observed in the increase of the uncertainties.

It is interesting to observe that for crystals~1 and~2 (which correspond to August measurements, see Table~\ref{tabla:QFvaluesAgo}), the QF results for BDs placed to the left of the beam (BD~$\geq$~9) are systematically higher than those placed to the right (BD~$\leq$~8). The same behaviour is observed for the two calibrations applied. The reason has not been understood yet. One possibility is that there was an error in BD positions measurements in August run. Another possibility is that the beam was not correctly aligned with the BD configuration, which would reduce the scattering angle value corresponding to BD~$\geq$~9 and increase the corresponding to BD~$\leq$~8.

Considering the independence of the sodium QF on the nuclear recoil energy (obtained when the second calibration method is applied) in the range of energies accessed in these measurements (from 10~keV to 80~keV), the weighted mean values of the QF for each crystal have been obtained. They are shown in Table~\ref{tabla:QFMean}. Additionally, the mean for all the crystals has been also obtained as (21.2~$\pm$~0.8)\%.

\begin{table}[h]
	\centering
	\begin{tabular}{|c|c|}
		\hline
		Crystal & Mean QF (\%) \\
		\hline
		1 & 20.4$\pm$0.7 \\
		2 & 21.0$\pm$0.8 \\
		3 & 22.1$\pm$0.8 \\
		4 & 21.4$\pm$0.9 \\
		5 & 21.6$\pm$1.1 \\
		\hline
		Weighted mean & 21.2$\pm$0.8 \\
		\hline
	\end{tabular} \\
	\caption{Mean QF values obtained for each crystal using the second NaI(Tl) energy calibration method.}
	\label{tabla:QFMean}
\end{table}

\subsection{QF of the iodine nuclei}\label{Section:QF_QFcal_I}

\begin{figure}[h!]
	\begin{center}
		\includegraphics[width=0.75\textwidth]{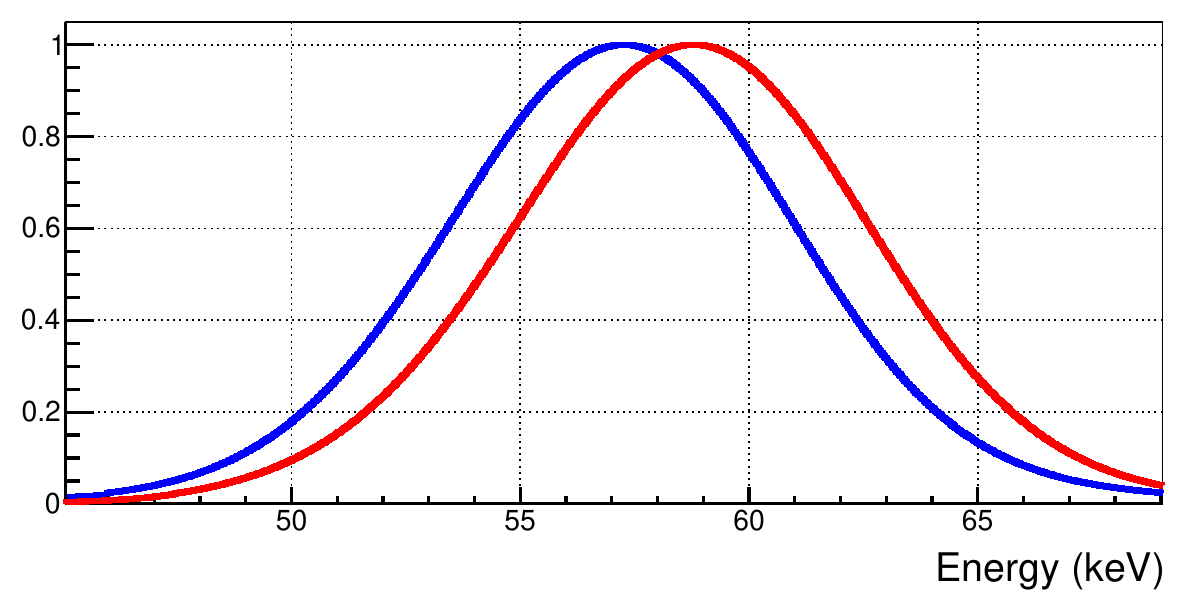}
		\caption{\label{IQF_Sim_57keV_Res}Energy shift expected in the $^{127}I$ inelastic peak for channel $\#$~0 and a QF of 10$\%$ (red line) compared to the gamma nominal energy value. The measured energy resolution for the 57.6 keV peak for the crystal~3 has been considered.}
	\end{center}
\end{figure}

As it was said, it was not possible to disentangle the iodine recoils from background for any channel, so a different strategy for the estimate of the iodine QF was followed. This was performed by studying the $^{127}I$ inelastic peak, which corresponds to the sum of the light produced by the energy depositions of the gamma (57.6~keV) and the iodine recoil, the latter quenched by the corresponding QF. As example, Figure~\ref{IQF_Sim_57keV_Res} shows the energy shift expected in the peak for channel $\#$~0 and a QF of 10$\%$ (red line) compared to the gamma nominal energy value (blue line), taking into account the energy resolution of the NaI(Tl) crystal~3.

In this case the ROI is centered around the 57.6~keV peak, therefore the proportional linear energy calibration (energy calibration method 1) presented in Section~\ref{Section:QF_Analysis_NaIcal_EnergyCal} was applied to the data. From the simulation, channels $\#$~8 and $\#$~9 have iodine recoil energies below 0.2~keV, that result in a negligible shift in electron equivalent energy, and then, they are used to build a reference for each crystal; channels $\#$~0 and $\#$~17, with the highest iodine recoil energies, above 10~keV, will be used for this analysis. The reference is built with events from channels $\#$~8 and $\#$~9 combined. After correcting gain drift (as explained in Section~\ref{Section:QF_Analysis_NaIcal_GainCorrection}) and calibrating in energy (energy calibration method 1, explained in Section~\ref{Section:QF_Analysis_NaIcal_EnergyCal}), the inelastic peak is fitted to a gaussian to obtain the position of the peak and the corresponding statistical uncertainty.

\begin{figure}[h!]
	\begin{center}
		\includegraphics[width=0.75\textwidth]{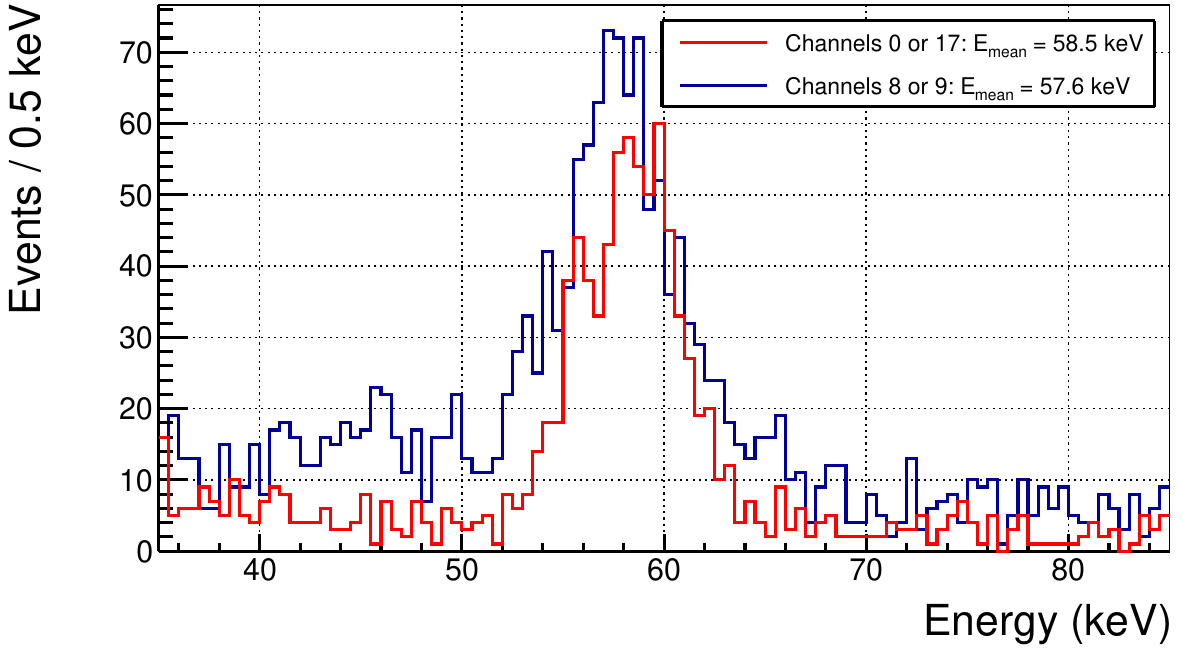}
		\caption{\label{QF_Iodine}Comparison between the $^{127}I$ inelastic peak for the reference channels (8 or 9, blue line) and the analysed channels (0 or 17, red line) for the measurements with the crystal~3.}
	\end{center}
\end{figure}

The difference between the mean energy obtained from the fit for each channel and that from the reference (referred to as \textit{Delta} variable) was calculated. The systematical uncertainty was estimated by changing the reference to only channel $\#$~8, instead of 8+9, and taking the difference between the corresponding delta values as uncertainty (systematical error $\#$~1). Moreover, channels $\#$~0 and $\#$~17 have a similar recoil energy in some of the crystals. In those cases, an additional systematical uncertainty was also estimated from the difference between the \textit{delta} values derived for both channels (systematical error $\#$~2). Results for the peak shift (\textit{Delta} variable) including statistical and systematical uncertainties at 1~$\sigma$ are presented in Table~\ref{tabla:IQF_errors} for crystals~2 and~3. The comparison between the $^{127}I$ inelastic peak for the reference channels (8 or 9) and the analysed channels (0 or 17) for the measurements with the crystal~3 is shown in Figure~\ref{QF_Iodine}. A difference in the mean energy of the distributions (lower than 1~keV) is observed.

\begin{table}[h!]
	\centering
	\begin{tabular}{|c|c|c|c|c|c|}
		\cline{3-6}
		\multicolumn{2}{c}{} & \multicolumn{4}{|c|}{Uncertainties (keV)} \\
		\hline
		Cr.~$\#$ & \textit{Delta} (keV) & Stat. & Sys. $\#$~1 & Sys. $\#$~2 & Total \\
		\hline
		2 & 0.73 & 0.24 & 0.29 & 0.01 & 0.38 \\
		3 & 0.93 & 0.26 & 0.01 & 0.02 & 0.26 \\
		\hline
	\end{tabular} \\
	\caption{Obtained \textit{Delta} values (see text) and their statistical and systematical uncertainties at 1$\sigma$ for crystals 2 and 3. Systematical errors are the difference of \textit{Delta} changing the reference channel (Sys. $\#$~1) and changing the analysed channel between those having similar recoil energy associated (Sys. $\#$~2).}
	\label{tabla:IQF_errors}
\end{table}

All of these uncertainties are combined and the corresponding \textit{Delta} variable estimates are used first, because of the high uncertainties, for the determination of upper limits at 90\% to the electron equivalent energy associated to the iodine recoil, taking into account only the parameter physical region (positive energies for the recoils). ULs for the QFs are obtained dividing the electron equivalent energy ULs by the mean iodine recoil energy corresponding to the channel, obtained from the simulation (Table~\ref{tabla:EnrIValues}). The UL derived at 90$\%$~C.L. at 11.7~keV and 13.8~keV for crystals 2 and 3 are 8.2$\%$ and 9.4$\%$, respectively. Results for crystals 1, 4 and 5 correspond to UL at the level of 20-30$\%$ because of lower statistics and poorer resolution, and they are not shown. 

\begin{figure}[h!]
	\begin{center}
		\includegraphics[width=\textwidth]{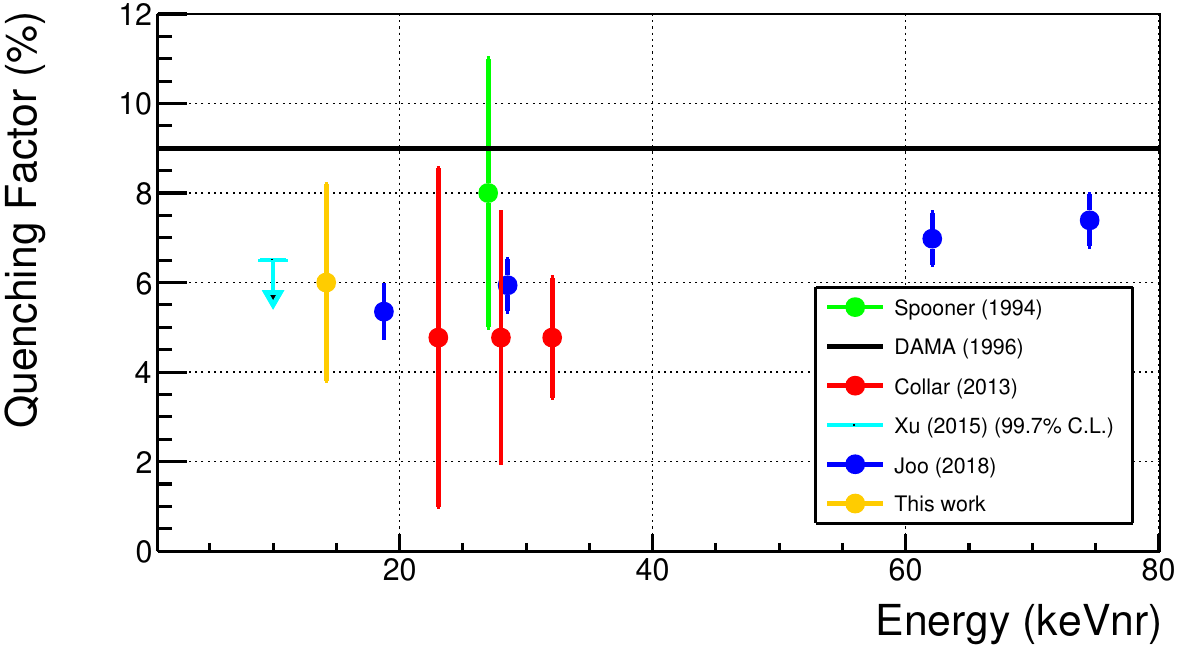}
		\caption{\label{QFIComparison}Iodine QF result for the combination of the values obtained with crystals~2 and~3  and the results presented by other experiments~\cite{Spooner,Bernabei:1996vj,Collar,Xu,Joo:2018hom}.}
	\end{center}
\end{figure}

However, we can also try to estimate the QF directly, as the ratio of the value of \textit{Delta} to the mean value of the iodine recoil energy for the corresponding channel and crystal obtained from the simulation (see Table~\ref{tabla:EnrIValues}). The uncertainties taken into account are those presented in Table~\ref{tabla:IQF_errors}. The values obtained for crystal~2 and~3 are~(5.1~$\pm$~2.7)\% and~(6.5~$\pm$~1.8)\%, respectively. As both measurements correspond to the same recoil energy (14.2~keV), they were combined together, obtaining a weighted mean of (6.0~$\pm$~2.2)\%. Figure ~\ref{QFIComparison} shows this value together with those obtained in previous measurements~\cite{Spooner,Bernabei:1996vj,Collar,Xu,Joo:2018hom}, showing a good agreement.

\section{Discussion and conclusions}\label{Section:QF_Discussion}

\fancyhead[RO]{\emph{\thesection. \nameref{Section:QF_Discussion}}}

The scintillation QF for sodium nuclei in five different NaI(Tl) crystals has been estimated, and a thorough analysis of possible systematic effects has been carried out. Following this analysis, the electron equivalent energy calibration method and the energy resolution modelling are the most relevant. To cope with this systematics, we have analysed the data considering two different linear energy calibrations (proportional and non-proportional) and convolving with two different resolution functions (with standard deviation constant for each channel and dependent on the square root of the energy). The results obtained with each resolution function were not compatible to each other within their statistical errors, and then, the corresponding systematical uncertainty was estimated from the difference between both estimates. On the other hand, the effect of the energy calibration in the results has been highlighted by presenting results separately for both calibration methods.

The sodium QF values obtained for each energy calibration applied showed very important differences. On the one hand, when a non-proportional lineal energy calibration in the ROI of the crystal signal is applied, compatible values for the 5~crystals were obtained from 10 to 80~keV, and no clear dependence with energy of the QF was observed. Considering this energy independence, the mean value of all the channels has been calculated, obtaining a value of (21.2~$\pm$~0.8)\%. On the other hand, when a proportional calibration is applied, an apparent decrease in QF at low energies, similar to that reported by other experiments is observed. This leads us to conclude that non-proportional behaviour in the detector response has to be taken into account, and further investigation is required to better understand the NaI(Tl) non-proportionality and its relevance on the QF determination.

Concerning the iodine QF results, it has not been possible to identify the nuclear recoil energy distributions in the experimental spectra, due to the low QF value and the low nuclear recoil energies at reach (less than 15~keV). However, it could be obtained by studying the $^{127}I$ inelastic peak, which corresponds to the sum of the light produced by the energy depositions of the gamma (57.6~keV) and the iodine recoil, the latter quenched by the corresponding QF. The UL at 90\% C.L. have been obtained for crystals~2 and~3 at 14.2~keV, obtaining values of 8.2$\%$ and 9.4$\%$, respectively. Moreover, the value of the QF combined of both crystals has been obtained as (6.0~$\pm$~2.2)\%. To obtain the value of this QF at higher nuclear recoil energies, possible future measurements should consider the possibility of working with more energetic neutrons or in a facility that allows to place the BD covering a higher scattering angle.

%% file: SiPMIntro.tex
\chapter{Silicon photomultipliers (SiPMs)} \label{Chapter:SiPM_Intro}
\fancyhead[LE]{\emph{Chapter \thechapter. \nameref{Chapter:SiPM_Intro}}}

Photo-Multiplier Tubes (PMTs) have been for long time the most efficient light detectors. However, they also have some disadvantages (as its large mass and its multiple-component structure) that make them less suitable in experiments that require low background. In addition, usually they consist of transparent media where charged particles can produce Cherenkov light. We study this effect in ANAIS-112 PMTs in Chapter~\ref{Chapter:OptSim}. Direct light emission from the PMTs has been also observed, for example by Double Chooz collaboration~\cite{DoubleChooz:2016ibm}, which was overviewed in Section~\ref{Section:Intro_Scintillators_PMTs}.

Recently, there has been a surge in the use of Silicon PhotoMultipliers (SiPMs) in physics experiments and many other applications~\cite{Buzhan:2003ur,Golovin:2004jt,Renker:2006ay,Danilov:2008fi,Simon:2018xzl}. They are solid-state Single-Photon sensitive devices based on Single-Photon Avalanche Diodes (SPAD) implemented on a common silicon substrate (see Section~\ref{Section:SiPM_PNJunction}). Their advantages in the detection of scintillation~\cite{Kovalchuk:2005cp,Laurenti:2008zz} are their high photon detection efficiency~\cite{Zappala:2016stq} (from 20\% to 60\%), immunity to magnetic fields, and operation at lower voltages (below 80~V, compared with the typical 1~kV of the PMTs). Additionally, SiPMs have a fast response time~\cite{Ronzhin:2010zzb} and a good level of radiation hardness~\cite{CMS:2008euu,Garutti:2018hfu}. Moreover, SiPMs are very compact, have low mass and can be made of very radiopure materials, which is very convenient for applications requiring low radioactive background, as direct dark matter searches (see Chapter~\ref{Chapter:Intro}).

The first implementation of the SiPMs in particle physics was in calorimeters for hadron colliders, being the CALICE Analog Hadron Calorimeter (AHCAL)~\cite{CALICESiPM:2006alf} the first detector prototype that used SiPMs, in 2006. They have been used as photon detectors of the scintillation light emitted by noble elements, as argon or xenon, both in gas or liquid phases~\cite{Chepel:2012sj,Gonzalez-Diaz:2017gxo}. These detectors are designed for diverse purposes, from the detection of the neutrinoless double beta decay in $^{136}Xe$, as NEXT~\cite{NEXT:2012zwy} and nEXO~\cite{Ostrovskiy:2015oja}, to the measurement of neutrino oscillations in DUNE experiment~\cite{DUNE:2020cqd,DUNE:2021hwx}, or dark matter detection in DarkSide-20k experiment~\cite{DarkSide-20k:2017zyg}. As ANAIS, these experiments require ultralow radioactive background, and to reach the radiopurity and performance desired they are developing their own SiPMs and read-out electronics. More recently, some experiments have started to use SiPMs in the detection of scintillation light from solid scintillators, as for example SoLid~\cite{SoLid:2017ema,SoLid:2018vak}, which expects to detect neutrino oscillations using diverse crystal scintillators. The use of SiPM for reading light from plastic scintillators is becoming also very popular to take advantage of the fast time response of these devices~\cite{Cattaneo:2014uya}. In less than 20 years after the first paper written about SiPMs they have become one of the most promising technologies for photon detection. They are applied in detectors aiming at many different applications as search for rare kaon decays in NA62 experiment~\cite{NA62:2017rwk}, Positron Emission Tomography (PET)~\cite{Otte:2005zz,Auffray:2015yla,J-PET:2016kal}, or muon imaging for archaeology, as MIMA does~\cite{Baccani:2018nrn}.

In this chapter, the working method and the architecture of SiPMs is presented (see Section~\ref{Section:SiPM_PNJunction}), as well as the main characteristics of these light detectors (see Section~\ref{Section:SiPM_Charac}) and the noise contributions in these devices (see Section~\ref{Section:SiPM_Noise}). Their use in the read-out of the scintillation light is overviewed in Section~\ref{Section:SiPM_Model}, besides modeling the light collection dependencies, as it will be used in the next two Chapters.

\section{The architecture of SiPMs} \label{Section:SiPM_PNJunction}
\fancyhead[RO]{\emph{\thesection. \nameref{Section:SiPM_PNJunction}}}

The SPADs in SiPMs are small (few micrometers long) silicon cells working in Geiger-mode. Photons absorbed in the silicon can excite the electrons from the valence to the conduction band in a process known as photoexcitation. The efficiency of this process depends on the bandgap of the silicon and the wavelength of the photon. Although this working method could be used with any semiconductor material, silicon is selected for these devices because it is relatively inexpensive, it has a high sensitivity to a wide range of wavelengths, and it is stable under a wide range of operating conditions.

\begin{figure}[h!]
	\begin{center}
		\includegraphics[width=\textwidth]{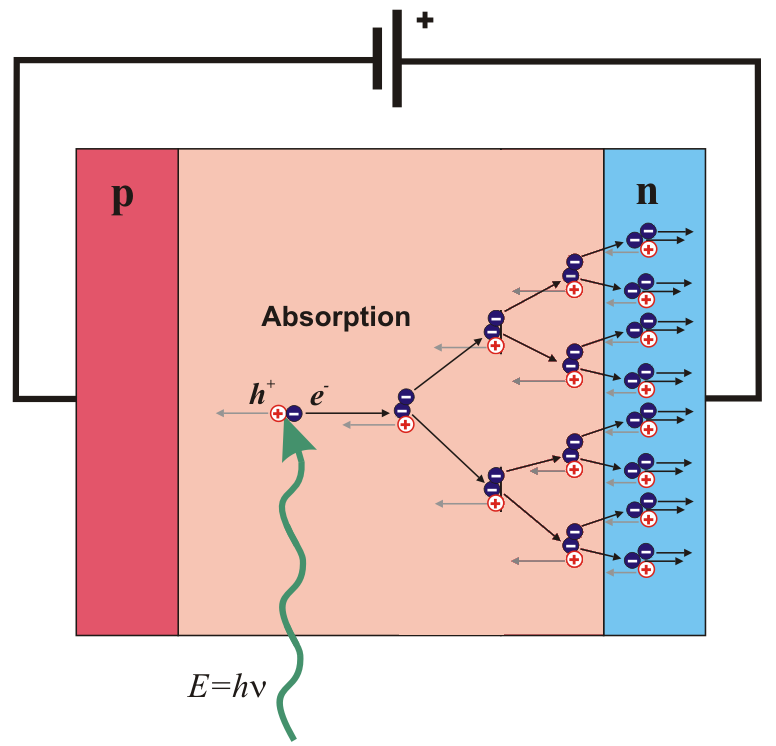}
		\caption{\label{PNJunction}Simplified scheme of a typical p-n junction and the generation of the avalanche.}
	\end{center}
\end{figure}

The structure of each SPAD is based on the p-n junction. It consists of two layers, one of which has an excess of electrons (n-type), and the other has an excess of holes (p-type). The excess of holes or electrons in the layers is typically achieved by introducing "dopants" into the semiconductor material. Typical dopants used for the p-layer are those from the group III, such as boron or aluminum, while for the n-layer are those from the group V, such as phosphorus or arsenic. A simplified scheme of a typical p-n junction is shown in Figure~\ref{PNJunction}. This junction is produced through the process of epitaxial growth, in which p-type and n-type silicon films are carefully layered on top of each other using Chemical Vapor Deposition (CVD) or Molecular Beam Epitaxy (MBE) techniques in a controlled environment.

Without an applied bias, the region between the p-type and n-type layers becomes depleted of charges due to the diffusion of electrons into the p-region and the recombination of these electrons with holes, as well as the diffusion of holes into the n-region and the recombination of these holes with electrons. This process results in the creation of a depletion layer that is free of charge carriers, as well as an electric field and voltage across the junction which opposes to the carriers diffusion, being an equilibrium reached. In the p–n junction electric charges can only flow in one direction; electrons (holes) can easily flow from n (p) to p (n) but not otherwise. When a forward bias is applied, the junction becomes more conductive and electrical current can flow through it more easily. However, when the junction is reverse-biased, the depletion zone expands and the electric field becomes stronger. This significantly reduces the flow of charge through the diode. When the bias exceeds a certain "breakdown voltage" the electric field becomes so strong that an electron-hole pair produced in the depleted layer is able to generate an avalanche of carriers. This is because each charge carrier acquires enough energy from the electric field to produce additional electron-hole pairs. In this condition, the response becomes non-linear and the device is working on Geiger mode. This process allows for the detection of a single photon, as the signal of a single carrier is highly amplified.

When a single photon is absorbed in one SPAD, it can produce an avalanche and generate a measurable current. However, once initiated, it will not stop. In order to quench the avalanche, the diode can be connected in serial with a resistor with a typical resistance of hundreds of~k$\Omega$. When the current generated in the SPAD flows through the quenching resistor, it produces a voltage drop across the diode, reducing the bias below the breakdown value and stopping the avalanche process. Then, the bias across the SPAD is restored and a new photon can be detected. The time required for the SPAD to recharge is called the recovery time.

SiPMs are an array of microcells built on a silicon wafer. Each microcell consists of a SPAD with a quenching resistor in series. They can have from hundred to several thousands of units per mm$^2$. They are connected in parallel, in such a way that the output signal of the SiPM is the sum of the charge released in each microcell. A single photon absorbed in one of them produces an avalanche which is confined to that unit, while all the other remain ready for detecting more photons. Their connections are usually made of gold or aluminum and are designed to minimize their lengths (and then, the resistance) by using a particular geometry and structure, such as a meander pattern or a spiral shape. They are coated with a dielectric or insulating layer to prevent electrical signals from leaking. These designs are also optimized to reduce the noise in the signal. The microcells are usually protected by a layer transparent to the photons being detected. It is usually made of silicon dioxide, which is also an excellent electrical insulator and a very stable and durable material. It is resistant to temperature changes and other environmental factors, which means that it can maintain its insulating properties over a wide range of conditions.

\begin{figure}[h!]
	\begin{subfigure}[b]{0.49\textwidth}
		\includegraphics[width=\textwidth]{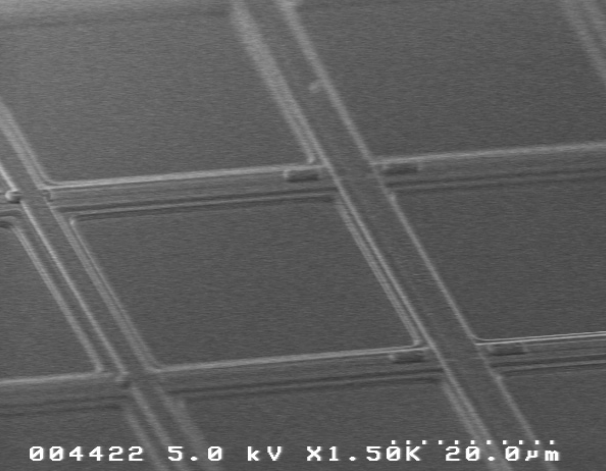}
	\end{subfigure}
	\begin{subfigure}[b]{0.49\textwidth}
		\includegraphics[width=\textwidth]{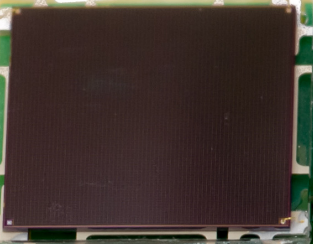}
	\end{subfigure}
	\caption{\label{Microcell}Close view of a microcell~\cite{OnsemiIntro} (left plot) and the 60035-4P-BGA SiPM, from SensL~\cite{ONSEMI} (right plot), with a surface of 6$\times$6~mm$^2$  and a FF of 67\%, having 35$\times$35~$\mu$m$^2$ microcells.}
\end{figure}

As it can be observed in Figure~\ref{Microcell}, only a portion of the SiPM area is light-sensitive, the so called active area. The ratio of active area to the total area of the SiPM is known as the Fill Factor (FF), and represents the probability of an incident photon on the SiPM reaching the active area of the device. It is an important performance parameter, because it directly affects the SiPM efficiency. It is determined by the manufacturing process of the SiPM, which requires that each microcell is separated from its neighbors to provide optical and electrical isolation, and also requires a certain amount of surface for the quench resistor and the connectors. The non-active surface for each microcell is more or less independent of its size, while the active area increases with the size of the microcells, and then, SiPMs having larger cells have higher FF.

SiPMs are often connected to a printed circuit board (PCB), commonly known as Front-End Board (FEB) in order to facilitate their integration into a larger system. These FEBs provide a stable and reliable platform for connecting the SiPM to other signal processing components, as for example preamplifiers. In general, the optimal preamplifier for the SiPM signal should provide high gain, while also having a low noise floor, low power consumption and a high frequency response.

\section{Key features of the SiPMs} \label{Section:SiPM_Charac}
\fancyhead[RO]{\emph{\thesection. \nameref{Section:SiPM_Charac}}}

The performance of SiPMs is determined by the following properties: breakdown voltage, gain, recharge time, photon detection efficiency (PDE) and dynamic range. These characteristics can be different for SiPMs having very similar design and manufacturing, and then, every unit has to be characterized individually.

\begin{figure}[h!]
	\begin{center}
		\includegraphics[width=0.5\textwidth]{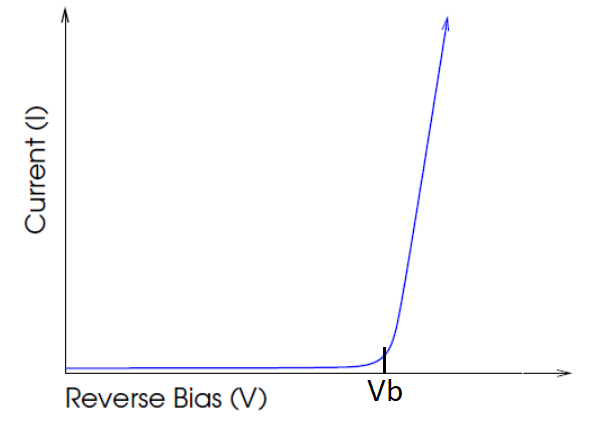}
		\caption{\label{QuenchingSiPM}I-V curve of a SiPM. The breakdown voltage, $V_b$, is the minimum operation voltage allowing the formation of the avalanches, and therefore the current can flow through the device, when it is reverse biased. Picture from~\cite{OnsemiIntro}.}
	\end{center}
\end{figure}

The breakdown voltage ($V_b$) of the SiPM is typically determined from its I-V curve, which represents the measured output current as a function of the applied voltage. The I-V curve should exhibit an abrupt current increase at $V_b$, indicating the point of breakdown. Figure~\ref{QuenchingSiPM} shows a typical I-V curve. Generally, $V_b$ is of the order of tens of volts and increases with the temperature~\cite{Dinu:2016hog}. Measurements with HAMAMATSU S13360 SiPMs presented in~\cite{NepomukOtte:2016ktf} showed an increase of the breakdown voltage with the temperature of 54.9~$\pm$~0.3~mV/K in a range of temperatures between -40 and 40~$^o$C. The difference between the bias voltage ($V_{bias}$) applied to the SiPM and the breakdown voltage is known as overvoltage ($V_{ov}=V_{bias}-V_{b}$).

When the microcell is reverse-biased above the breakdown voltage, it behaves as a capacitor whose capacitance $C$ increases with the surface of the microcell $S$ and with the capability of generating a carrier surface density in the silicon per unit voltage, $\kappa_e$ (which depends on the dopants concentration) as:
\begin{equation}
	C = S \cdot \kappa_e.
\end{equation}
For a typical microcell it can take values from~fF to a few~pF. Whenever an avalanche is produced, the microcell releases a highly uniform amount of charge. The gain of a SiPM, $G$, is therefore defined as the ratio of the charge generated by an activated microcell to the charge of a single electron $e$, and can be written as:
\begin{equation}\label{eq:Gain(OV)}
	G = \frac{V_{ov}\cdot C}{e} = \frac{V_{ov}\cdot \kappa_e \cdot S}{e}.
\end{equation}
It increases with the overvoltage and the surface of the microcell.

\begin{figure}[h!]
	\begin{center}
		\includegraphics[width=0.75\textwidth]{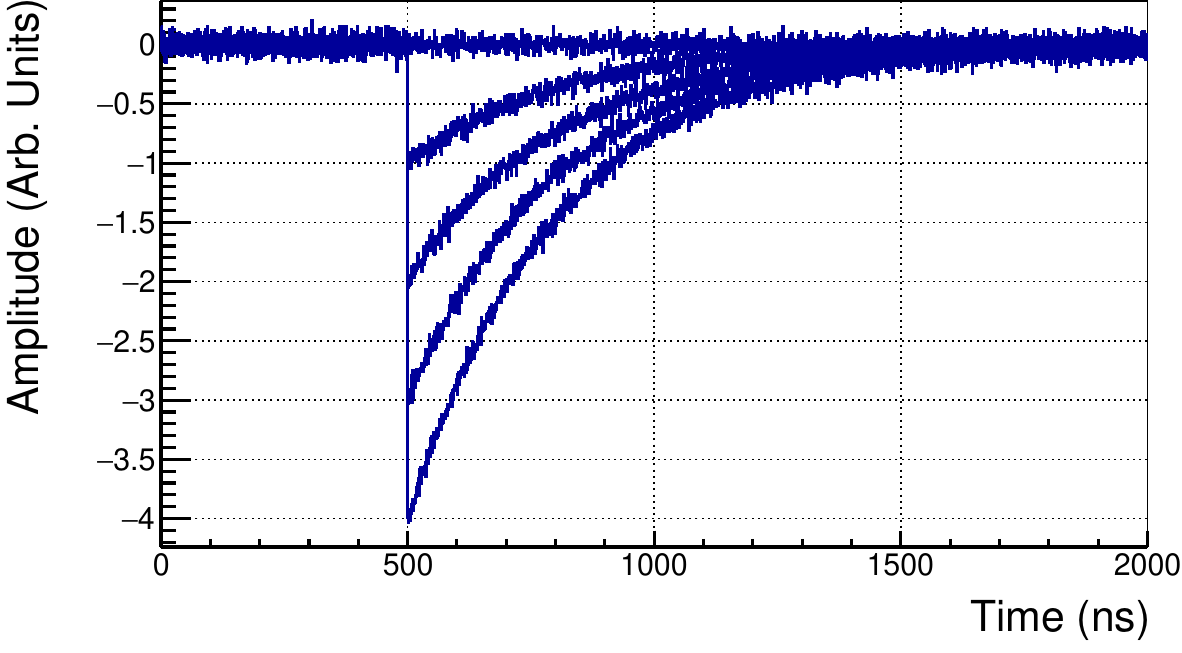}
		\caption{\label{SPE}Toy current pulses of a SiPM for a quenching resistor of 100~k$\Omega$ and a capacitance of 3~pF. The pulses correspond to a different number of avalanches (from~0 to~4).}
	\end{center}
\end{figure}

After an avalanche, the microcell has a recovery time ($t_{RC}$), related with the discharge of the capacitor through the quenching resistor as~\cite{Ronzhin:2010zzb}:
\begin{equation}
	t_{RC} = R_q \cdot C,
\end{equation}
being $R_q$ the resistance of the quenching resistor. As all the microcells are connected in parallel, the vias voltage is the same for all of them and the total capacitance of the SiPM is the sum of those from each microcell. The current that flows through the device in an avalanche follows an exponential decay with the time, being $t_{RC}$ its mean time. Figure~\ref{SPE} simulated toy current pulses produced by a different number of simultaneous avalanches. The simulation considers 100~k$\Omega$ for the quenching resistor and 3~pF for the microcell capacitance, giving as a result a recharge time of 300~ns. A white noise with a root mean squared of 10\% of the amplitude of the pulse corresponding to a single avalanche has been included.

\begin{figure}[h!]
	\begin{center}
		\includegraphics[width=0.75\textwidth]{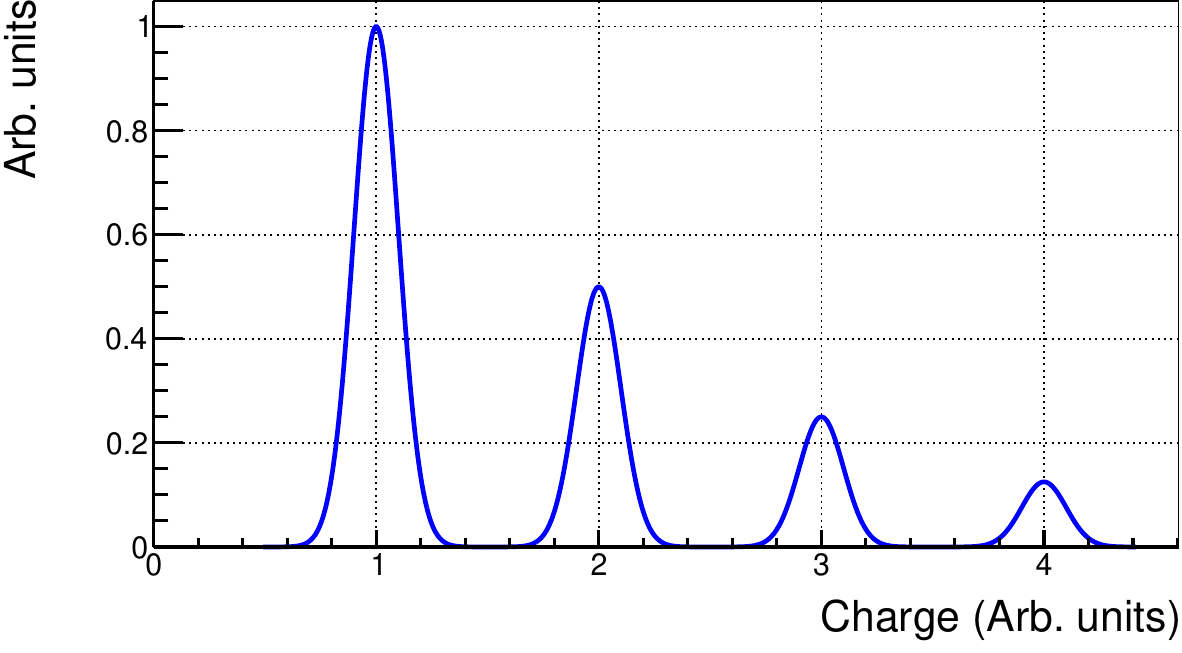}
		\caption{\label{plotGainArea}Typical charge spectrum in arbitrary units.}
	\end{center}
\end{figure}

By integrating the current pulse, a charge spectrum can be generated. The peaks corresponding to different number of detected photons can be identified in this signal, more or less easily depending on the noise conditions of the SiPM and the variability of the microcells properties, which produce dispersion in the SiPM signal. Figure~\ref{plotGainArea} shows a typical charge spectrum in arbitrary units. The separation between each pair of adjacent peaks is constant and represents the charge generated by a single fired microcell, i.e. a single photon interaction. It allows to calculate the gain as the total charge generated by a single photon. The charge generated by a single photon will be used, for instance, to calculate the number of photons detected, from the total charge generated. We will refer to the avalanche produced by a single photon interaction in the SiPM as photoelectron, $phe$, and the procedure to obtain the amplitude and the area of a single photoelectron signal (SPE) as the SPE calibration.

The Photon Detection Efficiency (PDE) of the SiPM is the probability that an incident photon produces an avalanche~\cite{Zappala:2016stq}. It can be factorized into three independent probabilities: that of the photon reaching an active volume of the device (a geometric efficiency, given by the FF), that of this photon producing an electron-hole pair (the quantum efficiency, $QE$), and that of this pair resulting in an avalanche ($\epsilon$), as
\begin{equation}\label{eq:PDE}
	PDE(V_{ov}) = FF \cdot QE\cdot\epsilon(V_{ov}).
\end{equation}
The avalanche initiation probability ($\epsilon$) increases with the overvoltage because it implies a higher electric field and then, higher kinetic energy transferred to the primary charge carriers produced, being easier for them to start the avalanche process.

\begin{figure}[h!]
	\begin{subfigure}[b]{0.39\textwidth}
		\includegraphics[width=\textwidth]{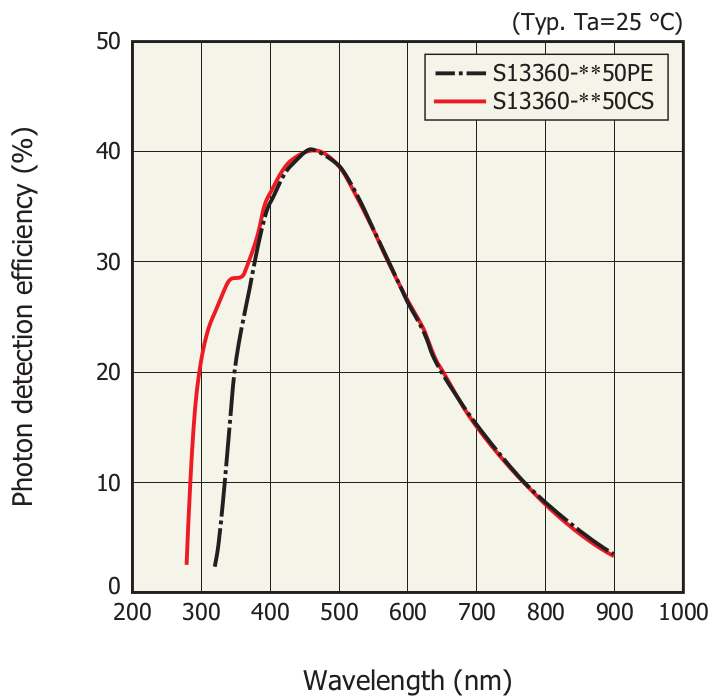}
	\end{subfigure}
	\begin{subfigure}[b]{0.59\textwidth}
		\includegraphics[width=\textwidth]{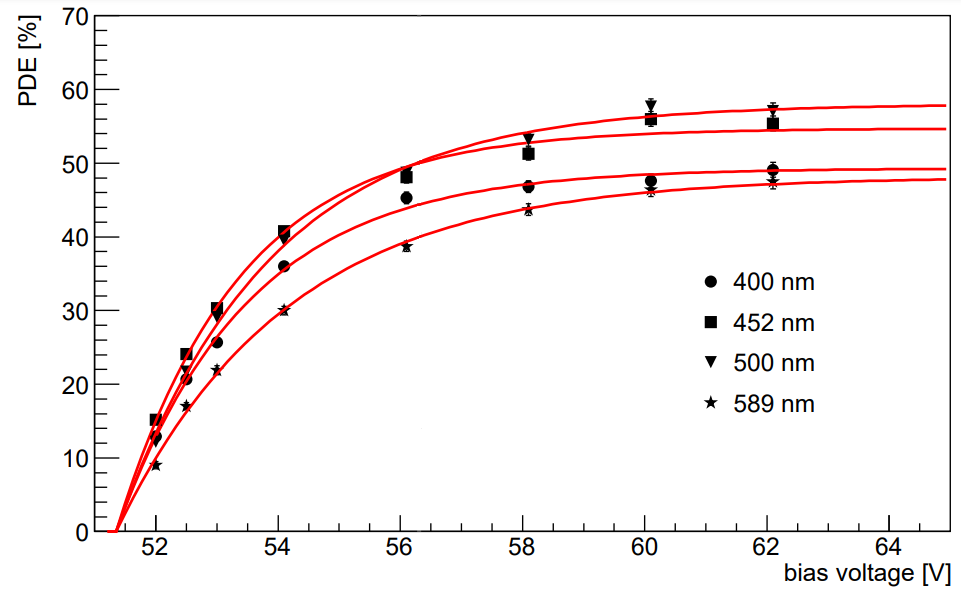}
	\end{subfigure}
	\caption{\label{S13360PDE}PDE of the SiPM from HAMAMATSU model S13360-6050PE. It is shown as a function of the incident photons wavelength in the left plot at an overvoltage of 2.7~V (from~\cite{Hamamatsu}) and as a function of the bias voltage for four different wavelengths in the right plot (from~\cite{NepomukOtte:2016ktf}).}
\end{figure}

The quantum efficiency, $QE$, depends strongly on the wavelength of the incident photon because of two different effects. First, the bandgap of the silicon is a threshold energy for the production of the electron-hole pairs, but above this threshold, the absorption probability still depends on the wavelength. Second, the probability of producing photoelectric absorption in the coating materials increases with the energy of the incident photon, and then absorption of the photons can occur before reaching the silicon. Due to all these factors, the typical shape of the PDE of SiPMs is peaked at a given wavelength, typically between 400 and 550~nm, with a maximum value between 20 and 50\%. As an example, Figure~\ref{S13360PDE} shows the PDE as a function of the wavelength of the incident photons for the SiPM model S13360-6050PE, from HAMAMATSU~\cite{Hamamatsu}, and the dependence with the bias voltage for four different wavelengths. The latter shows an asymptotic behaviour~\cite{NepomukOtte:2016ktf}.

The dynamic range of a SiPM is the range of incident radiation intensities over which the sensor provides an output current proportional to the number of photons. It has been modeled using different methods~\cite{Gruber:2013jia,Marano:2014zqa,Bretz:2016wzq}. For $N_{ph}$ photons arriving to the SiPM at the same time, the number of photoelectrons produced, $N_{phe}$, is the result of a binomial process and can be approximated by~\cite{Renker2009,Bretz:2020grl}:
\begin{equation}
	N_{phe} = N_{cell}\cdot\left(1-\exp{\left(\frac{-PDE \cdot N_{ph}}{N_{cell}}\right)}\right),
\end{equation}
where $N_{cell}$ is the number of cells of the SiPM that are illuminated by the incident light. The first term of the Taylor series for small $PDE \cdot N_{ph}/N_{cell}$ can be obtained for this function as:
\begin{equation}
	N_{phe} = PDE\cdot N_{ph}.
\end{equation}
When two or more photons result in avalanches in the same cell at times shorter than the recharge time, it is impossible to discriminate the resulting signal from a single-photon signal. Then, strong non-proportionality in the output can be observed in these devices and the linear behaviour is only valid for low intensity signals ($N_{ph} \ll N_{cell}$).

\section{Noise contibutions in the SiPMs} \label{Section:SiPM_Noise}
\fancyhead[RO]{\emph{\thesection. \nameref{Section:SiPM_Noise}}}

There are three different noise contributions in the SiPMs signals: dark count rate, the optical crosstalk and the afterpulses. The Dark Count Rate (DCR) is mainly caused by thermal excitations produced in the active volume of the cells, but it can also be produced by interactions of other particles able to produce electron-hole pairs. This can be achieved by radioactive decays inside the cell, environmental radiactivity or cosmic radiation. At room temperature the dominant effect is the thermal one. These signals are identical to those produced by photon-generated avalanches, which means that they act as a source of noise at the single photon level~\cite{Butcher:2017twm}. The DCR, neglecting the non-thermal contributions, depends on the number of thermally-generated electrons in a microcell per unit of time, $N_e$, multiplied by the probability of producing an avalanche, $\epsilon(V_{ov})$:
\begin{equation}
	DCR \propto N_e(T) \cdot \epsilon(V_{ov}).
\end{equation}
It increases with the overvoltage, as it was explained before for the $\epsilon(V_{ov})$ probability. The dependence of $N_e$ with the temperature is~\cite{Pagano:2012}:
\begin{equation}
	N_e(T) \propto T^2 \exp{\left(\frac{-E_i}{kT}\right)},
\end{equation}
where $E_i$ is the band-gap energy of the semiconductor and $k$ is the Boltzmann's constant. Typical values of the DCR range from 100~kHz/cm$^2$ to 10~MHz/cm$^2$ at room temperature~\cite{Hamamatsu} and is at the level of 1~Hz/cm$^2$ at the temperature of the liquid nitrogen, which shows the importance of working at low temperature for the operation of SiPMs in the very-low light signals regime. DCR can be a limiting factor in terms of triggering in the SiPM output at single-photon level, requiring to set a higher trigger threshold. However, DCR will be contributing to all the measurements, affecting to the proportionality with the number of photons illuminating the SiPM. The DCR was observed in~\cite{NepomukOtte:2016ktf} to increase an order of magnitude every 17.6~K for S13360 SiPMs from HAMAMATSU~\cite{Hamamatsu}. The increase of the DCR with the overvoltage is weaker and it depends strongly on the model of the SiPM.

\begin{figure}[h!]
	\begin{center}
		\includegraphics[width=0.75\textwidth]{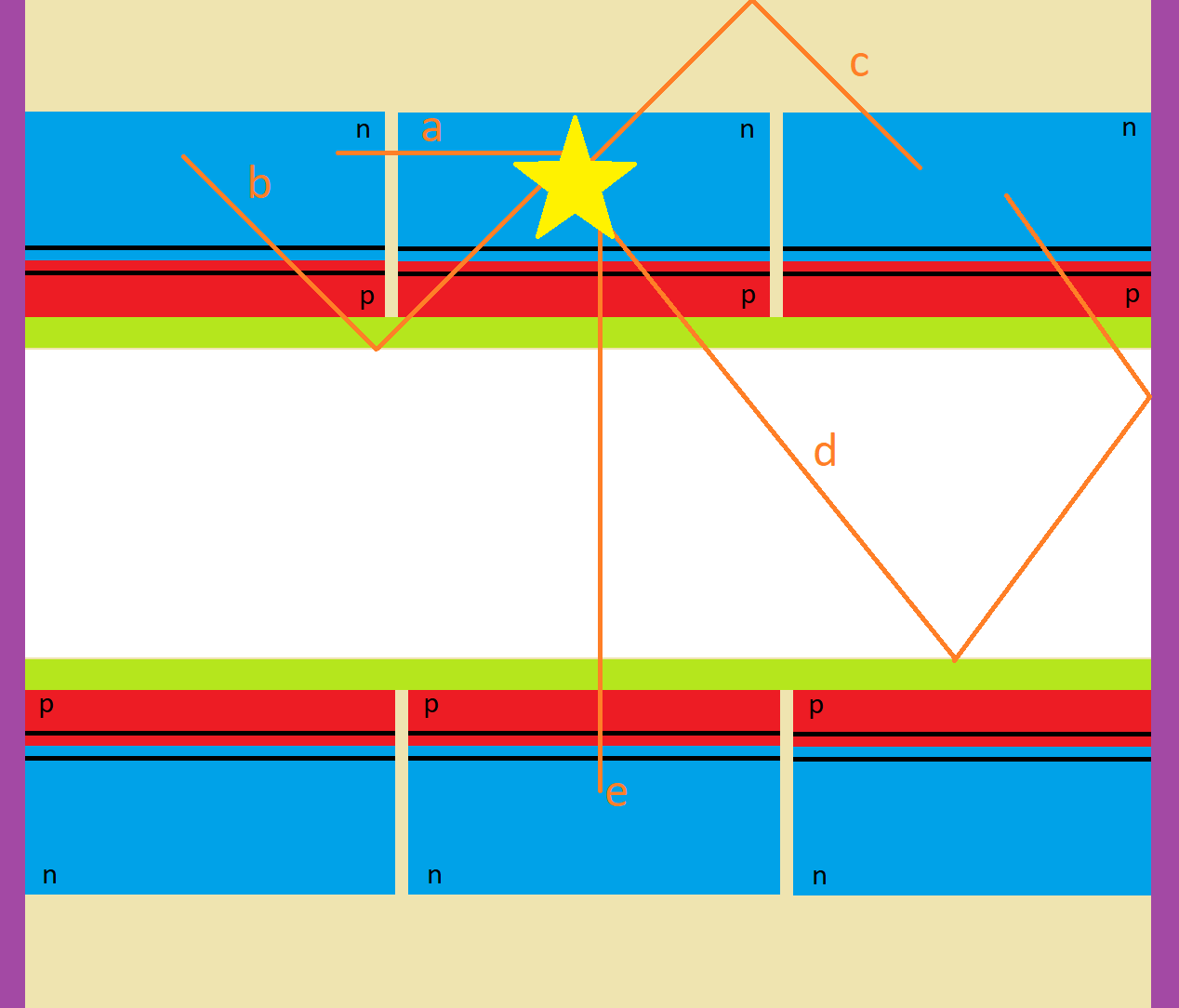}
		\caption{\label{CrossTalkPlot}Different processes contributing to optical CT. Secondary photons can reach neighboring microcells and produce CT, which can be internal if they reach directly a neighboring microcell (a), are reflected from the coating layer on top of the sensor (b) or from the bottom of the silicon substrate (c). It can also produce external CT by being reflected in the external medium where the SiPM is located (d) or feedback CT with other SiPMs in the same detector (e).}
	\end{center}
\end{figure}

While the avalanche is under development, accelerated charge carriers in the high-field region can emit photons that can initiate a secondary avalanche in a neighbouring microcell, a process which is called optical crosstalk (CT). These secondary photons are usually in the near infrared range and they can travel significant distances through silicon. Figure~\ref{CrossTalkPlot} shows how these secondary photons can affect neighbouring microcells: directly reaching other microcell, being reflected from the coating layer on top of the sensor, the bottom of the silicon substrate or in the external medium where the SiPM is located, or from feedback with other SiPMs in the same detector. The different processes can be classified as internal, external and feedback CT. The probability of this process is at a few\% level in typical SiPMs, and because all the CT processes before mentioned happen in very short time scales, any CT avalanche produced by a single incident photon contributes to the SiPM output signal and can be interpreted as an additional incident photon. Some models have been proposed to explain the CT and estimate the CT probability~\cite{Zappala:2016stq,Dinu:2016hog,Vinogradov:2011vr,Gallego:2013qua,NepomukOtte:2016ktf,Nagai:2018ovm,Boulay:2022rgb}. As in the case of the PDE and the DCR, the CT is a function of the avalanche initiation probability $\epsilon(V_{ov})$, and therefore it also increases with the overvoltage. Moreover, as shown in Figure~\ref{CrossTalkPlot}, it strongly depends on the geometry of the detector.

Although semiconductors in general, but silicon in particular, are manufactured with the highest quality standards, the presence of defects or impurities in the crystal lattice results in traps for the charge carriers. The trapped charged carriers may be released in a period of time ruled by the lifetime of the trap state, potentially initiating another avalanche and resulting in an afterpulse (AP) in the same microcell. If the trap lifetime is short, the AP will have little effect because the microcell will not be fully charged yet. However, if the lifetime of the trap is longer than the recharge time and the event is inside the integration time, these AP will affect the measurements in a similar way as the CT, multiplying the number of photoelectrons detected by a factor larger than one. The probability of an AP has typical values at few\% level~\cite{Butcher:2017twm}.

\section{SiPM modelization for scintillation detection} \label{Section:SiPM_Model}
\fancyhead[RO]{\emph{\thesection. \nameref{Section:SiPM_Model}}}

For the analysis presented in the next Chapters, it is important to model the performance of the SiPMs in the read-out of the scintillation light. After an energy deposition, a scintillator emits an amount of photons proportional to that energy with characteristics mean time and wavelength, and some of them will produce photoelectrons with a probability given by the PDE. Then, the total charge released by the device is used to estimate the energy deposition. Moreover, as it was seen in Chapters~\ref{Chapter:Intro} and~\ref{Chapter:ANAIS}, one of the parameters that determines the sensitivity of an experiment is the Light Colletion (LC), defined as the number of photoelectrons detected per unit of energy. This number depends on the collection of scintillation photons from the detector, but also on the three main noise effects of the SiPMs, commented in Section~\ref{Section:SiPM_Noise}. Therefore, to characterize the contribution of the scintillation light to the total LC, it is necessary to model the detector behaviour as a function of the overvoltage. The most simple model is the one that neglects the noise effects of SiPMs, being the number of photoelectrons detected if working in the linear regime:
\begin{equation}
	N_{phe}(V_{ov}) = N_{ph} \cdot PDE(V_{ov}),
\end{equation}
where $N_{ph}$ is the number of photons reaching the SiPM and $PDE$ is the photon detection efficiency (see Equation~\ref{eq:PDE}). DarkSide researchers developed a model that parameterizes the avalanche initiation probability ($\epsilon(V_{ov})$) as a function of the overvoltage~\cite{Boulay:2022rgb}:
\begin{equation}\label{eq:epsilon(V_{ov})}
	\epsilon(V_{ov}) = \epsilon_{max} \left[\xi\left(1-\exp{\left(\frac{-V_{ov}}{V_e}\right)}\right) + (1-\xi)\left(1-\exp{\left(\frac{-V_{ov}}{V_h}\right)}\right)\right],
\end{equation}
where $\xi$ and 1-$\xi$ are the probabilities of the avalanche to be initiated by the electron or by the hole, respectively, and they depend on the photon wavelength. The parameters $V_e$ and $V_h$ represent the mean energy needed for a carrier to produce secondary charges. By combining equations~\ref{eq:epsilon(V_{ov})} and~\ref{eq:PDE}, the $PDE(V_{ov})$ can be written as:
\begin{equation}
	PDE(V_{ov}) = PDE_{max}\cdot\frac{\epsilon(V_{ov})}{\epsilon_{max}},
\end{equation}
where
\begin{equation}
	PDE_{max} = FF \cdot QE\cdot\epsilon_{max}
\end{equation}
is the maximum value of the PDE that (assuming the model of $\epsilon(V_{ov})$ presented in the Equation~\ref{eq:epsilon(V_{ov})}) is achieved at $V_{ov}\rightarrow\infty$. Then, the number of photoelectrons increases with the overvoltage as:
\begin{equation}~\label{eq:Nphe(OV)}
	N_{phe}(V_{ov}) = N_{ph} \cdot PDE_{max}\cdot\frac{\epsilon(V_{ov})}{\epsilon_{max}}.
\end{equation}
It is possible to write this expression as a function of the maximum number of photoelectrons produced by scintillation light, $N_{phe}^{max}$, as it is directly:
\begin{equation}\label{eq:NpheMax}
	N_{phe}^{max} = N_{ph}\cdot PDE_{max}.
\end{equation}
Then, the Equation~\ref{eq:Nphe(OV)} becomes:
\begin{equation}\label{eq:Nphe(OV)2}
	N_{phe}(V_{ov}) = N_{phe}^{max}\cdot\frac{\epsilon(V_{ov})}{\epsilon_{max}}.
\end{equation}
Considering that the photoelectrons detected are produced by an energy deposition $E_o$, the LC can be written as:
\begin{equation}\label{eq:LC0(OV)}
	LC(V_{ov}) = \frac{N_{phe}(V_{ov})}{E_o} = \frac{N_{phe}^{max}(V_{ov})}{E_o} \cdot\frac{\epsilon(V_{ov})}{\epsilon_{max}} = LC^{scint}_{max}\cdot\frac{\epsilon(V_{ov})}{\epsilon_{max}},
\end{equation}
where $LC^{scint}_{max}$ is the maximum scintillation LC that the detector can achieve.

The contribution of the DC photoelectrons in the number of photoelectrons detected is the product of the DCR by the time integration window $t_I$:
\begin{equation}
	N_{DC}(V_{ov}) = DCR(V_{ov})\cdot t_I,
\end{equation}
and therefore the number of photoelectrons contributing to the signal will be larger than those directly produced by the scintillation:
\begin{equation}\label{eq:NpheMax(OV)}
	N_{phe}(V_{ov}) = N_{phe}^{max}\cdot\frac{\epsilon(V_{ov})}{\epsilon_{max}} + DCR(V_{ov})\cdot t_I.
\end{equation}
And the LC will increase accordingly:
\begin{equation}
	LC(V_{ov}) = LC^{scint}_{max}\cdot\frac{\epsilon(V_{ov})}{\epsilon_{max}} + DCR\cdot\frac{t_I}{E_o}.
\end{equation}

Similarly, the contributions of CT and AP to the number of photoelectrons have to be taken into account. Simple models for these contributions are presented next. First, the average number of secondary avalanches produced after a primary avalanche due to the optical CT, $\lambda_{CT}$, is the product of the number of carriers released in the avalanche (the gain, $G$), the probability of these carriers to emit the CT photons, $p_{ph}$, the probability of these photons to reach the neighbour cells, $p_{geo}$, the probability of being absorbed there and produce free carriers (which is given by the quantum efficiency of the silicon at the CT photons wavelength, $QE_{CT}$), and the probability of these carriers to produce avalanches, that is the initiation avalanche probability defined above. Then, in this model, the parameterization is:
\begin{equation}\label{eq:CT(OV)}
	\lambda_{CT}(V_{ov}) = G(V_{ov})\cdot p_{ph}\cdot p_{geo}\cdot QE_{CT}\cdot\epsilon(V_{ov}).
\end{equation}
The gain, $G$, is proportional to the overvoltage (see Equation~\ref{eq:Gain(OV)}). On the other hand, $p_{ph}$ and $p_{geo}$ do not depend on the overvoltage (the second one depends on the geometry of the detector and the photon wavelength). The optical photons that produce the CT are dominant in the red-infrared region~\cite{MIRZOYAN200998} and for these wavelengths the avalanches are mainly initiated by holes. That is why in $\epsilon(V_{ov})$ only the avalanches initiated by holes will be considered (i.e., $\xi = 0$). Then, $\lambda_{CT}$ can be parameterized as:
\begin{equation}
	\lambda_{CT}(V_{ov}) = \frac{C\cdot V_{ov}}{e}\cdot p_{ph}\cdot p_{geo}\cdot\epsilon_{max}\left(1-\exp{\left(\frac{-V_{ov}}{V_h}\right)}\right)
\end{equation}
All the parameters independent on the overvoltage can be unified to obtain:
\begin{equation}\label{eq:lambda(OV)}
	\lambda_{CT}(V_{ov}) = \xi_{CT} \cdot V_{ov} \cdot \left(1-\exp{\left(\frac{-V_{ov}}{V_h}\right)}\right).
\end{equation}
Due to the CT process, the number of primary avalanches produced in a SiPM photoelectrons produced in a SiPM (obtained in Equation~\ref{eq:NpheMax(OV)}) has to be multiplied by a factor $\mu_{CT}$, which behaves as an additional gain to obtain the total number of avalanches. This factor is the total number of avalanches produced by each primary avalanche:
\begin{equation}
	\mu_{CT}(V_{ov}) = \sum_{n=0}^{N_{cell}-1} \lambda_{CT}^{n}(V_{ov}).
\end{equation}
The sum is over all the microcells of the SiPM except the one that has produced the primary avalanche ($N_{cell} - 1$), but as this number is very large, it can be taken as infinity for simplicity. In that case, the gain $\mu_{CT}$ becomes:
\begin{equation}
	\mu_{CT}(V_{ov}) = \sum_{n=0}^{\infty} \lambda_{CT}^{n}(V_{ov}) = \frac{1}{1-\lambda_{CT}(V_{ov})},
\end{equation}
expression only valid for $\lambda_{CT} < 1$.

Concerning the AP, the average number of secondary avalanches ($\lambda_{AP}$) can be also obtained as the product of the total number of carriers produced in an avalanche, the probability of each carrier to be trapped, $p_{trap}$, the probability that the microcell is recharged when the carrier is released, $p_{rech}$, and the probability of this carrier to produce a secondary avalanche, $\epsilon(V_{ov})$. Additionally, as the secondary avalanche is produced some time after the primary one, there is a given probability that this AP-avalanche contributes to the signal, which can be seen as an efficiency, $E_{AP}$. Then, a possible parameterization is:
\begin{equation}
	\lambda_{AP}(V_{ov}) = G(V_{ov}) \cdot p_{trap} \cdot p_{rech} \cdot \epsilon(V_{ov}) \cdot E_{AP}.
\end{equation}
The efficiency $E_{AP}$ is just:
\begin{equation}
	E_{AP} = \frac{1}{\tau_{AP}} \left[1-\exp{\left(\frac{-t_I}{\tau_{AP}}\right)}\right],
\end{equation}
where $t_I$ is the integration time and $\tau_{AP}$ is the lifetime of the trap. $p_{trap}$ depends on the density of defects in the lattice, while $p_{rech}$ depends on the lifetime of the traps and the recharge time of the microcells. As they do not depend on the overvoltage, they can be unified to obtain the next relation:
\begin{equation}
	\lambda_{AP}(V_{ov}) = \xi_{AP} \cdot V_{ov} \cdot \epsilon(V_{ov}).
\end{equation}
In the same way as in the CT, the increase in the number of avalanches produced by the AP can be modeled as an additional gain $\mu_{AP}$ as:
\begin{equation}
	\mu_{AP}(V_{ov}) \approx \frac{1}{1-\lambda_{AP}(V_{ov})},
\end{equation}
expression only valid for $\lambda_{AP} < 1$. Finally, the total LC obtained at a given overvoltage considering the three main noise contributions of the SiPMs can be modeled as:
\begin{equation}\label{eq:LC(OV)}
	LC(V_{ov}) = \mu_{CT}(V_{ov})\cdot\mu_{AP}(V_{ov}) \cdot\left(LC^{scint}_{max}\cdot\frac{\epsilon(V_{ov})}{\epsilon_{max}}+DCR\cdot\frac{t_I}{E_o} \right).
\end{equation}

As commented before, the SiPM should produce an output current for a single photon with exponential shape and characteristic time $R_q\cdot C$. However, usually the output of the SiPMs is affected by parasite capacitances and resistances, and moreover, preamplified in the FEBs. Depending on the preamplifier band-width, for instance, the pulse shape of the SPE signal is modified, resulting in the response function of the SiPM to a single photon. Moreover, in the application of SiPMs to the readout of the scintillation light produced by an energy deposition, the output signal will be the convolution of the scintillation time profile and the time response of the SiPM.

The energy resolution achievable with SiPMs is strongly affected by the number of photoelectrons coming from noise effects (DC, CT and AP) that contribute to the signal integration window fixed by the typical scintillation time of the signal searched for. All these factors determine the resolution of the detector and depend on the overvoltage and the temperature. The PDE increases for higher overvoltages, and therefore also the LC and energy resolution, if the poissonian contribution dominates. However, the three main sources of noise in a SiPM (DCR, CT, and AP) also increase with the overvoltage, which degrade the resolution due to the inclusion of photoelectrons that are not produced by scintillation light. The combination of these opposite dependencies makes the resolution of the detector to be optimal at a given overvoltage, which depends on the system. Concerning the temperature, the PDE, CT and AP can change but no systematical dependence is observed in all the SiPMs. The most relevant factor that changes with the temperature is DCR, which decreases a factor~10 every 20~K (approximately, depending on the device). Therefore, for slow scintillation times (which require long time integration to acquire the signal completely), or for low-light events (needing large surfaces of SiPMs to increase the LC), the resolution is strongly improved by decreasing the temperature.

%% file: SiPM_frio2.tex
\chapter{Characterization of NaI(Tl) and undoped NaI crystal scintillation} \label{Chapter:SiPMStar2}

\fancyhead[LE]{\emph{Chapter \thechapter. \nameref{Chapter:SiPMStar2}}}

Being remarkable the results achieved by the ANAIS-112 experiment in the testing of the DAMA/LIBRA annual modulation result, experimental constraints prevent ANAIS from further sensitivity improvement. The Photomultiplier Tubes (PMTs) are one of the most important limiting factors, implying on the one hand the most relevant radioactive background contribution (after the NaI crystal itself), and on the other hand, being the possible origin of many spurious events observed in the experiment at very low energy. Replacing the PMTs by SiPMs will offer a large potential of sensitivity improvement: it could allow to increase the radiopurity of the setup, to increase the quantum efficiency for the light detection and to reduce the anomalous PMT-origin events. All of this will result in improvement in both, background and energy threshold. However, high DCR could endanger the required single-photon trigger and achieving an energy threshold below 1~keV. To profit from all the advantages of SiPMs, working at low temperatures, below 100~K, is then required.

The characterization of small crystals of pure NaI (from now on, NaI) and NaI(Tl) at low temperature was the objective of a research stage in the Gran Sasso National Laboratory (LNGS), in Italy, during the third year of my PhD. The group of Dr. Alessandro Razeto, has developed new models of SiPMs specifically designed for the detection of the scintillation of liquid argon (LAr) in the low-light regime to be applied in the DarkSide-20k experiment (to be installed in the LNGS)~\cite{DarkSide-20k:2017zyg}. They consist of an array of 6$\times$4 SiPMs, with a surface of 5$\times$5~cm$^2$, integrating appropriate front-end electronics for the read-out. They have shown excellent performance at 87~K~\cite{DIncecco:2017bau}. 

In Section~\ref{Section:SiPMSTAR2_Detector}, the NaI+SiPM array system used for the measurements is described. All of them were performed in Laboratory~$\#$7 of the LNGS, under the supervision of Dr. Alessandro Razeto in the STAR (SiPM Test bench in Argon) cryogenic system (described in Section~\ref{Section:SiPMSTAR2_Setup}), consisting of a cryostat and an argon recirculation circuit. Section~\ref{Section:SiPMSTAR2_GEANT4} describes the optical GEANT4 simulation of the experimental setup that I developed during my stage in LNGS, which showed to be very useful to choose optimal operating conditions and to obtain the averaged energies of the peaks used in the energy calibration. Moreover, I designed a procedure for the data taking using a $^{241}Am$ source for the characterization of the crystals at 87~K, presented in Section~\ref{Section:SiPMSTAR2_Analysis}. The results drawn from the measurements carried out during this stage, shown in Section~\ref{Section:SiPMSTAR2_Charac}, helped to improve the design of the prototype under development at the University of Zaragoza. Section~\ref{Section:SiPMSTAR2_Conclusions} includes some comments and outlook in this direction.

\section{NaI+SiPM detector} \label{Section:SiPMSTAR2_Detector}
\fancyhead[RO]{\emph{\thesection. \nameref{Section:SiPMSTAR2_Detector}}}

\begin{figure}[h!]
	\begin{center}
		\includegraphics[width=0.5\textwidth]{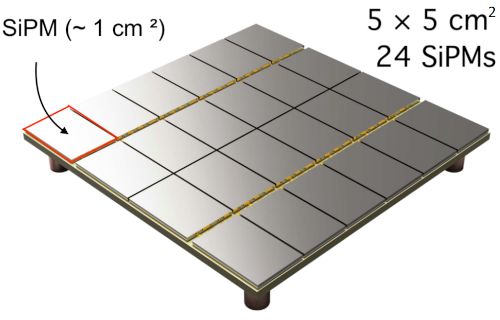}
		\caption{\label{DSTile}Picture of the SiPM array (DarkSide tile) used for these measurements. Image from~\cite{Walczak}.}
	\end{center}
\end{figure}

DarkSide has developed 6$\times$4 SiPM arrays~\cite{DIncecco:2017bau} (commonly known as tiles) together with a front-end read-out electronics~\cite{DIncecco:2017qta} to work in cryogenic and low electronic noise conditions. The SiPMs, model~NUVHD-LF, are produced by Fondazione Bruno Kessler~\cite{FBK} for DarkSide-20k. The surface of each SiPM is 11.9$\times$7.8~mm$^2$, and the array covers a 5$\times$5~cm$^2$ area with a Fill Factor of~89$\%$ (see Figure~\ref{DSTile}). They have a peak PDE of about 50$\%$ between 400–420~nm. A characterization of these SiPMs at liquid nitrogen temperature (77~K) can be found in~\cite{Acerbi:2016ikf}.

The SiPM array is mounted in the middle of a 6$\times$6~cm$^2$ PCB using conductive EPO-TEK EJ2189-LV~\cite{EPOTEK}, which is a silver-loaded epoxy that has been tested for mechanical and electrical stability after repeated thermal cycles in liquid nitrogen. The epoxy is deposited in the PCB in a 60~$\mu$m-thick film, creating a surface where the connectors of the SiPMs (made of aluminum) are placed. The anode and cathode of each SiPM are routed to a Samtec LSS-150-01-L-DV-A~\cite{Samtec} connector on the back of the PCB that fits with a Front-End Board (FEB).

\begin{figure}[h!]
	\begin{center}
		\includegraphics[width=0.5\textwidth]{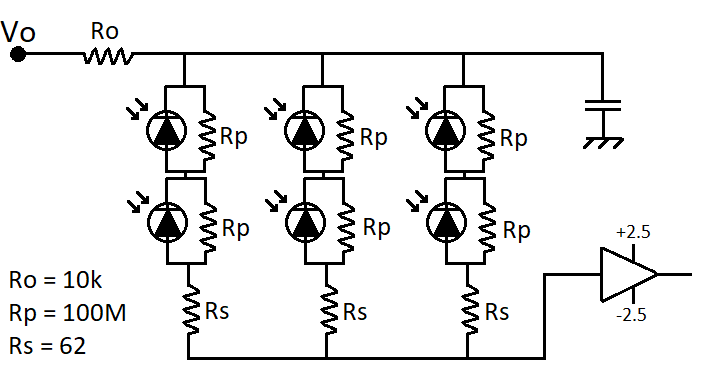}
		\caption{\label{PCBSiPM arrayScheme1ch}Connections scheme for one channel (3x2 SiPMs) of the SiPM array. SiPMs are connected in three parallel branches each one with two SiPM in series to optimize the SNR.}
	\end{center}
\end{figure}

Arranging SiPMs in series can reduce the noise by limiting the detector capacitance seen by the amplifier, but also attenuates the signal by a factor that is the number of SiPMs in series. The signal-to-noise ratio has to be optimized, and it is dependent on many factors, as the amplifier input impedance, the value of the series resistance ($R_s$) and the SiPM complex impedance. The configuration shown in Figure~\ref{PCBSiPM arrayScheme1ch} was chosen as the most convenient for the application in DarkSide experiment: 3~branches in parallel with 2~SiPMs in series each. This implies that the 24~SiPMs in the array are distributed into 4~channels. More information on this SiPM architecture can be found in~\cite{DIncecco:2017bau}. This circuit is printed in the FEB. The four output signals of the array, one per channel, are fed into independent Trans-Impedance Amplifiers (TIA), incorporating a LMH6629 preamplifier, which has a bandwidth of 30~MHz at liquid nitrogen temperature. The input voltage ($V_o$ in the figure) applied to the board is the same for all the channels. Because of the circuit configuration, the bias voltage applied to each SiPM is half of the voltage applied to the channel.

\begin{figure}[h!]
	\begin{center}
		\includegraphics[width=\textwidth]{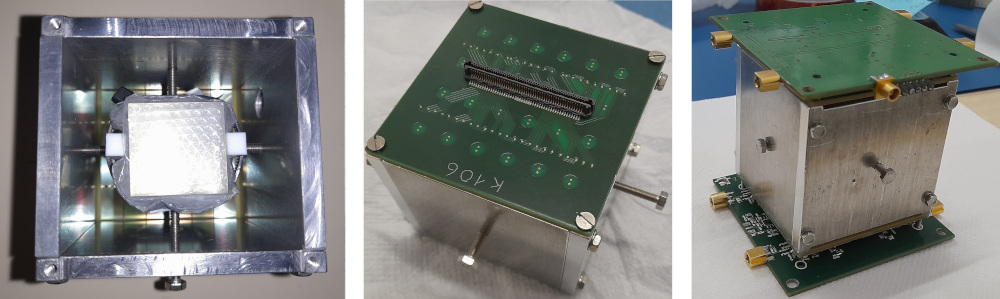}
		\caption{\label{ChamberPictures}Pictures of the detector built for these measurements. Left: view of the NaI crystal protected in a plastic bag in vacuum and fixed by four screws and teflon pieces. Middle: top array and the Samtec LSS-150-01-L-DV-A connector installed in the chamber. Right: the two front-end boards are onsite.}
	\end{center}
\end{figure}

Two of these arrays were used in the measurements to increase the LC and to reject DC events by requiring a signal coincidence. To hold the crystal and the tiles, an aluminum structure was designed, shown in Figure~\ref{ChamberPictures}. It was composed by two 6$\times$6$\times$0.5~cm$^3$ and two 6$\times$5$\times$0.5~cm$^3$ pieces hold together by eight screws, resulting in a experimental space of 6$\times$5$\times$5~cm$^3$. The two tiles were placed in the two faces not covered by the aluminum, and they were fixed to the structure by four screws each.

The crystals to test were cubic of 1" side, manufactured by Hilger Crystals~\cite{HilgerCrystals}. They were placed in the middle of the box and fixed by four screws with a diameter of 2.9~mm. To avoid the damage of the crystal during cooling procedures due to different expansion coefficients of the materials, four teflon pieces were placed between the screws and the crystal, distributing the stress applied to the crystal to hold it in place. In order to have a high LC, a high reflectivity of the inner surface for the aluminum box is required. To increase the reflectivity, the aluminum faces were polished. Moreover, the lateral crystal sides were covered by teflon reflecting tape, while top and bottom faces looked directly to the tiles.

\begin{figure}[h!]
	\begin{center}
		\includegraphics[width=\textwidth]{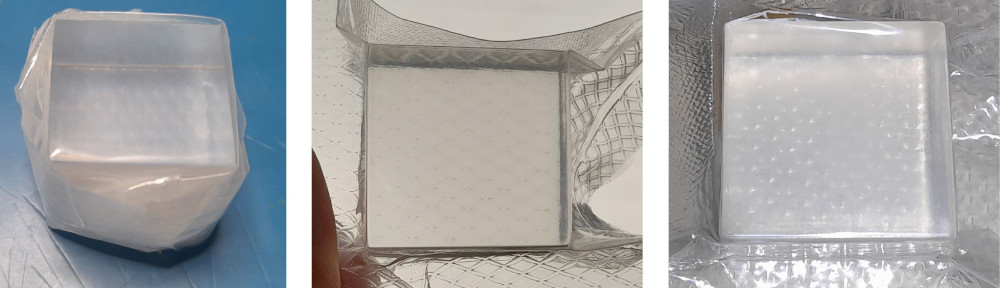}
		\caption{\label{NaIPictures}Pictures of the NaI crystal after encapsulation in plastic (left), before the cooling down procedure (center) and after warming up, one week after its encapsulation (right). A very good transparency of the crystal is observed by eye, and the encapsulation prevented the environmental humidity from damaging the crystal.}
	\end{center}
\end{figure}

NaI is a hygroscopic material, which means that absorbs very efficiently the water, and it degrades very fast when it is in contact with air with relative humidity $>$10\%. NaI crystals have to be tightly closed in dry atmosphere for any manipulation or operation. Two different methods for protecting the crystals from humidity were tried. First, the crystal was flushed with dry nitrogen gas while mounting the detector system, but the box was not tight and the crystal was damaged. Second, the crystal was encapsulated in a plastic bag in vacuum. It was done by introducing the crystal inside the plastic bag and using a system that reduces the internal pressure and heat-seals the plastic. Although the LC was expected to be reduced by the absorption and diffusion of the light in the plastic material, this encapsulation was effective in protecting the crystal, in the term of weeks after its encapsulation. The final encapsulation of the crystals is shown in Figure~\ref{NaIPictures}. It is observed by eye the good transparency of the crystal even after the cooling down and warming up procedures, one week after its encapsulation. However, for future crystal testing, a better protocol for the mounting of the detector system should be searched for in order to avoid unnecessary light losses.

\begin{figure}[h!]
	\begin{center}
		\includegraphics[width=0.75\textwidth]{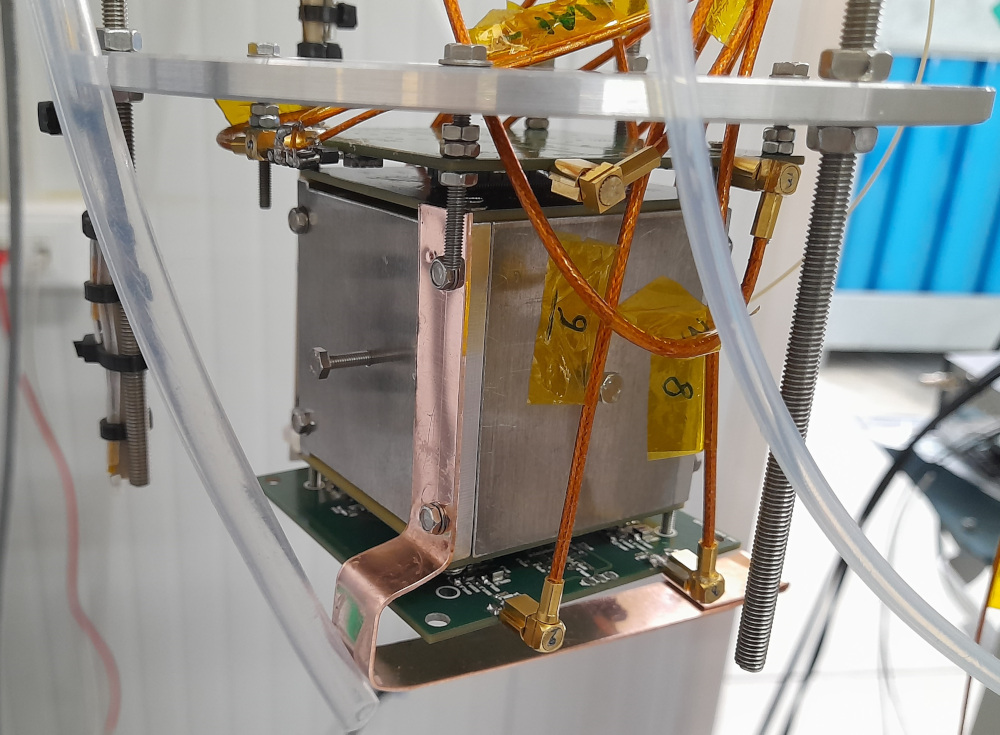}
		\caption{\label{ChamberCF}Picture of the detector attached to the copper cold plate in the cryostat.}
	\end{center}
\end{figure}

This detector system was designed to work immersed in LAr for cooling down to 87~K. Two holes in the aluminum structure allow the liquid filling the space around the crystal without bubbles. Moreover, to operate the detector without requiring the immersion in LAr, a copper plate (acting as a cold finger) was attached to the Al structure, allowing to keep the detector in thermal equilibrium with the LAr bath even in the case the LAr level was beneath the detector position in the cryostat (see Figure~\ref{ChamberCF}).

SPE calibration is essential to translate the acquired signal into number of photoelectrons, and then, to determine the LC if the corresponding energy deposition is well known. To do this calibration, a hole with a diameter of 1~mm was done in the middle of one of the faces of the aluminum chamber to introduce an optical fiber, that would be connected to a laser, for illuminating all the SiPMs. The laser used was a 400~mW Hamamatsu PLPC10196, which has a maximum emission at 405~nm and a pulse width of 60~ps.

\section{Experimental setup} \label{Section:SiPMSTAR2_Setup}
\fancyhead[RO]{\emph{\thesection. \nameref{Section:SiPMSTAR2_Setup}}}

\subsection{SiPM Test bench in Argon (STAR)} \label{Section:SiPMSTAR2_Setup_STAR}

\begin{figure}[h!]
	\begin{center}
		\includegraphics[width=0.75\textwidth]{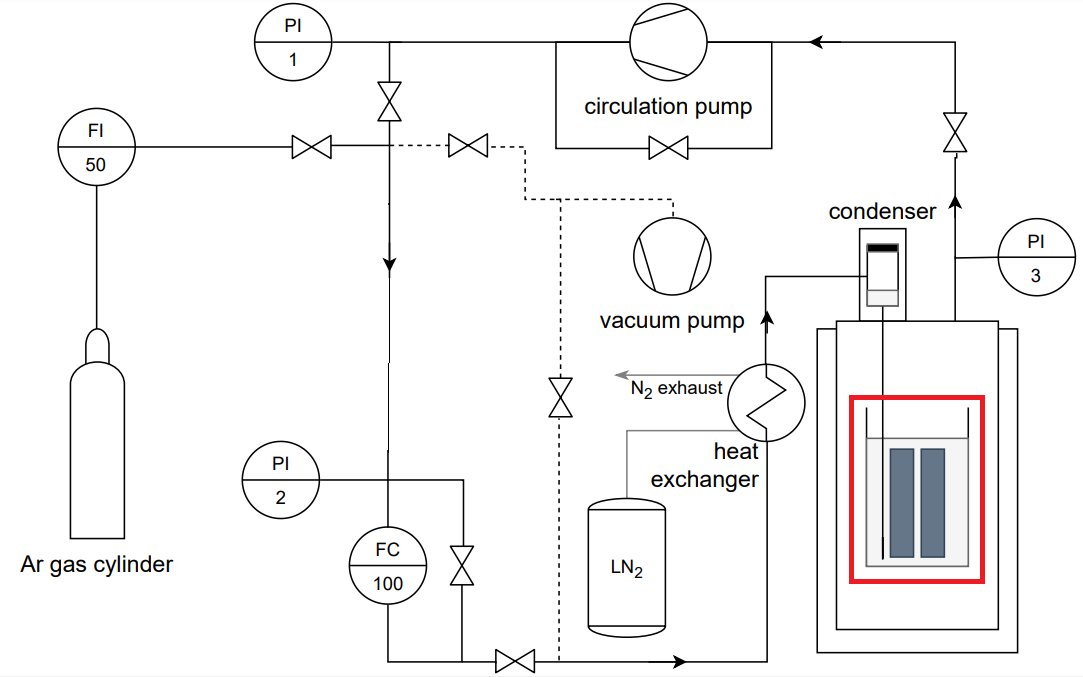}
		\caption{\label{STARScheme}Scheme of the STAR (SiPM Test bench in Argon) cryogenic system, from~\cite{Boulay:2021njr}. The red rectangle indicates the experimental space where the detector is placed.}
	\end{center}
\end{figure}

The measurements were carried out inside the cryogenic test setup called STAR (SiPM Test bench in Argon). An scheme of this system is shown in Figure~\ref{STARScheme}. It consists of a cryostat and an argon recirculation circuit~\cite{Boulay:2021njr}.

The cryostat is a vacuum insulated double jacket dewar with an inner diameter of 24~cm, a height of 93~cm and a CF250 top flange from CryoFAB~\cite{CryoFab}. It is equipped with an argon gas condenser cooled by a Cryomech~PT60 single-stage pulse tube cryocooler~\cite{CryoMech} capable of delivering 60~W at 80~K (corresponding to a nominal liquefaction rate of 6~l/min). A liquid nitrogen heat exchanger precools the input argon gas from room temperature to about 100~K before its introduction in the condenser to increase the speed of the condensation. The argon gas flow is promoted by a circulation pump (Metal Bellows MB-111) and regulated by a Sierra SmartTrack~100 gas flow controller.

\begin{figure}[h!]
	\begin{center}
		\includegraphics[width=\textwidth]{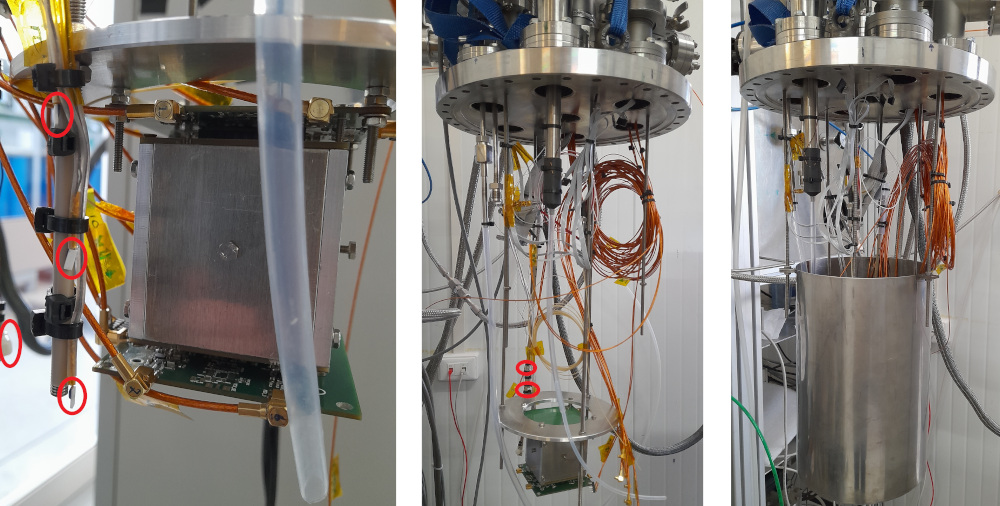}
		\caption{\label{CryostatPictures}Left picture: close view of the detector installed in the cryostat and the positions of the PT100 resistors (inside the red circles). The plastic tube in the foreground is connected to the condenser, and reaches the bottom part of the container. In the central picture, it can be seen how the detector is fixed to the top of the cryostat, and in the right picture the container that was filled with the LAr is shown.}
	\end{center}
\end{figure}

The liquefied argon is introduced inside the cryostat in a cylindrical container with a diameter of 19~cm and a length of 30~cm, which reduces the volume of LAr necessary to operate the system, and correspondingly the filling time to about 8~h. The detector was installed inside this container. It was fixed to the top of the cryostat through a metallic ring (seen in Figure~\ref{CryostatPictures}), leaving 3~cm between the bottom front-end board and the bottom surface of the container. In this gap, the end of the plastic tube that connects the condenser and the cryostat for the filling of the container was placed. The system allowed either to maintain up to 3~cm of LAr, while the rest of the container was filled with gaseous argon or to fill the container completely with LAr.

Connectors for the bias voltage of the SiPMs, the power supplies for the preamplifiers and the output signals were available at the top of the cryostat. Seven~PT100 resistors were installed close to the detector to monitor the level of LAr and the stability of the temperature during the measurements. The placement of the detector in the cryostat, the internal container and the positions of the PT100 resistors are shown in Figure~\ref{CryostatPictures}.

\subsection{Measurements schedule, electronic chain and DAQ} \label{Section:SiPMSTAR2_DAQ}

The measurements aimed at determining the LC and scintillation times of the NaI and NaI(Tl) crystals at different temperatures. For this goal, first, the SPE calibrations of the SiPMs using the laser were undertaken. Then, scintillation measurements with the crystal were carried out with an $^{241}Am$ external source (calibration measurements) and without external sources (background measurements). Data were taken both during the cooling down procedure and at LAr temperature. During the cooling down procedure, the temperature of the system changed between measurements but we tried to keep the overvoltage constant, while at LAr temperature different overvoltages were applied to the SiPMs.

\begin{figure}[h!]
	\begin{center}
		\includegraphics[width=\textwidth]{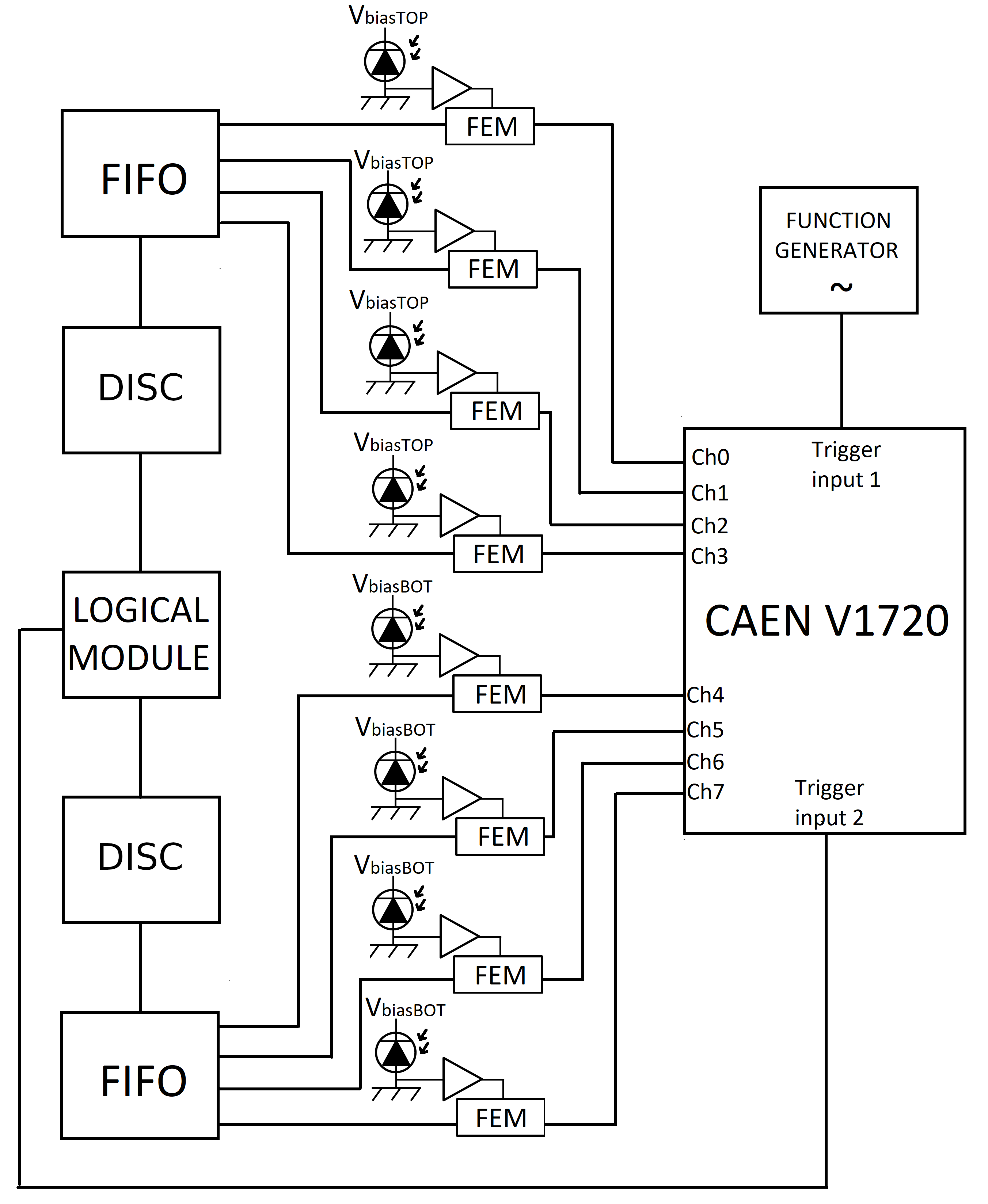}
		\caption{\label{ElectronicChainSTAR}Electronic chain designed for these measurements. Each channel (3$\times$2~SiPM array) is here represented as a single diode. A different voltage could be applied to top and bottom tiles, but it is the same for the four channels on each tile ($V_{biasTOP}$ or $V_{biasBOT}$). During SPE calibrations, the function generator biasing the laser was used for triggering (trigger input~1), while in the acquisition of scintillation light, a logic module required a logical AND coincidence between the signals of the two tiles for triggering (trigger input~2).}
	\end{center}
\end{figure}

To do these measurements, the electronic chain shown in Figure~\ref{ElectronicChainSTAR} was used. For SPE calibration the trigger was done by the function generator powering the laser (in Figure~\ref{ElectronicChainSTAR}, trigger input~1), defining a square pulse with an operation frequency of 400~Hz and a pulse width of 60~ps. On the other hand, a coincident signal between the two tiles was required to trigger during scintillation measurements to avoid DC events (in Figure~\ref{ElectronicChainSTAR}, trigger input~2).

As it was explained in Section~\ref{Section:SiPMSTAR2_Detector}, the architecture of the arrays in the tile provides four output signals (from now on, channels) per tile, which are preamplified at the FEB. Each one of the eight preamplified channels was connected to a Front-End NIM module (FEM) that divides the signal in two (see Figure~\ref{ElectronicChainSTAR}). These FEMs were specifically designed to be used in DarkSide-50~\cite{DarkSide:2017odo}. They also allow to amplify the signal (x10), something that was used to increase the SNR of the SPE calibrations at low overvoltages.

The two output signals of the FEM module for each channel are fed into the digitizer and to a Fan-In Fan-Out (FIFO) module for the trigger input~2. This module adds together the four signals of the same tile. The output signals of these FIFO modules were processed by two discriminator modules (DISC), which generated a NIM pulse when the signal was higher than a threshold that could be defined at will. Those NIM pulses entered into a logic coincidence module (setup in AND mode for a time window of 200~ns) that produces the DAQ trigger.

\begin{figure}[h!]
	\begin{center}
		\includegraphics[width=\textwidth]{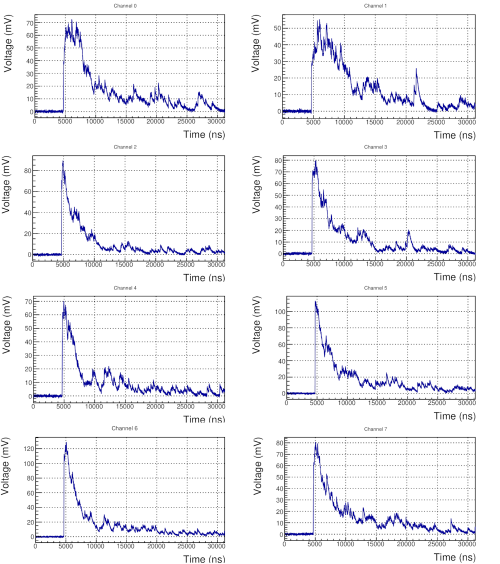}
		\caption{\label{8PulsesStar}Example of an event acquired in a background measurement with the NaI(Tl) crystal at LAr temperature and at an overvoltage of 5~V. The pulses for the eight channels are shown (channels [0,3] for top tile and [4,7] for bottom tile).}
	\end{center}
\end{figure}

An 8~Channel 12~bit 250~MHz 2~V digitizer model CAEN~V1720~\cite{CAENV1720} acquired the 8~signals (4~channels per tile) with an acquisition window and a pretrigger defined in a configuration file. As the scintillation time of the NaI and NaI(Tl) was expected to depend on the temperature, the digitization window was chosen for the crystals measurement as 60~$\mu$s (close to the limit allowed by the digitizer) trying to guarantee recording the whole pulse. In all the cases, the pretrigger was fixed as 15$\%$ of the record length. As an example, the 8~pulses of an event acquired in a background measurement with the NaI(Tl) crystal at LAr temperature and at an overvoltage of 5~V are shown in Figure~\ref{8PulsesStar}.

\begin{figure}[h]
	\begin{center}
		\begin{subfigure}[b]{0.75\textwidth}
			\includegraphics[width=\textwidth]{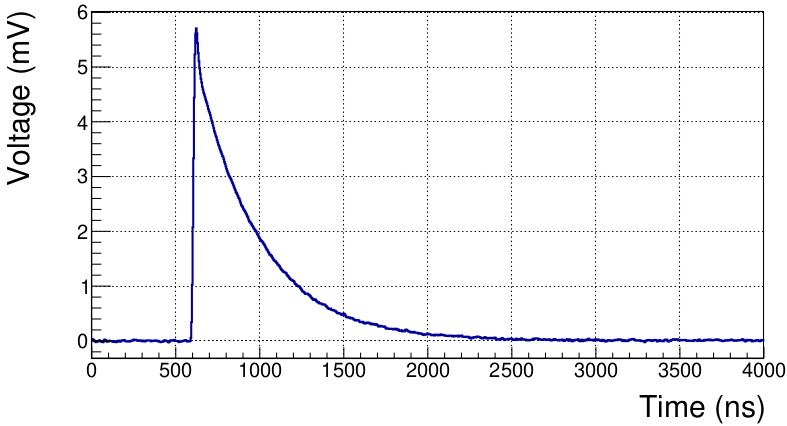}
		\end{subfigure}
		\caption{\label{AveragedPulsesStar}Average of $10^4$~pulses corresponding to a SPE for the channel~0 in the measurement with the NaI crystal at LAr temperature and at an overvoltage of 5~V. See Section~\ref{Section:SiPMSTAR2_Analysis_ScintTimes} for more information about how the pulses are selected for averaging.}
	\end{center}
\end{figure}

Figure~\ref{AveragedPulsesStar} shows the average of $10^4$~pulses of a SPE for one of the channels when the system is illuminated with the laser light and the DAQ triggered by the function generator powering the laser. The mean decay time of the SPE signal goes from 200~ns to 350~ns depending on the temperature, so that for measurements with the laser the digitization window was fixed as 4~$\mu$s during the characterization of the NaI crystal. To analyze possible changes due to the integration window, 6~$\mu$s were digitized in the characterization of the NaI(Tl) crystal.

For each event, the DAQ stores the value of the timestamp and the waveform of the 8~channels in a wav file. A decoder is then applied to this file to generate a ROOT file with the same information, while the analysis explained in Section~\ref{Section:SiPMSTAR2_Analysis} calculates additional parameters for each event.

\section{Optical simulation of the detector} \label{Section:SiPMSTAR2_GEANT4}
\fancyhead[RO]{\emph{\thesection. \nameref{Section:SiPMSTAR2_GEANT4}}}

An optical MC simulation was developed using the GEANT4 package~\cite{GEANT4:2002zbu}. The simulation addressed two complementary goals: analyzing the effect on the LC of the medium filling the container (liquid nitrogen, LN$_2$, liquid argon, LAr, or gaseous argon, GAr) and to choose the most convenient radioactive source from the two available ($^{241}Am$ and $^{133}Ba$) for the characterization of the NaI crystal properties by analyzing the corresponding efficiencies for the detection of the photopeaks. Additionally, it allowed to obtain the mean energies of the photopeaks used in the analysis.

The primary particles of this simulation were the isotopes $^{241}Am$ and $^{133}Ba$ decaying at rest from a point-like source placed close to the aluminum chamber but outside the cryostat. The $^{133}Ba$ emissions were presented previously in Tables~\ref{tabla:Ba133_Xray} and~\ref{tabla:Ba133_gamma} for x-rays and gammas, respectively, while the $^{241}Am$ x-ray emissions are presented in Table~\ref{tabla:Am241_Xray}. The main gamma emission of this isotope is at 59.5~keV with a branching ratio (BR) of 35.9\%, and it has another emission at 26.3~keV with a BR of 2.4\%. The x-ray emissions are not expected to reach the crystal, because they will be absorbed in the interposed materials. As the crystals used in the measurements have a small size (cubes of 1" side), after photoelectric absorption of the 59.5~keV photon, the x-rays of the K-shell of the iodine (which have energies between 28.3~and 32.3~keV) would escape easily, and therefore an escape peak around 29-30 keV will be observed.

\begin{table}[h!]
	\centering
	\begin{tabular}{|c|c|c|}
		\hline
		Energy (keV) & BR (\%) & Assignment \\
		\hline
		13.76 & 1.1 & Np L$_{\alpha2}$ \\
		13.95 & 9.6 & Np L$_{\alpha1}$ \\
		16.82 & 2.5 & Np L$_{\beta2}$ \\
		17.06 & 1.5 & Np L$_{\beta4}$ \\
		17.75 & 5.7 & Np L$_{\beta1}$ \\
		17.99 & 1.4 & Np L$_{\beta3}$ \\
		20.78 & 1.4 & Np L$_{\gamma1}$ \\
		\hline
	\end{tabular}
	\caption{X-ray emissions from the $^{241}Am$ source with a branching ratio (BR) higher than 1$\%$~\cite{Akovali:1994zfm}.}
	\label{tabla:Am241_Xray}
\end{table}

\begin{figure}[h!]
	\begin{center}
		\includegraphics[width=\textwidth]{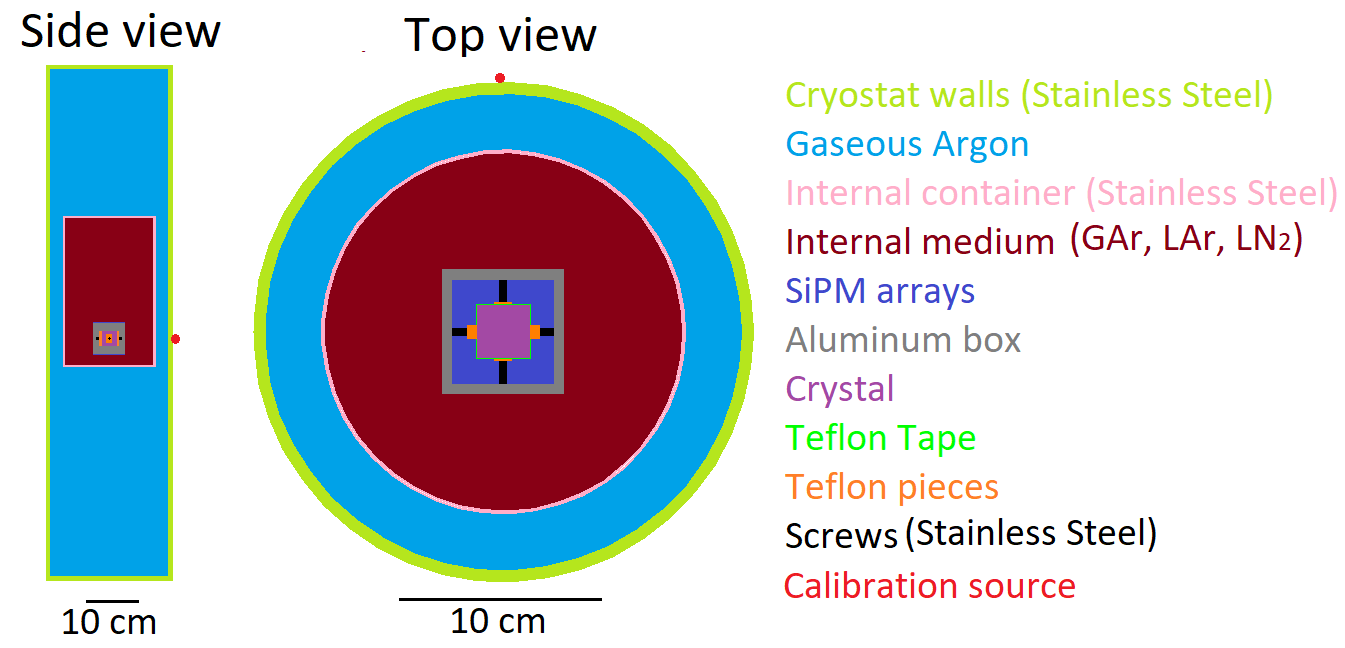}
		\caption{\label{CryostatGeant}Side and top views of the geometry defined in the GEANT4 simulation.}
	\end{center}
\end{figure}

\begin{figure}[h!]
	\begin{center}
		\includegraphics[width=\textwidth]{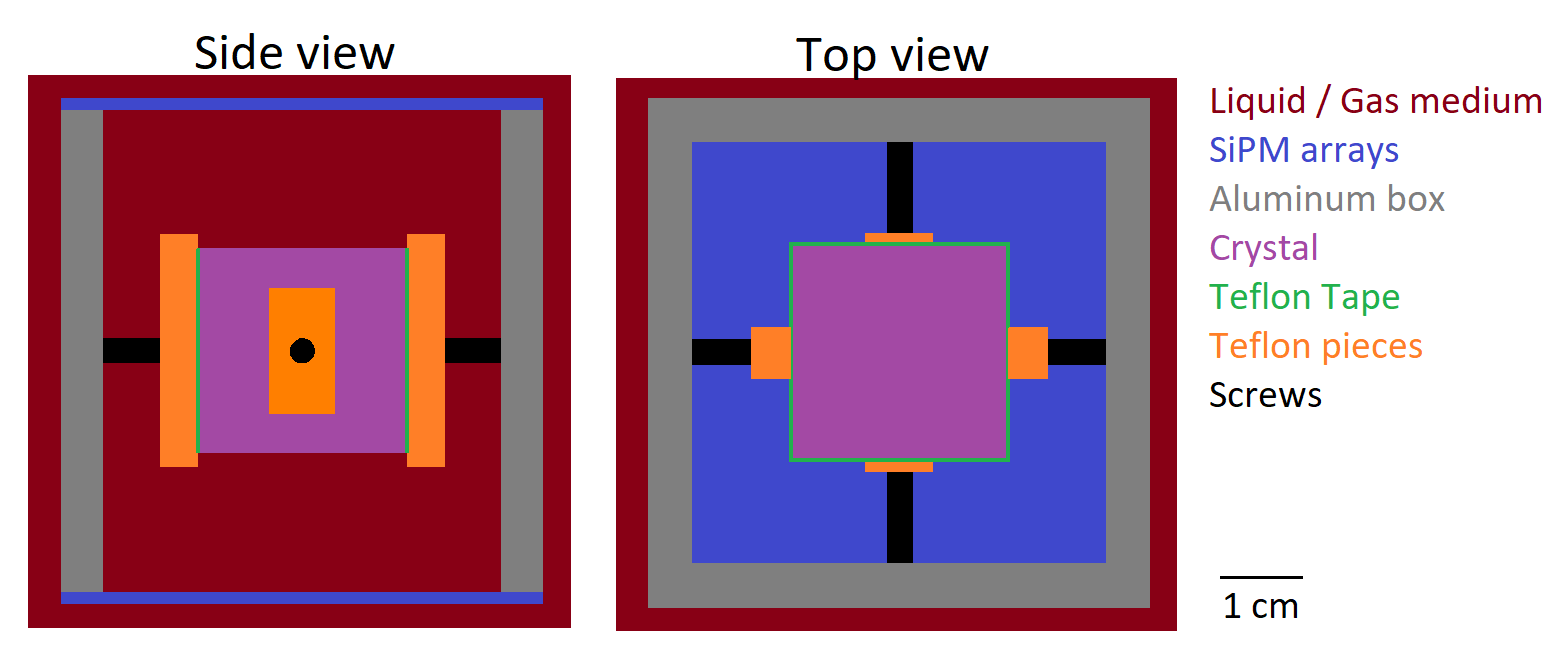}
		\caption{\label{ChamberGeant}Close side and top views of the geometry defined for the chamber containing the crystal and the SiPM tiles.}
	\end{center}
\end{figure}

The geometry implemented in the simulation included all the volumes and material media that could affect the LC or the energy depositions of the radiation before reaching the crystal (see Figures~\ref{CryostatGeant} and~\ref{ChamberGeant} and Sections~\ref{Section:SiPMSTAR2_Detector} and~\ref{Section:SiPMSTAR2_Setup} for the dimensions and positions of the different system components):

\begin{itemize}
	\item Cryostat walls (stainless steel)
	\item Medium inside the cryostat and outside the container (GAr or GN$_2$)
	\item Internal container walls (stainless steel)
	\item Internal medium (LAr, LN$_2$ or GAr)
	\item Aluminum box
	\item NaI/NaI(Tl) crystal
	\item SiPM arrays
	\item Teflon tape around the crystal
	\item Teflon pieces
	\item Screws to fix the crystal (stainless steel)
\end{itemize}

The teflon tape around the crystal was included assuming 0.5~mm thickness. The optical properties of the materials included in the simulation are presented in Table~\ref{tabla:OptPropSTAR}. As no information was found for the absorption lengths of LAr, LN$_2$ and GAr, they were chosen as 100~m (same as that used in Chapter~\ref{Chapter:OptSim} for the NaI(Tl) crystal) for the liquid media and infinity for the gaseous.

\begin{table}[h!]
	\centering
	\begin{tabular}{|c|c|c|}
		\hline
		Material & Refractive index & Absorption length \\
		\hline
		GAr & 1 & $\infty$ \\
		LAr & 1.23 & 100 m \\
		LN$_2$ & 1.205 & 100 m \\
		NaI / NaI(Tl) & $\sim$~1.85 (see Fig.~\ref{NaIPropSim}) & 100 m \\
		Teflon & 1.3 & 1 mm \\
		\hline
	\end{tabular}
	\caption{Refractive indices and absorption lengths for optical photons in the wavelength of interest for each material included in the optical simulation. Refractive indices have been obtained from~\cite{Sinnock:1969zz} for LAr, from~\cite{Johns:2011} for LN$_2$, from~\cite{Li1976RefractiveIO} for NaI and from~\cite{ScientificPolymer} for teflon. Absorption length of teflon was found in~\cite{Yang:2008}.}
	\label{tabla:OptPropSTAR}
\end{table}

\begin{figure}[h!]
	\begin{center}
		\begin{subfigure}[b]{0.49\textwidth}
			\includegraphics[width=\textwidth]{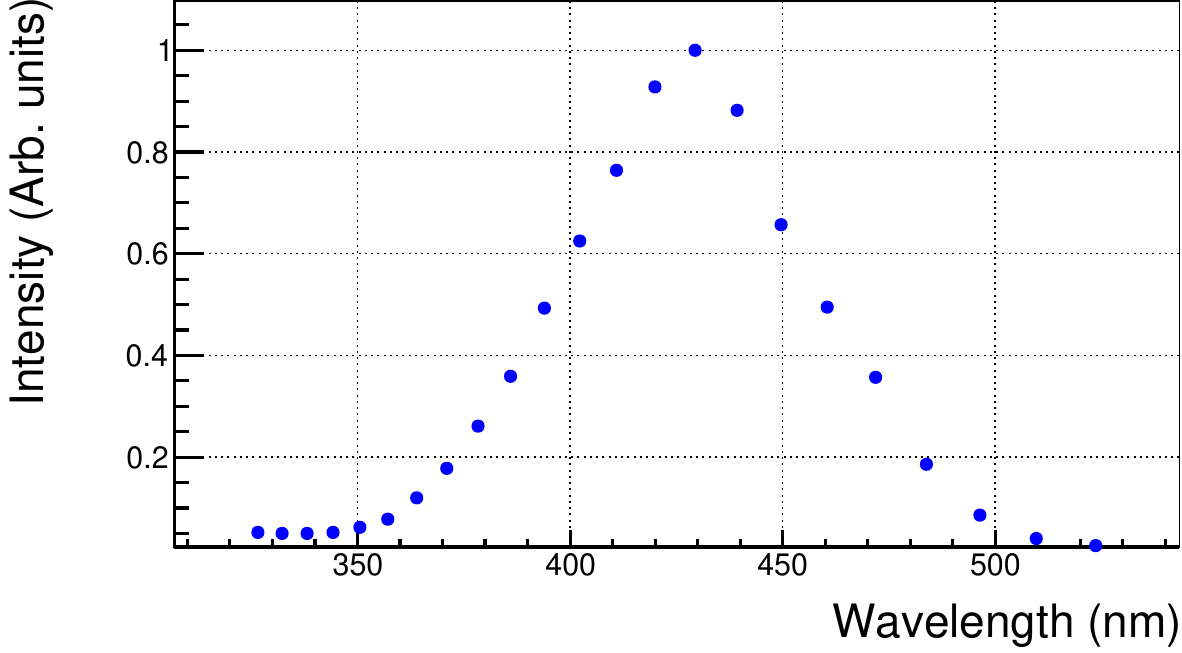}
		\end{subfigure}
		\begin{subfigure}[b]{0.49\textwidth}
			\includegraphics[width=\textwidth]{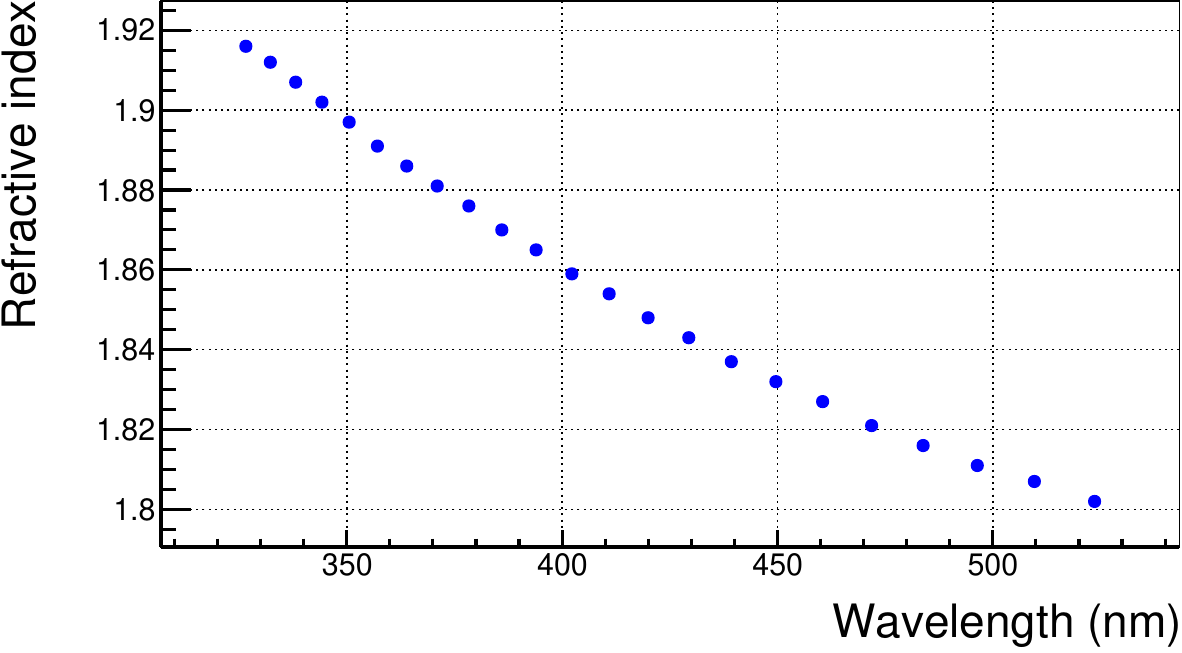}
		\end{subfigure}
		\caption{\label{NaIPropSim}Left plot shows the normalized emission spectrum~\cite{Kumar:2021}, and right plot presents its refractive index~\cite{Li1976RefractiveIO} as function of the wavelength for NaI(Tl) at liquid nitrogen temperature.}
	\end{center}
\end{figure}

The reflectivities of the aluminum and teflon surfaces were taken as~50 and~99$\%$, respectively, but the precise values could not be measured. The scintillation spectrum of the NaI(Tl) crystal at liquid nitrogen temperature was obtained from~\cite{Kumar:2021} and the refractive index from~\cite{Li1976RefractiveIO}. Both depend on the wavelength, and this dependence has been included in the simulation with the values shown in Figure~\ref{NaIPropSim}. The scintillation time was included as 230~ns and the light yield of the crystal as 40~photons/keV, nominal values at room temperature from Hilger Crystals~\cite{HilgerCrystals}. It is worth to note that these values are different at LAr temperature, and then, the corresponding scaling of the resulting LC should be applied.

The volumes defined as active for registering energy depositions were the crystal and the two SiPM arrays. For each event, the information stored was the energy deposition in the crystal and the number of optical photons that reached any of the arrays. A reduction in the number of optical photons of 50$\%$ was applied later to take into account the total PDE of the SiPMs.

This simulation could not reproduce the number of photons reaching the tiles, pointing to an incorrect modelling of the optical properties of the materials considered in the simulation. However, we rely on the simulation to evaluate the expected energy depositions in the crystals for different container filling conditions, and we expect our lack of knowledge of those optical properties does not prevent to draw some qualitative conclusions about the LC in different measurement conditions. The LC (corrected to take into account only the scintillation component as explained in Section~\ref{Section:SiPM_Model}) obtained for the 59.5~keV peak in the measurement presented in Section~\ref{Section:SiPMSTAR2_Charac_LAr} for NaI(Tl) was 143~phe/59.5~keV, while in the simulation with GAr, the result is much higher, 824~phe/59.5~keV.

\begin{figure}[h!]
	\begin{center}
		\includegraphics[width=0.75\textwidth]{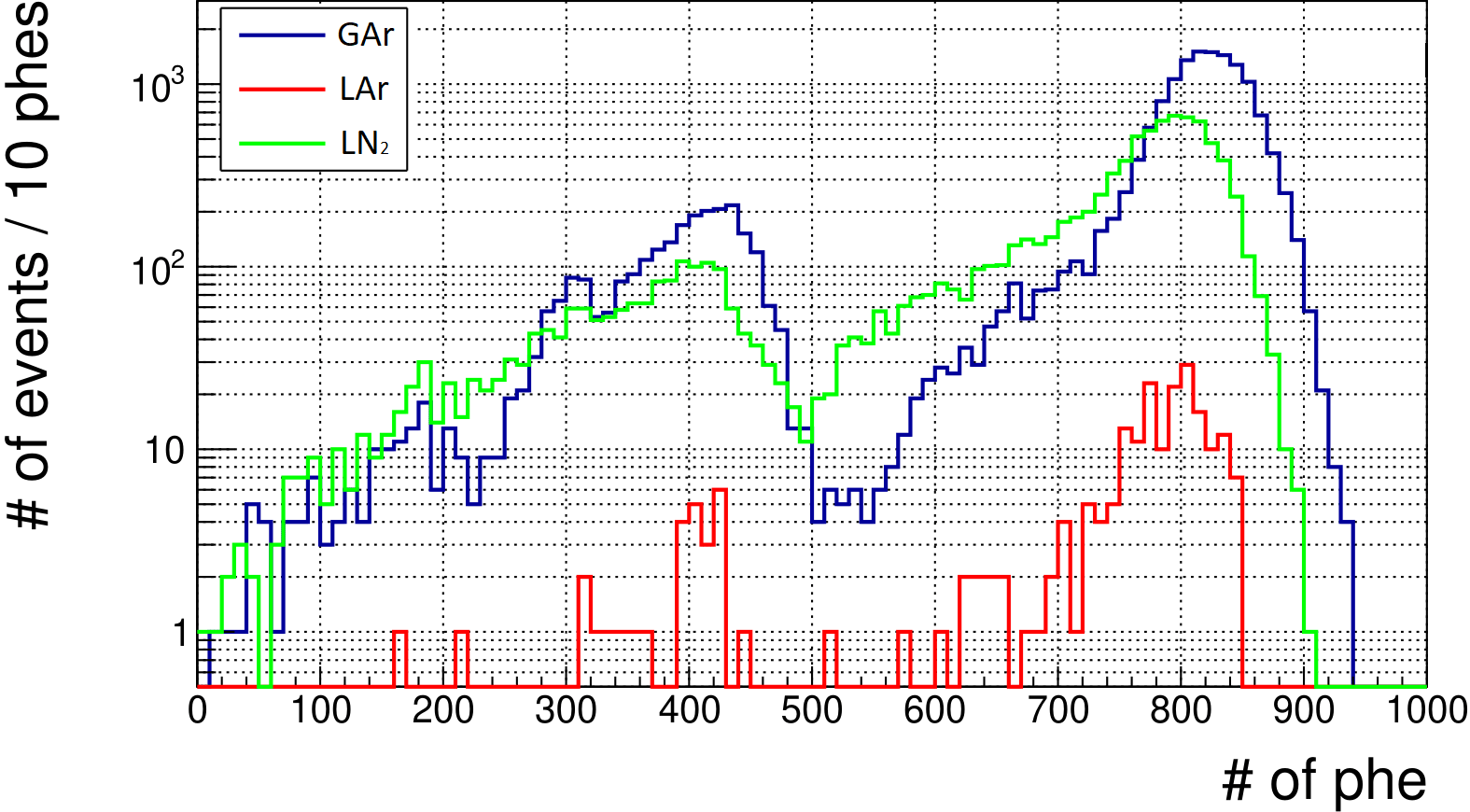}
		\caption{\label{Sim_Am241}Comparison of the $^{241}Am$ simulated spectra for the energy depositions in the NaI(Tl) crystal for the different filling materials.}
	\end{center}
\end{figure}

\begin{figure}[h!]
	\begin{center}
		\includegraphics[width=0.75\textwidth]{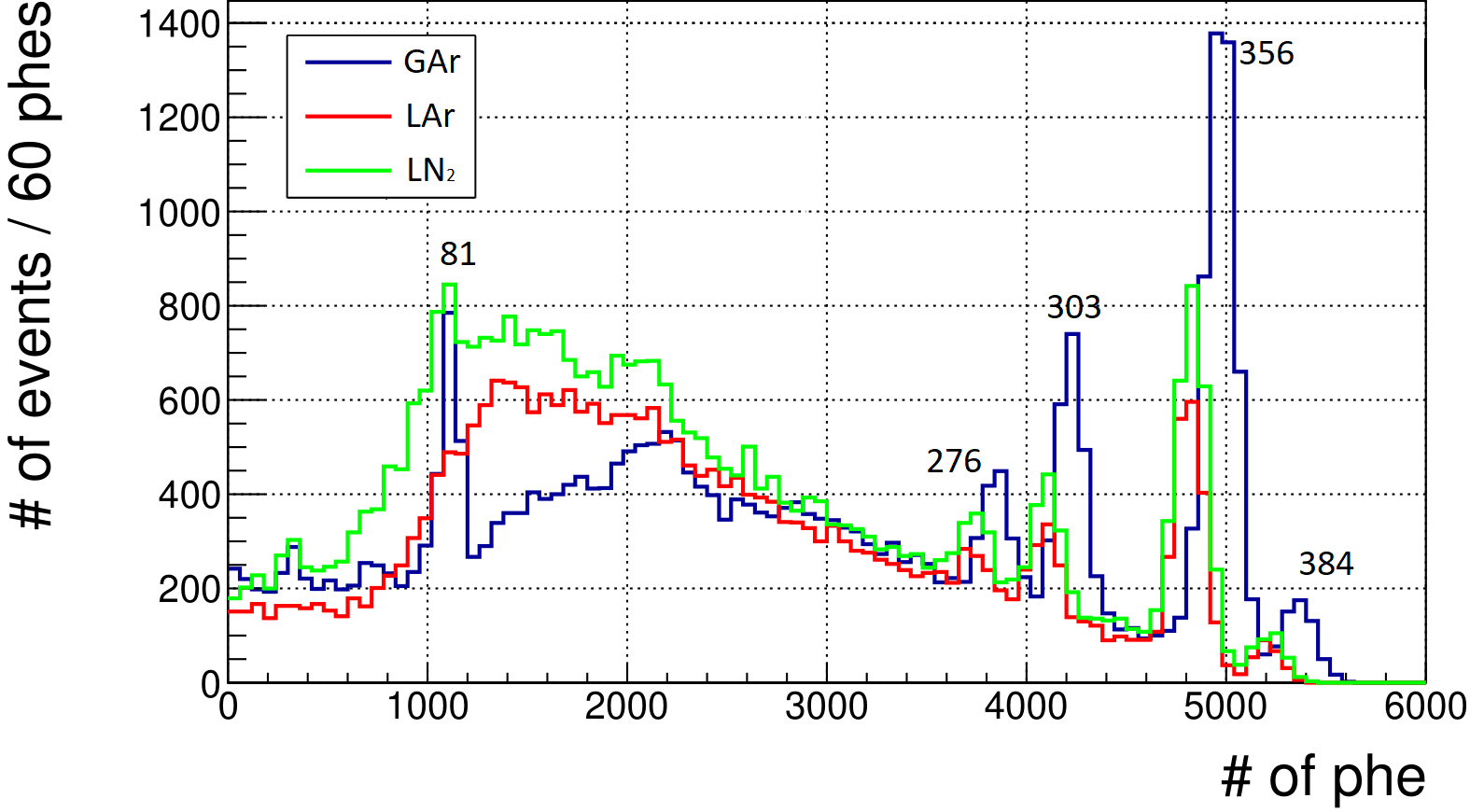}
		\caption{\label{Sim_Ba133}Comparison of the $^{133}Ba$ simulated spectra for the energy depositions in the NaI(Tl) crystal for the different filling materials.}
	\end{center}
\end{figure}

Six different simulations were run: using each radioactive source ($^{241}Am$ and $^{133}Ba$) and defining the three possible filling options (GAr, LAr and LN$_2$). In the simulations of the $^{241}Am$ source, 10$^9$ decays were simulated, while for the $^{133}Ba$ source, the number of decays was 10$^7$. The $^{241}Am$ source spectra obtained for the different filling materials are shown in Figure~\ref{Sim_Am241}, while for $^{133}Ba$ source are shown in Figure~\ref{Sim_Ba133}. The number of energy depositions per 10$^5$ events is shown in Table~\ref{tabla:SimSTAR}, as well as the density and atomic number of each material.

\begin{table}[h!]
	\centering
	\begin{tabular}{|c|c|c|c|c|}
		\hline
		Filling material & Density (g/cm$^3$) & Z & $\#$ events $^{133}Ba$ & $\#$ events $^{241}Am$ \\
		\hline
		GAr & 0.001 & 18 & 314 $\pm$ 2 & 166 $\pm$ 1 \\
		LAr & 1.399 & 18 & 280 $\pm$ 2 & 2 $\pm$ 1 \\
		LN$_2$ & 0.806 & 7 & 359 $\pm$ 2 & 97 $\pm$ 1 \\
		\hline
	\end{tabular}
	\caption{Properties of the three possible materials used for the filling of the chamber and the internal container, as well as the number of energy depositions in the NaI crystal per 10$^5$ simulated events for each radioactive source and each material.}
	\label{tabla:SimSTAR}
\end{table}

It can be concluded that although the $^{133}Ba$ source produces a higher number of energy depositions in the NaI(Tl) crystal per isotope decay than the $^{241}Am$ source, the high energy emissions of $^{133}Ba$ suffer from Compton interactions in the interposed materials, and can make more difficult the interpretation of the measured spectra, in particular in the case of using LAr in the filling. The reason is that this material has a higher density than the gaseous argon and higher atomic number than the nitrogen, and both things increase the absorption probability of the photons in the material. We use the results of these simulations to compare the LC obtained for the different filling media. In Figures~\ref{Sim_Am241}, and~\ref{Sim_Ba133} it can be observed that LC is slightly larger for the GAr while is very similar for both LAr and LN2. This can be probably explained by the higher light scattering probability in the liquid media, which increases the amount of photons reaching the aluminum and thus the probability of being absorbed before reaching the SiPMs. The simulation shows that the loss of light due to this process is around 5\%.

These results confirmed that the calibration run, independently from the source used, should not be done with the chamber filled with LAr. To ensure the stability of the system, we decided to cool down using nitrogen instead of argon, but after a first trial it was observed that the system had low efficiency for liquefying this gas, requiring too much time for cooling down. Instead, we decided to cool down the system using argon to fully fill the internal container and chamber and then wait 18~hours for reaching the base temperature. After that time, most of the LAr was removed and the system could be kept in stable conditions for the measurements, with some LAr in the bottom of the container (below the bottom SiPM array). The chamber and the upper part of the container were then filled with gaseous argon. A copper plate attached to the chamber and immersed in the LAr allowed to avoid the crystal warming-up due to Joule effect in the bottom SiPM array (as it was explained in Section~\ref{Section:SiPMSTAR2_Detector}). This effect in the top SiPM array is expected to heat only the gas above the chamber and therefore it should have a negligible effect on the detector. A picture of the final configuration used during the measurements was shown in Figure~\ref{ChamberCF}.

\begin{figure}[h!]
	\begin{center}
		\includegraphics[width=0.75\textwidth]{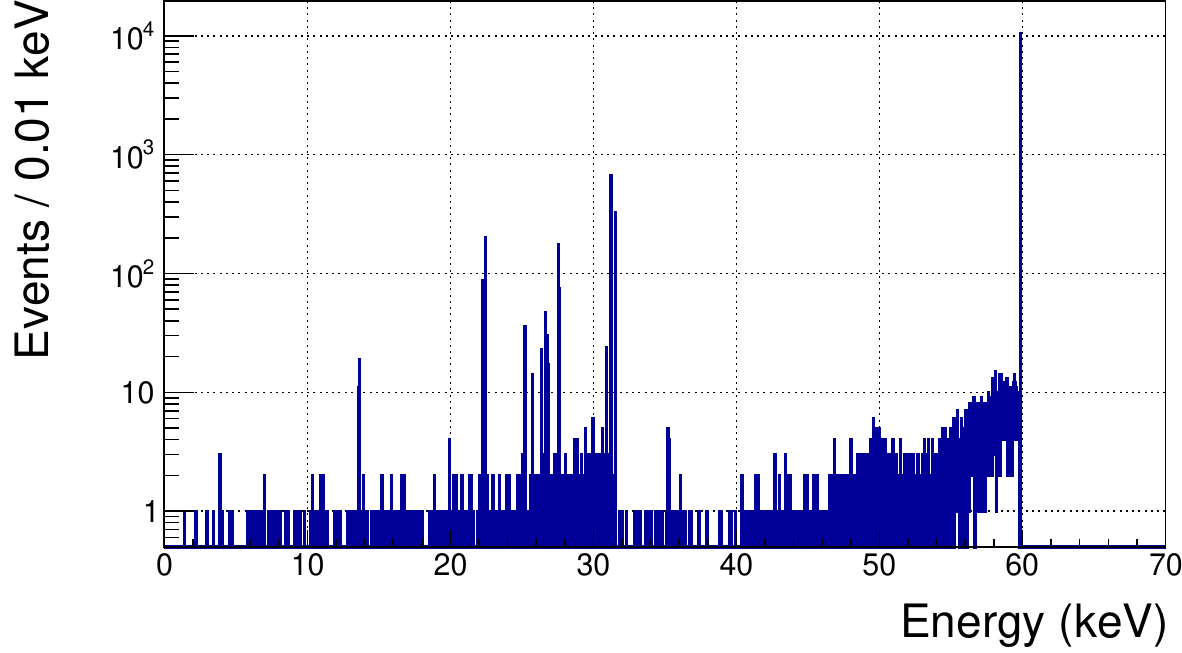}
		\caption{\label{EeeSimAm241}Spectrum of energy deposited in the crystal in the simulation of the $^{241}Am$ source when the chamber is filled with GAr.}
	\end{center}
\end{figure}

Once the $^{241}Am$ source was selected for the energy calibration, the mean energies of the two photopeaks observed in the spectrum were obtained. The spectrum of energy deposited in the crystal during the $^{241}Am$ calibration when the chamber is filled with GAr is shown in Figure~\ref{EeeSimAm241}. As expected, the main peak is that with an energy of 59.5~keV, and it is a single energy deposition. However, between 22 and 32~keV there are multiple lines contributing, due to the iodine x-ray escapes. The two main lines in that energy region have energies of 31.3 and 31.5~keV, but the calculated mean energy is 31.5~keV. This will be the energy considered for this peak in Section~\ref{Section:SiPMSTAR2_Charac}.

\section{Data analysis} \label{Section:SiPMSTAR2_Analysis}
\fancyhead[RO]{\emph{\thesection. \nameref{Section:SiPMSTAR2_Analysis}}}

The analysis of the data was done using ROOT package~\cite{Brun:1997pa}. The algorithm used for the calculation of the relevant variables for each event is explained in this section, as well as the method followed to obtain the LC and the scintillation times in each measurement carried out.

\subsection{Event analysis} \label{Section:SiPMSTAR2_EventAnalysis}

A first level analysis is applied to the wav files from each measurement (structure explained in Section~\ref{Section:SiPMSTAR2_DAQ}). This analysis is applied to the pulses recorded for each channel and to the addition of the eight channel pulses for the same event. In this analysis for each pulse, the baseline level is calculated using the first 100~ns of the pretrigger region, and the corresponding mean level and standard deviation are saved. The mean baseline of each pulse is subtracted from the waveform, and then, the pulse onset (\textit{t0}) is identified as the time when the waveform becomes higher than a threshold, defined as five times the standard deviation of the baseline. The maximum of the pulse is also saved (variable \textit{high}), and the corresponding time (\textit{tmax}).

The mean time of the pulse, $\mu$, is defined as
\begin{equation}\label{eq:mu}
	\mu = \frac{\sum_{t = t0}^{t = t_{acq}} V(t)\cdot (t-t0)}{\sum_{t = t0}^{t = t_{acq}}V(t)},
\end{equation}
where $V(t)$ is the value of the baseline-subtracted waveform at time $t$, and $t_{acq}$ is the end of the time acquisition window. This variable can be used for the analysis of the pulse shape, allowing to analyze effects as the loss of linearity in the behaviour of the detector or to select scintillation events.

To reduce the threshold effects introduced by the algorithm that looks for the pulse onset, a fixed integration window, independent from \textit{t0}, is used to build the \textit{area} variable. It was set differently for SPE and scintillation measurements. For the former, the integration window used was 2~$\mu$s from the hardware trigger position, fixed in the DAQ configuration file. Because the SPE mean time was of the order of 300~ns (see Section~\ref{Section:SiPMSTAR2_DAQ}), this range guarantees to integrate more than 99\% of the SPE area. For scintillation measurements, as the scintillation times depend on the scintillating material and the temperature, different digitization windows were used, with 15\% pretrigger region in all the cases (see Section~\ref{Section:SiPMSTAR2_DAQ}). For them, the integration window is chosen from 12\% to 100\% of the digitization window.

\subsection{Light collection measurements} \label{Section:SiPMSTAR2_Analysis_LC}

\begin{figure}[h!]
	\begin{center}
		\includegraphics[width=0.75\textwidth]{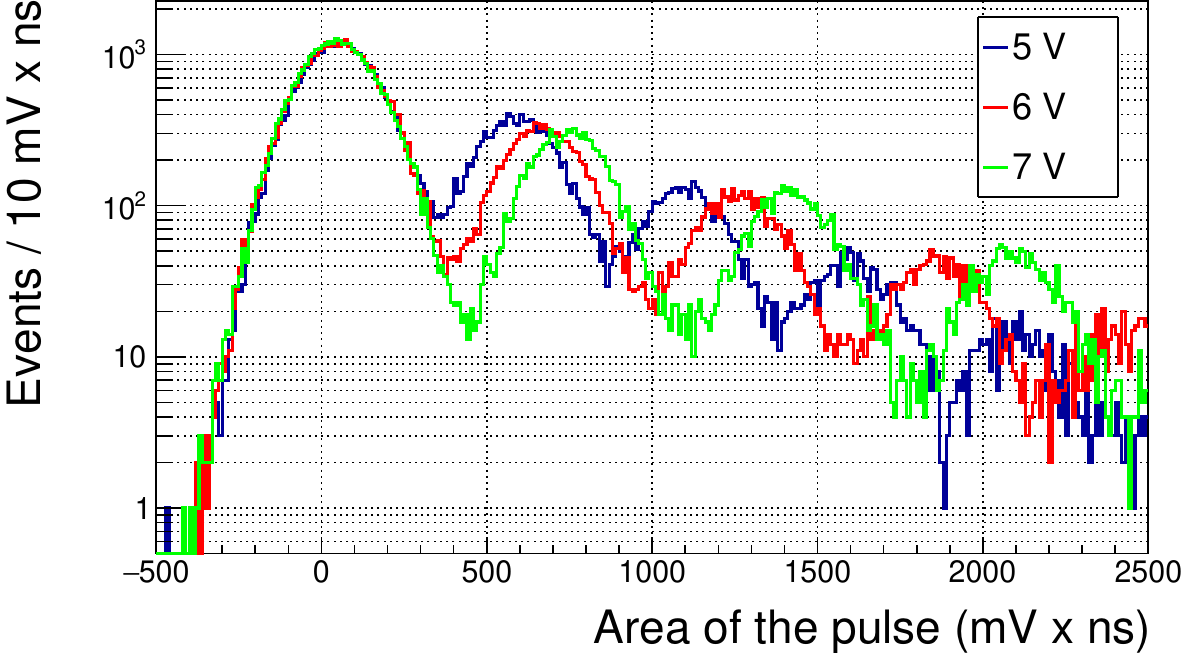}
		\caption{\label{SPECal_vs_OV}Pulse area spectra for SPE calibration of channel~7 for the measurement with the NaI(Tl) crystal at LAr temperature and at three different overvoltages. The peak with minimum area corresponds to the integral of the baseline, while the others include one or more phes.}
	\end{center}
\end{figure}

The process followed to calculate the LC in each condition (crystal, temperature and overvoltage) is explained in the following. First, the SPE area has to be estimated from the SPE calibrations. Figure~\ref{SPECal_vs_OV} shows the pulse area spectra for a SPE calibration of the same channel at three different overvoltages.

\begin{figure}[h!]
	\begin{center}
		\includegraphics[width=\textwidth]{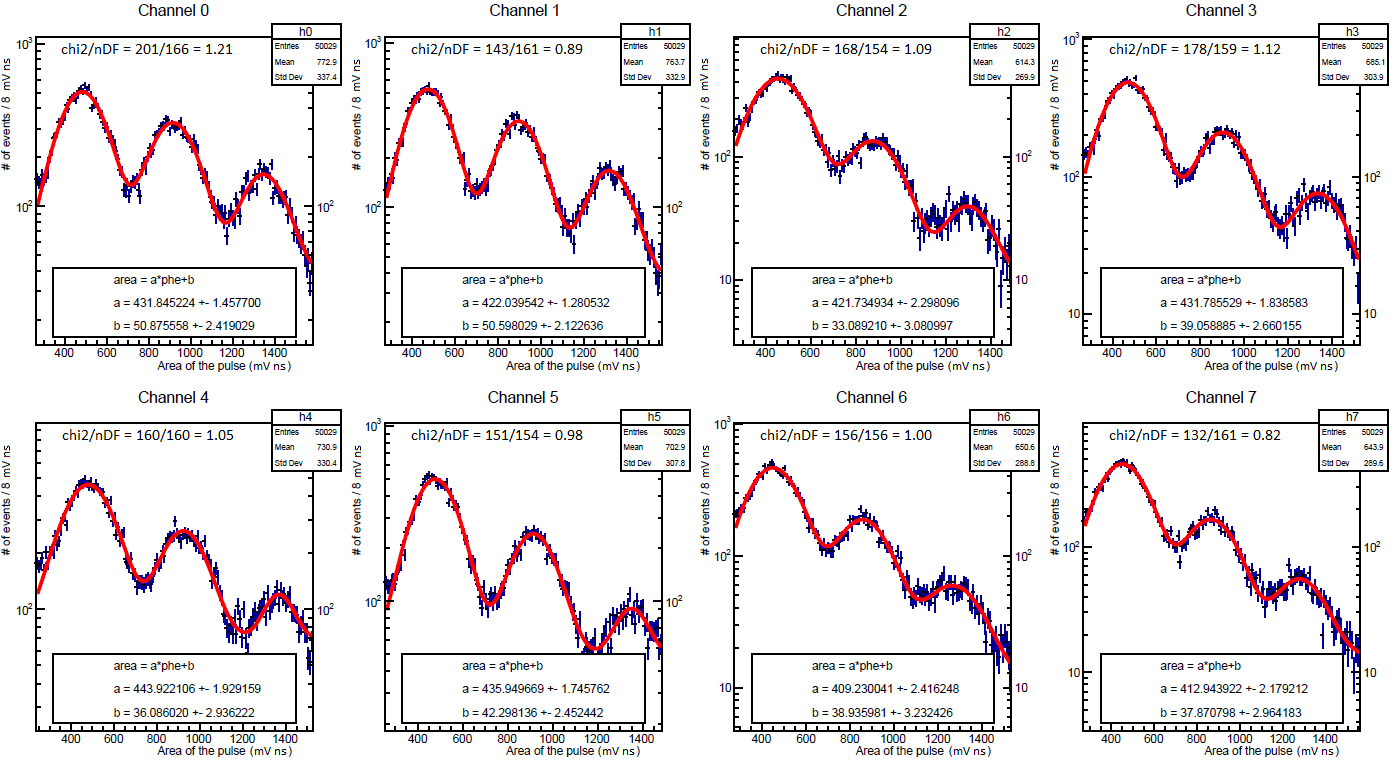}
		\caption{\label{SPECalSTAR}SPE calibration of the eight channels for the measurement with the NaI(Tl) crystal at LAr temperature and at an overvoltage of 5~V.}
	\end{center}
\end{figure}

The SPE area, $a$, can be obtained as the distance between consecutive peaks in the pulse area spectrum. Therefore, the first three peaks were fitted to three gaussians (the peak with minimum area observed in the pulse area spectrum in Figure~\ref{SPECal_vs_OV} was not fitted, as it corresponds to the integral of the baseline). Then, assuming linearity, and being $\mu$ the mean value of the area of the phe peaks:
\begin{equation}
	\mu = a\cdot N_{phe} + b,
\end{equation}
where $N_{phe}$ is the number of photoelectrons corresponding to the pulses with mean area $\mu$. The free parameters of this fitting procedure are $a$, $b$, and the standard deviation and area of the gaussians. An example of this calibration is shown in Figure~\ref{SPECalSTAR}.

\begin{figure}[h!]
	\begin{center}
		\begin{subfigure}[b]{0.49\textwidth}
			\includegraphics[width=\textwidth]{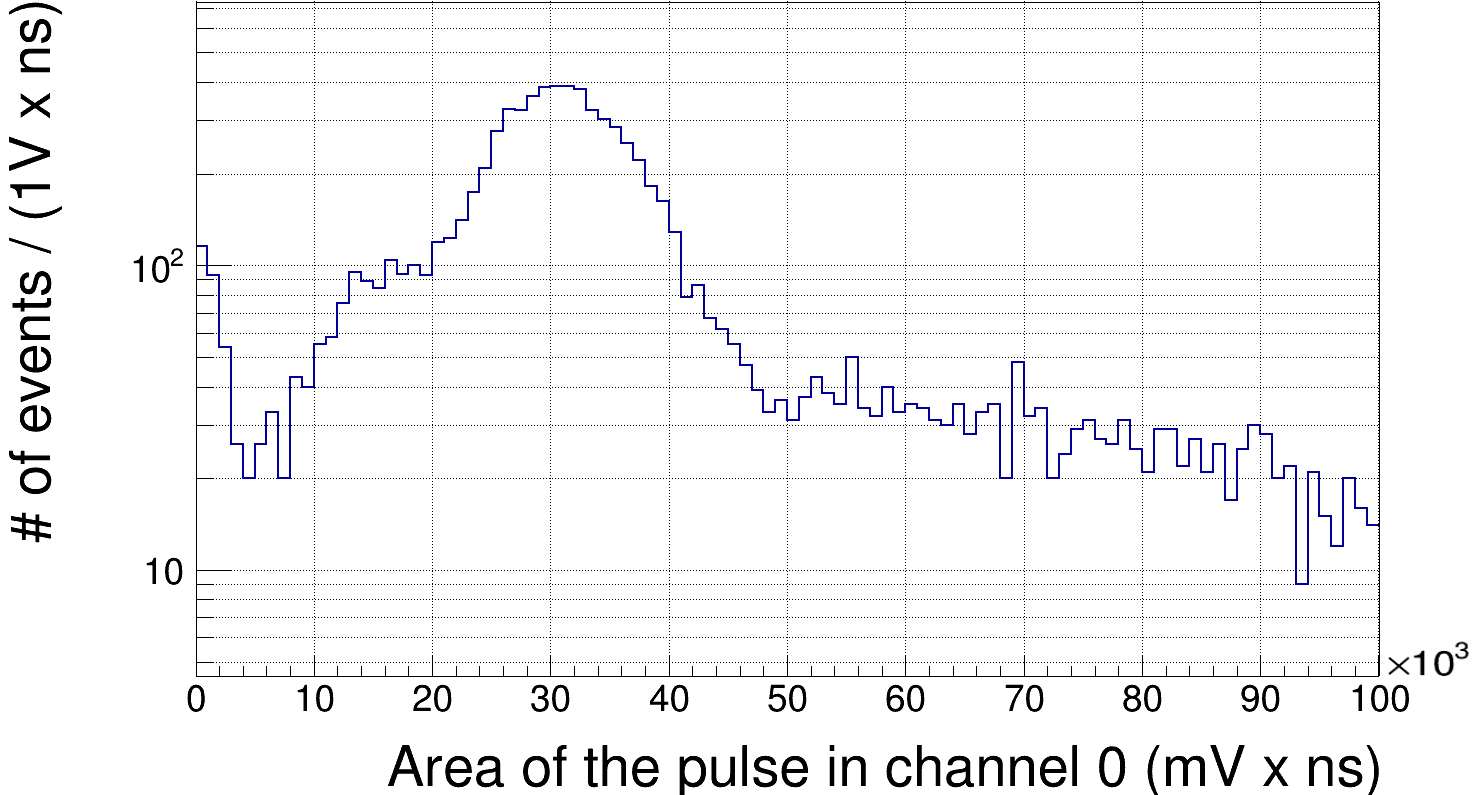}
		\end{subfigure}
		\begin{subfigure}[b]{0.49\textwidth}
			\includegraphics[width=\textwidth]{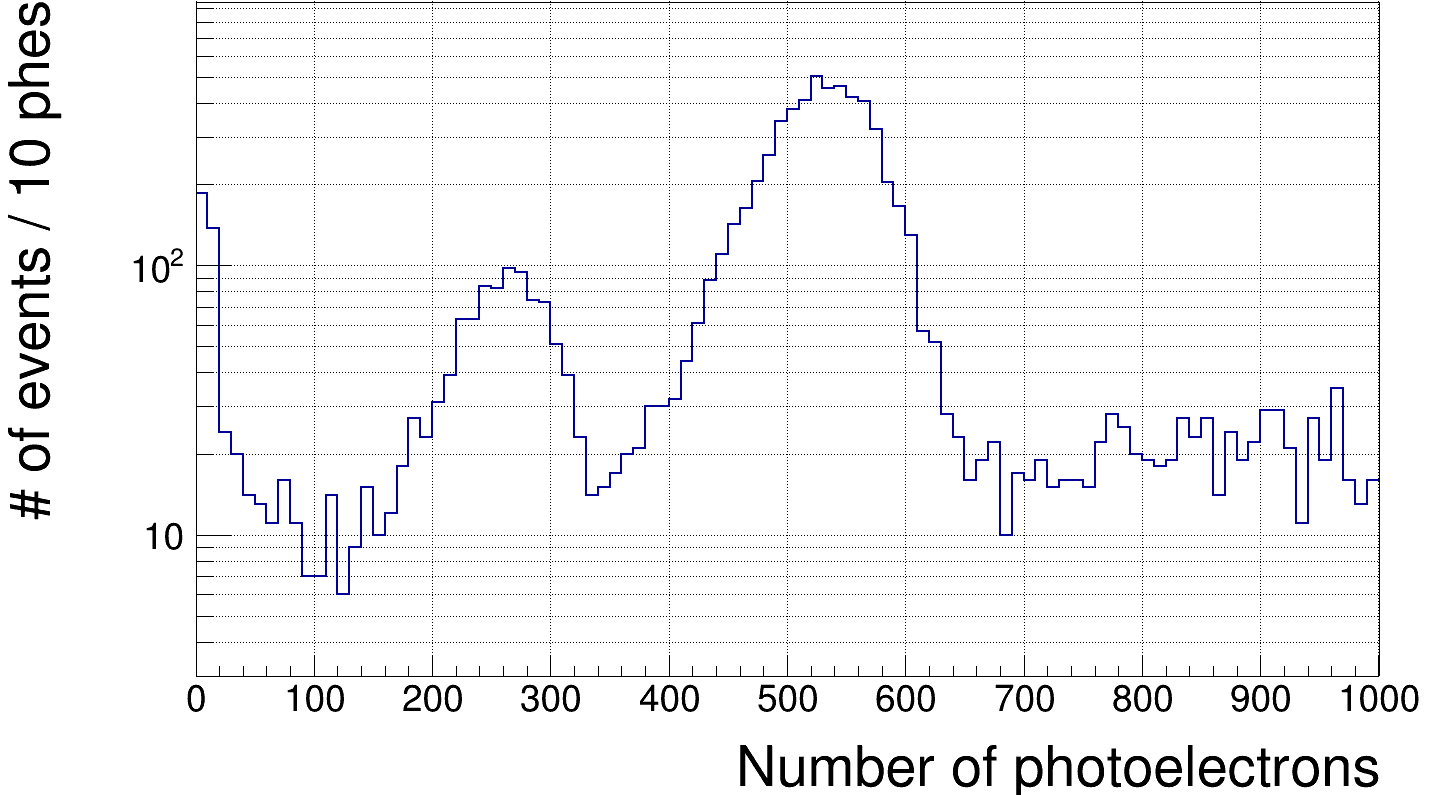}
		\end{subfigure}
		\caption{\label{AreaToLC}Measurement with the NaI crystal exposed to the $^{241}Am$ source at LAr temperature and at an overvoltage of 5~V. Left plot: Pulse area spectrum for a single channel (\#~0). Right plot: distribution of the total number of photoelectrons collected by all the channels. The resolution of the 59.5~keV peak (obtained as the ratio of its standard deviation to its mean) is 22\% for one channel and 9\% for the sum of eight channels.}
	\end{center}
\end{figure}

Then, when measuring scintillation signals, the pulse area for every channel is divided by the corresponding mean SPE area to convert it into number of photoelectrons collected by this channel. The number of photoelectrons corresponding to an event will be the sum of the photoelectrons measured in all the channels. As an example, Figure~\ref{AreaToLC} shows the spectra of the pulse area of a single channel and the distribution of the number of photoelectrons collected in all the channels for the measurement with the NaI crystal at LAr temperature and at an overvoltage of 5~V exposed to the $^{241}Am$ source. It is possible to observe the great improvement of the resolution obtained by adding all the channel signals. In the example shown, the resolution of the 59.5~keV peak (obtained as the ratio of its standard deviation to its mean) is 22\% for one channel and 9\% for the sum of eight channels.

\begin{figure}[h!]
	\begin{center}
		\includegraphics[width=0.75\textwidth]{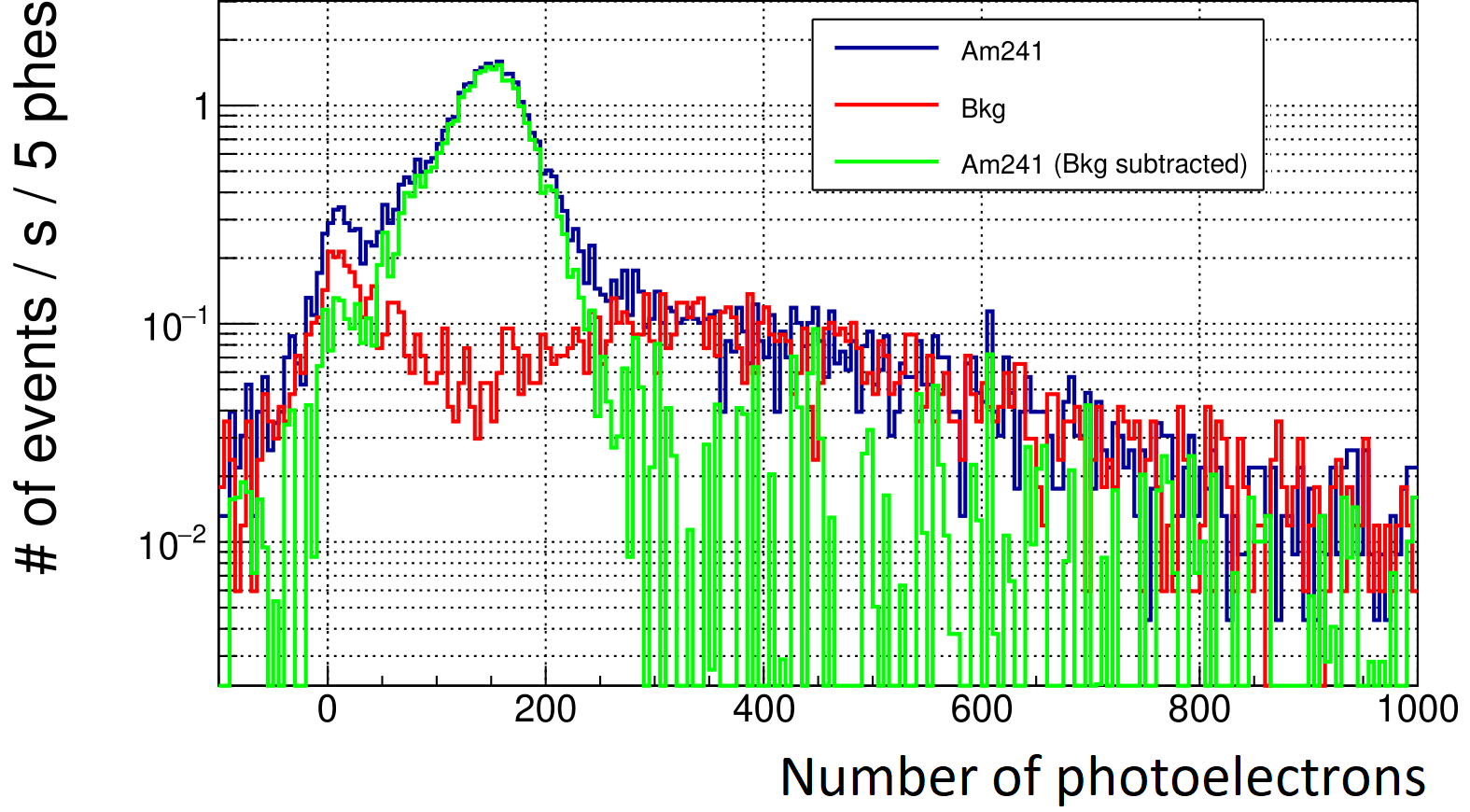}
		\caption{\label{CalBkgSubtraction}Comparison of the spectra for the calibration with $^{241}Am$ (blue line) and background (red line), and the result of subtracting the background from the calibration (green line) for measurements with NaI(Tl) at LAr temperature and at an overvoltage of 5~V.}
	\end{center}
\end{figure}

These calibration measurements have a non-negligible contribution from background events. Figure~\ref{CalBkgSubtraction} shows the comparison of the $^{241}Am$ calibration spectra and the background in number of photoelectrons for the NaI(Tl) crystal at LAr temperature and at an overvoltage of 5~V normalized to number of events per second. It can be observed a very good agreement between both measurements above the 59.5~keV peak from $^{241}Am$, pointing at the background origin from these energy depositions. If there is a background measurement available, it will be subtracted from the calibration spectrum (as shown also in Figure~\ref{CalBkgSubtraction}). 

The background subtracted $^{241}Am$ calibration spectra is then fitted to a function composed by a constant term and two Gaussians that account for the 59.5~keV and the 31.5~keV peaks (energies obtained from simulation, see Section~\ref{Section:SiPMSTAR2_GEANT4}), each of them with mean values ($\mu_{nphe}$) dependent on the energy of each peak, $E$, following the relation:
\begin{equation}
	\mu_{nphe} = LC \cdot E.
\end{equation}
The LC is assumed to be constant and the standard deviation of the peaks is considered to depend on the energy only by the poissonian distribution of the number of photoelectrons collected as:
\begin{equation} \label{eq:ResFitLC}
	\sigma = A \cdot \sqrt{E},
\end{equation}
The free parameters of the fit were $LC$, $A$ and the amplitudes of the two gaussians. The same procedure was followed if there were no background measurements.

\subsection{Scintillation time constants study} \label{Section:SiPMSTAR2_Analysis_ScintTimes}

The procedure followed to obtain the scintillation times of the crystals for each measurement is based on the modelling of both the scintillation signal ($s(t)$) and the response function of the SiPMs ($h(t)$) which have to be convolved and fitted to the data. The way each function was modeled is described next.

\begin{figure}[h!]
	\begin{center}
		\includegraphics[width=0.75\textwidth]{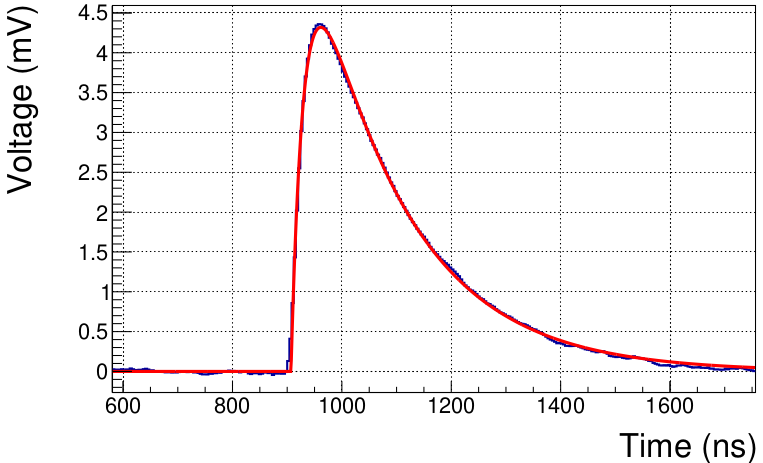}
		\caption{\label{SPEAveragedFit}Fit of the SPE to Equation~\ref{eq:SPE_Fit} for the measurement with the NaI(Tl) crystal at a temperature of 220~K and at an overvoltage of 3~V.}
	\end{center}
\end{figure}

\begin{figure}[h!]
	\begin{center}
		\includegraphics[width=0.75\textwidth]{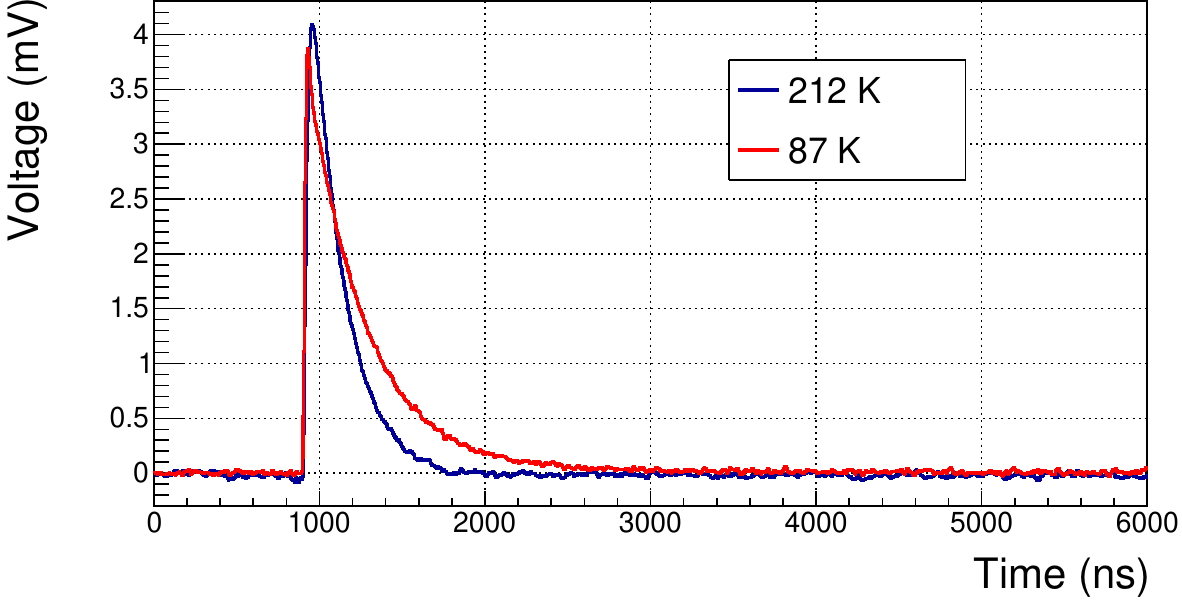}
		\caption{\label{STAR_SPEComparison}SPE for measurements with the NaI(Tl) crystal at a temperature of 212~K (blue line) and at 87~K (red line), both at an overvoltage of 3~V.}
	\end{center}
\end{figure}

The response function of the SiPMs is provided by the waveform of the SPE. To obtain it, the events of the first peak of the SPE area spectrum were selected using the SPE calibration measurements. The waveforms corresponding to those pulse area values within one standard deviation from the mean corresponding to the SPE peak were averaged. An example of this spectrum was shown in Figure~\ref{SPECalSTAR}, and an example of the averaged pulse for the SPE waveform is shown in Figure~\ref{AveragedPulsesStar}. The waveform of a SPE can be fitted to the function
\begin{equation}\label{eq:SPE_Fit}
	h(t) = \left\{ \begin{array}{lcc}
		0 & for & t \leq t_{PS} \\
		\\ h(t) = A_S\left(\exp{\left(\frac{t_{PS}-t}{\tau_{S1}}\right)}-\exp{\left(\frac{t_{PS}-t}{\tau_{S2}}\right)}\right) & for & t > t_{PS} \\
	\end{array}
	\right.,
\end{equation}
where $\tau_{S1}$ and $\tau_{S2}$ are the decay time and rise time of the pulse, respectively, $A_S$ is related with the pulse amplitude and $t_{PS}$ accounts for the initial time of the pulse in the waveform. These four parameters are left free in the fit. An example of this fit for the SPE waveform of the channel~0 at 220~K is shown in Figure~\ref{SPEAveragedFit}. Small differences in the SPE pulse shape at different temperatures were observed, as shown in Figure~\ref{STAR_SPEComparison}, between temperatures of 212~K and 87~K. Moreover, the same analysis was done for different channels in the same runs, obtaining compatible values with each other. As an example, Table~\ref{tabla:SPEFitComparison} shows the fit results for the eight channels of the same measurement (in this case, measurement at LAr temperature and an overvoltage of 5~V).

\begin{table}[h]
	\centering
	\begin{tabular}{|c|c|c|c|}
		\hline
		Channel & $A_S$ (mV) & $\tau_{S1}$ (ns) & $\tau_{S2}$ (ns) \\
		\hline
		0 & 5.54$\pm$0.22 & 344$\pm$29 & 9$\pm$6 \\
		1 & 5.24$\pm$0.26 & 354$\pm$27 & 9$\pm$5 \\
		2 & 5.36$\pm$0.46 & 407$\pm$21 & 12$\pm$6 \\
		3 & 6.04$\pm$0.24 & 350$\pm$30 & 8$\pm$4 \\
		4 & 5.31$\pm$0.31 & 367$\pm$24 & 18$\pm$10 \\
		5 & 5.53$\pm$0.47 & 409$\pm$17 & 15$\pm$8 \\
		6 & 6.07$\pm$0.37 & 384$\pm$24 & 17$\pm$8 \\
		7 & 5.74$\pm$0.32 & 370$\pm$22 & 8$\pm$6 \\
		\hline
		Mean & 5.61$\pm$0.31 & 377$\pm$24 & 11$\pm$6 \\
		\hline  
	\end{tabular} \\
	\caption{Results of the SPE pulse fits of the eight channels for the measurement at LAr temperature and an overvoltage of 5~V, and their weighted means.}
	\label{tabla:SPEFitComparison}
\end{table}

The scintillation signal $s(t)$ is defined as the sum of a number $n_{Exp}$ of exponential decay functions to take into account the possible presence of different scintillation components:
\begin{equation}
	s(t) = \left\{ \begin{array}{lcc}
		0 & for & t \leq t_P \\
		\\ \sum_{i=1}^{n_{Exp}} A_i \exp{\left(\frac{t_P-t}{\tau_i}\right)} & for & t > t_P \\
	\end{array}
	\right.,
\end{equation}
where $A_i$ and $\tau_i$ are the amplitude and decay time of the $i^{th}$ component and $t_P$ accounts for the initial time of the pulse in the waveform. The number of exponential functions used in the modeling depends on the measurements, as it is explained later. The scintillation function is numerically convolved with the response function of the SiPMs to obtain a PDF, $y(t)$.

\begin{figure}[h!]
	\begin{center}
		\begin{subfigure}[b]{0.49\textwidth}
			\includegraphics[width=\textwidth]{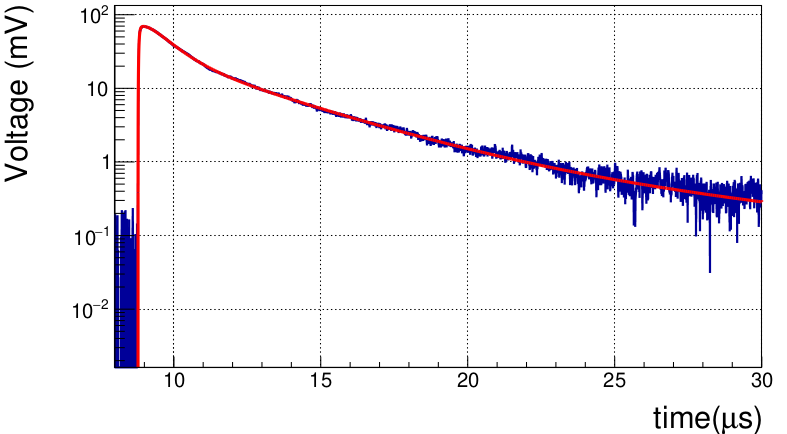}
		\end{subfigure}
		\begin{subfigure}[b]{0.49\textwidth}
			\includegraphics[width=\textwidth]{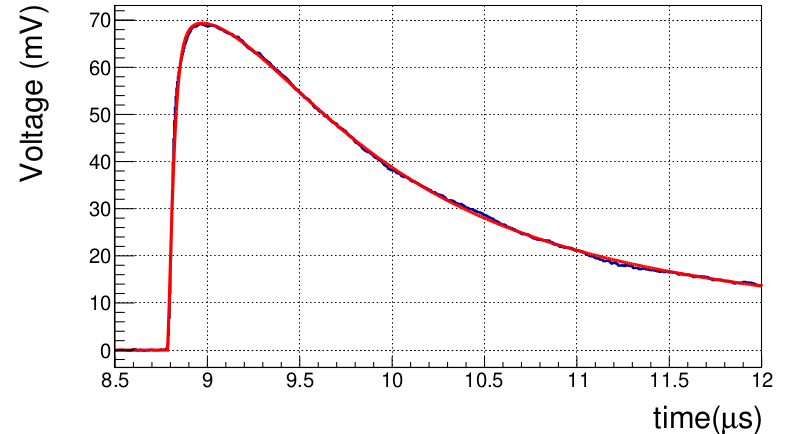}
		\end{subfigure}
		\caption{\label{ScintAveragedFit}Fit of the scintillation pulse for $^{241}Am$ events for the measurement with the NaI(Tl) crystal at around 100~K and at an overvoltage of 3~V. This fit considers four scintillation components. Left: Pulse and fit shown in the full digitization window in logarithmic scale. Right: zoomed view of the initial part of the pulse and fit, in linear scale, showing the goodness of the fit.}
	\end{center}
\end{figure}

The data to fit is the averaged pulse of scintillation, constructed using the events from the 59.5~keV peak of the measurements with the $^{241}Am$ source. The selected events have pulse areas from $\mu-\sigma$ to $\mu+\sigma$, where $\mu$ and $\sigma$ are the mean and standard deviation of this peak. Moreover, an additional selection in the onset time \textit{t0} of the pulse is applied: those events having $t_0$ within the range from $\langle$\textit{t0}$\rangle$-25~ns to $\langle$\textit{t0}$\rangle$+25~ns (being $\langle$\textit{t0}$\rangle$ the mean value) are used for averaging. Finally, this averaged pulse is fitted to $y(t)$, leaving free all the parameters included in the scintillation function ($A_i$, $\tau_i$ and $t_P$). As an example of this fit, Figure~\ref{ScintAveragedFit} shows that corresponding to the measurement with the NaI(Tl) crystal at around 100~K and at an overvoltage of 3~V.

The contribution of the $i^{th}$ scintillation component to the LC is calculated (to take into account the change in the scintillation times with temperature) as the average number of photons for the 59.5~keV peak. It is obtained as the ratio of the area of that component to that of the SPE:
\begin{equation}
	I_i = \frac{A_i \cdot \tau_i}{A_S \cdot \left(\tau_{S1}-\tau_{S2}\right)}.
\end{equation}
In measurements at LAr temperature, the relative contribution from each component to the LC is calculated as
\begin{equation}
	R_i = \frac{I_i}{\sum_{j=1}^{n_{Exp}} I_j}.
\end{equation}

\begin{figure}[h!]
	\begin{center}
		\includegraphics[width=0.75\textwidth]{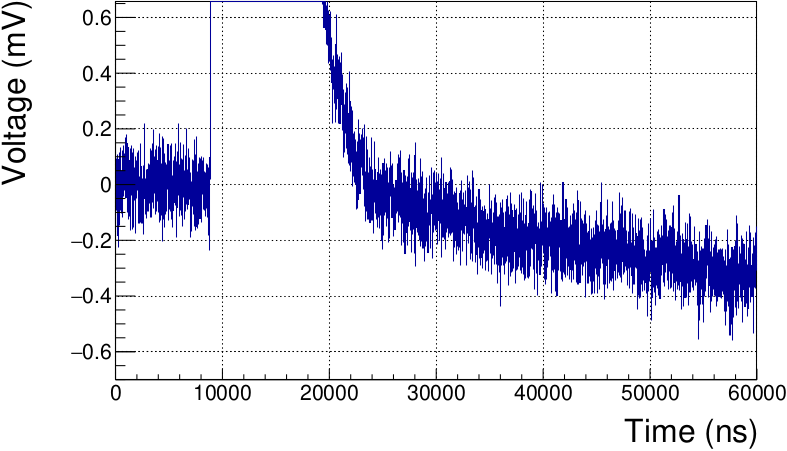}
		\caption{\label{PulseUndershoot}Undershoot observed in the averaged pulse of the measurements with the NaI crystal at LAr temperature and an overvoltage of 5~V.}
	\end{center}
\end{figure}

The procedure followed to select the number of scintillation components was based on the comparison between the fits using two, three or four components. More than four were not required to achieve successful fits for all the considered crystals and operation conditions. It is observed that in those cases where unnecessary scintillation components are included in the fit, some of the $A_i$ parameters converge to values compatible with zero. Therefore, the results presented for each pulse are those obtained in the fit with maximum number of components which have all of their amplitudes incompatible with zero at 1$\sigma$. The fit range used in this fitting procedure is from 8~$\mu$s (being close to 9~$\mu$s, the pretrigger time for the 60~$\mu$s digitization window, see Section~\ref{Section:SiPMSTAR2_DAQ}) to the time when the averaged pulse decreases below zero. The reason for this is that pulses with high amplitude show an undershoot, which is probably related with some saturation effect of the signal while processing. This undershoot cannot be reproduced by our signal modelling, requiring probably to include the front-end response. An example of this effect is shown in Figure~\ref{PulseUndershoot}. This effect changes the shape of the pulse, and it implies the most important limitation in the calculation of the scintillation times at present.

\section{Crystals characterization and results} \label{Section:SiPMSTAR2_Charac}
\fancyhead[RO]{\emph{\thesection. \nameref{Section:SiPMSTAR2_Charac}}}

\subsection{Characterization at LAr temperature} \label{Section:SiPMSTAR2_Charac_LAr}

\begin{figure}[h!]
	\begin{center}
		\includegraphics[width=\textwidth]{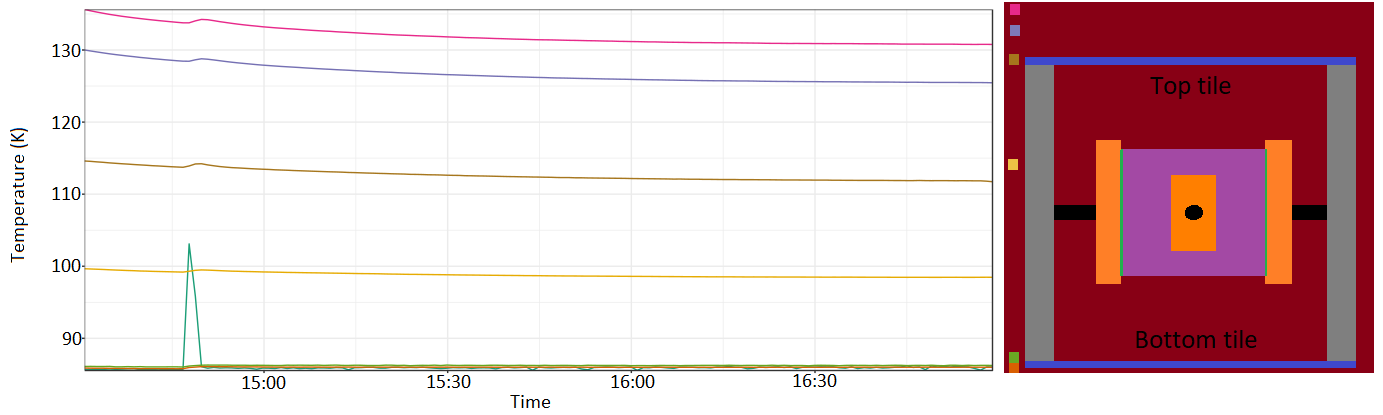}
		\caption{\label{TempNaI}Temperatures measured with the PT100 resistors inside the internal container during measurements at LAr temperature with the NaI crystal (left plot) and the PT100 resistors positions in the container (right plot).}
	\end{center}
\end{figure}

\begin{figure}[h!]
	\begin{center}
		\includegraphics[width=\textwidth]{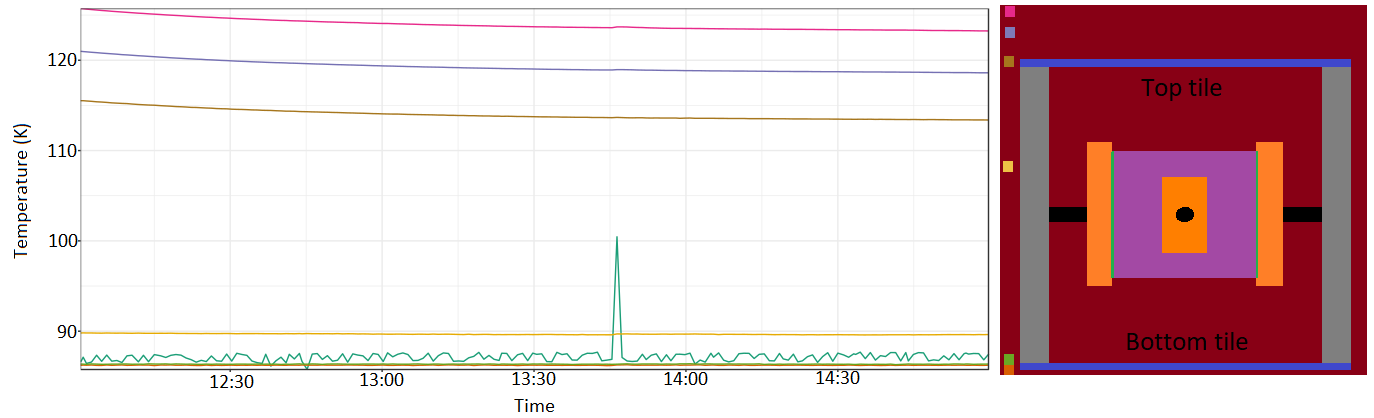}
		\caption{\label{TempNaITl}Temperatures measured with the PT100 resistors inside the internal container during measurements at LAr temperature with the NaI(Tl) crystal (left plot) and the PT100 resistors positions in the container (right plot).}
	\end{center}
\end{figure}

All the measurements carried out with each crystal at LAr temperature were done in the three hours period after removing the LAr from the aluminum chamber, but leaving 3~cm of LAr below the bottom tile, in contact with the copper plate (see Section~\ref{Section:SiPMSTAR2_Detector}). The temperature of the gas around the chamber was measured with the PT100 resistors, guaranteeing the stability of the system temperature while data taking. The temperatures measured by the PT100 were recorded, and they are shown in Figures~\ref{TempNaI} and~\ref{TempNaITl} for the measurements with NaI and NaI(Tl), respectively. Although the temperature of the crystal could not be directly measured, it was supposed to remain constant. As shown in Figures~\ref{TempNaI} and~\ref{TempNaITl}, the temperature of the gas in the container increases from bottom to top from 87~K up to about 130~K. However it remains quite stable while measuring. Being the crystal in thermal contact with the LAr in the bottom of the container, it is expected it reaches an equilibrium temperature close to 87~K, but in any case, not above 130~K. 

\begin{figure}[h!]
	\begin{center}
		\includegraphics[width=\textwidth]{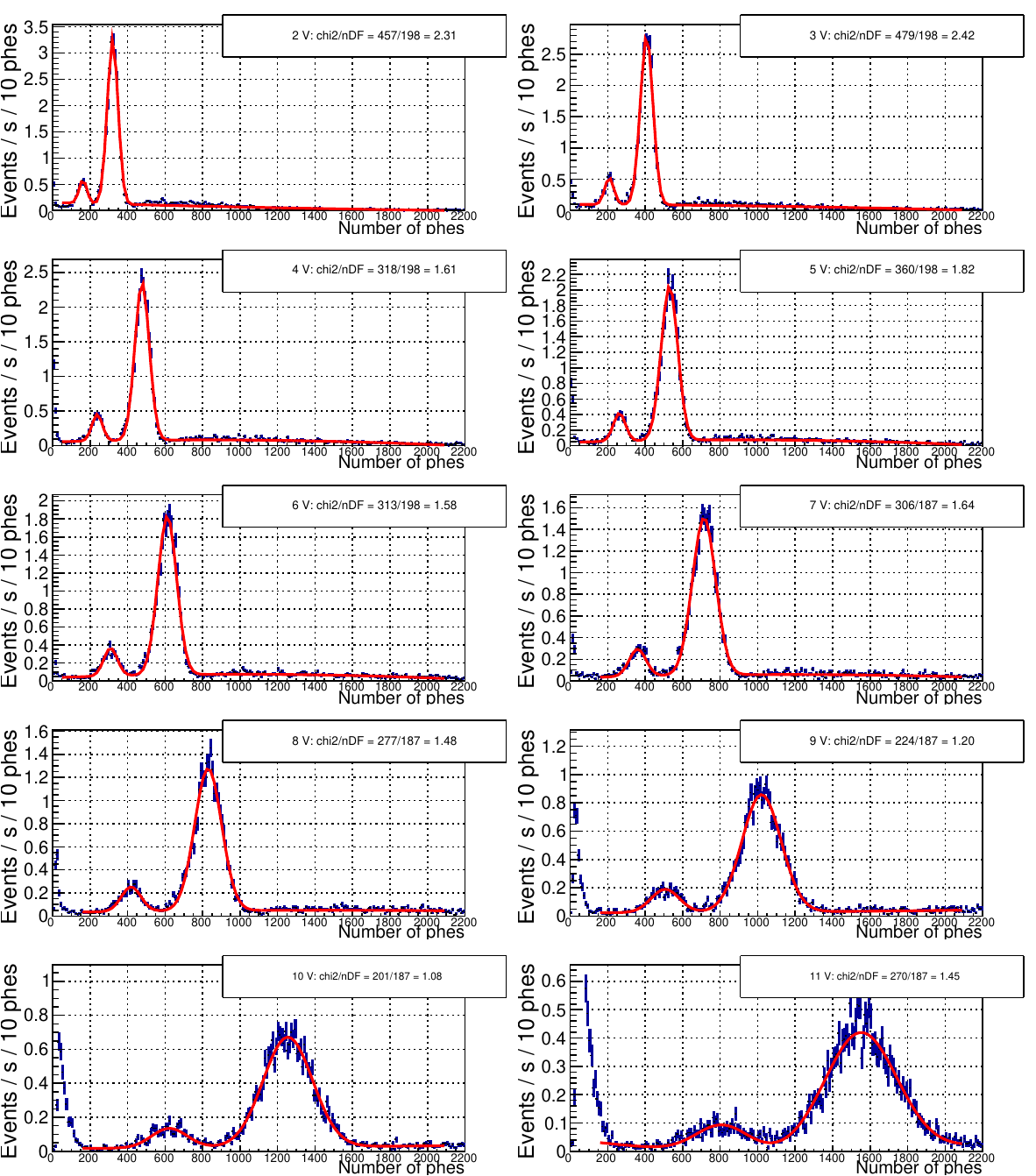}
		\caption{\label{Run2_Fits_Am241}Fits of the background subtracted $^{241}Am$ calibration spectra for the NaI crystal at LAr temperature at overvoltages in the range from 2~to 11~V.}
	\end{center}
\end{figure}

\begin{figure}[h!]
	\begin{center}
		\includegraphics[width=\textwidth]{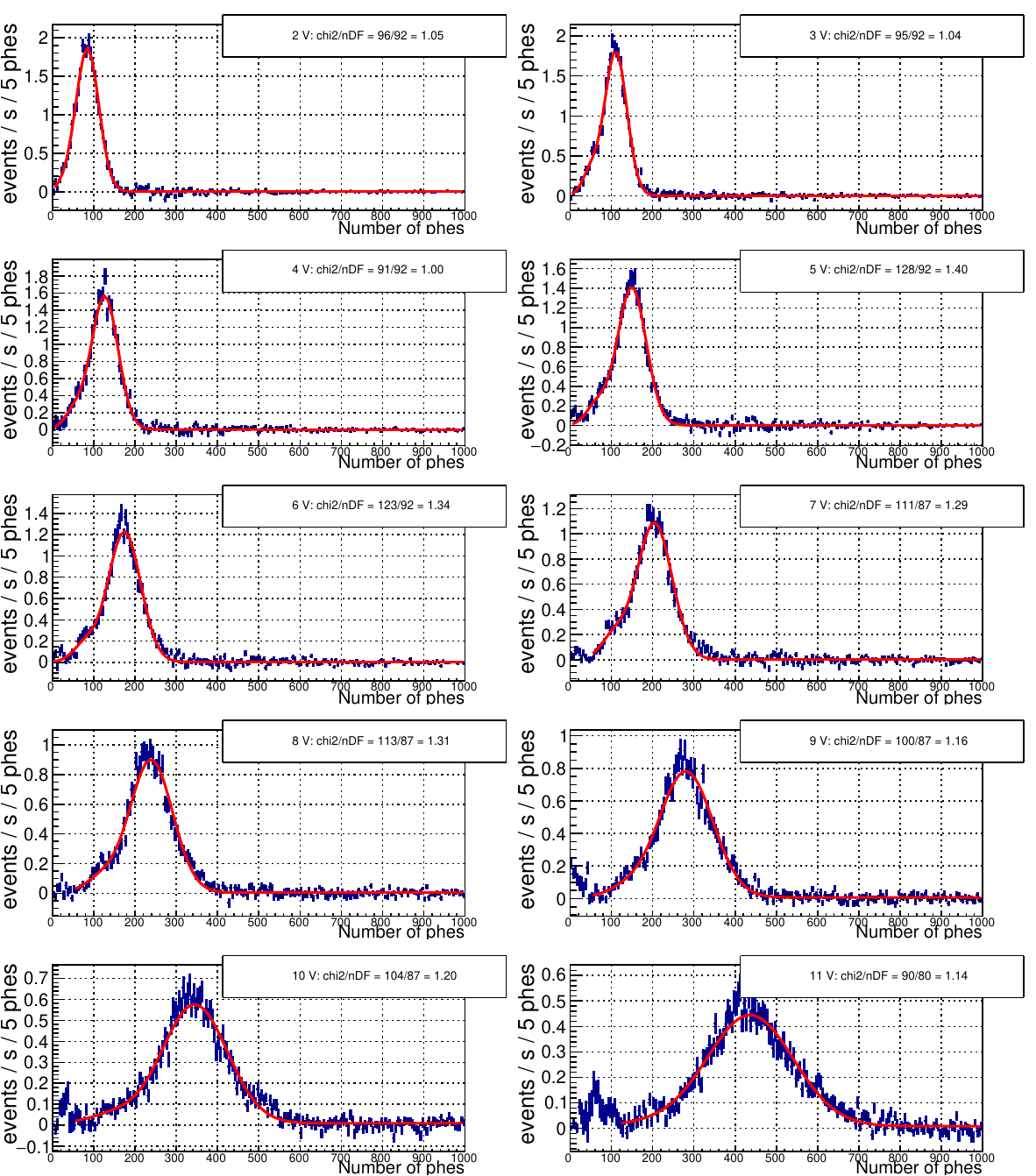}
		\caption{\label{Run3_Fits_Am241}Fits of the background subtracted $^{241}Am$ calibration spectra for the NaI(Tl) crystal at LAr temperature at overvoltages in the range from 2~to 11~V.}
	\end{center}
\end{figure}

\begin{figure}[h!]
	\begin{center}
		\begin{subfigure}[b]{0.49\textwidth}
			\includegraphics[width=\textwidth]{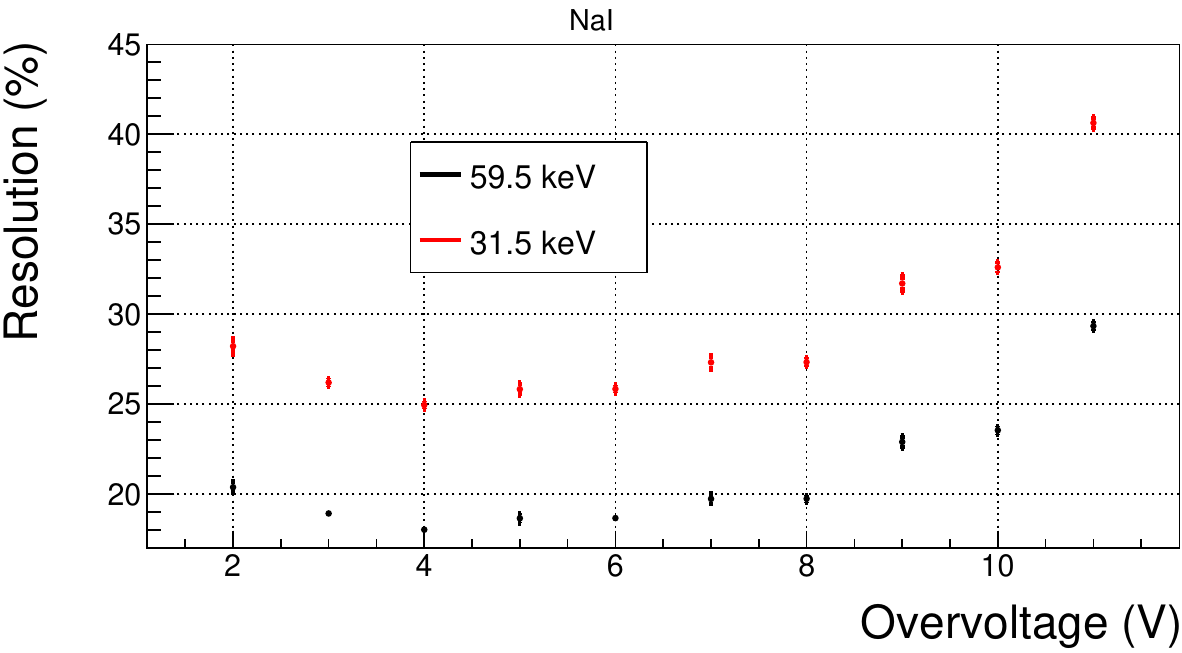}
		\end{subfigure}
		\begin{subfigure}[b]{0.49\textwidth}
			\includegraphics[width=\textwidth]{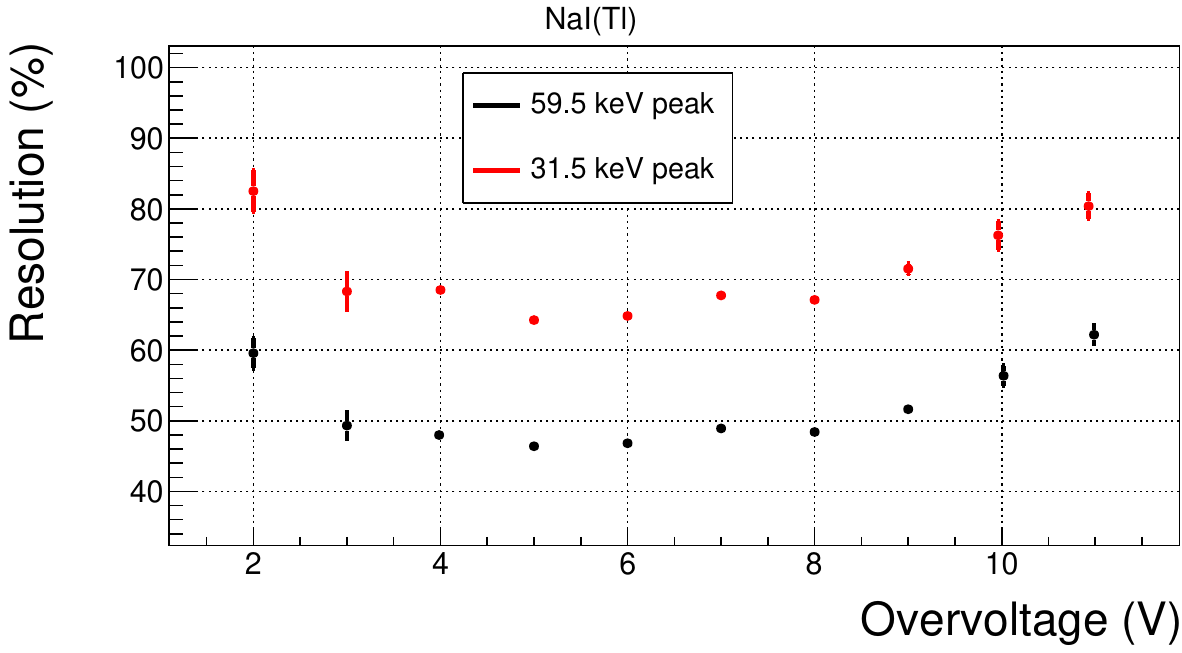}
		\end{subfigure}
		\caption{\label{ResolutionSTAR}Resolution of each peak as the FWHM from the fitted gaussian divided by the nominal energy, in percentage, for the measurements with NaI (left plot) and NaI(Tl) crystals (right plot).}
	\end{center}
\end{figure}

The program of measurements for each crystal consisted of 10 batches, changing the overvoltage applied to the SiPM arrays from 2~to 11~V in steps of 1~V. After the background subtraction of the $^{241}Am$ spectra, the two peaks were observed in the measurements with NaI, but it was not so clear in those of NaI(Tl), whose LC at this temperature was really low. The fits (whose procedure was explained in Section~\ref{Section:SiPMSTAR2_Analysis_LC}) are shown in Figures~\ref{Run2_Fits_Am241} and~\ref{Run3_Fits_Am241} for NaI and NaI(Tl) measurements, respectively, for the different overvoltage values. The fit results are also summarized in Table~\ref{tabla:NaIFits}. The values for the energy resolution of each peak were obtained as the FWHM from the gaussian peak divided by the energy, and they are shown in Figure~\ref{ResolutionSTAR} expressed as a percentage. This plot shows that the overvoltage that optimizes the resolution is around 4 or 5~V, depending on the crystal.

\begin{table}[h]
	\centering
	\begin{tabular}{|c|c|c|c|c|}
		\cline{2-5}
		\multicolumn{1}{c|}{} & \multicolumn{2}{|c|}{LC (phe/keV)} & \multicolumn{2}{|c|}{A (phe/$\sqrt{keV}$)} \\
		\hline
		Overvoltage (V) & NaI & NaI(Tl) & NaI & NaI(Tl) \\
		\hline
		2 & 5.56$\pm$0.05 & 1.84$\pm$0.65 & 0.67$\pm$0.01 & 1.96$\pm$0.07\\
		3 & 6.98$\pm$0.01 & 2.15$\pm$0.81 & 0.62$\pm$0.01 & 1.62$\pm$0.06 \\
		4 & 8.40$\pm$0.01 & 2.59$\pm$0.20 & 0.59$\pm$0.01 & 1.62$\pm$0.02 \\
		5 & 9.36$\pm$0.06 & 2.97$\pm$0.21 & 0.61$\pm$0.01 & 1.52$\pm$0.01 \\
		6 & 10.61$\pm$0.01 & 3.34$\pm$0.02 & 0.61$\pm$0.01 & 1.53$\pm$0.01 \\
		7 & 12.43$\pm$0.09 & 3.44$\pm$0.14 & 0.64$\pm$0.01 & 1.54$\pm$0.01 \\
		8 & 14.41$\pm$0.02 & 4.26$\pm$0.23 & 0.65$\pm$0.01 & 1.61$\pm$0.01 \\
		9 & 18.02$\pm$0.14 & 4.97$\pm$0.47 & 0.75$\pm$0.01 & 1.70$\pm$0.02 \\
		10 & 22.17$\pm$0.03 & 6.91$\pm$2.20 & 0.77$\pm$0.01 & 1.89$\pm$0.02 \\
		11 & 25.95$\pm$0.05 & 8.66$\pm$2.34 & 0.96$\pm$0.01 & 2.03$\pm$0.03 \\
		\hline
	\end{tabular} \\
	\caption{Results of the fit parameters $LC$ and $A$ for the measurements at LAr temperature of NaI and NaI(Tl) crystals, shown in Figures~\ref{Run2_Fits_Am241} and~\ref{Run3_Fits_Am241}.}
	\label{tabla:NaIFits}
\end{table}

To obtain the LC with origin in the scintillation in each measurement, the contribution of the CT has to be discounted from the LC shown in Table~\ref{tabla:NaIFits}. To do so, the behaviour of the SiPM was modeled as it was explained in Section~\ref{Section:SiPM_Model}. A parameterization of the LC, DC, CT and AP based on a model developed by DarkSide researchers~\cite{Boulay:2022rgb} allows to estimate the total scintillation LC (see Equation~\ref{eq:LC(OV)}). In the model used in this chapter, the AP and DC contributions to the LC have not been considered ($\xi_{AP} = 0$ and $DCR = 0$ in the Equation~\ref{eq:LC(OV)}), as they were found negligible in other applications of the same SiPM arrays~\cite{Boulay:2022rgb} using the same SiPM arrays. Moreover, the parameters $V_e$ and $V_h$ used in this model depend only on the SiPM arrays, and therefore they had been characterized previously using a similar chamber to that used in these measurements but designed to detect argon scintillation. The values determined for these parameters were 1~V and 5.4~V, respectively~\cite{Boulay:2022rgb}.

\begin{figure}[h!]
	\begin{center}
		\includegraphics[width=0.75\textwidth]{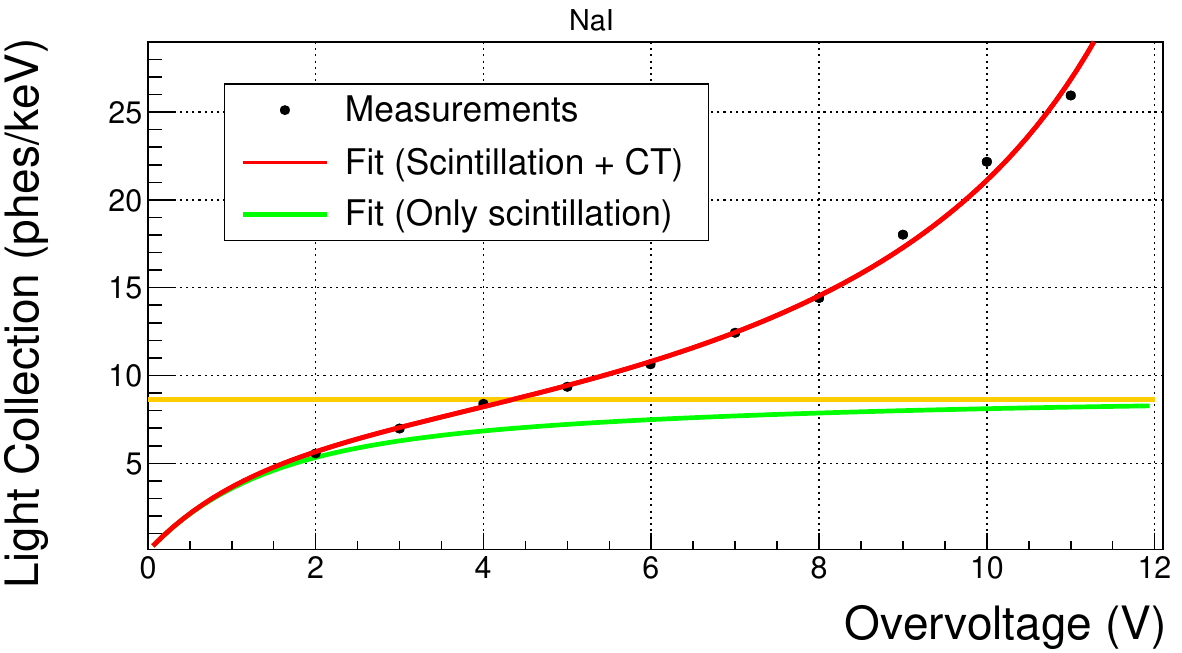}
		\caption{\label{Run2_LCVsOV}Fit of the LC values obtained at LAr temperature with the NaI crystal (black circles) to the Equation~\ref{eq:LC(OV)} (red line). Green line represents the LC from scintillation at each overvoltage ($LC^{scint}_{max}\cdot\epsilon/\epsilon_{max}$) and the orange line is the asymptotic LC of the detector, $LC^{scint}_{max}$.}
	\end{center}
\end{figure}

\begin{figure}[h!]
	\begin{center}
		\includegraphics[width=0.75\textwidth]{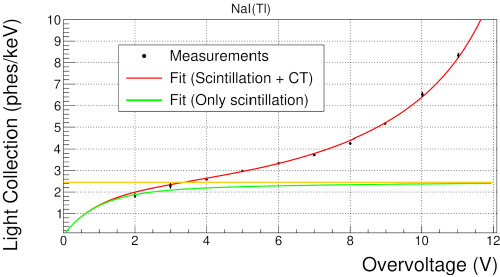}
		\caption{\label{Run3_LCVsOV}Fit of the LC values obtained at LAr temperature with the NaI(Tl) crystal (black circles) to the Equation~\ref{eq:LC(OV)} (red line). Green line represents the LC from scintillation at each overvoltage ($LC^{scint}_{max}\cdot\epsilon/\epsilon_{max}$) and the orange line is the asymptotic LC of the detector, $LC^{scint}_{max}$.}
	\end{center}
\end{figure}

However, $\xi$, $\xi_{CT}$ and $LC^{scint}_{max}$ depend on the properties of the scintillator crystal, the reflecting cavity, the optical coupling between the different components, etc., and they have to be obtained from our measurements. To do so, a fit of the previously obtained LC was done to the Equation~\ref{eq:LC(OV)}, and the contribution of the CT was removed to obtain the LC as a function of the overvoltage as $LC^{scint}_{max}\cdot\epsilon(V_{ov})/\epsilon_{max}$. The fits of the NaI and NaI(Tl) measurements are shown in Figures~\ref{Run2_LCVsOV} and~\ref{Run3_LCVsOV}, respectively, and the values obtained for $\xi$ and $\xi_{CT}$ resulting from those fits are shown in Table~\ref{tabla:etaResults}. They depend on the scintillation wavelength and on the characteristics of the detector. The different values obtained for the two crystals while all of the other setup components are the same, point at different wavelengths of the corresponding scintillation emissions (see Section~\ref{Section:SiPM_Model} for more details). The values obtained in~\cite{Boulay:2022rgb} for a cubic detector similar to that used in these measurements but using LAr as scintillator were 37\% for $\xi$ and 78~kV$^{-1}$ for $\xi_{CT}$, which are similar to the values obtained in these fits. After this correction, the LC for the scintillation of the NaI crystal at 87~K is more than three times larger than that for the NaI(Tl) crystal.

\begin{table}[h!]
	\centering
	\begin{tabular}{|c|c|c|c|}
		\hline
		Crystal & $\xi$ (\%) & $\xi_{CT}$ (kV$^{-1}$) & $LC^{scint}_{max}$ (phe/keV) \\
		\hline
		NaI & 60.35~$\pm$~1.51 & 68.3~$\pm$~0.1 & 8.72~$\pm$~0.02 \\
		NaI(Tl) & 80.24~$\pm$~3.79 & 69.6~$\pm$~0.3 & 2.41~$\pm$~0.12 \\
		\hline
	\end{tabular}
	\caption{Values of $\xi$, $\xi_{CT}$ and $LC^{scint}_{max}$ for NaI and NaI(Tl) measurements obtained from the fits shown in Figures~\ref{Run2_LCVsOV} and~\ref{Run3_LCVsOV}, respectively. Note that they depend on the wavelength of the scintillation photons, and being the same all the other detector components, these results can be interpreted as a hint of different wavelength emission in both crystals.}
	\label{tabla:etaResults}
\end{table}

\begin{figure}[h!]
	\begin{center}
		\includegraphics[width=0.75\textwidth]{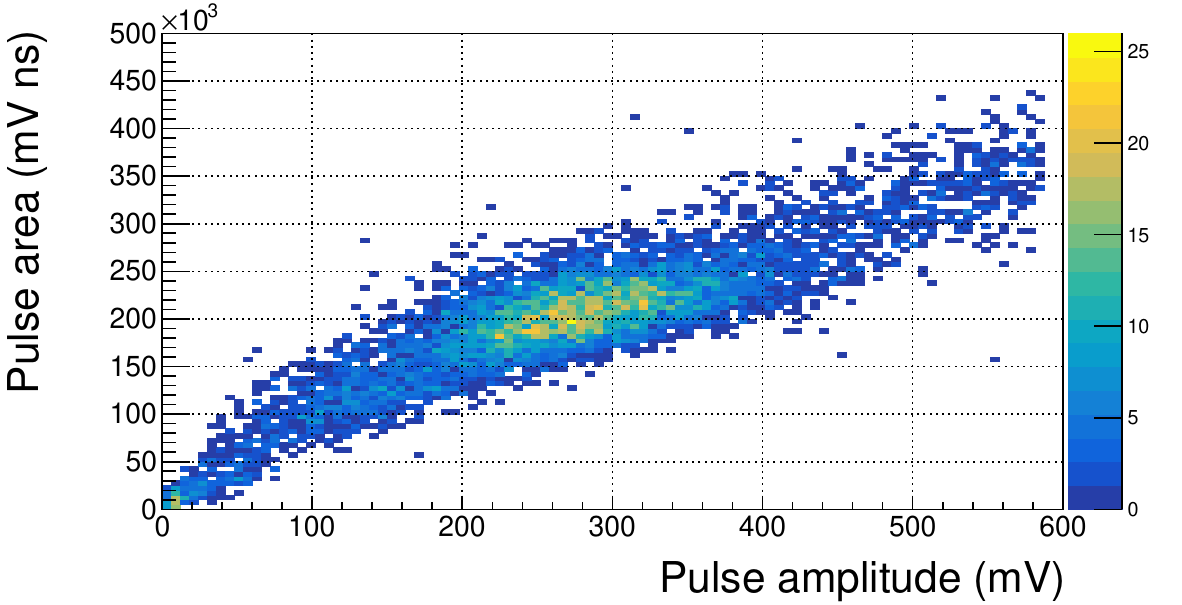}
		\caption{\label{PulseMinVsAmp}Scatter plot of the pulse area as a function of the pulse amplitude for the measurements with the NaI crystal at 5~V at LAr temperature.}
	\end{center}
\end{figure}

Scintillation time constants were analyzed by fitting the pulses as explained in Section~\ref{Section:SiPMSTAR2_Analysis_ScintTimes}. Two components were identified in the NaI crystal, while four components were needed for the NaI(Tl) crystal. Before fitting the pulses, it is interesting to analyze the effect of saturation/undershooting that we already commented in Section~\ref{Section:SiPMSTAR2_Analysis_ScintTimes}. The pulse area was plotted as a function of the pulse amplitude, as it is shown in Figure~\ref{PulseMinVsAmp} for the measurements with the NaI crystal at 5~V. It is clear that there is a pulse-shape dependence on the pulse amplitude and then, the obtained values of the scintillation components are expected to depend on the pulse amplitude, and hence, on the overvoltage.

The relative intensities and scintillation times of the four components obtained in each measurement with the NaI(Tl) crystal at LAr temperature are presented in Tables~\ref{tabla:ResultsIntScintNaI(Tl)} and~\ref{tabla:ResultsTimeScintNaI(Tl)}, respectively, while for the NaI crystal are shown in Table~\ref{tabla:ResultsScintNaI}. In those tables, mean values of the scintillation times and emission relative intensities are also calculated for each component. The uncertainties presented in those tables are standard deviations of the ten measurements, without considering systematical contributions, that could be much larger.

\begin{table}[h!]
	\centering
	\begin{tabular}{|c|c|c|c|c|}
		\hline
		OV (V) & $I_1$ (\%) & $I_2$ (\%) & $I_3$ (\%) & $I_4$ (\%) \\
		\hline
		2 & 5.5$\pm$1.1 & 15.4$\pm$1.8 & 56.0$\pm$2.3 & 23.1$\pm$0.8 \\
		3 & 6.8$\pm$1.8 & 16.9$\pm$1.5 & 53.4$\pm$2.1 & 22.9$\pm$0.7 \\
		4 & 7.0$\pm$1.7 & 18.9$\pm$1.3 & 45.6$\pm$1.7 & 28.5$\pm$1.2 \\
		5 & 6.8$\pm$1.4 & 18.6$\pm$1.1 & 47.4$\pm$1.5 & 27.2$\pm$0.9 \\
		6 & 6.5$\pm$1.2 & 18.2$\pm$0.9 & 48.5$\pm$1.2 & 26.7$\pm$0.7 \\
		7 & 6.2$\pm$0.3 & 18.1$\pm$0.9 & 49.6$\pm$1.1 & 26.1$\pm$0.4 \\
		8 & 5.3$\pm$0.4 & 18.1$\pm$0.7 & 50.5$\pm$0.8 & 26.1$\pm$0.4 \\
		9 & 5.9$\pm$0.3 & 18.1$\pm$0.7 & 51.4$\pm$0.8 & 24.7$\pm$0.3 \\
		10 & 6.0$\pm$0.6 & 21.6$\pm$0.6 & 50.7$\pm$0.8 & 21.7$\pm$0.3 \\
		11 & 7.7$\pm$4.3 & 20.9$\pm$1.0 & 52.9$\pm$2.3 & 18.5$\pm$0.8 \\
		\hline
		Mean & 6.0$\pm$0.7 & 18.7$\pm$1.2 & 50.3$\pm$1.2 & 24.4$\pm$0.5 \\
		\hline
	\end{tabular}
	\caption{Relative intensities of the four components identified in each measurement with the NaI(Tl) crystal. Mean and standard deviations are also shown.}
	\label{tabla:ResultsIntScintNaI(Tl)}
\end{table}

\begin{table}[h!]
	\centering
	\begin{tabular}{|c|c|c|c|c|}
		\hline
		OV (V) & $\tau_1$ (ns) & $\tau_2$ (ns) & $\tau_3$ (ns) & $\tau_4$ (ns) \\
		\hline
		2 & 1$\pm$1 & 539$\pm$68 & 2339$\pm$74 & 24469$\pm$1120 \\
		3 & 8$\pm$2 & 593$\pm$59 & 2510$\pm$68 & 34497$\pm$1908 \\
		4 & 16$\pm$2 & 659$\pm$47 & 2593$\pm$63 & 58107$\pm$4104 \\
		5 & 15$\pm$2 & 653$\pm$41 & 2552$\pm$52 & 51399$\pm$2728 \\
		6 & 13$\pm$2 & 622$\pm$32 & 2550$\pm$42 & 48396$\pm$2095 \\
		7 & 11$\pm$1 & 639$\pm$32 & 2512$\pm$41 & 41553$\pm$1369 \\
		8 & 7$\pm$1 & 590$\pm$24 & 2531$\pm$32 & 43726$\pm$1361 \\
		9 & 9$\pm$1 & 591$\pm$23 & 2492$\pm$30 & 39359$\pm$990 \\
		10 & 6$\pm$1 & 657$\pm$18 & 2539$\pm$27 & 35974$\pm$804 \\
		11 & 2$\pm$1 & 605$\pm$13 & 2409$\pm$19 & 29226$\pm$464 \\
		\hline
		Mean & 8$\pm$5 & 617$\pm$36 & 2497$\pm$71 & 36522$\pm$9717 \\
		\hline
	\end{tabular}
	\caption{Scintillation times of the four components identified in each measurement with the NaI(Tl) crystal. Mean and standard deviations are also shown.}
	\label{tabla:ResultsTimeScintNaI(Tl)}
\end{table}

\begin{table}[h!]
	\centering
	\begin{tabular}{|c|c|c|c|c|}
		\hline
		OV (V) & $I_1$ (\%) & $I_2$ (\%) & $\tau_1$ (ns) & $\tau_2$ (ns) \\
		\hline
		2 & 65.8$\pm$0.6 & 34.2$\pm$0.4 & 40$\pm$1 & 1833$\pm$18 \\
		3 & 64.1$\pm$0.4 & 35.9$\pm$0.4 & 37$\pm$1 & 1866$\pm$5 \\
		4 & 61.6$\pm$0.2 & 38.4$\pm$0.2 & 53$\pm$1 & 1861$\pm$6 \\
		5 & 62.3$\pm$0.3 & 37.7$\pm$0.2 & 36$\pm$1 & 1834$\pm$9 \\
		6 & 60.8$\pm$0.3 & 39.2$\pm$0.3 & 48$\pm$1 & 1845$\pm$9 \\
		7 & 63.7$\pm$0.4 & 36.3$\pm$0.2 & 50$\pm$1 & 1848$\pm$9 \\
		8 & 62.6$\pm$0.2 & 37.4$\pm$0.2 & 51$\pm$1 & 1854$\pm$7 \\
		9 & 59.7$\pm$0.2 & 40.3$\pm$0.1 & 52$\pm$1 & 1811$\pm$7 \\
		10 & 59.1$\pm$1.9 & 40.9$\pm$0.7 & 54$\pm$1 & 1803$\pm$10 \\
		11 & 56.7$\pm$0.1 & 43.3$\pm$0.1 & 62$\pm$1 & 1726$\pm$4 \\
		\hline
		Mean & 60.9$\pm$2.4 & 39.0$\pm$2.5 & 48$\pm$8 & 1822$\pm$39 \\
		\hline
	\end{tabular}
	\caption{Relative intensities and scintillation times of the two components identified in each measurement with the NaI crystal. Mean and standard deviations are also shown.}
	\label{tabla:ResultsScintNaI}
\end{table}

The scintillation of the crystals should not depend on the overvoltage applied to the SiPMs. However, in the results obtained for the NaI crystal, differences in the pulse shape are observed: the fast component becomes slower and its intensity decreases at high overvoltages while the slow component becomes faster and increase its intensity. In the NaI(Tl) crystal, it can also be observed such an effect for the slowest component, which reduces its intensity and becomes faster. We can conclude that in both crystals we observe pulse changes for the higher pulse amplitudes measurements, systematically producing faster time constants, which is easier to observe in the slower components.

This systematic effect should be better understood in future works, and as commented, is probably related with saturation in some step of the signal processing. However, we can conclude that the NaI(Tl) scintillation at LAr temperature has a very slow component that is not well fitted with our present digitizer configuration, which explains the large uncertainty in the fitted value. In addition, both crystals show a very fast component that is not easy to fit, because it is faster than the SPE, and then, it is affected by the deconvolution procedure applied in the fit, which also justifies the high dispersion in the values obtained for this parameter.

However, the results of the fits at different overvoltages behave in a consistent way, and are compatible with each other, allowing to draw some robust conclusions on the scintillation properties of both crystals at LAr temperature.

\subsection{Characterization at different temperatures} \label{Section:SiPMSTAR2_Charac_Cooling}

\begin{figure}[h!]
	\begin{center}
		\includegraphics[width=\textwidth]{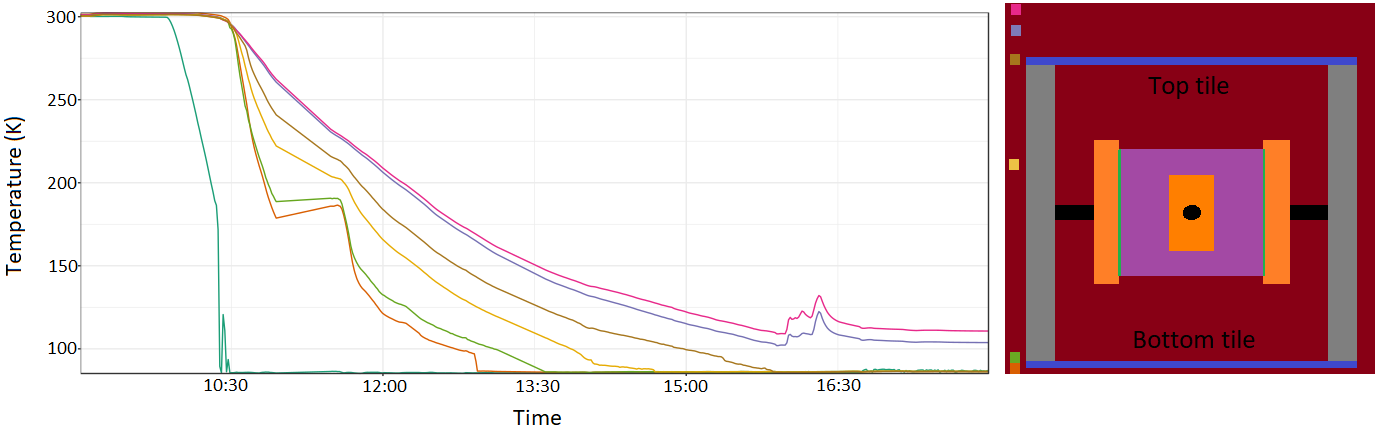}
		\caption{\label{TempNaI_Cooling}Temperatures measured with the PT100 resistors inside the internal container during the cooling down procedure with the NaI crystal (left plot) and the PT100 resistors positions (right plot).}
	\end{center}
\end{figure}

\begin{figure}[h!]
	\begin{center}
		\includegraphics[width=\textwidth]{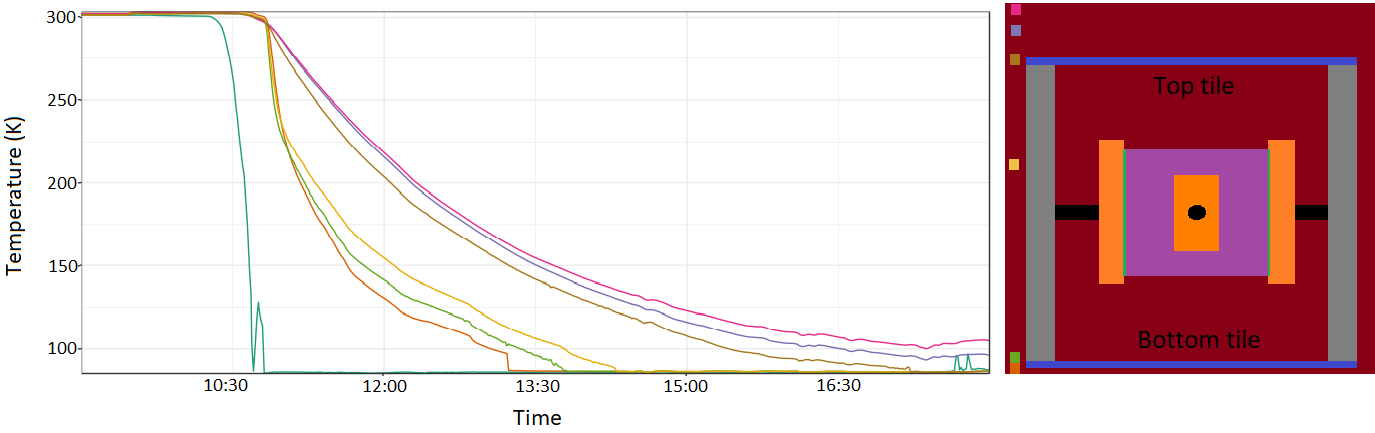}
		\caption{\label{TempNaITl_Cooling}Temperatures measured with the PT100 resistors inside the internal container during the cooling down procedure with the NaI crystal (left plot) and the PT100 resistors positions (right plot).}
	\end{center}
\end{figure}

During the cooling down of each crystal, some measurements were carried out with the purpose of observing the evolution of the scintillation properties with the temperature. The temperature of the gas around the chamber was under continuous monitoring with the PT100 resistors, as it is shown in Figures~\ref{TempNaI_Cooling} and~\ref{TempNaITl_Cooling} for NaI and NaI(Tl) crystals cooling down, respectively. The fast change observed in the gas temperature in few minutes and the temperature differences measured between PT100 resistors, depending on their position, was a clear indication that the system was not in equilibrium while cooling down.

Despite of this systematic uncertainty in the knowledge of the temperature, 25~measurements were carried out while cooling down for both crystals, NaI and NaI(Tl), in order to have some information about the evolution of the scintillation properties with the temperature. All the measurements followed the same procedure, and the temperature assigned to each of them is that obtained at the beginning of the measurement with the PT100 resistor placed closest to the chamber (shown in yellow in Figures~\ref{TempNaI_Cooling} and~\ref{TempNaITl_Cooling}) because the crystal temperature was not directly measured. All the measurements were carried out at the same overvoltage (3~V), requiring to measure the breakdown voltage as a first step, at every temperature. This is a crucial systematic problem behind these measurements, because as the temperature was continuously decreasing, the bias voltage setup at each measurement for the SiPM arrays would correspond to different overvoltages along the measurement. Two measurements after biasing the SiPMs were carried out: the SPE calibration (recording 5$\times$10$^4$ events) and the $^{241}Am$ source calibration (recording 10$^4$ events). In total, this procedure could take about 10~minutes (5~minutes of data taking and other 5 for preparing the measurements). No background measurements were taken.

\begin{figure}[]
	\begin{center}
		\begin{subfigure}[b]{0.3\textwidth}
			\includegraphics[width=\textwidth]{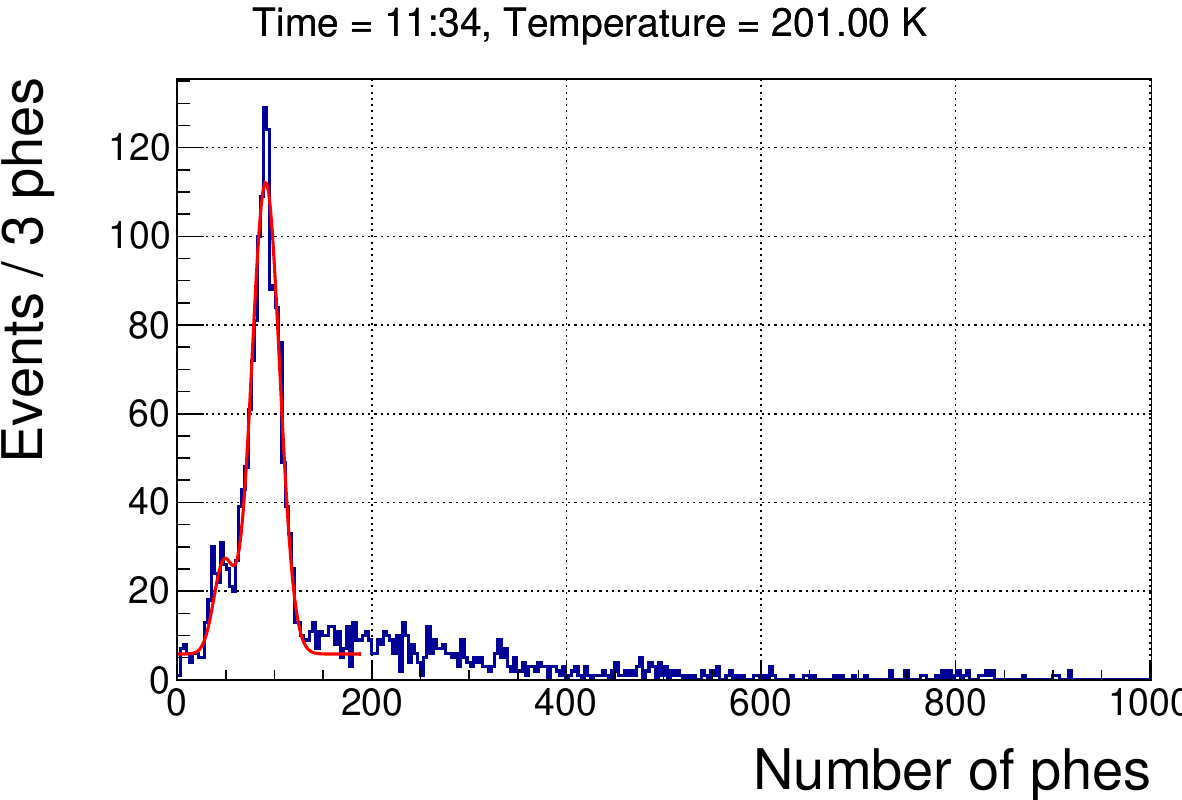}
		\end{subfigure}
		\begin{subfigure}[b]{0.3\textwidth}
			\includegraphics[width=\textwidth]{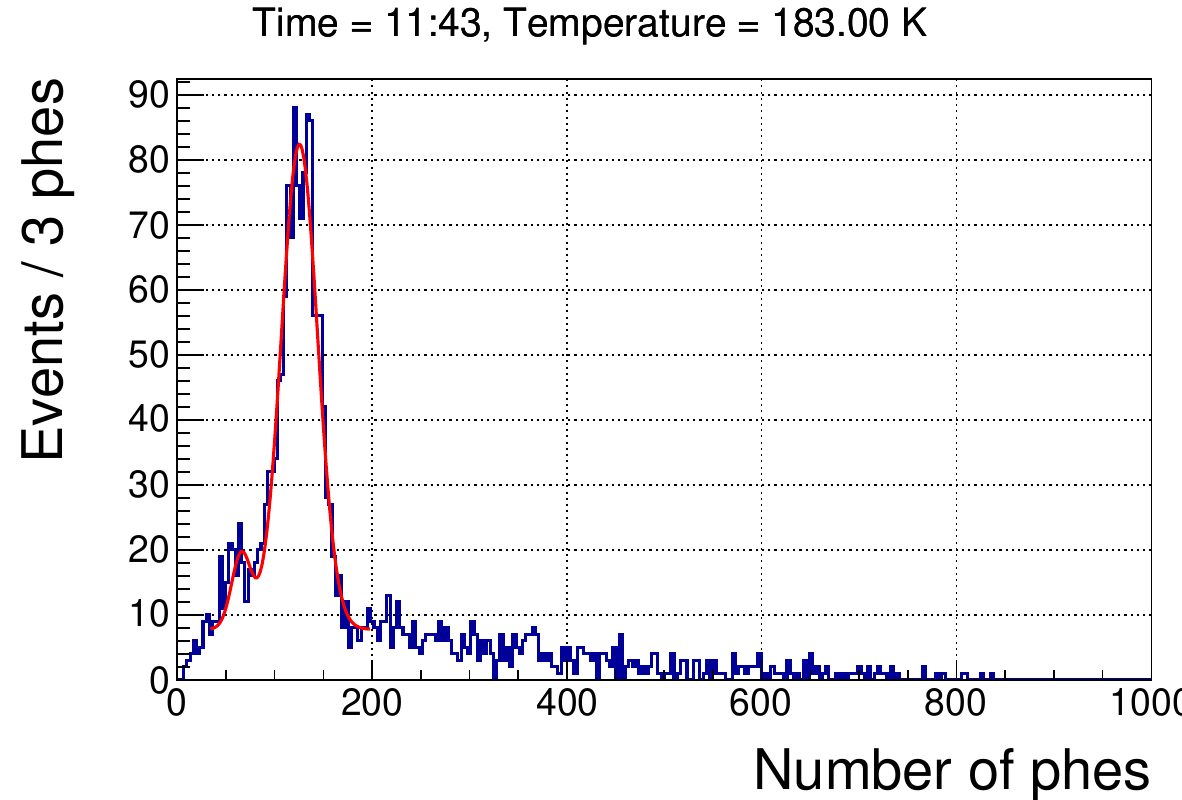}
		\end{subfigure}
		\begin{subfigure}[b]{0.3\textwidth}
			\includegraphics[width=\textwidth]{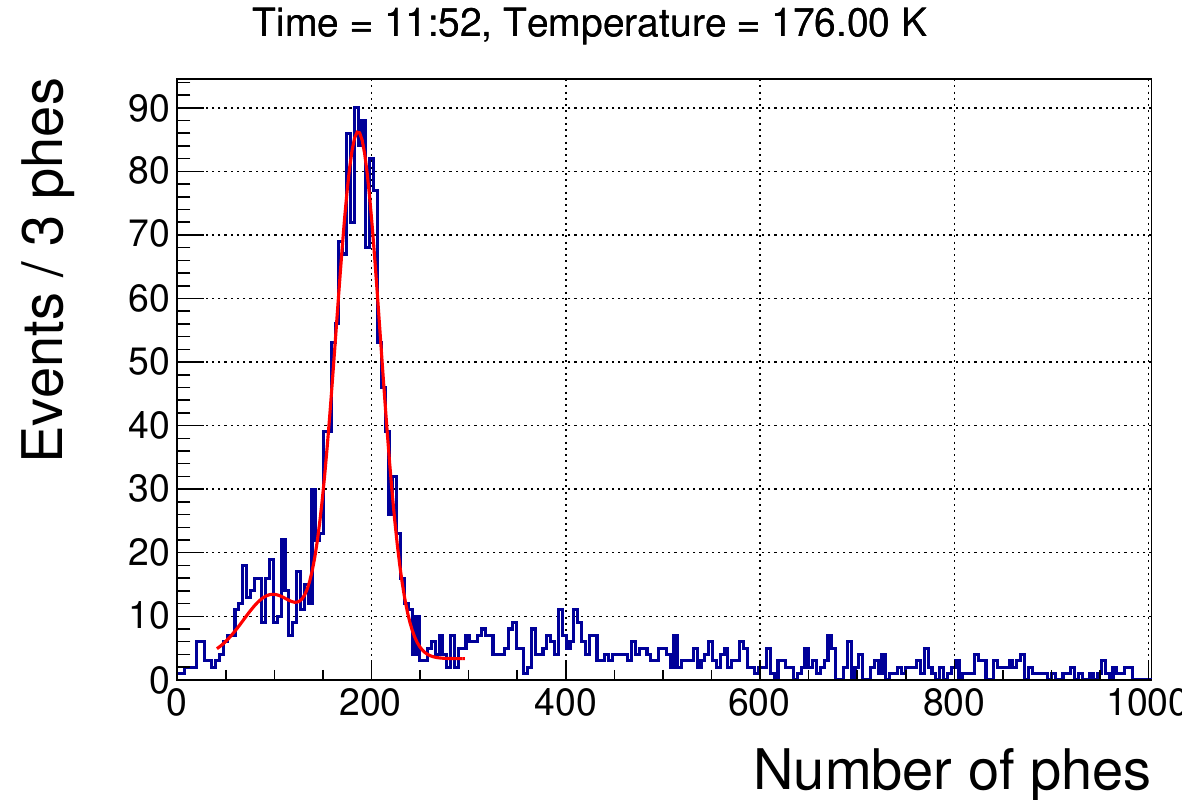}
		\end{subfigure}
		\begin{subfigure}[b]{0.3\textwidth}
			\includegraphics[width=\textwidth]{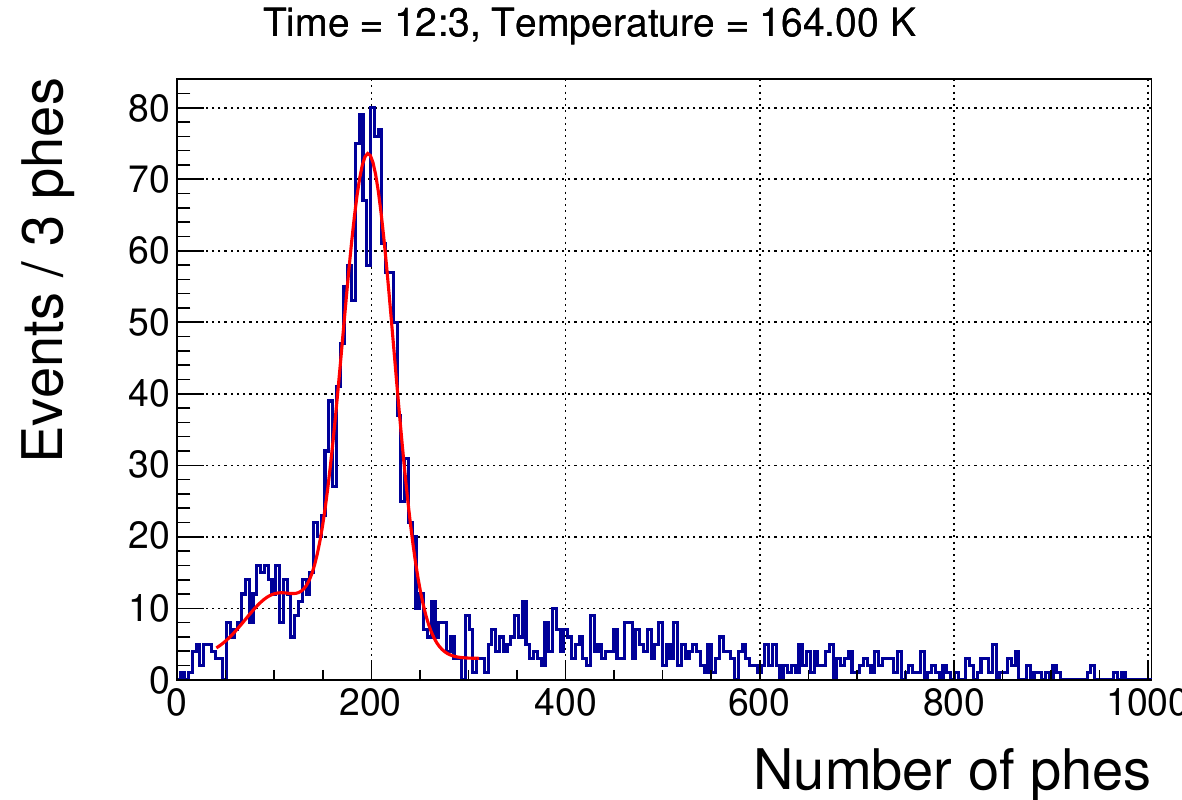}
		\end{subfigure}
		\begin{subfigure}[b]{0.3\textwidth}
			\includegraphics[width=\textwidth]{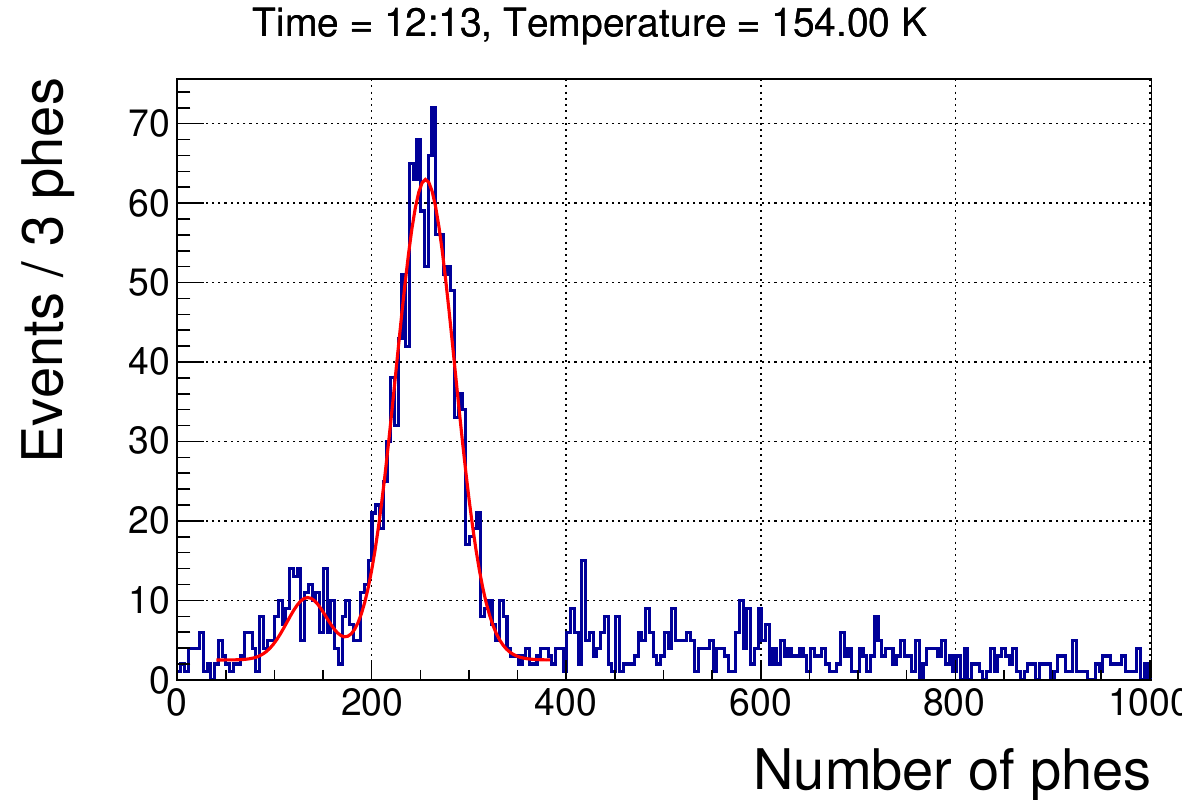}
		\end{subfigure}
		\begin{subfigure}[b]{0.3\textwidth}
			\includegraphics[width=\textwidth]{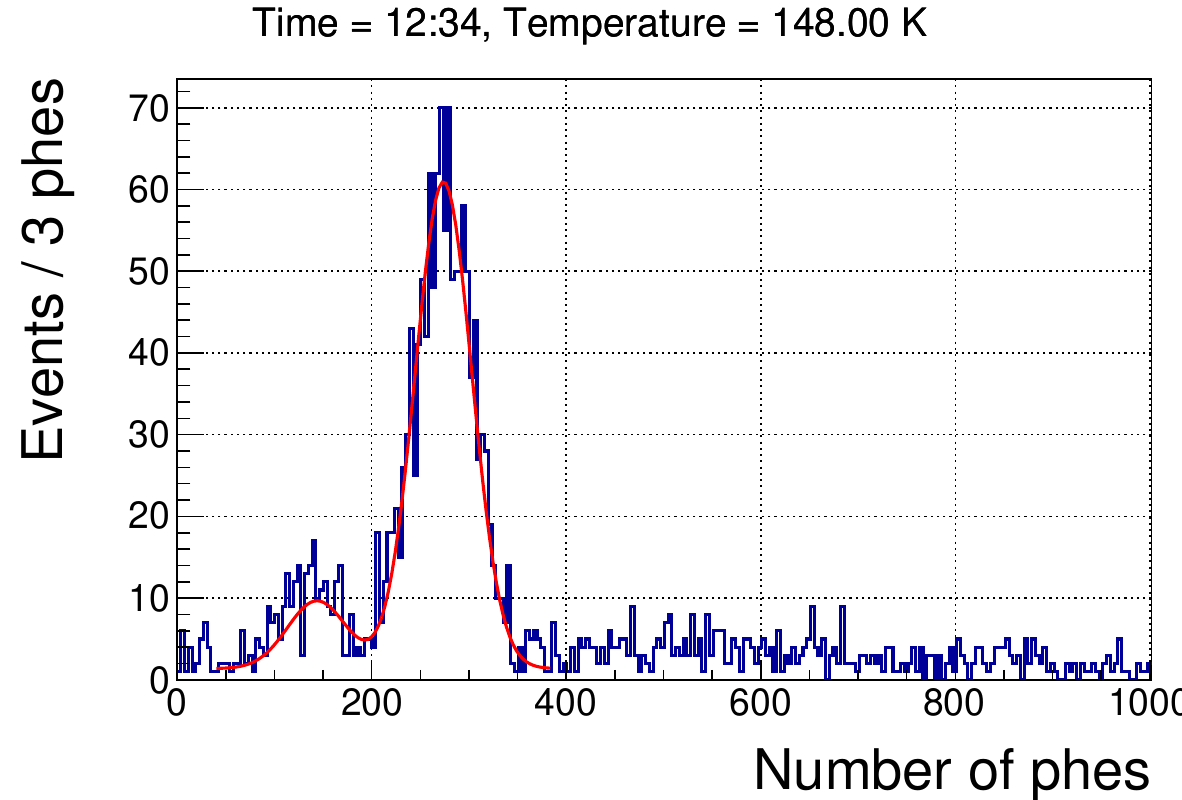}
		\end{subfigure}
		\begin{subfigure}[b]{0.3\textwidth}
			\includegraphics[width=\textwidth]{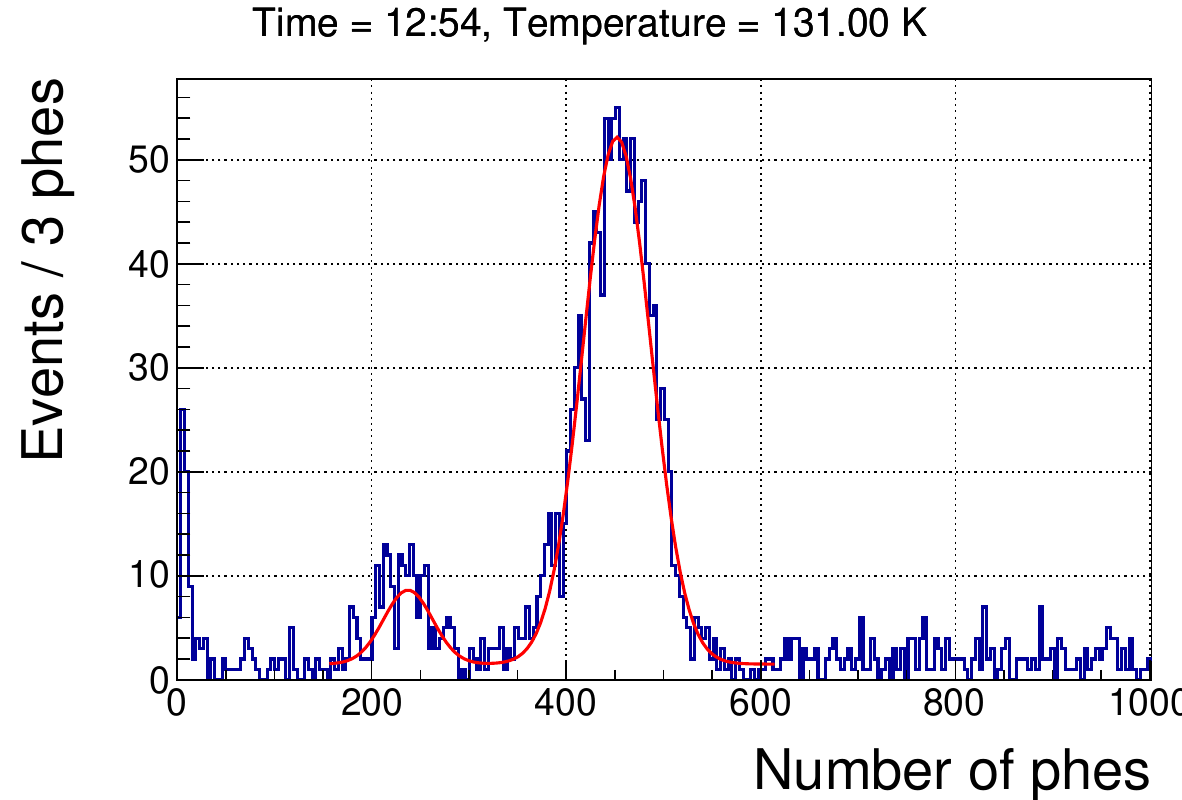}
		\end{subfigure}
		\begin{subfigure}[b]{0.3\textwidth}
			\includegraphics[width=\textwidth]{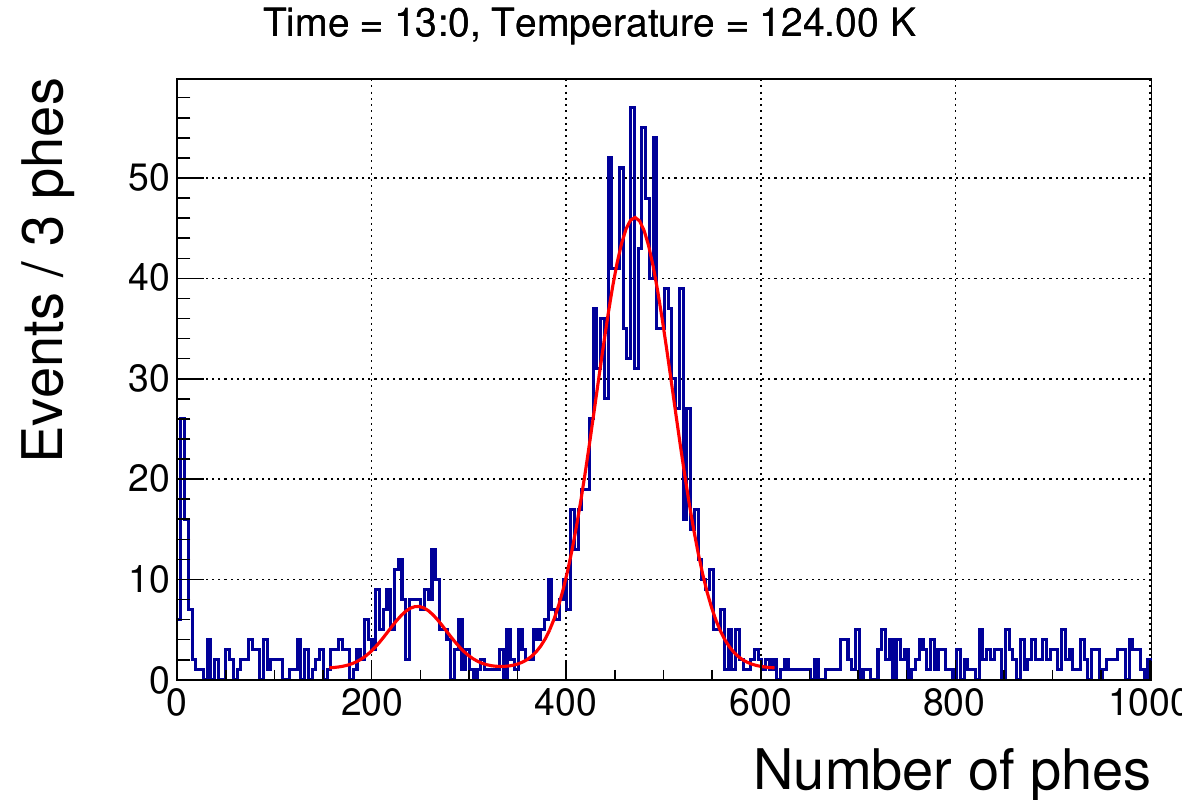}
		\end{subfigure}
		\begin{subfigure}[b]{0.3\textwidth}
			\includegraphics[width=\textwidth]{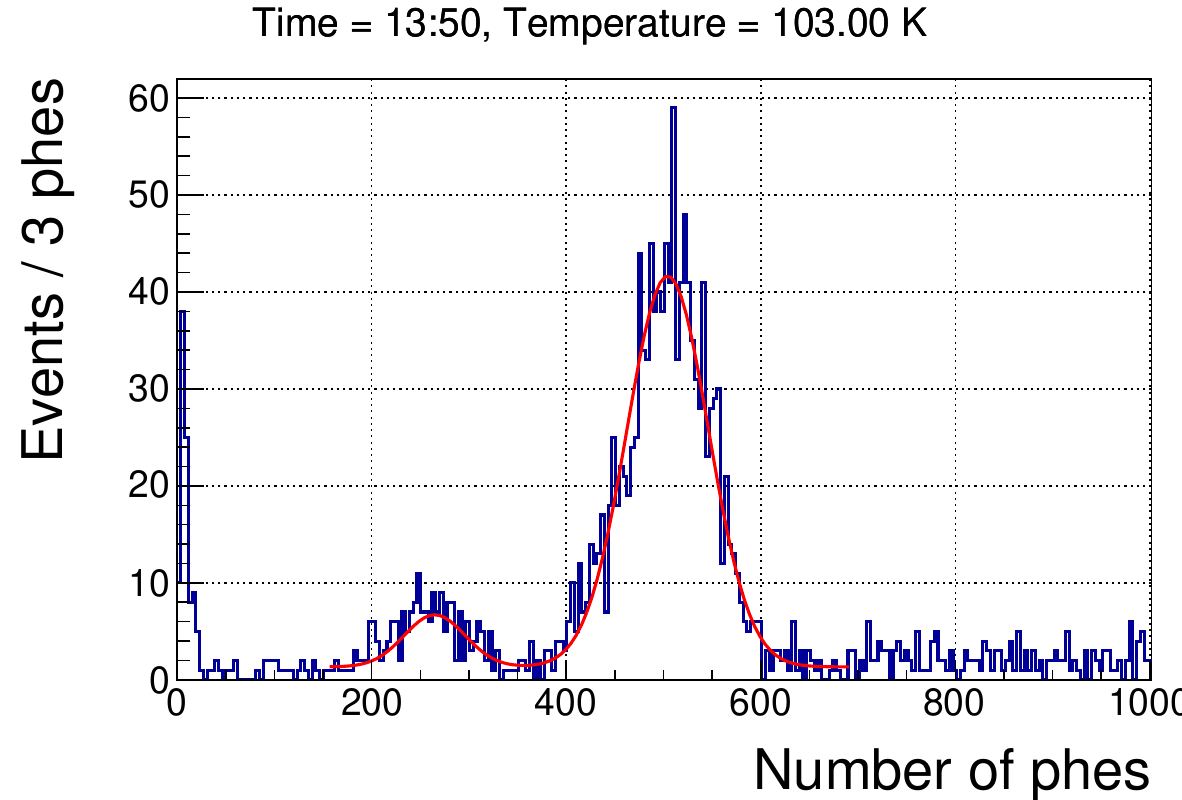}
		\end{subfigure}
		\begin{subfigure}[b]{0.3\textwidth}
			\includegraphics[width=\textwidth]{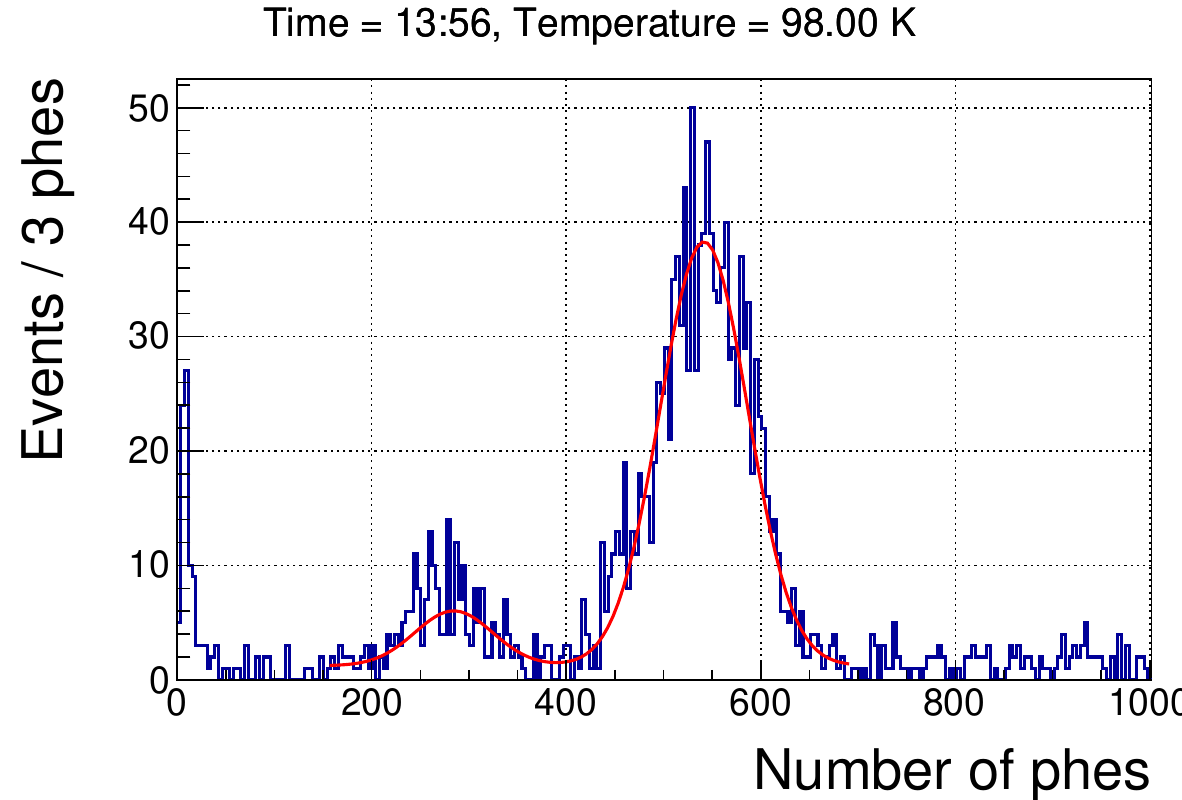}
		\end{subfigure}
		\begin{subfigure}[b]{0.3\textwidth}
			\includegraphics[width=\textwidth]{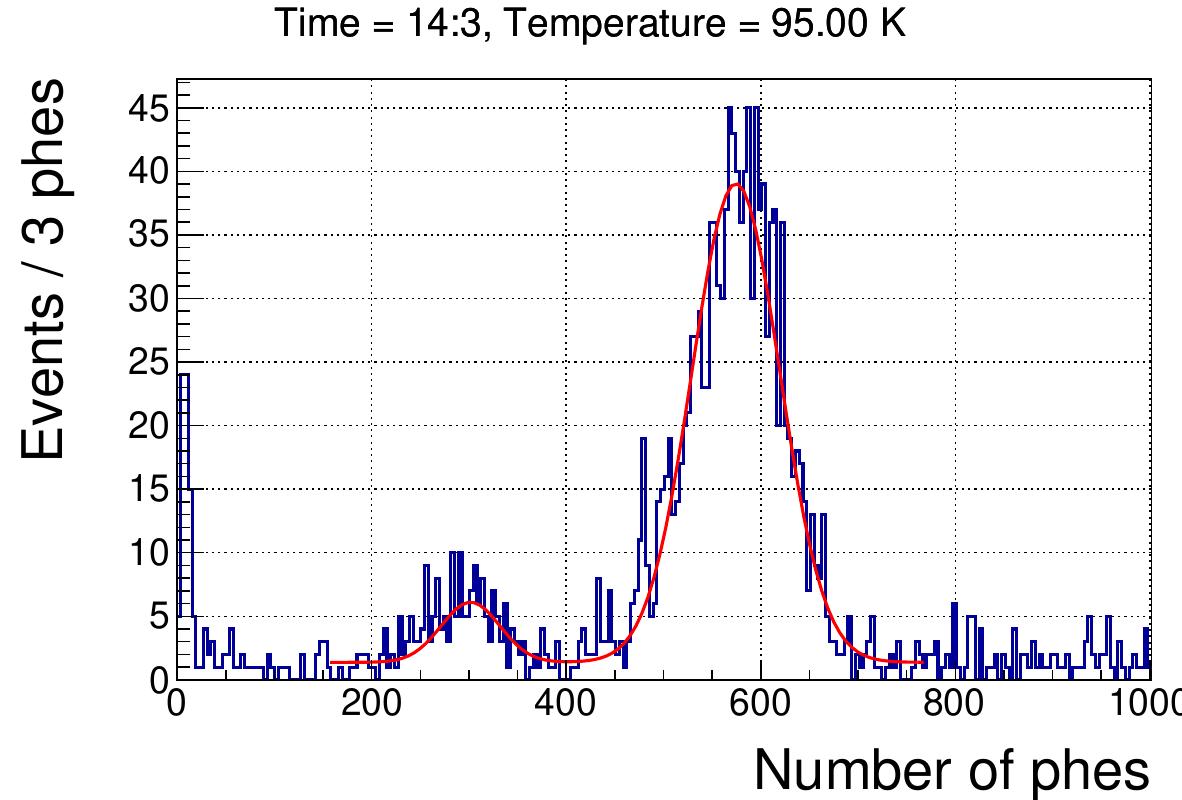}
		\end{subfigure}
		\begin{subfigure}[b]{0.3\textwidth}
			\includegraphics[width=\textwidth]{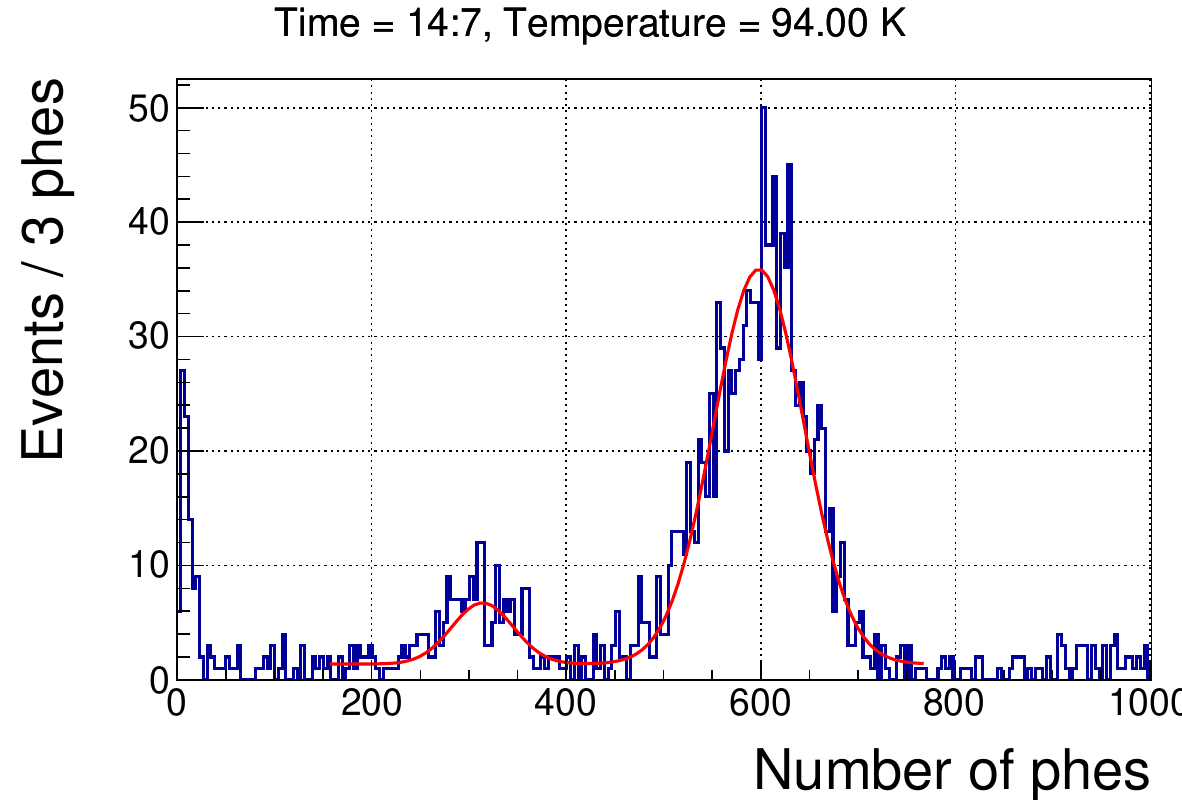}
		\end{subfigure}
		\begin{subfigure}[b]{0.3\textwidth}
			\includegraphics[width=\textwidth]{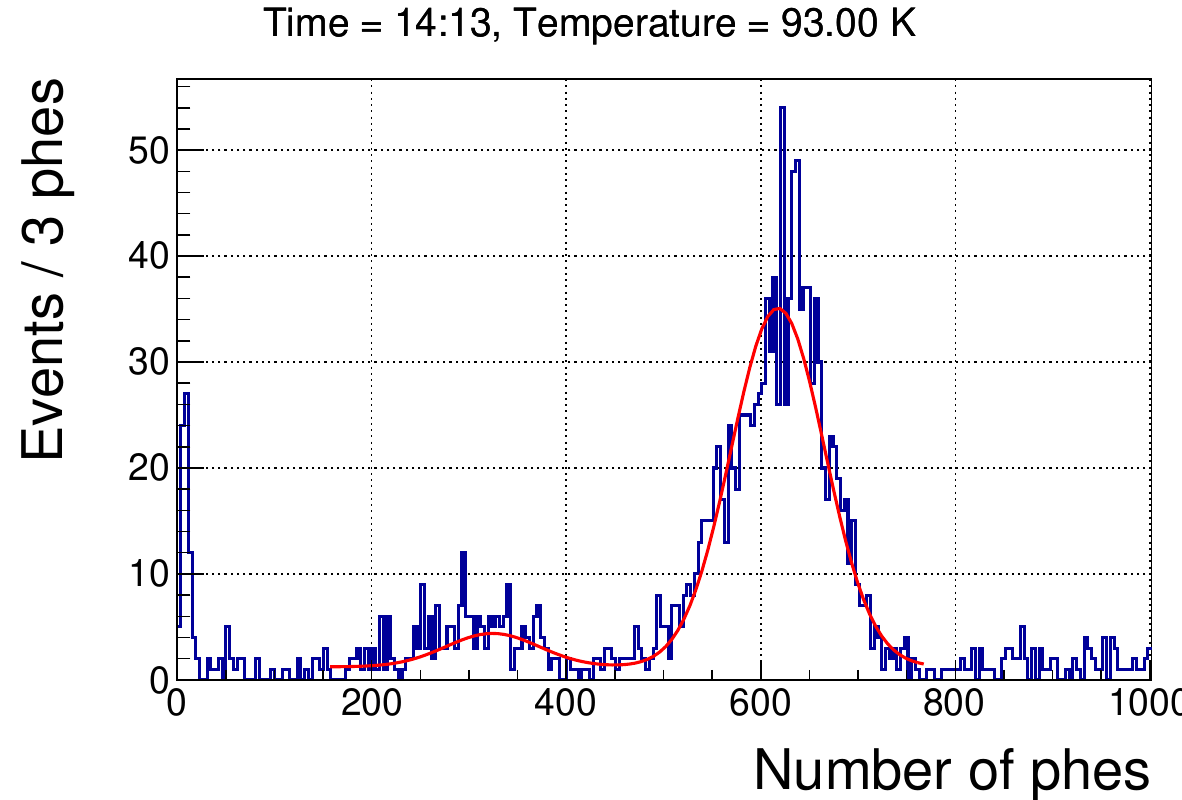}
		\end{subfigure}
		\begin{subfigure}[b]{0.3\textwidth}
			\includegraphics[width=\textwidth]{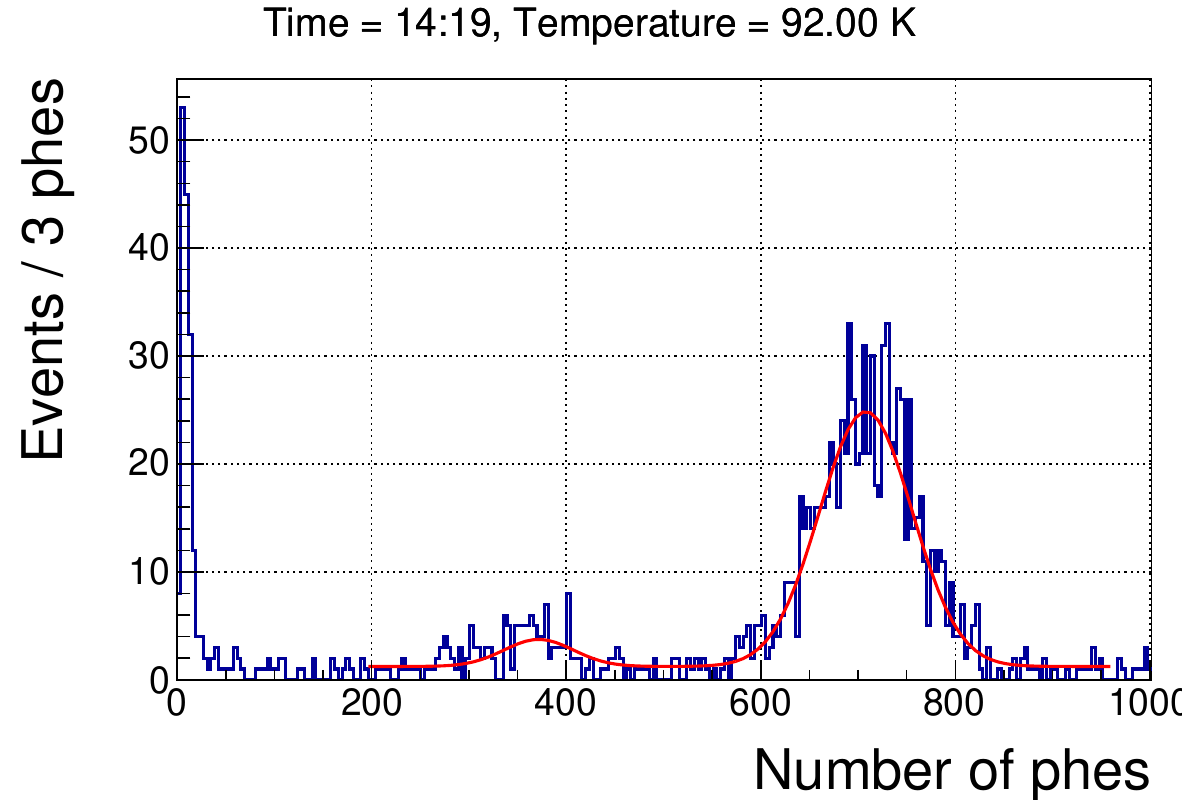}
		\end{subfigure}
		\begin{subfigure}[b]{0.3\textwidth}
			\includegraphics[width=\textwidth]{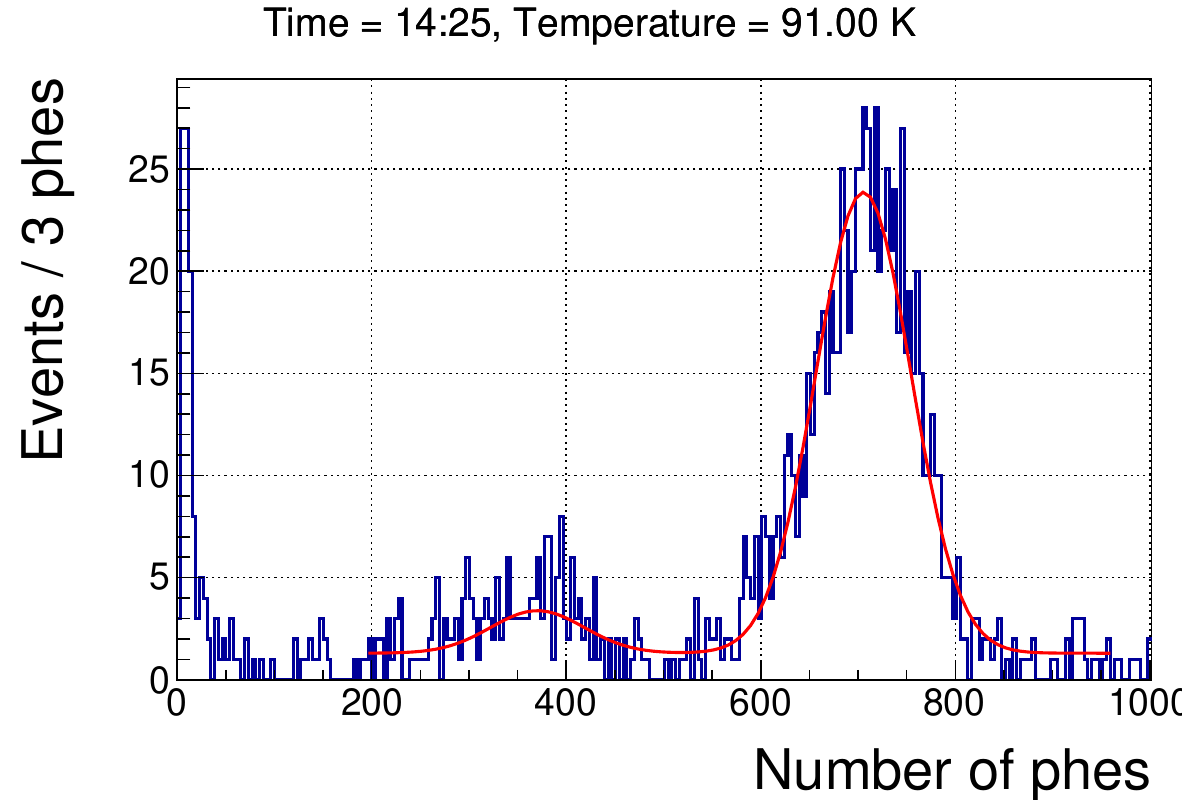}
		\end{subfigure}
		\begin{subfigure}[b]{0.3\textwidth}
			\includegraphics[width=\textwidth]{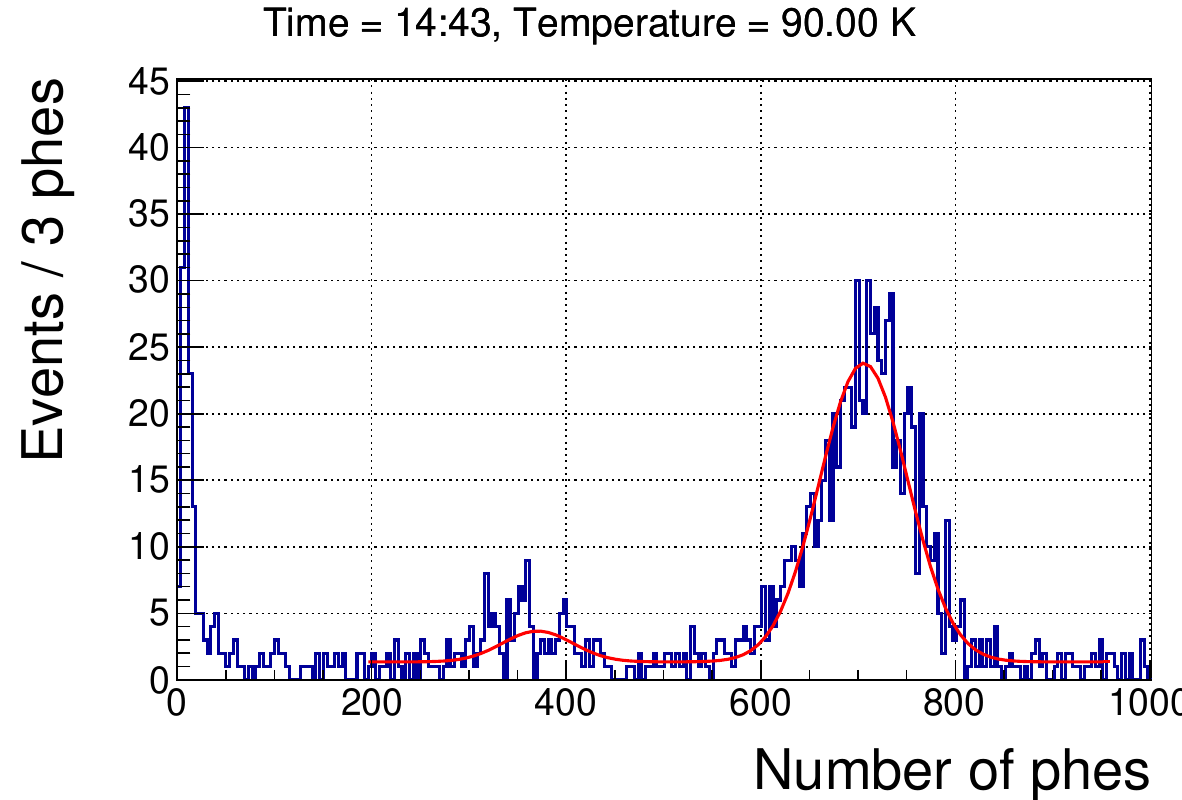}
		\end{subfigure}
		\begin{subfigure}[b]{0.3\textwidth}
			\includegraphics[width=\textwidth]{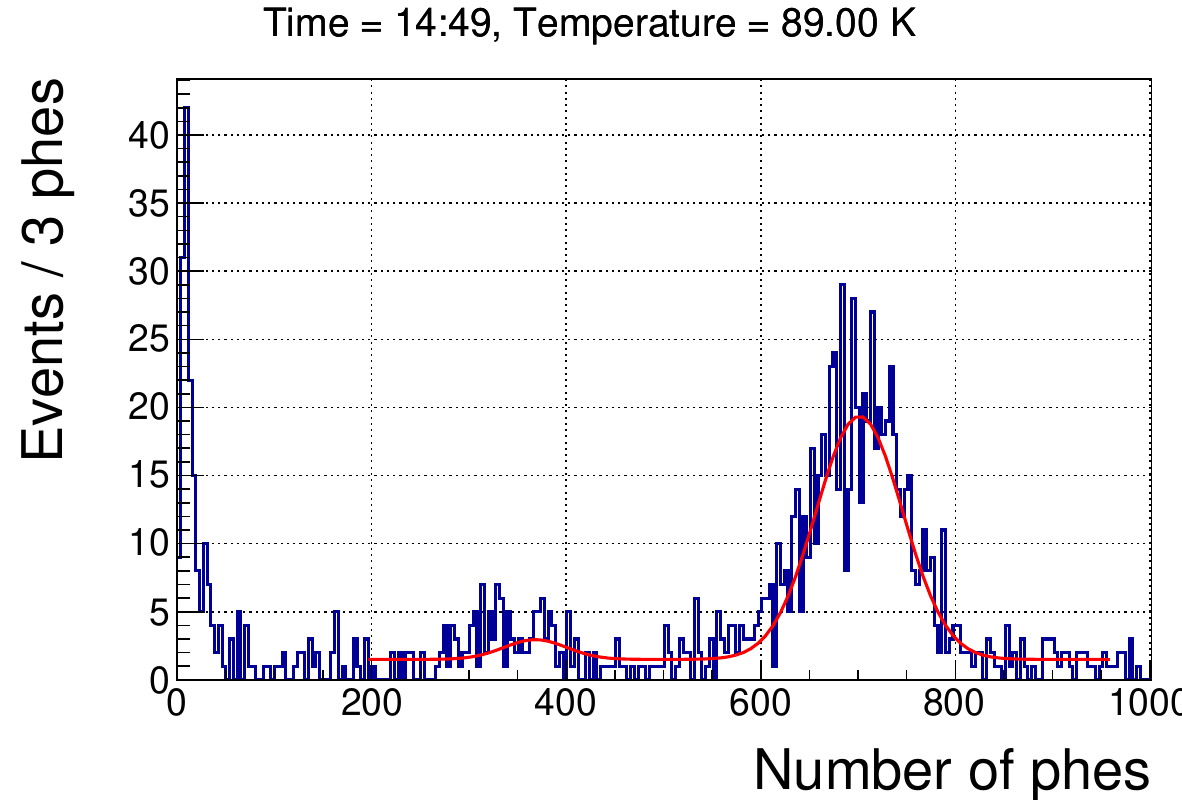}
		\end{subfigure}
		\caption{\label{FitsNaI_Cooling}Fits of the $^{241}Am$ calibration spectra of the measurements while cooling down the NaI crystal.}
	\end{center}
\end{figure}

\begin{figure}[]
	\begin{center}
		\begin{subfigure}[b]{0.3\textwidth}
			\includegraphics[width=\textwidth]{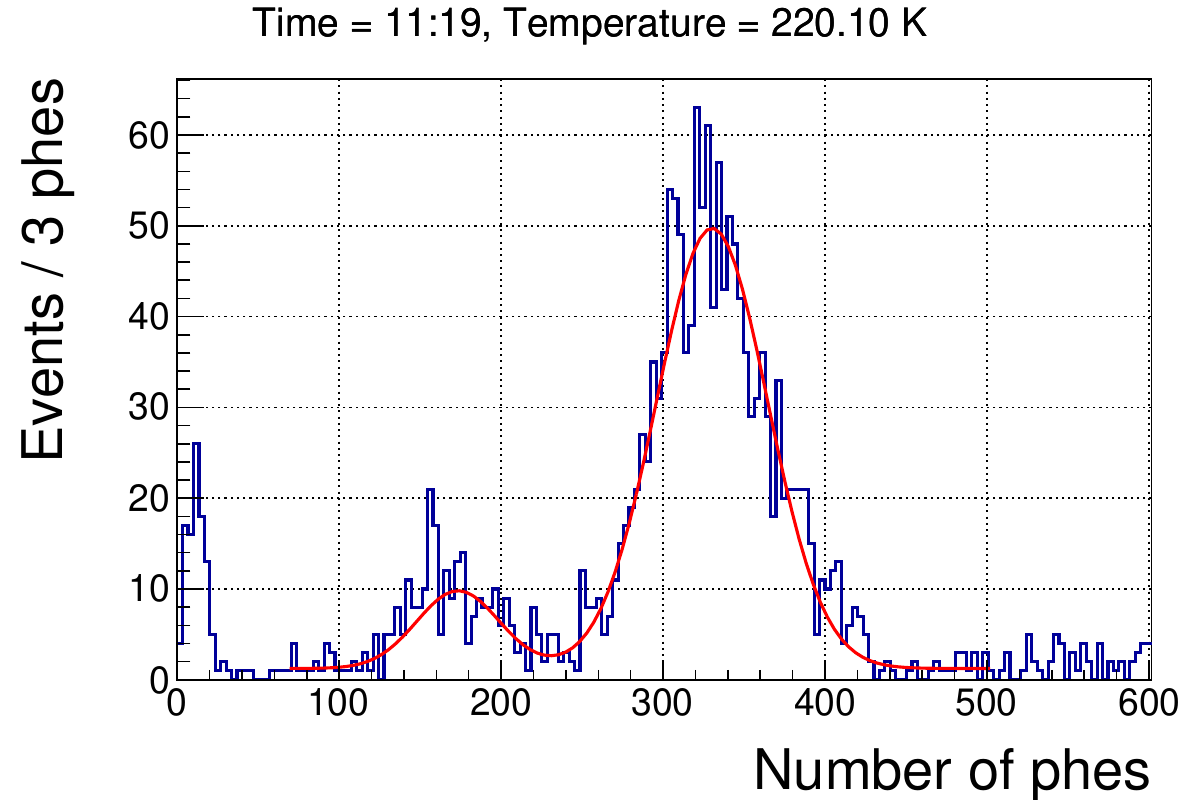}
		\end{subfigure}
		\begin{subfigure}[b]{0.3\textwidth}
			\includegraphics[width=\textwidth]{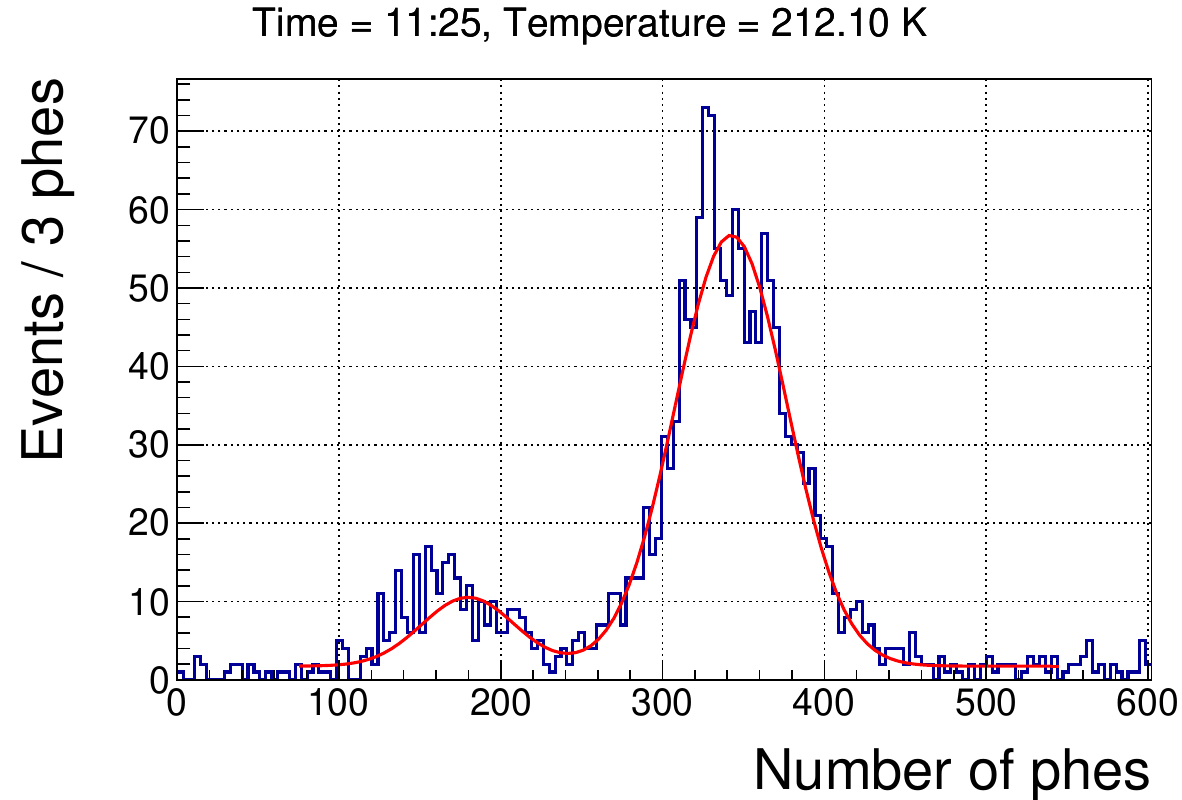}
		\end{subfigure}
		\begin{subfigure}[b]{0.3\textwidth}
			\includegraphics[width=\textwidth]{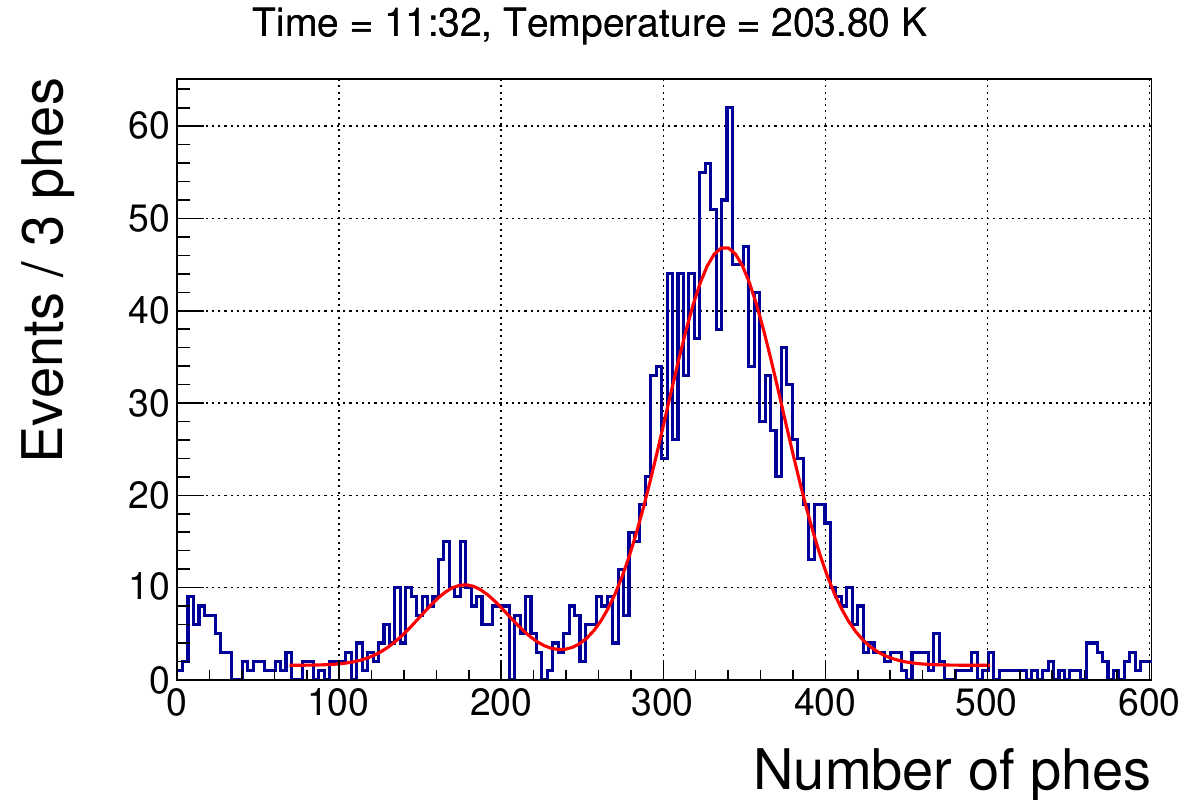}
		\end{subfigure}
		\begin{subfigure}[b]{0.3\textwidth}
			\includegraphics[width=\textwidth]{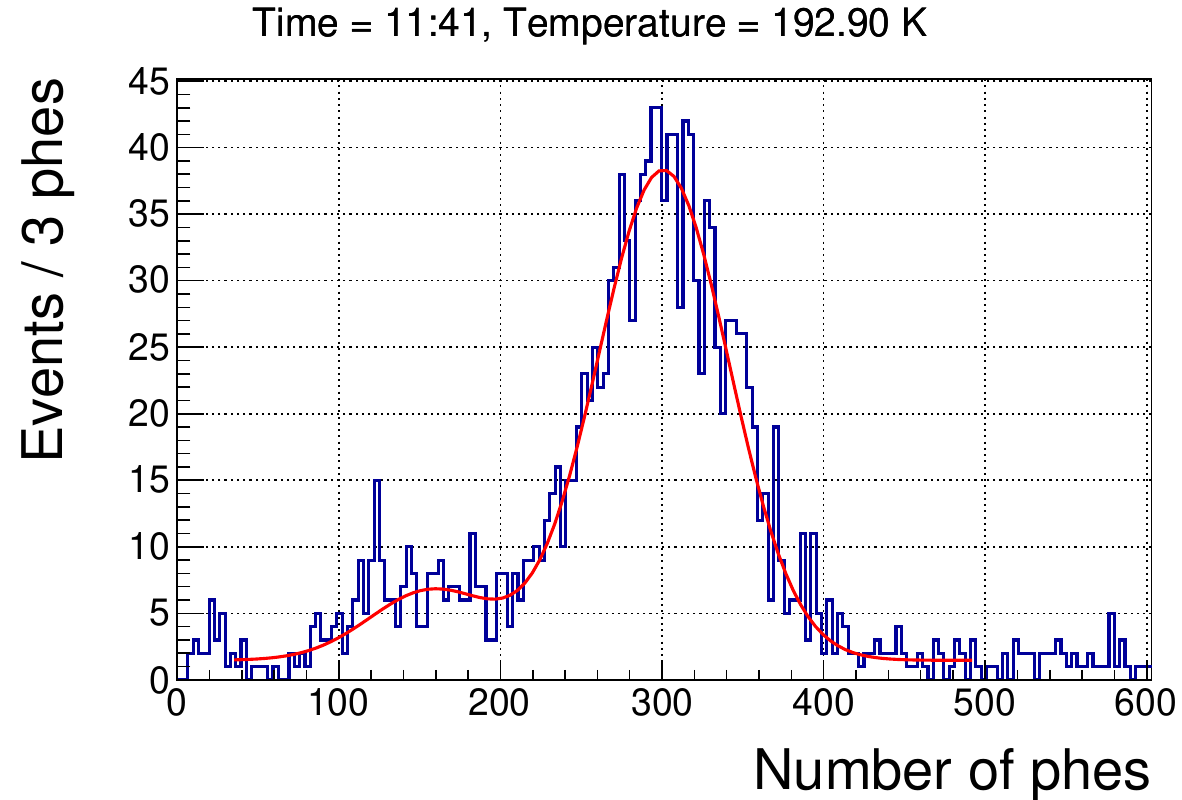}
		\end{subfigure}
		\begin{subfigure}[b]{0.3\textwidth}
			\includegraphics[width=\textwidth]{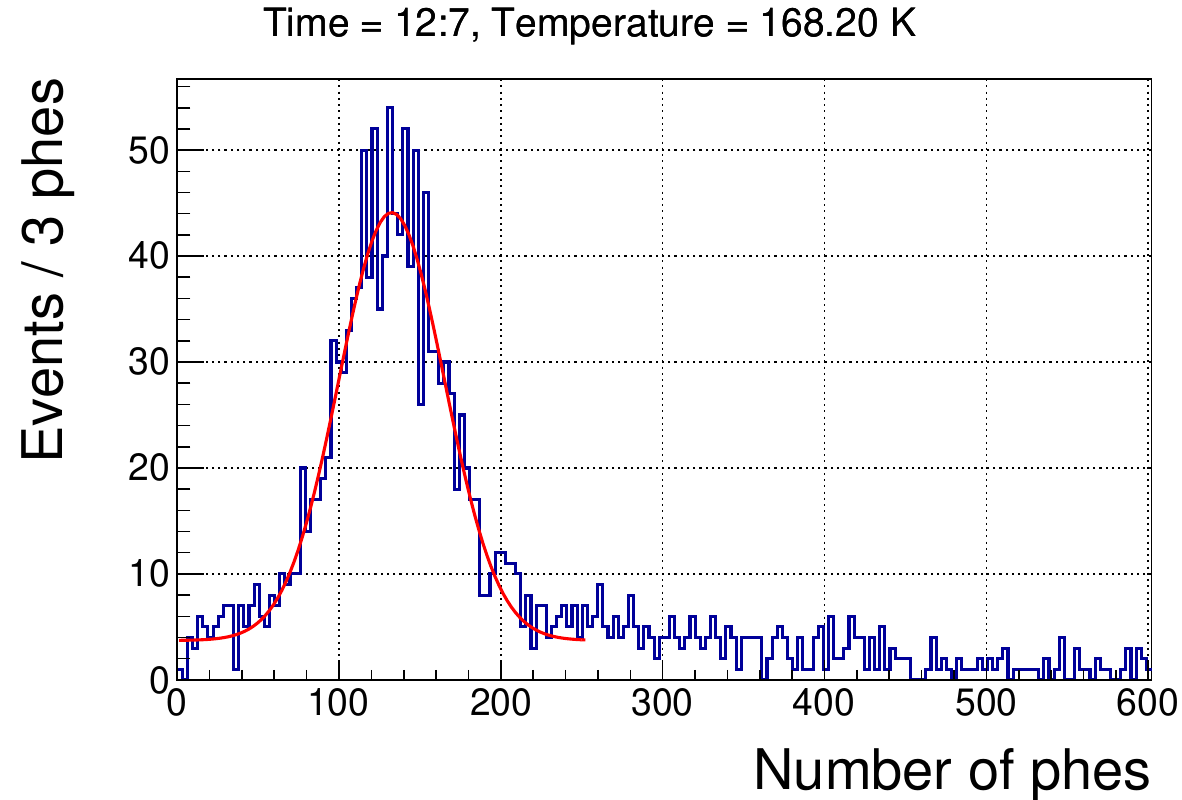}
		\end{subfigure}
		\begin{subfigure}[b]{0.3\textwidth}
			\includegraphics[width=\textwidth]{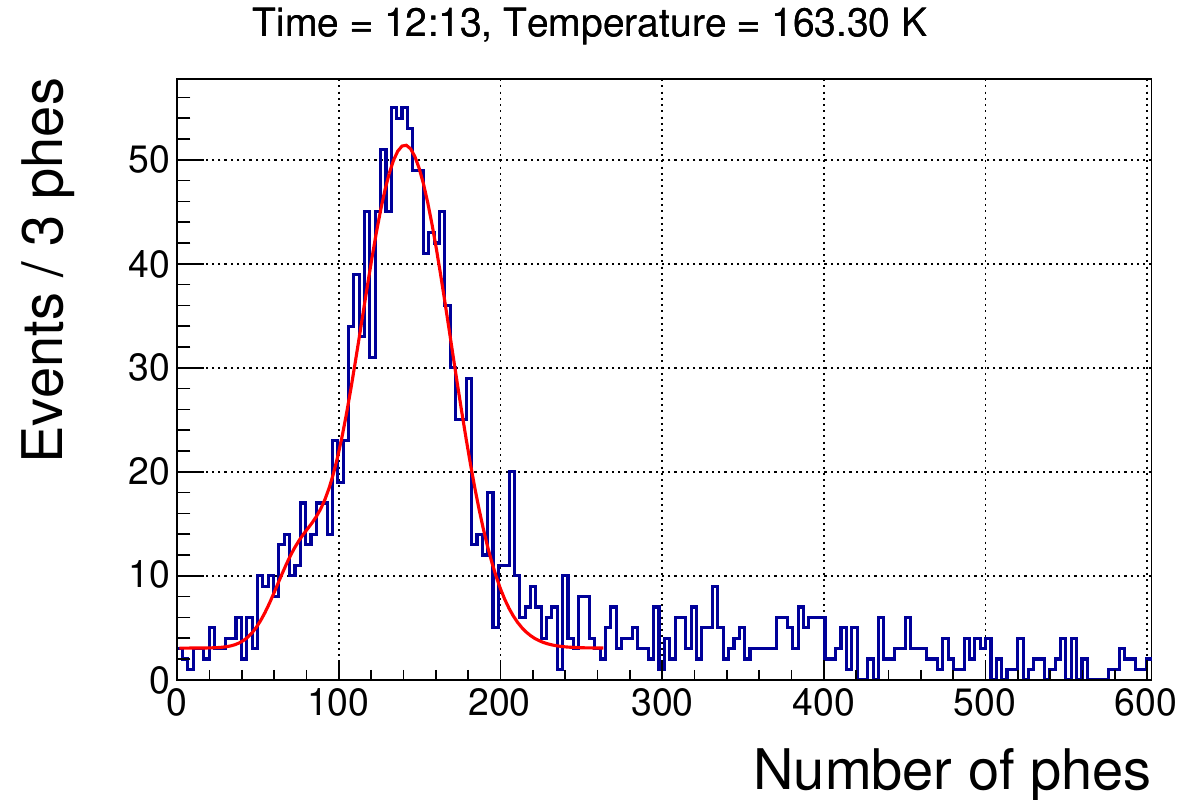}
		\end{subfigure}
		\begin{subfigure}[b]{0.3\textwidth}
			\includegraphics[width=\textwidth]{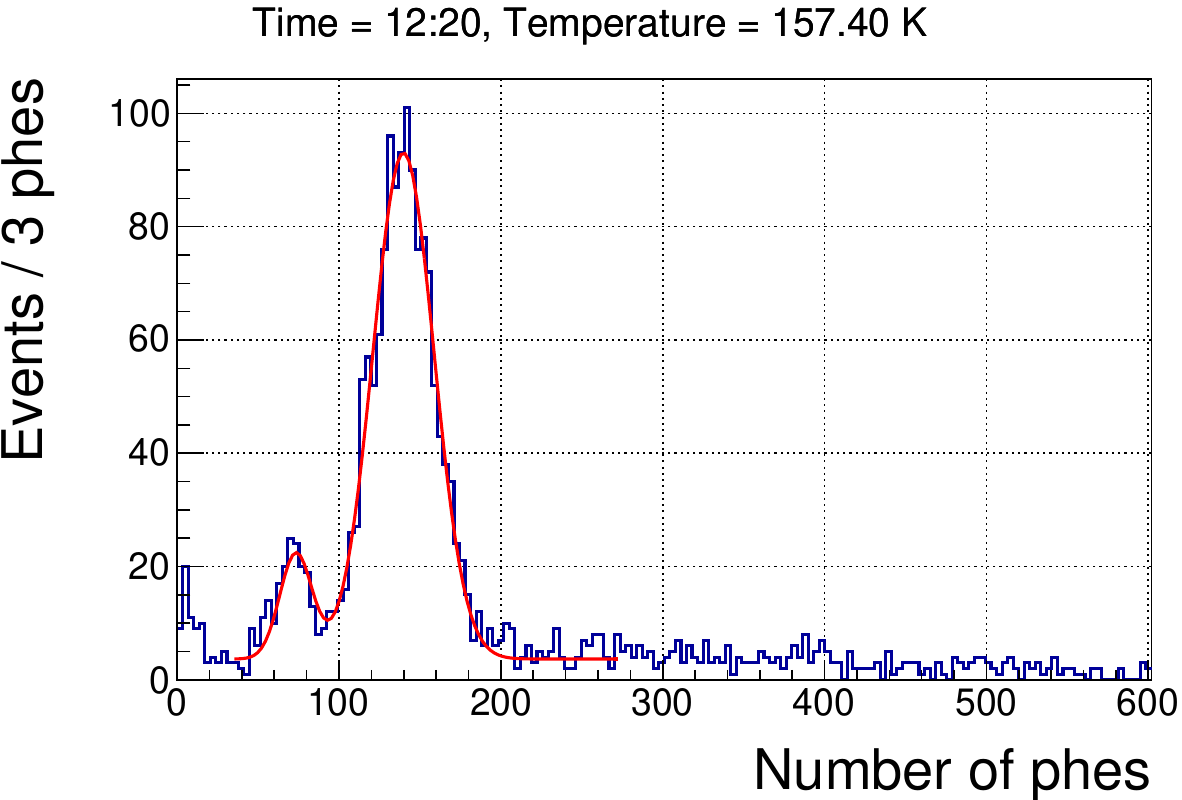}
		\end{subfigure}
		\begin{subfigure}[b]{0.3\textwidth}
			\includegraphics[width=\textwidth]{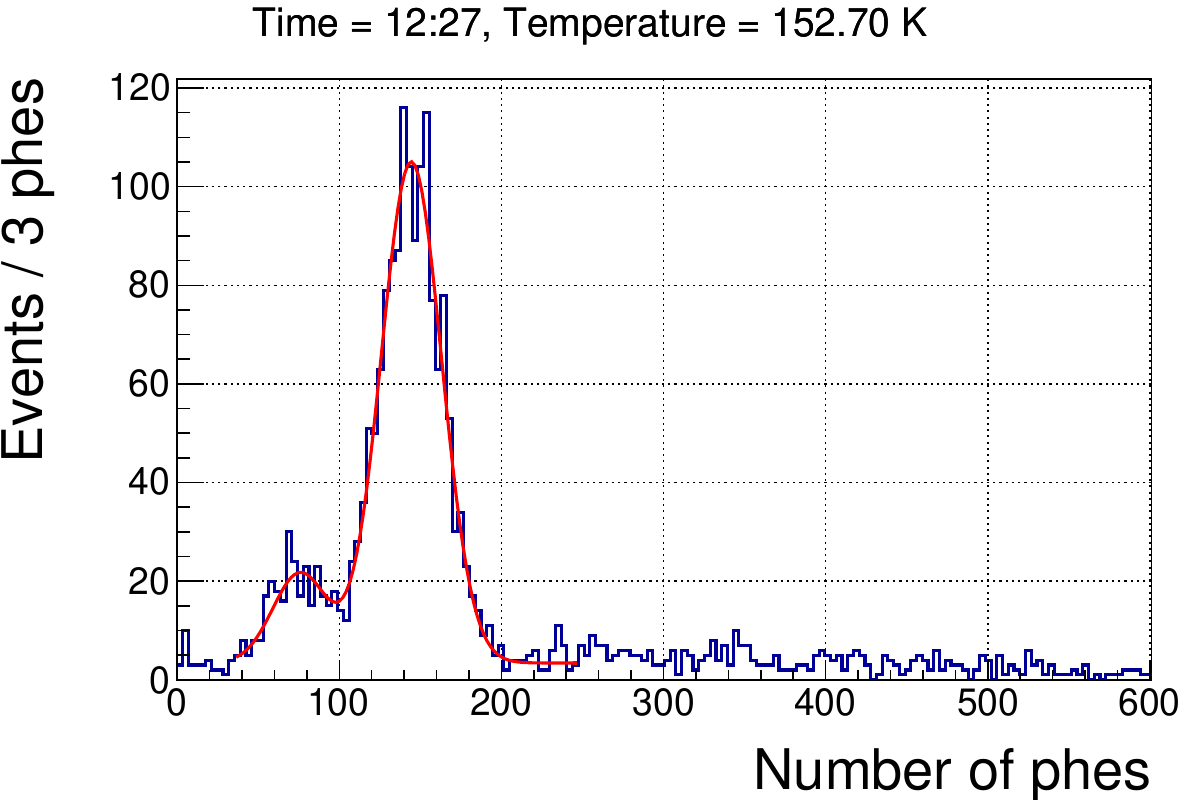}
		\end{subfigure}
		\begin{subfigure}[b]{0.3\textwidth}
			\includegraphics[width=\textwidth]{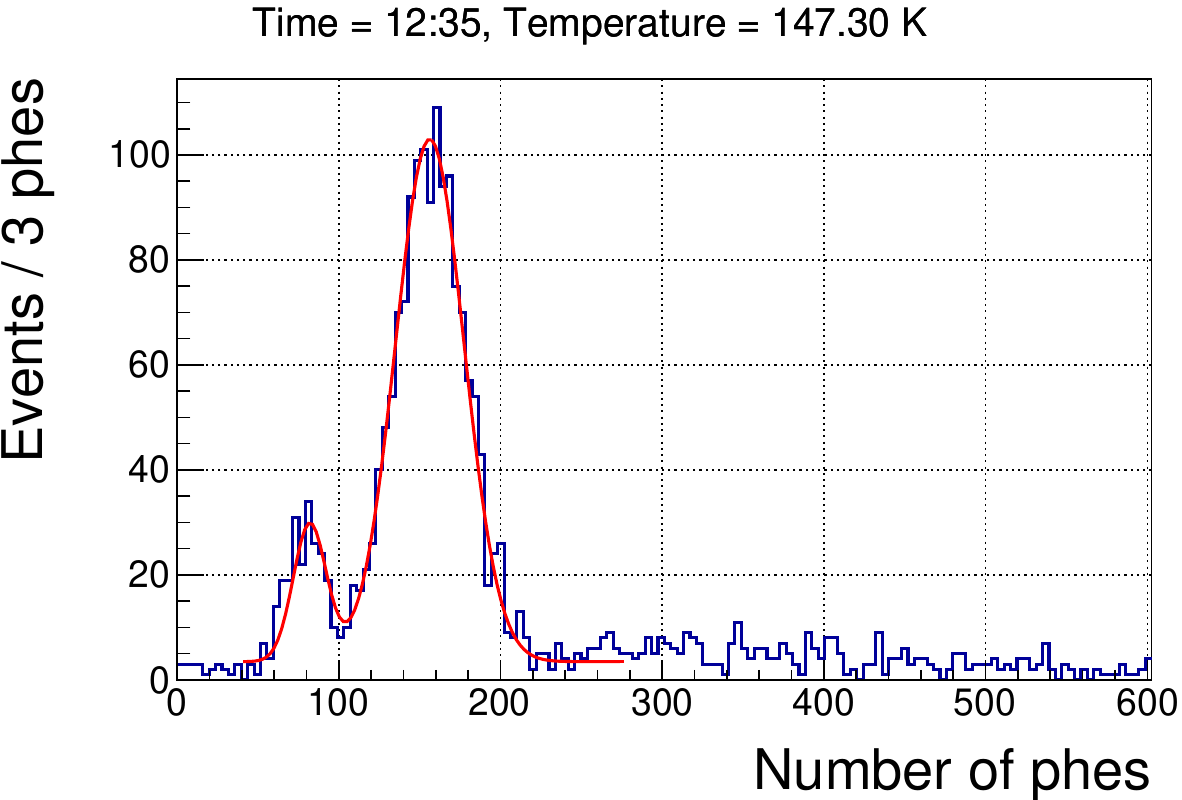}
		\end{subfigure}
		\begin{subfigure}[b]{0.3\textwidth}
			\includegraphics[width=\textwidth]{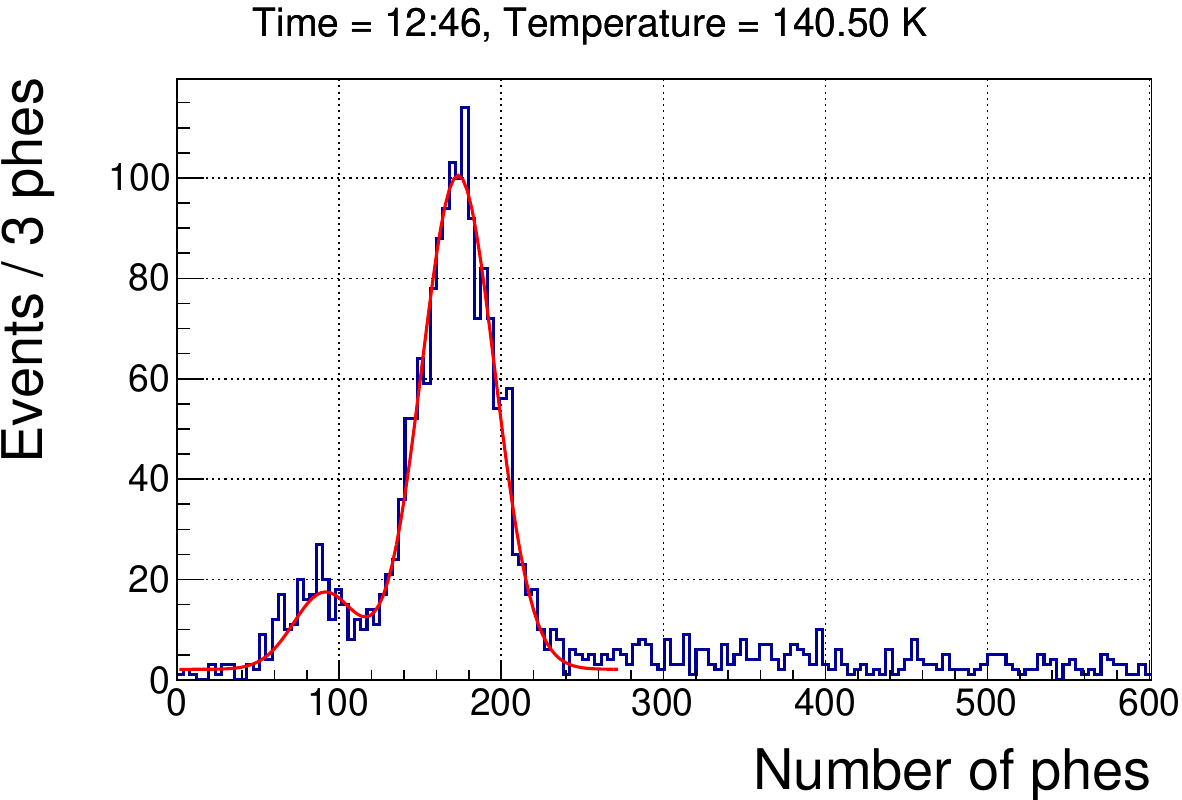}
		\end{subfigure}
		\begin{subfigure}[b]{0.3\textwidth}
			\includegraphics[width=\textwidth]{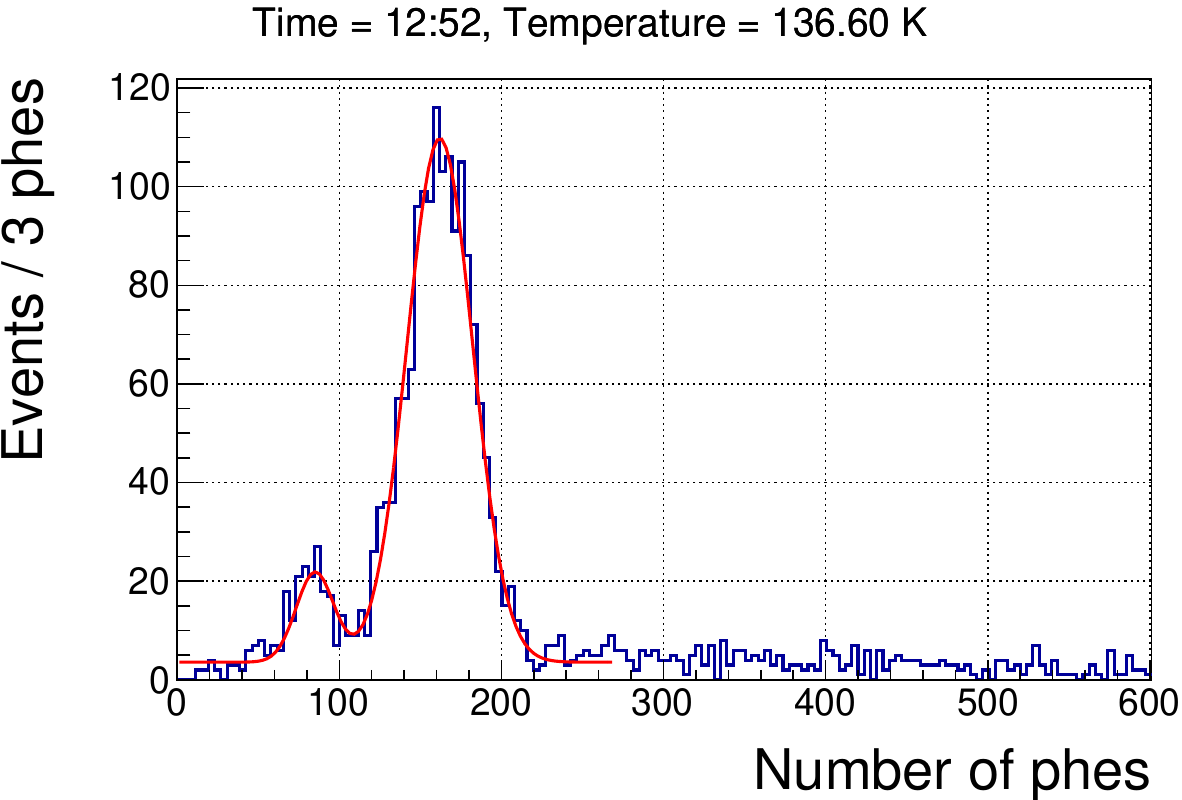}
		\end{subfigure}
		\begin{subfigure}[b]{0.3\textwidth}
			\includegraphics[width=\textwidth]{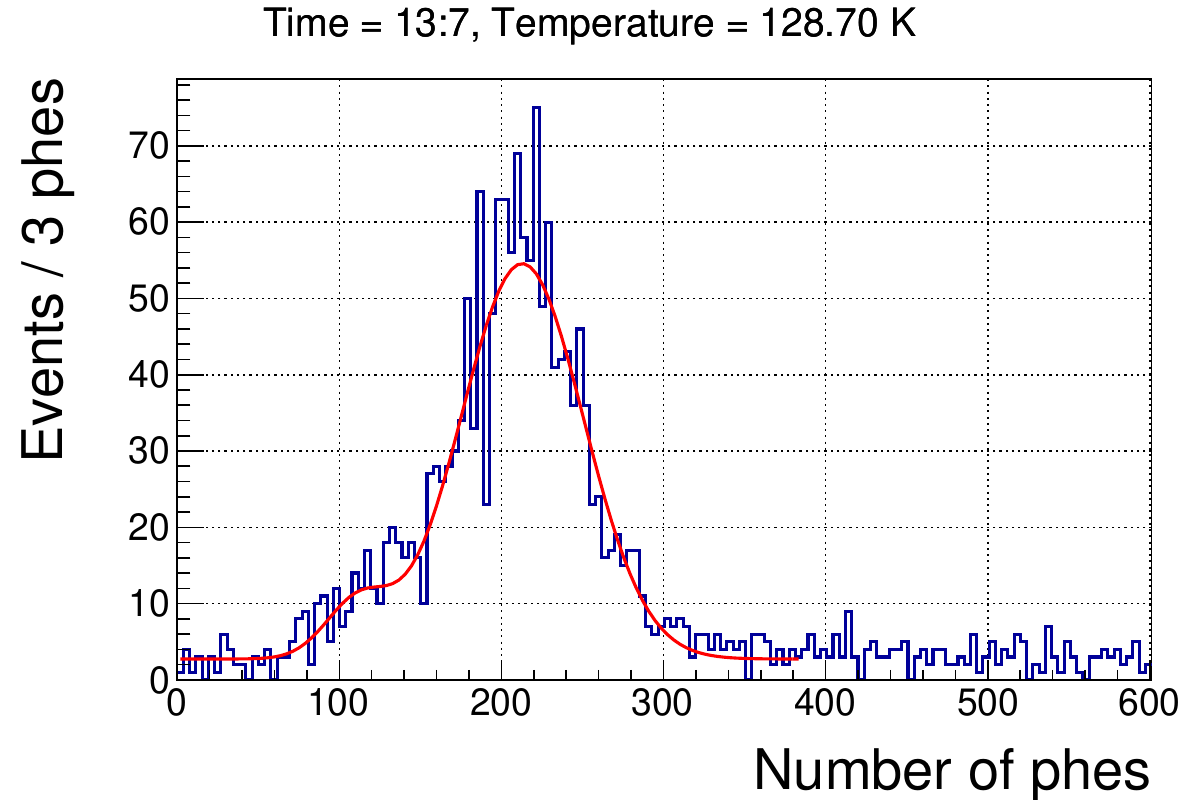}
		\end{subfigure}
		\begin{subfigure}[b]{0.3\textwidth}
			\includegraphics[width=\textwidth]{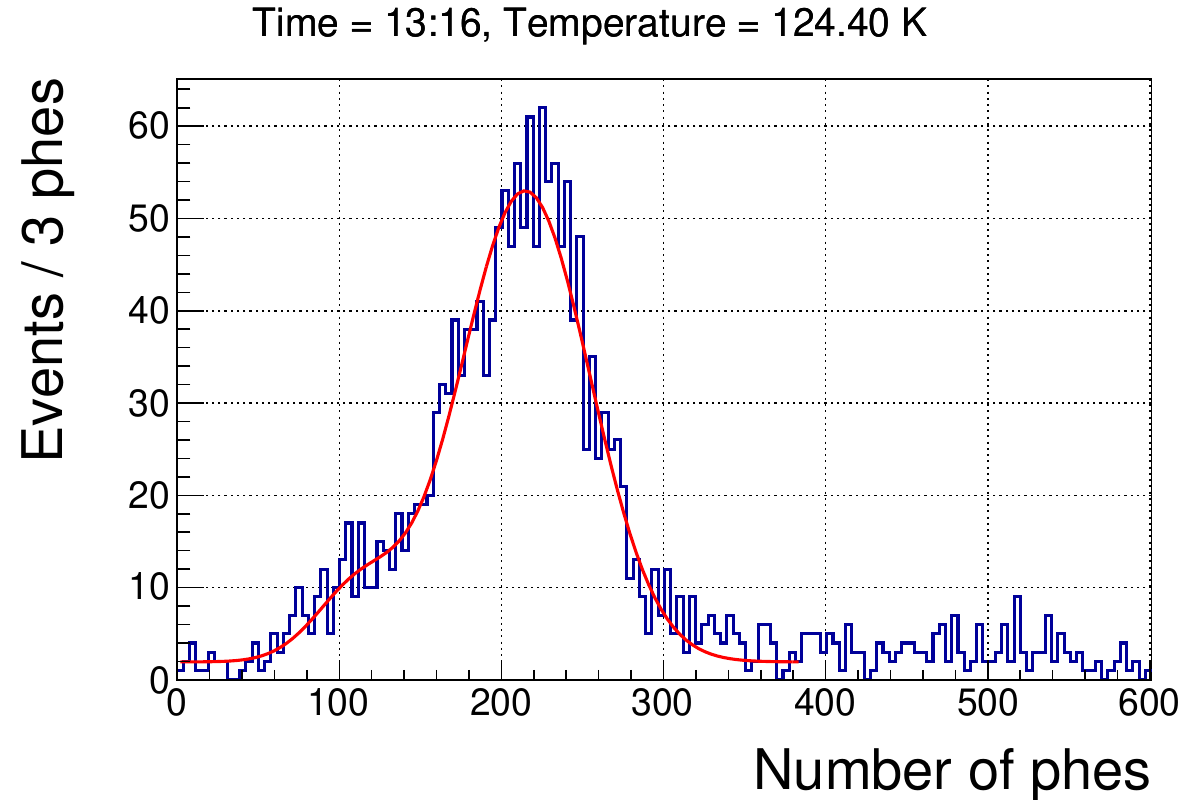}
		\end{subfigure}
		\begin{subfigure}[b]{0.3\textwidth}
			\includegraphics[width=\textwidth]{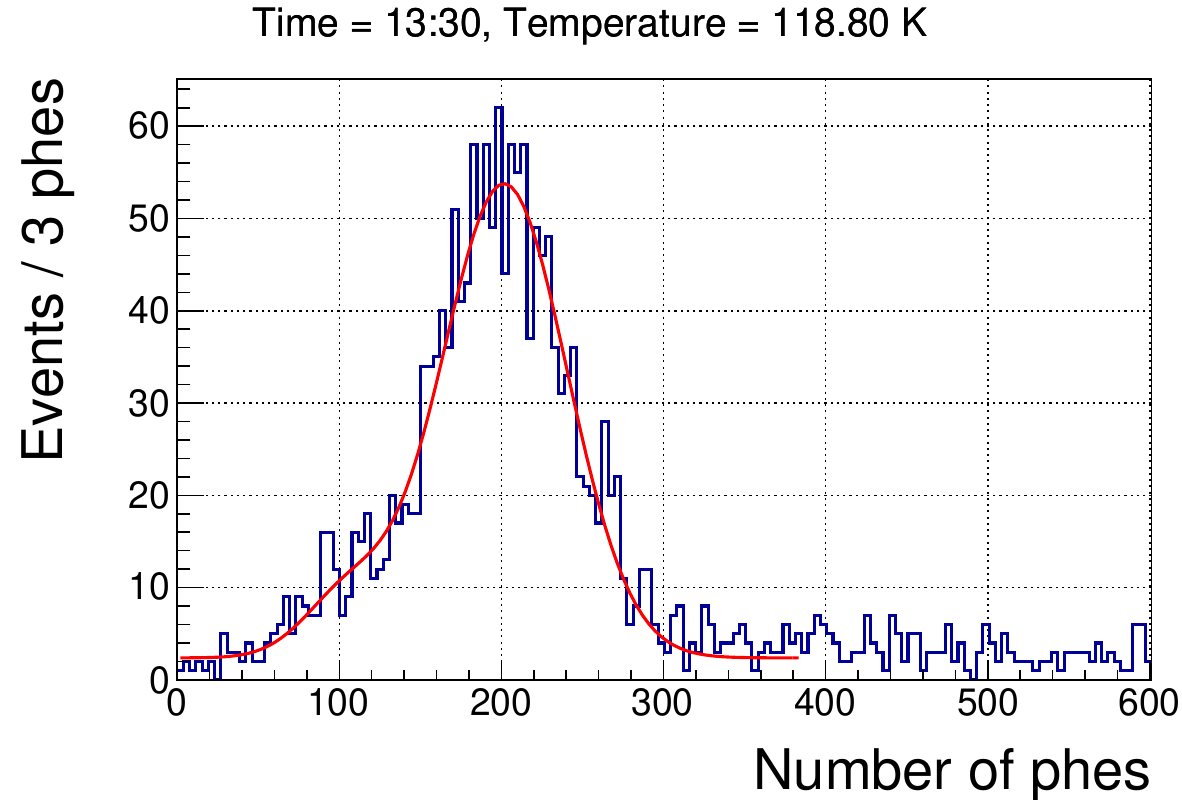}
		\end{subfigure}
		\begin{subfigure}[b]{0.3\textwidth}
			\includegraphics[width=\textwidth]{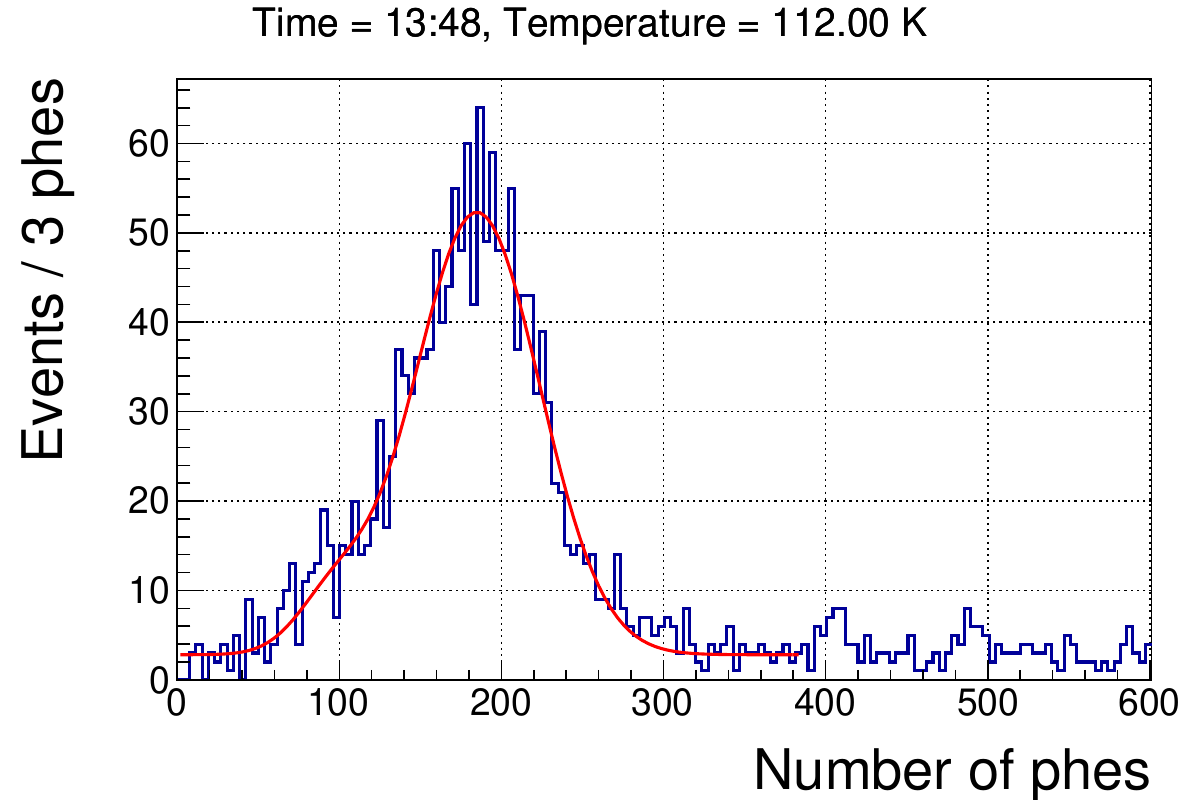}
		\end{subfigure}
		\begin{subfigure}[b]{0.3\textwidth}
			\includegraphics[width=\textwidth]{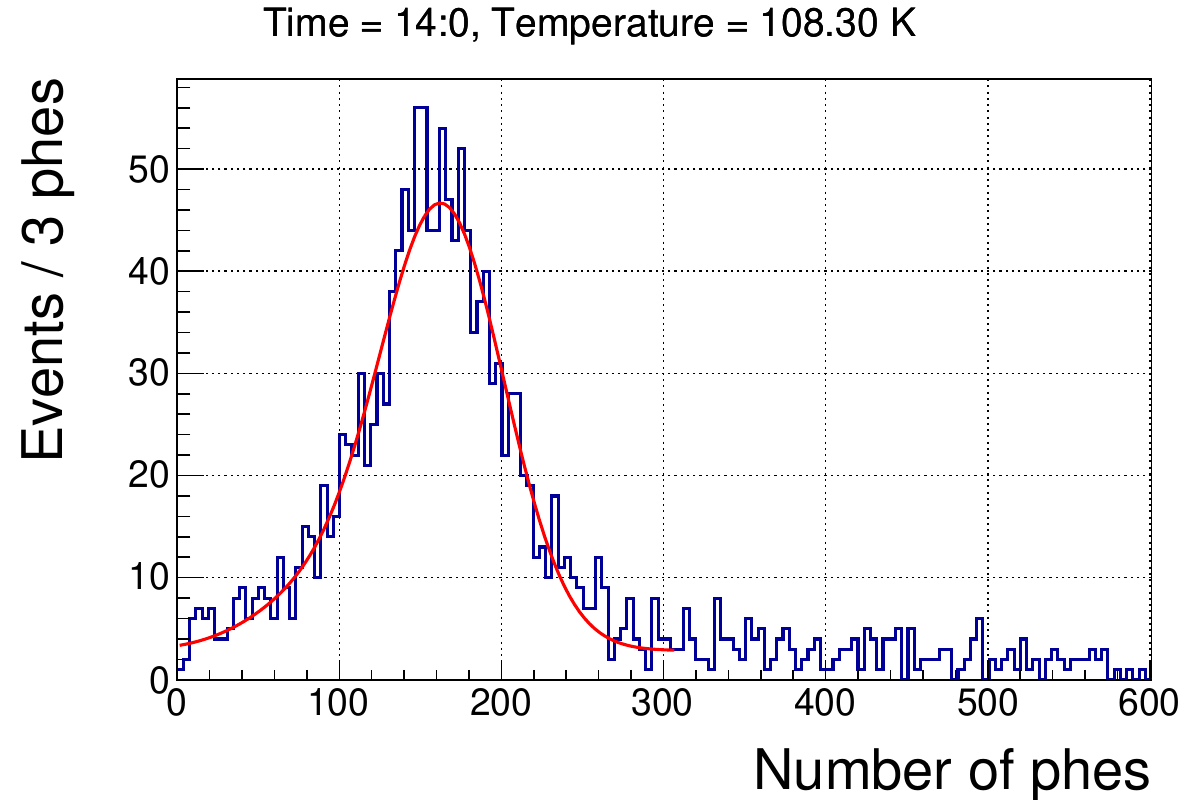}
		\end{subfigure}
		\begin{subfigure}[b]{0.3\textwidth}
			\includegraphics[width=\textwidth]{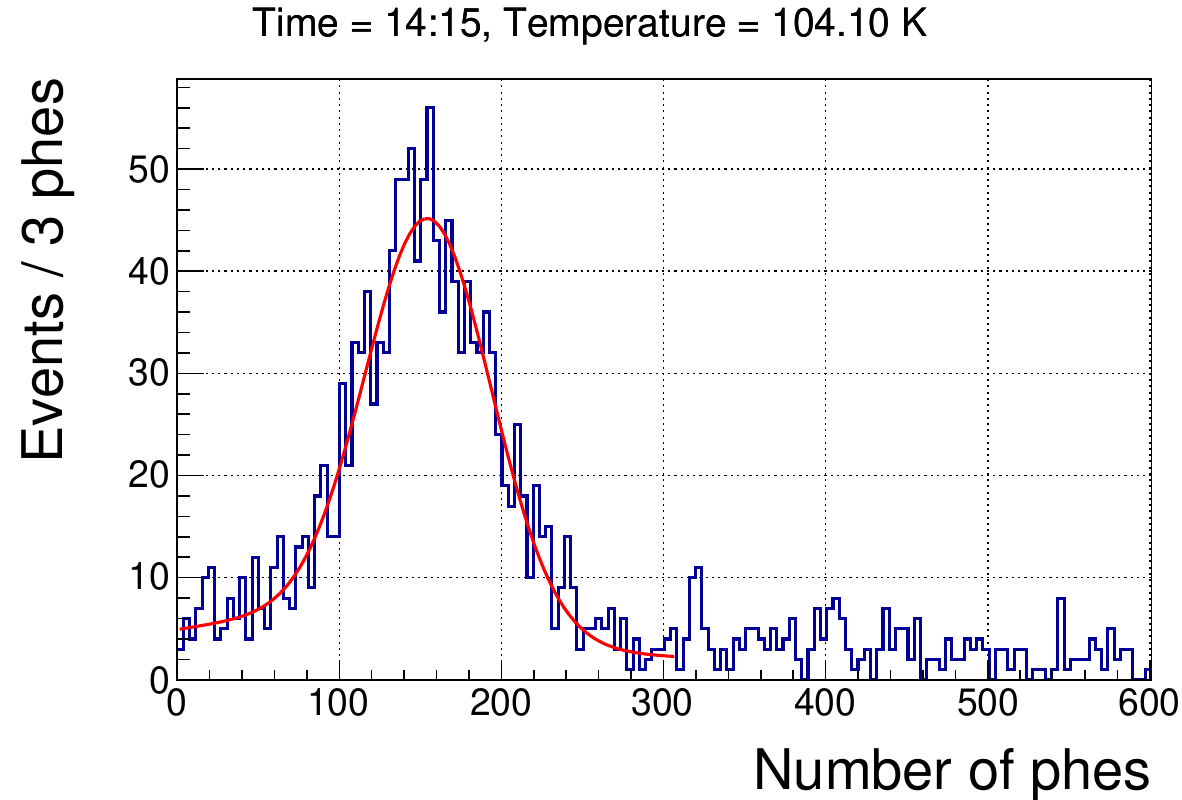}
		\end{subfigure}
		\begin{subfigure}[b]{0.3\textwidth}
			\includegraphics[width=\textwidth]{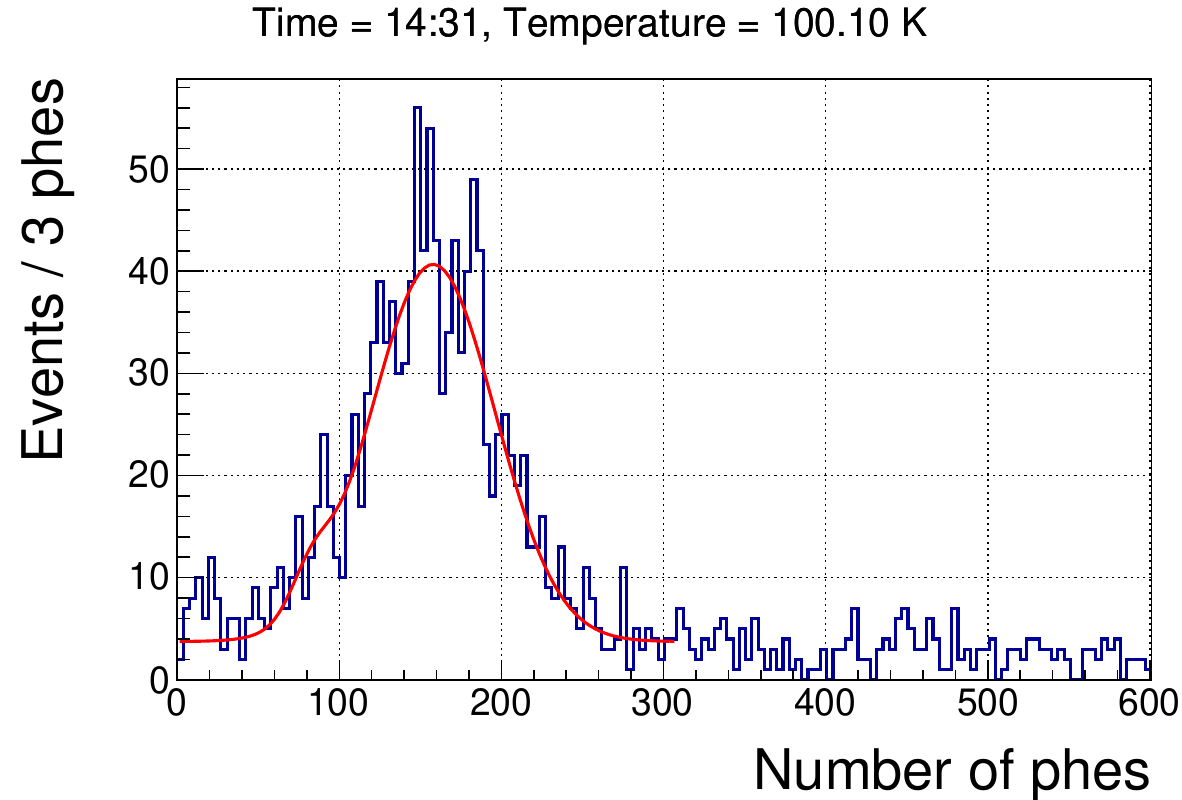}
		\end{subfigure}
		\begin{subfigure}[b]{0.3\textwidth}
			\includegraphics[width=\textwidth]{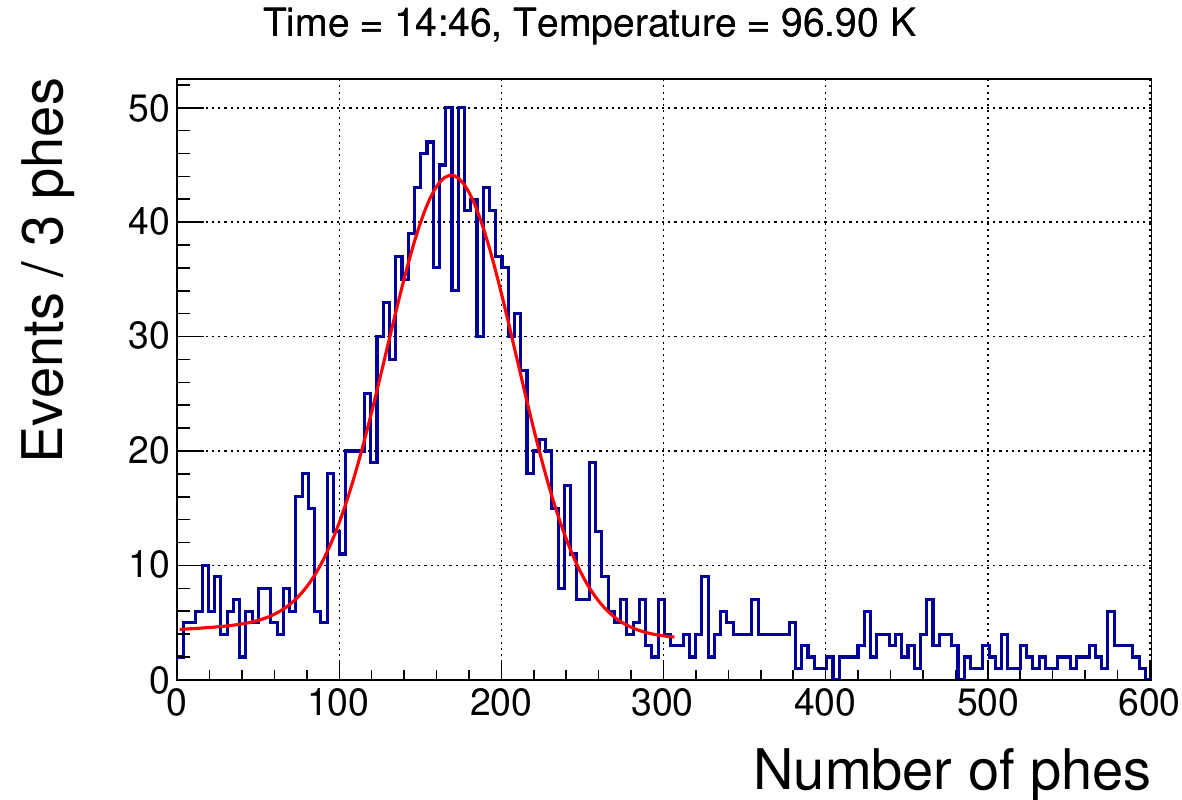}
		\end{subfigure}
		\begin{subfigure}[b]{0.3\textwidth}
			\includegraphics[width=\textwidth]{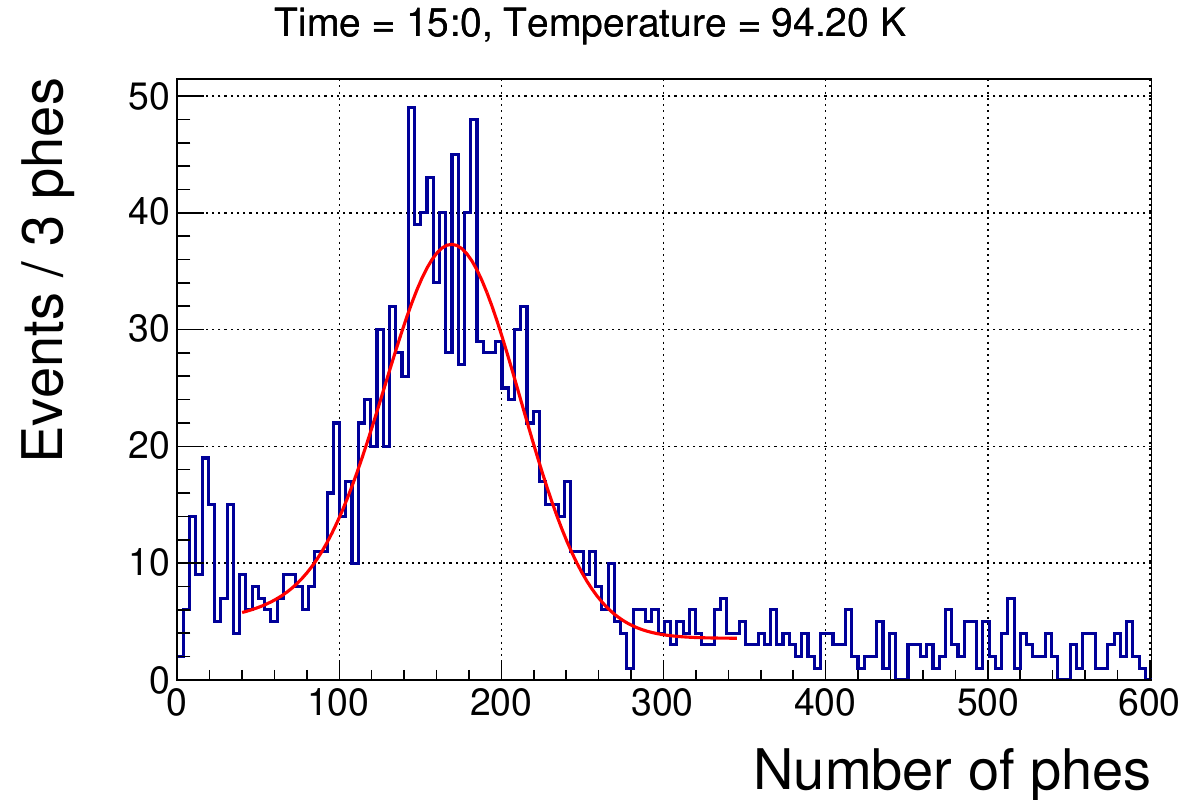}
		\end{subfigure}
		\begin{subfigure}[b]{0.3\textwidth}
			\includegraphics[width=\textwidth]{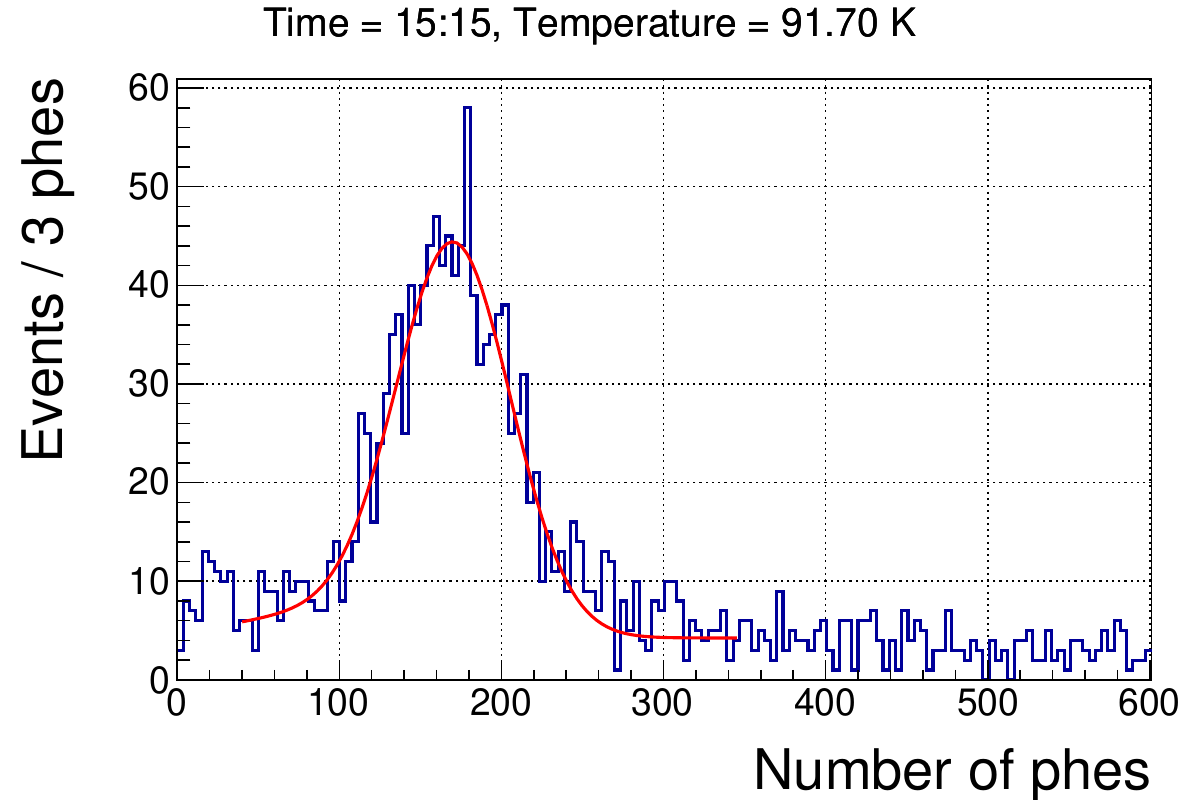}
		\end{subfigure}
		\caption{\label{FitsNaITl_Cooling}Fits of the $^{241}Am$ calibration spectra of the measurements while cooling down the NaI(Tl) crystal.}
	\end{center}
\end{figure}

The analysis of these data follows the procedure established in Section~\ref{Section:SiPMSTAR2_Analysis}: SPE area for each channel and temperature is estimated by fitting the first three peaks in the SPE calibration spectrum and used to translate the $^{241}Am$ calibration spectra into number of photoelectrons. The LC will be later estimated from the fitting of the $^{241}Am$ peaks. Figures~\ref{FitsNaI_Cooling} and~\ref{FitsNaITl_Cooling} show these fits for the measurements with NaI and NaI(Tl) crystals, respectively.

\begin{figure}[h!]
	\begin{center}
		\includegraphics[width=0.75\textwidth]{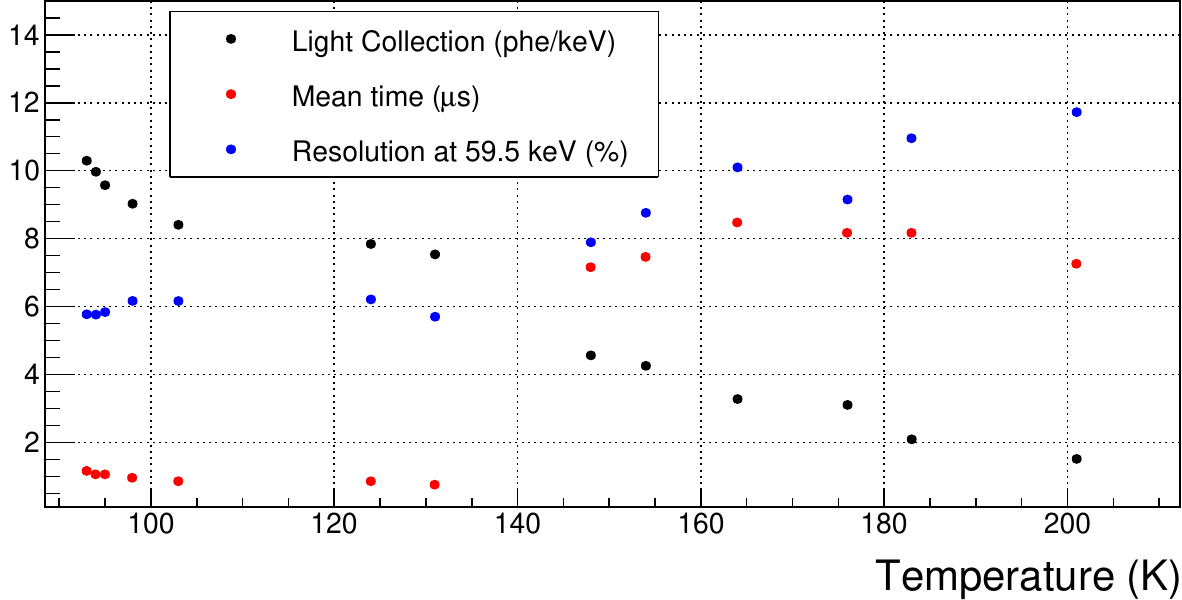}
		\caption{\label{LCMu(T)NaI}LC from the spectra fits of the NaI crystal shown in Figure~\ref{FitsNaI_Cooling}, mean times of the pulses and the resolution of the 59.5~keV peak as a function of the temperature.}
	\end{center}
\end{figure}

\begin{figure}[h!]
	\begin{center}
		\includegraphics[width=0.75\textwidth]{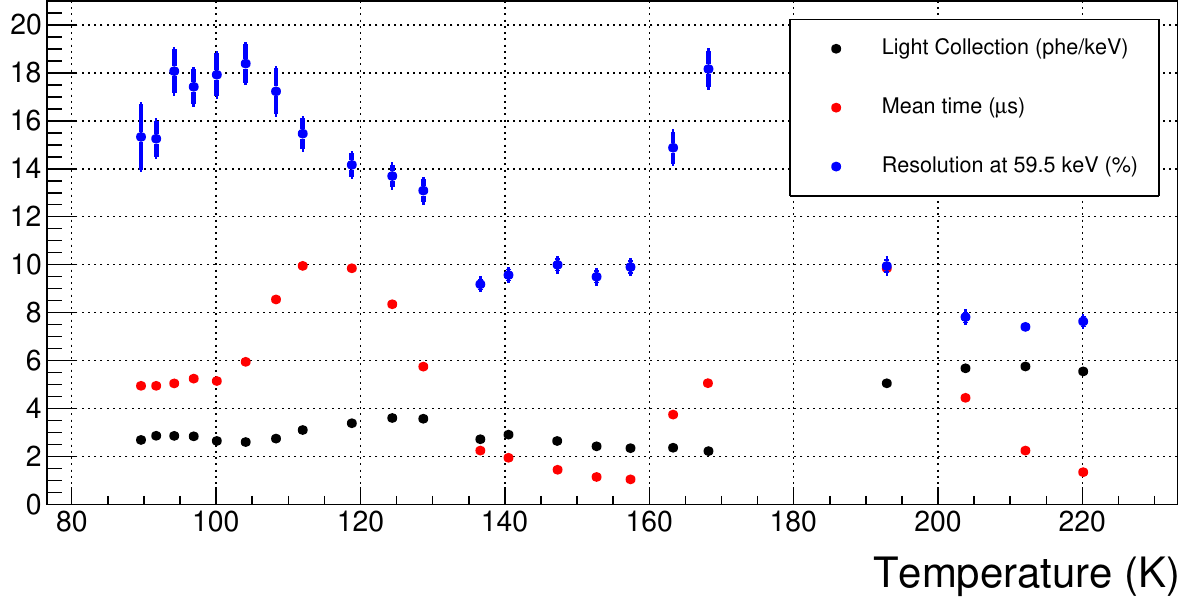}
		\caption{\label{LCMu(T)}LC from the spectra fits of the NaI(Tl) crystal shown in Figure~\ref{FitsNaITl_Cooling}, mean times of the pulses and the resolution of the 59.5~keV peak as a function of the temperature.}
	\end{center}
\end{figure}

Above 205~K, the LC in the NaI crystal was so low that the $^{241}Am$ peaks were not identified. Moreover, in the measurements with both crystals, the 59.5~keV photons were absorbed by the LAr when the chamber was almost full of this liquid, and therefore in those measurements the peak was not identified. Only those measurements where the 59.5~keV peak from $^{241}Am$ was identified have been used to estimate the LC and energy resolution. The mean time of the pulses corresponding to a given measurement, was calculated as the averaged value of this variable for 59.5~keV peak events, selected in a range between $\pm \sigma$ from the mean value in the pulse area distribution. The LC, the pulse mean times and the resolution as a function of the temperature are plotted in Figures~\ref{LCMu(T)NaI} and~\ref{LCMu(T)} for measurements with the NaI and NaI(Tl) crystals, respectively. Although the measurements were not taken in thermal equilibrium and the temperature of the crystal was not measured, Figures~\ref{LCMu(T)NaI} and~\ref{LCMu(T)} confirmed strong changes in the scintillation properties of both crystals with the temperature.

It is important to remark here the differences between LC and light yield. The first one can be affected by the change with temperature of the optical properties of all the other materials in the setup. Moreover, the observation of different time behaviour in the scintillation pulses points at different scintillation components which combined differently at each temperature, and that could be associated to different wavelengths in the emission. This would result in changes in the efficiencies for the detection of each component, according to the PDE of the SiPM arrays. Because of all of this, we will not draw any conclusion about the light yield in the following, but we will analyse the changes in the total LC of our system.

A constant increase of the LC was observed in the NaI crystal while the temperature decreased. The mean time of the scintillation pulses of this crystal remained stable (around 8~$\mu$s) during the first measurements above 120~K, but the pulses became suddenly much faster (around 1~$\mu$s) at temperatures close to that of LAr. Concerning the NaI(Tl) crystal, a decrease of the LC with respect to room temperature is observed down to 190~K, but below 170~K it remained stable at around 3~phe/keV. In this crystal, the mean time increased two times during the cooling down, showing a very unstable behaviour with the temperature, probably due to the switch between different scintillation components with different scintillation times, light yields and, possibly, wavelengths. The LC for this crystal at the lowest temperature measured is approximately three times lower than that of the NaI crystal, while the mean time of the pulses is around 4~times larger. Both properties indicate that the NaI crystal could be a much more suitable crystal for working at temperatures around 100~K.

\begin{figure}[h!]
	\begin{center}
		\includegraphics[width=0.75\textwidth]{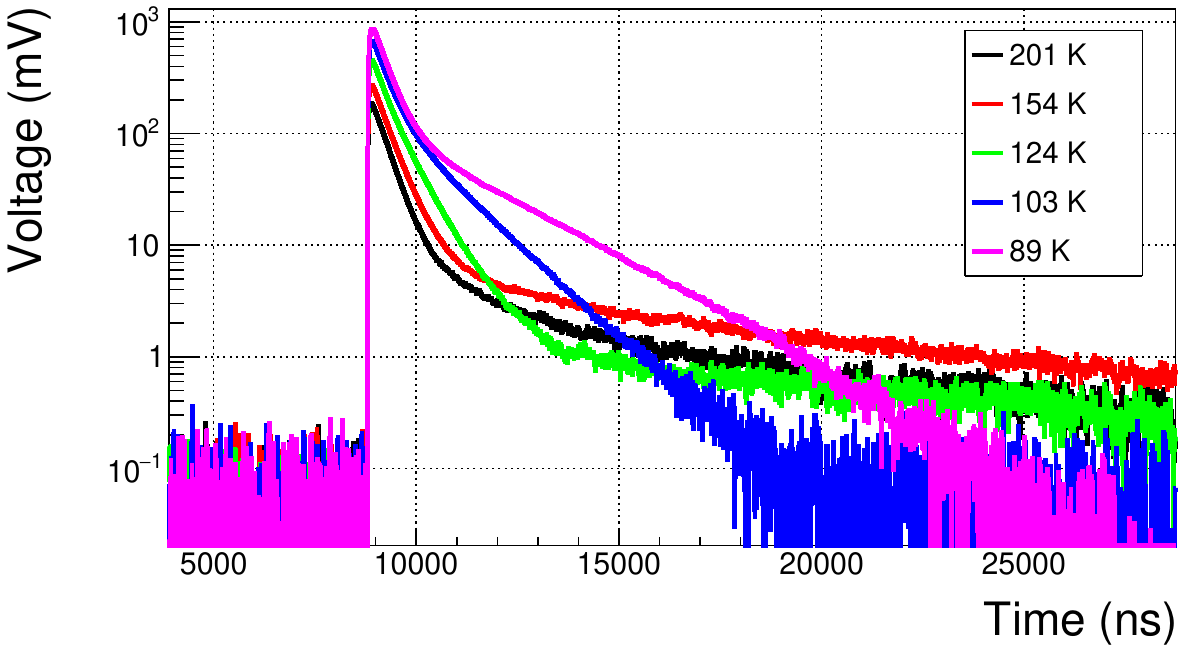}
		\caption{\label{PulsesNaI}Averaged pulses of 5 measurements at different temperatures with the NaI crystal selecting events in the 59.5~keV peak.}
	\end{center}
\end{figure}

\begin{figure}[h!]
	\begin{center}
		\includegraphics[width=0.75\textwidth]{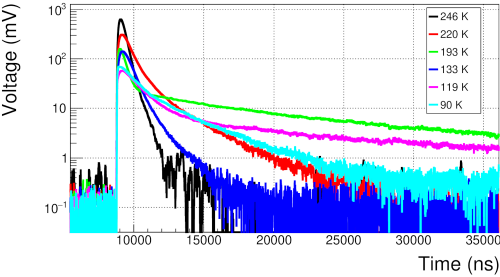}
		\caption{\label{PulsesNaI(Tl)}Averaged pulses of 5 measurements at different temperatures with the NaI(Tl) crystal selecting events in the 59.5~keV peak.}
	\end{center}
\end{figure}

Concerning the scintillation times, very clear changes were observed in the pulse shape while decreasing the temperature, as it is expected observing the mean times of the pulses presented in Figures~\ref{LCMu(T)NaI} and~\ref{LCMu(T)}. Examples of the different averaged pulses are shown in Figures~\ref{PulsesNaI} and~\ref{PulsesNaI(Tl)} for NaI and NaI(Tl) crystals, respectively.

It is possible to identify by eye in most of the measurements at least two scintillation components in the pulses. In the NaI(Tl) crystal the slow component has a low intensity at room temperature, but it increases at lower temperature, almost disappearing at 133~K, increasing again at 119~K and then decreasing again. Similar behaviour was measured in~\cite{Lee:2021jfx}, as it was shown in Figure~\ref{NaI(Tl)Scint}. It is worth to remind that these temperatures were not measured in the crystal. On the other hand, the NaI crystal presents an intense slow component at higher temperature, which decreases at lower temperatures. The steadily increase in LC while decreasing the temperature in NaI crystal is also clear in Figure~\ref{PulsesNaI}, while the contrary trend is observed for the NaI(Tl) crystal.

\begin{figure}[h!]
	\begin{center}
		\includegraphics[width=0.75\textwidth]{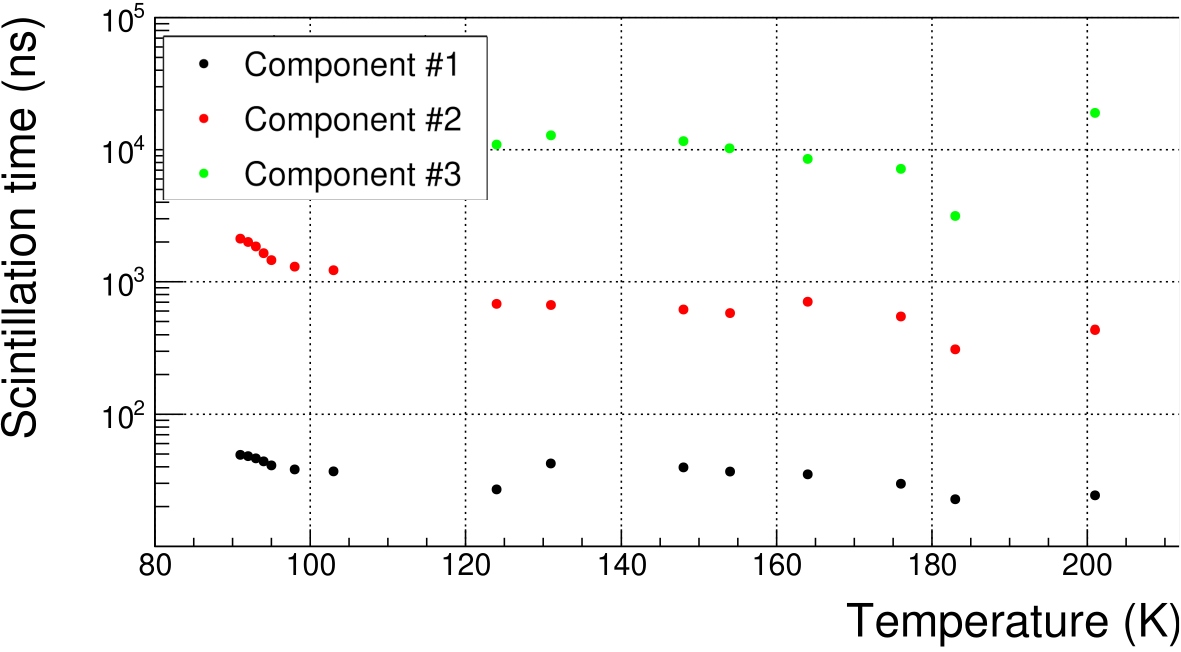}
		\caption{\label{NaICooling_ScintTimes}Scintillation time constants of each component obtained for all the measurements with the NaI crystal.}
	\end{center}
\end{figure}

\begin{figure}[h!]
	\begin{center}
		\includegraphics[width=0.75\textwidth]{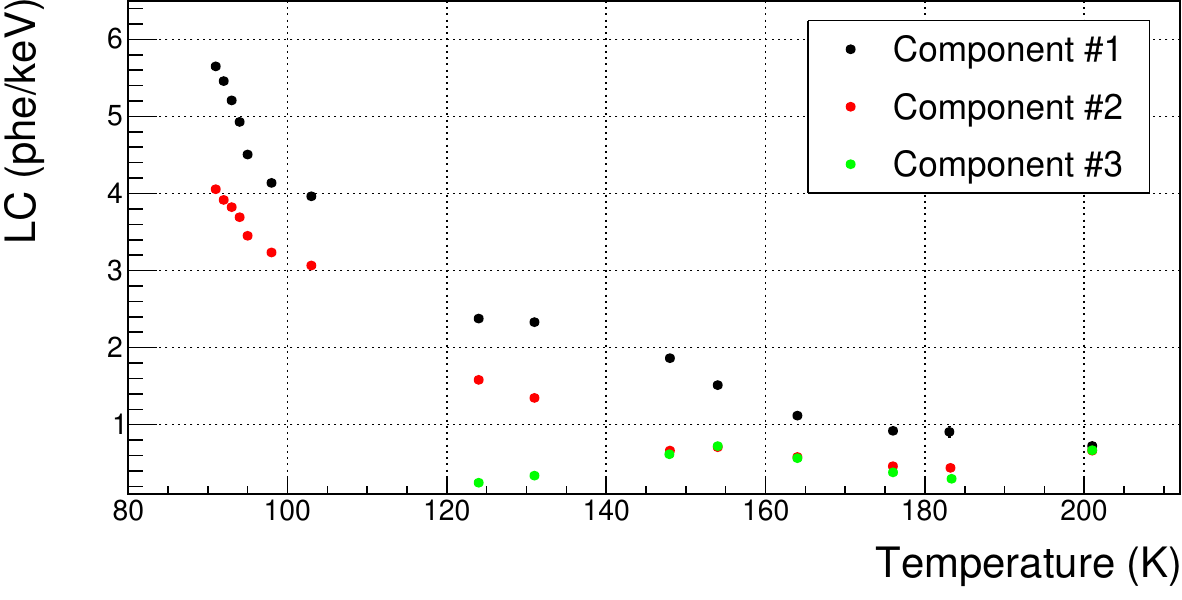}
		\caption{\label{NaICooling_LC}The contribution to the LC of each component obtained for all the measurements with the NaI crystal.}
	\end{center}
\end{figure}

All the pulses of the measurements where the 59.5~keV peak was identified were fitted as it was explained in Section~\ref{Section:SiPMSTAR2_Analysis_ScintTimes}. Figure~\ref{NaICooling_ScintTimes} and~\ref{NaICooling_LC} show the properties of the different scintillation components identified in the NaI crystal data: scintillation time constants and LC, respectively. In this crystal, three components were identified in the fits at temperatures above 120~K, and two components below that temperature. We will refer to them as fast, slow and very slow components. It can be observed in Figure~\ref{NaICooling_LC} how the very slow component LC starts to decrease below 150~K, being this trend compatible with the fact that only two components could be fitted below 120~K.

\begin{figure}[h!]
	\begin{center}
		\includegraphics[width=0.75\textwidth]{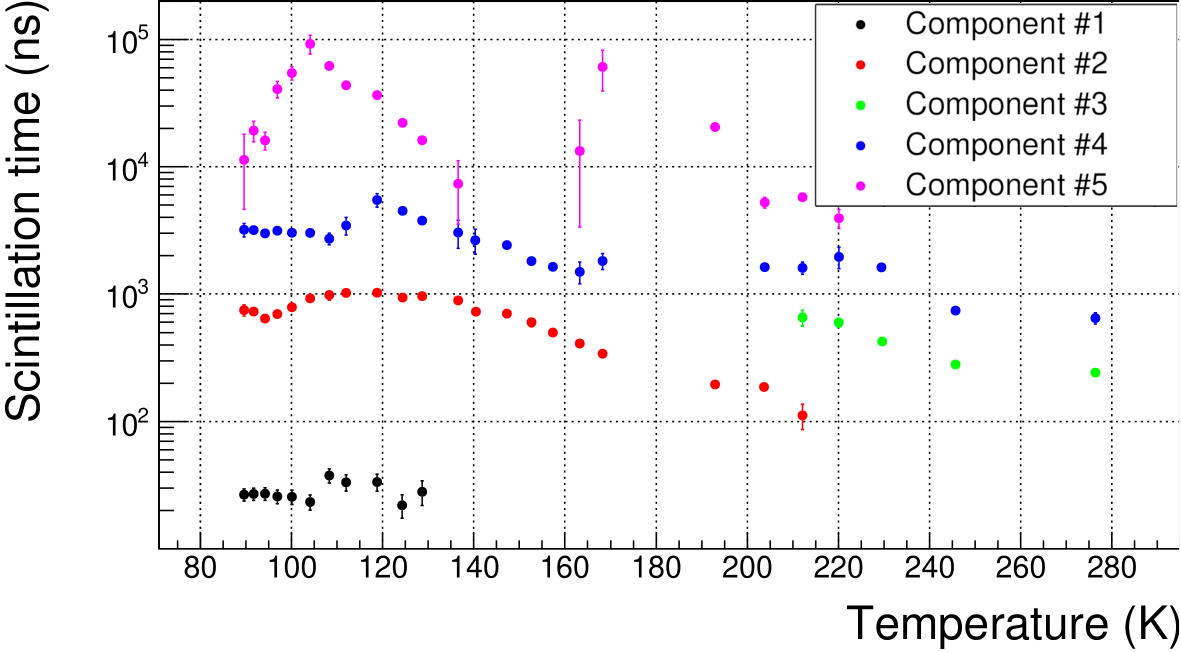}
		\caption{\label{NaITlCooling_ScintTimes}Scintillation time constants of each component obtained for all the measurements with the NaI(Tl) crystal.}
	\end{center}
\end{figure}

\begin{figure}[h!]
	\begin{center}
		\includegraphics[width=0.75\textwidth]{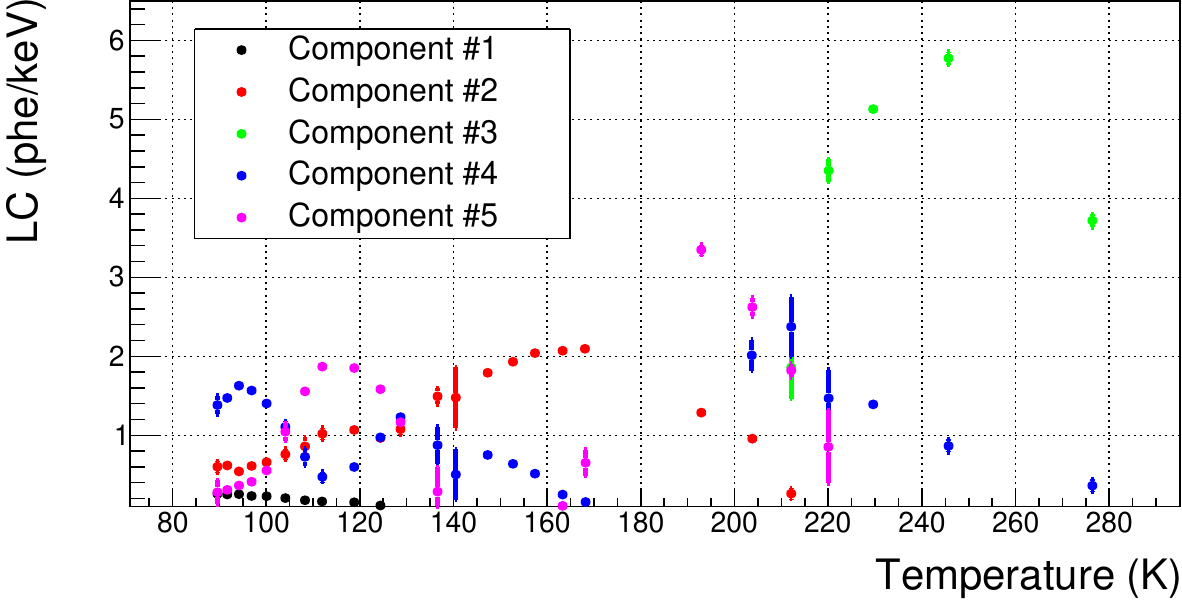}
		\caption{\label{NaITlCooling_LC}The contribution to the LC of each component obtained for all the measurements with the NaI(Tl) crystal.}
	\end{center}
\end{figure}

Figures~\ref{NaITlCooling_ScintTimes} and~\ref{NaITlCooling_LC} show the equivalent information for the measurements with the NaI(Tl) crystal. In this case, the shape of the pulse suffers from very drastic changes while the temperature decreases, as it was already shown in Figure~\ref{PulsesNaI(Tl)}. It implies the appearance and disappearance of different components along the measurements. The number of scintillation components considered has been different for each fit.  The component identification has been done considering as the same emission the one obtained with similar scintillation times when the number of components considered in the fit was changed. When this identification is done, it is possible to observe a tendency in the scintillation time constants of some components that remain quite stable independently on the number of components considered in the fit. One example is the measurement at 165~K with three components and that at 155~K with two components: these two components evolve with the temperature similarly and in a smooth way. This procedure allowed to identify five different components depending on their scintillation times.

It is worth remarking that the pulse fitting procedure followed has important limitations. In the measurements with NaI it was observed a clear undershoot for high amplitude pulses which is probably associated with saturation in some step of the readout electronics, as already commented. As all the pulses may be affected, both in the calculation of the pulse area, but also in the apparent scintillation times, the corresponding systematic contribution in our results can be important. It has been impossible to quantify, although it should affect more the highest total LC configurations/measurements and the highest externally amplified measurements. We can conclude, independently from this possible systematic, that the scintillation in NaI(Tl) depends strongly on the temperature, and it has several light emissions contributing from room temperature down to LAr temperature. On the other hand, NaI shows a smoother behaviour and a clear increase in LC when decreasing the temperature, being the best option to continue developing prototypes for working at the 100~K temperature range.

However, more work on understanding the readout electronics effect on the SiPM signal has to be carried out in order to work in high LC configurations, and of course, temperature control of the crystal and setup is mandatory in order to guarantee enough stability in the scintillation properties.

\section{Conclusions} \label{Section:SiPMSTAR2_Conclusions}
\fancyhead[RO]{\emph{\thesection. \nameref{Section:SiPMSTAR2_Conclusions}}}

The scintillation of the NaI and NaI(Tl) crystals has been compared in terms of LC and scintillation time behaviour at different temperatures down to the LAr temperature (87~K) in the same setup and LC system. The measurements with both crystals were carried out in the same conditions (both covered in plastic) and therefore their results are comparable. They confirm that at low temperature the pulses of the NaI crystal are much faster than those of the NaI(Tl) and that the LC with this crystal is three times larger. Both properties make the NaI crystal more interesting for working at these temperatures.

In order to improve the results presented in this chapter, for future work, SiPMs with faster response could be used, and better understanding of the saturation or band pass in the electronics could help to reduce pulse-shape dependencies on the overvoltage. The results obtained in this study have been useful to improve the design of detector prototypes using SiPMs and NaI or NaI(Tl). Future detectors must include a system (preferably an optical fiber) to perform the SPE calibration, either using a pulsed laser or a LED. To optimize the LC, they must include a reflecting surface around the scintillator and the number of components between the crystal and the SiPMs must be reduced. Moreover, the detector should allow the production of scintillation by placing a radioactive source close to the crystal, with the possibility of removing it to measure the background. Finally, this detector must be characterized in stable conditions of temperature, and to achieve this, the temperature of the crystal must be precisely monitored and controlled.

The analysis followed in these measurements has also allowed to establish a protocol for the characterization of the crystals. This procedure will be followed with the first NaI(Tl)+SiPM detector designed in the University of Zaragoza, as it is explained in the next Chapter.

%% file: SiPMsZgzNew2.tex
\chapter{Design of a NaI(Tl) detector prototype with SiPMs} \label{Chapter:SiPMZgz}

\fancyhead[LE]{\emph{Chapter \thechapter. \nameref{Chapter:SiPMZgz}}}

With the purpose of replacing the light readout of the ANAIS-112 experiment based on Photomultiplier Tubes (PMTs) by Silicon Photomultipliers (SiPMs), a R$\&$D program started in 2018. The first stage of this program aimed at the SiPM characterization in terms of the SPE calibration, pulse shape and DC rate, as well as designing convenient readout protocols, paving the way to the second stage: the application of SiPMs to the light readout of NaI(Tl) crystals and the design of a compact prototype.

A first batch of SiPMs was purchased to SensL (now Onsemi~\cite{ONSEMI}), and different PCBs were designed to connect them to a preamplifier. However, it was found that home-made PCBs induced a significant noise in the measurements, and no relevant results were obtained, being impossible to carry out their SPE calibration or DC rate characterization. Then, a single SiPM and an specific front-end board for the read-out were acquired from HAMAMATSU~\cite{SiPMHamamatsuManual}. In parallel, we started a collaboration within the Global Argon Dark Matter (GADM) program to characterize four DArTeyes~\cite{Acerbi:2016ikf,DIncecco:2017qta}, SiPMs designed to be used in the DArT experiment~\cite{DarkSide-20k:2020qfz} at the Canfranc Underground Laboratory. A DArTeye consists of a single 11.9$\times$7.8~mm$^2$ SiPM mounted in a board, specifically designed to work in low noise and background conditions and immersed in liquid argon, i.e. at $\sim$~87~K. The characteristics of these devices are presented in Section~\ref{Section:SiPMZgz_Setup_SiPM}.

SiPM characterization was done both in darkness and exposed to low intensity light signals. As light source, we used a controllable brightness LED that emits fast pulses at regular intervals in a simple but dedicated experimental setup (described in Section~\ref{Section:SiPMZgz_Setup}). The analysis applied to the data is explained in Section~\ref{Section:SiPMZgz_Analysis}. HAMAMATSU SiPMs and DArTeyes were characterized in terms of pulse shape, amplitude and area of the SPE and DC rate both in darkness and using the LED. In addition, the HAMAMATSU SiPM was applied to the readout of the scintillation light produced by a small NaI(Tl) crystal under particle excitation. The LC of the system, the resolution in energy and the scintillation time constant were determined at room temperature. All these measurements are summarized in Section~\ref{Section:SiPMZgz_CryoT}.

The next step was to design a compact NaI(Tl)+SiPM detector prototype (explained in Section~\ref{Section:SiPMZgz_Detector}) allowing the characterization at low temperature. This prototype consists of a 4$\times$4 array of SiPMs coupled to a cubic NaI(Tl) crystal of 1" side and covered by a reflective teflon tape. Both the SiPM array and the crystal are enclosed in a copper housing tightly closed and kept under dry atmosphere conditions. The housing design includes an optical fiber allowing to characterize the SPE using the light coming from an external LED. An eMUSIC board~\cite{eMUSICboardManual}, designed by SCIENTIFICA~\cite{SCIENTIFICA}, was used to control the power supply and process the SiPM output signals, in a versatile and simple way. Finally, an overview of the future work with this detector and future NaI(Tl)+SiPM detectors for the possible upgrade of the ANAIS experiment is presented in Section~\ref{Section:SiPMZgz_Conclusions}.

\section{SiPMs and Front-End electronics} \label{Section:SiPMZgz_Setup_SiPM}
\fancyhead[RO]{\emph{\thesection. \nameref{Section:SiPMZgz_Setup_SiPM}}}

Different SiPMs were acquired for their characterization. All of them have the maximum of the PDE at the wavelengths of the NaI(Tl) scintillation maximum light output at room temperature, peaked at 420~nm (see Section~\ref{Section:Intro_Scintillators_NaI(Tl)} and Figure~\ref{NaIEmission}). For the SiPM bias and readout, convenient Front-End Boards (FEB) were also required. Specific details of the two SiPMs used and their FEBs are given next.

\begin{figure}[h!]
\begin{center}
\includegraphics[width=\textwidth]{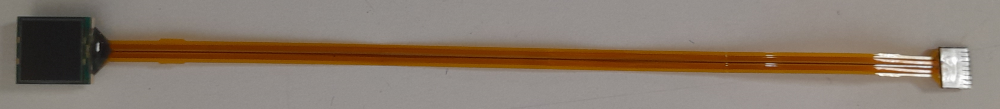}
\caption{\label{PictureS13360}Picture of the S13364~MPPC (to the left), the wire and the connector to the FEB.}
\end{center}
\end{figure}

The Multi-Pixel Photon Counter (MPPC) model S13364, from HAMAMATSU, integrates a 6$\times$6~mm$^2$ SiPM S13360-6050PE~\cite{SiPMHamamatsuManual} and a LM94021 Multi-Gain Analog Temperature Sensor~\cite{TemperatureSensorHamamatsu}. From now on, we will refer to this HAMAMATSU SiPM as MPPC. It is designed to be connected to a FEB also from HAMAMATSU model C13367-0459~\cite{MPPCModule}, which incorporates a power supply, a temperature compensation circuit and a preamplifier for the signal. Figure~\ref{PictureS13360} shows the S13364~MPPC and the connector. 

\begin{figure}[h!]
\begin{center}
\includegraphics[width=\textwidth]{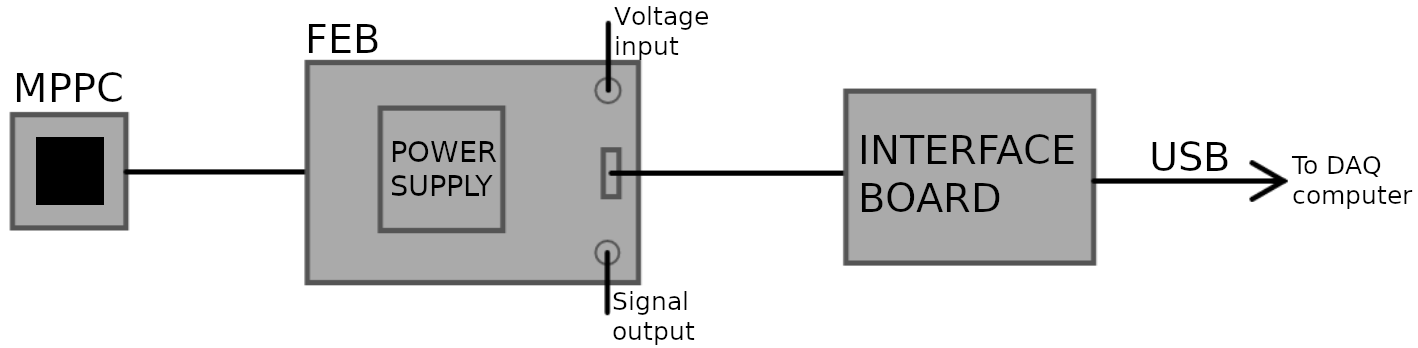}
\caption{\label{MPPCModuleScheme}Scheme of the connections between the different components of the system: the MPPC, the FEB, the interface board and the computer.}
\end{center}
\end{figure}

An external power supply provides an input voltage of $\pm$~5~V to the FEB. The internal power supply of the FEB is used to bias the SiPM (set using a control software at values between 40 and 90~V) and the preamplifier. An interface board connects the MPPC and the computer, allowing to read the current flowing through the SiPM, the temperature measured by the sensor, and switch on an "stable operation model", which modifies the SiPM bias as a function of the temperature in order to guarantee an stable overvoltage value. Although the latter is a very interesting option, it has not been used in the measurements shown in this chapter because the measurements were taken in stable temperature conditions. An scheme of the connections between the SiPM, the FEB, and the interface board can be seen in Figure~\ref{MPPCModuleScheme}.

\begin{figure}[h!]
\begin{center}
\includegraphics[width=0.5\textwidth]{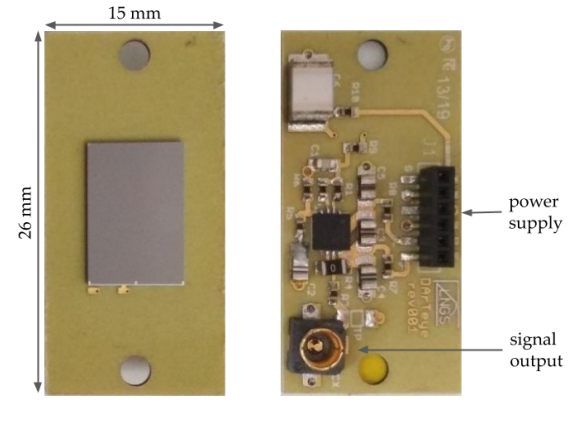}
\caption{\label{DArTeye}Front and back sides of the DArTeye module. Image from~\cite{ThesisEdgar}.}
\end{center}
\end{figure}

The DArTeye modules were specifically designed to maximise the radiopurity and simplify the connections for operation immersed in liquid argon. They integrate a SiPM and the readout electronics in a 15$\times$26~mm$^2$ board. The SiPMs used in the DArTeyes have the same design and characteristics as those used in Chapter~\ref{Chapter:SiPMStar2} (see Section~\ref{Section:SiPMSTAR2_Detector} for their specifications). A picture of this DArTeye module is shown in Figure~\ref{DArTeye}. The integrated preamplifier requires an input voltage supply of $\pm$~2.5~V, and the SiPM bias has to be provided by an external source. Four DArTeye modules (labeled as 2, 7, 9 and 10) were received at the University of Zaragoza for their characterization. Only three of them were finally characterized because the preamplifier of the $\#$~2 was damaged before the measurements at liquid nitrogen temperature.

\section{Experimental setup} \label{Section:SiPMZgz_Setup}
\fancyhead[RO]{\emph{\thesection. \nameref{Section:SiPMZgz_Setup}}}

This section describes first the two different test-benches designed to characterize the SiPMs to later explain the common elements of the setup and the DAQ system used in the measurements.

\subsection{Test benches for SiPM characterization} \label{Section:SiPMZgz_Setup_TestBenches}

Two different test benches were designed for the characterization of the MPPC and the DArTeyes.

\begin{figure}[h!]
\begin{center}
\includegraphics[width=\textwidth]{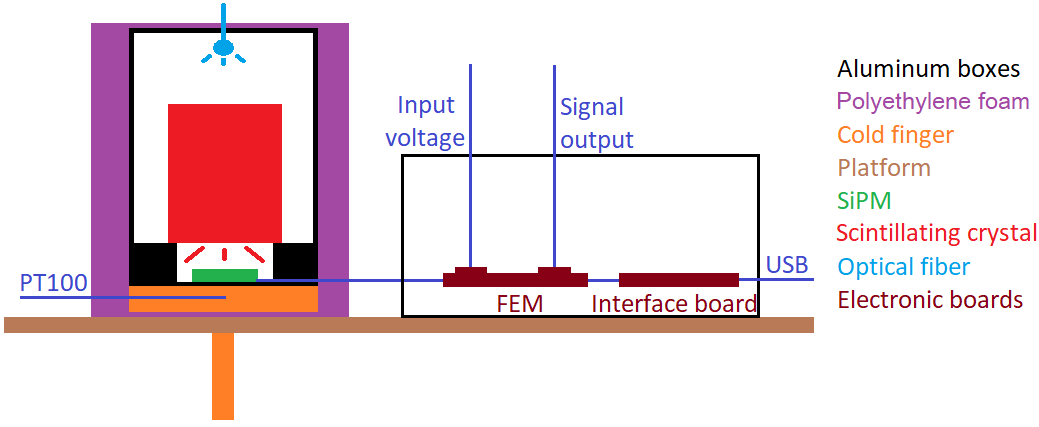}
\caption{\label{TestBenchHamamatsu}Scheme of the test bench used in the measurements with the MPPC.}
\end{center}
\end{figure}

For the MPPC characterization, a cylindrical aluminum box was designed to be light-tight and hold the SiPM in a position where it could be illuminated either by the optical fiber or by the scintillation of a NaI(Tl) crystal, which fitted inside the box. This box had a diameter of 6~cm, and was 6.3~cm long. The inner surface of the box was black painted to absorb possible external light. Both, the FEB and the interface board were installed inside a different 6$\times$9$\times$11.5~cm$^3$ aluminum box. Figure~\ref{TestBenchHamamatsu} shows an schematic representation of the system.

\begin{figure}[h!]
\begin{center}
\begin{subfigure}[b]{0.57\textwidth}
\includegraphics[width=\textwidth]{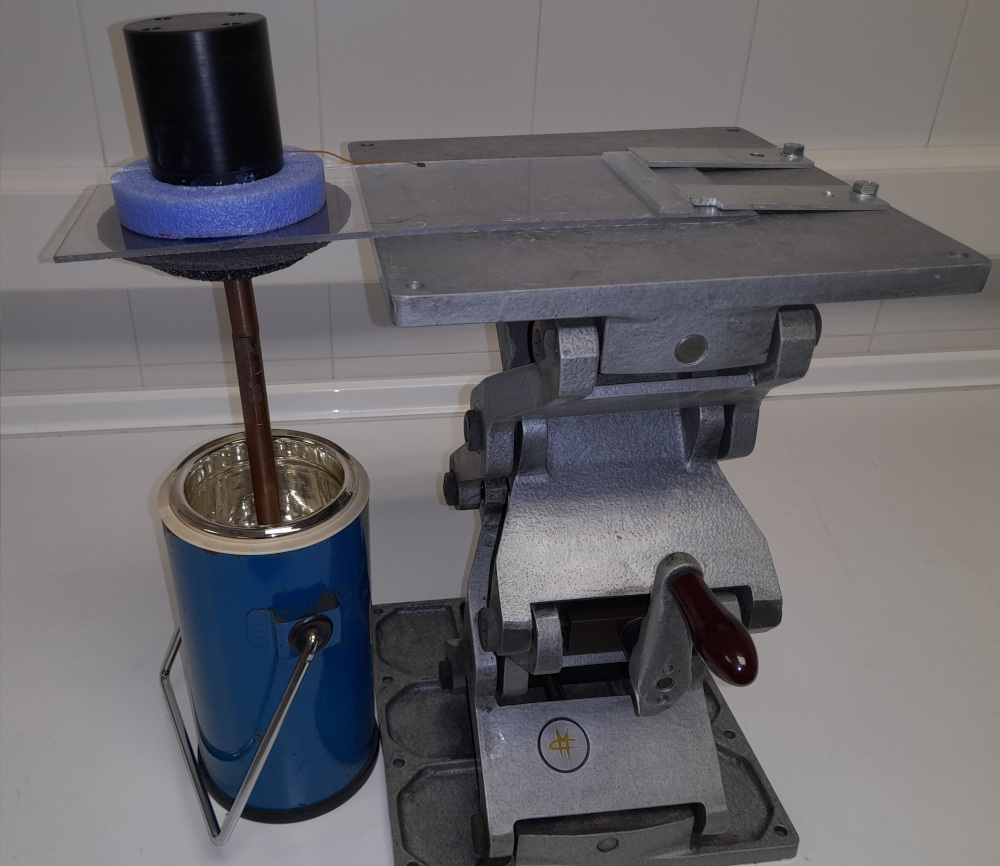}
\end{subfigure}
\begin{subfigure}[b]{0.42\textwidth}
\includegraphics[width=\textwidth]{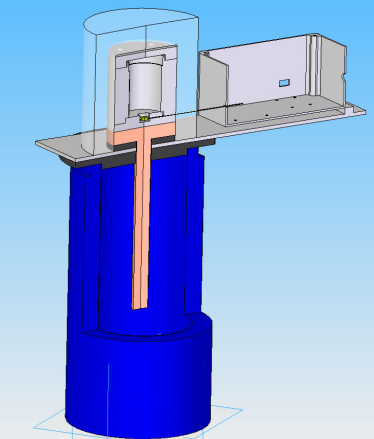}
\end{subfigure}
\caption{\label{HamamasuColdFinger}Picture (left) and scheme (right) of the cooling system for characterizing the MPPC down to 90~K.}
\end{center}
\end{figure}

To cool down the system, the cylindrical box was brought into contact with a copper cold finger that could be introduced in a dewar filled with liquid nitrogen. For this purpose, all the system (shown in Figure~\ref{TestBenchHamamatsu}) was placed on a platform that allowed to control the position of the cold finger inside the dewar, as it can be seen in Figure~\ref{HamamasuColdFinger}. To cross-check the temperature of the system, a PT100 resistor was placed in the cold finger, close to the aluminum box. To thermally isolate the system, the aluminum box was covered by a ring made of polyethylene foam. This system reached a temperature of 90~K.

\begin{figure}[h!]
\begin{center}
\includegraphics[width=0.75\textwidth]{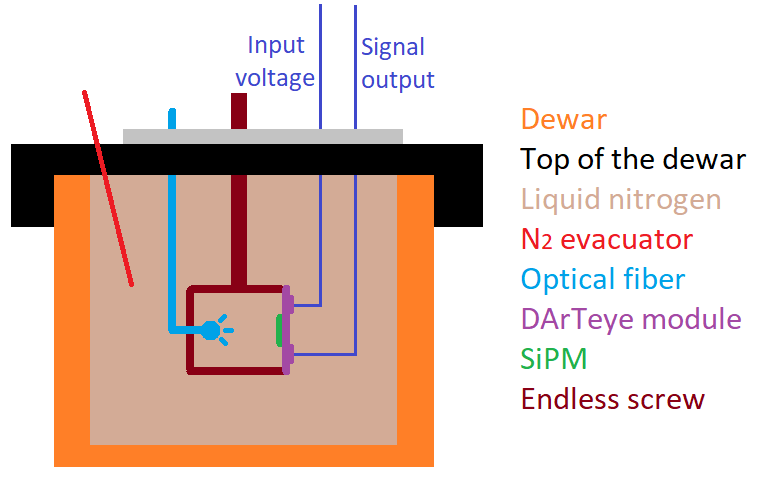}
\caption{\label{TestBenchDArT}Scheme of the test bench used in the measurements with the DArTeyes.}
\end{center}
\end{figure}

The test bench for the measurements with the DArTeyes was designed to introduce the modules directly on a dewar filled with liquid nitrogen. The DArTeye module was placed inside a plastic structure which holds an optical fiber for illuminating the SiPM and it was screwed to a metallic plate fixed to the top of the dewar. This metallic plate contained all the connectors for input voltages, output signal and the optical fiber. The complete test bench is shown in Figure~\ref{TestBenchDArT}.

\subsection{Common elements of the setup} \label{Section:SiPMZgz_Setup_Others}

The power supply used to bias the DArTeye was a Keithley~2200-72-1 programmable DC power supply~\cite{Keithley2200Manual,KeithleyWebPage}, which could be controlled from a computer. It can supply up to 72~V with a resolution of 1~mV. Between this power supply and the DArTeye module we connected an ammeter to measure I-V curves of the SiPMs and monitor the correct performance. A Keithley~2000-900-01 Multimeter~\cite{Keithley2000Manual} with a resolution of 10~nA was used for that purpose. To fed the preamplifiers of both the DArTeye and the MPPC, a TENMA~72-8000 DC power supply~\cite{TENMAManual} was used. This source also allows to monitor the behaviour of the system by providing access to the output current.

\begin{figure}[h!]
\begin{center}
\includegraphics[width=0.75\textwidth]{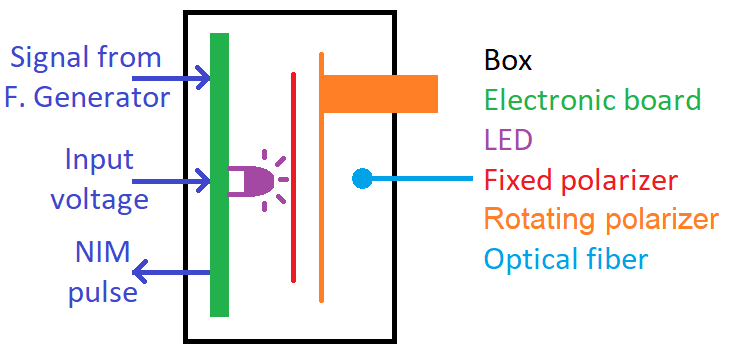}
\caption{\label{PicureLED}Scheme of the LED system used for the measurements.}
\end{center}
\end{figure}

The LED used was a HERO High Power Ultraviolet L400-5x0B LED~\cite{LEDManual}, and it was selected because its emission maximum is at 400~nm (with a bandwidth of 20~nm) close to the PDE maximum of the SiPMs and to the NaI(Tl) scintillation emission peak. This LED was placed inside a 12$\times$6.5$\times$4~cm$^3$ box facing an optical fiber. The amount of light that reached the optical fiber was controlled with two polarizing filters that could be rotated one respect to the other. The LED pulsing was set by a function generator Tektronix~AFG3102~\cite{GeneratorManual,TektronixWebPage} using a square function with a frequency of 1~kHz and width of 5~ns. The electronic board where the LED was connected required an input voltage and provided a NIM pulse synchronized with the signal from the function generator for triggering purpose. An scheme of the LED system is shown in Figure~\ref{PicureLED}.

\subsection{DAQ system} \label{Section:SiPMZgz_Setup_DAQ}

The DAQ system used for the readout of both, the MPPC and the DArTeye is based on a 4~channels 12-bit MATACQ32~board~\cite{MatacqManual}. Two different configurations of the electronic chain and trigger were implemented depending on the measurement purpose.

The MATACQ32 digitizer, connected in a CAEN VME~8100 Powered Crate~\cite{VMECrateManual}, is linked to the DAQ computer with an optical fiber using a VME~2718 Optical Link Bridge~\cite{VMEBridgeManual}. It has a dynamic range of 1V, sampling rate of 2~GS/s and records 1.26~$\mu$s (see Section~\ref{Section:ANAIS_DAQ}). It was configured to acquire negative pulses.

\begin{figure}[h!]
\begin{center}
\includegraphics[width=\textwidth]{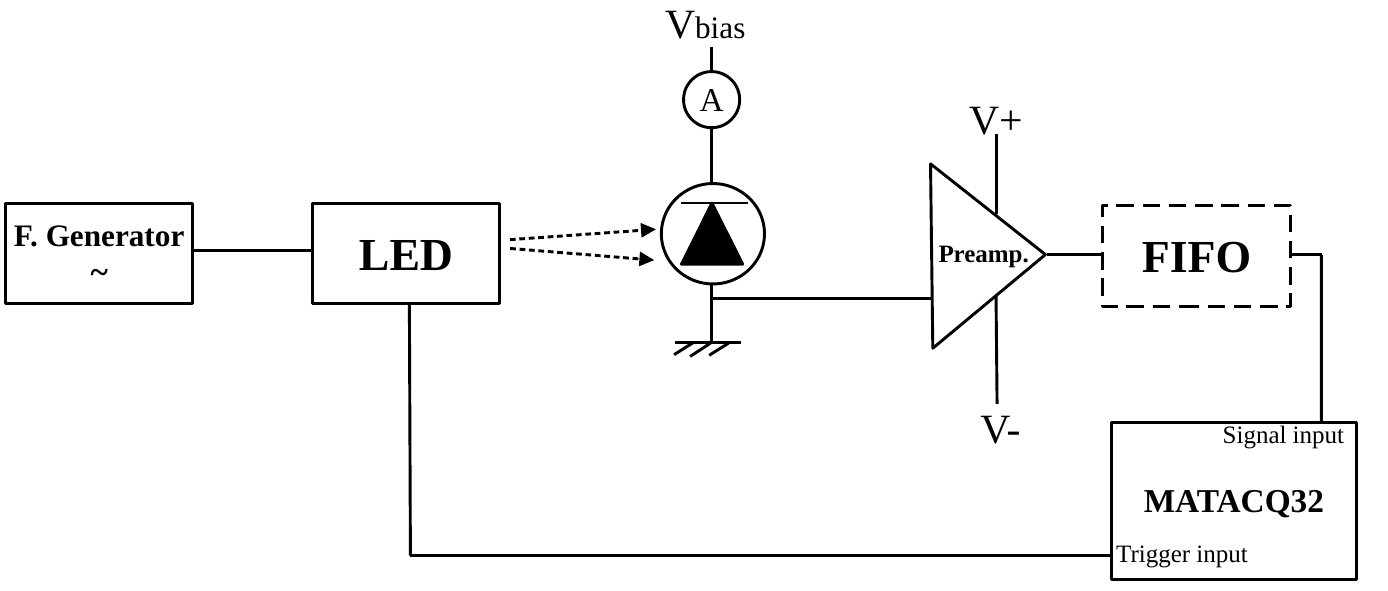}
\caption{\label{ElectronicChainSiPM_LED}Electronic chain of the measurements using the LED. Dashed arrows between the LED and the SiPM represent the optical fiber.}
\end{center}
\end{figure}

Two different electronic chains were used for the signal processing depending on the measurement purpose. In the measurements with LED illumination (shown in Figure~\ref{ElectronicChainSiPM_LED}) the trigger is set by the NIM pulse synchronized with the function generator. The preamplified output pulse can be positive or negative depending on the configuration of the FEB. Therefore, in case of positive pulses, a Fan-In Fan-Out (FIFO) module (CAEN Mod.625) is used to invert the signal that goes into the digitizer. In Figure~\ref{ElectronicChainSiPM_LED}, the dashed box of the FIFO module represents the possibility of inverting the signal.

\begin{figure}[h!]
\begin{center}
\includegraphics[width=\textwidth]{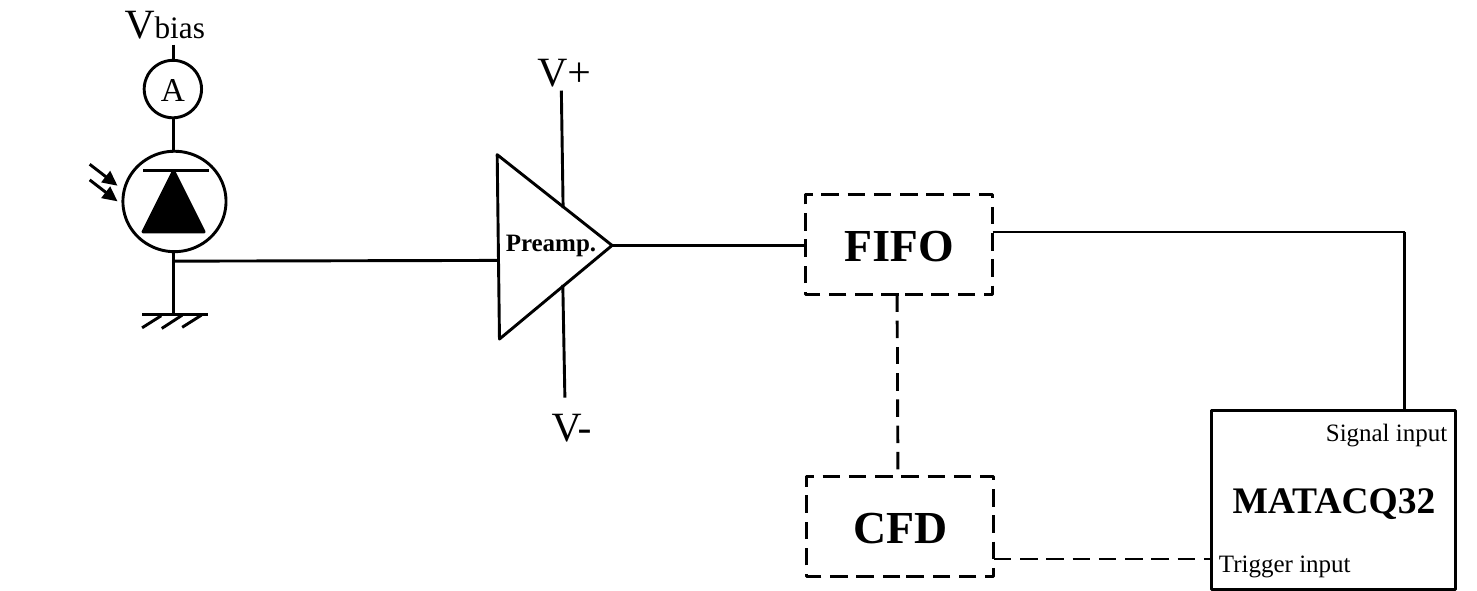}
\caption{\label{ElectronicChainSiPM_Dark}Electronic chain of the measurements in darkness or with a NaI(Tl) crystal as a light source.}
\end{center}
\end{figure}

In the measurements with the NaI(Tl) crystal as light source or in darkness (presented in Figure~\ref{ElectronicChainSiPM_Dark}) the trigger could be done internally by the digitizer (equipped with a threshold discriminator for every channel) or externally. In this last case, the signal of the preamplifier is connected to a FIFO module which divides the signal in two: one goes to the digitizer and the other to a Constant Fraction Discriminator (CFD) module (CAEN Mod.843) that generates a NIM pulse for triggering. In Figure~\ref{ElectronicChainSiPM_Dark}, the path followed to generate the trigger using the CFD is represented with dashed lines and boxes.

In all the data taking runs, the pretrigger time was set to 25\% of the digitization window (corresponding to 315~ns). As the scintillation time of the NaI(Tl) crystals at room temperature is approximately 230~ns, the posttrigger time window of 945~ns allows to collect about 98\% of the scintillation light. The digitizer stored the waveform for each event in a raw file. This file is later decoded, and its information is transcribed into a ROOT file, which is analyzed as explained in next section.

\section{Event analysis} \label{Section:SiPMZgz_Analysis}
\fancyhead[RO]{\emph{\thesection. \nameref{Section:SiPMZgz_Analysis}}}

\begin{figure}[h!]
\begin{subfigure}[b]{0.49\textwidth}
\includegraphics[width=\textwidth]{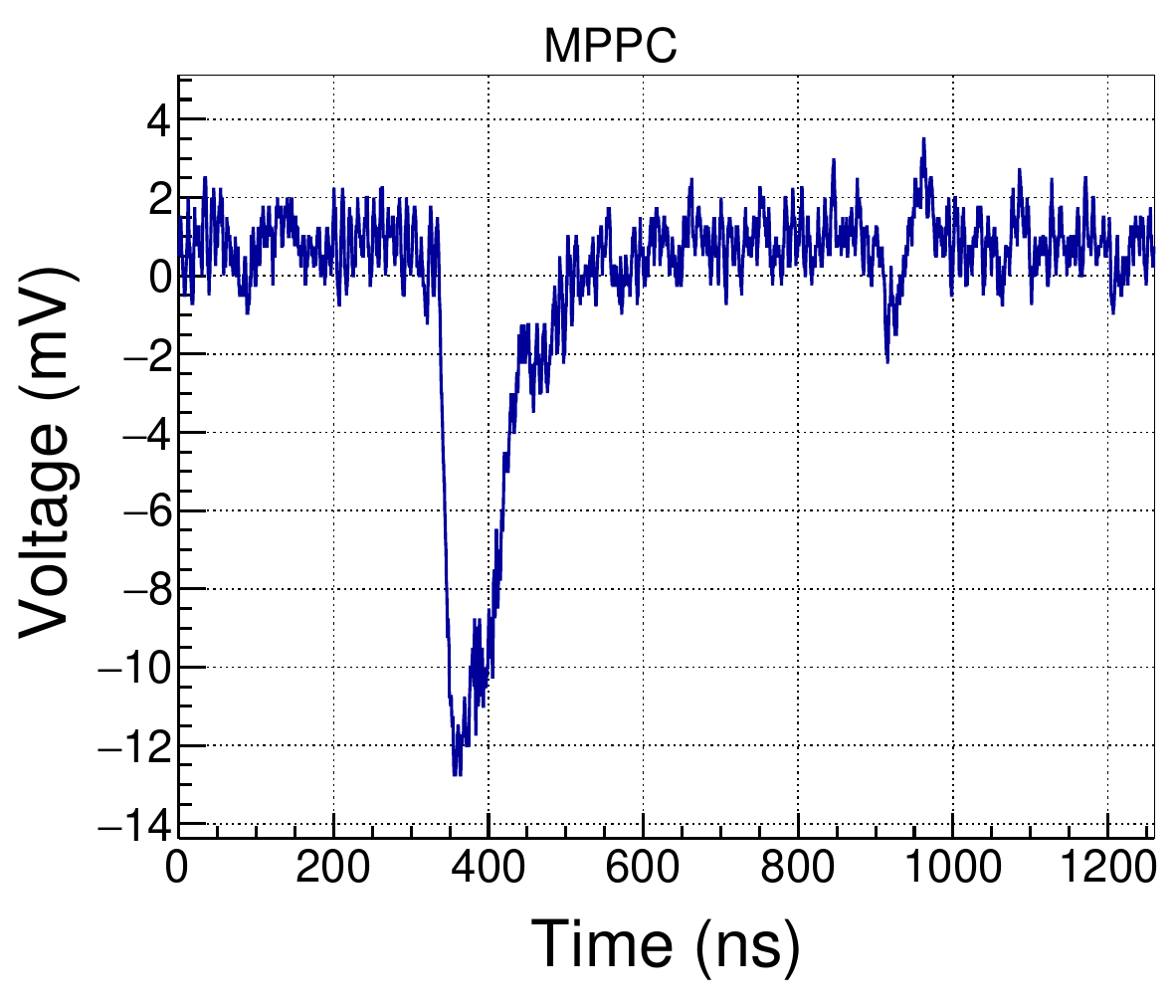}
\end{subfigure}
\begin{subfigure}[b]{0.49\textwidth}
\includegraphics[width=\textwidth]{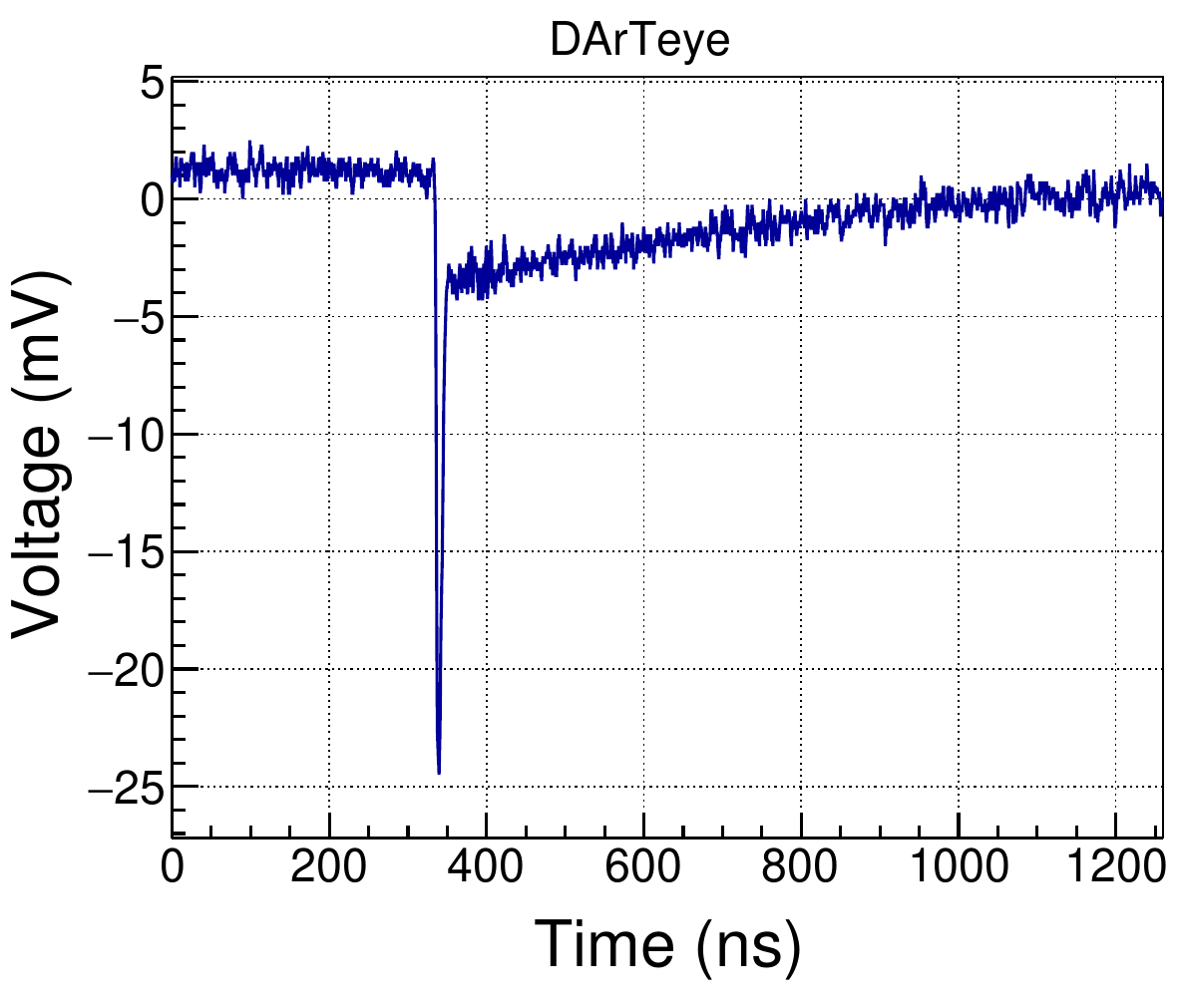}
\end{subfigure}
\caption{\label{exampleWaveforms}Example of acquired waveforms with the LED illumination and from the MPPC (left plot) and a DArTeye (right plot).}
\end{figure}

The algorithm used for the calculation of the relevant variables for each event is the same as that explained in Section~\ref{Section:SiPMSTAR2_EventAnalysis}, applied to only one channel. Figure~\ref{exampleWaveforms} shows an example of the pulses of a measurement with LED with similar illumination for both the MPPC and one DArTeye.

Variables used in this chapter are the pulse onset (\textit{t0}), its amplitude (\textit{high}), the time in the waveform when the pulse has the maximum amplitude (\textit{tmax}), the pulse mean time ($\mu$, defined in Equation~\ref{eq:mu}), and the \textit{area} variable, which integrates from \textit{t0} over the total acquisition window after the baseline subtraction. Moreover, the pulse-shape discrimination variable $p_1$ was used, being the ratio of two integrals of the pulse in different time windows
\begin{equation}\label{eq:p1}
p1 = \frac{\sum_{t = t0+100}^{t = t0+600} V(t)}{\sum_{t = t0}^{t = t0+600}V(t)},
\end{equation}
where $t0$ is the pulse onset, and $V(t)$ is the value of the baseline subtracted waveform at a given time, $t$. This variable, together with the mean time are used in ANAIS-112 experiment to reject non-bulk scintillation events, as it was explained in Section~\ref{Section:ANAIS_Filter}. Here, they also will be used for the selection of scintillation events and rejection of DC events.


\section{Results of the SiPM performance} \label{Section:SiPMZgz_CryoT}
\fancyhead[RO]{\emph{\thesection. \nameref{Section:SiPMZgz_CryoT}}}

\subsection{SiPM response in darkness and with LED illumination} \label{Section:SiPMZgz_CryoT_Characterization}

The MPPC was characterized both at room temperature (295~K) and at the lowest temperature achieved in the setup (90~K) and the DArTeyes were only characterized at the liquid nitrogen temperature (77~K). In order to let the system time enough for temperature stabilization, all the measurements at cryogenic temperatures were done at least one hour after reaching the base temperature.

The procedure involved several steps. Firstly, we measured the I-V curves in darkness to obtain the breakdown voltages. Then, we obtained the area and the amplitude distributions of the SPE at different overvoltages under LED illumination. Finally, the SPE waveforms were analyzed and the DC rates were determined in darkness.

\begin{figure}[h!]
\begin{center}
\begin{subfigure}[b]{0.49\textwidth}
\includegraphics[width=\textwidth]{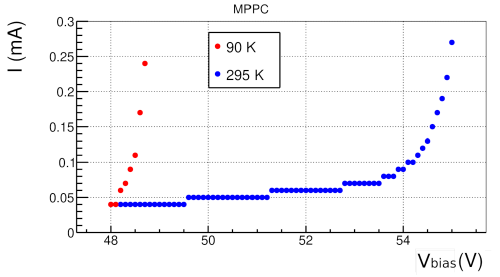}
\end{subfigure}
\begin{subfigure}[b]{0.49\textwidth}
\includegraphics[width=\textwidth]{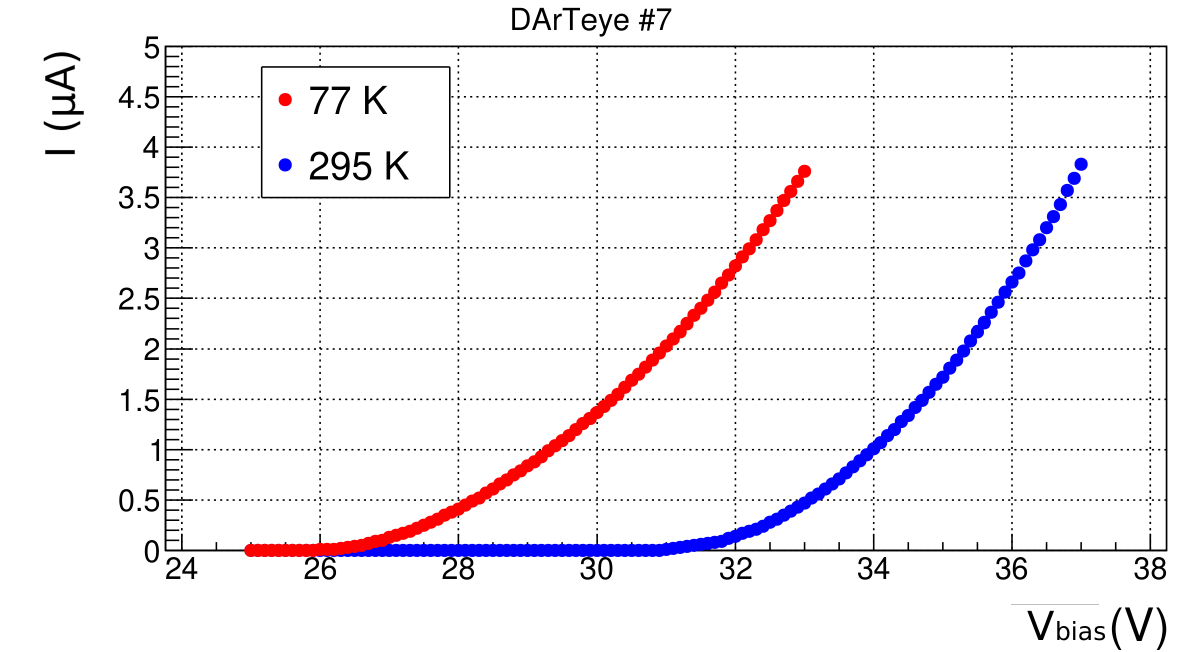}
\end{subfigure}
\caption{\label{IVcurves}I-V curves of the MPPC (left plot) and DArTeye~$\#$~7 (right plot). All DArTeyes had a similar behaviour. Breakdown voltages are summarized in Table~\ref{tabla:BreakdownVoltages}).}
\end{center}
\end{figure}

The I-V curves of the MPPC were measured with the software control tool, while for the DArTeyes curves the Keithley-2000 multimeter was used. The acquisition was performed in darkness, in steps of bias voltage of 0.1~V, starting from 25~V for the DArTeyes and from 48~V for the MPPC. The obtained curves are shown in Figure~\ref{IVcurves}.

The resolution in the measurement of the intensity in the control tool of the MPPC was 0.01~mA, as it is clearly noticeable in the left plot of Figure~\ref{IVcurves}. As the leakage current (that flowing at a bias voltage below the breakdown) of MPPC was 0.04~mA, it was difficult to estimate the breakdown voltage for these devices. At low temperature it was calculated as the bias voltage at which the current measured was higher than the leakage current, while at room temperature we took the bias voltage at which the current increased at a rate above 0.01~mA over a 1.5~V interval. For DArTeyes, having an stable and very low  leakage current (10~nA), the breakdown voltage was calculated as the voltage at which the intensity was higher than the leakage current. In this case, the current measurement exhibited a significantly improved resolution, on the order of 10~nA, as provided by the Keithley multimeter. All the DArTeyes have a similar behaviour. The breakdown voltages are shown in Table~\ref{tabla:BreakdownVoltages}.

\begin{table}[h!]
\centering
\begin{tabular}{|c|c|c|}
\hline
SiPM & V$_b$ at Low T. (V) & V$_b$ at Room T. (V) \\
\hline
MPPC & 48.2 V & 53.5 V \\
DArTeye~$\#$2 & - & 31.9 V \\
DArTeye~$\#$7 & 26.0 V & 31.5 V \\
DArTeye~$\#$9 & 26.0 V & 31.5 V \\
DArTeye~$\#$10 & 25.9 V & 31.1 V \\
\hline
\end{tabular}
\caption{Breakdown voltages at room and cryogenic temperature for the five SiPMs under study. Low temperature for MPPC module is 90~K and for DArTeyes is 77~K.}
\label{tabla:BreakdownVoltages}
\end{table}

The SPE was measured using the DAQ configuration shown in Figure~\ref{ElectronicChainSiPM_LED}, triggering with the LED. For the MPPC module, measurements both at room and low temperature were taken with overvoltages from 3 to 7~V, while DArTeyes were only measured at liquid nitrogen temperature at overvoltages from 3 to 10~V.

\begin{figure}[h!]
\begin{center}
\includegraphics[width=0.75\textwidth]{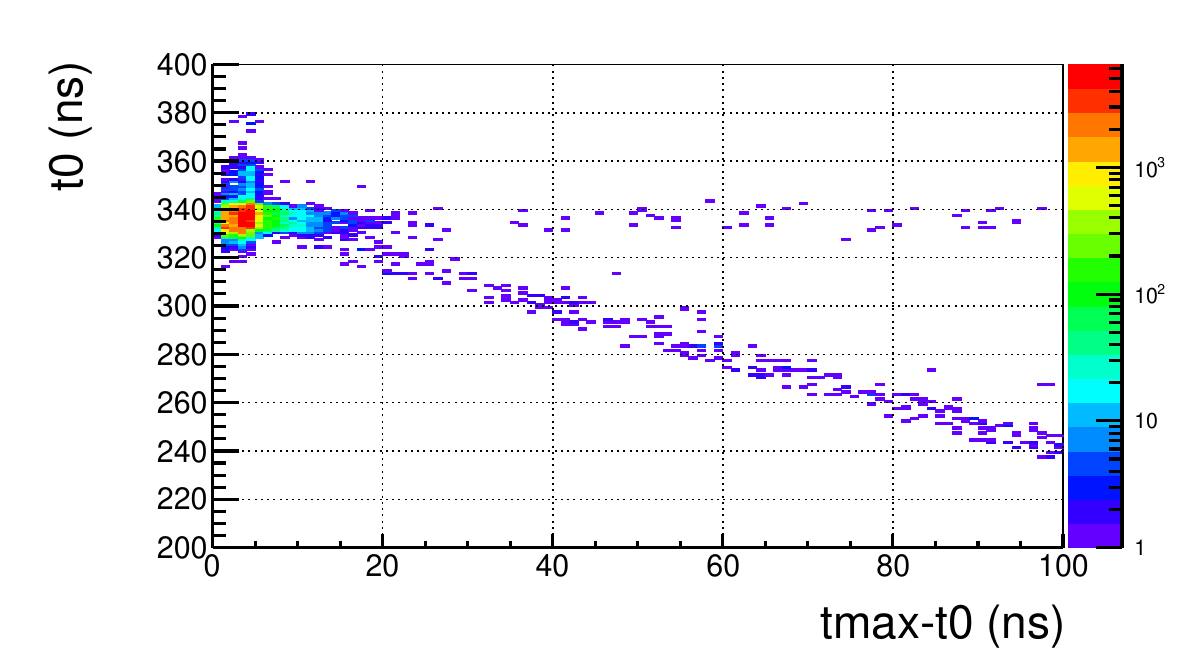}
\caption{\label{t0EventSelection}2D scattering plot of the t0 and tmax-t0 variables for the events acquired with the DArTeye~$\#$7 at an overvoltage of 5~V in the SPE calibration under LED illumination. Similar behaviour was observed in the measurements with the MPPC.}
\end{center}
\end{figure}

Before proceeding with the SPE calibration, the contribution of AP and DC events was reduced using \textit{t0} and \textit{tmax} variables (defined in Section~\ref{Section:SiPMZgz_Analysis}). As an example, the 2D scattering plot of \textit{t0} and \textit{tmax}-\textit{t0} variables for the measurement with the DArTeye~$\#$7 at an overvoltage of 5~V is shown in Figure~\ref{t0EventSelection}. Similar behaviour was observed in the measurements with the MPPC. It is possible to observe the main contribution with a \textit{t0} value close to the defined hardware trigger time (around 320~ns) and a small \textit{tmax}-\textit{t0} value, which corresponds to pulses with maximum value close to the trigger time. There are two other contributions: a population with \textit{t0} at the trigger time but a very distant maximum, corresponding to events which have DC pulses or AP after the trigger time with amplitudes higher than that of the triggered pulse, and another population with \textit{t0} lower than the trigger time, which are events with pulses in the pretrigger region. Both populations are removed by selecting \textit{t0} in the range from 320 to 360 ns, and a value of \textit{tmax}-\textit{t0} lower than 40~ns. However, this selection procedure is not removing those AP or DC events arriving after the trigger time that result in an amplitude lower than the pulse that triggered the DAQ.

\begin{figure}[h!]
\begin{center}
\begin{subfigure}[b]{0.49\textwidth}
\includegraphics[width=\textwidth]{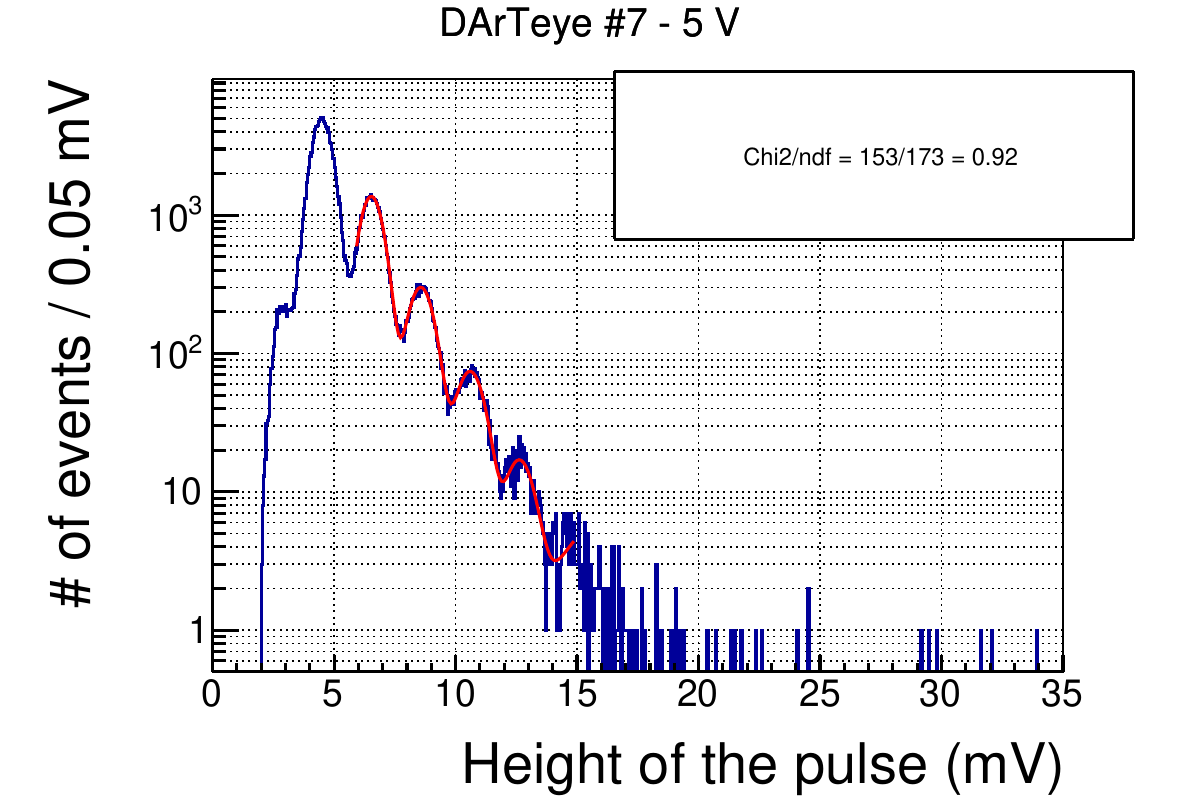}
\end{subfigure}
\begin{subfigure}[b]{0.49\textwidth}
\includegraphics[width=\textwidth]{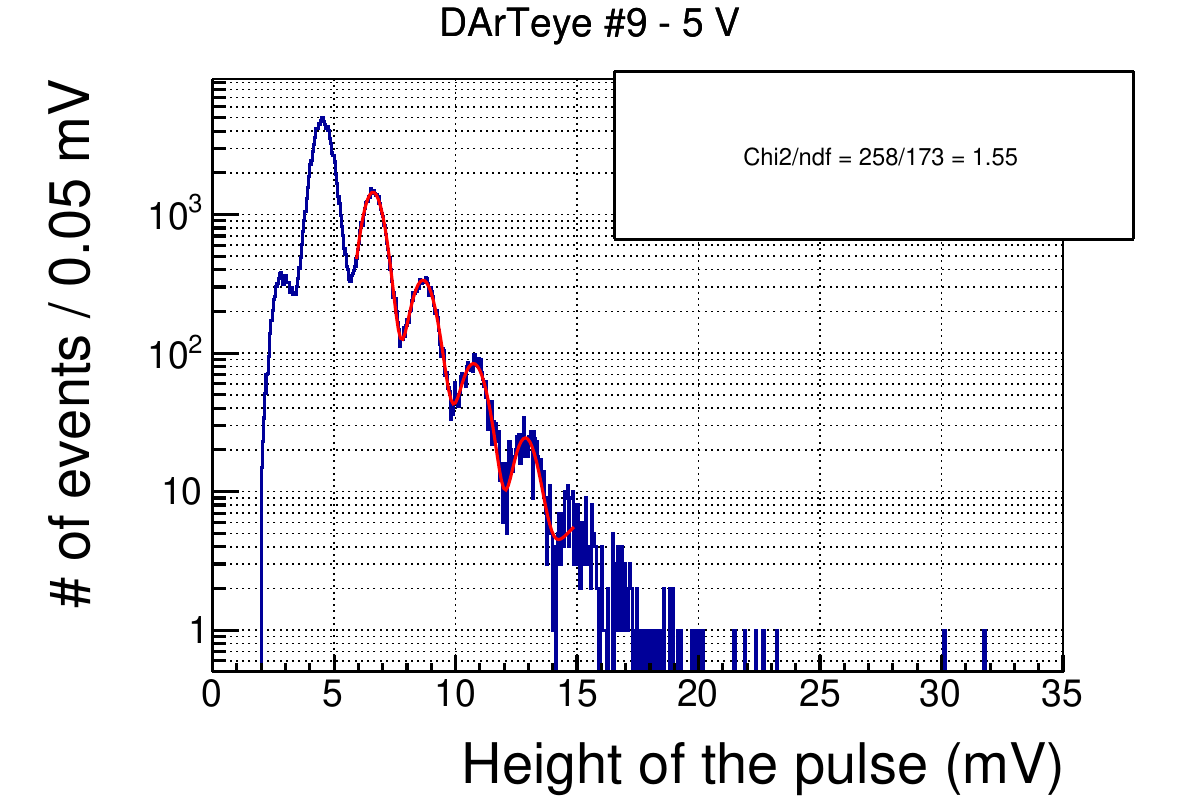}
\end{subfigure}
\begin{subfigure}[b]{0.49\textwidth}
\includegraphics[width=\textwidth]{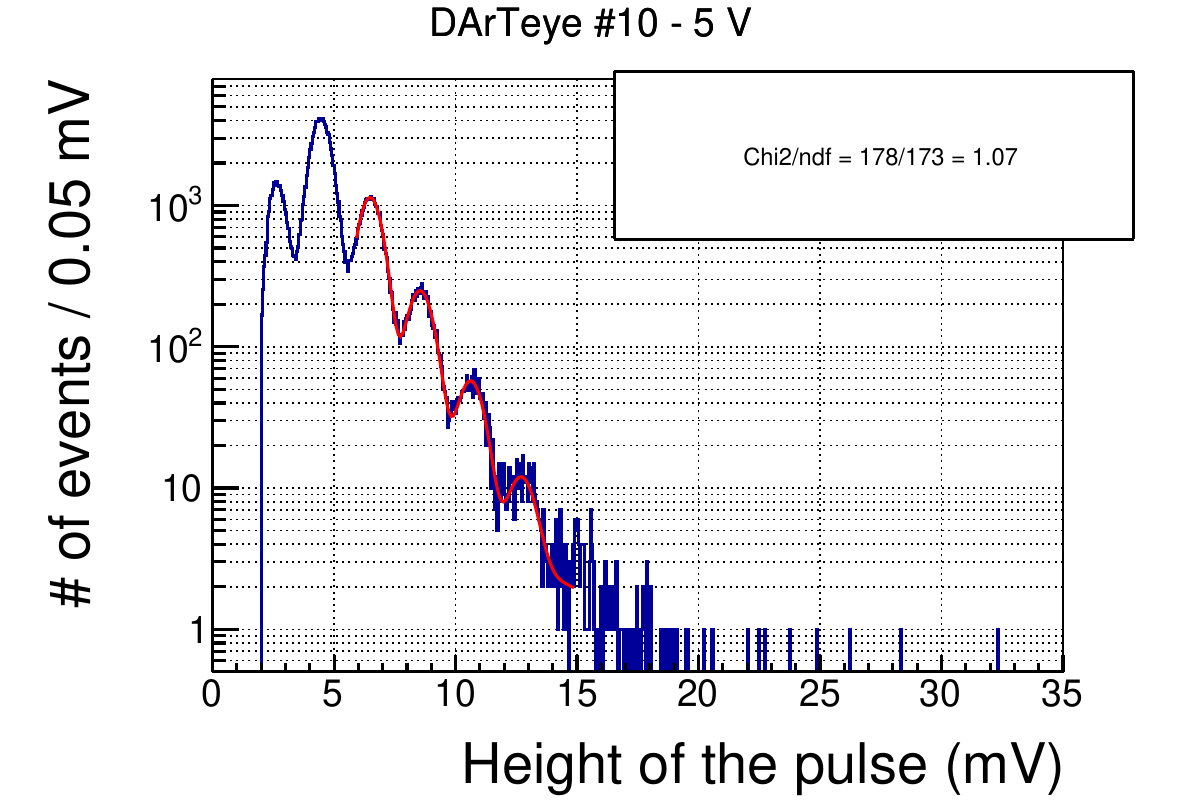}
\end{subfigure}
\begin{subfigure}[b]{0.49\textwidth}
\includegraphics[width=\textwidth]{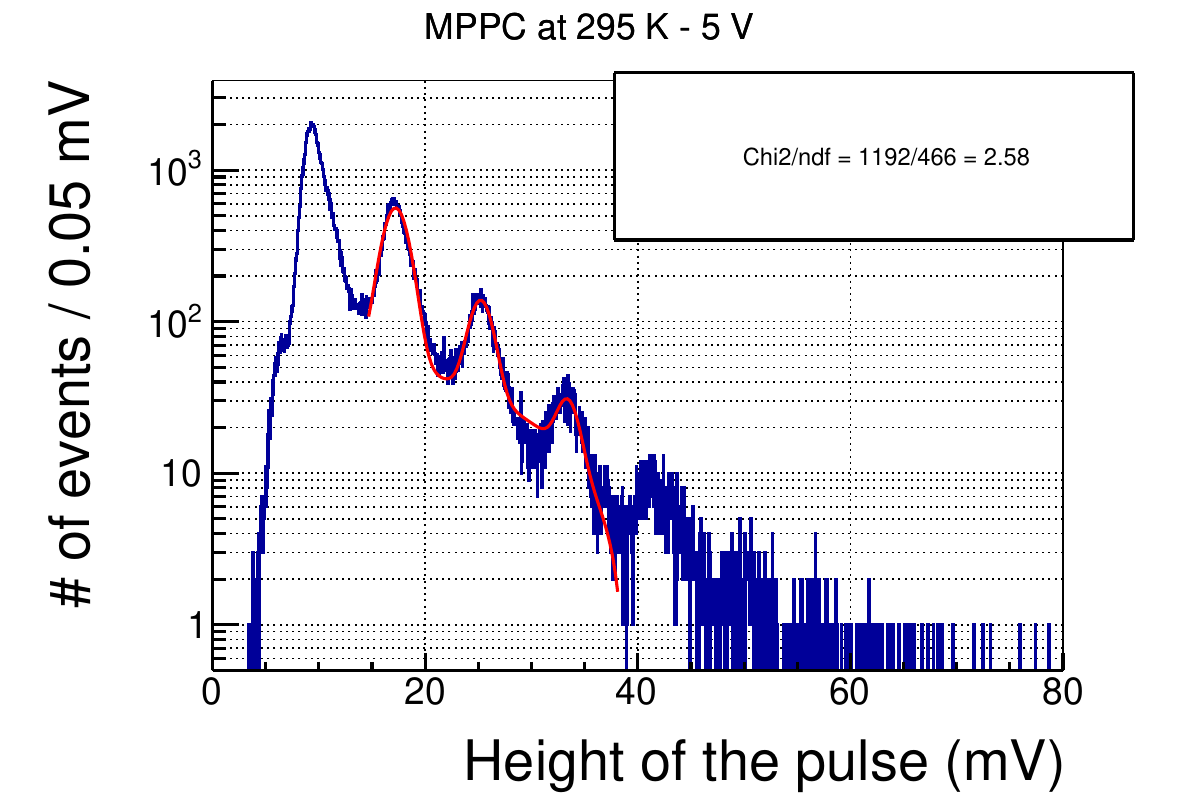}
\end{subfigure}
\begin{subfigure}[b]{0.49\textwidth}
\includegraphics[width=\textwidth]{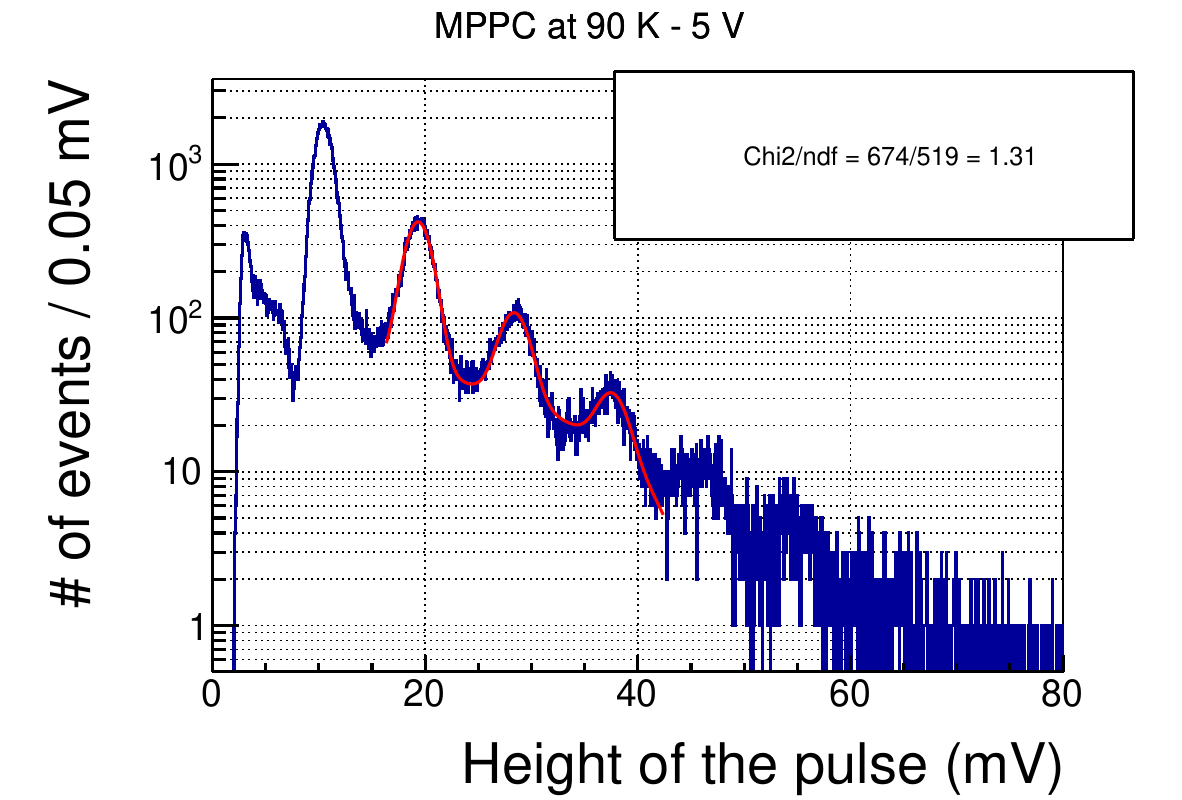}
\end{subfigure}
\caption{\label{HighFits}Plots of the height spectra and their corresponding fits for the SPE calibration under LED illumination for each SiPM at an overvoltage of 5~V.}
\end{center}
\end{figure}

After applying this selection, the pulse height spectrum was obtained in each measurement, as shown in Figure~\ref{HighFits} for the measurements at an overvoltage of 5~V. In these spectra, different pulses populations with similar height are clearly identified. Each one corresponds to a different number of photoelectrons in the pulse. They were fitted to a number of gaussian peaks plus a constant background. The number of peaks fitted depended on the measurement, but to avoid introducing noise or a peak affected by threshold effects, the first peak in the spectrum is not included in the fit, independently on the number of phe. The free parameters in the fit are the amplitudes and standard deviations of the gaussians, while for the means, a linear relation with the number of phes in the pulse is considered, $\mu = H_{spe}\cdot N_{phe} + c_0$. The association between each peak and its corresponding number of phes is that minimizing the independent term of the linear relation, $c_0$.

\begin{figure}[h!]
\begin{center}
\includegraphics[width=\textwidth]{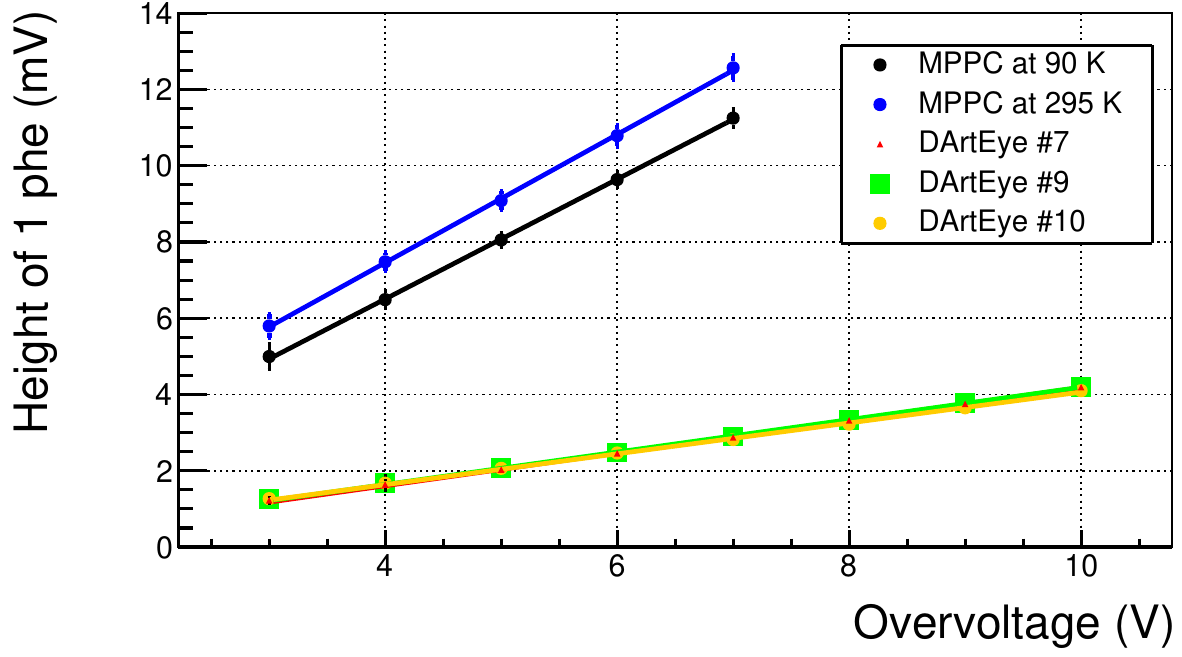}
\caption{\label{GainH}Fits of the mean height of a SPE, $H_{spe}$, as a function of the overvoltage applied for all the SiPMs.}
\end{center}
\end{figure}

This fit allows to obtain the mean height of a SPE, $H_{spe}$. The results at each overvoltage for each SiPM are shown in Table~\ref{tabla:SPECal_Height}. As the height of the SPE increases linearly with the overvoltage, a linear fit was done as $H_{spe} = c_{H1} \cdot V_{ov} + c_{H0}$. The fits are shown in Figure~\ref{GainH}, and the results in Table~\ref{tabla:Gains_vs_OV}. The calibration of the SPE amplitude will be important in the future to evaluate the threshold applied to trigger the acquisition of scintillation light in number of photoelectrons.

\begin{table}[h!]
\centering
\begin{tabular}{|c|c|c|c|c|c|}
\cline{2-6}
\multicolumn{1}{c|}{} & \multicolumn{5}{|c|}{Mean height (mV)} \\
\hline
OV (V) & MPPC 90~K & MPPC 295~K & De~$\#$7 & De~$\#$9 & De~$\#$10 \\
\hline
3 & 5.00$\pm$0.34 & 5.80$\pm$0.31 & 1.22$\pm$0.99 & 1.25$\pm$1.69 & 1.27$\pm$1.44  \\
4 & 6.49$\pm$0.23 & 7.48$\pm$0.24 & 1.64$\pm$1.17 & 1.68$\pm$0.92 & 1.67$\pm$2.01  \\
5 & 8.05$\pm$0.20 & 9.08$\pm$0.24 & 2.03$\pm$0.72 & 2.07$\pm$0.60 & 2.06$\pm$0.80  \\
6 & 9.64$\pm$0.22 & 10.79$\pm$0.28 & 2.46$\pm$0.54 & 2.48$\pm$0.49 & 2.47$\pm$0.70  \\
7 & 11.25$\pm$0.26 & 12.56$\pm$0.32 & 2.88$\pm$0.44 & 2.90$\pm$0.42 & 2.82$\pm$0.57  \\
8 & - & - & 3.32$\pm$0.39 & 3.34$\pm$0.38 & 3.22$\pm$0.50  \\
9 & - & - & 3.75$\pm$0.36 & 3.77$\pm$0.34 & 3.65$\pm$0.53  \\
10 & - & - & 4.19$\pm$0.36 & 4.20$\pm$0.35 & 4.10$\pm$0.53  \\
\hline
\end{tabular}
\caption{Mean height of the SPE, $H_{spe}$, at different overvoltages for each SiPM. "De" refers to DArTeye.}
\label{tabla:SPECal_Height}
\end{table}

\begin{figure}[h!]
\begin{center}
\begin{subfigure}[b]{0.49\textwidth}
\includegraphics[width=\textwidth]{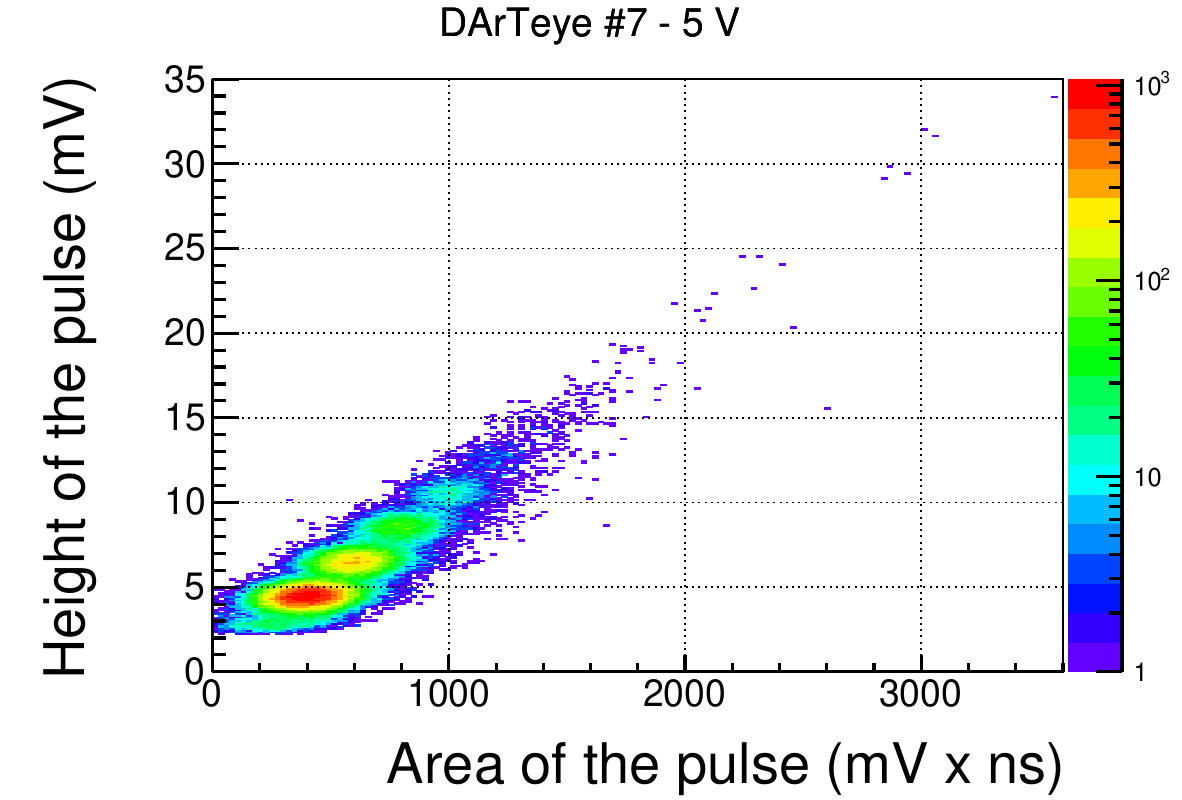}
\end{subfigure}
\begin{subfigure}[b]{0.49\textwidth}
\includegraphics[width=\textwidth]{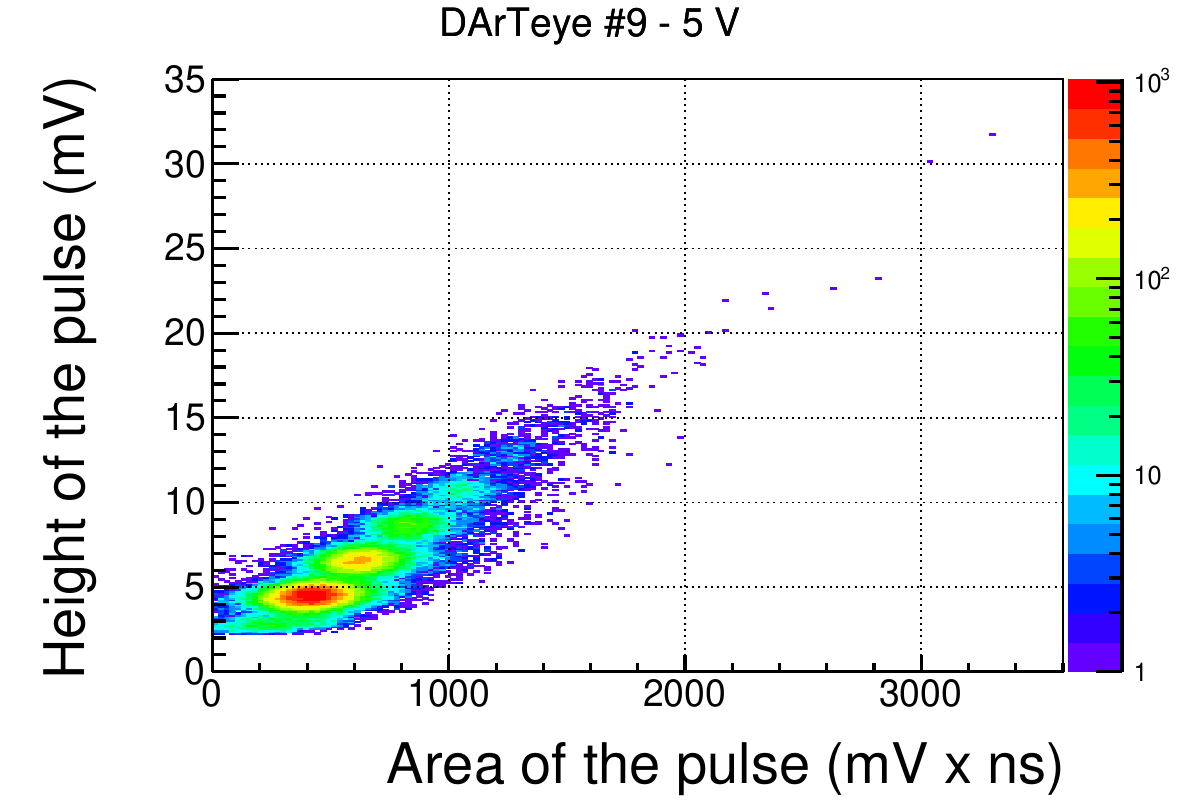}
\end{subfigure}
\begin{subfigure}[b]{0.49\textwidth}
\includegraphics[width=\textwidth]{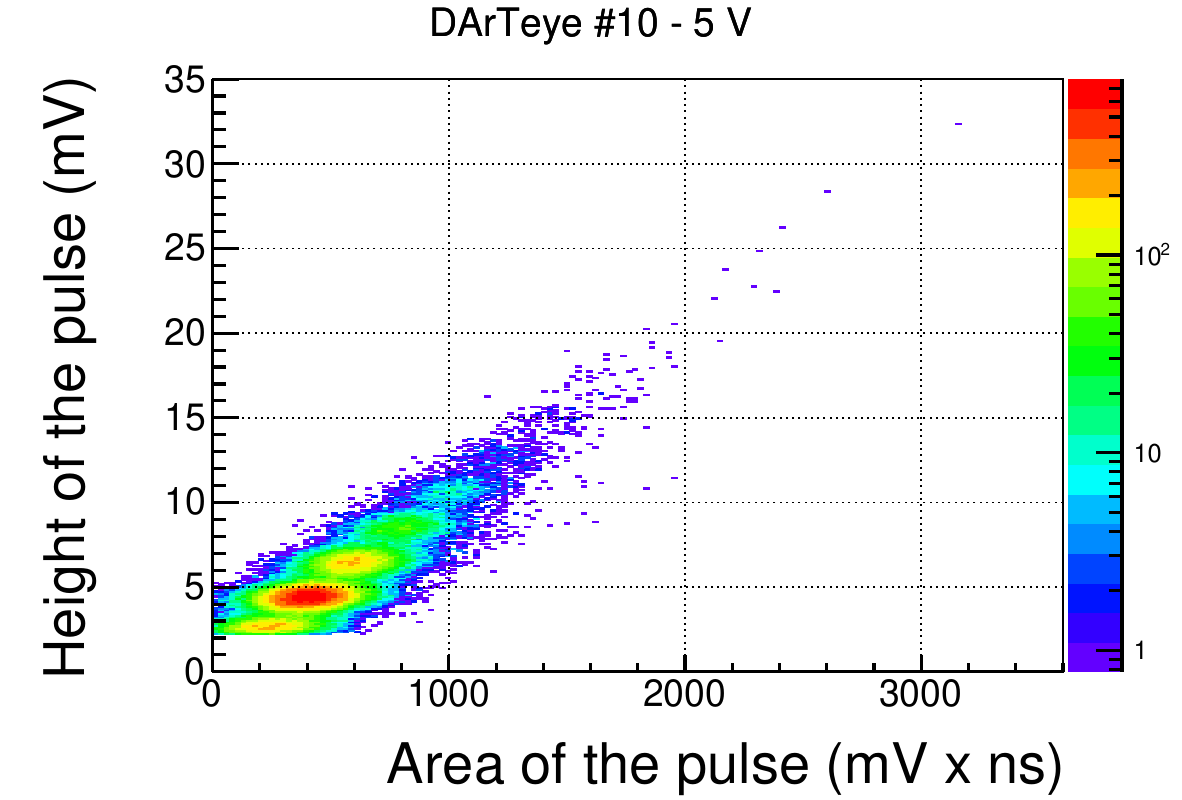}
\end{subfigure}
\begin{subfigure}[b]{0.49\textwidth}
\includegraphics[width=\textwidth]{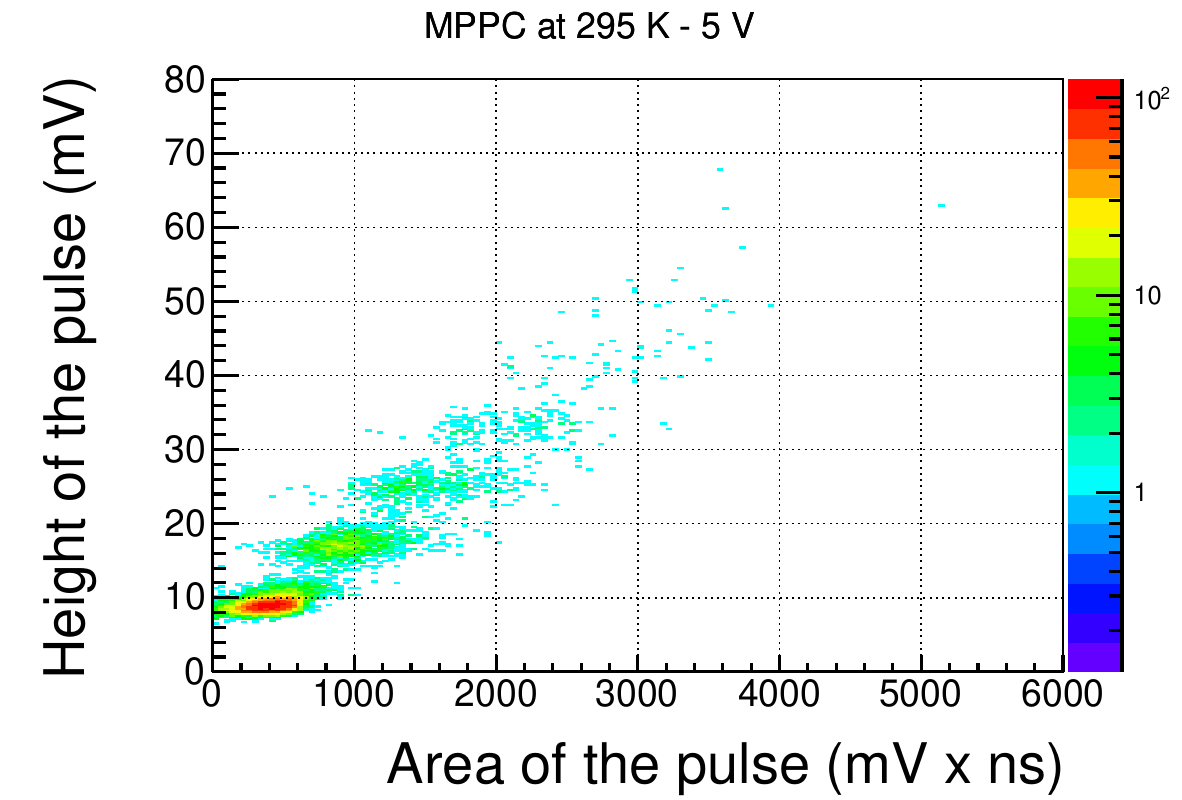}
\end{subfigure}
\begin{subfigure}[b]{0.49\textwidth}
\includegraphics[width=\textwidth]{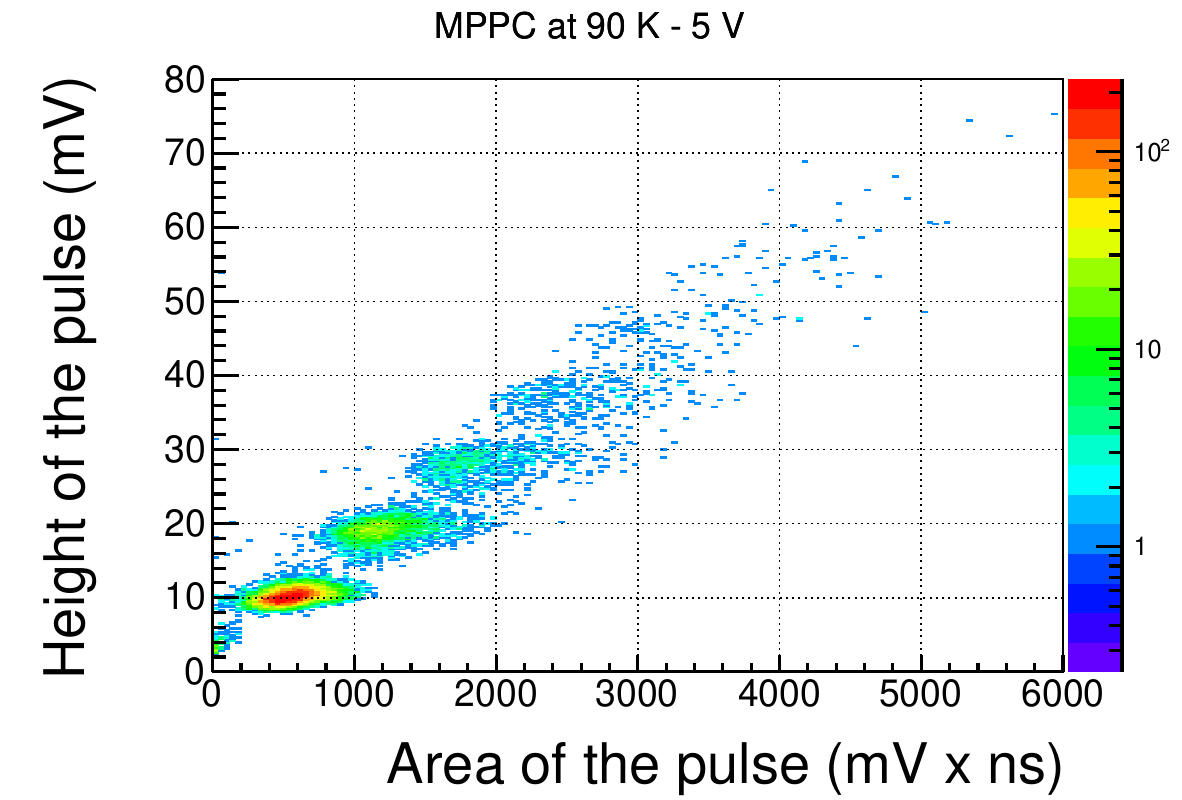}
\end{subfigure}
\caption{\label{SPECal_AreaHeight}Plots of the height vs pulse area in the SPE calibrations under LED illumination at an overvoltage of 5~V.}
\end{center}
\end{figure}

\begin{figure}[h!]
\begin{center}
\begin{subfigure}[b]{0.49\textwidth}
\includegraphics[width=\textwidth]{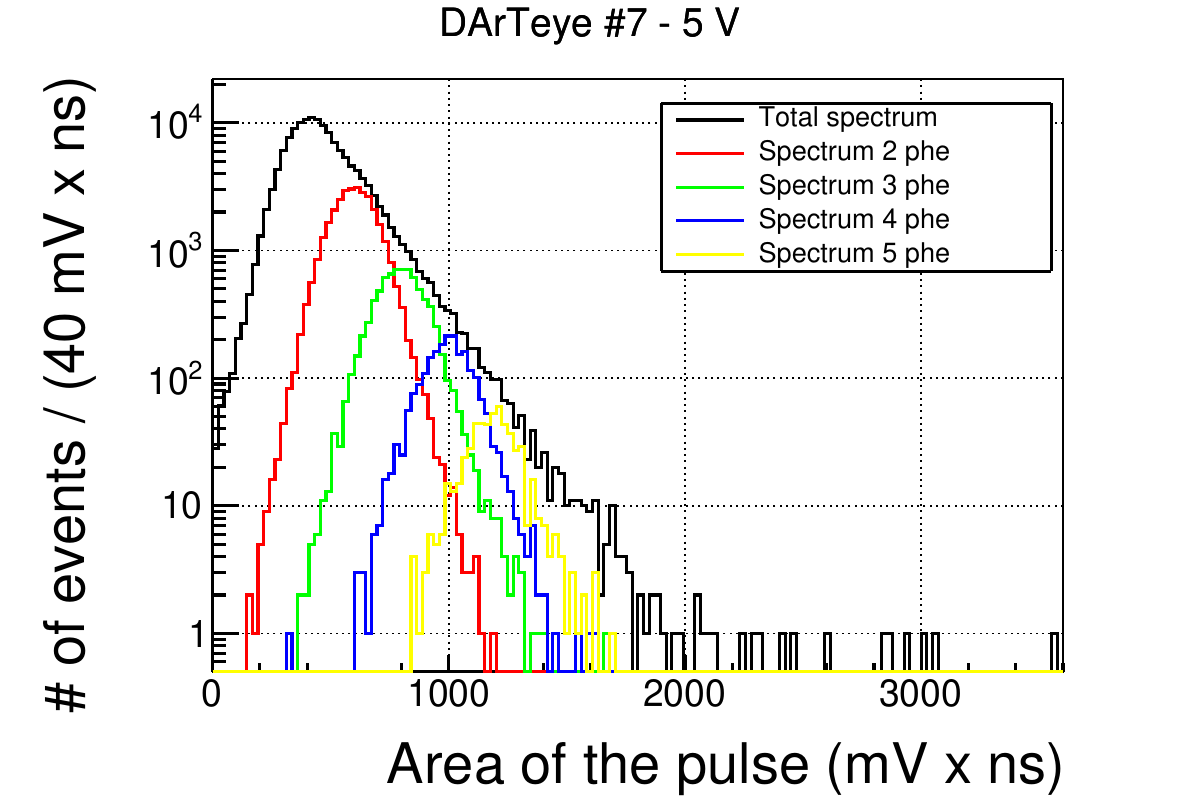}
\end{subfigure}
\begin{subfigure}[b]{0.49\textwidth}
\includegraphics[width=\textwidth]{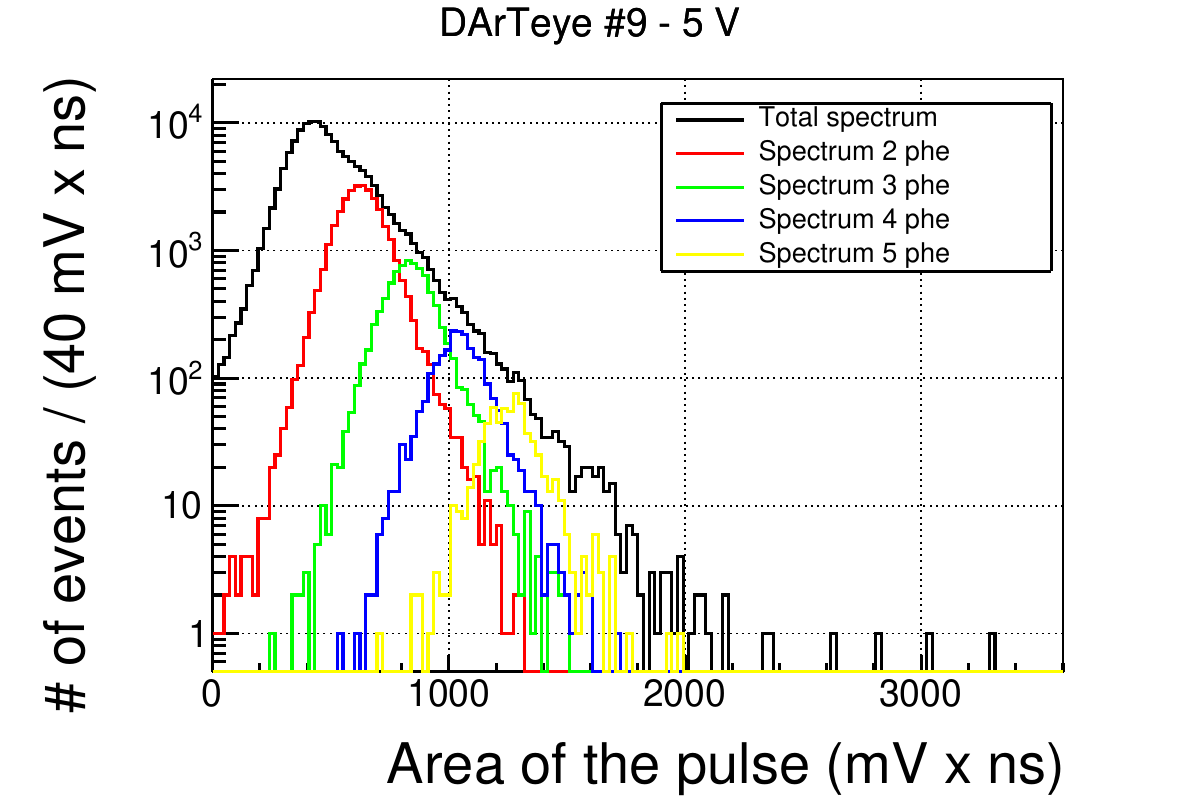}
\end{subfigure}
\begin{subfigure}[b]{0.49\textwidth}
\includegraphics[width=\textwidth]{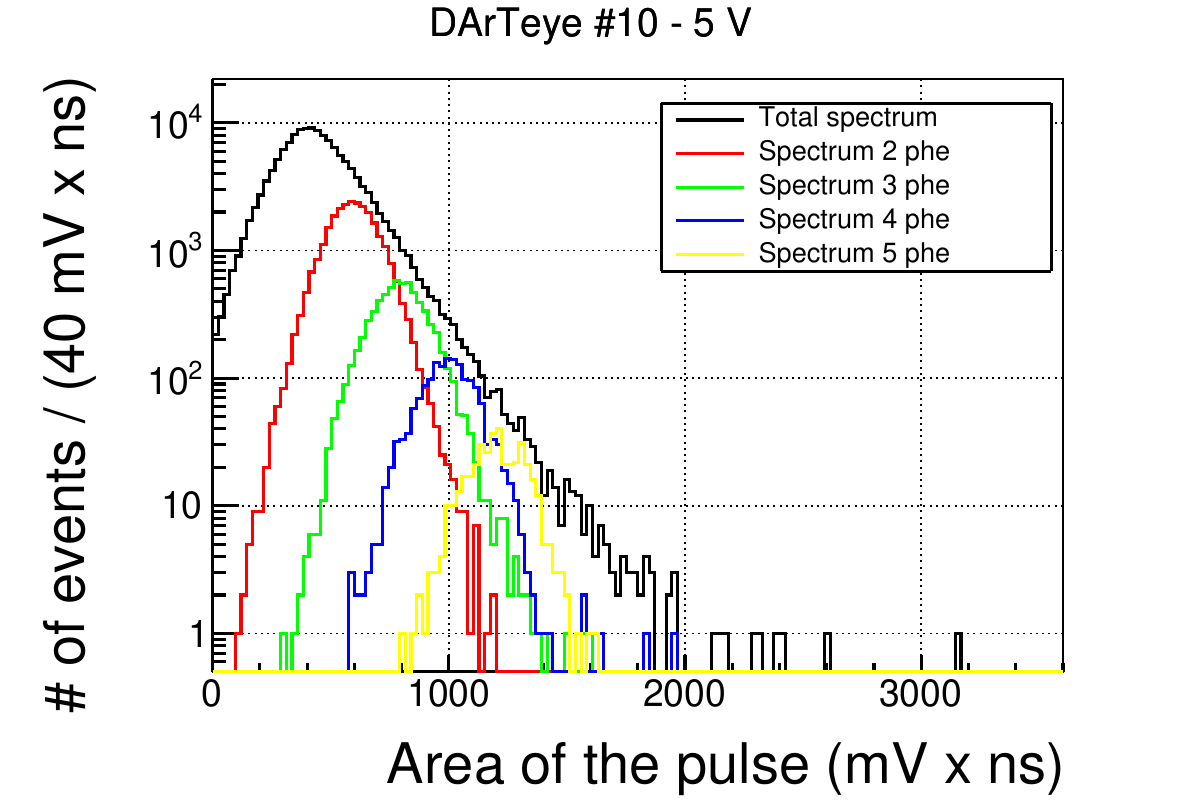}
\end{subfigure}
\begin{subfigure}[b]{0.49\textwidth}
\includegraphics[width=\textwidth]{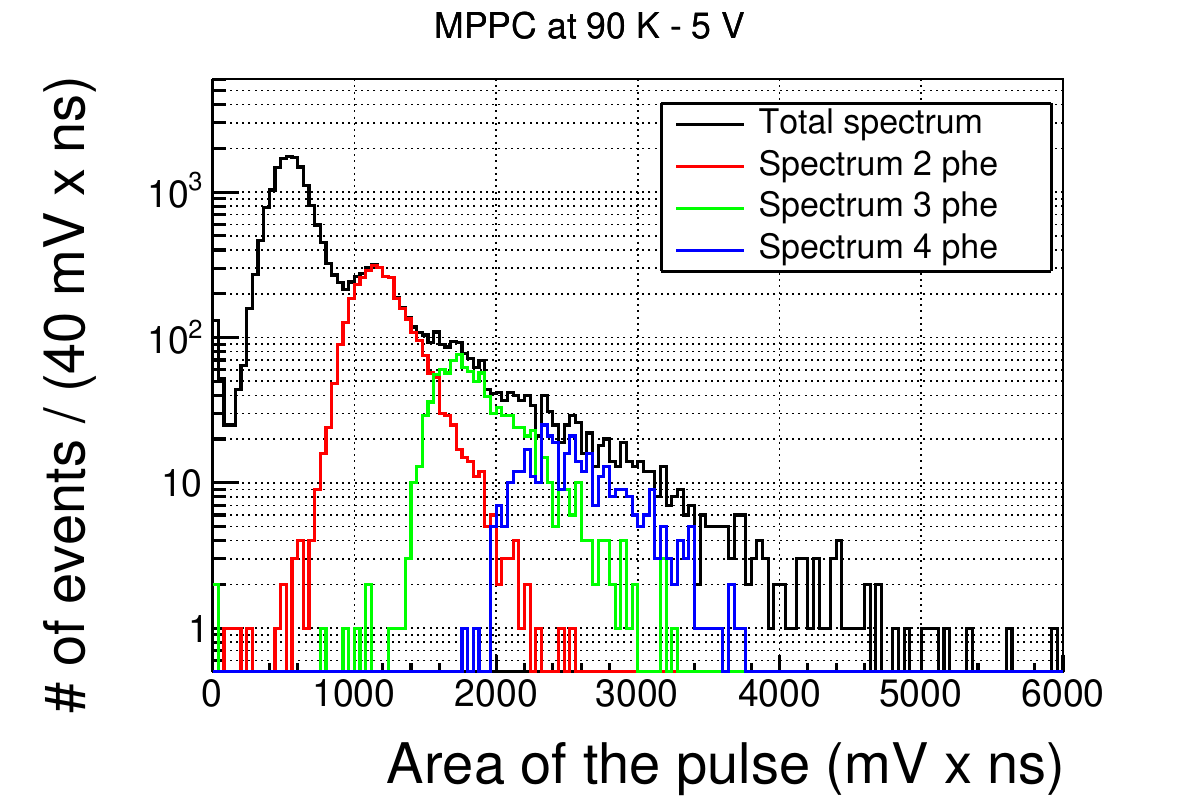}
\end{subfigure}
\begin{subfigure}[b]{0.49\textwidth}
\includegraphics[width=\textwidth]{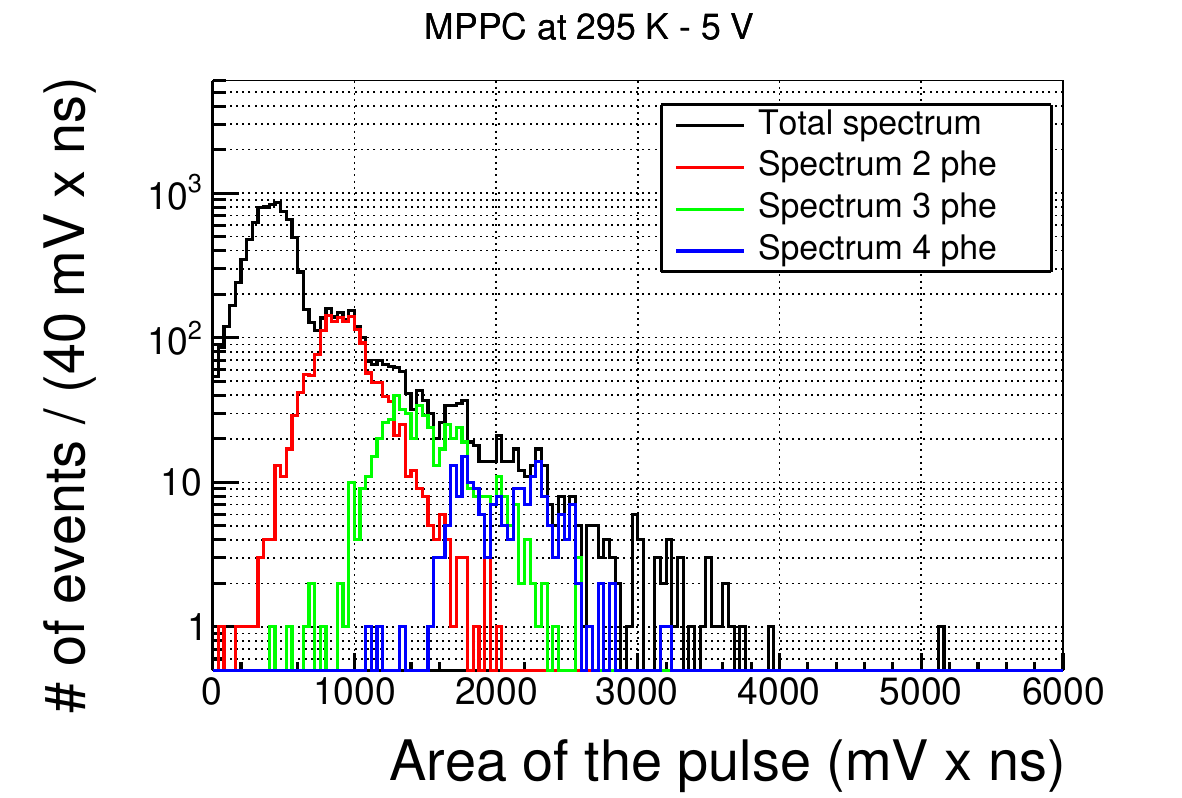}
\end{subfigure}
\caption{\label{AreaDistributions}Plots of the total area spectra and the corresponding spectra for the contributions of pulses with different number of phes, built by selecting in the height variable, in the SPE calibration under LED illumination. Measurements at an overvoltage of 5~V.}
\end{center}
\end{figure}

The distribution of pulse areas, on the other hand, does not allow to resolve the different populations. This can be observed in Figure~\ref{SPECal_AreaHeight}, where the height and the area of the pulses were plotted together in a scattering plot for each SiPM at an overvoltage of 5~V. The method followed to calibrate the area of the SPE was to select each population in the height spectra, taken as limits for each population the valley between consecutive peaks, and build the corresponding area distributions. An example of the area distributions obtained at an overvoltage of 5~V is shown in Figure~\ref{AreaDistributions}.

\begin{figure}[h!]
\begin{center}
\begin{subfigure}[b]{0.49\textwidth}
\includegraphics[width=\textwidth]{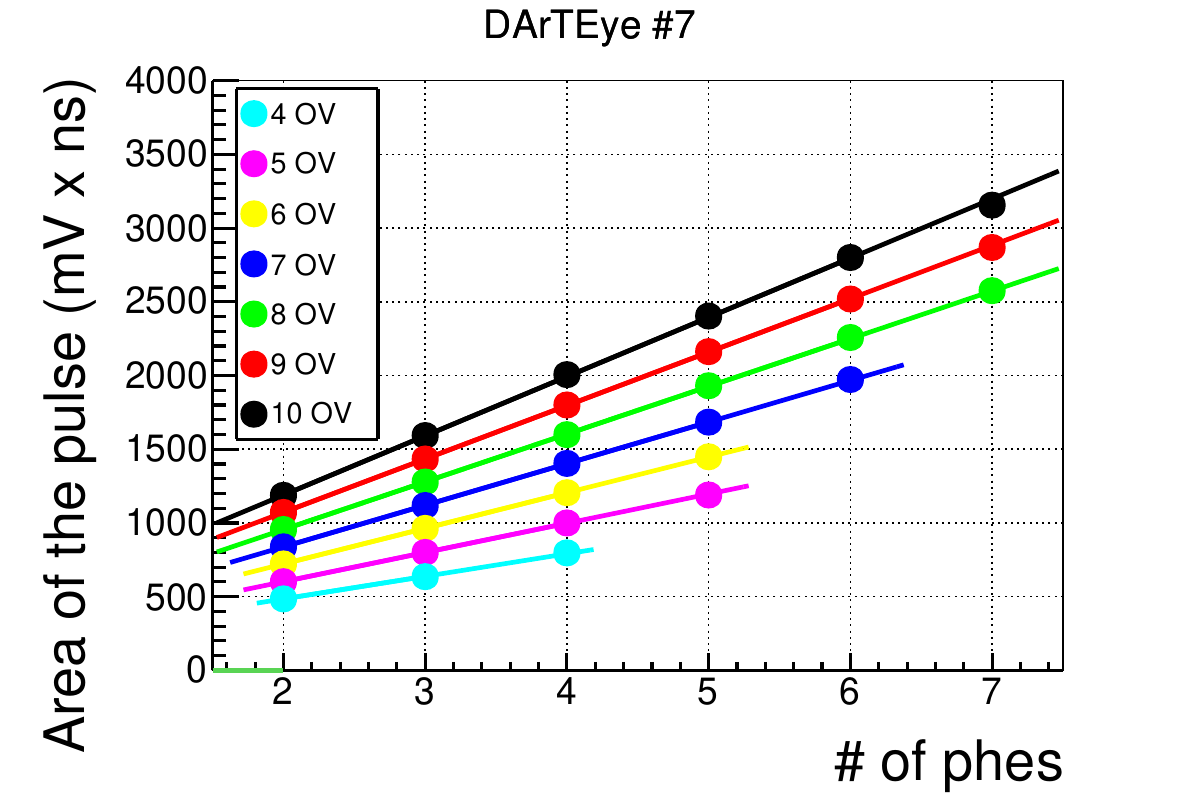}
\end{subfigure}
\begin{subfigure}[b]{0.49\textwidth}
\includegraphics[width=\textwidth]{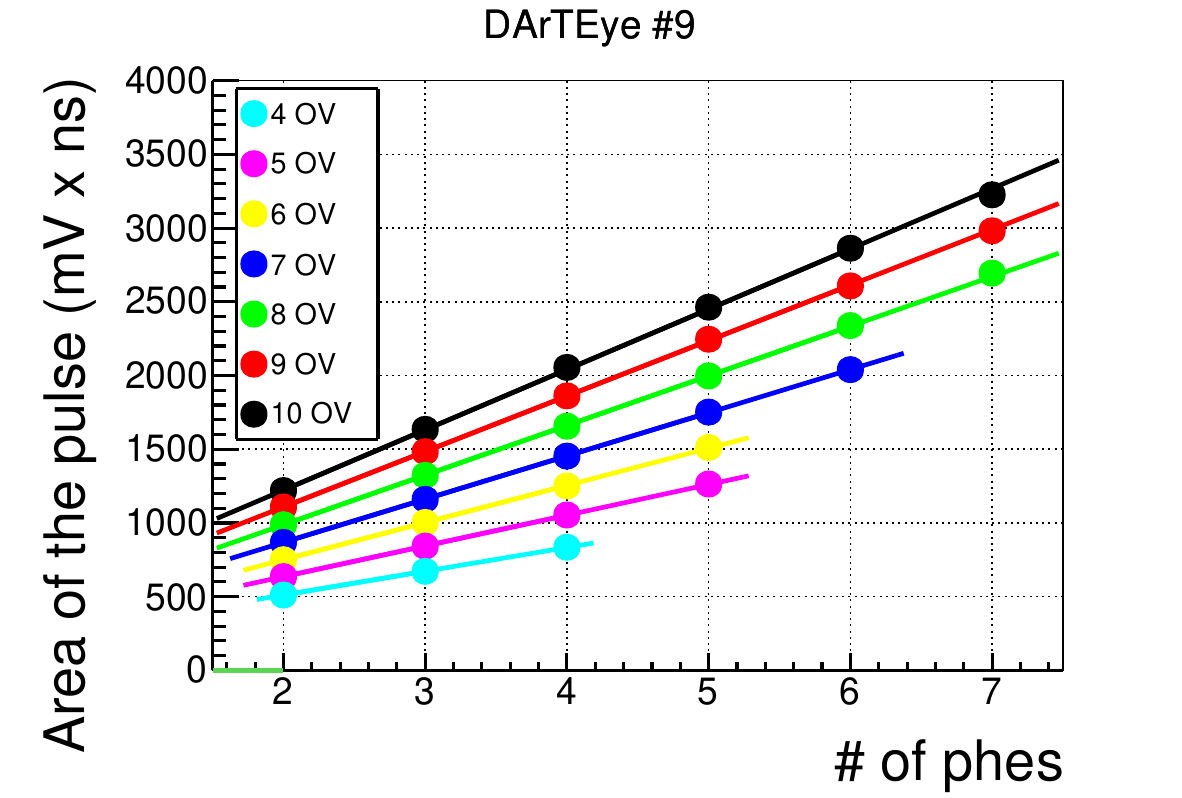}
\end{subfigure}
\begin{subfigure}[b]{0.49\textwidth}
\includegraphics[width=\textwidth]{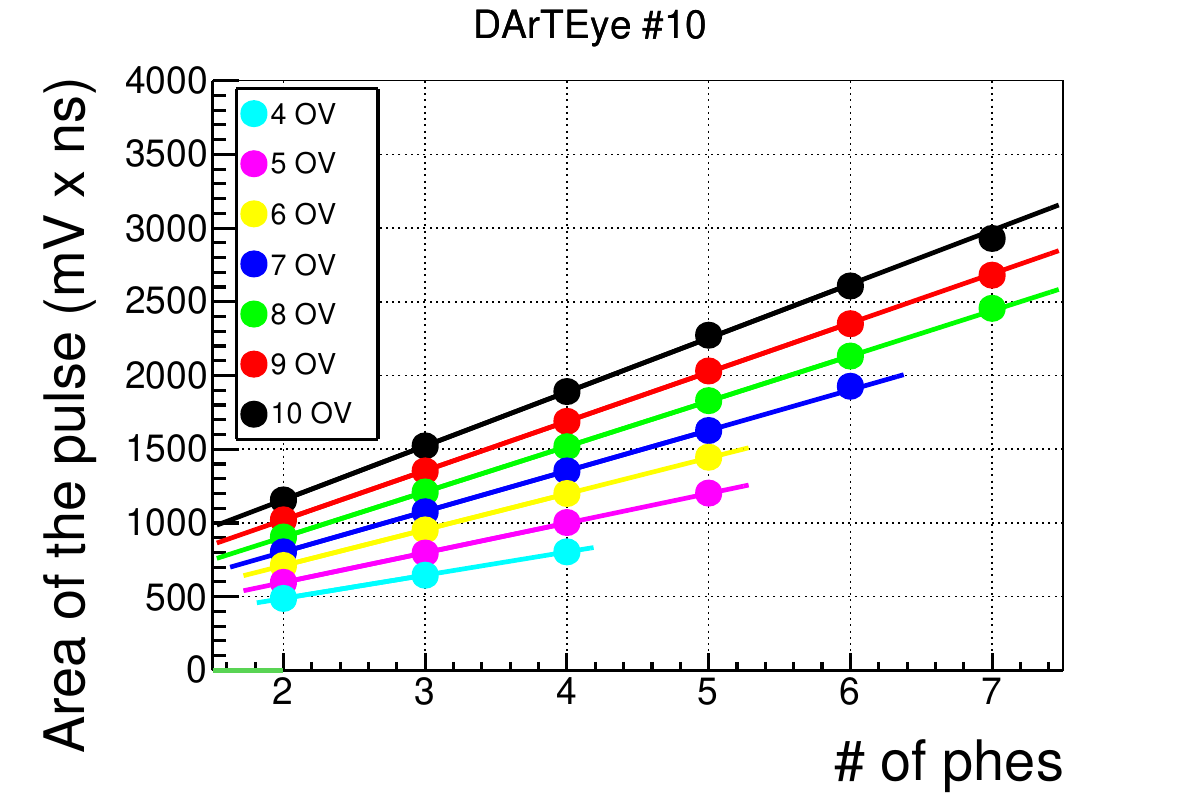}
\end{subfigure}
\begin{subfigure}[b]{0.49\textwidth}
\includegraphics[width=\textwidth]{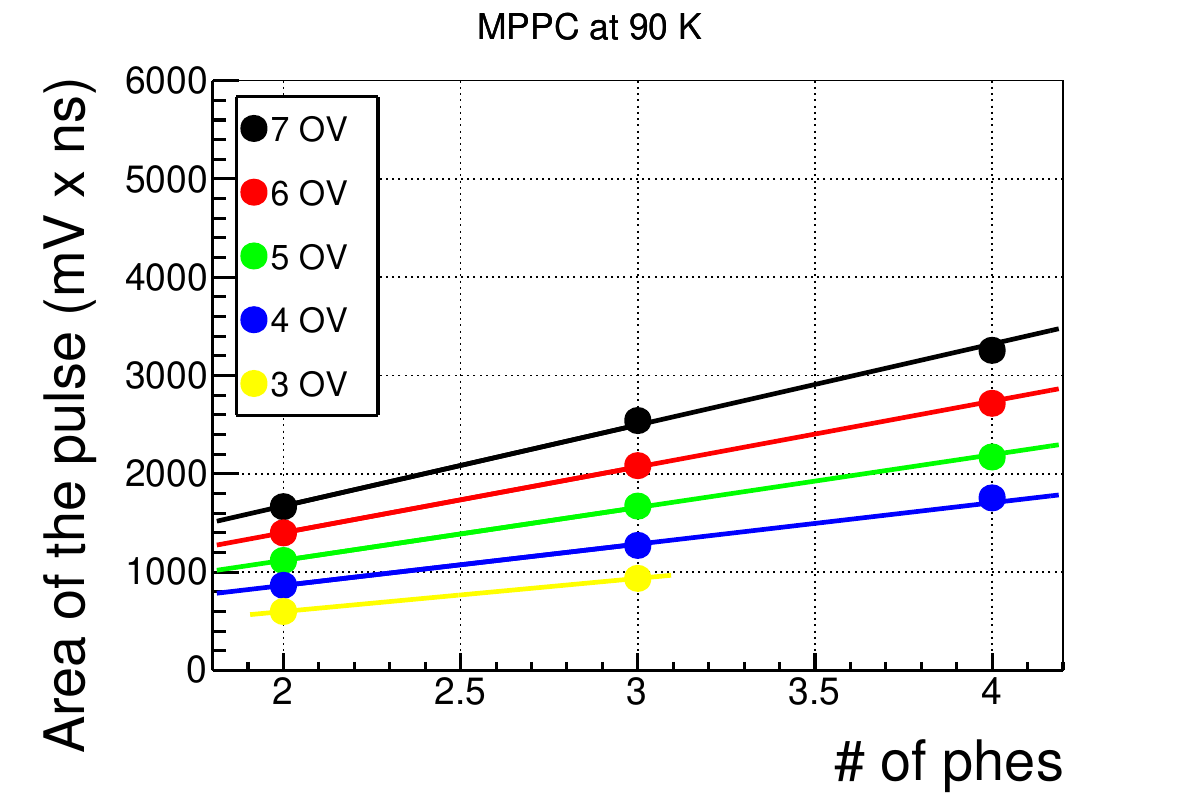}
\end{subfigure}
\begin{subfigure}[b]{0.49\textwidth}
\includegraphics[width=\textwidth]{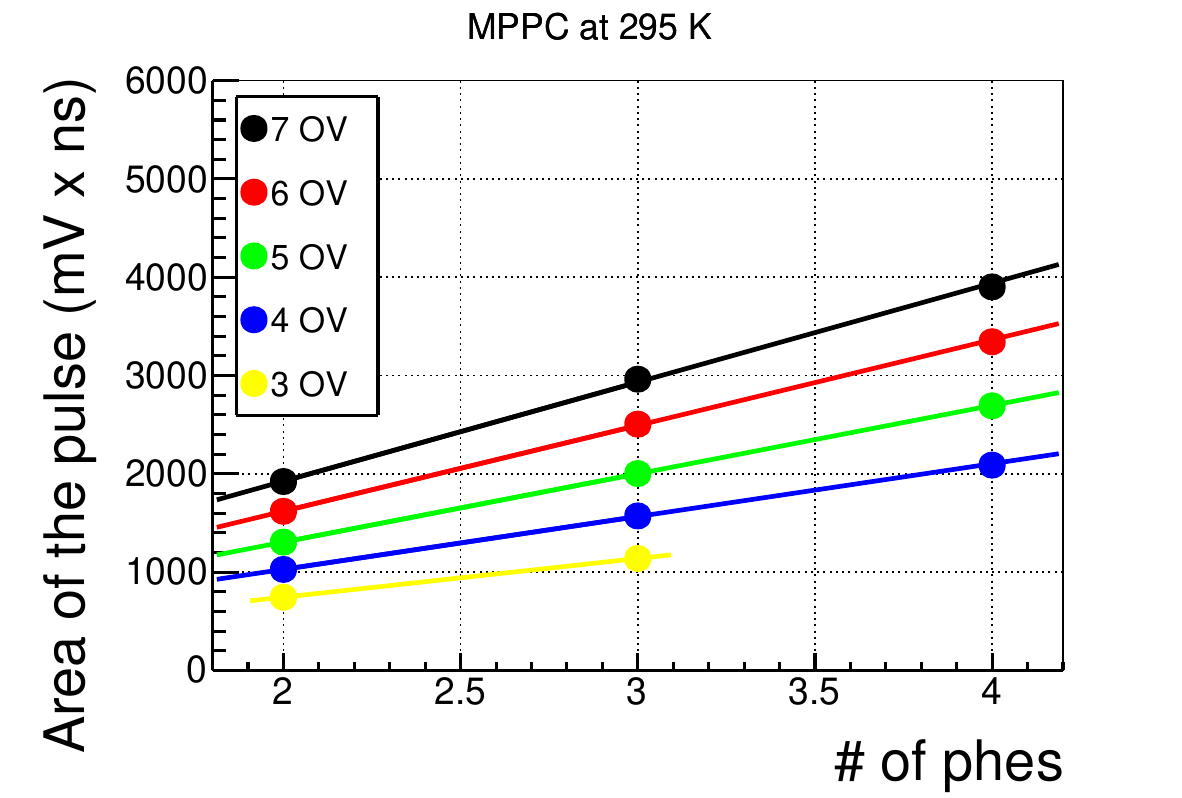}
\end{subfigure}
\caption{\label{SCECal_Area}SPE area calibrations under LED illumination.}
\end{center}
\end{figure}

\begin{figure}[h!]
\begin{center}
\includegraphics[width=\textwidth]{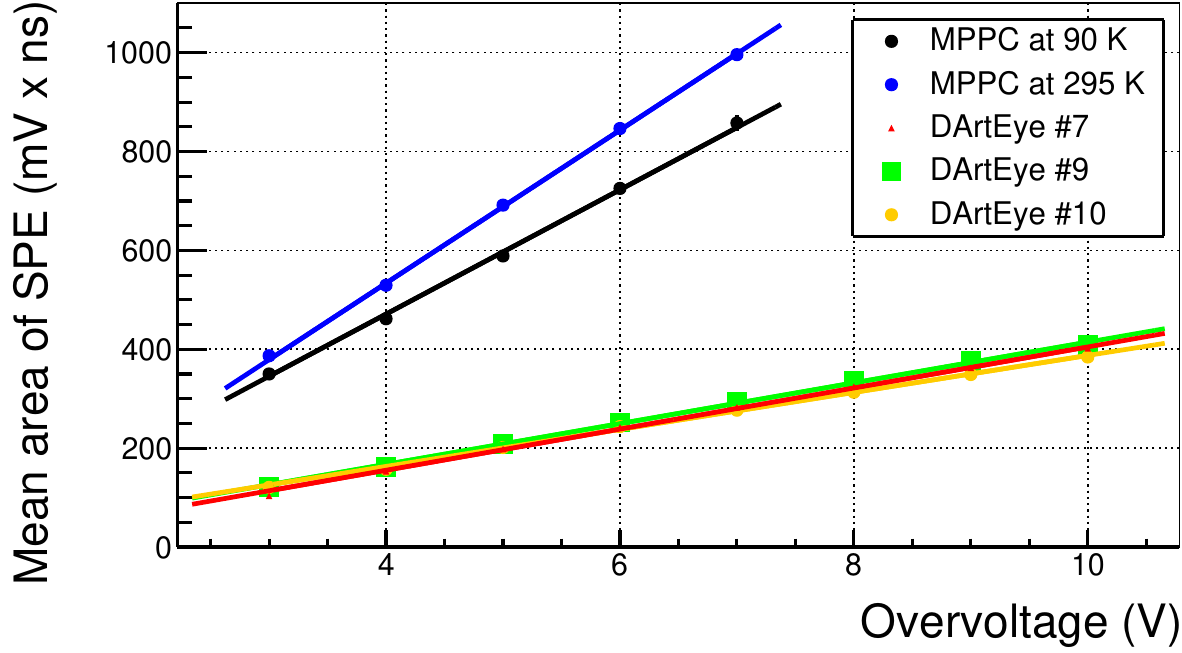}
\caption{\label{GainA}Fits of the mean area of the SPE, $A_{spe}$, as a function of the overvoltage applied for all the SiPMs in each temperature condition.}
\end{center}
\end{figure}

The pulse area for each population was linearly fitted as a function of the corresponding number of photoelectrons to obtain the mean area of the SPE, as it is shown in Figure~\ref{SCECal_Area}. The mean area of the SPE as a function of the overvoltage for each module is shown in Table~\ref{tabla:SPECal_Area}. As the mean area of the SPE is observed to increase linearly with the overvoltage, a linear fit was done following the relation $A_{spe} = c_{A1} \cdot V_{ov} + c_{A0}$. The fits are shown in Figure~\ref{GainA}, and the results in Table~\ref{tabla:Gains_vs_OV}.

\begin{table}[h!]
\centering
\begin{tabular}{|c|c|c|c|c|c|}
\cline{2-6}
\multicolumn{1}{c|}{} & \multicolumn{5}{|c|}{Mean Area (mV ns)} \\
\hline
OV (V) & MPPC 90~K & MPPC 295~K & De.~$\#$7 & De.~$\#$9 & De.~$\#$10 \\
\hline
3 & 349$\pm$6 & 386$\pm$9 & 103$\pm$1 & 120$\pm$1 & 120$\pm$1  \\
4 & 461$\pm$10 & 529$\pm$5 & 152$\pm$1 & 161$\pm$1 & 157$\pm$1  \\
5 & 588$\pm$9 & 691$\pm$7 & 198$\pm$1 & 208$\pm$1 & 201$\pm$1  \\
6 & 725$\pm$11 & 846$\pm$8 & 242$\pm$1 & 251$\pm$1 & 243$\pm$1  \\
7 & 857$\pm$13 & 995$\pm$10 & 282$\pm$1 & 292$\pm$1 & 276$\pm$1  \\
8 & - & - & 324$\pm$1 & 336$\pm$1 & 313$\pm$1  \\
9 & - & - & 362$\pm$1 & 376$\pm$1 & 348$\pm$1  \\
10 & - & - & 401$\pm$1 & 408$\pm$1 & 384$\pm$1  \\ 
\hline
\end{tabular}
\caption{Mean area of the SPE, $A_{spe}$, at different overvoltages for each SiPM. "De" refers to DArTeye.}
\label{tabla:SPECal_Area}
\end{table}

\begin{table}[h!]
\centering
\begin{tabular}{|c|c|c|c|c|}
\hline
Module & $c_{H0}$(mV) & $c_{H1}$(mV/V) & $c_{A0}$(mV$\cdot$ns) & $c_{A1}$(mV$\cdot$ns/V) \\
\hline
MPPC at 295 K & 0.73$\pm$0.48 & 1.68$\pm$0.09 & -85$\pm$14 & 155$\pm$3 \\
MPPC at 90 K & 0.22$\pm$0.46 & 1.57$\pm$0.09 & -32$\pm$14 & 126$\pm$3 \\
De. $\#$7 & -0.11$\pm$0.08 & 0.43$\pm$0.01 & -11$\pm$1 & 42$\pm$1 \\
De. $\#$9 & -0.06$\pm$0.08 & 0.43$\pm$0.01 & 2$\pm$1 & 41$\pm$1 \\
De. $\#$10 & -0.02$\pm$0.10 & 0.41$\pm$0.01 & 13$\pm$1 & 37$\pm$1 \\
\hline
\end{tabular}
\caption{Parameters obtained in the fits of the mean height and mean area of the SPE as a function of the overvoltage shown in Figures~\ref{GainH} and~\ref{GainA}, respectively. "De" refers to DArTeye.}
\label{tabla:Gains_vs_OV}
\end{table}

\newpage

\begin{figure}[h!]
\begin{center}
\begin{subfigure}[b]{0.49\textwidth}
\includegraphics[width=\textwidth]{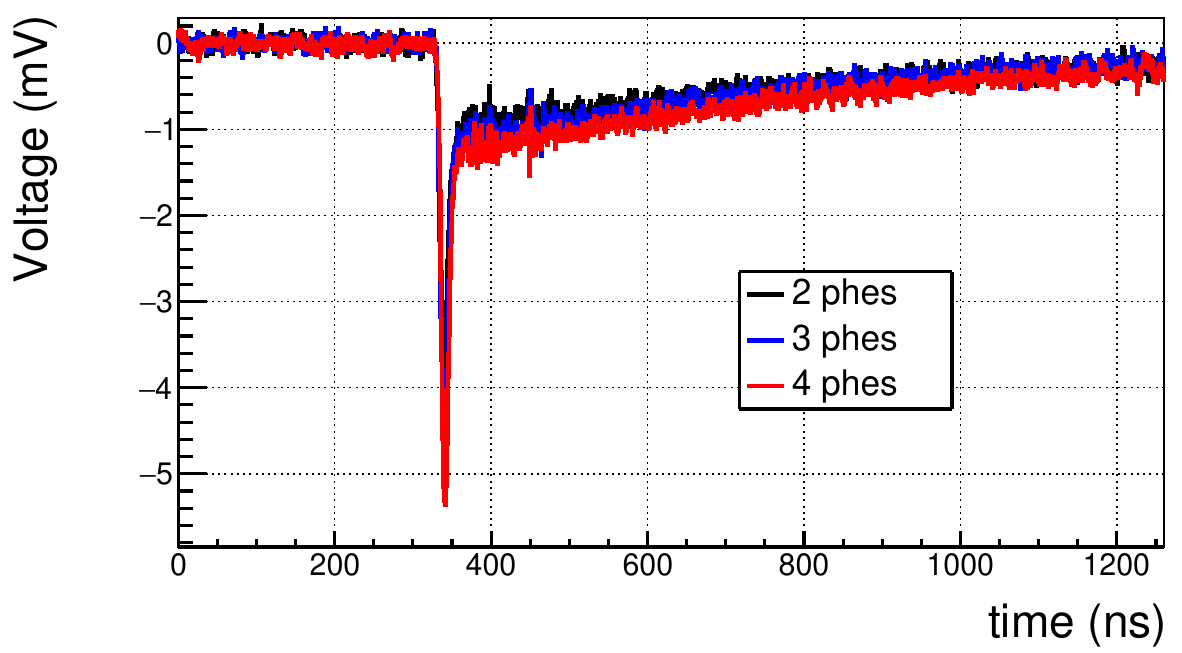}
\end{subfigure}
\begin{subfigure}[b]{0.49\textwidth}
\includegraphics[width=\textwidth]{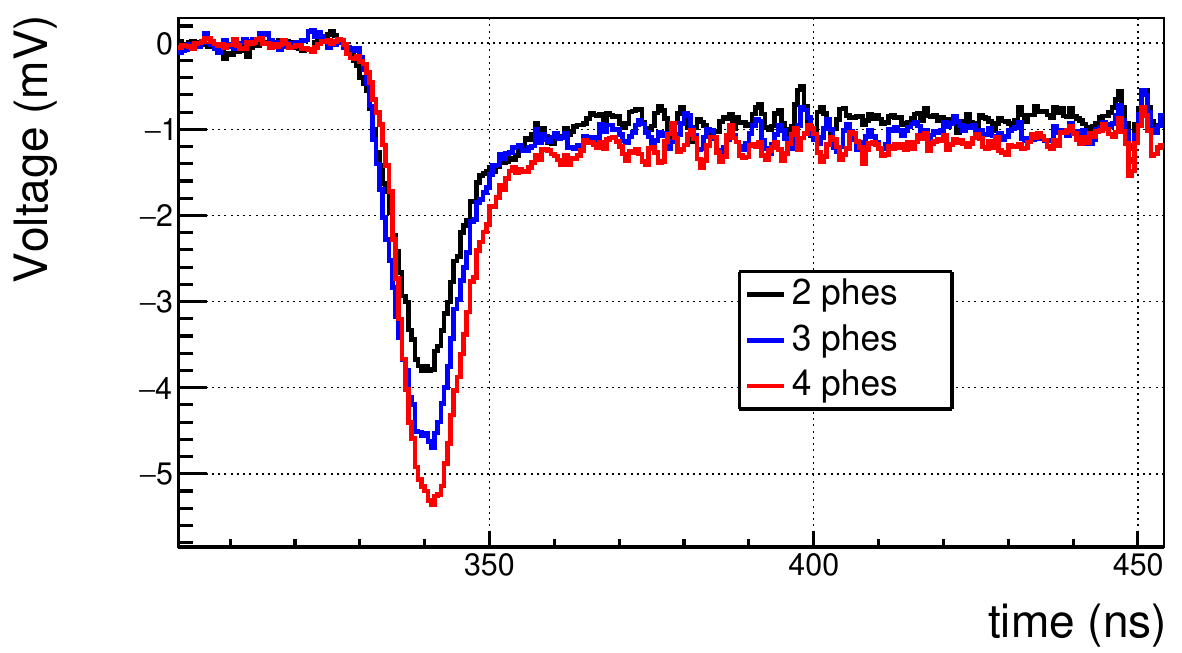}
\end{subfigure}
\caption{\label{AveragedPulsesDArT}Averaged pulses of 1000 events corresponding to 2, 3 and 4~photoelectrons for measurements with DArTeye~$\#$~7 at an overvoltage of 4~V. Right plot shows a closer view of the maximum of the pulse.}
\end{center}
\end{figure}

Averages of the pulses were obtained for the measurements with each DArTeye from the SPE calibration. 1000 events of each one of the populations corresponding to 2, 3 and 4~photoelectrons were selected using the information of their height. They are shown for DArTeye~$\#$~7 at an overvoltage of 4~V in Figure~\ref{AveragedPulsesDArT}. The response of these devices has two different components: a very fast one with a decay time of $\sim$~10~ns and a slow one with a decay time of 600~ns, which is the main contribution to the pulse area. Although these SiPMs are the same as those used in Chapter~\ref{Chapter:SiPMStar2}, the differences in the electronics of the FEBs can explain the important changes in the shape of the pulses.

\begin{figure}[h!]
\begin{center}
\includegraphics[width=0.75\textwidth]{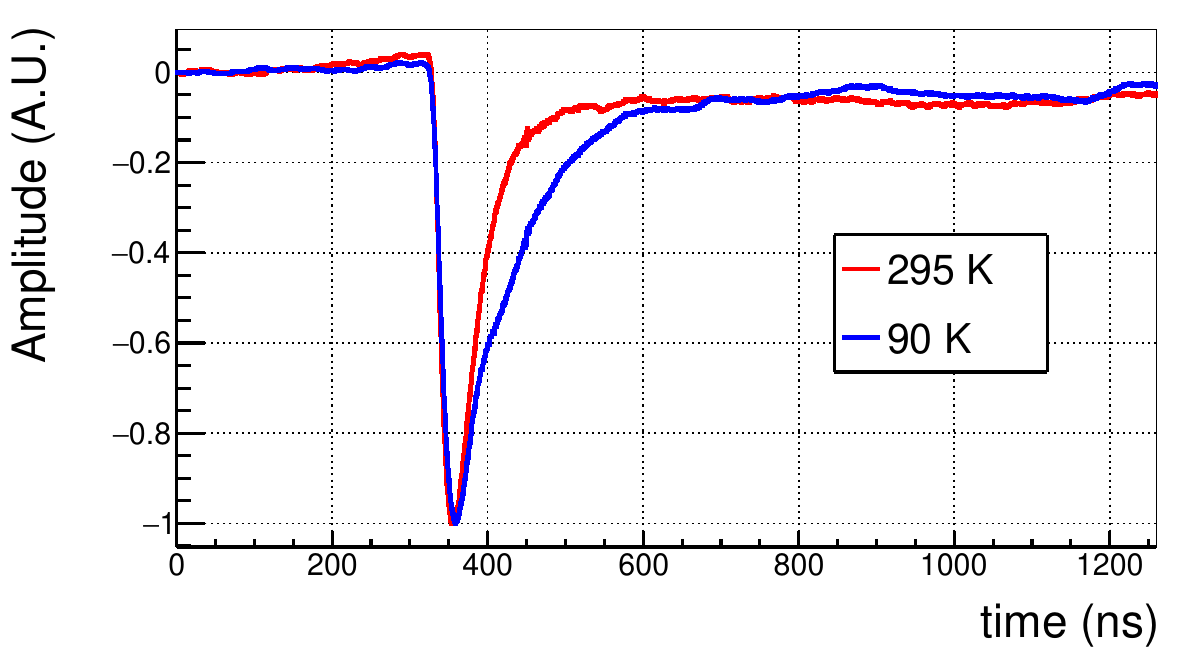}
\caption{\label{AveragedPulsesMPPC}Averaged pulses (normalized to minimum) of 1000 events corresponding to 2~photoelectrons for MPPC at an overvoltage of 4~V at the two different temperatures.}
\end{center}
\end{figure}

The same analysis was done for the measurements with the MPPC, but comparing the shape at each temperature. 1000~events were selected from the population corresponding to 2~photoelectrons at an overvoltage of 4~V. The results are shown in Figure~\ref{AveragedPulsesMPPC}. The response of the MPPC becomes slower at low temperatures. Their mean times have been obtained as 91~ns at 295~K and 106~ns at 90~K. Although most of the pulse is inside the acquisition window, the pulse is not recovering the initial baseline level, pointing at an additional slow component whose mean time cannot be calculated. It clearly affects the pulse area obtained and therefore it can be a source of non-proportionalities.

\begin{figure}[h!]
\begin{center}
\includegraphics[width=0.75\textwidth]{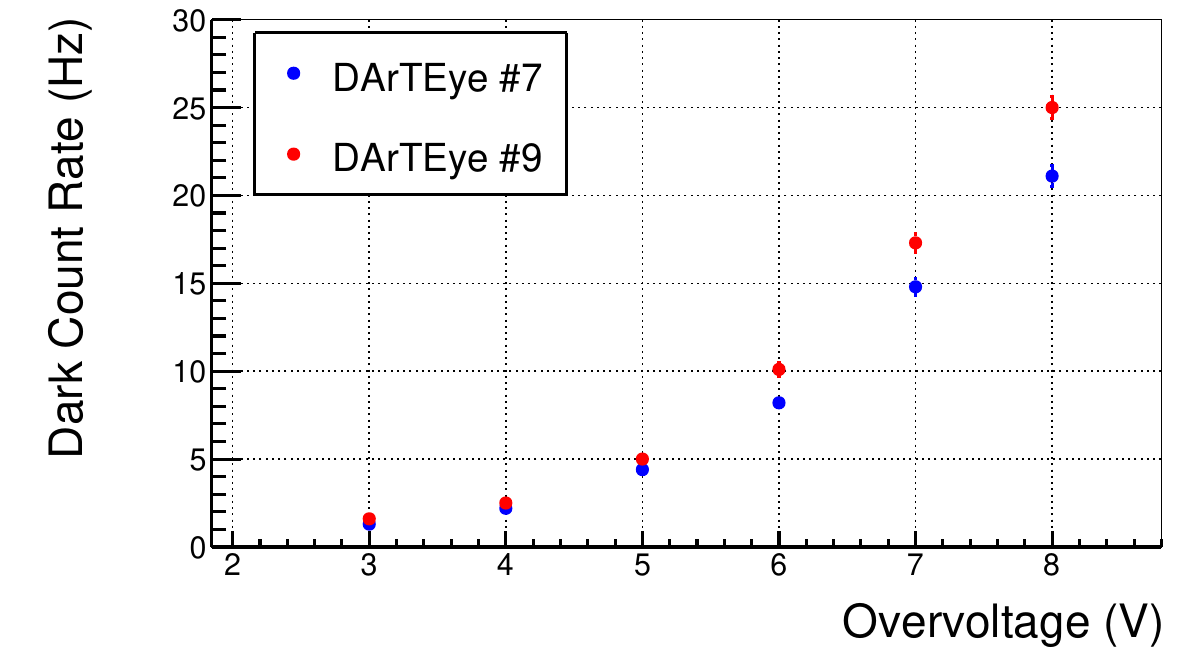}
\caption{\label{DCrates}DC rates of the DArTeyes~$\#$~7 and~9 at 77~K as a function of the overvoltage applied.}
\end{center}
\end{figure}

To measure the DC rate, the threshold of the CFD was fixed lower than the mean height of a SPE, and a Dual Counter Timer 2071A module~\cite{CounterManual} was connected to the CFD output in the configuration shown in Figure~\ref{ElectronicChainSiPM_Dark}. Measurements of one minute long were taken in darkness to obtain the DC rate at overvoltages from 3 to 8~V. In the measurements at room temperature the rate measured by the counter did not increase with the overvoltage, pointing to a saturation of the system at around 400~kHz. For that reason, the measurements were done only at low temperatures. The DC rates of the DArTeyes~$\#$~7 and~9 are shown in Figure~\ref{DCrates}. As it was expected, DC rates increase with the overvoltage, with values of the order of few~Hz. It means that the DC rate observed at room temperatures is strongly reduced at cryogenic temperatures (as it was described in Section~\ref{Section:SiPM_Charac}). The measurements for the MPPC and the DArTeye~$\#$~10 modules are not shown because oscillations in the baseline with amplitudes higher than the threshold applied to the CFD were present during the measurements.

Once characterized the SPE response and the DC, the next step was to analyze the capability to measure the scintillation light of the NaI(Tl) with the MPPC.

\subsection{MPPC response to NaI(Tl) scintillation light} \label{Section:SiPMZgz_CryoT_Characterization_NaI}

In order to measure the scintillation using the MPPC at room temperature, a NaI(Tl) crystal was placed in the test bench. This crystal had a cylindrical shape with equal diameter and length (1"), was encapsulated in aluminum and had one optical window made of quartz. The measurements were done under irradiation with a $^{137}Cs$ source, which has a single gamma emission at 661.7~keV. First, 1.5$\times$10$^5$~events were acquired at an overvoltage of 3~V (close to the value recommended by the manufacturer) obtaining the spectrum and expressing it in number of photoelectrons using the SPE calibration obtained in Section~\ref{Section:SiPMZgz_CryoT_Characterization}. Then, the averaged pulse for the events corresponding to the photoelectric peak is obtained. The same procedure is followed for overvoltages from 3 to 6~V, obtaining in each case the resolution and the LC of the detector. Finally, an analysis of the procedure to reject DC events is done.

\begin{figure}[h!]
\begin{center}
\includegraphics[width=0.75\textwidth]{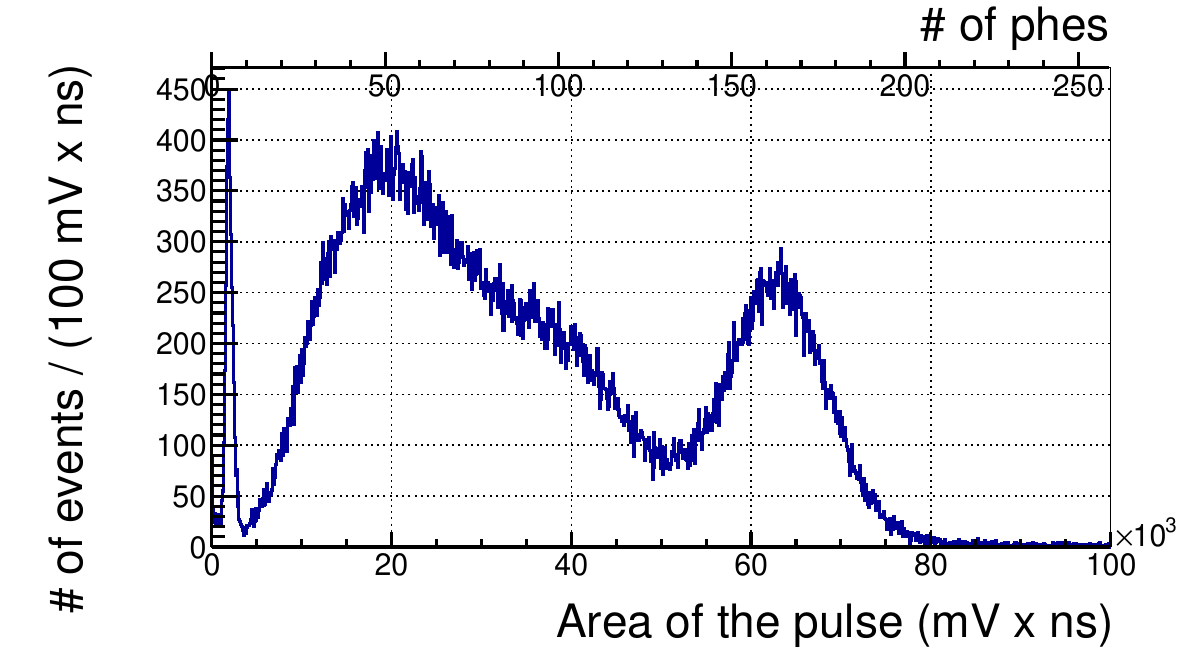}
\caption{\label{Cs137Spectrum}$^{137}Cs$ spectrum obtained using the MPPC as light sensor for the scintillation of a NaI(Tl) crystal at room temperature (295~K) and at an overvoltage of 3~V. Bottom x-axis shows the area of the pulse, while top axis corresponds to number of photoelectrons.}
\end{center}
\end{figure}

\begin{figure}[h!]
\begin{center}
\includegraphics[width=0.75\textwidth]{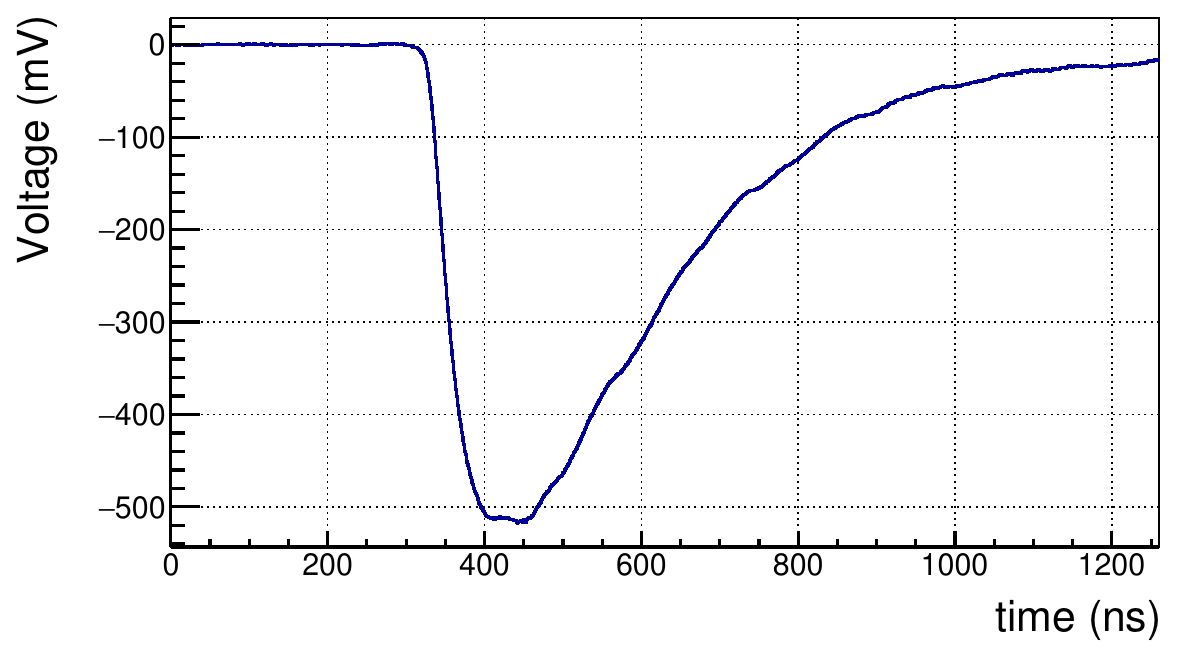}
\caption{\label{NaIPulseRT}Averaged pulse of the events corresponding to the photoelectric peak of the $^{137}Cs$ at an overvoltage of 3~V using the MPPC as light sensor for the scintillation of a NaI(Tl) crystal at room temperature.}
\end{center}
\end{figure}

For the first characterization, the configuration shown in Figure~\ref{ElectronicChainSiPM_Dark} was used, fixing the threshold in the CFD 5~times higher than the mean height of the SPE at room temperature, in order to reduce the triggering of the DC events. Figure~\ref{Cs137Spectrum} shows the corresponding spectrum of pulse areas, also converted into number of phes. To obtain the averaged pulse of the events corresponding to the photoelectric peak, a selection in the pulse area was applied from~55000 to~70000~mV$\cdot$ns. The obtained pulse is shown in Figure~\ref{NaIPulseRT}, the corresponding mean time is 287~ns. The difference between this value and the main scintillation time of the NaI(Tl) can be produced (at least partially) by the slow time component of the SPE observed in Figure~\ref{AveragedPulsesMPPC}.

\begin{figure}[h!]
\begin{center}
\includegraphics[width=\textwidth]{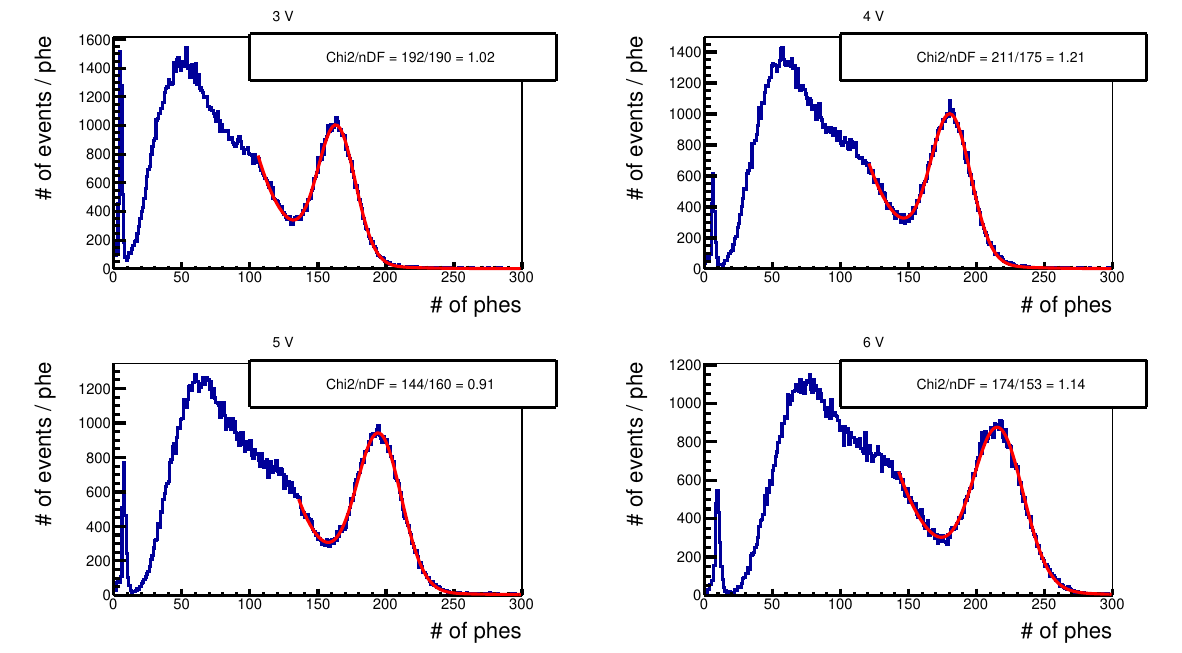}
\caption{\label{Cs137Fits}$^{137}Cs$ measured spectra for overvoltages from 3 to 6~V and the corresponding fits of the photoelectric peak using the MPPC as light sensor for the scintillation of a NaI(Tl) crystal at room temperature.}
\end{center}
\end{figure}

\begin{figure}[h!]
\begin{center}
\includegraphics[width=0.75\textwidth]{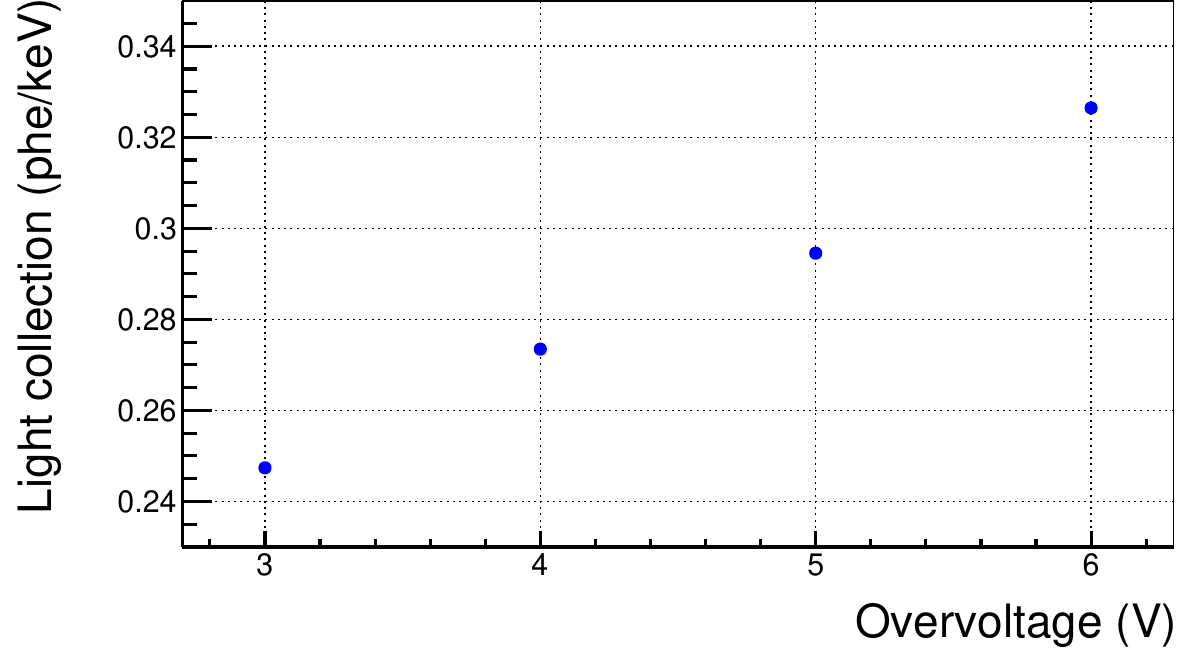}
\caption{\label{NaILy_vs_OV}LC of the NaI(Tl)+SiPM system at room temperature at 661.7~keV as a function  of the overvoltage.}
\end{center}
\end{figure}

\begin{figure}[h!]
\begin{center}
\includegraphics[width=0.75\textwidth]{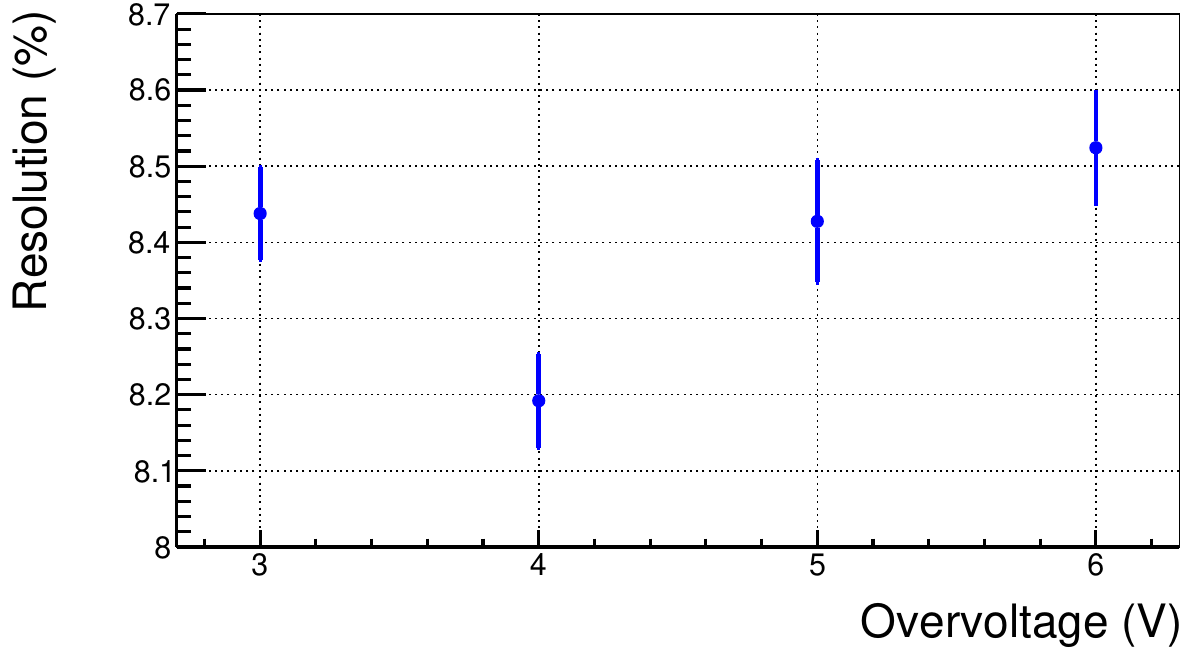}
\caption{\label{NaIRes_vs_OV}Resolution of the NaI(Tl)+SiPM system at room temperature at 661.7~keV as a function  of the overvoltage.}
\end{center}
\end{figure}

Similar measurements were taken with the $^{137}Cs$ source at overvoltages from 3 to 6~V. In each spectrum, the photoelectric peak was fitted to a gaussian with area, mean and sigma free over an exponential decay as background, as it shows the Figure~\ref{Cs137Fits}. From each fit, the resolution was obtained as the ratio of the sigma to the mean of the gaussian, and the LC as the mean number of photoelectrons of the photopeak divided by its energy. The obtained LC and energy resolution are summarized in Figures~\ref{NaILy_vs_OV} and~\ref{NaIRes_vs_OV}, respectively.

Both LC and resolution  exhibit a similar behaviour vs. the overvoltage as observed in the measurements done in Chapter~\ref{Chapter:SiPMStar2}, and follow the model of the PDE and CT developed in Chapter~\ref{Chapter:SiPM_Intro}. However, a fit to Equation~\ref{eq:LC(OV)} was not attempted due to the limited number of data points available. There was a plan for these dedicated measurements but unfortunately the power supply incorporated in the FEB was damaged before those measurements. This analysis was stopped until a detector with a higher LC was available (see Section~\ref{Section:SiPMZgz_Detector}).

\begin{figure}[h!]
\begin{center}
\includegraphics[width=0.75\textwidth]{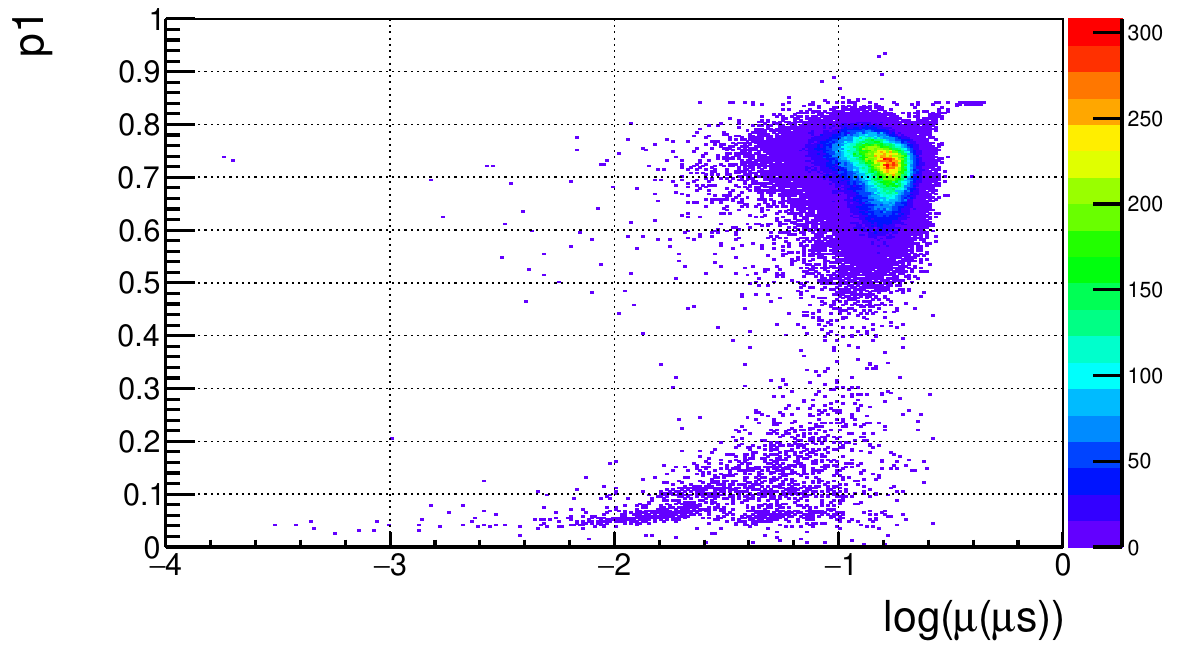}
\caption{\label{p1_vs_mu_MPPC}Scattering plot of log($\mu$) versus $p_1$ variables for the measurement with the $^{137}Cs$ source at an overvoltage of 6~V.}
\end{center}
\end{figure}

Low light events can be observed in the four spectra of Figure~\ref{Cs137Fits} as a peak at low number of photoelectrons. An analysis of the pulse shape was done to analyze this contribution using $\mu$ and $p_1$ variables, the same used in ANAIS experiment. Figure~\ref{p1_vs_mu_MPPC} shows the scattering plot of both variables for measurement at an overvoltage of 6~V.

\begin{figure}[h!]
\begin{center}
\includegraphics[width=0.75\textwidth]{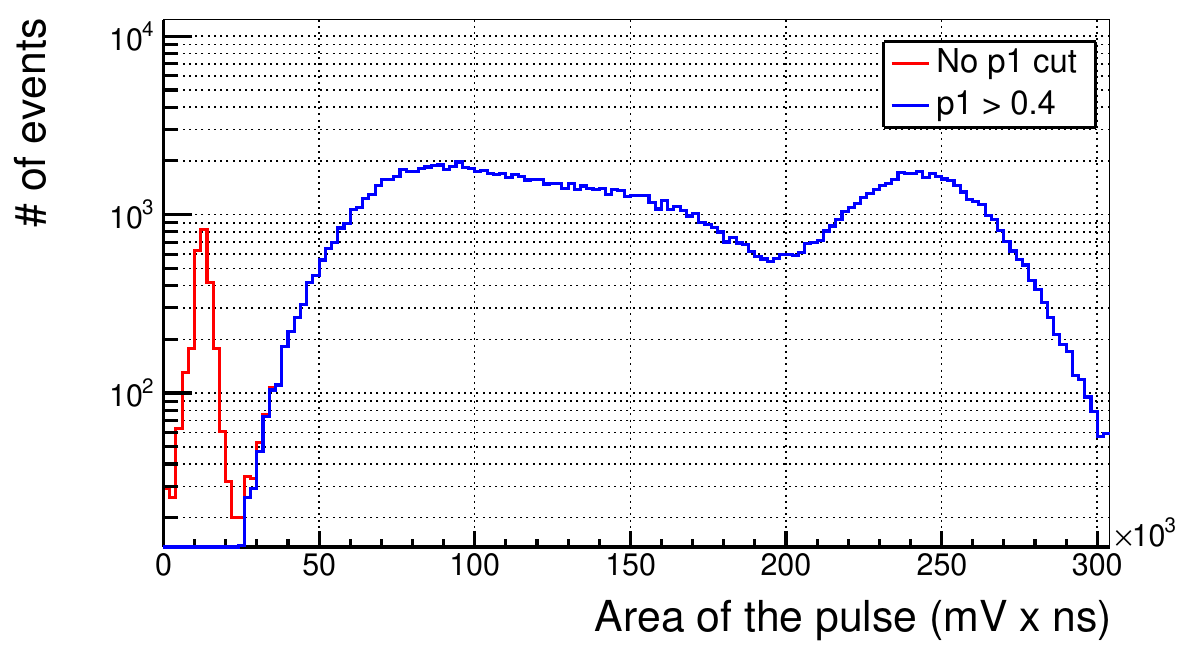}
\caption{\label{MPPC_p1Cut}Comparison of the spectra of the measurement with the $^{137}Cs$ source at an overvoltage of 6~V with (blue spectrum) and without (red spectrum) $p_1 > 0.4$ selection.}
\end{center}
\end{figure}

The scintillation events correspond to $p_1 >$~0.4. The other population of events can be DC events or Cherenkov light produced in components of the detector as the quartz window of the crystal. It is possible to observe that a simple cut in the $p_1$ variable above~0.4 (similar to that applied in ANAIS-112 experiment, see~\cite{Amare:2018sxx}) is enough to reject those events. The comparison of both spectra (with and without this cut) is seen in Figure~\ref{MPPC_p1Cut}.

The measurements presented in this section suggest that SiPMs can be applied to detect scintillation light from NaI(Tl) at room temperature, as it was observed in Chapter~\ref{Chapter:SiPMStar2}. Again, a significant decrease in the DC rate was observed at lower temperatures, which could lead to improved resolution and lower energy thresholds for scintillation detection in NaI(Tl). The LC obtained in the measurements with NaI(Tl) was clearly low compared with its light yield (40 photons/keV) because the system was not designed to be efficient in the LC. A new design for a NaI(Tl)+SiPM detector is required to improve LC and ease the operation at low temperature. The coupling between the SiPM and crystal has to be improved by avoiding the encapsulation of the crystal, which implies keeping a dry atmosphere in the interior of the detector system, which should be tightly closed to prevent that ambient humidity damages the crystal.

\section{Compact NaI(Tl)+SiPM detector} \label{Section:SiPMZgz_Detector}
\fancyhead[RO]{\emph{\thesection. \nameref{Section:SiPMZgz_Detector}}}

In this section, the design of a compact NaI(Tl)+SiPM detector is shown, as well as the new experimental setup developed for its characterization. Finally, the measurements planned are also summarized.

\subsection{Detector design} \label{Section:SiPMZgz_Detector_Design}

\begin{figure}[h!]
\begin{center}
\includegraphics[width=0.5\textwidth]{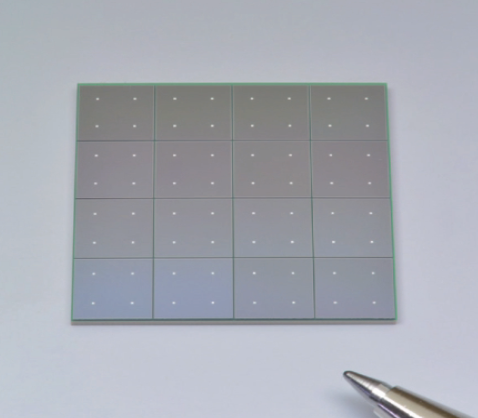}
\caption{\label{ArrayMPPC}Picture of the SiPM array model S13361-6050PE~\cite{SiPMHamamatsuArrayManual}.}
\end{center}
\end{figure}

The SiPMs used in the previous section (MPPCs HAMAMATSU model S13360-6050PE) were purchased in a 4$\times$4 array (model named S13361-6050PE~\cite{SiPMHamamatsuArrayManual}), with a surface of 25$\times$25~mm$^2$. A picture of this array can be seen in Figure~\ref{ArrayMPPC}. To bias each cell and extract its signal, the input and output voltage lines of each SiPM are grouped in two connectors on the backside of the array. This allows to apply a different bias voltage to each SiPM and to read each one of them independently. The bias/read-out configuration is determined by the FEB used.

\begin{figure}[h!]
\begin{center}
\includegraphics[width=0.75\textwidth]{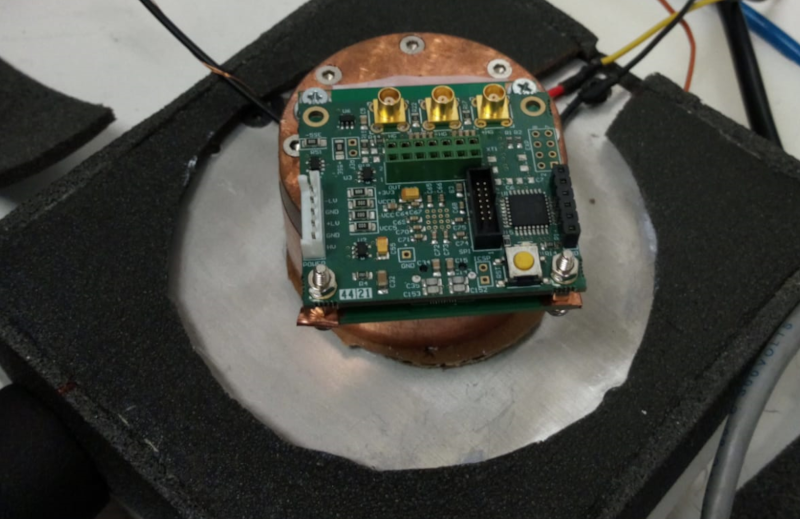}
\caption{\label{eMusicPicture}Picture of the eMUSIC board installed for the readout and biasing of the SiPM array in the NaI+SiPM prototype.}
\end{center}
\end{figure}

For the measurements with the prototype, an eMUSIC~\cite{eMUSICboardManual} board designed by SCIENTIFICA~\cite{SCIENTIFICA} was selected (it is shown in Figure~\ref{eMusicPicture}). It requires the connection of the power supply for the bias voltage of the SiPM and a $\pm$~6.5~V voltage for biasing the preamplifier and other electronics components. To connect the SiPM array to the eMUSIC, an interface board was designed in the University of Zaragoza. The eMUSIC is controlled with a computer using a software specifically designed with this purpose. It has some interesting features that make it a very versatile board: the 16 single SiPMs of the array can be measured in eight different channels, being the signal of each channel the sum of two SiPMs. Moreover, the software allows to enable or disable the read-out of any of these channels. Additionally, an RC filter can be applied to the output signal of the preamplifier to modify the shape of the pulses, and the values of the resistance and capacitance can be modified at will. Finally, a different offset in the bias voltage can be applied to each channel to homogenize the gain of the SPE in case of small differences in the breakdown voltage between the individual SiPM units. 

\begin{figure}[h!]
\begin{center}
\includegraphics[width=\textwidth]{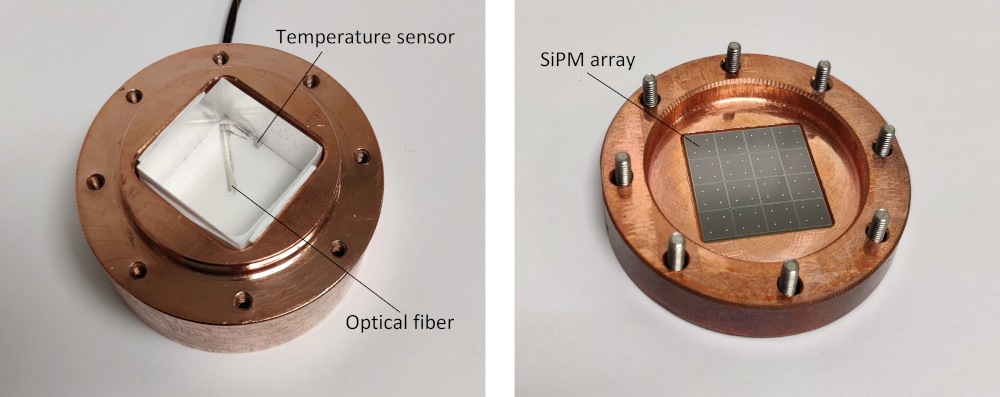}
\caption{\label{CompactDetector}Pictures of the two sides of the NaI(Tl)+SiPM detector. See the text for detailed information of the detector structure. Due to the hygroscopic character of the NaI(Tl) crystal, a cubic piece of methacrylate was used instead of the NaI(Tl) crystal for taking the left picture.}
\end{center}
\end{figure}

The detector designed consists of a cylindrical structure made of copper with a diameter of 6~cm and a length of 3.5~cm, as seen in Figure~\ref{CompactDetector}. This copper structure has two pieces, which hold the NaI(Tl) cubic crystal (1" side), manufactured by Hilger Crystals~\cite{HilgerCrystals}, and the SiPM array (see Figure~\ref{ArrayMPPC}). After sealing both pieces, the SiPM array lays on top of the crystal. This copper piece has built-in the connectors required to led in the bias voltage for the array and to take out the output signals to the readout electronics. The detector design should guarantee the tightness against water for a long time. This is achieved thanks to a kapton seal between the two copper pieces, closed under the pressure of 8~screws. The connectors were sealed using epoxy resin. Tightness tests were carried out filling the detector with Helium at overpresure, and approaching a leak detector to the structure joints. Such tests were also done at liquid nitrogen temperature, after the immersion of the detector in a dewar. In both cases, the concentration of Helium measured by the leak detector was compatible with the atmospheric value.

The coupling between the SiPM and the NaI(Tl) crystal is a critical point. Ideally, it should be done by using a convenient interface with intermediate refractive index between the two surfaces to be coupled, in order to increase the transmission of the light by reducing the internal total reflection. Silicone grease or silicone pads are typically used for this goal at room temperature. However, SiPMs could be damaged in the process of coupling and the behaviour at low temperature of these materials had not been tested, which led us to avoid the use of any optical coupling. This results in poorer light transmission, because there is air between the crystal and the SiPM. On the other hand, avoiding optical coupling media reduces the possible radioactive background contribution and emission of spurious light in that material.

To allow the calibration of the SPE response of the SiPM array with the LED, an optical fiber was installed inside the copper cylinder (see Figure~\ref{CompactDetector}), letting the crystal between the optical fiber and the SiPM array. In order to homogeneously distribute the light on the surface of the SiPM array, the optical fiber was laid parallel to the SiPM and its surface was scratched in several lines so that the light could exit perpendicular to the direction of the fiber. This or a similar system will be used in the future to extract the scintillation light of the crystal to analyze the emission spectrum at different temperatures using an spectrometer. To measure the temperature of the crystal, a PT100 resistor was put in contact with the crystal with thermal paste. Finally, trying to maximize the LC of the system the NaI(Tl) crystal was completely covered by teflon tape, except the crystal side facing to the SiPM array. The surface below the crystal was also covered by teflon but allowing the contact of the PT100 sensor to the crystal and the light transmission of the optical fiber (see left picture in Figure~\ref{CompactDetector}).

\begin{figure}[h!]
\begin{center}
\includegraphics[width=\textwidth]{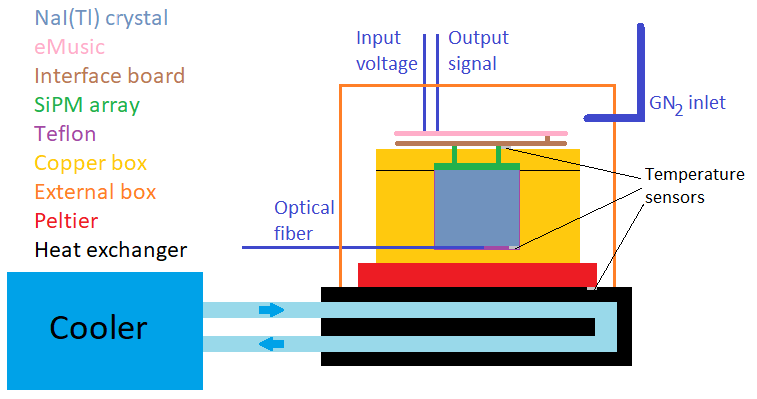}
\caption{\label{SetUpSchemeCompact}Scheme of the setup used in the measurements with the compact NaI(Tl)+SiPM detector. The polyethylene foam covering the aluminum box is not drawn to allow a more simplified image of the setup.}
\end{center}
\end{figure}

A cooling system was designed to characterize the detector until temperatures of around -40$^o$C. An scheme of the complete system is presented in Figure~\ref{SetUpSchemeCompact}. It consists of a metallic surface that cools down through the Peltier effect and on which the detector is placed in contact with thermal paste. This metal surface is in thermal contact with a heat exchanger filled with a refrigerant liquid which is connected to a cooler in a closed circuit. The temperature is monitored in four different positions of the system: in the crystal, in the SiPMs array, in the heat exchanger and in the cooler. The temperature of the detector is controlled by selecting the operation voltage of the Peltier system and the temperature of the refrigerant liquid in the cooler. To avoid condensation around the detector, all the system is placed inside a cylindrical box made of aluminum and flushed with nitrogen gas (GN$_2$ in the figure). Input voltage and output signal connectors are placed on the top of that box. Finally, the aluminum box is covered by 5~mm-thick thermal insulator made of polyethylene foam.

An 8-channel 14-bit 250~MS/s digitizer model DT5725 made by CAEN was purchased for these measurements, and the DAQ electronic chains used both under LED illumination or for scintillation measurements are the same as those presented in Section~\ref{Section:SiPMZgz_Setup}.

\subsection{Planning of the measurements} \label{Section:SiPMZgz_Detector_Meas}

The measurements with this detector started during the last months of my PhD, in the framework of the degree's thesis of Laura Navarro, under my supervision, and the PhD of Jaime Apilluelo. In the following, an overview of the measurements and analysis scheduled is provided.

As it was explained in Chapter~\ref{Chapter:SiPM_Intro}, the two most relevant operation conditions that determine the behaviour of the detector are the temperature and the overvoltage, and therefore the measurements protocol was designed to characterize some properties of the detector as a function of these observables.

First, as the eMUSIC board allows to apply different filters to the SiPM output signal, it is important to analyze the effect of this filtering in terms of SPE pulse mean time and gain. Fast SPE reduce the integration time required to acquire the complete pulse, thus improving the resolution of the SPE in the amplitude and area spectra by reducing the DC contribution.

Moreover, fast response also implies an easier reconstruction of the scintillation pulse shape, and at low energies, a better identification of single phe in the signal pulse, enabling the development of event discrimination filtering procedures, as those developed for ANAIS (explained in Section~\ref{Section:ANAIS_Filter}). On the other hand, higher gain increases the resolution of the SPE in the amplitude and area spectra and the SNR (both in the SPE calibration and in the scintillation measurements).

Another relevant measurement is the DCR. As the scintillation light is always present in the measurements with this prototype, two methods can be followed for that characterization. The first one consists of calculating the baseline without considering any DC event and subtracting it from the pulse. The integral of the signal in the pretrigger region, $a_{pT}$, divided by the area of the SPE will provide the number of DC in the pretrigger time. Therefore, the DCR can be calculated as follows:
\begin{equation}
DCR = \frac{a_{pT}}{a_{SPE}\cdot t_{pT}},
\end{equation}
where $t_{pT}$ is the pretrigger time, in seconds. The other method is to apply a peak-finding algorithm similar to the one used in the ANAIS-112 analysis (see Section~\ref{Section:ANAIS_Analysis}) to the acquired pulses. The DCR is then calculated as the number of peaks found in the pretrigger region (which will be considered as photoelectrons) divided by the pretrigger time. This method requires a highly efficient phe identification and very low noise level.

A very important characterization of the detector is the LC, which is obtained by measuring the scintillation signal corresponding to a known energy and dividing by the previously calibrated SPE area. The dependence of the LC with the overvoltage was modeled in the Chapter~\ref{Chapter:SiPM_Intro}, and it is described by the Equation~\ref{eq:LC(OV)}. If the AP probability and the DCR are measured independently of the LC measurement, they can be included in the model as fixed parameters. This measurement can be done channel by channel to analyze its distribution among them, and it should be done at different temperatures in order to check the validity of this model. Moreover, this test should be repeated several times at room temperature, after one or several cooling cycles to analyze the possible effect of mechanical and thermal stress while warming and cooling on the optical properties of the system.

The first preliminary measurements carried out from room temperature down to -32~$^o$C using a $^{133}Ba$ source show a good LC and resolution. At 81~keV, the maximum LC reached by the detector, $LC^{scint}_{max}$ after discounting the CT contribution, is between 6 and 7~phe/keV, which is not too far from the LC of each PMT in the ANAIS-112 experiment, which has a mean value of 7.22~phe/keV. The optimal energy resolution in the current conditions of the detector was reached at overvoltages from 5 to 6~V with values between 8 and 9\% for the 81~keV peak.

\section{Future work with NaI(Tl)+SiPM detectors} \label{Section:SiPMZgz_Conclusions}
\fancyhead[RO]{\emph{\thesection. \nameref{Section:SiPMZgz_Conclusions}}}

In this section, the future work planned with NaI+SiPMs detectors is overviewed. In the short term, the previously explained characterizations will be done from room temperature to the minimum temperature reachable by the currently in operation system. It is planned to design a new electronic board that will be connected to the eMUSIC board to read the eight channels of the SiPM array independently with the 8-channel digitizer available. It will allow to perform the SPE calibration channel by channel and also to discriminate DC events by requiring coincidence among different channels. Also in the short term, it is expected to carry out a measurement of the emission spectrum of the NaI(Tl) crystal in the temperature range down to -40~$^o$C using a spectrometer.

In the middle term, a cryocooler (that has been already purchased) is going to be received in the next months. It will allow to characterize the detector in a wide range of temperatures down to about 80~K. This system will be used for the characterization of a similar prototype to that explained in section~\ref{Section:SiPMZgz_Detector} but using pure NaI crystal as scintillator. As shown in Chapter~\ref{Chapter:SiPMStar2} this material has a higher light emission than NaI(Tl) at low temperature, and a faster scintillation, being a quite interesting target for the ANAIS+ purposes. 

In the medium-long term an improved detector can be constructed. To increase the LC, four of the six faces of the crystal will be covered by SiPM arrays. Moreover, convenient materials as transparente silicone grease can be tried to increase the light transmission between SiPM array and crystal. Before, its behaviour at low temperature should be checked. In the longer term, the objective is to develop a demonstrator in the 1 kg mass range incorporating in the design all the knowledge acquired. This demonstrator could be characterized in terms of background and general performance assessment in a bath of liquid argon at the Canfranc Underground Laboratory.

%% file: Conclusions.tex
\setcounter{secnumdepth}{0}
\chapter*{Summary and conclusions}\label{Chapter:Conclusions}
\fancyhead[LE]{\emph{\nameref{Chapter:Conclusions}}}
\addcontentsline{toc}{chapter}{Summary and conclusions}
\markboth{Summary and conclusions}{Summary and conclusions}

Numerous astronomical and cosmological observations (such as the galaxy rotation curves, the anisotropy spectrum of the cosmic microwave background or gravitational lenses) suggest the existence of a type of matter called "Dark Matter" that makes up the 26\% of the mass-energy density of the universe. This type of matter does not interact electromagnetically (then it does not emit or absorb light), and it can only have weak interactions with the rest of the matter. These requirements rule out the possibility that this matter is made up of particles from the Standard Model of Particle Physics, and candidates beyond this model have to be searched for, being Axions and WIMPs among the most successful. The understanding of the nature of this dark matter is at present one of the most important challenges faced by the Physics, in the frontier from cosmology, astrophysics and particle physics.

The DAMA/LIBRA experiment, which uses NaI(Tl) scintillator crystals and is located at LNGS, is the only experiment that has observed an annual modulation in its detection rate consistent with the signal that would produce dark matter particles distributed in the galactic halo following the most standard model. However, this result has not been confirmed by any other experiment, but it has neither been ruled out in a way independent of the particle model and halo model. For this model-independent testing of DAMA/LIBRA using the same target material is an important asset. This is the goal of ANAIS-112 and COSINE-100 experiments. The ANAIS-112 experiment, which has been taking data since August 3rd, 2017 at the LSC, consists of a 3$\times$3 matrix of cylindrical NaI(Tl) crystals of 12.5~kg each (thus reaching a total mass of 112.5~kg). The scintillation of each crystal is read by two HAMAMATSU R12669SEL2 Photomultiplier Tubes (PMTs), with high Quantum Efficiency (QE). ANAIS-112 modules have shown to have a high light collection, above 14~ph/keV in 7 out of the 9~modules. This results in a low energy threshold of the experiment. Although triggering efficiency is good below 1~keV, the analysis threshold is set at 1~keV, allowing to explore the DAMA/LIBRA signal.

The works included in this memory correspond to developments carried out in parallel to the ANAIS-112 data taking and analysis, but sharing the same ANAIS goals. First, the knowledge of the response of NaI(Tl) crystals to nuclear recoils is required in order to compare ANAIS and DAMA/LIBRA results. Second, the origin of some of the spurious events populations found in ANAIS-112 could be analysed by developing convenient optical simulations of the system. Finally, the possibility of improving the sensitivity of NaI scintillators applied to dark matter searches by replacing the Photomultiplier Tubes by SiPMs provides very interesting opportunities that are under research in the frame of a new project ANAIS+.

One of the main goals of my PhD dissertation has been the analysis of the measurement of the scintillation quenching factor of $Na$ and $I$ nuclei in NaI(Tl) crystals. This parameter is defined as the ratio between the amount of light emitted by the crystal after a given energy deposition by a nuclear recoil and that emitted in an electronic recoil for the same deposited energy. Since the ANAIS-112 experiment is calibrated in energy using x-ray and gamma-ray sources, which interact through electronic recoils, knowing this factor is essential to interpret the measured energy depositions in terms of nuclear recoil energy spectra produced by dark matter particles. Although it has been measured since the 1990s for both $Na$ and $I$, the values obtained by different experiments are not compatible with each other. In order to measure this factor and analyze the possible systematics that affect the measurements, 5~crystals produced by Alpha Spectra grown from powder with different quality, were characterized in the same experimental set-up.

These measurements were carried out in August and October 2018 at the Triangle Universities Nuclear Laboratory (TUNL) within a collaboration between members of the COSINE-100, COHERENT and ANAIS-112 experiments. In them, protons were directed against a LiF target, producing the reaction $^7Li(p,n)^7Be$, which releases monoenergetic neutrons with an energy of approximately 1~MeV. These neutrons were targeted at the NaI(Tl) crystal, where they could scatter, producing a nuclear recoil. The scattered neutron was later detected by one of the neutron detectors (BDs) that surrounded the crystal allowing to determine the nuclear recoil energy by measuring the corresponding neutron scattering angle. On the other hand, the light emitted by the crystal was measured by a PMT coupled to the crystal, being the same PMT used for all the measurements. The measurements were designed to minimize the effect of different systematics as, for instance, reducing the background produced by multiple scattering using small crystals, reducing the possible effects of channeling by rotating periodically the crystals and avoiding threshold effects by triggering the system with the neutron signal in the BDs.

The QF calculation requires selecting single scattering of a neutron in the NaI(Tl) and the detection of the scattered neutron in a BD. This implies to develop an analysis protocol which rejects multiple hits in the BDs, and which selects neutron events in the BDs (both by the pulse shape and the time of flight). To avoid the threshold effect in the crystal spectrum, the integration window of its signal was fixed taking into account the time of flight of the neutrons from the crystal to the BDs. Moreover, a simulation of the experiment using the GEANT4 package allowed to obtain the nuclear recoil energy distributions from the initial energy of the neutrons and the positions of the various elements of the system, as well as the average energies of the peaks observed in the electron equivalent energy calibration of the crystal with a source of $^{133}Ba$. This simulation has allowed to analyze the different systematic effects.

The stability of the NaI(Tl) response was analyzed and the gain drift corrected using the inelastic peak of $^{127}I$ as reference. Moreover, to analyze the effect that the calibration of the crystal in electron equivalent energy can have on the resulting QF estimate, it was decided to apply two different calibration procedures and compare the results obtained in each case. The first method consisted of applying a proportional calibration using a single energy as reference, which being followed by many of the previous experiments allowed for a direct comparison of the results. However, the inelastic peak of $^{127}I$ is outside the region of interest for the QF determination and this method is not considering the non-proportional light yield of NaI(Tl). In the second method, a non-proportional linear response was considered using three energy peaks identified in the measurements made with the $^{133}Ba$ source.

The QF value of sodium corresponding to each triggered BD was obtained by fitting the energy calibrated spectrum to a PDF built by adding three contributions: the iodine nuclear recoil signal obtained from the simulation, the background obtained from the experimental data, and the signal of the nuclear recoil of sodium. The latter is obtained from the simulation applying a gaussian convolution (to take into account the energy resolution of the detector) and converted into electron equivalent energy using the QF. This PDF has a free parameters the QF of the sodium nuclei, the energy resolution of the recoil spectrum and the scale factor of each contribution: sodium recoils, iodine recoils and background. Two different energy resolution models were considered: constant with the energy and a poissonian energy dependence. Then, the QF results derived from this work include the systematical uncertainties from: electron equivalent energy calibration of the NaI(Tl), energy resolution, uncertainties in positions of the different detectors, and quenching factor of iodine. These systematics affecting our analysis have been incorporated into the final uncertainties for the derived Na-QF values, but the energy calibration. We have decided to present results for both calibration procedures separately, as we will comment later.

Since the iodine recoil spectrum could not be disentangled from the background, a different process was followed to calculate its QF. The interaction of the neutrons with the crystal produced an inelastic recoil in the iodine nuclei, and the energy measured in those events was the sum of the energy of the gamma emitted after de-excitation of the iodine nucleus and that of the nuclear recoil. The recoil energy can be obtained from the different positions of the peak for the different scattering angles. The combination of the results from crystals~2 and~3 allows to obtain a QF value of (6.0~$\pm$~2.2)\%.

The energy calibration used for the conversion of the light signal into electron equivalent energy is, according to our results, a critical issue in the determination of the QF, and then, in the comparison among results from different measurements. First, we obtain compatible results for the five crystals measured in both calibration scenarios, with the proportional and the non-proportional calibration. However, both are not compatible with each other. While for the proportional calibration a clear decrease of the QF when the nuclear recoil energy decreases is observed, for the non-proportional case, constant with energy QF is obtained (see Figures~\ref{QF_57} and~\ref{QF_Ba133}, respectively). This is one of the most important results derived in this work: the confirmation of the relevance of the energy calibration procedure applied in the QF estimates. This allows to understand some of the disagreements among previous measurements and highlights the importance to progress further in the understanding of the conversion from energy deposited into light in NaI(Tl), considering for instance, non-proportional behaviour, possible surface effects, etc.

Summarizing our results, we can conclude that our measurement is compatible with a constant QF for sodium of (21.2~$\pm$~0.8)\%, considering an energy calibration non-proportional but linear in the ROI, derived using $^{133}Ba$ calibration data, that could be affected by surface effects. On the other hand, our measurement is compatible with a QF decreasing at low energy when considering proportionality and using energy depositions in the bulk. More research should be carried out in order to establish more clearly a preference for one or the other model. 

An optical simulation of one of the ANAIS modules has been developed in the frame of my PhD work. The goal of this simulation is to improve the understanding of the light signal building, taken into account all the optical processes following the energy deposition by the particles in the different components of the module and later, even modelling the light sensor and in subsequent developments, the electronic chain. This simulation is the starting point for future developments. 

Since the optical processes involved in the detector from the emission of light after an energy deposit to the collection of photons in the photocathode of the PMT (considering propagation, reflection, diffusion and absorption) play a fundamental role in the construction of the detector signal, an optical simulation is necessary for a deeper understanding of the response of the system to light signals with different origin. With this objective, a MC simulation has been developed using the GEANT4 package with the geometry of an ANAIS module. This simulation generates the output signal of the detector taking into account the response of the ANAIS-112 PMTs. The same analysis applied to the experimental pulses has been applied to the simulated events, allowing for comparison between the most relevant analysis variables: pulse area, pulse shape parameters, numbers of peaks in each PMT pulse, etc. 

The optical simulation developed incorporates the basic geometry of ANAIS-112 modules, but an improved design of the PMT with respect to previous simulations. In addition, the relevant optical properties of all the component materials, and the characteristic features of the NaI(Tl) scintillation  have been incorporated into the code. This effort is still ongoing, and following the works of my PhD, a more realistic model of the ANAIS-112 module should be developed. This improved model should consider, for instance, the incorporation of the slow scintillation component of NaI(Tl), a better photocathode model considering its transparency, introducing some spatial effects in the light propagation as inhomogeneities in the Tl concentration that should produce a dispersion in the light yield, etc.

This work focused on the simulation of the events populations used in ANAIS-112 experiment for the energy calibration of the ROI. These events are produced by the decays of $^{22}Na$ and $^{40}K$ contaminant isotopes distributed homogeneously in the NaI(Tl) crystal and that of $^{109}Cd$ isotopes in the calibration source. These optical simulations have produced some preliminary results, summarized below.

A non-proportional energy calibration is obtained when considering the pulse area relation with the deposited energy in the crystal as observed in the experiment, when same digitization parameters are considered (2~GHz sampling rate and 1~$\mu$s integration window). However, this effect is not observed for a different configuration (1~GHz sampling rate and 2~$\mu$s integration window). Further understanding of the non-proportionality origin would be very relevant for improving in the future the calibration of ANAIS-112 experiment in the ROI. 

The pulse shape variables have not been correctly reproduced by the simulation because only the fast scintillation component of the NaI(Tl) has been considered. However, we report on a decrease of the mean time for the simulated low energy pulses, an effect that should be further investigated, as it is also observed in the experimental data of ANAIS-112.

We have analyzed the effect that changing the signal sampling rate from 2~GHz to 1~GHz would have on the resolution of the detectors, increasing at the same time the digitization window from 1.26~$\mu$s to 2.52~$\mu$s. No effect of this modification on the energy resolution is observed. The energy resolution resulting from the simulation is slightly better than the experimentally determined in ANAIS-112 modules. We have been able to identify and quantify the contribution to this energy resolution of the SER area distribution, and then, to conclude that some other effect not yet included in the simulation is responsible of an additional contribution to the experimental energy resolution. This could be related with inhomogeneities in the scintillation properties of the crystal and/or in the light propagation.

PMTs are a very important component of the ANAIS-112 modules. They are quite transparent and they are contaminated with isotopes from the natural chains and $^{40}K$ (see Table~\ref{table:ActivityPMTs}). This implies that light emission can take place in the ANAIS-112 PMTs by Cherenkov effect initiated by electrons or positrons in the PMTs borosilicate. Simulation of this light emission should produce relevant outputs to evaluate the capability of the filtering protocols applied in ANAIS-112 to fight this kind of events. The simulations results allow to conclude that all the Cherenkov events produced by the decay of $^{40}K$ isotopes present in the borosilicate of the PMTs are symmetric regarding the number of peaks identified in the pulses of each PMT, so they may not be the cause of asymmetric events. Moreover, these events are rejected because they are very fast and they have a low number of peaks. The simulation allows to conclude that modifying the selection criteria from a minimum of 5~peaks in each PMT to 3~the efficiency for the detection of the $^{22}Na$ events at 0.9~keV would improved from 40\% to 70\%, while rejecting all the Cherenkov events. However, the asymmetric events present in the data would not be rejected effectively.

This simulation also is able to explain the trigger rate of ANAIS-112 experiment, dominated by events with 1~peak in each PMT. The Cherenkov contribution previoulsy explained, resulting from $^{40}K$ and isotopes in the natural chains at the PMTs is able to generate such a rate after applying the ANAIS-112 trigger condition. Better quantification of this rate of Cherenkov events, including the natural chains in the simulation, is a natural step to follow as continuation of this work. 

ANAIS-112 sensitivity improvement is limited in several directions by the PMTs used as light sensors. PMTs could be responsible of the asymmetric events identified in ANAIS-112 and the blank module, very difficult to reject and obliging to impose very restrictive asymmetry cuts, as commented above. In addition, PMTs contaminations in $^{40}K$ and natural chains are a main contributor to the background model. Trying to boost the sensitivity of the NaI(Tl) scintillators applied to rare event searches it is mandatory to look for alternatives to the PMTs. In particular, in this work we have explored the implementation of Silicon Photomultipliers (SiPMs) as light sensors. 

SiPMs are solid-state detectors made up of multiple Single-Photon Avalanche-Diodes (SPADs) operating in Geiger mode and connected in parallel on a common silicon substrate. Their quantum efficiency for NaI(Tl) scintillation wavelengths is comparable to that of PMTs, but they have the advantage of being much lighter and can easily be made from materials with low radioactive content. In my PhD work the goal was to build prototypes of undoped NaI and NaI(Tl) crystals coupled to SiPMs for the light readout. To achieve this goal I have characterized first different SiPMs, later I carried out measurements at different temperatures down to 87~K at the LNGS under the supervision of Dr.~A.~Razeto using a very simple detector design, to finally contribute to the design and building of a prototype incorporating all the know-how acquired after the measurements at LNGS. This prototype will be characterized along the next months in a facility under commissioning at the University of Zaragoza to evaluate the performance improvement at reach.

These detectors have a large dark current (between 1~and 10~MHz/cm$^2$) at room temperature, which makes it impossible to reach a low energy threshold and a good resolution with these devices under these conditions. Since this dark current follows an exponential dependence on temperature, working at very low temperature (100~K range) is mandatory. However, in this temperature range working with NaI(Tl) and undoped NaI crystals is a challenge. The hygroscopic character of NaI crystals makes difficult the building and operation of the detectors because of the thermal and mechanichal stresses the system suffer, and the scintillation properties of both NaI(Tl) and undoped NaI have to be better understood in that temperature range.

In the research stage I carried out at LNGS, a detector prototype was designed using both NaI(Tl) and undoped NaI crystals, reading their scintillation with arrays of SiPMs whose models were specifically designed to be used in the reading of liquid argon scintillation in the DarkSide-20k experiment. The measurements were carried out in the LNGS laboratory~7, in the STAR cryogenic system, a test bench consisting of a cryostat and an argon recirculation circuit. Inside the cryostat, the detector could be immersed in liquid argon or kept at a temperature close to that of the condensation of this element (around 87~K).

An optical simulation of the system was designed using the GEANT4 package, based on the optical simulation of the ANAIS module presented above to analyze the effect that the medium (liquid or gaseous) surrounding the detector had on both light collection and the detection efficiency of the photopeaks, as well as the suitability of the source chosen for the characterizations ($^{241}Am$ or $^{133}Ba$). This simulation showed that although the $^{133}Ba$ source would produce a higher number of energy deposits in the NaI(Tl) crystal, the high energy gammas of $^{133}Ba$ would suffer from Compton interactions in the interposed materials, which would make interpretation of the measured spectra difficult, particularly in the case of using liquid argon as cryostat filler, due to its high density. This analysis allowed us to identify that the optimal choice consists in using the $^{241}Am$ source and filling the chamber with gaseous argon.

Based on these results, a measurement protocol was established. The procedure followed to characterize the scintillation light collection by the SiPMs for each operating temperature and voltage was the same. First, measurements were taken with a pulsed laser and the charge released by the device in each avalanche was characterized. Afterwards, calibration and background measurements were taken, and the charge spectra obtained in each case was normalized with the time exposure. This allowed to subtract the background spectrum, thus having a cleaner spectrum of $^{241}Am$. Finally, the light collection was determined as the ratio between the average charge produced at a given energy and the charge released in each avalanche.

The light collected by the SiPM has always a contribution of the optical crosstalk produced in SiPMs that ranges from 5\% to 50\%, depending on the system and the applied voltage. This crosstalk contribution was modeled in~\cite{Boulay:2022rgb} as a function of the operation voltage, allowing to derive the light collection by fitting the measurements to this model. The scintillation time behaviour of both crystals at the different temperatures was also searched for. First, the SER was modelled and later, convolved with several exponential scintillation components for fitting the measured pulses.

The results obtained confirmed, as expected, that the light collected for the undoped crystal (NaI(Tl)) increases (decreases) while cooling down from room temperature to liquid argon temperature. At this temperature, approx. 87~K, the light collection measured for the undoped crystal is almost three times that from the NaI(Tl). It is expected that the undoped NaI emits in the UV where the SiPM is not very efficient, so it is reasonable to think that there is more light than that detected, and other ways to increase the light collection efficiency will have to be found in the future.

It can be concluded from this analysis that NaI(Tl) scintillation times depend strongly on the temperature and that this material has several scintillation components with a complicated behaviour while the undoped NaI crystal shows a smoother dependence and scintillation times are 4~times faster than those of NaI(Tl) at liquid argon temperature. However, our results cannot be considered as direct estimates of the scintillation times because the pulse shapes measured are affected by the electronics, resulting in a signal undershoot that introduces a systematic contribution in all the intended fits. Better understanding of the possible effects of the electronic readout chain in the pulse shapes would help to improve this kind of analysis in the future. As main conclusion from these measurements and analysis, because of the higher light collection and faster scintillation, the undoped NaI seems a better target to work at temperatures in the 100~K range. 

The next steps of the research consisted of characterizing SiPMs in terms of gain, time response and dark current, to subsequently design the first compact NaI+SiPMs prototype detector developed at the University of Zaragoza. A SiPM and a specific Front-End Board (FEB) were purchased from HAMAMATSU. In parallel, we started a collaboration within the Global Argon Dark Matter (GADM) program to characterize four DArTeyes, composed of a SiPM on an electronic board and designed to be used in the DArT experiment, at the LSC. These SiPMs were characterized both in the dark and under illumination with a pulsed LED in simple experimental setups. Those setups allowed to work both at room temperature and at that of liquid nitrogen. All SiPMs were characterized in terms of pulse shape, photoelectron amplitude and area, and dark current rate at different voltages. In addition, the HAMAMATSU SiPM was used to read the scintillation of a small NaI(Tl) crystal, allowing to determine the light collection, energy resolution, and scintillation time constant at room temperature.

A compact NaI(Tl)+SiPM detector prototype was designed to allow its characterization at different temperatures. This prototype consists of an array of 4$\times$4 SiPMs attached to a 1" side cubic NaI(Tl) crystal and covered by reflective Teflon tape. Both the SiPM array and the crystal are enclosed in a housing of copper tightly closed and maintained in dry atmosphere conditions. The design of the housing includes an optical fiber that allows to characterize the photoelectron using the light coming from an external LED. An eMUSIC board, designed by SCIENTIFICA, is used to control the supply and process the output signals of the SiPM, in a versatile and simple way.

First measurements carried out from room temperature down to -32~$^o$C using a $^{133}Ba$ source show good resolution and high light collection. At 81~keV, the scintillation light collection is between 6 and 7~phe/keV, which is not too far from that obtained by each PMT in the ANAIS-112 experiment. In the near future this detector will be characterized at temperatures down to 100~K inside a cryostat that has already been purchased.

In conclusion, this memory summarizes different lines of research I have been following along the last four years. They are quite different, but they share a common framework, the ANAIS experiment, and a common goal, from a better understanding of the response of NaI-NaI(Tl) scintillator detectors we could progress towards a sensitivity improvement of this detection technique. This memory encloses the analysis of the scintillation quenching factor in NaI(Tl) crystals, where a very important systematic related with the energy calibration has been found, the develoment of an optical simulation of the ANAIS-112 module to improve the understanding of the optical processes taking part in its signal generation, and the investigation into the use of SiPMs as an alternative to PMTs for detecting the NaI scintillation light.

%% file: Conclusiones.tex
\setcounter{secnumdepth}{0}
\chapter*{Resumen y conclusiones}\label{Chapter:Conclusions}
\addcontentsline{toc}{chapter}{Resumen y conclusiones}
\markboth{Resumen y conclusiones}{Resumen y conclusiones}

Numerosas observaciones astronómicas y cosmológicas (como las curvas de rotación de galaxias, el espectro de anisotropía del fondo cósmico de microondas o las lentes gravitatorias) apuntan a la existencia de un tipo de materia denominada "Materia Oscura" que constituye el 26\% de la densidad de masa-energía del universo. Este tipo de materia no interactúa electromagnéticamente (no emite ni absorbe luz), y solo puede tener interacciones débiles con el resto de la materia. Estos requisitos descartan la posibilidad de que esta materia esté formada por partículas del Modelo Estándar de la Física de Partículas, por lo que hay que buscar candidatos más allá de este modelo, estando los axiones y los WIMPs entre los más exitosos. La comprensión de la naturaleza de esta materia oscura es en la actualidad uno de los retos más importantes a los que se enfrenta la Física, en la frontera entre la cosmología, la astrofísica y la física de partículas.

El experimento DAMA/LIBRA, que utiliza cristales de centelleadores de NaI(Tl) y está ubicado en Laboratorio Nacional del Gran Sasso (LNGS), es el único experimento que ha observado una modulación anual en su tasa de detección consistente con la señal que producirían partículas de materia oscura distribuidas en el halo galáctico siguiendo el modelo más estándar. Sin embargo, este resultado ni ha sido confirmado por ningún otro experimento, ni ha sido descartado de forma independiente del modelo de partículas y del modelo de halo. Para poner a prueba el resultado de DAMA/LIBRA de forma independiente de dichos modelos, es necesario utilizar el mismo material blanco. Este es el objetivo de los experimentos ANAIS-112 y COSINE-100. El experimento ANAIS-112, que lleva tomando datos desde el 3 de agosto de 2017 en el Laboratorio Subterráneo de Canfranc (LSC), consiste en una matriz 3$\times$3 de cristales cilíndricos de NaI(Tl) de 12,5~kg cada uno (alcanzando así una masa total de 112,5~kg). El centelleo de cada cristal convertido en una señal eléctrica mediante dos Tubos Fotomultiplicadores (PMTs) HAMAMATSU R12669SEL2, con alta Eficiencia Cuántica (QE). Los módulos de ANAIS-112 han demostrado tener una alta recolección de luz (LC), por encima de 14~phe/keV en 7 de los 9~módulos, que posibilita un bajo umbral en energía. Los módulos de ANAIS-112 han probado ser sensibles a depósitos energéticos inferiores a 1~keV con alta eficiencia, pero el umbral de análisis se ha establecido en este valor, lo que permite explorar la señal DAMA/LIBRA.

Los trabajos incluidos en esta memoria corresponden a desarrollos realizados en paralelo a la toma y análisis de datos de ANAIS-112, pero compartiendo los mismos objetivos de ANAIS. En primer lugar, se requiere un mejor conocimiento de la respuesta de los cristales de NaI(Tl) a los retrocesos nucleares para poder comparar los resultados de ANAIS y DAMA/LIBRA. En segundo lugar, el origen de algunas de las poblaciones de sucesos espurios que se observan en ANAIS-112 podría analizarse mediante el desarrollo de simulaciones ópticas del detector. Finalmente, la posibilidad de mejorar la sensibilidad de los centelleadores de NaI aplicados a la búsqueda de materia oscura mediante la sustitución de los Tubos Fotomultiplicadores (PMTs) por Fotomultiplicadores de Silicio (SiPMs) ofrece oportunidades muy interesantes que se encuentran bajo investigación en el marco de un nuevo proyecto ANAIS+.

Uno de los principales objetivos de mi tesis doctoral ha sido la medida del Factor de \textit{Quenching} (QF) de centelleo de los núcleos de $Na$ y $I$ en cristales de NaI(Tl). Este parámetro se define como la relación entre la cantidad de luz emitida en el cristal después de un depósito de energía por un retroceso nuclear y la emitida en un retroceso electrónico que deposite exactamente la misma energía. Dado que el experimento ANAIS-112 está calibrado en energía utilizando fuentes de rayos-X y rayos-gamma, que producen retrocesos electrónicos, conocer este factor es esencial para interpretar las depósitos de energía medidas en términos de espectros de energía de retroceso nuclear producidos por partículas de materia oscura. Aunque se ha estudiado experimentalmente este factor desde la década de 1990 tanto para $Na$ como para $I$, los valores obtenidos por diferentes experimentos no son compatibles entre sí. Con el fin de medir este factor y analizar los posibles efectos sistemáticos que afectan las mediciones, se caracterizaron en el mismo montaje experimental 5~cristales producidos por Alpha Spectra y crecidos a partir de polvo de NaI de diferentes calidades.

Estas medidas se llevaron a cabo en agosto y octubre de 2018 en el Triangle Universities Nuclear Laboratory (TUNL) dentro de una colaboración entre miembros de los experimentos COSINE-100, COHERENT y ANAIS-112. En ellas, aces de protones eran acelerados y dirigidos contra un blanco de LiF, produciéndose la reacción $^7Li(p,n)^7Be$, que libera neutrones monoenergéticos con una energía de aproximadamente 1~MeV. Estos neutrones eran dirigidos hacia el cristal de NaI(Tl), donde podían dispersarse con uno de sus núcleos produciendo un retroceso nuclear. El neutrón era detectado posteriormente por uno de los detectores de neutrones (BD) que rodeaban el cristal, lo que permitió determinar la energía de retroceso nuclear midiendo el ángulo de dispersión correspondiente. Por otro lado, la luz emitida por el retroceso nuclear en el cristal se midió mediante un PMT al que estaba acoplado. Se utilizó el mismo PMT para todas las medidas. El experimento se diseñó para minimizar los diferentes efectos sistemáticos, como, por ejemplo, la dispersión múltiple (se utilizaron cristales de pequeño tamaño), el \textit{channeling} (los cristales se rotaron periódicamente) y efectos de umbral (los BDs disparaban la adquisición).

El cálculo del QF requiere seleccionar la dispersión de un neutrón en el NaI(Tl) y la detección de ese neutrón en un BD. Esto implica desarrollar un protocolo de análisis que rechace coincidencias múltiples en los BD y que seleccione eventos de neutrones en esos detectores (tanto por la forma del pulso como por el tiempo de vuelo). Para evitar efectos de umbral, se fijó la ventana de integración de la señal en el cristal de NaI(Tl) teniendo en cuenta el tiempo de vuelo de los neutrones desde el cristal hasta los BDs. Además, una simulación del experimento con el paquete GEANT4 permitió obtener las distribuciones de energía de retroceso nuclear a partir de la energía inicial de los neutrones y las posiciones de los distintos elementos del sistema, así como las energías medias de los picos observados en la calibración de energía equivalente de electrón del cristal con una fuente de $^{133}Ba$. Esta simulación permitió además analizar otros efectos sistemáticos.

Se analizó la estabilidad de la respuesta del detector de NaI(Tl) y se corrigió la deriva de ganancia usando como referencia el pico asociado a la desexcitación del $^{127}I$ tras una dispersión inelástica del neutrón en 57.6~keV. Además, para analizar el efecto que la calibración del cristal en energía equivalente de electrón puede tener sobre la estimación del QF resultante, se decidió aplicar dos procedimientos de calibración diferentes y comparar los resultados obtenidos en cada caso. El primer método consistió en aplicar una calibración proporcional usando una sola energía como referencia, la del pico en 57.6~keV, un procedimiento que al ser seguido por muchos de los experimentos anteriores para medir el QF permitió una comparación directa de los resultados. Sin embargo, esta energía está fuera de la región de interés para la determinación de QF y este método no tiene en cuenta que la producción de luz del NaI(Tl) no es proporcional a la energía en este rango energético. En el segundo método, se consideró una respuesta lineal no proporcional utilizando tres picos de energía identificados en las mediciones realizadas con la fuente $^{133}Ba$.

El valor del QF del sodio correspondiente a cada ángulo de dispersión se obtuvo ajustando el espectro calibrado de energía a una PDF construida sumando tres contribuciones: la señal de retroceso nuclear de yodo obtenida de la simulación, el fondo obtenido de los datos experimentales y la señal de la retroceso nuclear del sodio. Esta última se obtiene de la simulación aplicando una convolución gaussiana (para tener en cuenta la resolución energética del detector) y es convertida en energía equivalente de electrón mediante el QF. Esta PDF tiene como parámetros libres el QF de los núcleos de sodio, la resolución en energía de los retrocesos nucleares y el factor de escala de cada contribución: retrocesos de sodio, retrocesos de yodo y fondo. Se consideraron dos modelos diferentes de resolución energética: uno constante con la energía y otro con una dependencia energética poissoniana. Los resultados del QF derivados incluyeron las incertidumbres sistemáticas de la calibración de energía equivalente de electrón del NaI(Tl), de la resolución en energía, de las incertidumbres en las posiciones de los diferentes detectores y del QF de los núcleos de yodo. Estos sistemáticos que afectan nuestro análisis se han incorporado a las incertidumbres finales para los valores derivados de QF del sodio.

Dado que el espectro de retroceso del yodo no se pudo identificar adecuadamente, se siguió un proceso diferente para calcular su QF. La interacción de los neutrones produce dispersiones inelásticos en los núcleos de $^{127}I$, como ya hemos comentado, y la energía depositada en esos sucesos es la suma de la energía del fotón emitido tras la desexcitación del núcleo de yodo y la del retroceso nuclear. La energía de retroceso se puede obtener a partir de las diferentes posiciones del pico para los diferentes ángulos de dispersión. La combinación de los resultados para los cristales~2 y~3 permitió obtener un valor para este QF de (6,0~$\pm$~2,2)\%.

La calibración de energía utilizada para la conversión de la señal de luz en energía equivalente de electrones es, según nuestros resultados, un tema crítico en la determinación del QF y, por lo tanto, en la comparación entre los resultados de diferentes experimentos. En primer lugar, en este trabajo hemos obtenido resultados de QF compatibles para los cinco cristales medidos en ambos escenarios de calibración, con la calibración proporcional y no proporcional. Sin embargo, estos resultados no son compatibles entre ellos. Mientras que para la calibración proporcional se observa una clara disminución del QF cuando disminuye la energía de retroceso nuclear, para el caso no proporcional se obtiene un QF constante con la energía (ver Figuras~\ref{QF_57} y~\ref{QF_Ba133}, respectivamente). Este es uno de los resultados más importantes derivados de este trabajo: la confirmación de la relevancia del procedimiento de calibración de energía aplicado en las estimaciones del QF. Esto permite comprender algunos de los desacuerdos entre las mediciones anteriores y destaca la importancia de avanzar más en la comprensión de la conversión de energía depositada en luz de centelleo en NaI(Tl), considerando por ejemplo, un comportamiento no proporcional, efectos superficiales, etc.

Resumiendo nuestros resultados, podemos concluir que nuestra medida es compatible con un QF constante para el sodio de 21.2~$\pm$~0.8\%, considerando una calibración de energía no proporcional sino lineal en la ROI, basada en los datos de calibración de $^{133}Ba$, que podrían verse afectados por efectos de superficie. Por otro lado, nuestra medida es compatible con un QF decreciente a baja energía cuando se considera la proporcionalidad y se utilizan depósitos de energía distribuidos homogéneamente en el volumen del cristal. Se deben realizar más investigaciones para establecer más claramente una preferencia por uno u otro modelo.

Se ha desarrollado una simulación óptica de uno de los módulos de ANAIS en el marco de mi trabajo de doctorado. El objetivo de esta simulación es mejorar la comprensión de la construcción del proceso de generación de la señal, teniendo en cuenta todos los procesos ópticos que siguen al depósito de energía por parte de las partículas en los diferentes componentes del módulo, y modelando el comportamiento del sensor de luz. Esta simulación sirve como punto de partida para futuros desarrollos, en los que, por ejemplo, se podría incorporar a la simulación el efecto de toda la cadena de procesado electrónico de la señal.

Dado que los procesos ópticos que intervienen en el detector desde la emisión de luz tras un depósito de energía hasta la recogida de fotones en el fotocátodo del PMT (considerando propagación, reflexión, difusión y absorción) juegan un papel fundamental en la generación de la señal del detector, una simulación óptica es necesaria para una comprensión más profunda de la respuesta del sistema a señales de luz con diferente origen. Con este objetivo se ha desarrollado una simulación de MC utilizando el paquete GEANT4 con la geometría de un módulo de ANAIS. Esta simulación genera la señal de salida del detector teniendo en cuenta la respuesta de los PMTs de ANAIS-112. El mismo análisis aplicado a los pulsos experimentales se ha aplicado a los eventos simulados, permitiendo la comparación entre las variables de análisis más relevantes: área de pulso, parámetros de forma de pulso, número de picos en cada pulso de PMT, etc.

La simulación óptica desarrollada incorpora la geometría básica de los módulos de ANAIS-112, pero un diseño mejorado del PMT respecto a simulaciones anteriores. Además, se han incorporado al código las propiedades ópticas relevantes de todos los materiales de los componentes del módulo y los rasgos característicos del centelleo de NaI(Tl). Este esfuerzo aún está en curso para desarrollar un modelo más realista del módulo de ANAIS-112. Este modelo mejorado debería considerar, por ejemplo, la incorporación de la componente de centelleo lento de NaI(Tl), un mejor modelo de fotocátodo teniendo en cuenta su transparencia, la introducción de algunos efectos espaciales en la emisión y propagación de la luz como inhomogeneidades en la concentración de Tl que deberían producir una dispersión en el rendimiento lumínico, etc.

Este trabajo se centró en la simulación de las poblaciones de eventos utilizadas en el experimento ANAIS-112 para la calibración energética en la región de interés. Estos eventos son producidos por la desintegración de isótopos contaminantes de $^{22}Na$ y $^{40}K$ distribuidos homogéneamente en el cristal de NaI(Tl) y de isótopos de $^{109}Cd$ en la fuente de calibración. Estas simulaciones ópticas han producido algunos resultados preliminares, que se resumen a continuación.

Se ha obtenido una calibración de energía no proporcional al considerar la relación del área del pulso con la energía depositada en el cristal, como se observa en el experimento, cuando se consideran los mismos parámetros de digitalización (frequencia de muestreo de 2~GM/s y ventana de integración de 1~$\mu$s). Sin embargo, este efecto no se observa para una configuración diferente (frequencia de muestreo de 1~GM/s y ventana de integración de 2~$\mu$s). Una mayor comprensión del origen de la no proporcionalidad sería muy relevante para mejorar en el futuro la calibración del experimento ANAIS-112 en la ROI.

La simulación no ha reproducido correctamente las variables de forma de pulso porque solo se ha considerado el componente de centelleo rápido del NaI(Tl). Sin embargo, se ha observado una disminución del tiempo medio de los pulsos de baja energía simulados, un efecto que debe investigarse más a fondo, ya que también se observa en los datos experimentales de ANAIS-112.

Hemos analizado el efecto que tendría cambiar la frecuencia de muestreo de la señal de 2~GM/s a 1~GM/s sobre la resolución de los detectores, aumentando al mismo tiempo la ventana de digitalización de 1,26~$\mu$s a 2,52~$\mu $s. No se observa ningún efecto de esta modificación sobre la resolución energética, pero se ha observado que la resolución obtenida en la simulación es ligeramente mejor que la determinada experimentalmente en los módulos de ANAIS-112. Hemos podido identificar y cuantificar la contribución a esta resolución de la distribución del área de la SER y concluir que algún otro efecto aún no incluido en la simulación es responsable de una contribución adicional a la resolución energética experimental. Esto podría estar relacionado con falta de homogeneidad en las propiedades de centelleo del cristal y/o en la propagación de la luz.

Los PMTs son un componente muy importante de los módulos de ANAIS-112. Sus ventanas ópticas son transparentes y están contaminadas con isótopos de las cadenas naturales y $^{40}K$ (ver Tabla~\ref{table:ActivityPMTs}). Esto implica que la emisión de luz puede tener lugar en los PMTs de ANAIS-112 por efecto Cherenkov iniciado por electrones o positrones en el borosilicato de los PMTs. La simulación de esta emisión de luz debería producir resultados relevantes para evaluar la capacidad de los protocolos de filtrado aplicados en ANAIS-112 para combatir este tipo de sucesos. Los resultados de las simulaciones permiten concluir que todos los eventos Cherenkov producidos por la desintegración de los isótopos de $^{40}K$ presentes en el borosilicato de los PMTs son simétricos en cuanto al número de picos identificados en los pulsos de cada PMT, por lo que no pueden ser la causa de los sucesos asimétricos observados en ANAIS-112. Además, estos eventos son rechazados porque son muy rápidos y tienen un bajo número de picos. La simulación permite concluir que modificando el criterio de selección de un mínimo de 5~picos en cada PMT a 3, la eficiencia de detección de los eventos de $^{22}Na$ a 0.9~keV mejoraría de un 40\% a un 70\%, mientras que seguirían rechazándose todos los eventos Cherenkov. Sin embargo, los eventos asimétricos presentes en los datos no serían rechazados de manera efectiva.

Esta simulación también puede explicar el ritmo de disparo del experimento ANAIS-112, dominado por eventos con 1~pico en cada PMT. La contribución de sucesos Cherenkov explicada anteriormente, resultante de la desintegración de $^{40}K$ y de isótopos de las cadenas naturales en los PMTs, es capaz de generar ese ritmo tras aplicar la condición de disparo de ANAIS-112. Una mejor cuantificación de este ritmo de eventos de Cherenkov, incluyendo la simulación de las cadenas naturales, es un paso natural a seguir como continuación de este trabajo.

La mejora de la sensibilidad de ANAIS-112 está limitada en varias direcciones por los PMTs utilizados como sensores de luz. Los PMTs podrían ser los responsables de los eventos asimétricos identificados en ANAIS-112 y el módulo \textit{Blank}, que son muy difíciles de rechazar y que obligan a imponer cortes muy restrictivos en la asimetría de los pulsos, como se ha comentado anteriormente. Además, las contaminaciones de $^{40}K$ y de otras cadenas naturales en los PMTs son una de las principales contribuciones al fondo del experimento. Para aumentar la sensibilidad de los centelleadores de NaI(Tl) aplicados a búsquedas de sucesos raros, es obligatorio buscar alternativas a los PMTs. En particular, en este trabajo hemos explorado la implementación de Fotomultiplicadores de Silicio (SiPMs) como sensores de luz.

Los SiPM son detectores de estado sólido formados por múltiples diodos de avalancha de un único fotón (SPAD) que funcionan en modo Geiger y están conectados en paralelo en un sustrato de silicio común. Su eficiencia cuántica para las longitudes de onda de centelleo del NaI(Tl) es comparable a la de los PMTs, pero tienen la ventaja de ser mucho más ligeros y pueden fabricarse fácilmente con materiales con bajo contenido radiactivo. En mi trabajo de doctorado, el objetivo era construir prototipos de cristales de NaI(Tl) y NaI sin dopar acoplados a SiPM para la lectura de luz. Para lograr este objetivo, primero caractericé diferentes SiPMs, luego realicé mediciones a diferentes temperaturas hasta 87~K en el LNGS bajo la supervisión del Dr.~A.~Razeto usando un diseño de detector muy simple, para finalmente contribuir al diseño y construcción de un prototipo incorporando todo el conocimiento adquirido tras las mediciones en LNGS. Este prototipo se caracterizará a lo largo de los próximos meses en la Universidad de Zaragoza para evaluar la mejora de rendimiento que permite alcanzar.

Los SiPMs tienen una gran corriente oscura (entre 1~y 10~MHz/cm$^2$) a temperatura ambiente, lo que hace imposible alcanzar un umbral de baja energía y una buena resolución con estos dispositivos en estas condiciones. Dado que esta corriente oscura sigue una dependencia exponencial con la temperatura, es obligatorio trabajar a muy baja temperatura (en torno a 100~K). Sin embargo, en este rango de temperatura, trabajar con NaI(Tl) y cristales de NaI sin dopar es un desafío. El carácter higroscópico de los cristales de NaI dificulta la construcción y el funcionamiento de los detectores debido a las tensiones térmicas y mecánicas que sufre el sistema. Además, las propiedades de centelleo tanto del NaI(Tl) como del NaI sin dopar deben comprenderse mejor en ese rango de temperatura.

En la estancia de investigación que realicé en LNGS, se diseñó un prototipo de detector utilizando tanto cristales de NaI(Tl) como de NaI sin dopar, leyendo su centelleo con matrices de SiPMs diseñados específicamente para ser utilizados en la lectura del centelleo del argón líquido en el experimento DarkSide-20k. Las mediciones se realizaron en el laboratorio \#~7 del LNGS, en el sistema criogénico STAR, un banco de pruebas compuesto por un criostato y un circuito de recirculación de argón. Dentro del criostato, el detector podría sumergirse en argón líquido o mantenerse a una temperatura cercana a la de condensación de este elemento (alrededor de 87~K).

Se diseñó una simulación óptica del sistema utilizando el paquete GEANT4, basado en la simulación óptica del módulo de ANAIS presentado anteriormente para analizar el efecto que el medio (líquido o gaseoso) que rodea al detector influyó tanto en la recolección de luz como en la eficiencia de detección de los fotopicos. Además, también sirvió para analizar la idoneidad de la fuente elegida para las caracterizaciones ($^{241}Am$ o $^{133}Ba$). Esta simulación mostró que aunque la fuente de $^{133}Ba$ produciría una mayor cantidad de depósitos de energía en el cristal de NaI(Tl), las gammas de alta energía de $^{133}Ba$ sufrirían interacciones Compton en los materiales interpuestos entre la fuente y el cristal, lo que dificultaría la interpretación de los espectros medidos, particularmente en el caso de tener el criostato lleno de argón líquido, debido a su alta densidad. Este análisis nos permitió identificar que la elección óptima consistía en utilizar la fuente de $^{241}Am$ y llenar la cámara con argón gaseoso.

En base a estos resultados, se estableció un protocolo de medida. El procedimiento seguido para caracterizar la recolección de luz de centelleo por parte de los SiPM para cada temperatura y voltaje de operación fue el mismo. En primer lugar, se tomaron medidas con un láser pulsado y se caracterizó la carga liberada por el dispositivo en cada avalancha. Posteriormente, se tomaron medidas de calibración y de fondo, y se normalizaron los espectros de carga obtenidos en cada caso con el tiempo de exposición. Esto permitió restar el espectro de fondo, teniendo así un espectro más limpio de $^{241}Am$. Finalmente, la recolección de luz se determinó como la relación entre la carga promedio producida a una energía dada y la carga liberada en una avalancha.

La luz recogida por los SiPM tiene siempre una contribución del \textit{crosstalk} óptico producido en los SiPM que oscila entre el 5\% y el 50\%, dependiendo del sistema y la tensión aplicada. Esta contribución del \textit{crosstalk} se modeló en~\cite{Boulay:2022rgb} en función del voltaje de operación, y usando este modelo se ha obtenido la luz de centelleo recogida en nuestro sistema ajustando las medidas a este modelo. También se estudió el comportamiento del tiempo de centelleo de ambos cristales a diferentes temperaturas. Primero, se modeló la respuesta de una sola avalancha y luego se convolucionó con varias componentes de centelleo exponenciales para ajustar los pulsos medidos.

Los resultados obtenidos confirmaron, como se esperaba, que la luz recolectada por el cristal no dopado (NaI(Tl)) aumenta (disminuye) mientras se enfría desde temperatura ambiente hasta la temperatura del argón líquido. A esta temperatura, la recolección de luz medida para el cristal sin dopar es unas tres veces mayor que la del NaI(Tl). Se espera que el NaI sin dopar emita en la región de luz ultravioleta donde el SiPM no es muy eficiente, por lo que es razonable pensar que la relación de luz emitida es aún mayor que la obtenida de la luz detectada, y habrá que encontrar otras formas de aumentar la eficiencia de recolección de luz en el futuro.

De este análisis se puede concluir que los tiempos de centelleo de NaI(Tl) dependen fuertemente de la temperatura y que este material tiene varios componentes de centelleo con un comportamiento complicado mientras que el cristal de NaI sin dopar muestra una dependencia más suave y los tiempos de centelleo son 4~veces más rápidos que los de NaI(Tl) a la temperatura del argón líquido. Sin embargo, nuestros resultados no pueden considerarse estimaciones directas de los tiempos de centelleo porque las formas de los pulsos medidos se ven afectadas por la electrónica, lo que da como resultado una deformación del pulso que introduce un efecto sistemático en todos los ajustes. Una mejor comprensión de los posibles efectos de la cadena de lectura electrónica en las formas de los pulsos ayudaría a mejorar este tipo de análisis en el futuro. Como principal conclusión de estas mediciones y análisis, debido a la mayor recolección de luz y al centelleo más rápido, el NaI sin dopar parece un mejor objetivo para trabajar a temperaturas en el rango de 100~K.

Los siguientes pasos de la investigación consistieron en caracterizar los SiPMs en términos de ganancia, tiempo de respuesta y corriente oscura, para posteriormente diseñar el primer prototipo de detector compacto de NaI+SiPMs desarrollado en la Universidad de Zaragoza. Se compró un SiPM y una tarjeta interfaz (FEB) de HAMAMATSU. Paralelamente, iniciamos una colaboración dentro del programa Global Argon Dark Matter (GADM) para caracterizar cuatro DArTeyes, compuestos por un SiPM en una placa electrónica y diseñados para ser utilizados en el experimento DArT, en el LSC. Estos SiPMs se caracterizaron tanto en la oscuridad como bajo iluminación con un LED pulsado en configuraciones experimentales simples. Esas configuraciones permitieron trabajar tanto a temperatura ambiente como a la de nitrógeno líquido. Todos los SiPMs se caracterizaron en términos de forma de pulso, amplitud y área de fotoelectrones, y tasa de corriente oscura a diferentes voltajes. Además, el SiPM de HAMAMATSU se utilizó para leer el centelleo de un pequeño cristal de NaI(Tl), lo que permitió determinar la recolección de luz, la resolución de energía y la constante de tiempo de centelleo a temperatura ambiente.

Se diseñó un prototipo de detector compacto de NaI(Tl)+SiPM que permite su caracterización a diferentes temperaturas. Este prototipo consta de una matriz de 4$\times$4 SiPM unidos a un cristal de NaI(Tl) cúbico de 1" de lado y cubierto por una cinta reflectante de teflón. Tanto la matriz de SiPM como el cristal están encerrados en una carcasa de cobre herméticamente cerrada y mantenida en condiciones de atmósfera seca. El diseño de la carcasa incluye una fibra óptica que permite caracterizar el fotoelectrón utilizando la luz proveniente de un LED externo. Además, se utiliza una placa eMUSIC (diseñada por SCIENTIFICA) para controlar la alimentación y procesar las señales de salida del SiPM, de forma versátil y sencilla.

Las primeras mediciones realizadas desde la temperatura ambiente hasta -32~$^o$C usando una fuente de $^{133}Ba$ muestran una buena resolución y una alta recolección de luz. A 81~keV, la recolección de luz de centelleo está entre 6 y 7~phe/keV, que no se aleja mucho de la obtenida por cada PMT en el experimento ANAIS-112. En un futuro próximo, este detector se caracterizará a temperaturas de hasta 100~K dentro de un criostato que ya se ha comprado.

En conclusión, esta memoria resume diferentes líneas de investigación que he ido siguiendo a lo largo de los últimos cuatro años. Son bastante diferentes, pero comparten un marco común, el experimento ANAIS. Todas siguen un objetivo común: mejorar la comprensión de la respuesta de los centelleadores de NaI y NaI(Tl) y avanzar hacia una mejora de la sensibilidad de esta técnica de detección. Esta memoria engloba el análisis del QF en cristales de NaI(Tl) (en el que se ha encontrado un sistemático muy importante relacionado con la calibración en energía), el desarrollo de una simulación óptica de un módulo de ANAIS-112 para mejorar la comprensión de la procesos que intervienen en la generación de su señal, y la investigación del uso de SiPMs como alternativa a los PMTs para la detección de la luz de centelleo del NaI.

%% file: ANEXO1.tex
\chapter{Neutron interactions in the NaI(Tl) crystal} \label{Chapter:Anexo_NeutronNaI}
\fancyhead[LE]{\emph{Appendix}}
\fancyhead[RO]{\emph{A. Neutron interactions in the NaI(Tl) crystal}}

When a neutron interacts with a nucleus of the crystal, different processes can take place: 

\begin{itemize}
    \item \textbf{Elastic scattering:} The energy lost by the neutron is completely transferred to the nuclear recoil.
    \item \textbf{Inelastic scattering:} The energy lost by the neutron is partially transferred as kinetic energy to the nucleus, being the rest of the energy invested in the excitation of some accessible nuclear level above the ground state. The nucleus will de-excite later by emitting one or more gammas.
    \item \textbf{Radiative capture:} The neutron is absorbed by the nucleus, reaching an excited state. The nucleus will de-excite after a $\gamma$ emission.
    \item \textbf{Spallation:} Part of the energy of the neutron is invested in the ejection of other nucleons or groups of nucleons from the nucleus. This process is very unlikely for neutrons with an energy of the order of 1~MeV.
\end{itemize}

The cross-section for each process depends strongly on the target nucleus and the neutron energy. As the neutron energy produced at TUNL facilities and used in this work was of the order of 1~MeV, the values of the cross-sections for each interaction can be compared for the two isotopic components of the target: $^{23}Na$ and $^{127}I$. They are obtained from TENDL-2019 nuclear data library~\cite{Koning:2019qbo}, and summarized in Table~\ref{tabla:CrossSections}.

\begin{table}[h!]
    \centering
        \begin{tabular}{|c|c|c|c|c|}
             \hline
             & \multicolumn{2}{|c|}{$^{23}Na$} & \multicolumn{2}{|c|}{$^{127}I$} \\
             \hline
             Process & $\sigma$ (mbarn) & Final nuclei & $\sigma$ (mbarn) & Final nuclei \\
             \hline
             Elastic scattering & 2700 & $^{23}Na$ & 5100 & $^{127}I$ \\
             Inelastic scattering & 440 & $^{23}Na^*$ & 970 & $^{127}I^*$ \\
             Radiative capture & 0.23 & $^{24}Na^*$ & 55 & $^{128}I$ \\
             Spallation (n,$\alpha$) & 0 & - & 2.2$\cdot$10$^{-9}$ & $^{123}Sb$ \\
             Spallation (n,p) & 0 & - & 9.6$\cdot$10$^{-11}$ & $^{126}Te$ \\
             Spallation (n,2$\alpha$) & 0 & - & 2.4$\cdot$10$^{-18}$ & $^{119}In$ \\
             \hline
         \end{tabular} \\
    \caption{Interaction cross-sections for some selected channels for 1~MeV neutrons in $^{23}Na$ and $^{127}I$ nuclei and the final states after the neutron interactions. Values from TENDL-2019 nuclear data library~\cite{Koning:2019qbo}.}
    \label{tabla:CrossSections}
\end{table}

In the inelastic scattering, the nucleus can reach different energy states and then emit one or more $\gamma$ particles in the de-excitation. In the case of $^{23}Na$, there is only one possible nuclear excitation (with an energy of 440~keV, and a half-life of 1.14~ps~\cite{ShamsuzzohaBasunia:2021nun}), because the next excited state has an energy of 2076~keV, higher than the beam neutron energy. However, the first 9 excited energy levels of $^{127}I$ are below 1~MeV, and they can be reached in the inelastic scattering process. The cross-sections of the production of each excitation~\cite{Koning:2019qbo}, the probabilities of reaching a given level following an inelastic scattering, the corresponding level energies and the half-life of each level are presented in Table~\ref{tabla:IodineScattering}~\cite{Hashizume:2011dst}.

\begin{table}[h!]
    \centering
        \begin{tabular}{|c|c|c|c|c|}
             \hline
             Level~$\#$ & Energy (keV) & $\sigma$ (mbarn) & Level prob. ($\%$) & T$_{1/2}$ (ps) \\
             \hline
             1 & 57.6 & 240 & 24.74 & 1950 \\
             2 & 203 & 150 & 15.46 & 387 \\
             3 & 375 & 55 & 5.67 & 31 \\
             4 & 418 & 160 & 16.49 & 3.3 \\
             5 & 618 & 84 & 8.66 & $<$~2.1 \\
             6 & 629 & 120 & 12.37 & 1.9 \\
             7 & 651 & 68 & 7.01 & 3.9 \\
             8 & 717 & 26 & 2.68 & 3.0 \\
             9 & 745 & 67 & 6.91 & 2.8 \\
             \hline
         \end{tabular} \\
    \caption{Level scheme of $^{127}I$ isotope. Only levels below 1 MeV are listed. Cross-section values are extracted from TENDL-2019 nuclear data library~\cite{Koning:2019qbo}, while energies and half-life of the states are obtained from~\cite{Hashizume:2011dst}.}
    \label{tabla:IodineScattering}
\end{table}

The half-life of these states are of the order of ns or ps, which means that the light produced by the nuclear recoil will be added to that produced by the gamma. All these lines will be present during all the beam-on measurements, but they can be difficult to identify because the small size crystal implies a low efficiency for the photoelectric absorption of gammas above 100~keV.

\begin{figure}[h!]
    \begin{center}
        \includegraphics[width=\textwidth]{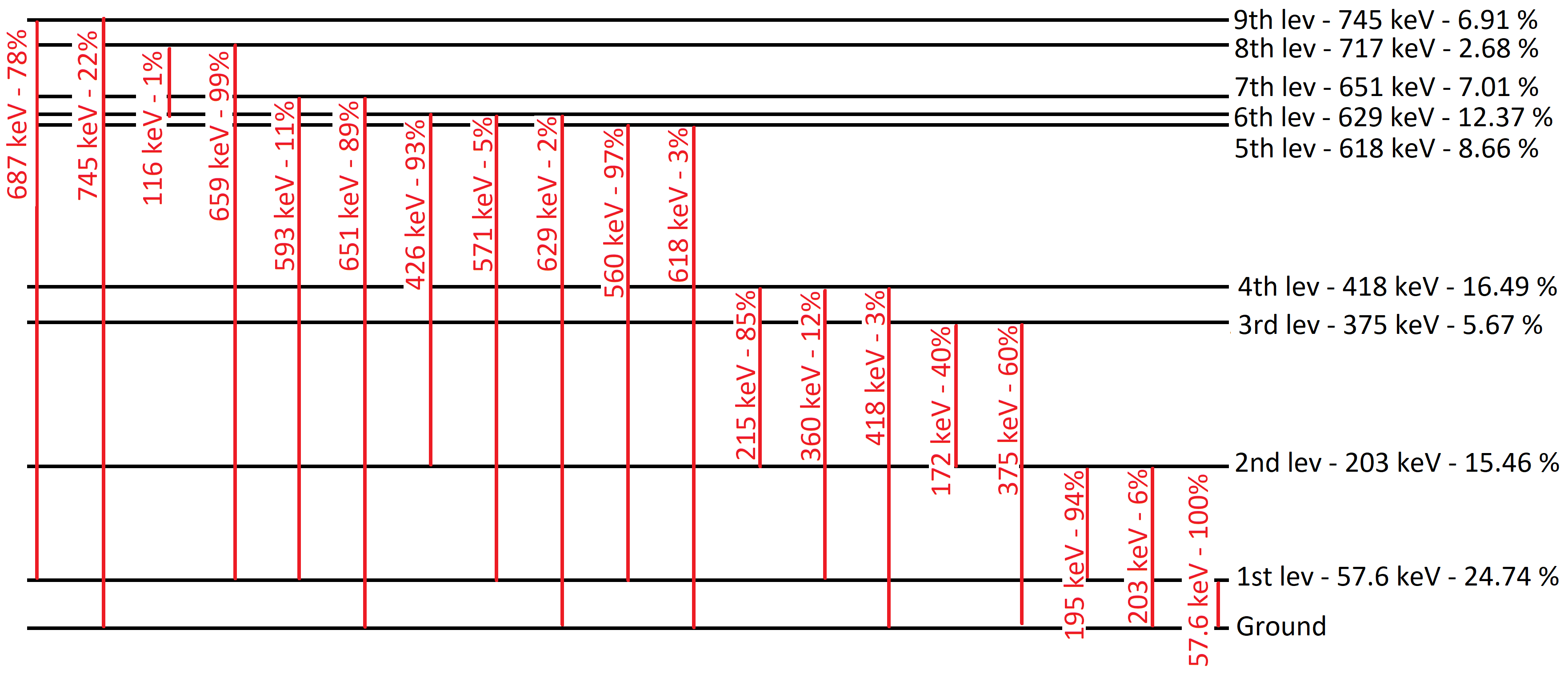}
        \caption{\label{InelasticScatLines}Level scheme of $^{127}I$ isotope. Black percentages represent the probability of exciting that level after a neutron of 1~MeV scatters inelastically $^{127}I$, and red percentages are the probability of emission of the corresponding gammas. Level energies and emission energies are also shown (black and red text, respectively).}
    \end{center}
\end{figure}

%% file: ANEXO2.tex
\chapter{Technical specifications of detector components} \label{Anexo_Specifications}
\fancyhead[LE]{\emph{Appendix}}
\fancyhead[RO]{\emph{B. Technical specifications of detector components}}

\begin{table}[h!]
    \centering
        \begin{tabular}{|l|l|}
            \hline
             Wide effective area & 23~mm~×~23~mm \\
             Spectral range & 300~to~650~nm \\
             Peak wavelength & 400~nm \\
             Photocathode material & Ultra bialkali \\
             Window material & Borosilicate glass \\
             Dynode structure & Metal channel \\
             Dynode stages & 12 \\
             Supply voltage between anode and cathode & -1000~V \\
             Average anode output current & 0.018~mA \\
             Typical luminosity & 0.135~mA/lm \\
             Typical blue sensitivity index (CS 5-58) & 15.5 \\
             Typical quantum efficiency & 43~$\%$ \\
             Typical cathode radiant sensitivity & 130~mA/W \\
             Typical gain & 4.8~$\cdot$~10$^5$ \\
             Typical dark current & 2~nA \\
             Typical rise time & 1.3~ns\\
             Typical transit time & 5.8~ns\\
            \hline
         \end{tabular} \\
    \caption{Specifications of Hamamatsu H11934-200-10 ultra-bialkali PMT~\cite{PMTNaIQFManual}.}
    \label{tabla:pmtNaI}
\end{table}

\begin{figure}[h!]
\begin{center}
\includegraphics[width=\textwidth]{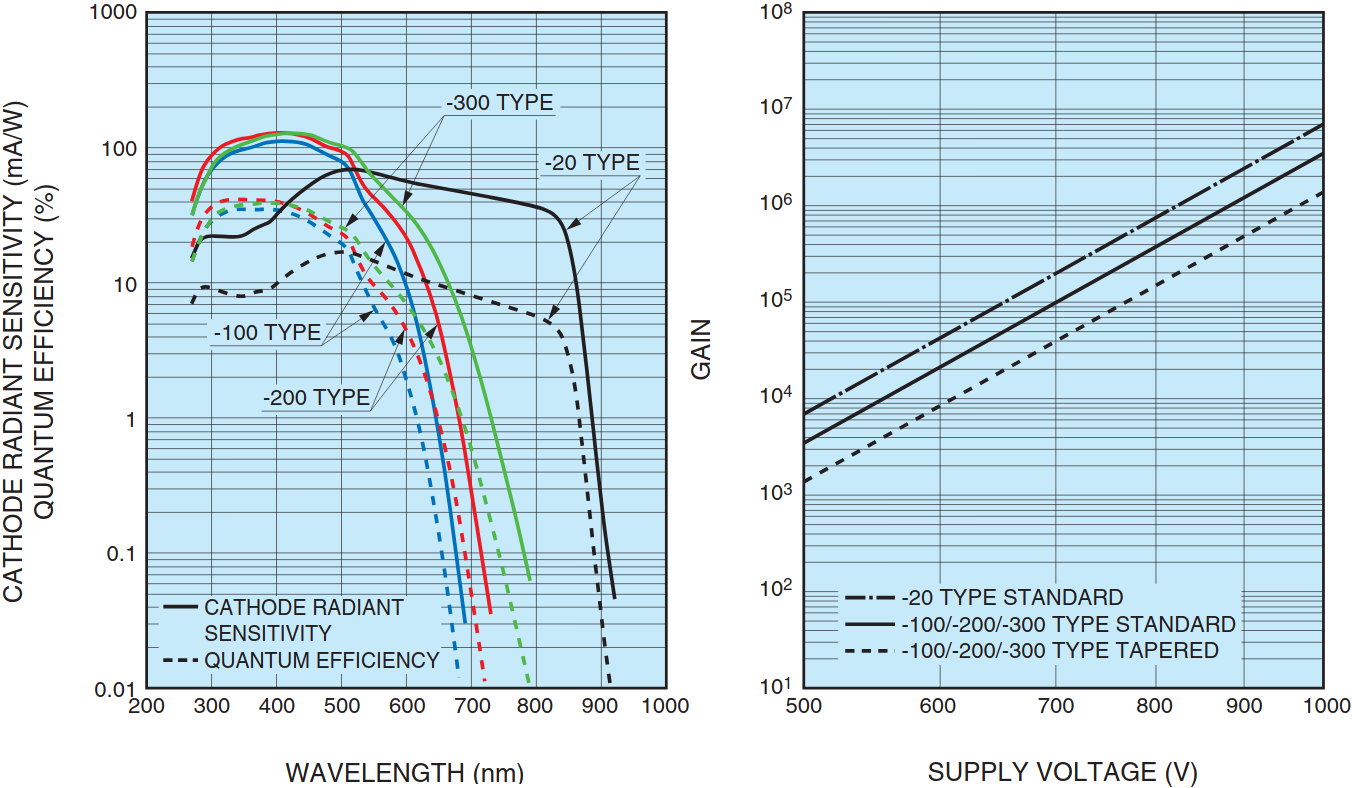}
\caption{\label{pmtNaISpec}Typical spectral response (left plot, red line) and gain (right plot, continuous line) of the Hamamatsu H11934-200-10 ultra-bialkali PMT~\cite{PMTNaIQFManual}.}
\end{center}
\end{figure}

\begin{table}[h!]
    \centering
        \begin{tabular}{|l|l|}
            \hline
             Scintillation yield & 12300~photons/MeV~e- \\
             Wavelength of maximum emission & 424~nm \\
             Decay time, short component & $\sim$~3.5~ns \\
             Average decay time for electronic recoils & $\sim$~10~ns \\
             Average decay time for nuclear recoils & $\sim$~20~ns \\
             Bulk light attenuation length & $>$~1~m \\
             Specific gravity & 0.959 \\
             Refractive index & 1.57 \\
            \hline
         \end{tabular} \\
    \caption{Specifications of EJ-309 from manufacturer~\cite{EJ-309:Eljen}.}
    \label{tabla:LSSpec}
\end{table}

\begin{figure}[h!]
    \begin{center}
        \includegraphics[width=0.75\textwidth]{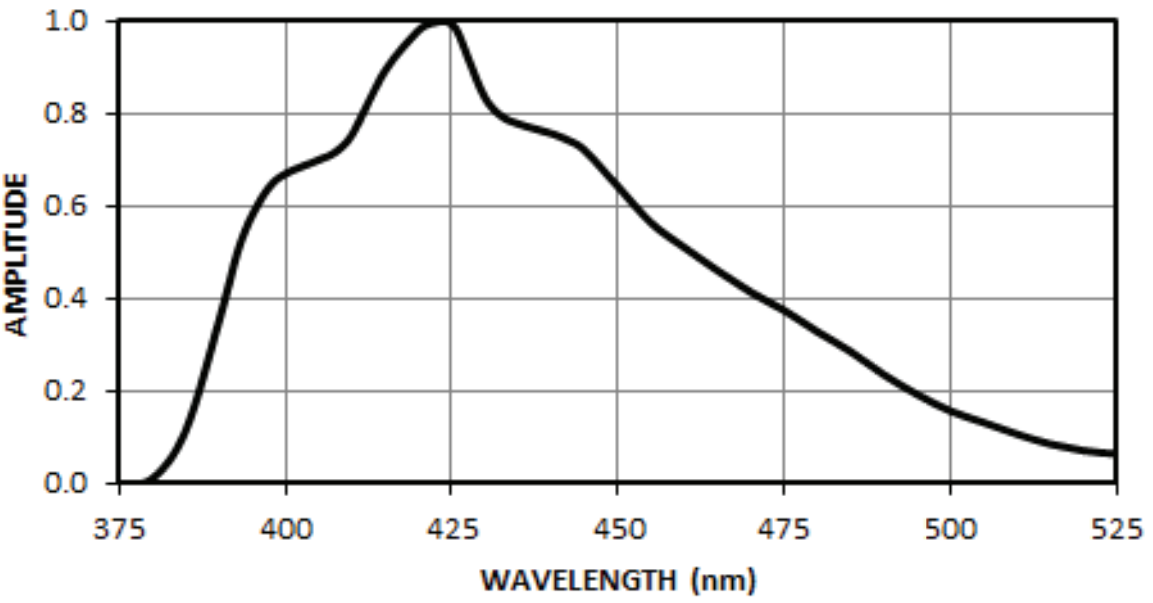}
        \caption{\label{LSSpec}EJ-309 emission spectrum, data from manufacturer~\cite{EJ-309:Eljen}.}
    \end{center}
\end{figure}

\begin{figure}[h!]
    \begin{center}
        \includegraphics[width=\textwidth]{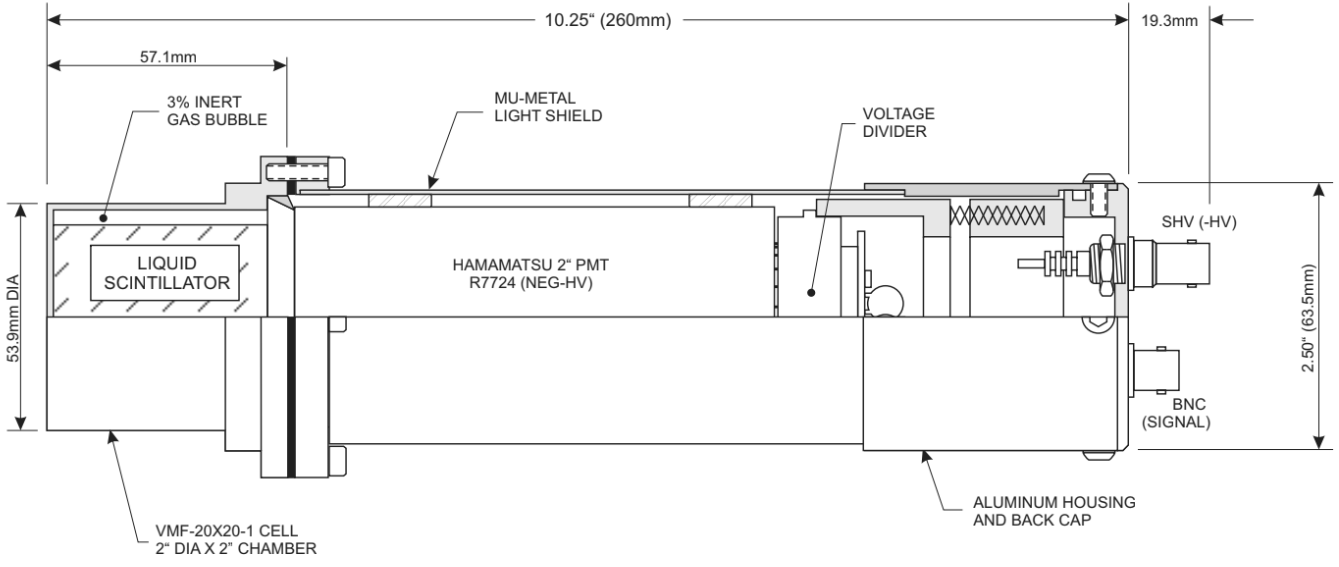}
        \caption{\label{BDSpec}Specifications of a BD, with the EJ-309 scintillator optically coupled to a PMT R7724 (NEG-HV)~\cite{PMT_LS:Hamamatsu}.}
    \end{center}
\end{figure}

\begin{table}[h!]
    \centering
        \begin{tabular}{|l|l|}
            \hline
             Minimum effective area & 46~mm (diameter) \\
             Spectral range & 300~to~650~nm \\
             Peak wavelength & 420~nm \\
             Photocathode material & Super bialkali \\
             Window material & Borosilicate glass \\
             Dynode structure & Linear-focused \\
             Dynode stages & 10 \\
             Supply voltage between anode and cathode & -1750~V \\
             Average anode output current & 0.02~mA \\
             Typical luminosity & 0.3~mA/lm \\
             Typical blue sensitivity index (CS 5-58) & 10.5 \\
             Typical quantum efficiency & 26~$\%$ \\
             Typical cathode radiant sensitivity & 85~mA/W \\
             Typical gain & 3.3~$\cdot$~10$^6$ \\
             Typical dark current & 6~nA \\
             Typical rise time & 2.1~ns\\
             Typical transit time & 29~ns\\
            \hline
         \end{tabular} \\
    \caption{Specifications of Hamamatsu R7724 PMT~\cite{PMT_LS:Hamamatsu}.}
    \label{tabla:PMT_BD_Spec}
\end{table}

\begin{figure}[h!]
\begin{center}
\includegraphics[width=\textwidth]{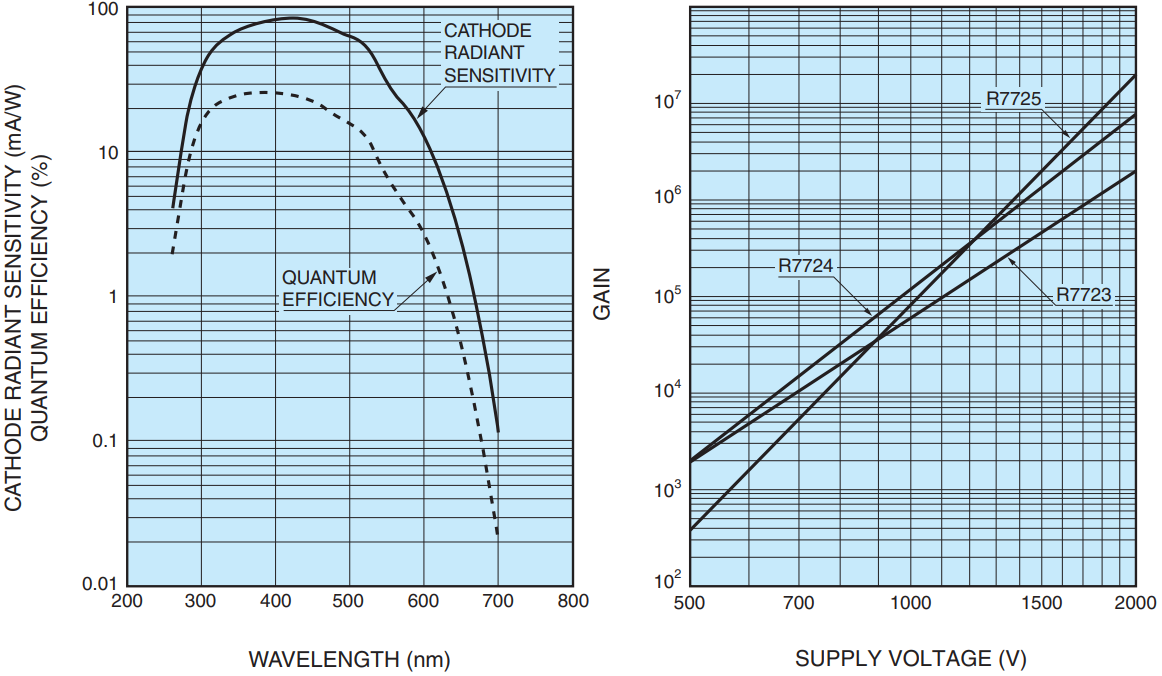}
\caption{\label{PMT_BD_Spec}Typical spectral response (left plot, red line) and gain (right plot, continuous line) of the Hamamatsu R7724 PMT~\cite{PMT_LS:Hamamatsu}.}
\end{center}
\end{figure}

\begin{figure}[h!]
  \begin{subfigure}[b]{0.49\textwidth}
    \includegraphics[width=\textwidth]{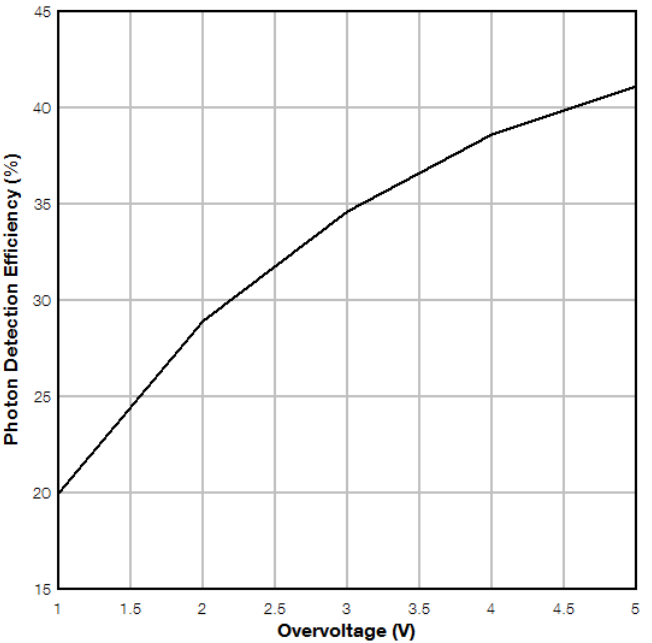}
  \end{subfigure}
  \begin{subfigure}[b]{0.49\textwidth}
    \includegraphics[width=\textwidth]{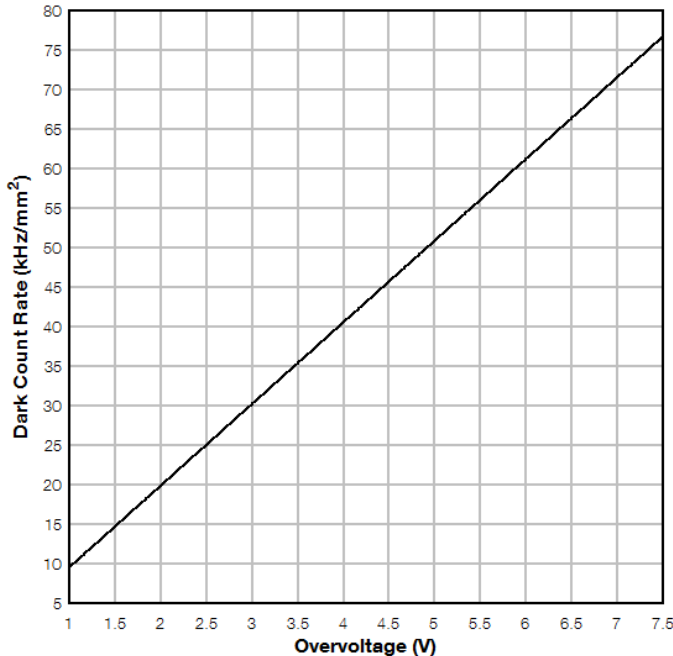}
  \end{subfigure}
  \caption{\label{SensLSpecifications1}Typical behaviour of the PDE and DC rate of the ARRAYC-60035-4P-BGA SiPM as a function of the overvoltage. Left plot shows the total PDE, and right plot is the Dark Count rate~\cite{ONSEMI}.}
\end{figure}

\begin{figure}[h!]
\begin{center}
\includegraphics[width=0.75\textwidth]{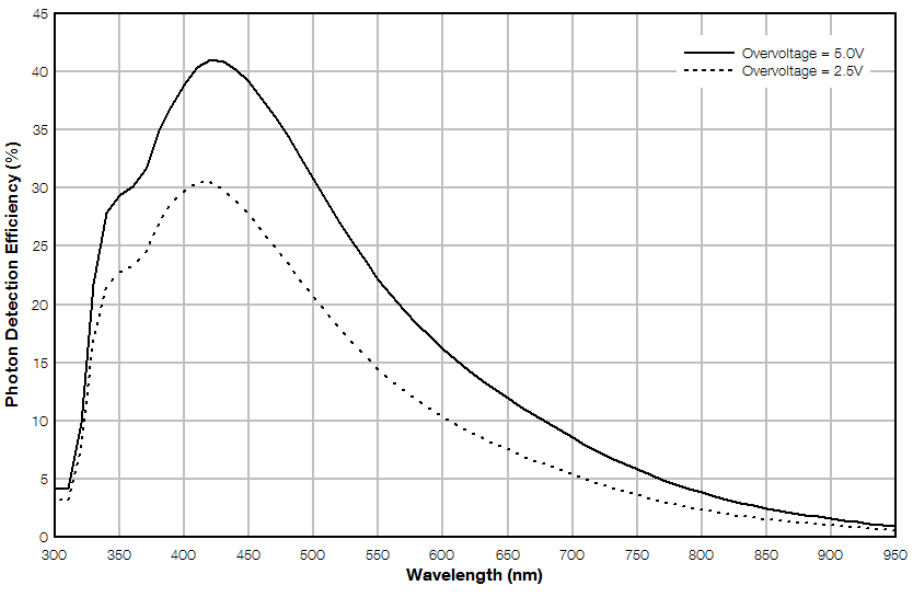}
\caption{\label{SensLSpecifications2}PDE of the ARRAYC-60035-4P-BGA SiPM as a function of the light wavelength at two different overvoltages~\cite{ONSEMI}.}
\end{center}
\end{figure}

\begin{table}[h!]
    \centering
        \begin{tabular}{l l}
             Length of each cell & 6 mm \\
             Microcell size & 50 $\mu$m \\
             Geometrical Fill Factor & 74 $\%$ \\
             Breakdown voltage & 53 V \\
             Recommended overvoltage & 3 V \\
             Spectral Range & 270 to 900 nm \\
             Maximum PDE wavelength & 450 nm \\
             PDE at 3 OV & 40 $\%$ \\
             Typical DC rate at 3 OV & 2 MHz \\
             Maximum DC rate at 3 OV & 6 MHz \\
             CrossTalk at 3 OV & 3 $\%$ \\
         \end{tabular} \\
    \caption{Specifications of S13360-6050PE SiPM.}
    \label{tabla:S13360Spec}
\end{table}

\begin{table}[h!]
    \centering
        \begin{tabular}{l l}
             Surface of each cell & 11.9 x 7.8 mm$^2$ \\
             Microcell size & 25 $\mu$m \\
             Breakdown voltage at 77 K & 26 V \\
             Breakdown voltage at 300 K & 31.5 V \\
             Spectral Range & 270 to 750 nm \\
             Maximum PDE wavelength & 400-420 nm \\
             PDE at room temperature & $>$ 50 $\%$ \\
             DC rate at 4 OV at 300 K & $\sim$ 1 MHz \\
             DC rate at 4 OV at 77 K & $\sim$ 10 Hz \\
             CrossTalk at 4 OV & 10 $\%$ \\
             Afterpulse at 4 OV at 300 K & 1 $\%$ \\
             Afterpulse at 4 OV at 77 K & 8 $\%$ \\
         \end{tabular} \\
    \caption{Specifications of SiPM of the DArTeye.}
    \label{tabla:DArTSpec}
\end{table}

\begin{figure}[h!]
\begin{center}
  \begin{subfigure}[b]{0.47\textwidth}
    \includegraphics[width=\textwidth]{SiPM_Figuras/DArTeye.png}
  \end{subfigure}
  \begin{subfigure}[b]{0.52\textwidth}
    \includegraphics[width=\textwidth]{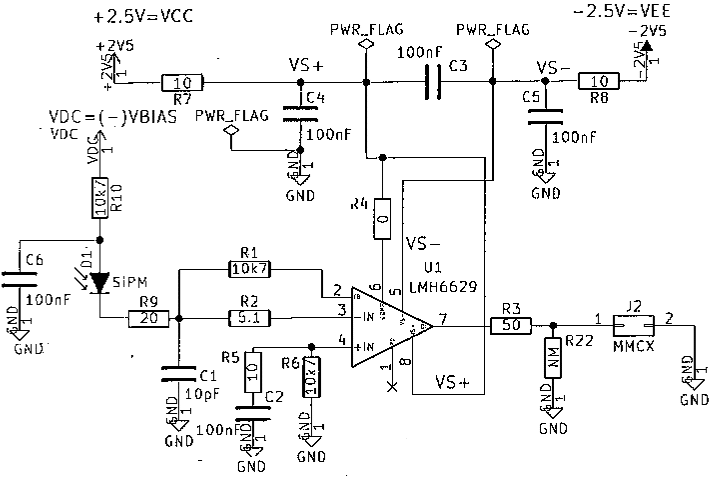}
  \end{subfigure}
\caption{\label{DArTeye2}Left picture: front and back sides of the DArTeye PCB. Right picture: scheme of the printed circuit in the DArTeye module. Images and information from~\cite{ThesisEdgar}.}
\end{center}
\end{figure}

%% file: Agradecimientos.tex
\chapter*{Agradecimientos}\label{Agradecimientos}

Con las siguientes palabras quiero agradecer a todas aquellas personas que me han acompañado en esta aventura que ha sido para mí obtener el título de doctor en física.

En primer lugar, quiero agradecer a mis directoras de tesis Marisa Sarsa y María Martínez por la enorme cantidad de consejos que me han dado y por la forma de entender la física que me han transmitido. La perspicacia y perseverancia de Marisa han sido todo un ejemplo para mí y creo firmemente que me ha sido mi mayor referente a la hora de entender cómo desenvolverme en el mundo científico. Las ayudas, consejos y perspectivas de María y su entrega a la hora de trabajar han hecho que este camino sea mucho más agradecido y me han permitido disfrutarlo plenamente.

Quisiera también agradecer a mi compañero de despacho y de viajes a Canfranc, Alfonso Ortiz de Solórzano, por la pasión que transmite en todo lo que hace y por cómo consigue que todo lo que se hace en el laboratorio sea más divertido y emocionante. Su compañía en el despacho siempre alegraba los días, y las múltiples visitas diarias de personas preguntando por él llenaban de vida el entorno de trabajo.

También quiero dedicar un agradecimiento a Julio Amaré por su entrega en el laboratorio y por sus amplísimos conocimientos en múltiples ramas de la física, pero sobretodo por lo mucho que me ha ayudado a desarrollar gran parte de esta tesis, trabajando duro junto a Alfonso para solucionar las dificultades propias del trabajo en el laboratio.

Una parte importante de mi motivación para el camino que estoy siguiendo se la debo a varios profesores que tuve durante mis años de estudiante de carrera y que me supieron transmitir la belleza de la física de partículas: las clases de Jorge Puimedón, Theopisti Dafni, Igor García, Eduardo García y Jose Ángel Villar fueron las asignaturas más bonitas y de las que guardo mejor recuerdo de toda la carrera. También mi trabajo de fin de grado dirigido por Gloria Luzón y Javier Galán. Muchas gracias por hacerme partícipe de esta historia.

Agradecer a Susana Cebrián por sus consejos y su compañía en las prácticas de laboratorio. A Javier Mena por su inestimable ayuda con los temas informáticos... y su facilidad para arreglar los ordenadores cuando mi trabajo colapsaba el sistema. A Iván Coarasa, por sus consejos con los procedimientos a seguir en el trabajo a realizar como doctorando. También a Carmen Pérez y Jaime Ruz por los buenos ratos tomando café o hablando en los pasillos, y su forma de ser tan abierta que siempre me saca una sonrisa.

Las experiencias de laboratorio tanto en el LSC como durante mi estancia en el LNGS han sido muy enriquecedoras. Por un lado, me han hecho darme cuenta de la facilidad con la que me puedo desenvolver en el laboratorio y, por otro lado, me han permitido conectar con personas con una forma de pensar y de amar la ciencia muy cercana a la mía. Es por eso que quiero agradecer a todos los trabajadores del LSC y a los compañeros de los experimentos DArT y DarkSide que la vida me ha permitido conocer. De todos ellos, quiero agradecer especialmente a Alessandro Razeto por recibirme tan abiertamente en su despacho y transmitirme ese gran conocimiento que él posee de Fotomultiplicadores de Silicio, así como del trabajo con estos detectores a temperaturas criogénicas.

No quiero olvidarme de los alumnos cuyos trabajos de fin de grado y master he codirigido: A Marta, Diego, Víctor y Laura. También a Ainara. Veros aprender ha sido una experiencia muy enriquecedora. También, como no, a los compañeros que han ido llegando con los años: Tamara y Jaime, por transmitirme lo agradecido que es formar parte de un equipo de trabajo.

Agradecer a mis amigos del Consejo de Sabios por estar recorriendo a mi lado este camino, así como compartimos juntos todas las cosas que enriquecen nuestra vida. También a mis amigos de El Callejón y a todos mis familiares, que me han visto crecer con el paso de los años y estar cada día un poco más cerca de mis objetivos. Por encima de todo, quiero agradecer a mis padres por transmitirme desde pequeño esa curiosidad por la naturaleza, el Universo y la vida, así como por enseñarme a ser mejor cada día. Quiero que esta tesis sea un referente más para que nunca olvidemos que siempre debemos perseguir nuestros sueños, y que cada caída es precisamente una fuente de motivación para aprender, levantarnos, y crecer mucho más.

%% file: Acknowledgements.tex
\chapter*{Acknowledgements}\label{Acknowledgements}

This work has been financially supported by the Ministerio de Economía y Competitividad and the European Regional Development Fund (MINECO-FEDER) under Grant No. FPA2017-83133-P; the Ministerio de Ciencia e Innovación – Agencia Estatal de Investigación under Grant No. PID2019-104374GB-I00; the Consolider-Ingenio 2010 Programme under Grants No. MultiDark CSD2009-00064 and No. CPAN CSD2007-00042, the Laboratorio Subterráneo de Canfranc (LSC) Consortium and the Gobierno de Aragón and the European Social Fund (Group in Nuclear and Astroparticle Physics, GIFNA). I would like to acknowledge the use of Servicio General de Apoyo a la Investigación-SAI, Universidad de Zaragoza, and the technical support from LSC, GIFNA and LNGS staff.

This PhD dissertation has been possible thanks to the Aid for Predoctoral Contract for the Training of Doctors 2018 of the Ministerio de Ciencia e Innovación – Agencia Estatal de Investigación, reference PRE2018-083789, and the research stage at the Laboratori Nazionali del Gran Sasso to the the Ministerio de Ciencia e Innovación – Agencia Estatal de Investigación Subprograma Estatal de Movilidad.